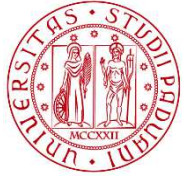

# Exoplanets through extreme optics:
# from PLATO to SHARK-NIR

**Coordinatore:** Ch.mo Prof. Giampaolo Piotto

**Supervisore:** Ch.mo Prof. Roberto Ragazzoni

**Co-Supervisore:** Ch.mo Prof. Jacopo Farinato

**Dottorando:** Gabriele Umbriaco

Padova, Novembre 2019

# Table of contents

# Index of figures

# Index of tables

# Abstract

In the last years, the Observatory of Padova (Istituto Nazionale di Astrofisica – Osservatorio Astronomico di Padova) and the University of Padova have been involved massively in projects dedicated to the exoplanets search, both ground, and space-based.

SHARK-NIR, the acronym of "System for coronagraphy with High order Adaptive optics from R to K band – Near-Infrared", is an instrument designed to search and characterize the young exo-planetary system and star-forming regions in the NIR domain, in coronagraphic direct imaging and spectroscopic mode. It has been selected for the 2nd generation Large Binocular Telescope (LBT) instruments, and it will take advantage of the excellent performance of the LBT adaptive optics system, which is delivering an eXtreme Adaptive Optics (XAO) correction, necessary for SHARK-NIR to achieve the best possible coronagraphic performance, which is mandatory to detect faint planets orbiting around bright stars.

CHEOPS, the acronym of "CHaracterising ExOPlanet Satellite", is the first mission dedicated to the characterization of small-size transiting planets using ultrahigh precision photometry on bright stars already known to host planets. It will allow an accurate determination of the radii of transiting planets, for which the mass has already measured from ground-based spectroscopic surveys. It will also provide precision radii for new planets discovered by the next generation of ground-based transits surveys (Neptune-size and smaller).

PLATO, the acronym of "PLAnetary Transits and Oscillations of stars", is a mission of Medium size satellites proposed for the European Space Agency in the program "Cosmic Vision", with the target to detect and characterize exoplanets utilizing their transit on a bright star. The overall instrumental layout proposed by the Plato Payload Consortium consists of a multi-telescope concept instrument, composed by several tens of telescope units, for which it has developed an all-refractive optical solution. These devices are characterized by a very large Field of View (more than 20 degrees on one side) with an optical quality that fits most of the energy into a single sensor pixel. Such a goal can be achieved in a variety of solutions, some including aspheric elements as well.

The activities concerning my Ph.D. have been exploited both in the framework of the space projects and in the field of ground instrumentation. For the PLATO project, I participated in the Assembly, Integration, and Verification (AIV) of the Telescope Optical Unit prototype, to validate the AIV procedure and the telescope optical performance in-flight conditions. Concerning SHARK, my activity has been performing optical alignment and qualification of the instrument.

# Riassunto

L'Osservatorio di Padova (Istituto Nazionale di Astrofisica) e l'Università di Padova negli ultimi anni sono stati coinvolti massicciamente in progetti dedicati alla ricerca di pianeti extrasolari, sia con strumenti e telescopi da terra che dallo spazio.

SHARK-NIR, che sta per "System for coronagraphy with High order Adaptive optics from R to K band – Near-Infrared", è uno strumento disegnato per cercare e caratterizzare sistemi solari giovani e regioni di formazione stellare nel dominio di lunghezze d'onda del vicino infrarosso. La tecnica devota all'osservazione è quella spettroscopica e dell'immagine diretta. Questo strumento ottico è stato selezionato per la seconda generazione di dispositivi per il Large Binocular Telescope (LBT), con il vantaggio di sfruttare le eccellenti prestazioni del sistema di ottica adattiva di LBT. La correzione di ottica attiva estrema di LBT (XAO), è il requisito necessario di SHARK-NIR per ottenere la migliore cronografia attualmente disponibile con LBT ed è obbligatoria quando l'obiettivo è studiare pianeti poco luminosi che orbitano attorno a stelle brillanti.

CHEOPS, che sta per "CHaracterising ExOPlanet Satellite", è la prima missione spaziale dedicata alla caratterizzazione di piccoli pianeti già noti attorno a stelle brillanti tramite fotometria ad altissima precisione. Si otterranno accurate misure del raggio dei pianeti per i quali la massa è già nota da campagne spettroscopiche con telescopi da terra. Inoltre si conosceranno con precisione i raggi dei nuovi pianeti scoperti dalle campagne di osservazione da terra di nuova generazione basate sulla tecnica dei transiti, fino a pianeti di dimensioni di Nettuno o inferiori.

PLATO, che sta per "PLAnetary Transits and Oscillations of stars", è una missione proposta per il programma di nuovi satelliti di medie dimensioni "Cosmic Vision" dell'Agenzia Spaziale Europea. Il telescopio è focalizzato alla ricerca e caratterizzazione di eso-pianeti attorno a stelle brillanti e vicine al nostro Sole. Il progetto proposto dal consorzio PLATO consiste in un telescopio multiplo, composto da decine di telescopi singoli uguali, per i quali si è sviluppata una soluzione ottica totalmente rifrattiva. Ogni singolo telescopio ha un grande campo di vista (fino a 20 gradi) e una qualità ottica tale da concentrare la maggior parte dell'energia raccolta in un singolo elemento del sensore di immagini. Un tale scopo è raggiungibile applicando una molteplicità di soluzioni, tra cui anche l'uso di elementi ottici asferici.

In questa tesi descriverò le attività svolte sia nell'ambito di un progetto spaziale che di uno strumento per telescopio a terra. Il progetto PLATO è stato trattato nell'ambito dell'integrazione, assemblaggio e verifica (AIV) del prototipo della singola unità ottica del telescopio, allo scopo di validare la procedura completa di AIV e le prestazioni in condizioni di volo. Riguardo lo strumento SHARK-NIR si spiegheranno le attività svolte per l'allineamento ottico e la qualificazione finale.

# List of acronyms

| | |
|---|---|
| ADC | Atmospheric Dispersion Corrector |
| ADI | Angular Differential Imaging |
| AGN | Active Galactic Nuclei |
| AIT | Assembly Integration Test |
| AIV | Assembly Integration Verification |
| AO | Adaptive Optics |
| ASM | Adaptive Secondary Mirror |
| BB | Bread Board |
| BRL | Back Reflected Light |
| BRR | Back Reflected Rings |
| CC | Corner Cube |
| CCD | Charged Couple Device |
| CHEOPS | CHaracterising ExOPlanets Satellite |
| CI | Classical Imaging |
| CMM | Coordinate Measuring Machine |
| CS | Confocal Sensing |
| CTE | Coefficient of Thermal Expansion |
| DBI | Dual-Band Imaging |
| DLAs | Dumped Ly-α systems |
| DLAs | Dumped Ly-α systems |
| DM | Deformable Mirror |
| EE | Ensquared Energy |
| ESA | European Space Agency |
| FEE | Front End Electronics |
| FFT | Fast Fourier Transform |
| FLAO | First Light AO system |
| FoV | Field of View |
| FPA | Focal Plane Assembly |
| FPM | Focal Plane Mask |
| FQPM | Four Quadrant Phase Mask |
| FSS | FEE Support Structure |
| FWHM | Full Width Half Maximum |
| GPI | Gemini Planet Imager |
| GSE | Ground Support Equipment |
| GUI | Graphics User Interface |
| HFFS | High Frequencies in Fourier Space |
| HZ | Habitable Zone |
| INAF-OAPD | Istituto Nazionale di Astrofisica – Osservatorio Astronomico di Padova |
| IPAG | Institut de Planétologie et d'Astrophysique de Grenoble |
| IWA | Inner Working Angle |
| LBT | Large Binocular Telescope |
| LESIA | Laboratoire d'études spatiales et d'instrumentation en astrophysique |
| LMIRCAM | Large Binocular Telescope mid-infrared camera |
| LND | Leonardo S.p.A. |
| LoG | Laplacian of Gaussian |
| LOP | Long-Duration Observation Phase |
| LoS | Line-of-Sight |
| LS | Lyot Stop |
| LSS | Long Slit Spectroscopy |
| NCPA | Non-Common Path Aberrations |
| ND | Neutral Density Filter |
| NIR | Near Infra-Red |
| NPF | Northern Plato Field |

| OD | Optical Density Filter |
| OWA | Outer Working Angle |
| OWA | Other Working Angle |
| PBS | Pellicle Beam Splitter |
| pCMM | portable Coordinate Measuring Machine |
| PIRR | Preliminary Instrument Requirement Review |
| PLATO | PLAnetary Transits and Oscillations of stars |
| PtV | Peak to Valley |
| RoC | Radius of Curvature |
| SAG | Sagitta |
| SHARK | System for coronagraphy with High order AO for R and K band |
| SOP | Step-and-stare Observation Phase |
| SOUL | Single conjugated adaptive Optics Upgrade for LBT |
| SP | Shaped Pupil |
| SPHERE | Spectro-Polarimetric High-contrast Exoplanet REsearch instrument |
| SR | Strehl Ratio |
| TDS | Test Detector System |
| TOU | Telescope Optical Unit |
| TT | Tip Tilt |
| TVC | Thermal Vacuum Chamber |
| UBE | University of Bern |
| VOSA | Virtual Observatory Sed Analyzer |
| WFS | WaveFront Sensing |
| XAO | eXtreme Adaptive Optics |

# 1 General introduction

AIV (Assembly, Integration, and Verification) is the engineering discipline that verifies to a very high level of confidence and probability that the hardware will perform the desired scientific project. The AIV process includes the assembly of many components, parts, and manufactured subsystems, followed by the integration and test of the flight spacecraft or of a prototype very close to the final one. There are many processes and tasks that contribute to the verification goal, as the development of procedures, test of components, error budget, software code, and documents delivery too.

In the Ph.D. I was involved in the AIV of two different projects, designed for the exoplanet research. I was first involved in the space mission is PLATO (Planetary Transits and Oscillations of stars) of ESA (Rauer et al., 2016), which will detect many exoplanets down to Earth-size in the habitable zone, with an extensive sky coverage, larger than the previous space telescopes. The targets are major stars nearby ($m_V$ ≤13) the Solar System, that open the doors of the science of habitable planet near us. PLATO will be composed of a mosaic of 26 Telescope Optical Units (TOU) to collect light as a telescope of 1-meter class. I present in this work the AIV of a TOU prototype, which was under the responsibility of INAF in the PLATO Consortium, aimed to deliver to the industry a robust procedure for the optics alignment to be performed in the warm condition in order to achieve best optical performance at the working temperature of -80°C. The optical alignment is challenging both for the large field of view (FoV) of each TOU, near 20 degrees in radius, and for the tight requirements in both manufacturing and alignment for each single optical component.

After this I perform optical alignment qualification of SHARK-NIR (System for coronagraphy with High order Adaptive optics from R to K band), a ground infrared instrument for Large Binocular Telescope (LBT), exploiting challenging science from exoplanet to extragalactic topics through eXtreme Adaptive Optics (XAO). The optical layout comprises, among the others, four off-axis parabolas, shaped pupil masks, coronagraphic mask, a phase mask, and phase diversity techniques. All this work was done in the environment of international cooperation with the ESA PLATO Consortium and the LBT SHARK Team. This thesis mainly focuses on technical aspects and challenges posed by these modern high-performance instruments.

## 1.1 Research of Exoplanets

Started approximately in the late 1980s, exoplanetology has up to now unveiled the main gross bulk characteristics of planets and planetary systems. In the future, it will benefit from more telescopes and advanced space missions. These instruments will dramatically improve performances in terms of photometric precision, detection speed, multipixel imaging, high-resolution spectroscopy, allowing to go much deeper in the knowledge of planets, where PLATO (Planetary Transits and Oscillations of stars) play a crucial role. The main goal of the PLATO mission is to detect terrestrial exoplanets in the habitable zone of solar-type stars and to characterize their bulk properties.

Searching and characterization of planets beyond the solar system have been a long quest for observational astronomy. Interestingly, the very first extra-solar planets (hereafter exoplanets) were not found around a main-sequence star, but rather a neutron star using the change in pulsar timing (Wolszczan and Frail, 1992). In the following text, we describe the techniques for detecting exoplanets, in particular, focused on the analysis of exoplanets with Earth-size in our galaxies.

### 1.1.1 Techniques for detecting exoplanets

Most exoplanets have been discovered using the radial velocity (RV) technique, which measures the host star's reflex motion around the system's center of mass, or with the transit method, which targets the regular and temporal dimming of a star caused by a transiting planet.

**The RV technique** is used principally in ground-based telescopes, requiring spectrograph with a resolving power of $R>10^5$ to detect a radial shift of the spectral lines up to cm/s. This feature can be induced by an unseen planet and it scales with the planet mass, the square-root of the orbital distance, and the sine of the orbital inclination. Massive planets, like hot-Jupiter mass and relatively closed orbits, are simple to detect producing radial velocity around 200 m/s, i.e. the case of HD217107 (Fischer et al., 1999). In **Errore. L'origine riferimento non è stata trovata.** the radial velocity distribution of confirmed Hot-Jupiters are reported. Earth-mass planets in an Earth orbit around a solar-mass star induces a radial velocity of 9 cm/s. Radial velocity searches aim for planets in our immediate galactic neighborhood, up to 100 light-years from Earth.

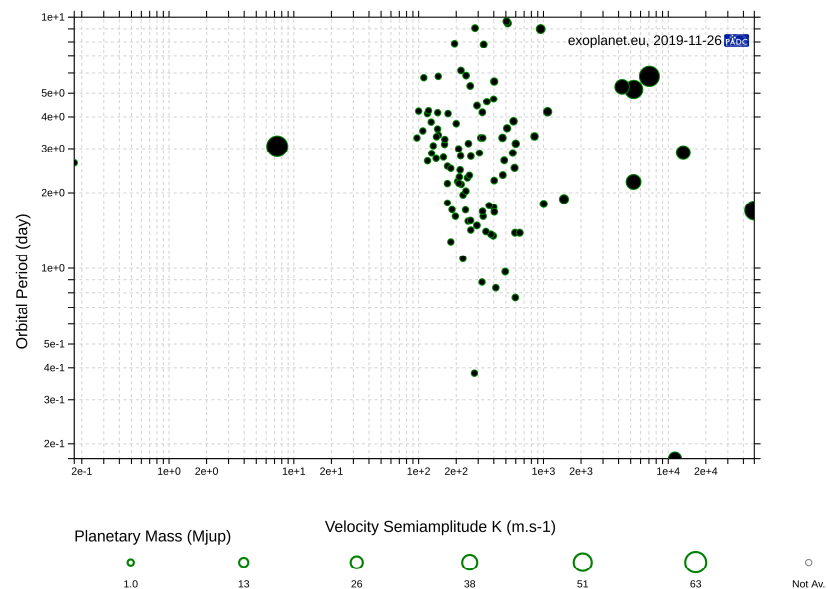

*Figure 1. The plot shows the Hot-Jupiters distribution based on the radial velocity measurements, from exoplanet.eu (26 Nov 2019)*

**The transit method** depends on wide-field CCD photometry to monitor the flux of thousands of stars simultaneously. The transit amplitude scales with the square of the ratio of the planet to star radius, it corresponds to 1% for a Jupiter-size planet transiting a solar-type star and around 0.01% for an Earth-size planet. The transit time varying from one to a few hours for a close-in planet, to half a day for a planet in an Earth-like orbit. At this time transit surveys are strongly biased to planets with short orbital periods, because large orbits size means long orbiting period, i.e. to detect a planet similar to the Earth the minimum detection time is one year. Ground-based transit surveys have found hundreds of hot and warm gas-giant planets. The CoRoT and NASA Kepler space mission have probed down to rocky planet sizes, although mostly for relatively faint and distant stars. In particular, Kepler has resulted in thousands of discoveries, reaching down to Earth-size planets in Earth-size orbits around solar-type stars (Petigura et al., 2013). Transit photometry can potentially detect planets at a distance of hundreds of light-years.

**High-contrast imaging** is currently entirely performed by ground-based telescopes and depends on adaptive optics (AO) techniques that can correct the stellar wavefront from atmospheric turbulence. In particular, to exploit high contrast imaging, ground-based telescopes must be equipped with extreme adaptive optics (XAO) systems optimized to acquire high-contrast images of the immediate surrounding of nearby bright stars. This is done using a coronagraph, suppressing the starlight as much as possible, assisting a state-of-the-art wavefront control. Planets relatively bright and far enough from their host star are the most accessible to this technique, and current discoveries are mostly limited to young (10-100 Myr) self-luminous gas giant planets orbiting at tens of AU. The achieved contrast concerning the host star and the faint planet is typically $10^{-4}$, $10^{-5}$.

The age of a planet is a fundamental parameter affecting the contrast, as very young exoplanets exhibit a high luminosity and thus they are easier to detect, while mature objects, aged billion years or more, are dimmer and below the current detection limits.

For against in the infrared band, the young planets (Marley et al., 2007) have higher luminosity in systems with M-class host stars with respect to object of other class. For this reason the exoplanet detection in this epoch is focused on searching objects around M-class stars.

**Microlensing** is an astronomical effect predicted by Einstein's General Theory of Relativity, when the light emanating from a star passes very close to massive object on its way to an observer on Earth, the gravity of the intermediary massive object will slightly bend the light rays from the source star, causing the two stars to appear farther apart than they normally would. This effect was used by Sir Arthur Eddington in 1919 to provide the first empirical evidence for General Relativity during a solar eclipse. For exoplanets, a foreground star with accompanying the planet can act as a gravitational lens of a background star and significantly boost its flux. Although this technique does not provide precise measurements of a planet's mass and orbit, a large ensemble can provide statistics on planets in relatively wide (4-10 AU) orbits that are otherwise inaccessible with the other methods. Microlensing surveys are currently conducted by ground-based telescopes mostly monitoring very wide star fields towards the bulge. Microlensing can find planets orbiting stars near the center of the galaxy, thousands of light-years away.

**Astrometry** measures the reflex motion of the star in the plane of the sky, and the signals are inversely proportional with distance, making the technique most sensitive to the closest stars, as the last results of the ESA Gaia Mission (Kervella et al., 2019) achieved in synergy with other high accuracy catalogs. Interestingly, the observational bias on orbital radius is opposite to those for the RV and transit techniques, generating a more significant signal for planets in wider orbits, with a limit set by the total mission duration. Astrometry opens the possibility to identify long period orbital companions otherwise inaccessible.

**Red dwarf opportunity** (Alibert and Benz, 2017) is a chance for finding planets around red dwarfs star since stellar masses are up to an order of magnitude smaller than that of solar-mass stars and the radial velocity signals are correspondingly larger. Moreover, the stellar radii are up to a factor ten smaller, the corresponding transit signal for a given radius planet can be up to two orders of magnitude larger. Their luminosities can be a thousand times lower meaning that temperate planets can be found in smaller orbits meaning that even small, rocky planets are accessible through both the transit and the radial velocity technique.

## 1.1.2 Planets detected in our galaxies

A key point for the search for extrasolar planets similar to the Earth is to define their distance from the host star, which is commonly called habitable zone (HZ) (Kopparapu et al., 2013). The habitable zone position is defined by several parameters, including the brightness, spectral class, the mass, age and density of the host star. Until now, most of the discovered planets have dimensions, mass, and distance from the star that are not compatible with the habitable zone. To locate the earth-like exoplanets discovered in the HZ we start selecting updated catalogs of exoplanets:

- Nasa Exoplanet Archive, 3949 exoplanets,
  https://exoplanetarchive.ipac.caltech.edu/

- The Exoplanet Orbit Database, 5748 exoplanets, http://exoplanets.org/

- The Extrasolar Planets Encyclopaedia, 4133 exoplanets, http://exoplanet.eu/

- Open Exoplanet Catalogue, 3793 exoplanets,
  http://www.openexoplanetcatalogue.com/

I used the Exo-MerCat (Alei et al., 2019), that compare and merge the previous catalogs listed, providing excellent uniformity among database, more effective association, fixing error on previous databases, provide a link to stellar sources archives. The catalog maintained update once a week and accessible by Virtual Observatory resource (ivo://ia2.inaf.it/catalogs/exomercat), include candidates and confirmed exoplanets.

Starting from this catalog, it was necessary to compute the host stellar physical parameter such as mass and $T_{eff}$, to define the habitable zone for each planet. I have done the computation by VO Sed Analyzer (VOSA) (Bayo et al., 2008), a web tool that helps to work with large numbers of sources, and we determine stellar physical parameters (Stassun et al., 2017). The main steps on the computation are:

- uploading the Exo-MerCat on VOSA;

- check the coordinates of the sources with GAIA DR2 catalog with a radius of search below 10 arcseconds;

- make a VO photometry combining observation on star coordinates with 2MASS, DENIS, WISE, Tycho-2, Gaia DR2, and Galex observations;

- by VOSA information on "quality of photometry" we automatically identify bad photometric points and remove them from the Chi-square fitting;

- we use the "BT-Settl-CIFIST" Model grid of theoretical spectra, a cloud model, valid across the entire stellar parameter range and using the Caffau et al. (2011) solar abundances(Baraffe et al., 2015);

- We select for stellar fit $T_{eff}$: 2600-7200K and log(g): 4.0-5.0 dex (Sun: G2 V star Teff = 5777K, log(g) = 4.43).

The habitable zone was calculated (Kopparapu et al., 2013) with a python code by John Armstrong selecting HZ fluxes for stars with 2600 K < T_eff < 7200 K.

The diagram Hertzsprung–Russell of all stars in the catalog with the data of luminosity and $T_{eff}$ computed with VOSA is shown in Figure 2.

Figure 3 shows the distribution of exoplanets discovered without a selection of mass or radius from the previous Exo-MerCat catalog with the data of stellar masses calculated. The two images in the plot are the position of planets Venus and Earth.

Figure 4 shows the selection of exo-planets with radius below twice that of Earth and mass from one to ten Earth-mass, produce only a few planets in the HZ, especially coming from discoveries with the technique of transit. The number of planets in HZ (43) is in agreement with the independent analysis from Planetary Habitability Laboratory (http://phl.upr.edu/projects/habitable-exoplanets-catalog) at the date of August 2019.

In Figure 3 and Figure 4 there are many detections of exoplanets near the host stars, and a broad gap of objects in the HZ.

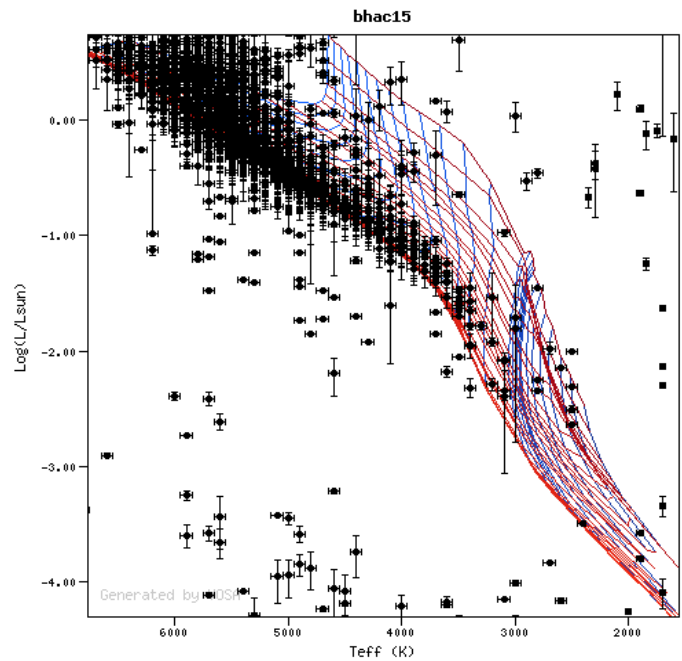

*Figure 2. Hertzsprung–Russell diagram for the star in the Exo-MerCat catalog computed on 31 July 2019 by using VOSA.*

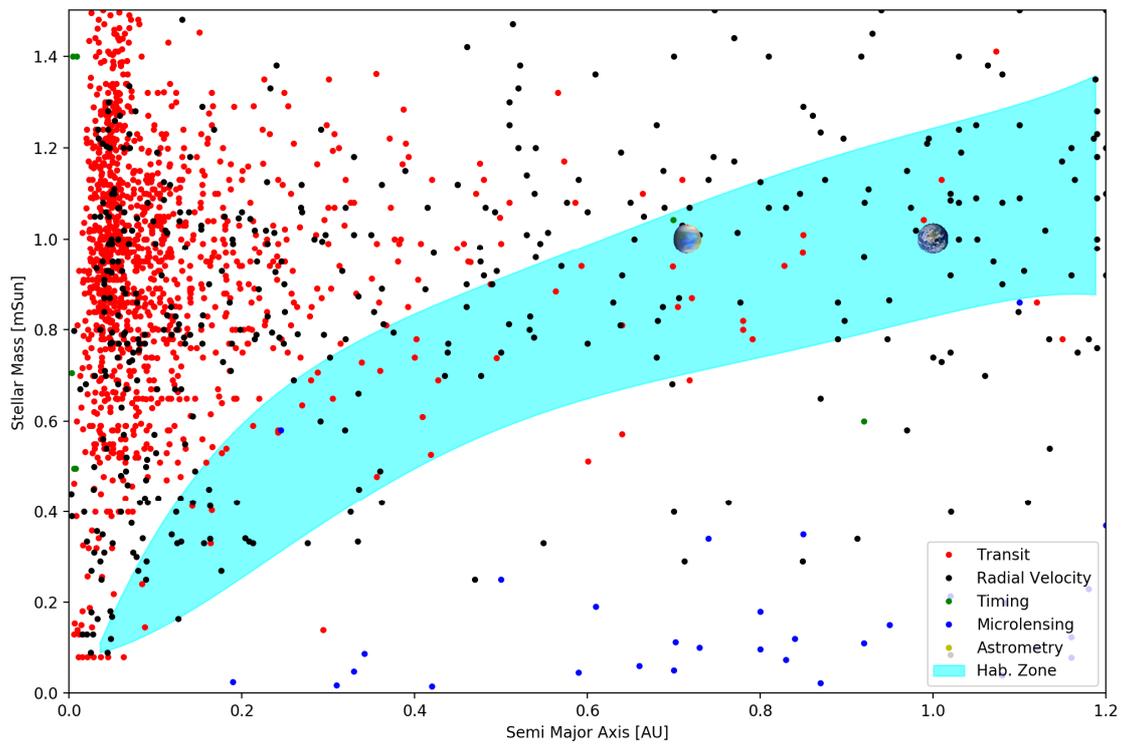

*Figure 3. All planets discovered until 31 July 2019, from exoplanets.eu catalog, selected by host star mass and with semi-major axis below 1 A.U., without distinction in mass and radius. In light blue is highlighted the habitable zone.*

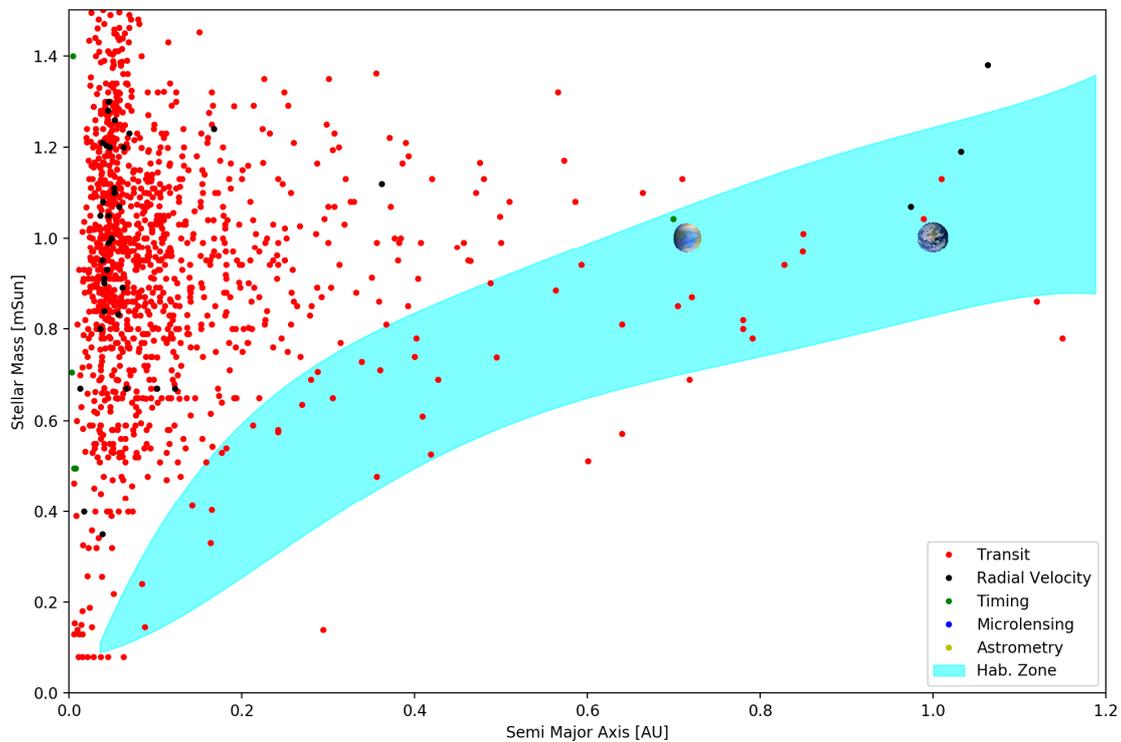

*Figure 4. The plot show the distribution of the know Super-Earth exoplanets (1< M planet ≤ 10ME or R planet ≤ 2RE ) for different host star masses in comparison with the position of the habitable zone in light blue.*

Plato survey will cover this gap, and with ground telescope follow-up, the new exoplanet discovered will also be characterized in life sustainable parameter, locating them with high precision in the HZ.

The direct image of an exoplanet in the HZ (Turnbull et al., 2012) depends on the distance from the Earth and orbital parameters, as orbital radius and eccentricity, representing a new challenge for science and technology. As an example, Figure 5 shows the angular extent of the HZ of the Hipparcos Catalog (ESA 1997) for stars within 30 pc (parallax > 33.33 mas, 2350 stars). With a 4m space telescope plus 50m starshade located at the Earth-Sun L2 point, as the New Worlds Observer NASA's mission, an exo-Earth could be detected and characterized with separation from their star of around 65 mas. The difference in flux from star and planet represent another challenge to overcome because the contrast can reach $10^{-10}$ for an Earth planet in the HZ at 30 pc from the Solar System. In Figure 6 is plotted the fractional planet brightness as a function of the angular separation, for planets in the habitable zone of star within 30pc. As seen from a distance, Jupiter and Venus are the most detectable planets in the solar system, each of them being several times brighter than the Earth. Now, an Earth-like planet could be detected within 30pc with space telescope of minimum 4m of diameter and sophisticate coronagraphic techniques, with detection limits of $4 \times 10^{-11}$ in contrast and angular HZ size of 65 mas. The dashed line in Figure 6 represents the detection space of the New Word Observer, a NASA proposed space telescope, for exoplanets in the habitable zone.

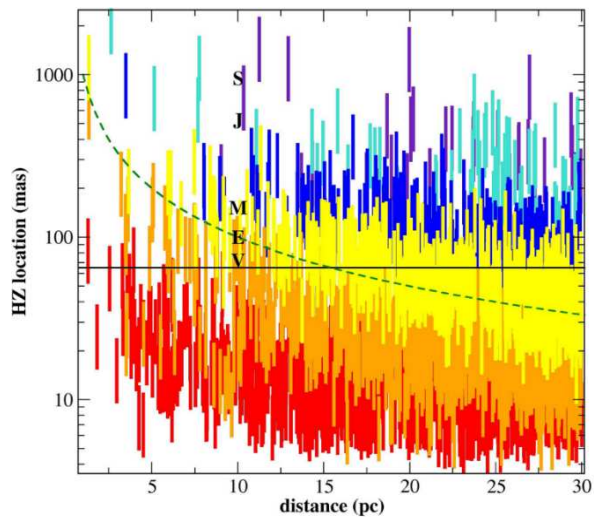

*Figure 5. Habitable zone locations (0.75 – 1.8 AU for the Sun, seen face-on, scaled for stellar luminosity and distance) for Hipparcos stars within 30 pc, colors by spectral types (M-red, K-orange, ... B-white). 65 mas is shown for reference, as is the Earth's angular separation from the Sun as a function of distance (dashed line). The locations of the Sun's planets are scaled for a distance of 10 pc (from Turnbull et al., 2012).*

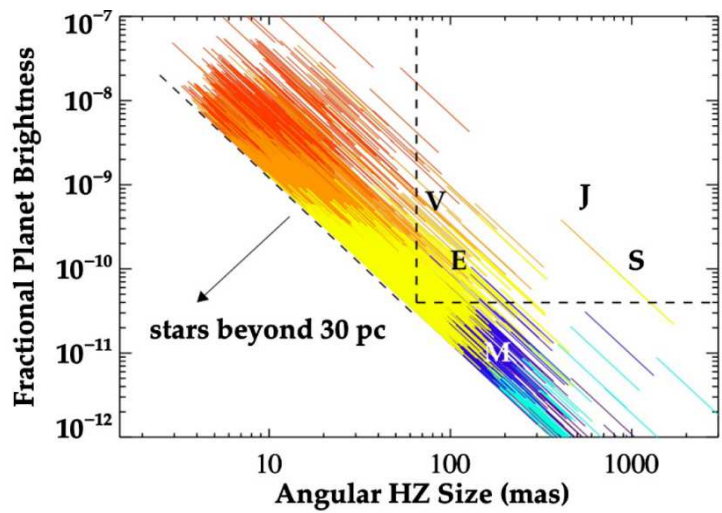

*Figure 6. Fractional planet brightness as a function of angular separation, for planets seen in reflected starlight throughout the habitable zones of stars within 30 pc. Colors represent the stellar spectral type, assuming the systems are seen pole-on. The Sun's planets are plotted for a distance of 10 pc. The dashed line represents the discovery space for a space telescope of 4m in diameter with starshade occulting mask, (from Turnbull et al., 2012).*

### 1.1.3   Space missions for detecting planets

A large variety of space-based telescopes are working on exoplanet research, helped by the absence of atmosphere. The telescopes cover a large variety of diameter sizes and instruments for the science of exoplanets, looking for hot-Jupiter planets to earth-size planets around a variety of host stars. In the next future space, missions will characterize exoplanets discovered and figure out most exciting cases, like Earth-size planets in the habitable zone.

*Table 1. A list of a space mission for exoplanets studies.*

| Name | Launch date | Mission goals |
|---|---|---|
| **HST** | Start: 24 April 1990 End: ongoing | Hot-Jupiter studies, and transit spectroscopy of exoplanets |
| **MOST** | Start: June 30, 2003 End: March 2019 | First spacecraft dedicated to the study of asteroseismology |
| **SPITZER SPACE TELESCOPE** | Sart: Aug. 25, 2003 End: 30 Jan. 2020 | Infrared characterization and discovery of exoplanets |
| **EPOXI** | Start: July 21, 2005 End: August 8, 2013 | Characterized planets and fly-by of comet |
| **SWEEPS** | Start: 2006 End: | Based on the HST, a short 7-day mission looking for exoplanets |
| **COROT** | Start: December 27, 2006 End: November 2, 2012 | Mission to look for exoplanets using the transit method |
| **Kepler** | Start: March 7, 2009 End: August 15, 2013 | Mission to look for large numbers of exoplanets using the transit method |
| **K2** | Start: November 18, 2013 End: October 30, 2018 | After the reaction wheels failed on Kepler, this mission was created |
| **Gaia** | Start: December 19 End: 2013- Ongoing | Map 1 trillion of astronomical objects in the Milky Way |

| **ASTERIA** | Start: November 2017<br>End: Ongoing | CubeSat, technology demonstrator |
|---|---|---|
| **TESS** | Start: April 18, 2018<br>End: Ongoing | To search for new exoplanets; rotating so by the end of its two-year mission it will have observed stars from all over the sky. It is expected to find at least 3,000 new exoplanets. |
| **CHEOPS** | Start: 2019 | To learn more about how exoplanets form, probe atmospheres, and characterize super-Earths. 20% of the time will be open to community use. |
| **JWST** | Start: March 2021 | To study atmospheres of known exoplanets and find some Jupiter-sized exoplanets, and Earth-size planets with Starshade occulting mask. |
| **PLATO** | Start: 2026 | To search for and characterize rocky planets around stars like our own. |
| **ARIEL** | Start: 2028 | Observe exoplanets using the transit method, study and characterise the planets' chemical composition and thermal structures |
| **WFIRST** | Start: 2020 | To search for and study exoplanets while studying dark matter. It is expected to find about 2,500 planets. |

# 2 PLATO

## 2.1 General description of Plato space telescope

PLATO (Planetary Transits and Oscillations of stars) is a mission of the Cosmic Vision Programme(Wilson et al., 2005) M3 selected from ESA for launch in 2026. The main goal of the PLATO mission is to detect terrestrial exoplanets in the habitable zone of solar-type stars and to characterize their bulk properties.

In this chapter, I describe the science that PLATO can achieve, describing the payload and the observing strategy, and defining the Telescope Optical Unit optical and mechanical design and mission requirements.

## 2.1.1   Science with Plato: how unique is our Solar System?

PLATO's primary goal is to detect terrestrial exoplanets in the habitable zone of solar-type stars and to characterize their bulk properties. By providing key information required to determine habitability, PLATO will provide the answer to one of the most important questions in modern astrophysics: how common are worlds like ours, and are they suitable for the development of life?

PLATO planet finder is based on ultra-high precision photometry (Rauer et al., 2016) from space of a large number of stars, more than 400000 stars brighter than $m_V = 13$, with long term scale of at least five years. PLATO will combine the photometric results to determine habitability such:

- planet detection and radius determination from photometric transits

- determination of planet masses from ground-based radial velocity follow-up

- determination of accurate stellar masses, radii, and ages from asteroseismology

- identification of bright targets for atmospheric spectroscopy

The PLATO mission will characterize thousands of rocky, icy or gaseous planets, ranging from giants to Earth-twins, by providing accurate measurements of their radii (at 3% precision), masses (better than 10% precision), and ages (10% precision). This will revolutionize our understanding of planet formation, and of the evolution of planetary systems. Also, important properties of host stars, such as chemical composition and activity levels, will be measured by PLATO and the associated ground-based follow-up for a large sample of systems. It is expected that the final catalog will include between 300,000 and 1,000,000 of high precision stellar light curves.

PLATO will provide key information needed to determine the habitability of these undoubtedly diverse new worlds: planet radii, mean densities, stellar irradiation levels, and planetary system architecture. Understanding planet habitability is a truly multi-disciplinary endeavor. It requires knowledge of planetary composition, to distinguish terrestrial planets from non-habitable gaseous mini-Neptunes, and of the atmospheric properties of planets.

With PLATO will help to understand the uniqueness of our Solar System because while the structure and mass distributions of bodies in our Solar System are well known, we only have indirect and partial knowledge of its formation and evolution. In fact, to place our system in a proper context, we must look to other systems, and study their architectures and composition. From current observations, it has become obvious that the bulk compositions of exoplanets can differ substantially from those of Solar System planets and this must be indicative of the formation process.

PLATO also provides ways to constrain the interior models of exoplanets because many confirmed exoplanets fall into classes that are unknown from our Solar System (Buchhave et al., 2014). Now we have new categories of exoplanets like "hot Jupiters", "mini-Neptunes", or "super-Earths" (rocky planets with masses below 10 MEarth). It came as a surprise that gaseous planets can be as small (or light) as a few Earth radii (or masses). As a result, many of the smallest (or lightest) exoplanets known today cannot be classified as either rocky (required for habitability) or gaseous, because their mean densities remain unknown for lack of mass or radius measurements. PLATO will be unique in providing vital constraints for planetary interior models.

Through accurate measurements of stellar ages, PLATO will study the evolution of planetary systems and how their host stars. Using accurate radius and mass measurements, we will determine how planets form and evolve by observationally building evolutionary tracks for gaseous exoplanets as functions of stellar properties. For example, giant gas planets cool and contract (Fortney et al., 2005), a process which can last up to several billion years, and PLATO will study this evolution. Over time, terrestrial planets lose their primary hydrogen atmospheres, develop secondary atmospheres, and may develop life (Madhusudhan et al., 2016) PLATO will provide key data on terrestrial planets at intermediate orbital distances, including in the habitable zones of solar-like stars with different ages, allowing us to study Earth-like planets at different epochs.

Planets discovered around the bright PLATO stars ($m_V$= 4 to 11 mag) will be prime targets for spectroscopic transit follow-up observations of their atmospheres, for example using JWST or E-ELT. Small planets with low mean density are particularly interesting, as they are likely to have a primordial hydrogen atmosphere. Small planets with high densities are likely to be terrestrial planets with secondary atmospheres. The PLATO catalog will, therefore, play a key role in identifying small planet targets of interest at intermediate orbital distances. It will also provide information on planetary albedos and the stratification of planetary atmospheres. The close-in planets found around stars of different types and ages will provide a huge sample to study the interaction between stars and planets due to, for example, stellar winds or tides.

The intrinsic luminosity of red giant stars allows us to probe distances of up to 10kpc into our Galaxy, and to determine accurate ages from asteroseismology. Red giants can thus be efficiently used to map and date the Galactic disc. These data will complement the information on distances and chemical composition obtained by the Gaia mission. In addition, asteroseismic ages provided for PLATO targets can be compared to age determinations by other means, for example, to calibrate gyro chronology (Brown, 2014), the method of estimating stellar ages based on their rotational rates.

## 2.1.2   Payload concept

The instrument concept is based on a multi-telescope approach, involving four sets of six Normal Cameras monitoring stars fainter than $m_V=8$ at a cadence of 25s, plus two Fast Cameras observing extremely bright stars in the V range with magnitudes range from 4 to 8.

The 24 Normal Cameras see Figure 7, are arranged in four sub-groups of 6 Cameras. All 6 Cameras of each sub-group have nominally the same FoV of 19 degrees, and the line of sight of the four sub-groups is tilted from the payload mean line of sight, along which the two Fast Cameras are pointing. This particular configuration allows surveying a very large field, with various parts of the field monitored by 24, 12 or 6 Normal Cameras. This strategy optimizes both the number of targets observed at a given noise level and their brightness.

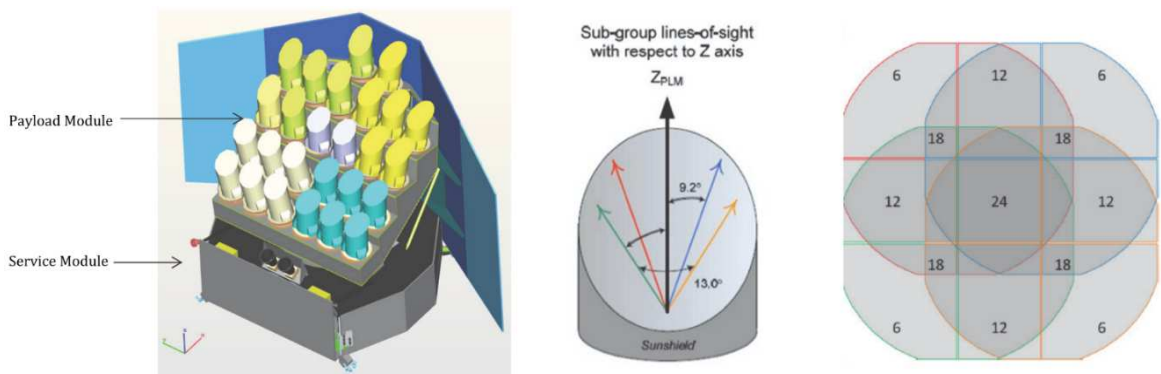

*Figure 7. On the left: spacecraft configuration proposed by the company OHB System AG, with the Payload Module and Service Module separated. Center: the overlapping line-of-sight for each group of six cameras. On the right: the resulting field-of-view of all the Telescope Optical Units with the overlapping zones The telescope is based on a fully dioptric design, working in an extended visible light range. It has been designed to be able to observe a very large field, with a sufficiently large pupil diameter (European Space Agency, n.d.).*

The cameras are mounted on an optical bench, which provides structural and thermo-elastic stability to ensure their relative line of sight do not drift.

### 2.1.3 Observing strategy

The sky coverage of PLATO achieves 40% of the whole sky (European Space Agency, n.d.) by implementing two different observing strategies for the (~2,750 deg$^2$) PLATO FOV, plotted in Figure 8:

- Long-Duration Observation Phase of at least two years (LOP), divided on Southern Plato Field (SPF) and Northern Plato Field (NPF);

- Step-and-stare Observation Phase (SOP);

The Long-Duration Observation Phase is required for the detection of planets with periods long enough to orbit in the habitable zone of solar-type stars. The Step-and-stare Observation Phase is intended to include the repointing of the spacecraft to observe more transits of long-period Earth-like planets detected in the Long-Duration Observation Phase of large numbers of targets in different areas of the sky. The LOP will be observed in one field for up to 4 years, and the SOP duration is between 2 and 5 months.

During long observations, the spacecraft must maintain the same line-of-sight (LoS) towards one field for up to several years. However, the spacecraft shall be periodically re-pointed to ensure the solar arrays points towards the Sun. This is achieved by rotating the spacecraft around the LoS by 90° roughly every three months, as shown in Figure 9.

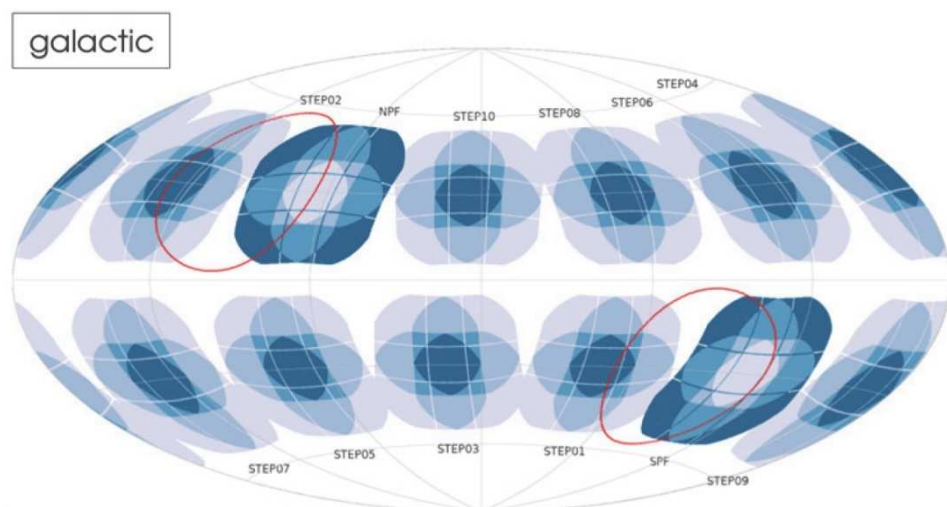

*Figure 8. The Aitoff projection of the two preliminary Long-Duration fields (SPF, NPF) and ten step-and-stare fields (STEP01–STEP10), all centered at |b| = 30°, in the Galactic reference frame. The red line is the Long-Duration pointing requirement limit at |β| = 63°. The Long-Duration fields are color-coded on an inverted scale (European Space Agency, n.d.).*

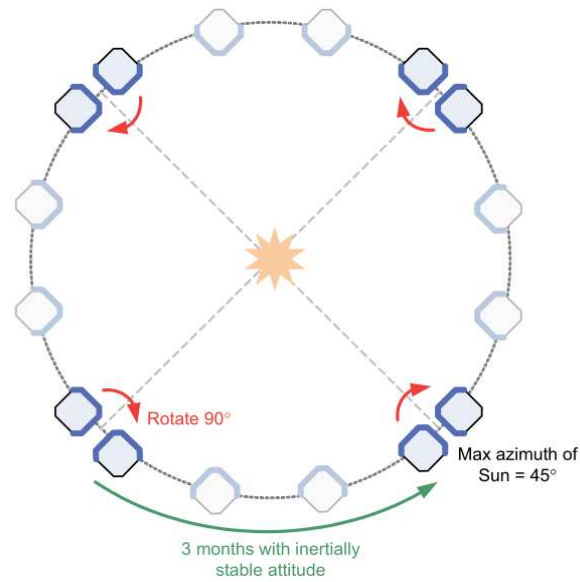

*Figure 9. The spacecraft rotation around payload line-of-sight during one orbit.*

### 2.1.4   Observation requirements

PLATO aims at obtaining high precision, continuous time series of photometric measurements of a large sample of stars. The drivers for performance are the high photometric efficiency of the instrument including optics transmission and vignetting, CCD quantum efficiency, CCD charge transfer efficiency, low particulate and molecular contamination) and the high photometric stability of the instrument such as pointing performance and thermal stability.

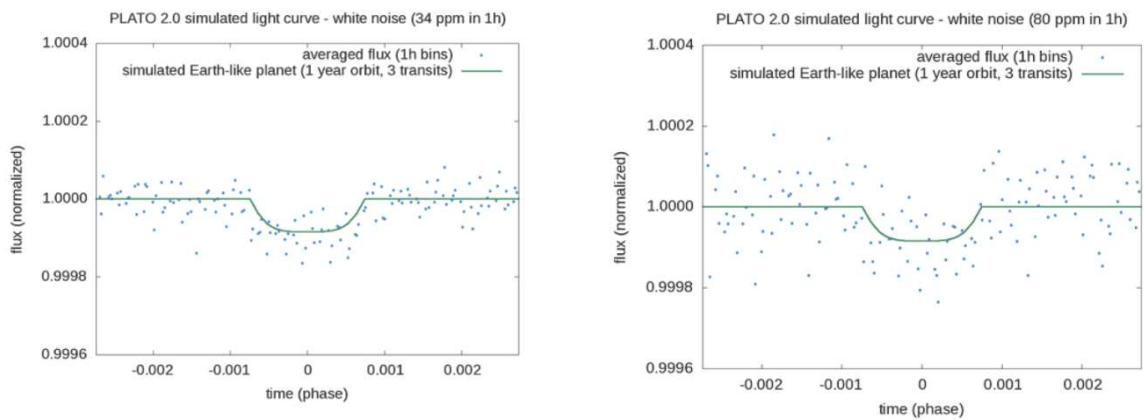

*Figure 10. Simulated light curves for planet detection with PLATO. The data correspond to three simulated transit of an Earth-sized planet orbiting around a Sun-like star with a period of 1 year. Left: noise level of 34 pp, in 1h. Right, a noise level of 80 ppm in 1h.*

PLATO's noise requirements have been set to allow for the detection and characterization of planets and stars with high accuracy. Figure 10 shows an example of the simulated signal averaging three transits for our benchmark study case, an Earth around a Sun. It can be seen that an Earth can be detected already with a noise level of 80 ppm in 1 hour, whereas highly accurate characterization requires 34 ppm in 1 hour as discussed below (European Space Agency, n.d.).

## 2.1.5   TOU optical design and requirements

Each of the 26 Cameras of PLATO includes a Telescope Optical Unit (TOU), based on a fully dioptric design, working in an extended visible light range. It has been designed to be able to observe a very large field, with respect to sufficient pupil diameter.

This results in 1037 deg$^2$ effective field-of-view for the normal cameras and 613 deg$^2$ field-of-view for each of the two fast cameras.

Besides providing star brightness measurements for bright stars, the fast cameras also work as fine guidance sensors for the attitude control system of the spacecraft. In addition, they allow measurements of stars in two spectral bands. For this purpose, one of the fast cameras is equipped with a blue filter, while the other one with a red bandpass filter. The TOU optical configuration consists of 6 lenses, plus one window, placed at the entrance of the telescope, protecting from radiation and thermal shocks. The first surface of the first lens contains even aspherical terms (K, a4, a6), while the second surface is flat to facilitate the interferometric surface measurements during the aspheric manufacturing.

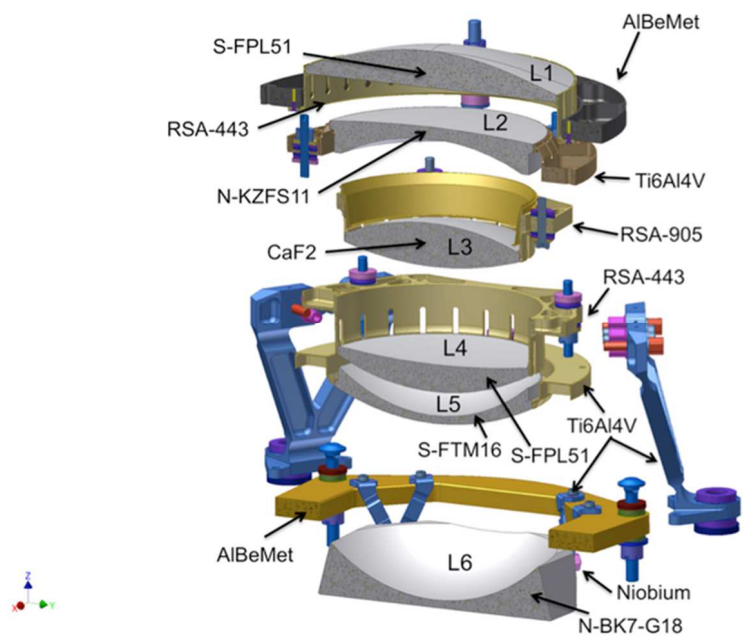

*Figure 11. The 3D model of the TOU structure.*

There is no difference in the TOU optical design (Magrin et al., 2010) for normal and fast telescopes, but the latter will include coating filters on the optical surface. In Table 2 are shown the alignment requirements for the prototype, which have to be compared to the final error budget, which will consider all the error sources affecting the final precision achieved after the alignment. A layout of the design is shown in Figure 12, and the general performance and parameters of the baseline optical configuration are summarized in Table 3.

All the other lenses are standard spherical surfaces. A well-defined mechanical aperture close to the first surface of the third lens is the optical system stop and guarantees a real entrance pupil diameter of 120 mm. This configuration provides a corrected field-of-view up to 18.9° accepting slightly degraded image quality, as well as a ~14% vignetting, at the edge of the field. The nominal field distortion is 3,89% in the max FoV, see grid distrotion in Figure 13. This values is tipical for such optical configuraion with large >FoV.

The prototype optics used in the laboratory consist of a set of six lenses with different radii of curvature and materials as close as possible to the final design.

The chromatic aberrations due to the use of refractive optics are well corrected as shown in the spot diagrams of reported in Figure 14 and Figure 15, for FoV of 0, 2, 4, 6, 8, 10, 12, 13, 13.6 degree and working wavelength of PLATO TOU of 500nm, 650nm, 800nm, 950nm. The resulting polychromatic ensquared energy is within specifications, i.e. 90% EE within 2x2 pixels. The correction at 950nm is slightly worse than the other wavelengths because the optical optimization takes into account the relative weight between the various wavelengths, based on the average spectral type of the target stars and the quantum efficiency of the CCD detector.

*Table 2. The prototype alignment tolerances for the lenses*

| Lens | Decenter | Tilt | Focus |
|------|----------|------|-------|
| L1 | ± 22 µm | ± 44 arcsec | ± 15 µm |
| L2 | ± 22 µm | ± 44 arcsec | ± 30 µm |
| L3 | ± 22 µm | ± 44 arcsec | ± 40 µm |
| L4 | ± 22 µm | ± 44 arcsec | ± 20 µm |
| L5 | ± 22 µm | ± 44 arcsec | ± 20 µm |
| L6 | ± 42 µm | ± 44 arcsec | ± 20 µm |

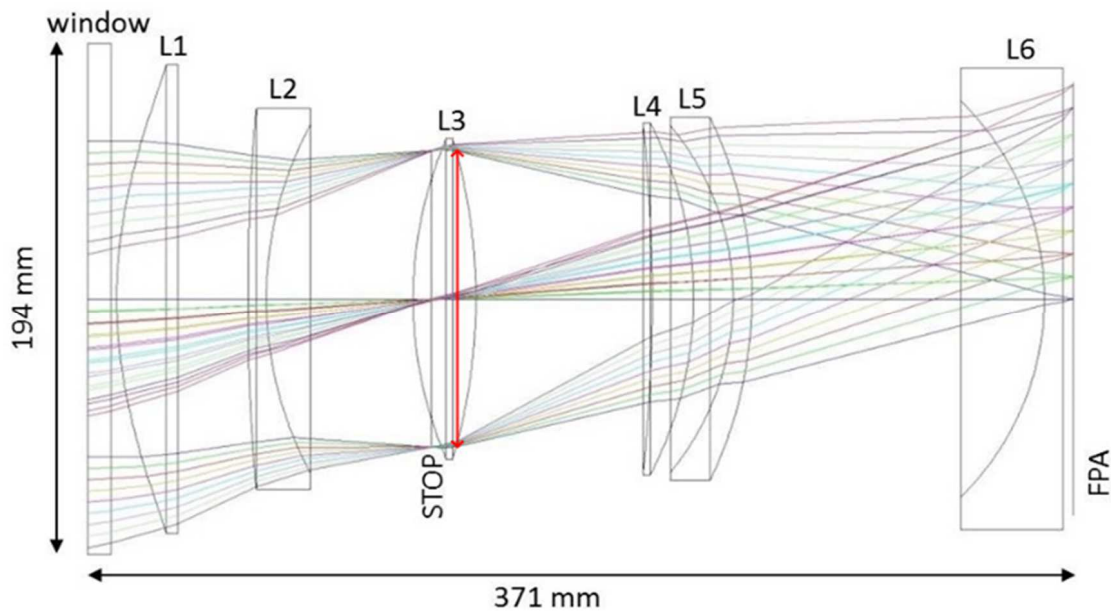

*Figure 12. The optical design of the TOU Prototype, L3 is a convex CaF2 lens, the reference for the alignment of the TOU. Between the stop and L3 is located the entrance pupil diameter of 120mm, in red.*

*Table 3. Design parameters of the TOU.*

| Summary table of the TOU | |
|---|---|
| **Spectral range:** 500 – 1000 nm | **Maximum Vignetting (FOV edge):** 13.55% |
| **Entrance Pupil Diameter:** 120 mm | **Plate scale:** 15 arcsec/pixel |
| **Effective Focal Length:** 247.533 mm @ 700 nm | **Pixel size:** 18 micron |
| **Working F/#:** 2.058 @ 700 nm | **CCD format:** 4510x4510 (x4) pixel$^2$ |
| **TOU optical FoV radius:** 18.8876 deg | **FPA size:** 164.776 mm |
| **CAMERA effective FoV area:** 1042 deg$^2$ | **Optical elements weight:** 5420 g |
| **Image quality criterion:** 1 x 1 px$^2$ < 90% Enclosed energy < 2×2 px$^2$ | **Nominal Working Temperature:** -80°C |
| **Maximum Field Distortion (FOV edge):** 3.89% | **Nominal Working Pressure:** 0 atm |

*Figure 13. Field curvature and distortion computed with Zemax for the optical design of the TOU prototype.*

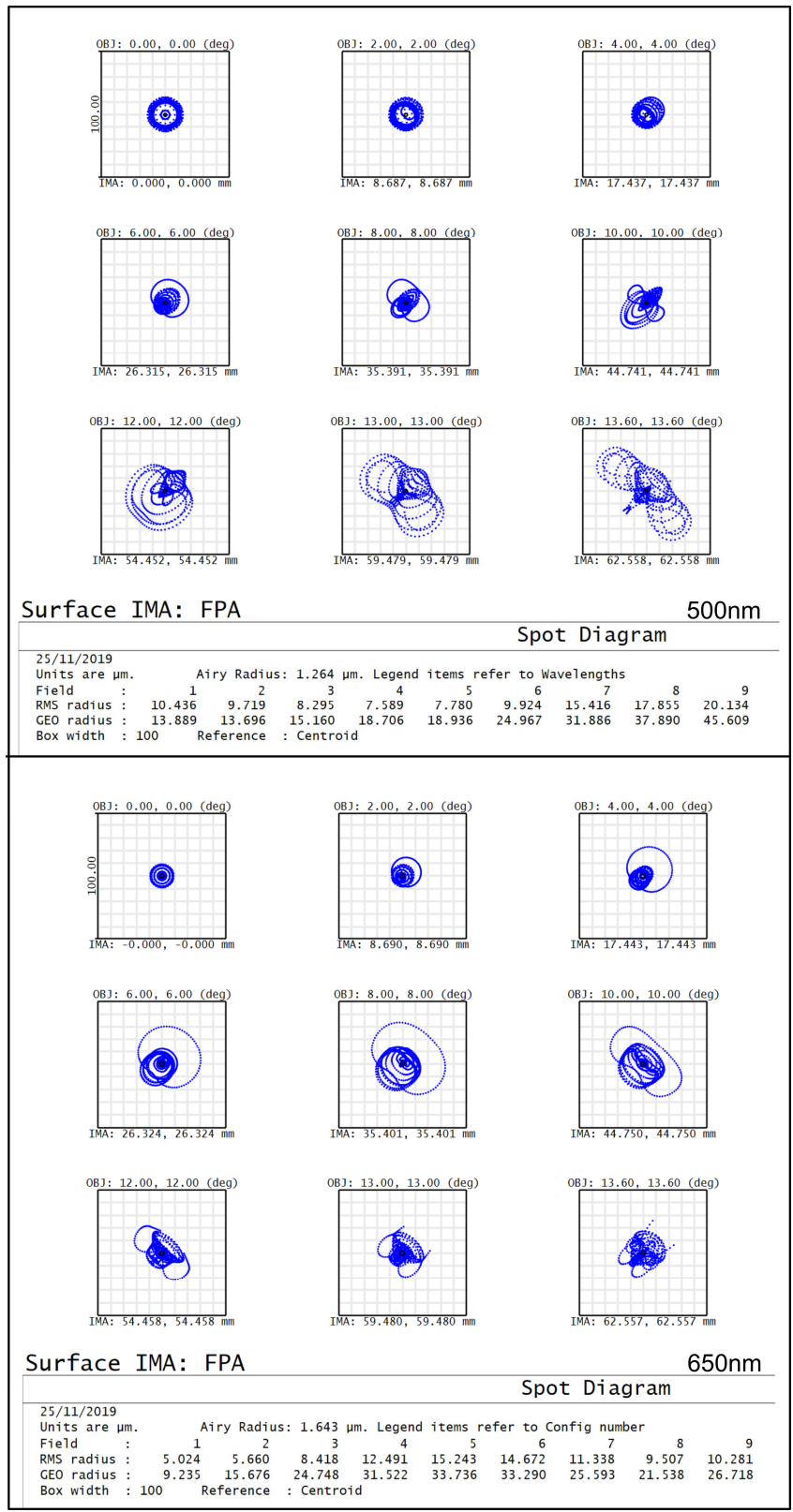

Figure 14. Spot diagram analysis computed with Zemax for all FOV and wavelength of 500nm and 650nm.

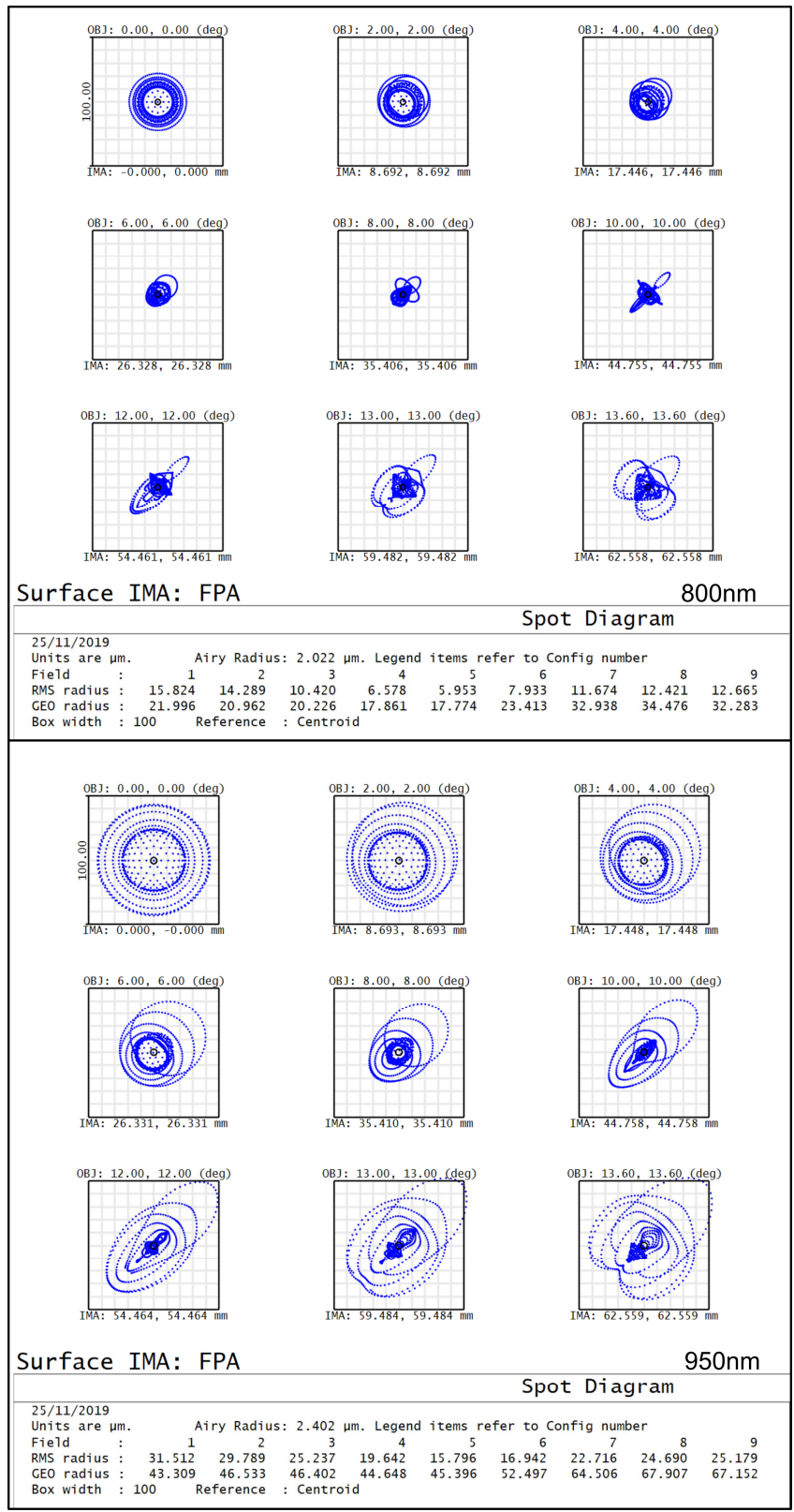

Figure 15. Spot diagram analysis computed with Zemax for all FOV and wavelength of 800nm and 950nm.

## 2.1.6  TOU mechanical design

The TOU main structure consists of a machined tube with all the interface planes, threads, and holes necessary to mount the other components.

The temperature gradients along the TOU shall be limited to a few Kelvin to guarantee the necessary optical performance. Due to the stringent mass requirement, only very limited thermal hardware can be implemented, and therefore the heat dissipated in the focal plane assembly is removed through the mechanical structure and radiated by the baffle into deep space. Since the integration is carried out at room temperature (25°C), while the TOU will be operated at about -80°C, the design must allow the compensation of the thermal stress between the different materials used inside TOU and between the TOU units and the optical bench. Low density and high stiffness materials are necessary to cope with the launch environment and limited mass budget.

The prototype mechanical structure (Figure 16) is equivalent to the TOU final structure in terms of thermal behavior, mimicking the relative movements of the lenses when going from "warm" to "cold" conditions. The purpose of such a prototype is to perform the final alignment procedure and test the on-axis and off-axis optical performance of the system. There is also the requirement to characterize the time needed to perform the full AIV. The final tests are done both in warm (~25°C) and cold (-80°C) conditions.

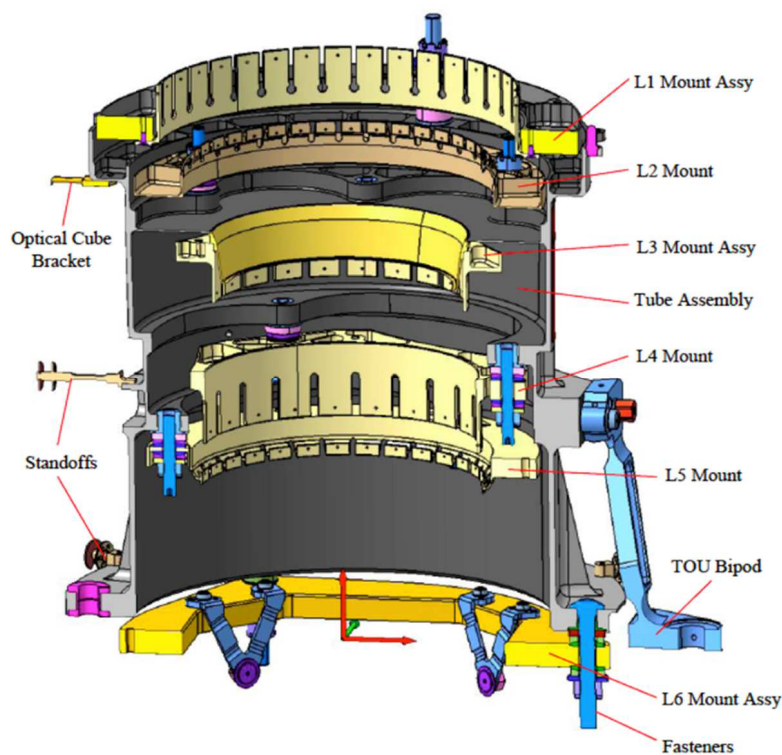

*Figure 16. Structure with highlighted subcomponents*

In the final TOU, the heat dissipated by the CCD needs to be transported through the TOU structure, which therefore must be made of a material with high thermal conductivity. Besides, the large temperature difference between integration and operation requires a design able to accommodate the dimensional changes of the assembled components without leading to unacceptable mechanical stresses. The TOU is quasi-static mounted to the Camera Support Structure with three bipods.

## 2.2  The TOU Prototype Optical Design

The TOU team built a prototype of the TOU, mainly with the purpose of making an assessment of the foreseen assembly and alignment procedure. The prototype optics consist of a set of custom made lenses as close as possible to the final design lenses, differing only for the absence of the AR coating, not performed for schedule and cost issues.  The prototype mechanical structure is equivalent to the TOU final structure in terms of thermal behavior (to mimic the relative movements of the lenses when going from "warm" to "cold" conditions) and interfaces to the FPA, but realized with rods and interface plates, which allow to connect the lenses mounts, which are identical (at the moment of the prototype lenses purchase) to the final TOU ones instead. The purpose of such a prototype is to perform the final alignment procedure and test - both in warm and in cold conditions - the on-axis and off-axis optical performance of the system, to validate in-flight environment the "warm" AIV. In Figure 17 there is a view of the overall optomechanical layout of the TOU prototype.

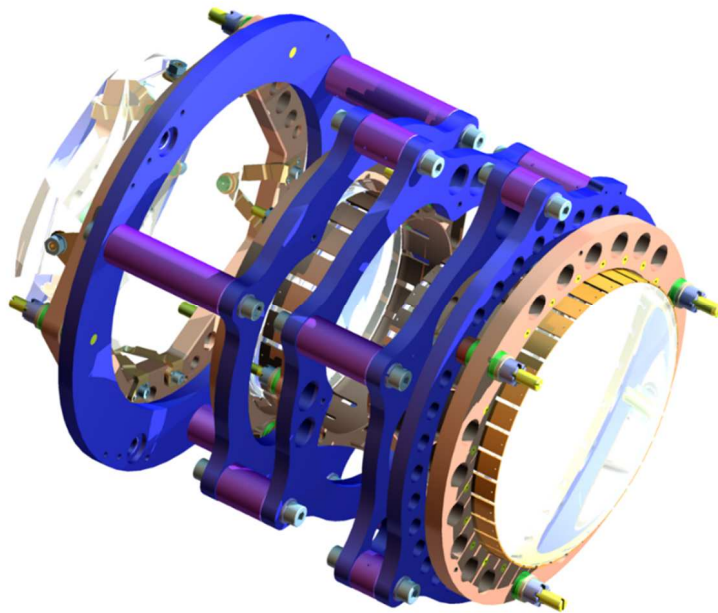

*Figure 17. The TOU Prototype*

To reduce the cost of the lenses and to relax the manufacturing tolerances we adopted a multi-step adaptation of the design in cooperation with Thales SESO, the company selected for the manufacturing of the prototype lenses:

- **Tool adaptation**: the radii of curvature of the lenses have been optimized to match the test plates available at Thales SESO. This step allowed to speed up the manufacturing and reduce the cost of the lenses with only a small decrease in the nominal optical performance. The thickness and distance between lenses, together with the asphericity coefficient of L1 have been used to recover the optical performance during the optimization process.

- **Melt adaptation**: after receiving the glass melt data from the glass manufacturer (index of refraction at different wavelengths), the design was re-optimized by using the measured index of refraction of the glasses.

- **Coefficient of Thermal Expansion (CTE) adaptation**: the distance between lenses at warm conditions has been optimized applying the CTE map of the prototype structure provided by UBE. This step was necessary to confirm the length of the mechanical components.

- **Post manufacturing adaptation**: in order to relax manufacturing tolerances, we agreed to accept decreased specifications regarding central thickness and radius of curvature. These relaxed specs were compensated re-optimizing the air space between lenses once the measured parameters of the lenses (radii and thickness) were inserted in the model.

For all the optimization steps mentioned above, the design was optimized looking at the optical performance at cold conditions (T=-80°C). Since all measures are taken at warm conditions (T=20°C), we used the thermal modeling capabilities of the ray-tracing software Zemax to transform the model into cold conditions. Note that, because of schedule reasons, the CTE adaptation has been done before the post-manufacturing adaptation and the last adjustment of lens positions relies on the use of shims to compensate manufacturing errors both of the mechanics and optics.

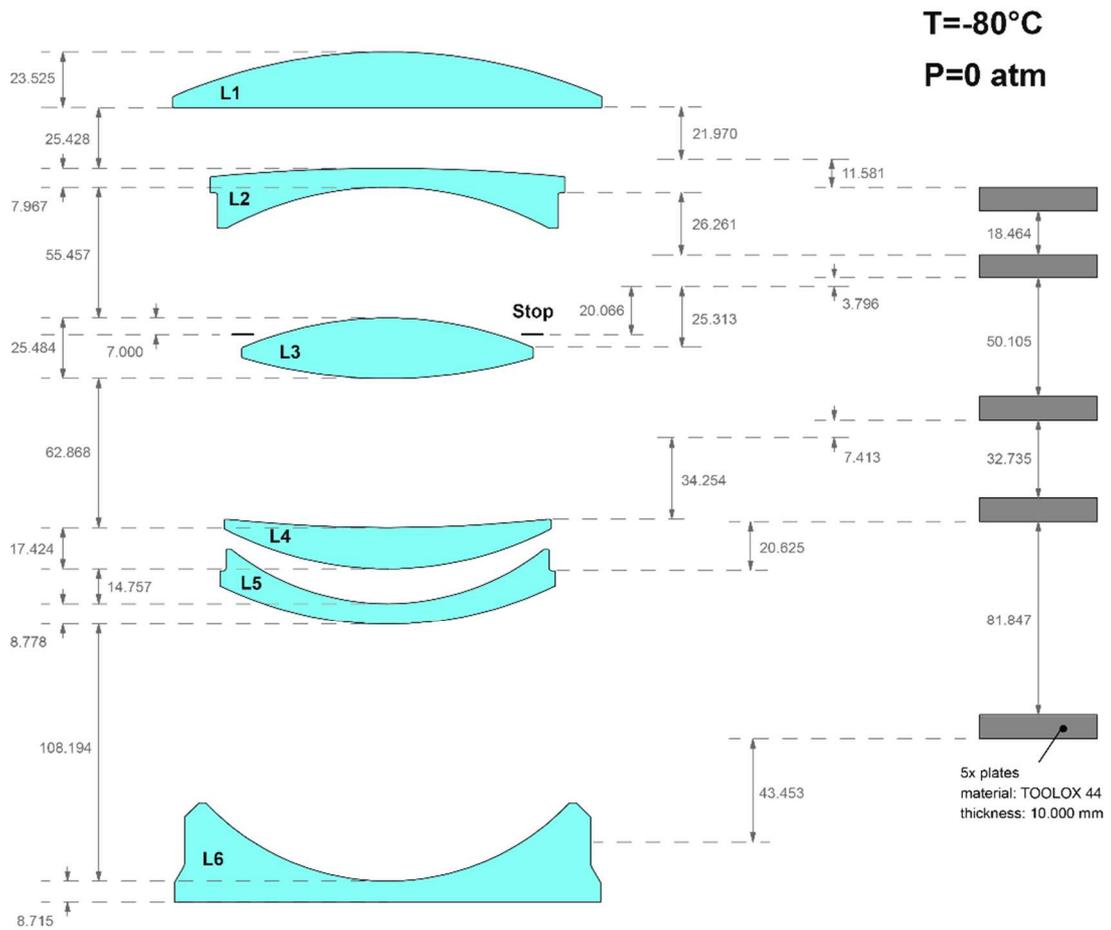

*Figure 18. Lens distances (on the left side) and thickness of the mechanical spacers (on the right side) after CTE adaptation at cold conditions.*

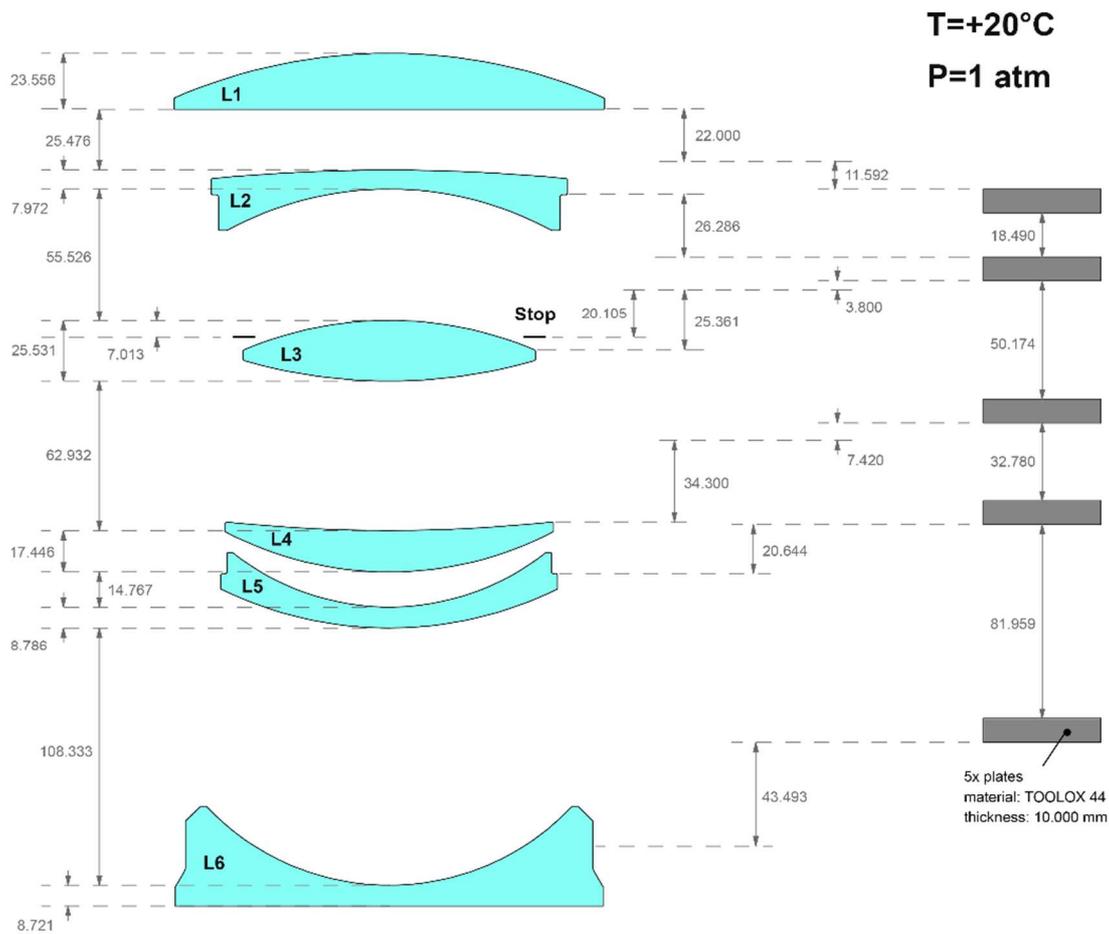

*Figure 19. Lens distances (on the left side) and thickness of the mechanical spacers (on the right side) after CTE adaptation at warm conditions.*

Figure 18 and Figure 19 show the output of the CTE adaptation process and they give the nominal thickness (respectively at cold and warm conditions) required for every mechanical spacer in order to have the lenses at the correct distance. For the optical prescription of the final optical design refer to the following section.

Table 4 summarizes the optical prescription of the final design (warm condition) obtained after the four adaptation processes. The measurement uncertainty for the lens parameters are the following:

- Central thickness: +/- 30 μm

- Radius of curvature: +/- 0.02%

The nominal performance of the Prototype evaluated in terms of ensquared energy (EE) is 90% within a pixel of 18 μm and is plotted in Figure 20.

*Table 4. Optical prescription of the PLATO 2.0 Prototype at T=20°C.*

| Surface name | Radius of Curvature | Distance to Next Surf. | Half Diameter | Aspheric coefficients | Material |
|---|---|---|---|---|---|
| L1 S1 | 185.81 | 23.772 | 89.775 | K = -3.857<br>a4 = 2.9171e-8<br>a6 = -3.9479e-12 | S-FPL51 |
| L1 S2 | Infinity | 24.990 | 89.775 | | |
| L2 S1 | 769.48 | 8.100 | 74.23 | | N-KZFS11 |
| L2 S2 | 140.705 | 62.410 | 67.7 | | |
| STOP | Infinity | -7.013 | 56.31 | | |
| L3 S1 | 154.34 | 25.610 | 60.99 | | CAF2 |
| L3 S2 | -217.28 | 63.347 | 60.99 | | |
| L4 S1 | -638.54 | 17.420 | 66.4 | | S-FPL51 |
| L4 S2 | -146.835 | 14.825 | 68.435 | | |
| L5 S1 | -104.48 | 8.880 | 65.35 | | S-FTM16 |
| L5 S2 | -158.75 | 108.580 | 70.235 | | |
| L6 S1 | -104.395 | 8.660 | 76.1 | | N-BK7 |
| L6 S2 | Infinity | 4.694 | 89.21 | | |

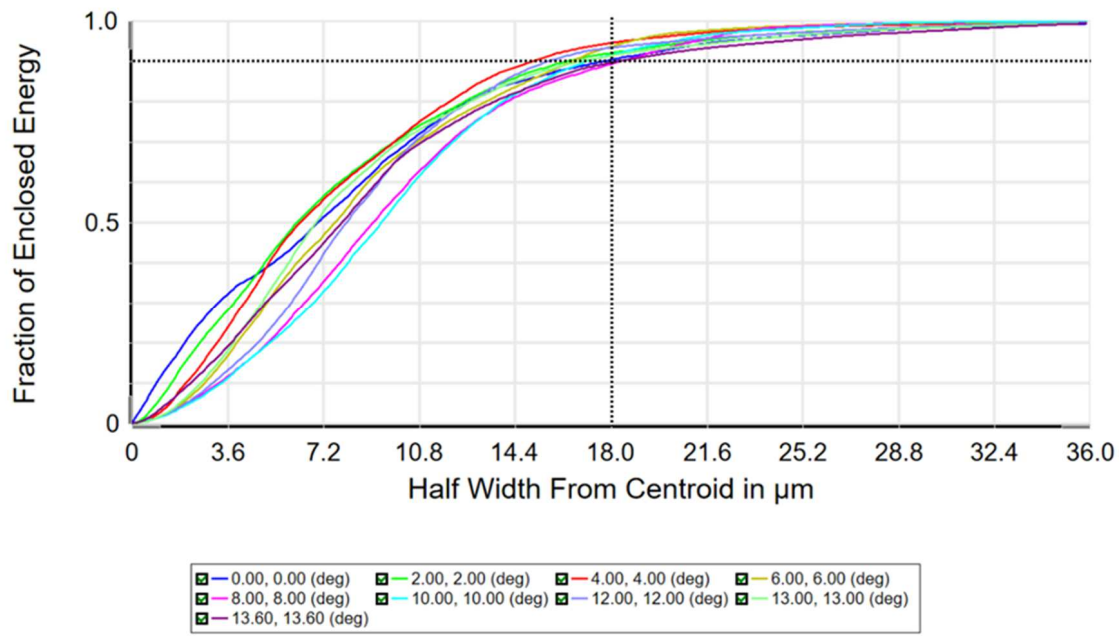

*Figure 20. The nominal fraction of EE as a function of the semi-width of the square for 9 field positions at cold conditions. The horizontal dotted line indicates 90% of EE and the vertical one shows the position of one pixel of PLATO. It matches the image quality criterion 1 x 1 px² < 90% Enclosed energy < 2×2 px².*

### 2.2.1   The TOU Prototype Mechanical Design

The mechanical design of the prototype structure has been performed by the University of Bern (UBE) following these guidelines:

- the mechanics of the prototype is a simplified structure, representative in terms of thermal expansion coefficient and Young's modulus, to the final TOU mechanical structure in AlBeMet, a metal alloy matrix composite material made by beryllium and aluminum used for weight advantage;

- the lenses mounts are identical to the flight design ones both in shape (a part from the adaptation to the optical design) and in materials, but the 2 mounts foreseen to be (partly) in AlBeMet (L1 and L6), which has been replaced by TOOLOX 44, which is a whilst Iron and Nickel alloy with nearly identical CTE and Young's moduli to the AlBeMet ones.

- the mechanics holding the mounts is a very simplified one, done by interface plates and columns holding the plates in a way that the distance between the mounts, at the final working nominal temperature (-80°C), is the nominal one. It differs from the flight model one for thermal conductivity, specific heat, and density. For these reasons, while tests at a given temperature are significant, the prototype is not particularly suitable for testing when the temperature gradient occurs.

- the interfaces between the holding plates and the mounts are identical to the final TOU one (see Figure 21), meaning that the concept is unchanged (shims and spherical washers / oversized holes), and lens frames attachment points are at the same location.

- The prototype has an optomechanical interface on the detector side identical to the final structure, in order to possibly integrate and test, in a second moment if needed, the FPA.

The resulting structure is shown in Figure 22, where all the materials used for the mounts are also specified, while some additional information on the prototype mechanics is also reported in Figure 21.

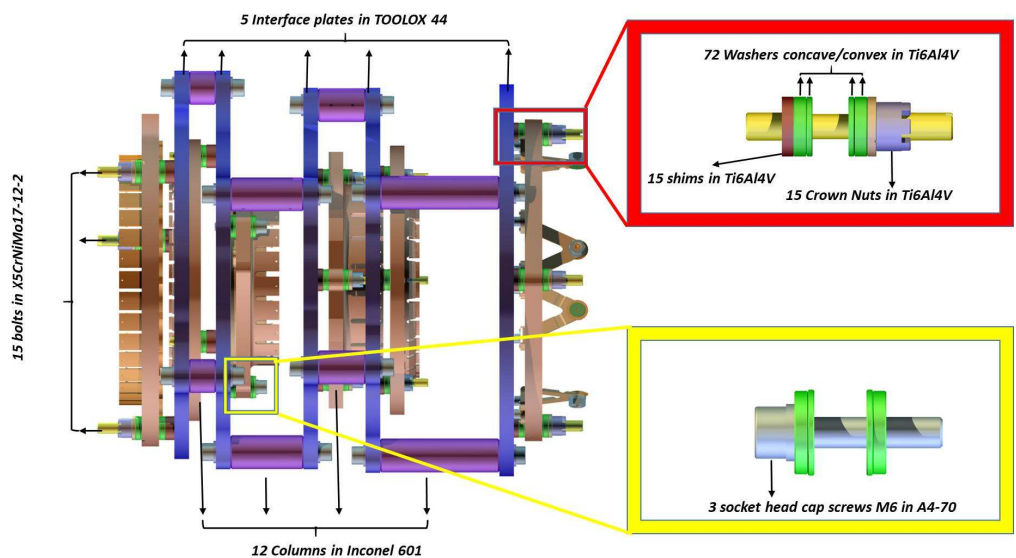

*Figure 21. The interfaces between the mounts and the mechanical structure holding plates of the prototype and some more details about the structure*

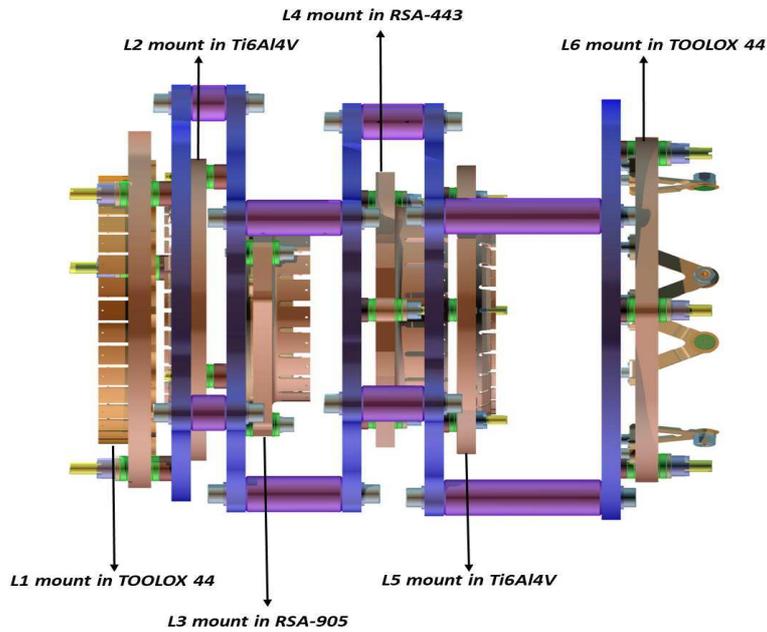

*Figure 22. The materials used for the prototype mounts.*

In Figure 23 we show on the right side a view of the overall prototype optomechanical structure, while on the left side a cross-section of the prototype.

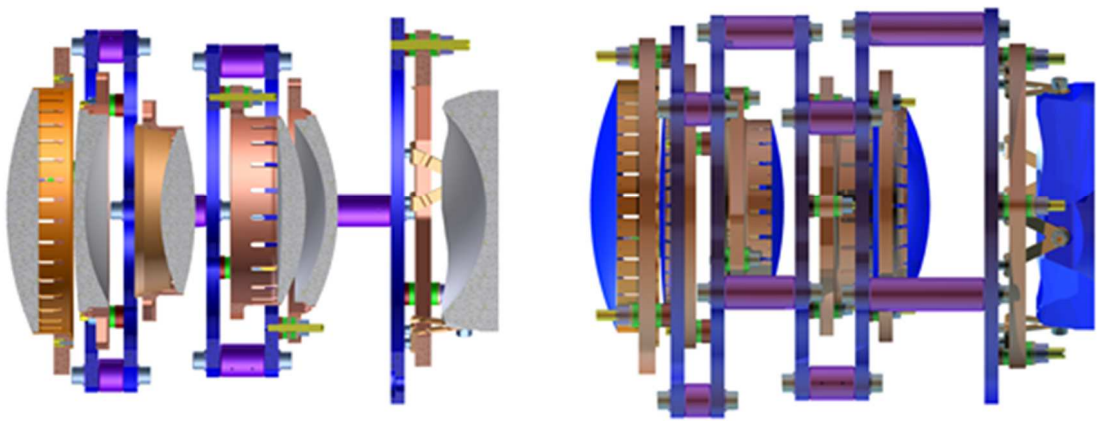

*Figure 23. The prototype optomechanical structure seen from the side, on the left: cross-section, on the right: complete prototype.*

## 2.3  TOU strategy of alignment study

The TOU development follows these three prototyping steps in the framework of TOU construction.

A **"pre-Breadboard"** structure, see Figure 24, was built from the TOU Team with off-the-shelf lenses used to test Ground Support Equipment (GSE) and procedures to be used with the breadboard.

A "**Breadboard**" (BB) structure, Figure 25, was built to test alignment in warm and cold environment before the Preliminary Instrument Requirement Review  (PIRR), achieving the optical performance only on-axis, in Figure 26 are plotted the results for the EE both for working and ambient temperature.

Finally, a "**Prototype**" structure was built after the PIRR, during the Phase B1 (Preliminary Definition of the ESA mission), obtained from the breadboard, replacing the L1 and L2 lenses, to assess the final cold optical quality in the full FoV.

My work was mainly devoted to the AIV phase of the Prototype.

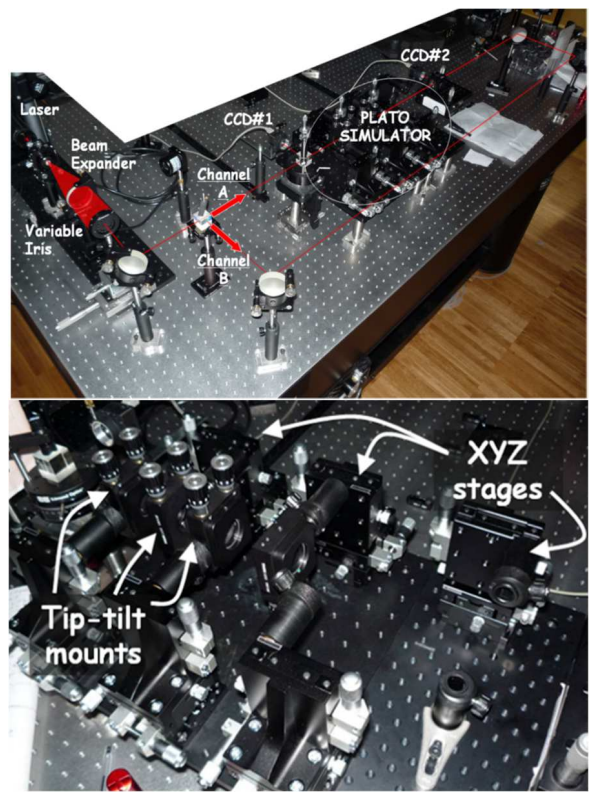

*Figure 24. Pre-BreadBoard setup in the INAF-OAPd laboratory from the TOU Team. Right the detail on the simulation of the tip-tilt and decenter mounts of the lenses.*

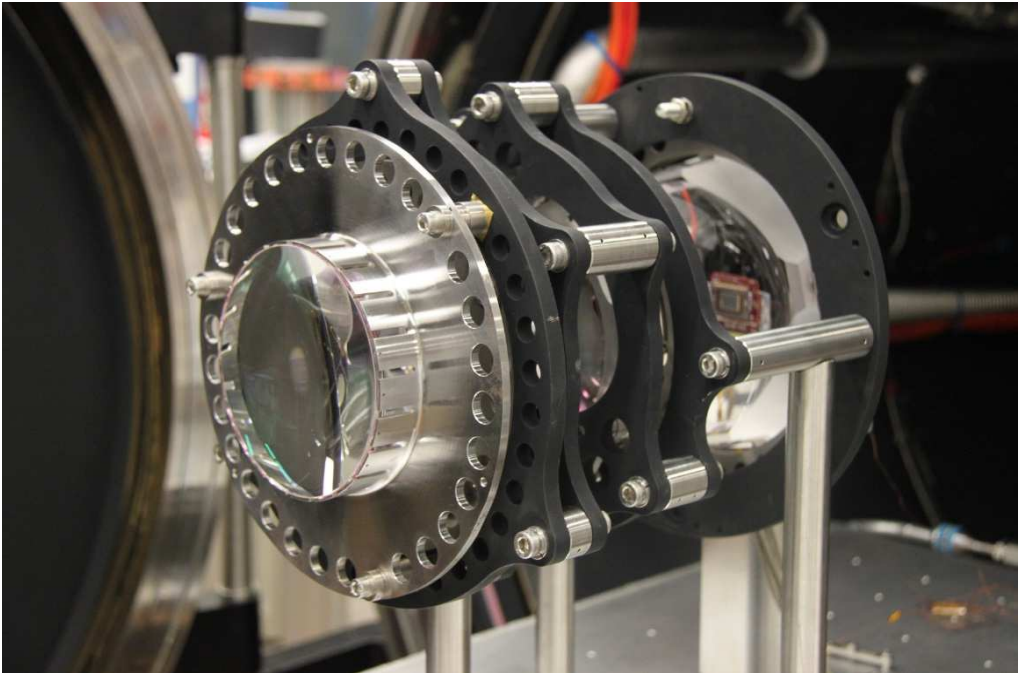

*Figure 25. The Breadboard of TOU mounted and aligned from the TOU Team.*

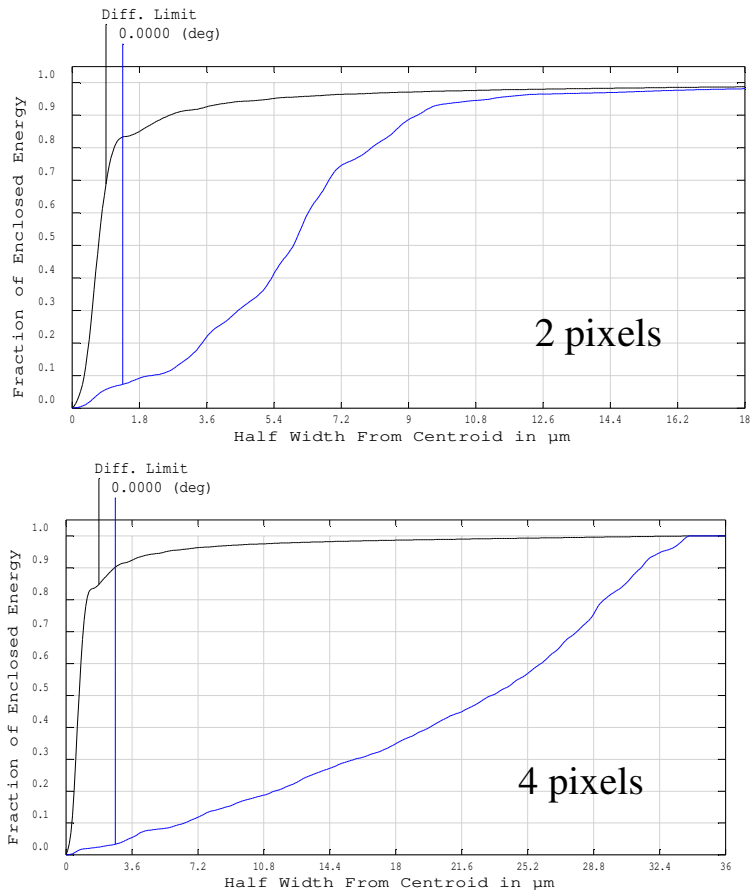

*Figure 26. The Enclosed Energy for the on-axis field of the breadboard in the case of working temperature (on the left) and ambient temperature (on the right).*

The purposes of making a prototype to validate the overall AIV process, as already mentioned, are the following:

- the identification of an AIV strategy in co-operation with possible industrial partners, performed at ambient temperature but ensuring the proper optical quality when testing the system in a cold environment

- the check of the time requested for the overall AIV process, due to the high number of telescopes to be assembled, integrated and tested during the manufacturing phase

- the exploitation of the assembly and integration process step by step, offering in this way time to address and solve possible problems encountered in the early prototypes, or simply to optimize the overall AIV procedure, also in term of time needed to accomplish it

Overall, the scope is to consolidate a development plan ensuring that the full set of TOUs, including the spares, can be successfully assembled, aligned, and tested accordingly to the schedule.

Due to the impossibility to have a test detector as large as the final FPA, a smaller detector will be used and moved along the FoV with a motorized linear stage to span the complete TOU field, when necessary.

The AIV strategy defined for PLATO TOU Prototype takes advantage of the previous experience acquired with the alignment and verification of the breadboard.

## 2.3.1   Alignment concept.

The Alignment concept that we present here is derived from the one used to align the Bread Board (BB), modified to accomplish the alignment of the lenses also along the optical axis, i.e. in focus. In fact, we briefly recall that the BB had a modified optical design including a spherical lens instead of the aspherical one (L1), giving performance only on-axis (the aspheric lens is the one which is needed to reach the required off-axis performance); also the design of L2 had to be slightly modified to achieve a good optical performance on-axis.

Thus, the basic idea behind the AIV concept is the following:

- the L3 mount is just positioned in the structure in its nominal position, acting as a reference for all the following lenses alignment;

- an optical beam is materialized and aligned to L3;

- every other lens is then inserted and aligned in centering, tip-tilt and focus with respect to L3, using the optical beam previously aligned to L3 itself.

In Figure 27 the Prototype alignment concept is shown. We now describe the procedure we will use for the alignment of the lenses decenter, tip-tilt, and defocus.

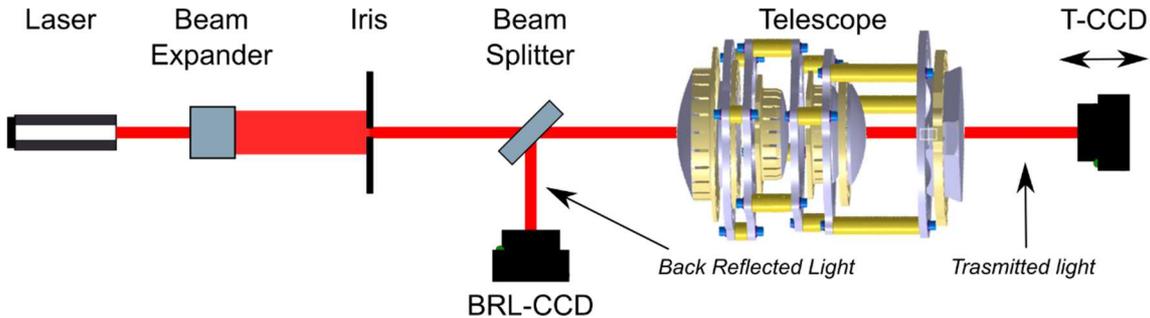

*Figure 27. The prototype alignment concept.*

## 2.3.2   The lenses centering and tip-tilt alignment

The idea for centering the lenses is to use the back-reflected light coming back from the previously aligned lens and passing through the newly inserted lens. Each lens is transmitting most of the light, while a small percentage is back-reflected, showing the typical pattern of rings due to the diffraction of a circular aperture.

For the BB, the transmitted spot on the CCD, T-CCD in Figure 27,  was used only as a sanity check of the alignment performed by using the Back Reflected Light (BRL), since the sensitivity to the decenter on the T-CCD was too low. Thus, also for the prototype, we do not foresee to initially use the transmitted spot to perform the alignment, even if the sensitivity to tilt and decenter of each observable will be carefully checked for each lens to identify the more precise way to perform the alignment.

The alignment of each lens in tip-tilt will be achieved by proper shimming between each lens mount and the prototype structure reference plane. The decenter correction was performed with the transmitted beam. After a series of iterations of tip-tilt and decenter corrections, the lenses were aligned.

## 2.3.3   The Focus Alignment

The idea concerning the focus alignment of the lenses is to install the T-CCD on a linear stage which can move the CCD along the optical axis, in a way that we can position the T-CCD on the nominal focus position for each intermediate configuration of installed lenses.

The other idea is to use an interferometer to measure the defocus power of the different combinations of the lenses following the insertion order, see Figure 28. By using a ray-tracing software, Zemax, we calculate the intermediate focal position, the dynamic range of the interferometer to ensure a correct measure for all lenses. To do this, we need an interferometer co-aligned with the laser beam reference because a single ray from the flat beam of the interferometer does not have sufficient intensity to be used itself.

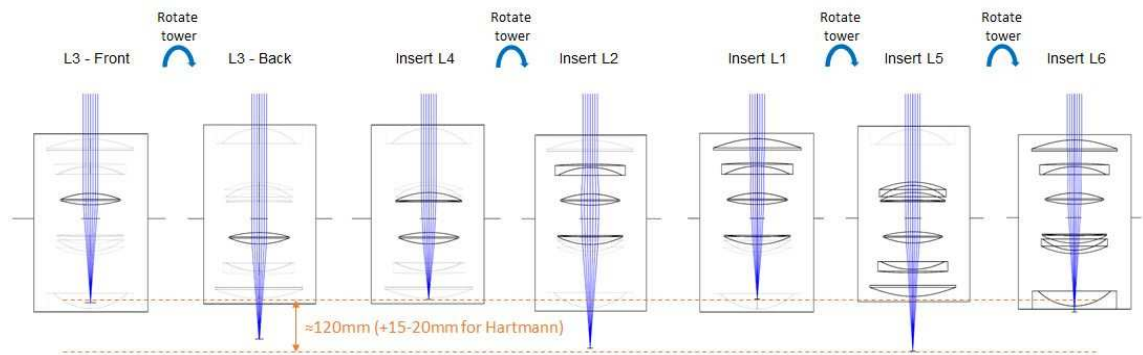

*Figure 28. The lenses insertion order to allow their alignment along the optical axis*

Since L3 is the first lens to be installed into the prototype, and it is, for this reason, the reference to which all the other lenses are aligned, of course, we do have some constraints in the lenses insertion order. We thus identified the optimal sequence of lenses insertion, given the obvious constraint just mentioned (after L3, we must insert either L2 or L4, and so forth), following these guidelines:

- the optimization of the visibility of the Back Reflected Rings (BRR, simulated with FRED Photon Engineering) since they are the observables used for the alignment;

- the minimization of the travel required for the T-CCD to be in focus in the various configuration of inserted lenses;

- the minimization of the number of rotations required to the prototype for the lenses insertion, since the rotation of the structure required each time a re-alignment of the beam to the prototype;

In Figure 28 we show the optimal sequence that we devised, from which it is clear that a total of four rotation of the prototype is needed, and about 150mm of travel is required for the linear stage moving the T-CCD along the optical axis on the transmitted beam side. It is obvious that such a stage must have a very small wobble of the moving carriage (holding the T-CCD), not to introduce additional errors in the focus measurements. Particularly, Pitch and Yaw must be minimized (Roll is orthogonal to the optical axis and thus not causing troubles), and we computed that they should be below 10 arcsec over the whole travel range not to introduce an additional error in the focus measurement larger than 5μm.

The focus alignment procedure is thus the following:

1. with L3 alone in the prototype, move the stage with the T-CCD till finding the best focus position, and record the correspondent linear stage encoder value. This operation has to be performed with L3 in both orientations with respect to the optical axis;

2. insert the next lens (L4 accordingly to the order shown in Figure 28) and adjust the T-CCD linear stage position in the nominal position computed through Zemax in the configuration L3-L4. The last operation has to be performed by moving in a relative way the T-CCD linear stage of the nominal amount given by Zemax, with respect to the best focus found at item #1;

3. move L4 along Z till reaching the best focus on T-CCD in the configuration L3-L4;

4. compute the right shimming to be applied to put L4 in the right focus position (also considering the tilt to be applied, compensated by using shims too);

5. apply the shims and tighten the lens mount, and repeat the same procedure for all the prototype lenses.

Alternatively, in the sequence, we can use the Zygo interferometer to check and correct the T-CCD determination in-focus position.

## 2.3.4   Confocal sensing and low-coherence interferometry

The PLATO lenses have to be gently manipulated in particular L3, because it is made of CaF2 glass, for this reason, various non-contact optical sensing (Berkovic and Shafir, 2012) techniques were used to measure distances of an object and related parameters. The z-axis position of the PLATO lens may also be achieved with Confocal Sensing (CS, Jordan et al., 1998), employing both monochromatic and polychromatic light sources. CS method measures the position of the lens by non-contact technique, it is based on the beam reflected from the surface of a lens, flat or curved. In the PLATO-case we propose to realize an optical head, getting a focused beam from the collimated beam of the interferometer, that moves in z-axis below the TOU. The head, composed by a microscope objective in a precise translation stage, collect the back-reflected light from the lens surface. We test the sensibility of this method with the optical bench sketched in Figure 29 and realized in Figure 30. A beam expander with spatial filter creates a reference beam of monochromatic light of 633nm of wavelength. A scanning lens is mounted in a piezoelectric translation stage capable of 6.9nm steps per counts. Near the focus of the lens, we mounted the lens under test of which we want to measure the position. The role of the folding mirror "FM" was of reflecting off-axis the unwanted light. The back-reflected light from the surface of the test lens is collected by a second lens, in the focus of which we placed a single-mode fiber optics, to erase any speckle generated by the coherent source. The fiber optic output is entirely captured in a CCD with a 12-bit dynamic range. We use two different plano-convex scanning lenses with 100mm and 38mm of focal length, and f-number 4 and 1.5 respectively. The sensitivity is directly proportional to the numerical aperture of the lens because the reflected light is correlated to the size of the sampling PSF and the deep focus of the scanning lens. The experiment was performed moving the translation stage with constant velocity and recording images continuously on the CCD, the scanning time is around 200 seconds, collecting an image every 22µm of translation of the scanning lens. The lens under test is plano-convex of 50mm of diameter with its curved face turned towards the incoming ray, with a focal length of 150mm.

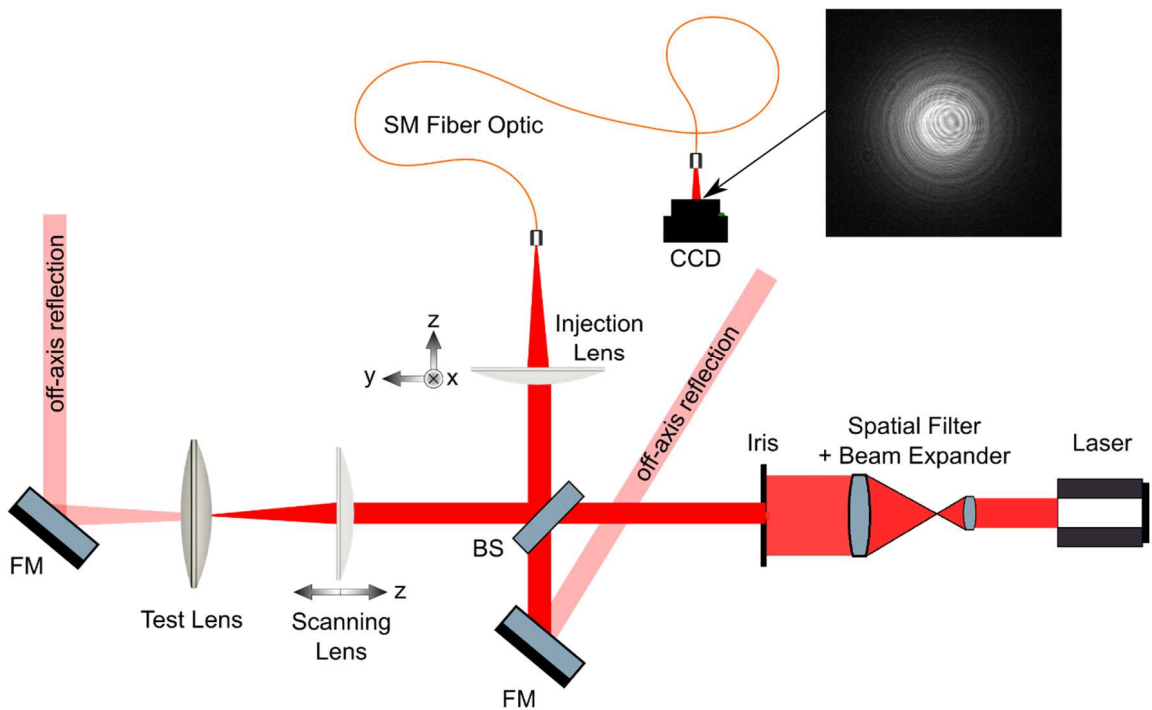

*Figure 29. The optical layout to simulate confocal sensing. A laser beam is focused on the surface of the test lens by the scanning lens, mounted on a precision translation stage. The back-reflected light is collected in a CCD by a fiber optic, (see the image in the box).*

Figure 31 and Figure 32 show the result for the lens of 100mm and 38mm of focal length, which is also the working distance. The signal increases when the focus is near the surface of the lens under test, and we estimate the position of the maximum by a Full-Width Half Maximum from which we deduce the standard deviation, FWHM = 2.355 std. As expected the standard deviation decrease with the focal lens.

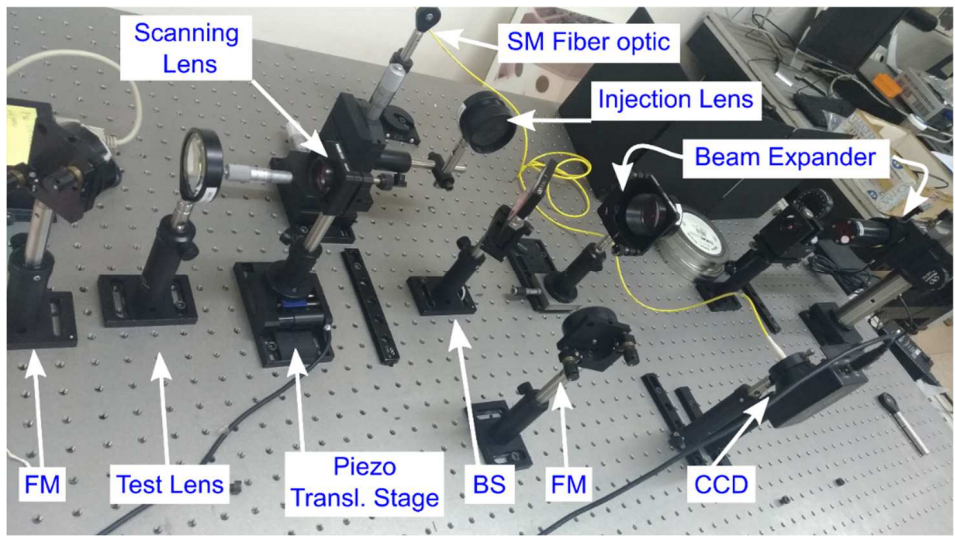

*Figure 30. The image of the optical bench used to simulate confocal sensing.*

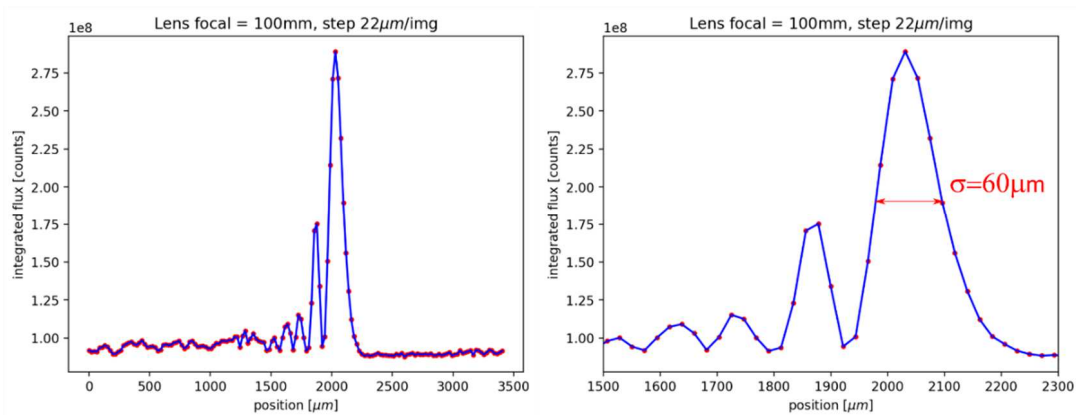

*Figure 31. The back-reflected signal for the lens with a focal length of 100mm. When the scanning lens focus is near the first surface of the lens under test the signal rising, 60 µm is an estimation of the error on defining the maximum.*

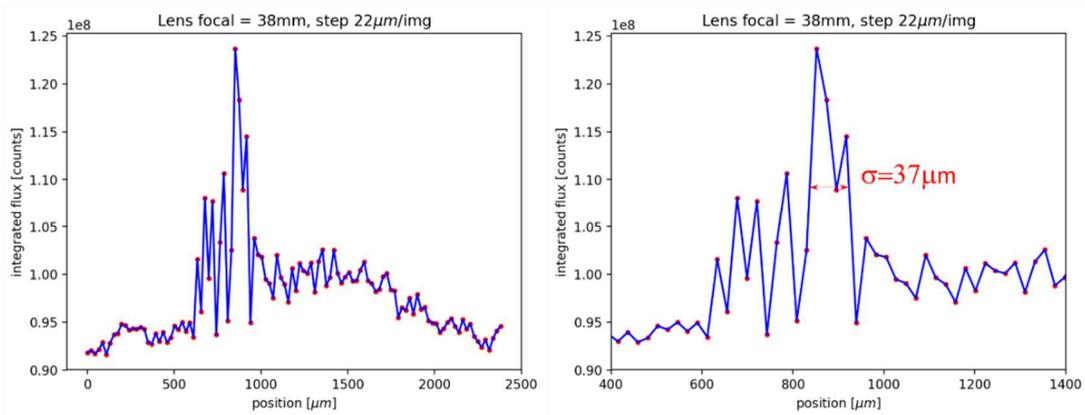

*Figure 32. The back-reflected signal for the lens with a focal length of 38mm. When the scanning lens focus is near the first surface of the lens under test the signal rising. 37 µm is an estimation of the error in defining the maximum.*

We try to simulate this situation also utilizing the L3 of PLATO as a scanning lens on measuring the distance of the L4 lens, mounting the setup as shown in Figure 33 and Figure 34.

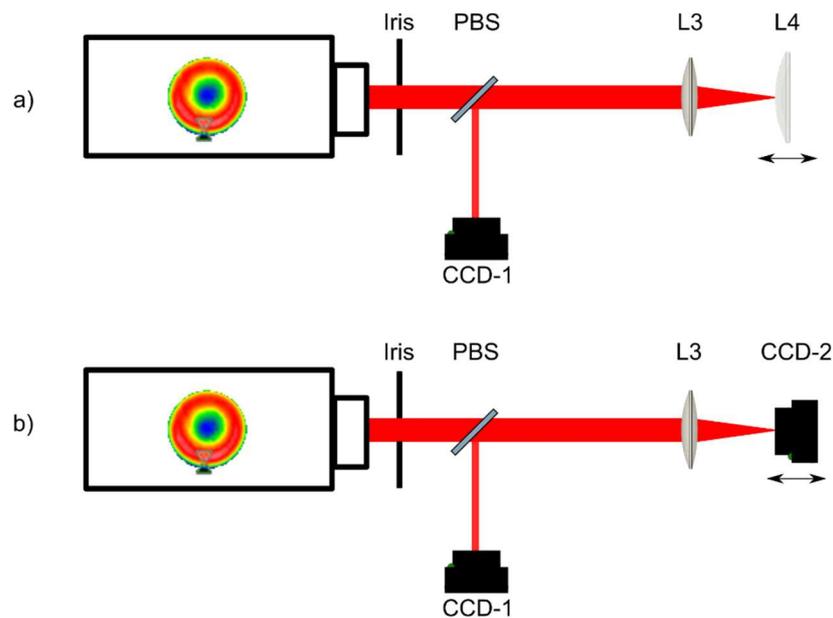

*Figure 33. The optical layout of the confocal test. a) focusing L3 on the surface of L4. b) focusing L3 on the CCD-2 optical window.*

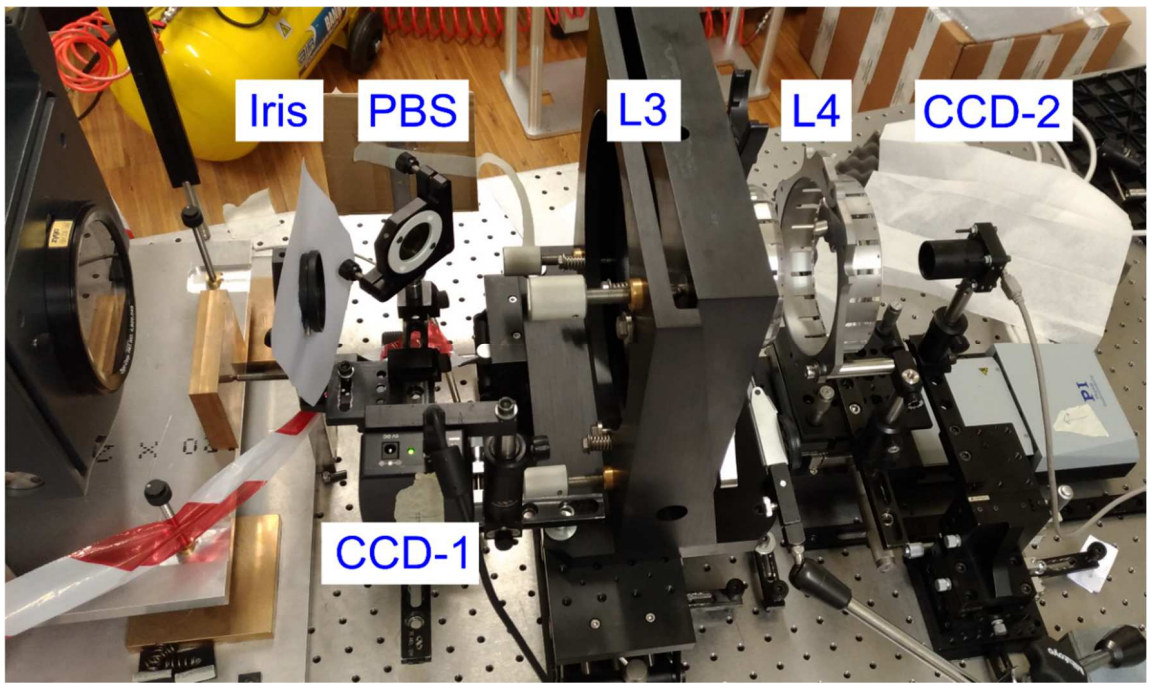

*Figure 34. The optical setup of the confocal test, L3 is fixed, and L4 is on a translation stage. CCD-1 collects the back-reflected light from the L4 surface or CCD-2 when we remove L4 from the optical axis.*

We use a Physik Instrumente M-511.DG1 translation stage, with a design resolution of 0,033 µm/count driven with a velocity of 10000 counts/sec, with a travel range of 101,6mm. We continuously move the stage and collect 910 images on CCD-1 for 95,97 seconds, corresponding to a displacement of L4 lens of 31,67mm, with the focus of the L3 passing through the surface of L3 near the middle travel range. Each image is acquired in an incremental position of 0,035 µm. In Figure 35, we plot the averaged flux in the frame of the CCD vs. the L4 position; from 183 to 202 µm the signal increase, corresponding to reflection at the edge of the L4 lens. This method reaches the accuracy of the order of 20 µm on localizing L4. Naturally, this is only an example of the power of this technique and is limited to the light back-reflected only from one surface.

To apply this technique for all the lenses of PLATO we need to investigate the correct propagation of the light in the other lenses, also calculating the power lens-to-lens. For completeness, in Figure 36 it is shown the result of the confocal sensing when the focus of L3 is located in the optical window of a CCD. In this case, we double-check the position of the optical window with the interferogram of the back-reflected light. Equal spaced fringes appear, but this is not the case of PLATO.

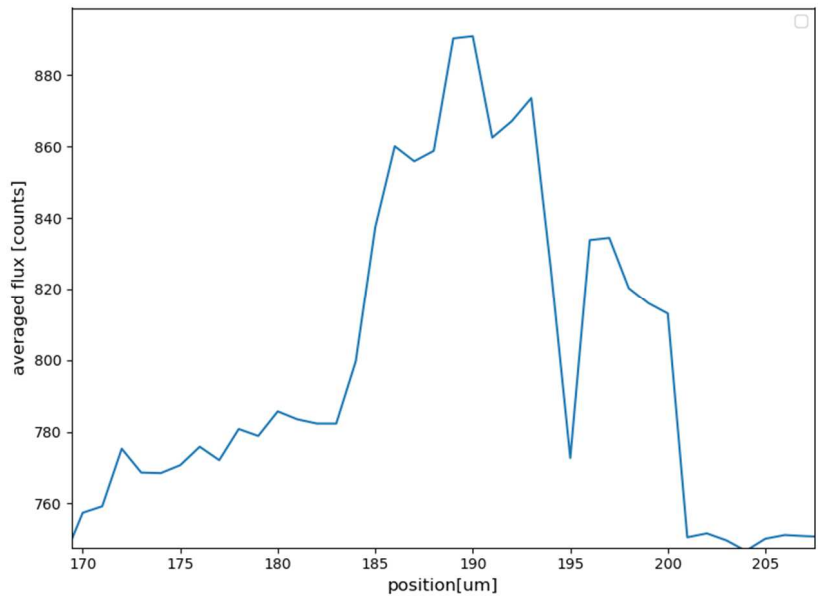

*Figure 35. The averaged flux on CCD-1 frame through the range of 31,67mm. In the positions between 183-202 µm the signal id increasing, corresponding to the edge of L4 lens, with a maximum error of around 20 µm.*

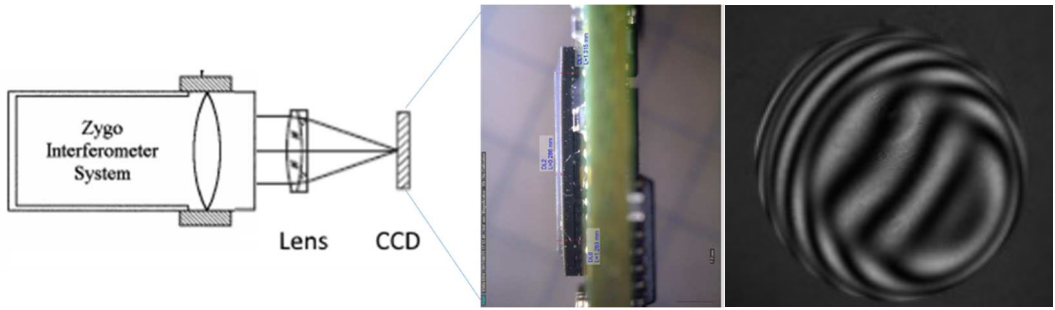

*Figure 36. Confocal sensing of the optical windows f the CCD-2.*

I anticipate here that Leonardo Company, after the delivery of the documents related to the AIV process, is studying the TOU AIV procedure with a Trioptics centering machine working on five axes, that uses also a similar principle of the confocal sensing. The centering machine made centration measurement in reflection and transmission. To perform the measurement in reflection, follow Figure 37, the head lens is focused on the center of curvature of the lens surface being tested. The resulting reflected image of the reticle is observed using the CCD camera integrated into the measurement head and analyzed with the software. If there is a centration error, the observed image describes a circle while the sample rotates on the reference axis. The center of the described circle is on the reference axis. The radius of the circle is proportional to the centration error and describes the distance from the center of curvature of the lens surface to the reference axis. If the centration error is described as an angle, this is called a surface tilt error when measuring in reflection. As a result, the shift of the center of curvature of the lens surface was determined with respect to the axis of rotation. The data reduction software uses this data to calculate the spatial position of the optical axis.

In low-coherence interferometry, the light from a low-coherence light source is divided by a beam splitter, one towards the lenses to be measured (the object beam) and one towards a movable mirror (reference beam), see Figure 38. The object beam illuminates the lens system along its optical axis. A fraction of the incoming light beam is back-reflected at each surface of the sample. This light is superimposed on a photodetector with the light from the reference arm. The light in the reference arm is varied to follow the attenuation of the reflected ray. The length of the reference arm is varied through a movable mirror and measured by using an encoder. When the resulting intensity of the back-reflected reference beam and object beam are analyzed as a function of the change in length/delay of the reference arm then interference patterns are always observed when the optical path lengths coincide in the two interferometer arms, the difference of the group delay is zero. Thus, the position of each surface of the sample can be determined mathematically, from interference peak positions. The alignment accuracy is declared to reach x/y/z <1 μm and $\theta_x, \theta_y$ < 2arcsec, about ten times better than the TOU budget requirement.

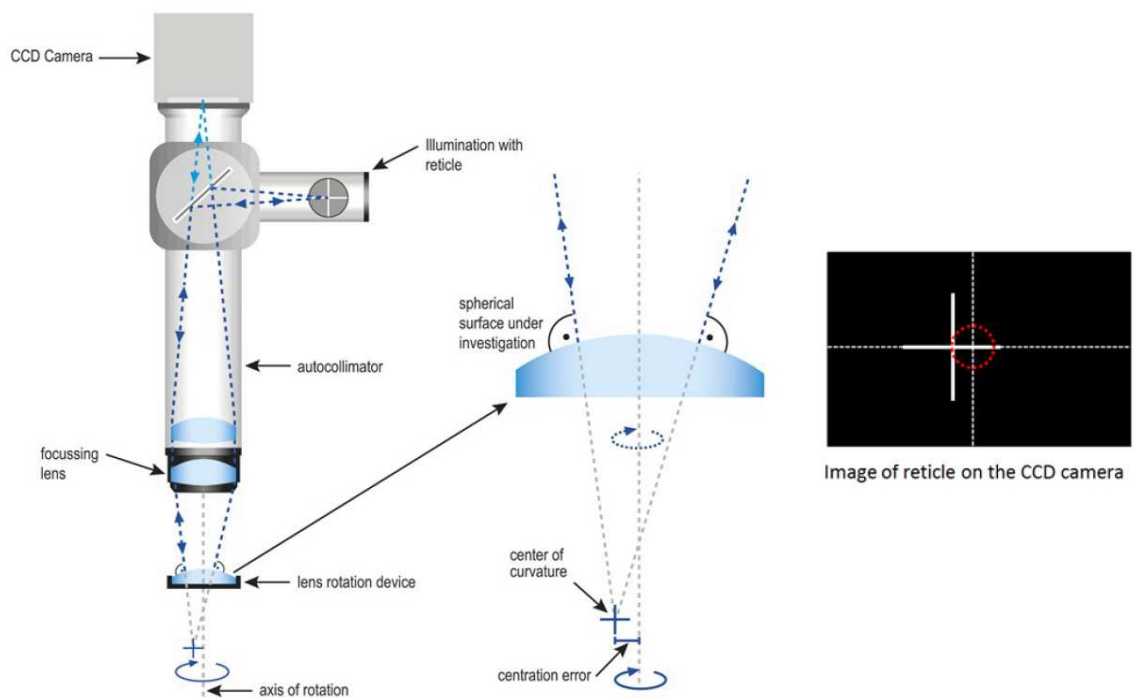

*Figure 37. Measurement of the centration error in reflection mode with autocollimator and reticle, from Trioptic product catalog.*

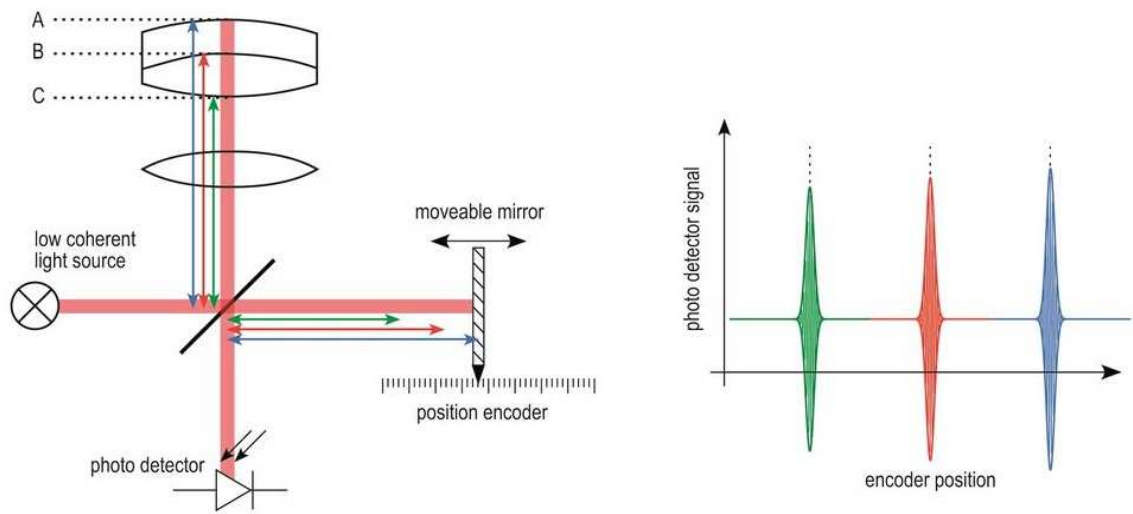

*Figure 38. Operating principle of a low-coherent interferometer, from Trioptics web site.*

## 2.3.5   The overall assembly and alignment concept

Summarizing what reported in the previous sections, the overall AIV procedure concept can be summarized as follows:

1. L3 is the first lens inserted into the prototype and coarsely aligned only in decentre to be close to the mechanical axis of the structure;

2. the laser beam (that will be used as reference beam for the alignment) is aligned to L3 in a way to minimize the tilt of the back-reflections and to have the back-reflections of the alignment beam superimposed to the ongoing alignment beam itself (iterative process between tilt and decenter);

3. move the stage with the T-CCD till finding the best focus position, and record the correspondent linear stage encoder value in that prototype orientation;

4. rotate the prototype and move the stage with the T-CCD till finding the best focus position, and record the correspondent linear stage encoder value in that prototype orientation;

5. go back to the prototype orientation of item #2, and insert L4;

6. set the focus position of L4 to the nominal one;

7. align in centering and TT L4 as described in Sec. 2.3.2;

8. align in focus L4 as described in Sec. 2.3.3;

9. repeat the previous items until having aligned all the lenses.

### 2.3.6 The Prototype Verification concept

Both "warm" and "cold" tests on the Prototype are planned to be performed either in Padova laboratory or at Leonardo, which is equipped with a suitable climate chamber, and they will be done by a Test Detector System (TDS).

Two different performance tests in a "warm" condition will be performed in Padova laboratory:

- A standard interferometric test (see Figure 39), performed with an F/1.5 transmission sphere to achieve the proper prototype input F/# and a reference flat mirror with tip-tilt capabilities;

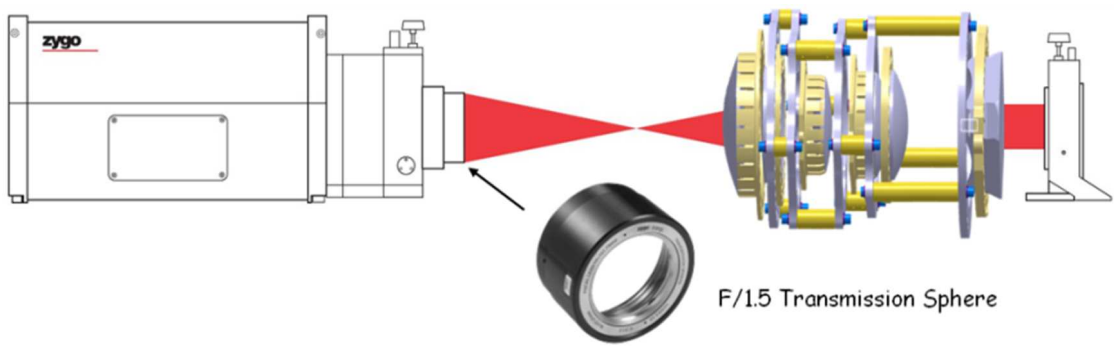

*Figure 39. The warm interferometric test of the BB*

- A PSF test, done with monochromatic light. A 200mm flat folding mirror, equipped with tip-tilt capabilities, will be used to simulate the light coming from off-axis objects to check the PSF over the FoV. The PSF shape of off-axis objects can be recorded by moving the TDS over the complete FoV by using motorized linear stages.

Since the PSF off-axis, at ambient temperature, will be very large, a Hartmann mask may be used to assess the optical quality which can be achieved in warm conditions.

The prototype "Cold" tests in vacuum have the purpose of double-checking the on and off-axis performance in the final working conditions, at the operating temperature of about -80º.

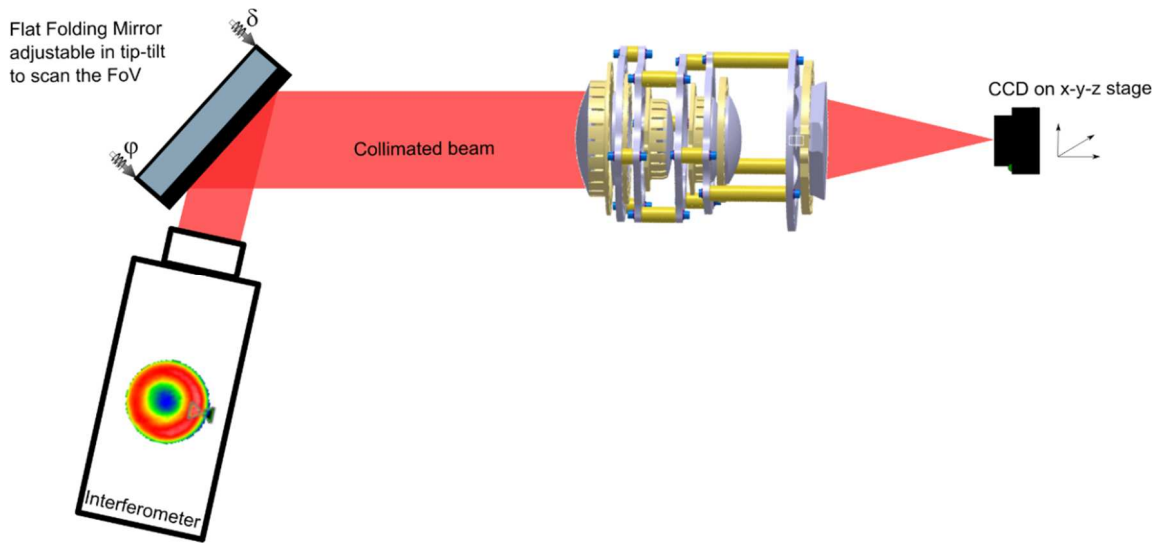

*Figure 40. The warm interferometric test setup.*

This test is conducted at Leonardo, where a climate chamber is made available. It is very important the involvement of the industry already in this phase in order to conceive a commonly agreed strategy both for the alignment and for the test to be performed on the TOU in each phase.

The selection of the test to be performed in "cold" conditions has been made taking into account a few problems to be faced, mostly due to the low testing temperature and vacuum conditions (Hearn, 1999). The temperature difference between the inside and outside parts creates axial and radial gradients within the window, usually this effect curves the flat window as a meniscus lens, with a convexity facing outwards the TVC, so-called "shrinking effect". The gradient relative to the atmospheric pressure curves the window as a meniscus lens, with a convexity facing inwards the TVC. The optical window assumes an optical power if these two gradients are not well balanced.

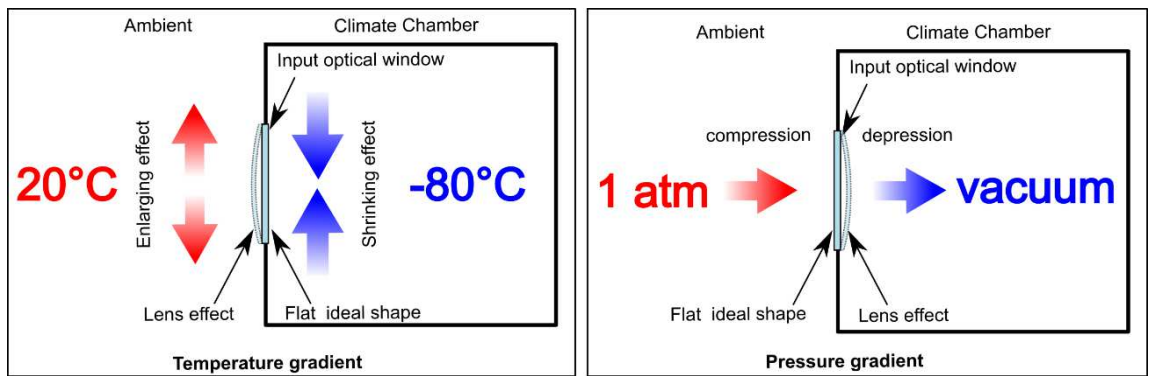

*Figure 41. Lens effect at the entrance of the Climate Chamber from temperature and pressure gradient.*

In fact:

- the optical input window of the TVC is affected by aberrations (the so-called "lens" effect, see Figure 41) of a combination of temperature and pressure gradient;

- if the test requires auxiliary optical components inside the chamber, they will also be affected by aberrations and particular care shall be given to their mounts to ensure the proper temperature compensation.

For these reasons, we have outlined a strategy for the "cold" test, which is the following:

- to collect information concerning the Input Optical Window aberrations;

- to minimize the lens effect of the Input Optical Window, it is better to use parallel beams setup at the climate chamber entrance;

- it is better not to use additional optics inside the climate chamber;

- if additional optics inside the chamber is needed, it is better to use flat mirrors instead of concave ones, again to minimize the "lens effect".

With these criteria, we have identified some tests to be performed on the prototype:

- A Hartmann test, as described in Figure 42;

- A PSF test, done with monochromatic light, on-axis, and off-axis. A 500mm flat folding mirror, equipped with tip-tilt capabilities, will be used to simulate the light of off-axis objects. The PSF shape of off-axis objects can be recorded by moving the TDS over the complete FoV with motorized linear stages. Of course, an additional couple of linear stages in XY configuration, working in cold conditions are required.

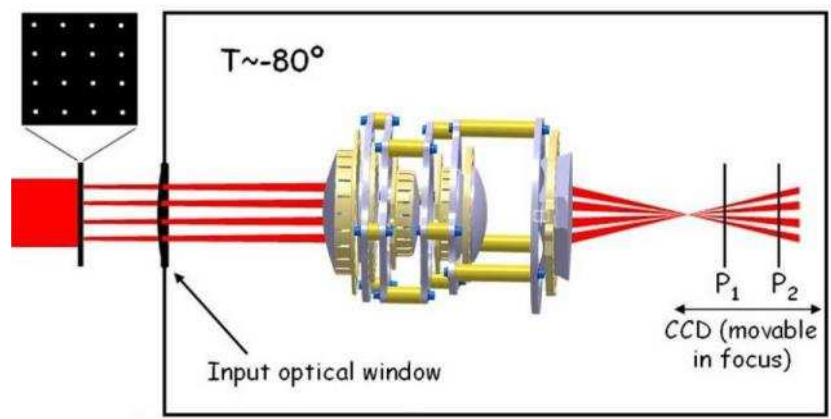

*Figure 42. Prototype "cold" test, performed by a Hartmann mask*

Both the selected tests present the advantages of requiring a parallel beam in input, not asking for any additional optical component inside the chamber and requiring only 1 motorized axis in cold.

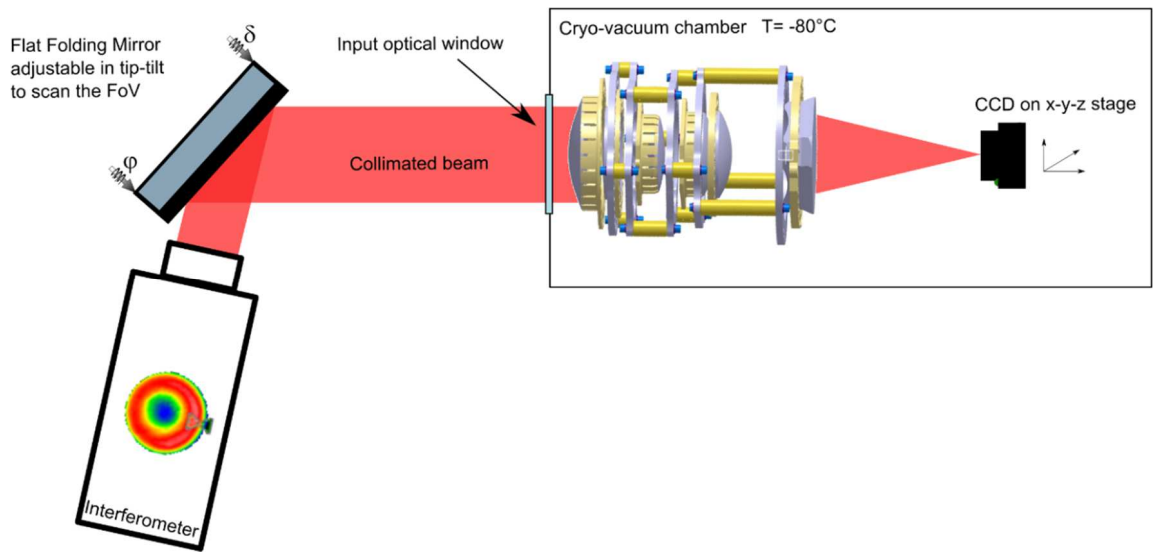

*Figure 43. The cold PSF test setup.*

The source is the collimated beam generated by an interferometer, equipped with a beam expander and previously aligned in auto-collimation with a reference flat mirror.

To align the collimated beam to the prototype optics, the following procedure will be used:

- a small, collimated laser beam, coming from outside the climate chamber, is aligned with respect to the prototype optics looking at the back-reflected spots, recorded on one of the test cameras used for the prototype AIV. The chamber optical window, being flat, will produce back-reflections easily identifiable, which will be neglected during this alignment;

- the position of the re-imaged spot in the TDS, inside the chamber, is recorded;

- the laser beam is removed and replaced with the collimated beam coming from the interferometer, equipped with a beam expander to obtain a 300mm wide beam;

- a flat mirror is used as a reference to auto-collimate the interferometer;

- the interferometer beam is tip-tilted, with a 500mm flat folding mirror, toward the climate chamber optical window and adjusted until the focused spot reaches the recorded position on the TDS.

In summary, the test that will be performed on the prototype are:

- interferometric test in warm conditions;

- optical quality check over the FoV in warm conditions, PSF and Hartmann test;

- Hartmann test in cold conditions;

- PSF test over the FoV in cold conditions.

## 2.3.7   The Ground Segment Equipment

The various GSEs will be used to provide these actions:

- the lens insertion into the main mechanical structure always from the top;

- possibility to align each lens but L3 in decentering, tilt and focus with a sensitivity of 2 μm in decentering and focus and 5'' in tilt;

- isolation of the back-reflected spots coming from a single lens;

- a stiff setup bench is allowing the injection of a laser beam from the top towards the TOU, see Figure 48.

The concept is then to have a frame with the possibility to hold the TOU mechanical structure in a vertical position in both the orientation (L1-L6 and L6-L1). This structure is thus provided of two pivoting points located close to the position of L3, to minimize the rigid shift of the TOU along the optical axis when the rotation occurs. This GSE also provides the possibility to lock the TOU in the desired position and adjustment screws to fine-tune the rotation of the TOU in order to have exactly 180° rotation between L1-L6 and L6-L1 orientations (see Figure 44) within the accuracy provided by the adjustment screw (~10 arcsec).

The need to have the TOU always oriented in the vertical position with respect to the gravity is to simplify the insertion and the alignment of the lenses inside the TOU.

Since inserting the lenses by hand into the TOU structure would be difficult or very dangerous, especially for the lenses located toward the middle of the tube, like L4, we designed a special GSE allowing to handle the lenses always from outside of the mechanical structure of the TOU.

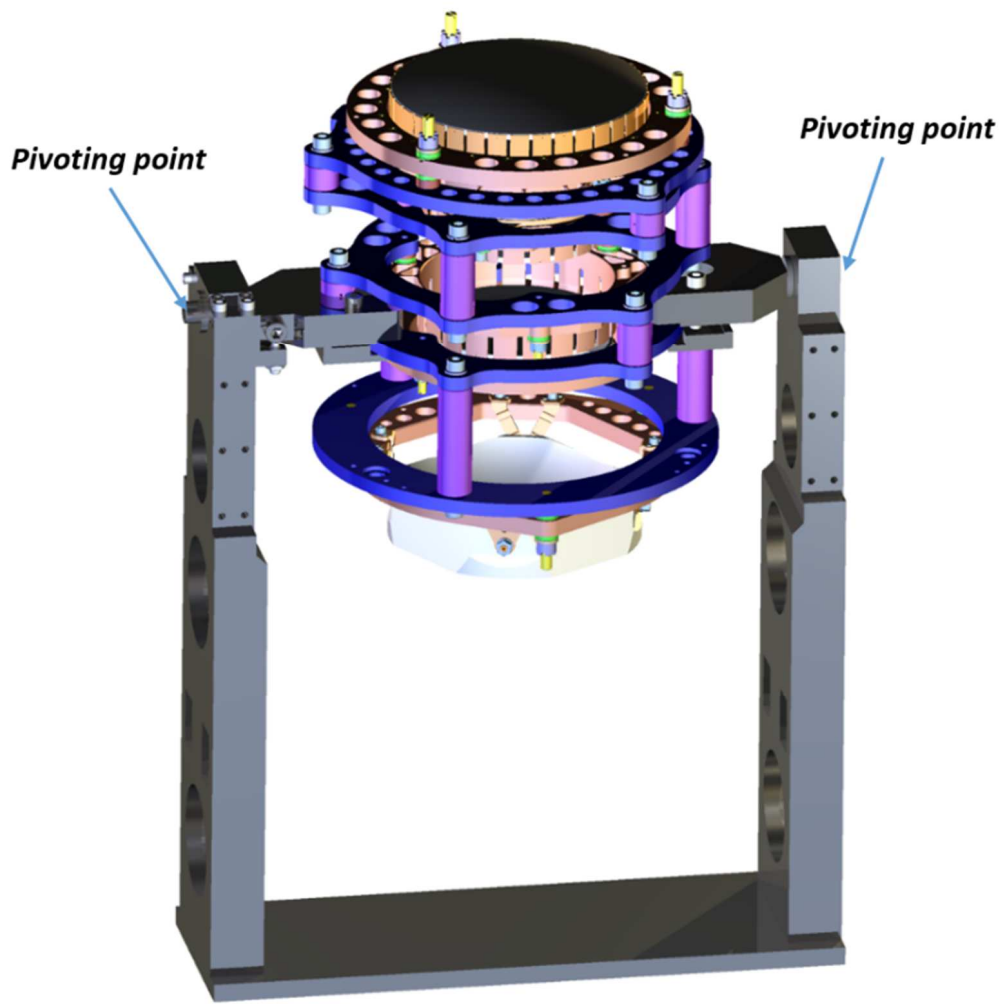

*Figure 44: The GSE allowing to keep the TOU Prototype structure in the vertical position, both in L1-L6 and L6-L1 orientation. The pivoting points are located close to the position of L3 so that when changing the orientation, the TOU maintains more or less the same position along the gravity axis.*

These micrometers are always kept in correspondence of the fixing holes of the lens mount which is being aligned at that moment, to simplify the computation of the thicknesses needed for the shims to be inserted under the lens mount. For this reason, the metal ring, from now on called "**manipulator**" (shown in Figure 45), provides several interface holes for the micrometers, in order to locate them always in the correct position, and several interface holes for the rods. In this way, the manipulator is common to all the lenses, but L3.

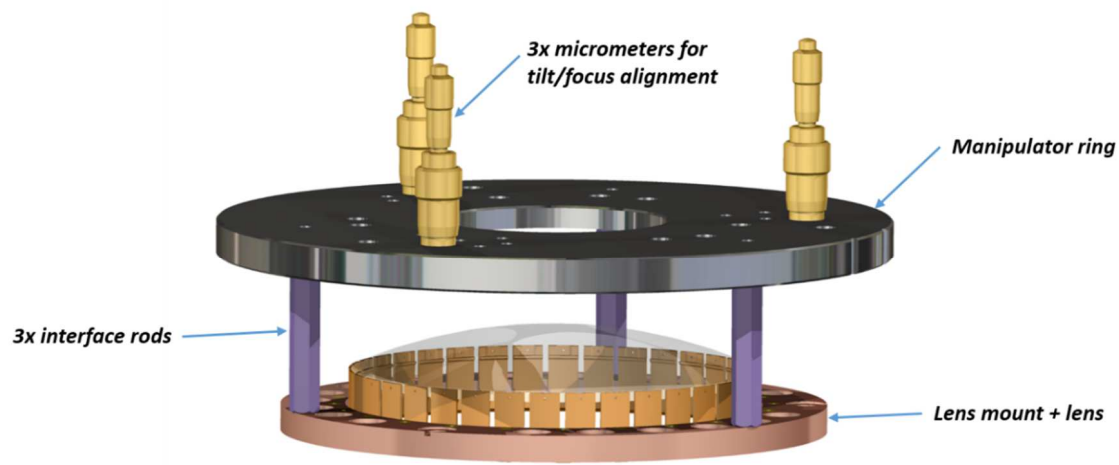

*Figure 45. The manipulator used for the insertion and tilt/focus alignment of the lenses into the TOU mechanical structure. Several holes are provided in the manipulator, to use it for the insertion/alignment of all the lenses.*

Once the lens mount leans on the mechanical structure of the TOU, acting on the 3 micrometers for tilt/focus adjustment we slightly lift the lens. The micrometers are now pushing on a custom-designed rectified flange. Since the dimension and the position of this rectified flange would not allow the insertion of the lenses, this flange must be removable and repositionable. For this reason, the rectified flange has pins and it is screwed to the GSE fixed structure. The sequence of a lens insertion can be seen in Figure 46. As a first step, the lens is inserted from the top keeping it from the interface rods (panel 1). Once the lens mount is correctly leaning on the TOU mechanical structure, install the rectified flange on the GSE fixed part and screw it in position. The rectified flange has an opening to allow the interface rods to protrude above it (panel 2). Finally, insert the manipulator ring and screw it to the interface rod (panel 3).

When the lens alignment setup is correctly installed, acting on the micrometers installed onto the manipulator ring the lens is manipulated in tilt/focus, while for the decenter alignment there are 2 micrometers, displaced by 90°, on the fixed part of the GSE and pushing against the manipulator ring. A pre-load block, displaced by 135° from the two micrometers, counter-acts the action of the centering micrometers, with a strength sufficient to pull the system back when one of the centering micrometers is rotated backward.

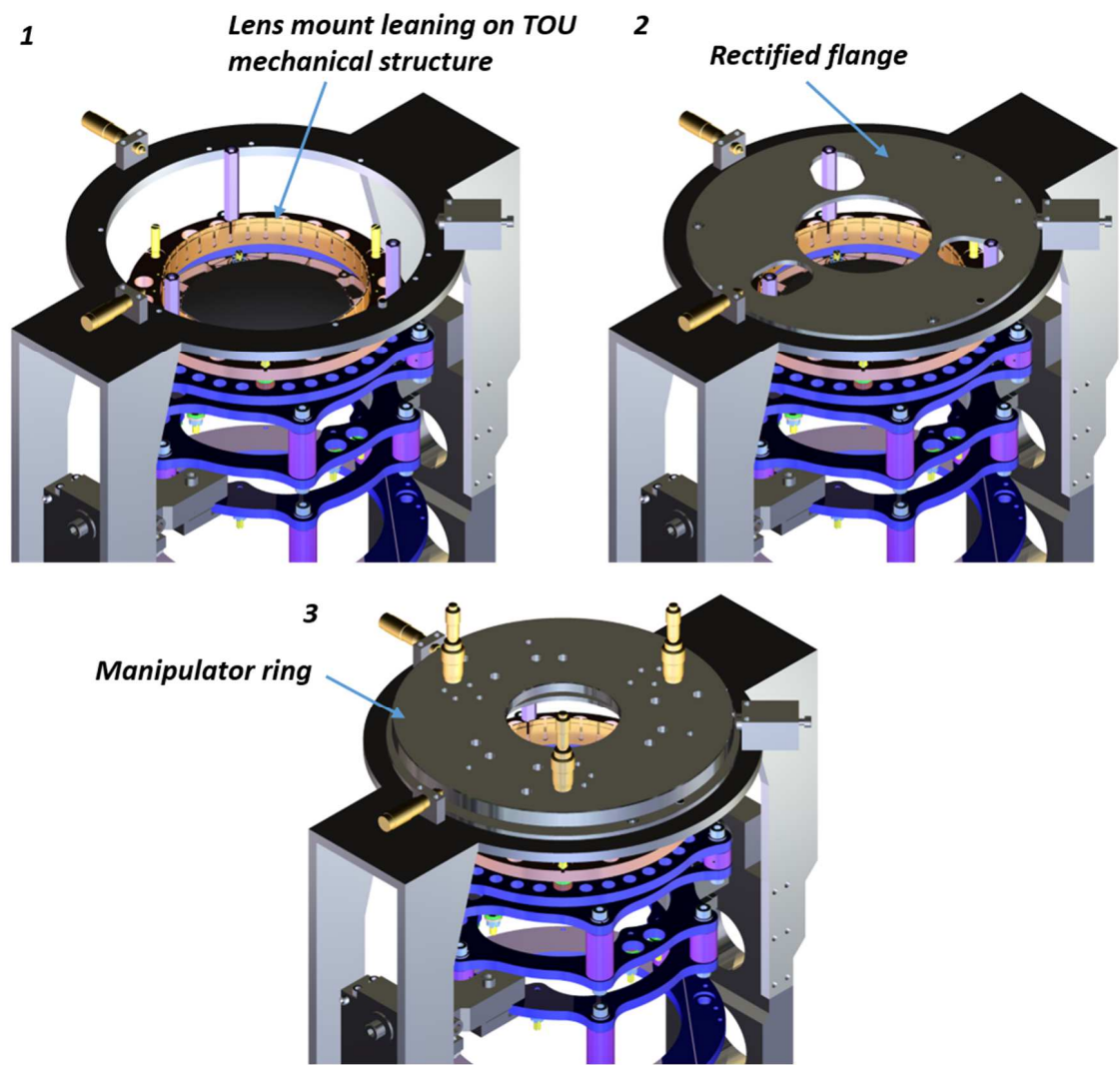

*Figure 46: The lens insertion strategy. As a first step, the lens is inserted from the top keeping it from the interface rods (panel 1). Once the lens mount is correctly leaning on the TOU mechanical structure, install the rectified flange on the GSE fixed part and screw it in position. The rectified flange has an opening to allow the interface rods to protrude above it (panel 2). Finally, insert the manipulator ring and screw it to the interface rod (panel 3). Now the setup for the lens alignment is ready.*

We actually verified during the alignment of the prototype that the strategy of aligning in centering the lenses by means of such a manipulator, with long interface rods connected to the lenses, has problems related to the flexures of the rods themselves when trying to shift the lens leaning on the shims. This introduces spurious tilt on the lenses and highly non-linear movements of the lens in decentering, making the alignment procedure very difficult. For this reason, in the end, we decided to align the lens in centering by using a precision hammer and acting on the side of the lens mount. In principle, this would not be possible with the flight model tube, as it is completely closed, and for this reason, it would be really useful to foresee some small openings in the tube allowing to access the lens mounts from the side.

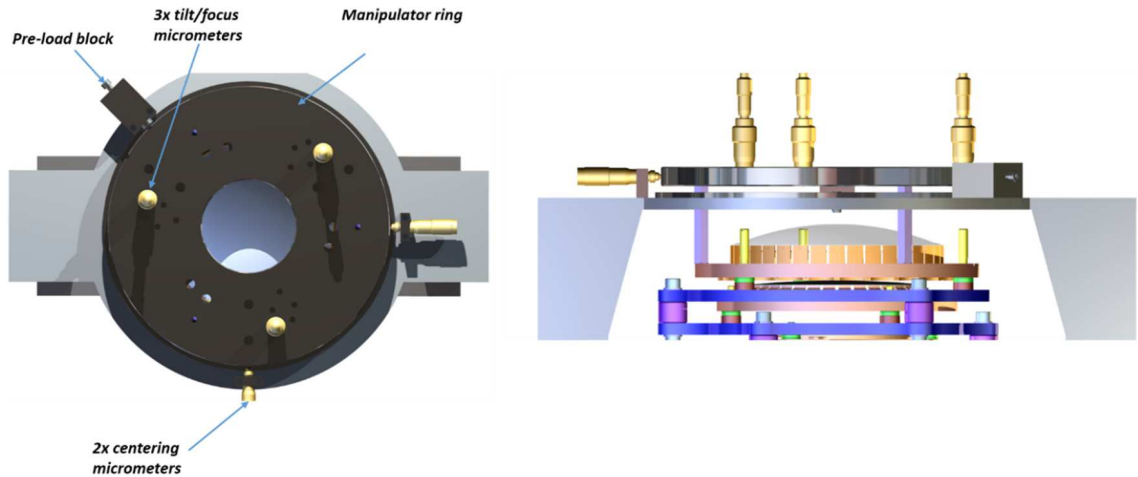

*Figure 47. The manipulator view from the top and from the side.*

On the top of the GSE is also fixed a small setup bench, hosting some folding mirror/beam splitter to send a laser beam towards the TOU and 2 CCDs; one for the analysis of the back-reflected light and the other to check the stability of the alignment beam.

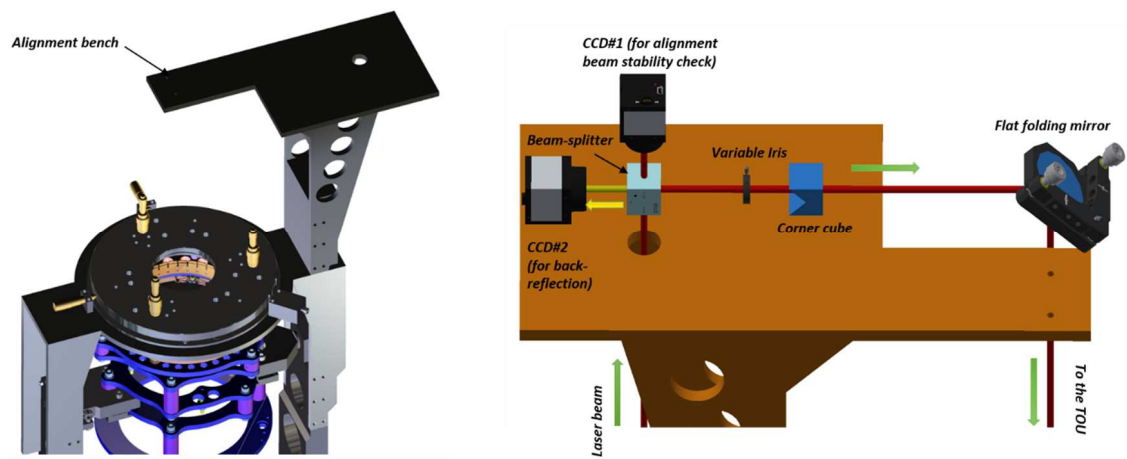

*Figure 48. The alignment bench located on the top of the TOU AI GSE. A collimated laser beam passes through a hole in the bench and half of the light is folded by a beam splitter parallel to the alignment bench towards a flat-folding mirror.*

A linear stage with a travel range of 150 mm and an accuracy positioning better than 1 μm has been installed on the lower part of the fixed part of the GSE. The linear stage, an Aerotech PRO165SL-150, carried a small CCD, without housing, for the analysis of the TOU transmitted spot. Alternatively, it can carry a spherical mirror, for the interferometric measurements. Since the travel range of the linear stage is longer than the required one, in order to avoid accidental collision between the CCD and the lenses or the mechanical structure of the TOU a security screw is installed on the fixed part of the GSE in order to block the stage before any possible collision.

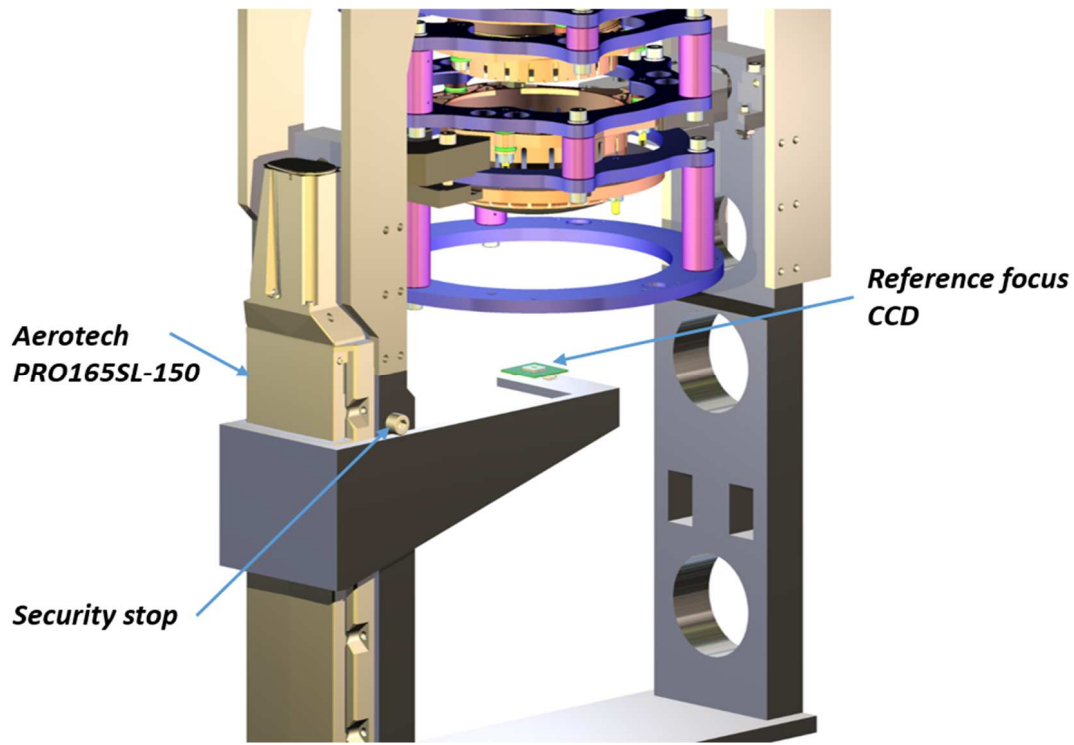

**Reference focus CCD**

**Aerotech PRO165SL-150**

**Security stop**

*Figure 49. A linear stage carrying a CCD has been installed on the lower fixed part of the TOU GSE. The travel range of this stage was sufficient to cover the range in which the focus of the TOU system was moving each time a new lens was inserted in the system, as depicted in Figure 28. The accuracy of the linear stage was better than 1 µm.*

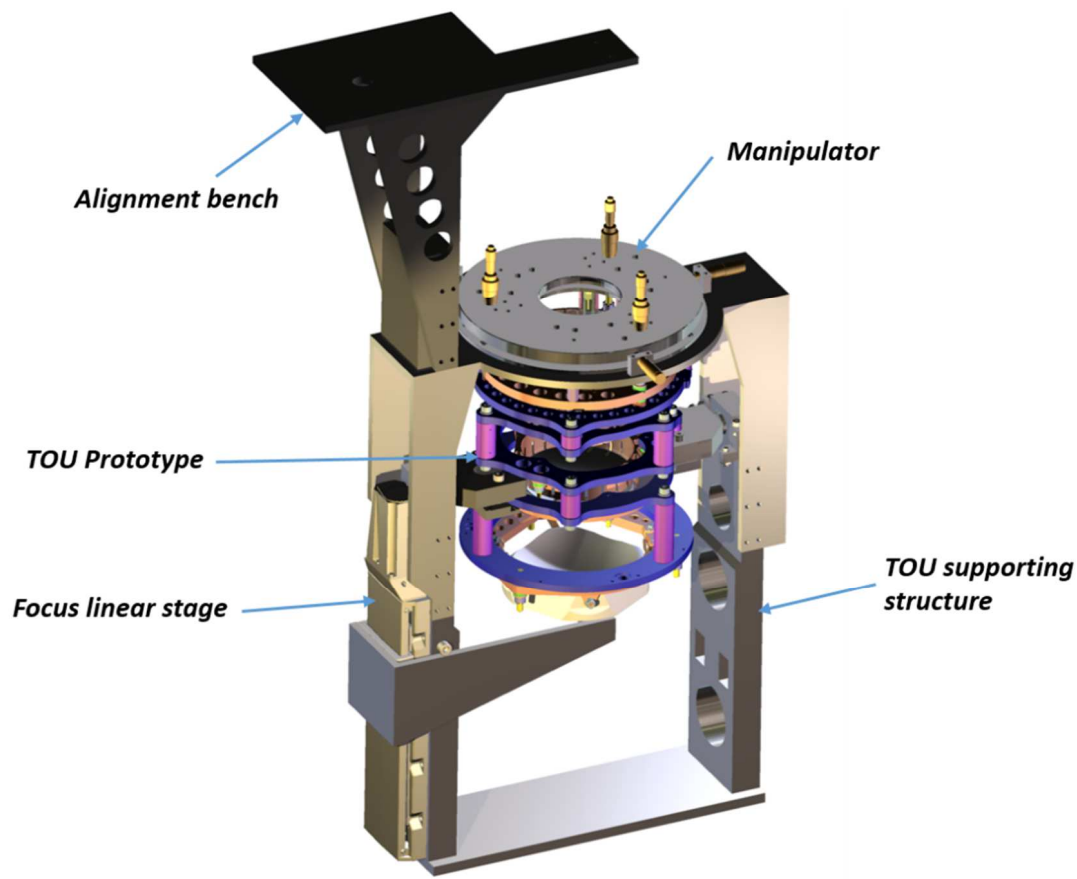

*Figure 50. A sketch of the full GSE used for the TOU Prototype AI phase. The whole structure was located onto an optical bench where other optical components delivered a collimated laser beam towards the alignment bench of the TOU Prototype GSE.*

## 2.3.8 Mechanical tools

We use two elongation tools provides by UBE to close the lens mount into the TOU mounts. When we torque a normal screw, torsional stress is introduced to the parts involved. Because of optical performance requirements, introducing tension into the lens mounts needs to be avoided. We place the elongation tool over the screw used to block the mount. At the top, the elongation tool has a screw barrel, named "screw assy", composed by several fine screws. By evenly increasing the torque at the screw assy, the bolt of the mount is elongated, until a specific torque value.

The crown nut is tightened until it touches the washer of the mount. At this point evenly decreasing the torque at the screw assy the tension of the elongation tool is transferred to the washer, see Figure 51. The goal is to avoid any torsional stresses during the fixing procedure. During this operation, the lens is monitored in tip-tilt and decenter. We use digital torque wrenches to apply the correct torque following Table 5.

.

*Table 5. Torque at the "screw assy" for 0.1% of elongation.*

|    | Bolt elongation [mm] | Bolt force [N] | Necessary torque [Nm] |
|----|----------------------|----------------|------------------------|
| L1 | 0.029                | 4880           | 1.1                    |
| L2 | 0.019                | 2850           | 0.5                    |
| L4 | 0.024                | 2850           | 0.5                    |
| L5 | 0.015                | 2850           | 0.5                    |
| L6 | 0.026                | 4880           | 1.1                    |

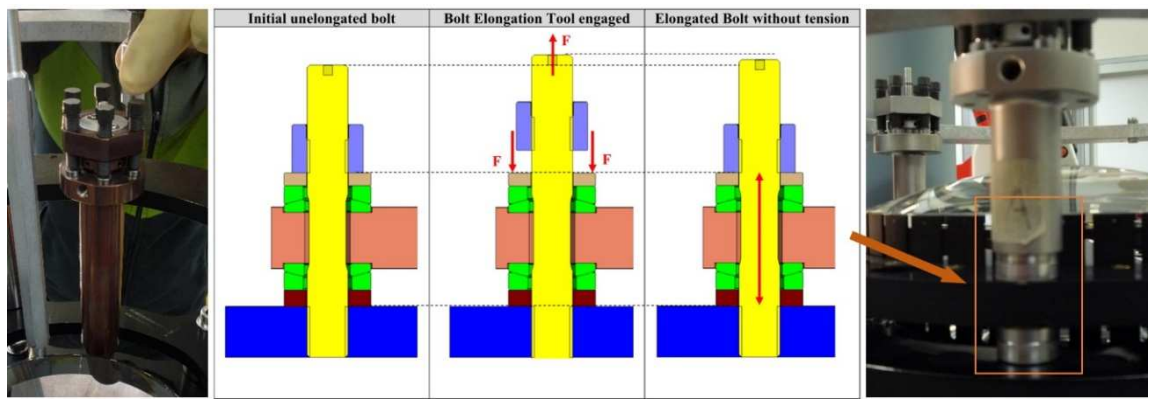

*Figure 51. On the left: mechanical tools for the alignment. Right: the image of the elongation tools inserted over the screw assy. On the center: the tool initially incorporates the screw barrel and the bolt is not elongated, then the tool is engaged and the nut rises. The washers are gently aligned over the mount and we screw the nut. Finally, the elongation tools are removed and a torque force compresses the nut and washers. Courtesy Tymothy Bandy (UBE).*

## 2.3.9   Washer stability

The lens mounts are locked in the TOU mount by screws, bolts, and washers in special material for avionics. During the alignment process the optical mount will be locked and unlocked few times, and, when the lens reaches the final position, we need that the locking procedure doesn't move the lens. To achieve this we pay attention to using a spherical washer and test the position stability after several lock/unlock cycles.

A test setup (Figure 52) was built in order to measure the variations of tip-tilt, decenter and defocus by using the L1 lens and mount of the PLATO breadboard. The autocollimator (λ = 633 μm, Model = CONEX-LDS, Sensitivity: down to 0.01 μrad) was used for tip-tilt measurements by a flat mirror (d = 100 mm) glued to the L1 mount with thermal glue. Two dial gauges (1 and 2 in Figure 52)  with precision 2μm was used for decenter Two dial gauges (3 and 4 in Figure 52) with a precision of 1μm was used for defocusing. The test was performed with the three pre-tensioning tools for M6 bolts, each assigned to the three bolt appropriately marked, by using the digital torque screwdriver setting a force from 0.10 N to 0.50 N (and reverse), with a step of 0.10 N each.

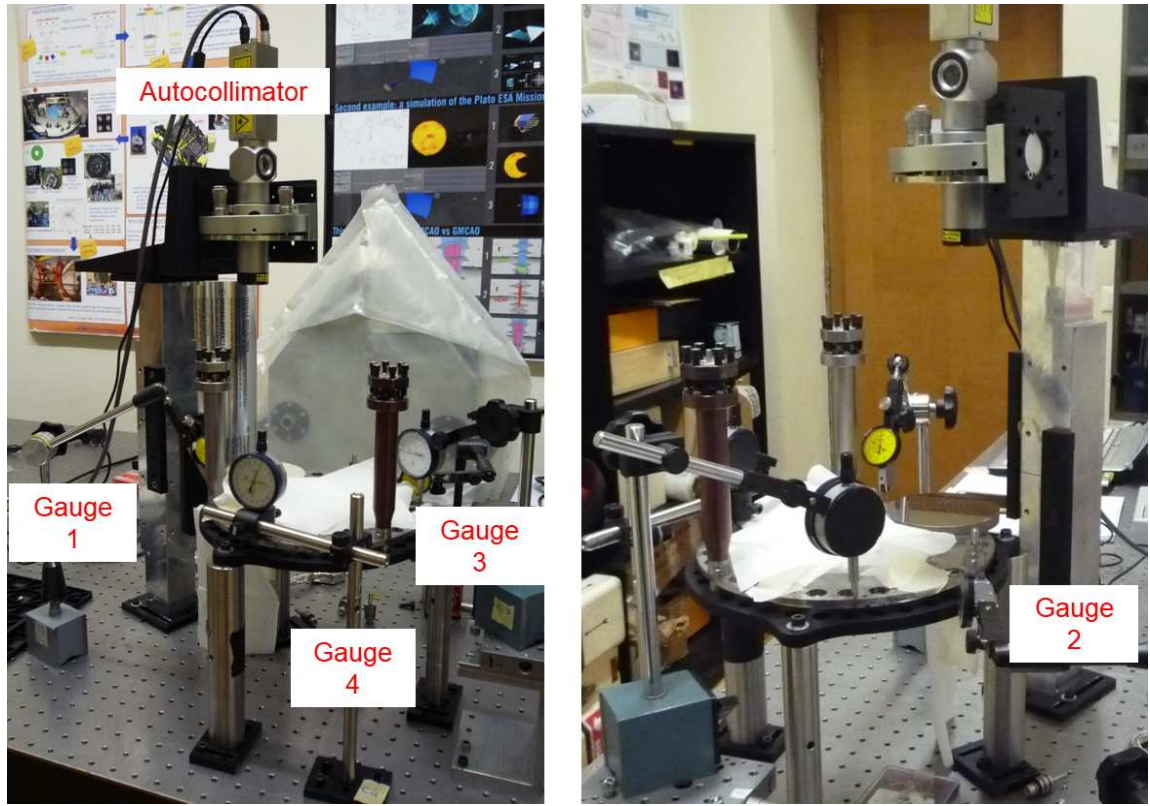

*Figure 52. The setup for testing the PLATO washers, three gauges, and an autocollimator is used.*

To measure such a movement we used a procedure that mimics the real assembly method with shaking of the spherical washers, the movement of the torque tools, the insertion of the crown nut by hand, the screwing and unscrewing procedure. We also tested a normal washer, and in total, we perfume 29 cycles. In Table 6 and Table 7 we summarize the results of the 29 runs performed with the spherical washers and 14 runs performed with the normal washers, to assess the tilt repeatability with different references:

Ref 1.  is the initial position of the washers

Ref 2.  is after shaking washers

Ref 3.  is after manually tightening the nuts

Ref 4.  after placing the 3 tools

Ref 5.  is after screwing at 0.10 N

Ref 6.  is after screwing at 0.20 N

*Table 6. Summary of the results on the tilt repeatability with spherical washers over 29 tests performed; reference positions are described in the text.*

| Tip-Tilt Spherical | Ref 1 [arcsec] | Ref 2 [arcsec] | Ref 3 [arcsec] | Ref 4 [arcsec] | Ref 5 [arcsec] | Ref 6 [arcsec] |
|---|---|---|---|---|---|---|
| Max(PtV$_x$) | 41,05 | 41,05 | 41,05 | 19,6 | 13,61 | 9,28 |
| Max(PtV$_y$) | 38,16 | 38,16 | 34,65 | 13,82 | 9,28 | 5,78 |
| Max(PtV$_{xy}$) | 56,05 | 56,04 | 53,72 | 23,98 | 16,48 | 10,93 |
| Mean(PtV$_x$) | 13,28 | 11,63 | 10,28 | 6,83 | 5,53 | 3,88 |
| Mean(PtV$_y$) | 12,28 | 11,37 | 9,3 | 4,88 | 3,65 | 2,4 |
| Mean(PtV$_{xy}$) | 18,48 | 16,26 | 13,86 | 8,39 | 6,63 | 4,56 |
| Stddev(PtV$_x$) | 7,78 | 7,97 | 7,95 | 4,92 | 3,54 | 2,27 |
| Stddev(PtV$_y$) | 7,12 | 7,08 | 6,84 | 3,42 | 2,33 | 1,42 |
| Stddev(PtV$_{xy}$) | 7,5 | 7,5 | 7,4 | 4,17 | 2,94 | 1,85 |

*Table 7. Summary of the results on the tilt repeatability with normal washers over 14 tests performed; reference positions are described in the text.*

| Tip-Tilt Normal | Ref 1 [arcsec] | Ref 2 [arcsec] | Ref 3 [arcsec] | Ref 4 [arcsec] | Ref 5 [arcsec] | Ref 6 [arcsec] |
|---|---|---|---|---|---|---|
| Max(PtV$_x$) | 59,61 | 56,93 | 52,19 | 24,13 | 16,29 | 9,08 |
| Max(PtV$_y$) | 40,43 | 35,27 | 30,53 | 14,02 | 7,22 | 2,89 |
| Max(PtV$_{xy}$) | 72,03 | 66,45 | 60,46 | 27,91 | 17,82 | 9,52 |
| Mean(PtV$_x$) | 20,79 | 19,61 | 17,68 | 9,22 | 6,53 | 4,34 |
| Mean(PtV$_y$) | 32,35 | 30,42 | 23,99 | 9,24 | 4,2 | 1,69 |
| Mean(PtV$_{xy}$) | 38,45 | 36,2 | 29,8 | 13,05 | 7,76 | 4,67 |
| Stddev(PtV$_x$) | 14,92 | 13,66 | 12,4 | 5,4 | 4,3 | 3,02 |
| Stddev(PtV$_y$) | 5,13 | 3,63 | 4,85 | 3,71 | 1,02 | 0,97 |
| Stddev(PtV$_{xy}$) | 10,03 | 8,65 | 8,63 | 4,56 | 2,66 | 2 |

*Table 8. Summary of the results on the focus repeatability with spherical washers over 10 test performed; Ref 1 is the initial position, Ref 2 is after shaking washers, Ref 3 is after manually tightening the nuts, Ref 4 is after screwing at 0.10 N and Ref 5 is after screwing at 0.20 N.*

| Defocus Spherical | Ref 1 [µm] | Ref 2 [µm] | Ref 3 [µm] | Ref 4 [µm] | Ref 5 [µm] |
|---|---|---|---|---|---|
| Max(PtV) | 19 | 17 | 12,5 | 4 | 2,5 |
| Mean(PtV) | 11 | 9,4 | 6,9 | 2 | 1,4 |
| Stddev(PtV) | 4,3 | 3,6 | 2,7 | 0,9 | 0,6 |

In Table 8 there is a summary of the results obtained when we measured the focus by using the two dial gauges. The values were obtained averaging the two measures. The defocus is always less than 20 µm for the spherical. Considering the Mean(PtV), the focus repeatability becomes very good (of the order of 2 µm) from reference 4.

Table 9 resumes the results obtained when we measured the decenter by using the two dial gauges. The values were obtained averaging the two measures. The mean of the measured decenter starts to be within 10µm from reference 3 (manual bolts tightening) and from there, there is no difference between normal and spherical washers. With both normal and spherical washers, from reference 3 (manual tightening of the bolts) there is still a not negligible movement of about 8µm on average.

*Table 9. Summary of the results on the decenter repeatability with the two tested setups, spherical washers on a flat plane and normal washers on a tilted plane, reporting just the Max(PtV) and the Mean(PtV) (yellow background). Ref 1 is the initial position, Ref 2 is after shaking washers, Ref 3 is after manually tightening the nuts and Ref 4 is after screwing at 0.15 N*

| Decenter | Ref 1 [µm] | Ref 2 [µm] | Ref 3 [µm] | Ref 4 [µm] |
|---|---|---|---|---|
| Max(PtV) Spherical (10 run) | 107,5 | 34 | 21,5 | 3 |
| Max(PtV) Normal (8 run) | 72 | 16 | 15 | 2,5 |
| Mean(PtV) Spherical (10 run) | 48,8 | 15,95 | 8,2 | 1,45 |
| Mean(PtV) Normal (8 run) | 31,19 | 9,19 | 8,25 | 1,5 |

In conclusion, the tilt repeatability for spherical washers is of the order of 15-20 arcsec PtV on average over 44 full tightening processes, and it appears to be not very repeatable in direction. The tilt repeatability can be improved (10 arcsecs PtV) by gently tightening the bolts manually first and checking the tip-tilt with the fixing GSEs inserted on the bolts before starting the tightening process. The focus changes of 5um PtV on average (over 10 test), always in the same direction is very repeatable.

## 2.4 Alignment observables and simulations

We identify as observables the features created by the reflected and transmitted incident wavefront of an on-axis laser beam (Figure 53). The initial plane wavefront encounters the first lens surface (1) and is reflected (wavefront named "a" in blue color) and transmitted (wavefront "b" in green color). The wavefront is curved both from the geometry of the first surface and from the index of refraction. Airy's spot is generated in points A and B by the emerging wavefront a" and b'. Far away from the back-reflected wavefront, a and b creates an interference pattern, Newton's rings, where the relationship of constructive interference is achieved by reflected wavefront a' and b'. In Figure 53 right, the line profile at the center of Newton's Rings and Airy spot is plotted.

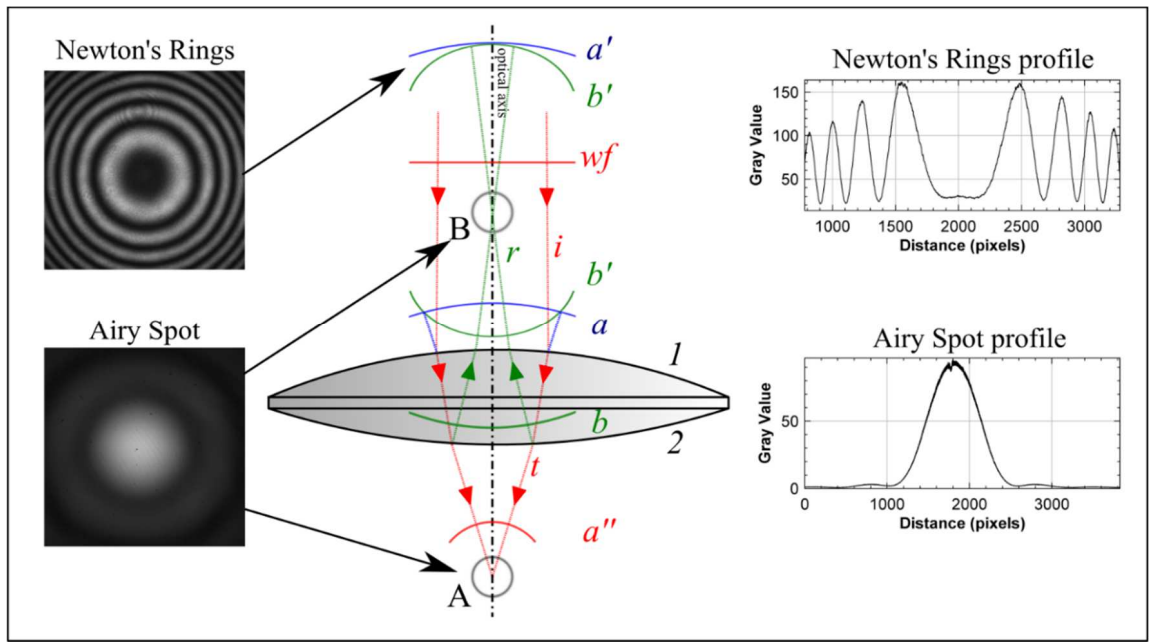

*Figure 53. On the top: laser light back-reflected from the lens surface 1 and 2 creates an interference pattern, the Newton's Rings. On the bottom: near point A and B the wavefront converges and create an Airy spot. On the right: plot of the central lines profiles of the resulting intensity maps.*

We use these interference shapes to define the optical axis of the third (L3) lens (Farinato, 2010), which is the reference for the alignment in tip-tilt, decenter (x-y) and focus (z) of the other lenses. The same interference shapes are used for aligning the other lens. We use a laser beam to realize the geometry of Figure 53. Moving the beam decenter and tilt, when Newton's Rings line profile becomes symmetrical, we know that the ray passes through the vertex of the lens, see Figure 54.

The reference observables on L3 are:

- the center of the principal Newton's ring: this defines the tip-tilt of the laser ray with respect to the vertexes of the lens;

- the symmetry, in terms of intensity, of the all maximum and minimum on Newton's rings: this defines the decenter of the laser ray with respect to the center of the lens.

We have developed algorithms to determine precise x-y decenter and tip-tilt measurements, with accuracy to match the design tolerances. Due to the double pass of the light in the front windows of the CCD, all spot laser images have an interference pattern superimposed. We decide to remove it by spatially filtering with an automatic selection in the power spectrum frequencies of this image. Even though fringes move over time due to temperature changing in the CCD, the corresponding spatial frequencies appear fixed in the 2dFFT image, see Figure 81. The laser beam is used as reference for the alignment of all the other lenses, by analyzing their transmitted and back-reflected spots (An et al., 2016).

The other lenses alignment will be performed looking at the Airy spots and/or Newton rings (from now on we will call them Back Reflected Rings - BRR), when visible, on a test camera collecting the light reflected back by the lenses (from now on BRR-CCD). On a second test camera, the transmitted spot position is monitored too (from now on, T-CCD). We emphasize that the lens tilt is mostly (but not only) acting on the position of the BRR, while the lens decenter is mostly (but not only) acting on the uniformity of the BRR. Instead, the transmitted spot position is mostly affected by lens decenter.

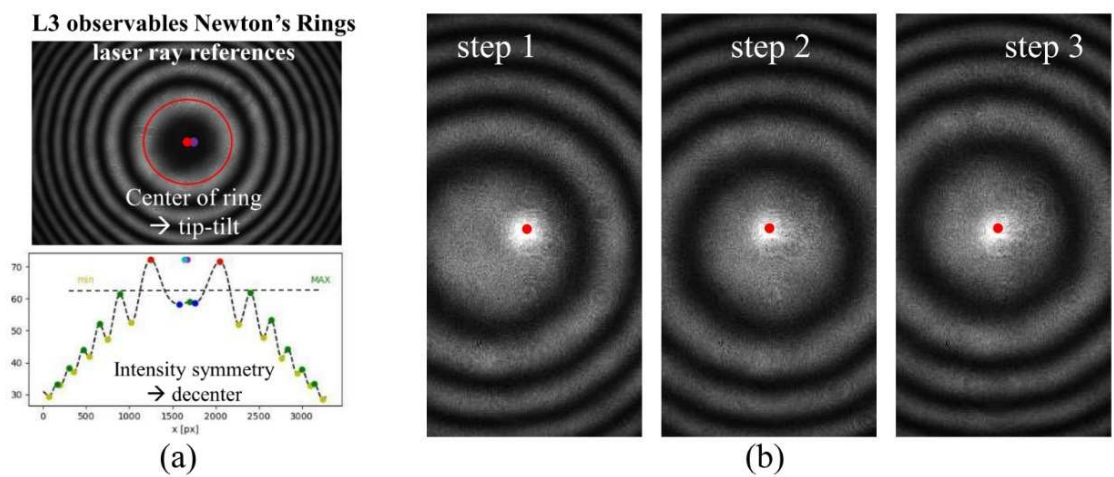

Figure 54. Panel a) the L3 observables defines the reference for tip-tilt and decenter. Panel (b) the loop of the procedure to align the laser concerning L3 tip-tilt, red spot is the reference position of the laser ray for every step. We finish the iteration for decentering and tilting when the tip-tilt and decenter is below the optical design tolerance.

## 2.4.1 Simulation of Newton's Rings

Ray tracing software is normally used to design an optical system, to find the best performance in terms of Encircle Energy, or MTF or any other parameter, providing geometrical analysis of the rays. The intensity map in the focal plane, or in intra and extra focal position, need software capable of calculating the propagation of a ray, including diffraction analysis; to achieve this, we did consider these possibilities. The open-source Opticspy python module for optics application, computes the diffraction, the interference and the Zernike's coefficients by optics propagation. The open-source PROPER (Krist, 2007) library is written in IDL and translated to Python and Matlab, it is a versatile routine for simulating optical propagation in the near and far-fields exploiting Fourier-based Fresnel and angular spectrum methods. The PRYSM open-source library for physical and first-order modeling of optical systems and analysis of related data. FRED Optical Engineering Software, a commercial software by Photon Engineering, simulates the propagation of light through any optomechanical system by raytracing. The TOU Team simulated with FRED the outline of Newton's rings to help the coding for real alignment case. We search for Newton's ring back-reflected by the TOU during the lenses insertion order, as previously described in Figure 46: L3, L3L4, L3L4L2, L3L4L2L1, L3L4L2L1L5, L3L4L2L1L5L6. Figure 55 resumes the simulated shapes of the lenses in the order of insertion in the TOU, applying a software matrix of 256x256 dots in front of at the simulated TOU illuminated by a 633nm source. In Figure 56 the simulation is limited to the back-reflected light from the single-lens as if the other lenses were obscured. At every alignment step, we consider only the last lens inserted (back-reflections from the following lenses are masked by a piece of optical paper).

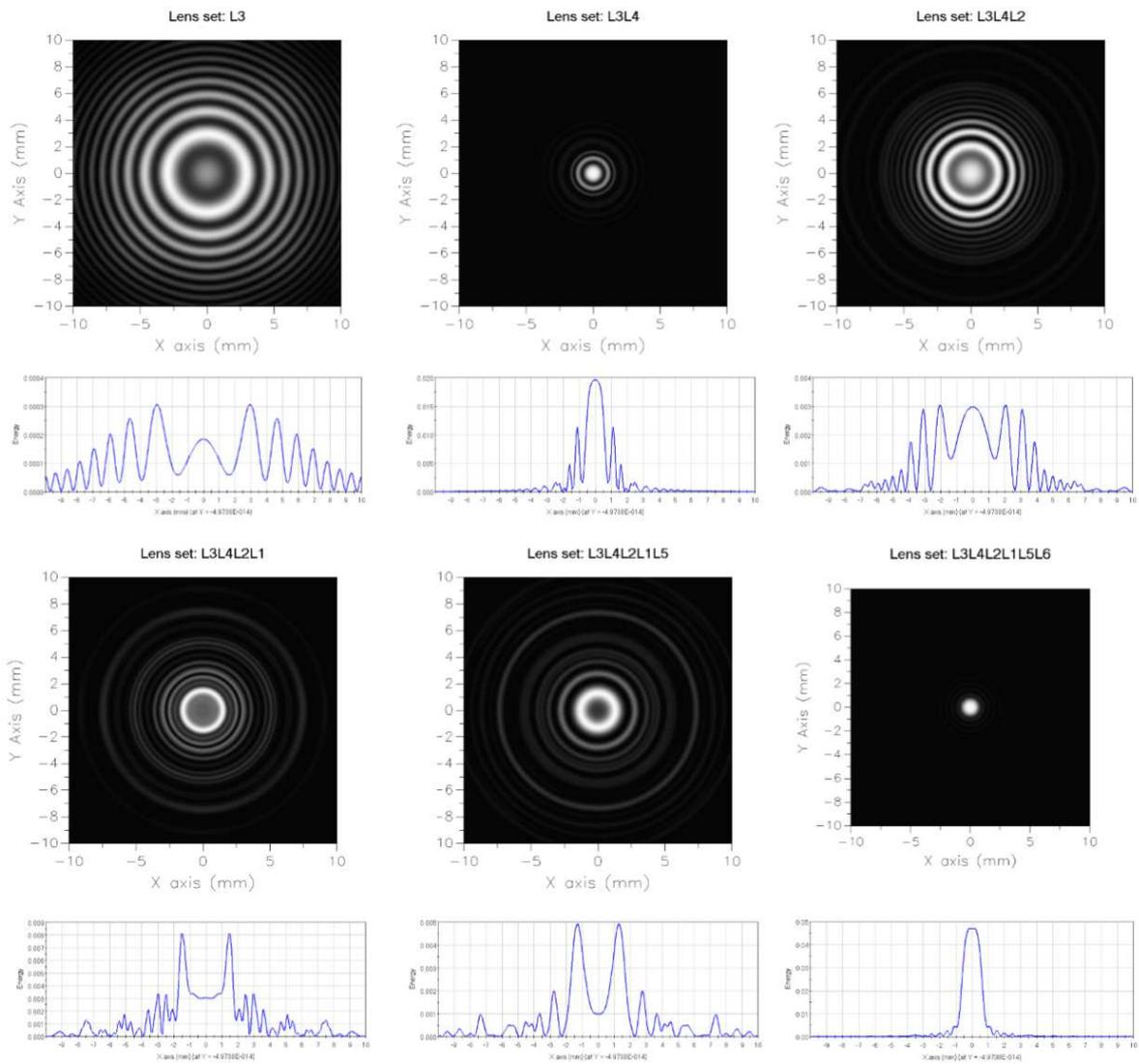

*Figure 55. The FRED simulation of Newton's rings by back-reflected ray from the TOU with insertion lens order.*

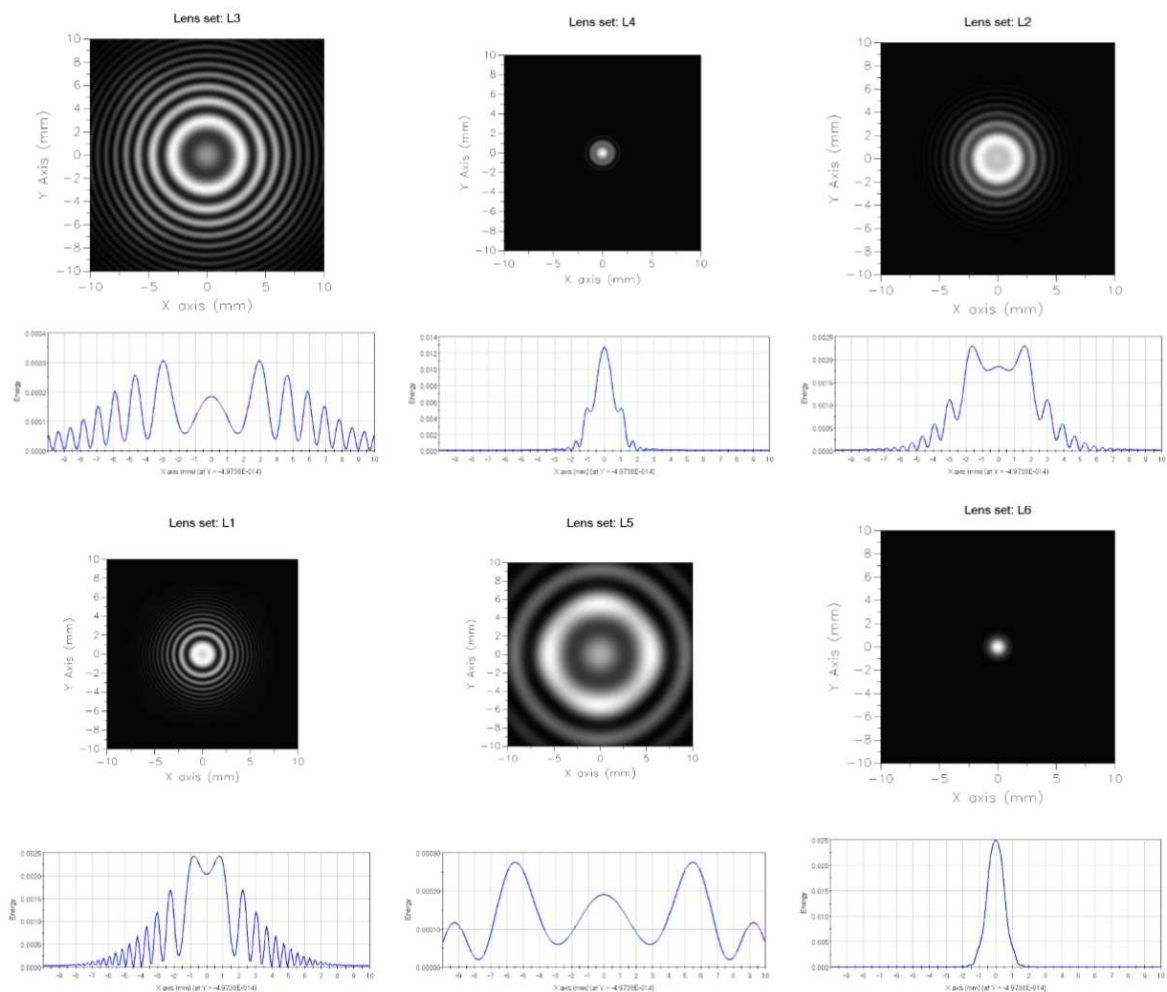

*Figure 56. Simulation of the Newton rings generated by each lens during the alignment procedure. At every alignment step, we consider only the last lens inserted (back-reflections from the following lenses are masked by a piece of optical paper). Simulations are based on gaussian beamlets ray-tracing with the software FRED by Photon Engineering.*

## 2.4.2   Defocus and Zernike coefficient

The interferometer, Zygo FlashPhase GPI is equipped with a Flat Reference lens λ/20 PtV and it is mounted in parallel to the laser system, co-aligned with respect to the laser optical path. Its 22 mm beam illuminates the GSE in common path with the laser by a folding mirror (FM2) mounted in a precision motorized translation stage. The role of the interferometer is to perform a double pass optical test to evaluate the precise position of a spherical reference mirror (SM1).

The Graphical User Interface of the software Metro Pro, see Figure 57, was configured to take more interferograms that are post-processed to achieve a good estimate of aberrations coefficients in 37 Zernike terms. Zernike polynomials (Goodwin and Wyant, 2006) were derived by Fritz Zernike in 1934. They express the wavefront data since they are of the same form as the types Zernike Polynomials of aberrations often observed in optical tests. These polynomials are a complete set of two variables, ρ and θ', that are orthogonal in a continuous fashion over the unit circle. It is important to note that Zernikes are orthogonal only in a continuous fashion and that in general, they will not be orthogonal over a discrete set of data points. Note that θ' is the angle counterclockwise from the $x_P$ axis.

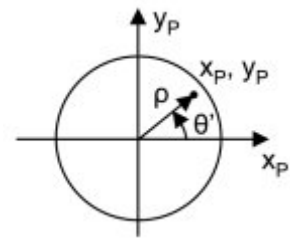

$$x_p = \rho\,cos\theta' \quad y = \rho\,sin\theta' \quad \rho = \sqrt{x_p + y}$$

There are several common definitions for the Zernike polynomials, so care should be taken that the same set is used when comparing Zernike coefficients. One of the convenient features of Zernike polynomials is that their simple rotational symmetry allows the polynomials to be expressed as products of radial terms and functions of angle, r(ρ)g(θ'), where g(θ') is continuous and repeats itself every 2π radians. The coordinate system can be rotated by an angle without changing the form of the polynomial. Each Zernike term is referenced by a single number or by two subscripts, n and m, where both are positive integers or zero.

$$g(\theta' + \alpha) = g(\theta')g(\alpha)\,;\ g(\theta') = e^{\pm im\theta}$$

The radial function, r(ρ), must be a polynomial in ρ of degree n and contain no power of ρ less than m. Also, r(ρ) must be even if m is even and odd if m is odd. A list of the first 37 polynomials is shown in Table 10, according to the Zygo notation (also known as "fringe" or "Arizona" Zernike polynomials).

*Table 10. The Zernike Polynomials Table. In this table, φ = polar coordinate angle, and ρ = radius (normalized to 1 at the edge of the aperture). The numbers in columns m and n are the indices for Zernike polynomials.*

| n | m | Term# | Polynomial | Aberration |
|---|---|-------|------------|------------|
| 0 | 0 | 0 | 1 | Piston or Bias |
| 1 | +1 | 1 | ρ cosθ' | Tilt x |
|   | -1 | 2 | ρ sinθ' | Tilt y |
|   | 0 | 3 | 2ρ²- 1 | Defocus or Power |
| 2 | +2 | 4 | ρ² cos 2θ' | Astigmatism X |
|   | -2 | 5 | ρ² sin 2θ' | Astigmatism Y |
| | +1 | 6 | $(3\rho^2 - 2)\rho \cos\theta'$ | Coma X |
| | -1 | 7 | $(3\rho^2 - 2)\rho \sin\theta'$ | Coma Y |
| | 0 | 8 | $6\rho^4 - 6\rho^2 + 1$ | Primary Spherical |
| 3 | +3 | 9 | $\rho^3 \cos 3\theta'$ | Trefoil X |
| | -3 | 10 | $\rho^3 \sin 3\theta'$ | Trefoil Y |
| | +2 | 11 | $(4\rho^2 - 3)\rho^2 \cos 2\theta'$ | Secondary Astigmatism X |
| | -2 | 12 | $(4\rho^2 - 3)\rho^2 \sin 2\theta'$ | Secondary Astigmatism Y |
| | +1 | 13 | $(10\rho^4 - 12\rho^2 + 3)\rho \cos\theta'$ | Secondary Coma X |
| | -1 | 14 | $(10\rho^4 - 12\rho^2 + 3)\rho \sin\theta'$ | Secondary Coma Y |
| | 0 | 15 | $20\rho^6 - 30\rho^4 + 12\rho^2 - 1$ | Secondary Spherical |
| 4 | +4 | 16 | $\rho^4 \cos 4\theta'$ | Tetrafoil X |
| | -4 | 17 | $\rho^4 \sin 4\theta'$ | Tetrafoil Y |
| | +3 | 18 | $(5\rho^2 - 4)\rho^3 \cos 3\theta'$ | Secondary Trefoil X |
| | -3 | 19 | $(5\rho^2 - 4)\rho^3 \sin 3\theta'$ | Secondary Trefoil Y |
| | +2 | 20 | $(15\rho^4 - 20\rho^2 + 6)\rho^2 \cos 2\theta'$ | Tertiary Astigmatism X |
| | -2 | 21 | $(15\rho^4 - 20\rho^2 + 6)\rho^2 \sin 2\theta'$ | Tertiary Astigmatism Y |
| | +1 | 22 | $(35\rho^6 - 60\rho^4 + 30\rho^2 - 4)\rho \cos\theta'$ | Tertiary Coma X |
| | -1 | 23 | $(35\rho^6 - 60\rho^4 + 30\rho^2 - 4)\rho \sin\theta'$ | Tertiary Coma Y |
| | 0 | 24 | $70\rho^8 - 140\rho^6 + 90\rho^4 - 20\rho^2 + 1$ | Tertiary Spherical |
| 5 | +5 | 25 | $\rho^5 \cos 5\theta'$ | Pentafoil X |
| | -5 | 26 | $\rho^5 \sin 5\theta'$ | Pentafoil Y |
| | +4 | 27 | $(6\rho^2 - 5)\rho^4 \cos 4\theta'$ | Secondary Tetrafoil X |
| | -4 | 28 | $(6\rho^2 - 5)\rho^4 \sin 4\theta'$ | Pentafoil X |
| | +3 | 29 | $(21\rho^4 - 30\rho^2 + 10)\rho^3 \cos 3\theta'$ | Pentafoil Y |
| | -3 | 30 | $(21\rho^4 - 30\rho^2 + 10)\rho^3 \sin 3\theta'$ | Secondary Tetrafoil X |
| | +2 | 31 | $(56\rho^6 - 105\rho^4 + 60\rho^2 - 10)\rho^2 \cos 2\theta'$ | Secondary Tetrafoil Y |
| | -2 | 32 | $(56\rho^6 - 105\rho^4 + 60\rho^2 - 10)\rho^2 \sin 2\theta'$ | Tertiary Trefoil X |

| | | | |
|---|---|---|---|
| +1 | 33 | $(126\rho^8 - 280\rho^6 + 210\rho^4 - 60\rho^2 + 5)\rho \cos\theta'$ | Tertiary Trefoil Y |
| -1 | 34 | $(126\rho^8 - 280\rho^6 + 210\rho^4 - 602 + 5)\rho \sin\theta'$ | Quaternary Astigmatism X |
| 0 | 35 | $252\rho^{10} - 630\rho^6 + 560\rho^6 - 210\rho^4 + 30\rho^2 - 1$ | |

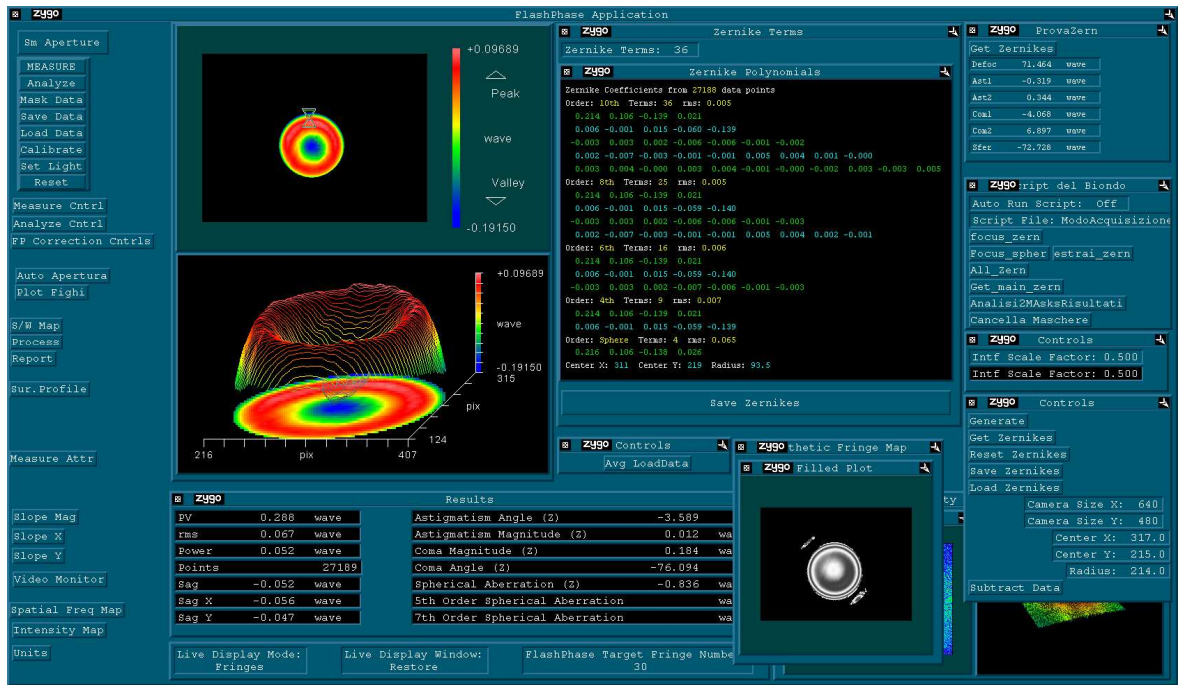

*Figure 57. The Graphics User Interface of Metro Pro software of the Zygo interferometer, in this case, shows the analysis of L3 lens under the reference beam with an aperture of 20mm.*

For the correct analysis of the aberration, the interferometer applies an aperture over the lens diameter. By default, the aperture was placed in the center with respect to the lens center, and the diameter was chosen by 100% or 95% of the lens diameter, or less for example in the GSE case. A test to measure the accuracy of the interferometric analysis was carried out in section 2.6.3. To display the wavefront, we used python modules from PRYSM that fit also the aberration measured with the interferometer, see Figure 60, without the necessity to use the MetroPro software of Zygo interferometer.

The dimension of the aperture affects the focus position in the TOU, and we could calculate the precise focal position along the z-axis. To assess the impact of the mask diameter and positioning, we made a simulation figuring out the error in the z-axis position due to a diameter error of the beam of 300 μm, and to a decenter of 500 μm. In Figure 58, we can see that the maximum error for an aperture of 22mm (GSE case) is 20 μm for the L2 lens. In Figure 59 are shown the errors for a decentering of the beam, this varies with the square of the decentering amount and, in this case, the most sensitive lens is L5 that has an error in the z-axis of about 8µm. These results demonstrate that the system is more sensitive to errors in the diameter of the beam to be analyzed than in decenter, and this will be considered in the error budget.

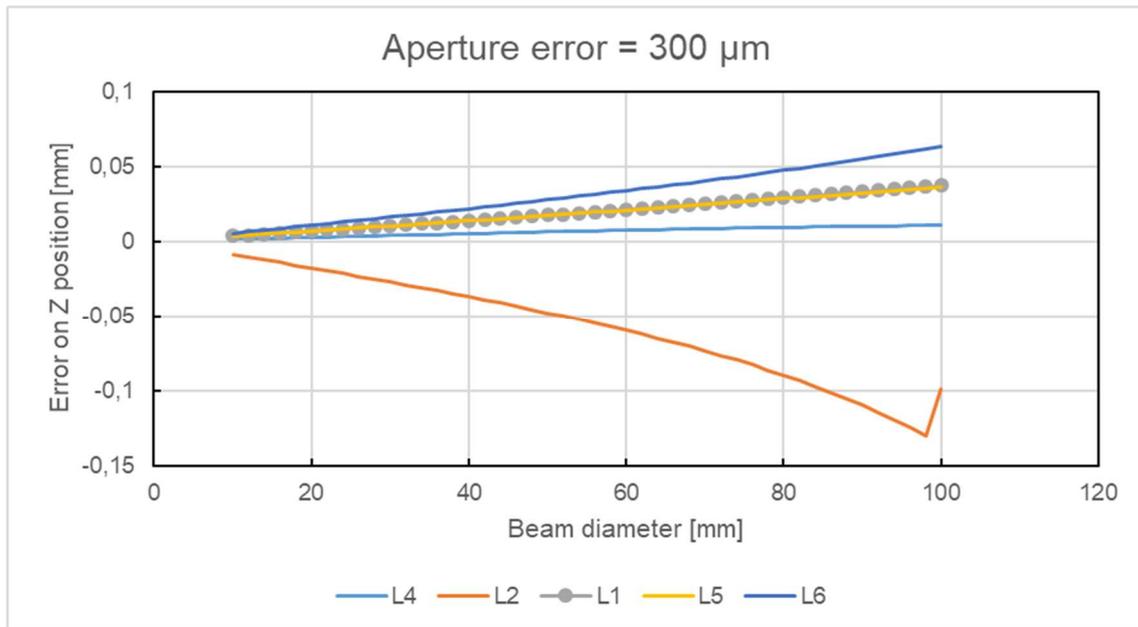

*Figure 58. The simulation figuring out the error in the z-axis position due to a diameter error of the beam mask of 300 µm.*

To measure the best focus of L3 we acquired a sweep in focus measuring the Zernike power term. The defocus zero-position is found by the linear least-squares fitting, both for orientation 0 deg and 180 deg. L3 was illuminated with a beam of 22mm in diameter, and it was measured by a double pass interferometer test adding a spherical mirror below L3. Figure 61 shows the results of the sweep.

Figure 60 shows an example of interferograms analysis by using the software PRYSM, applied to L3 after the alignment.

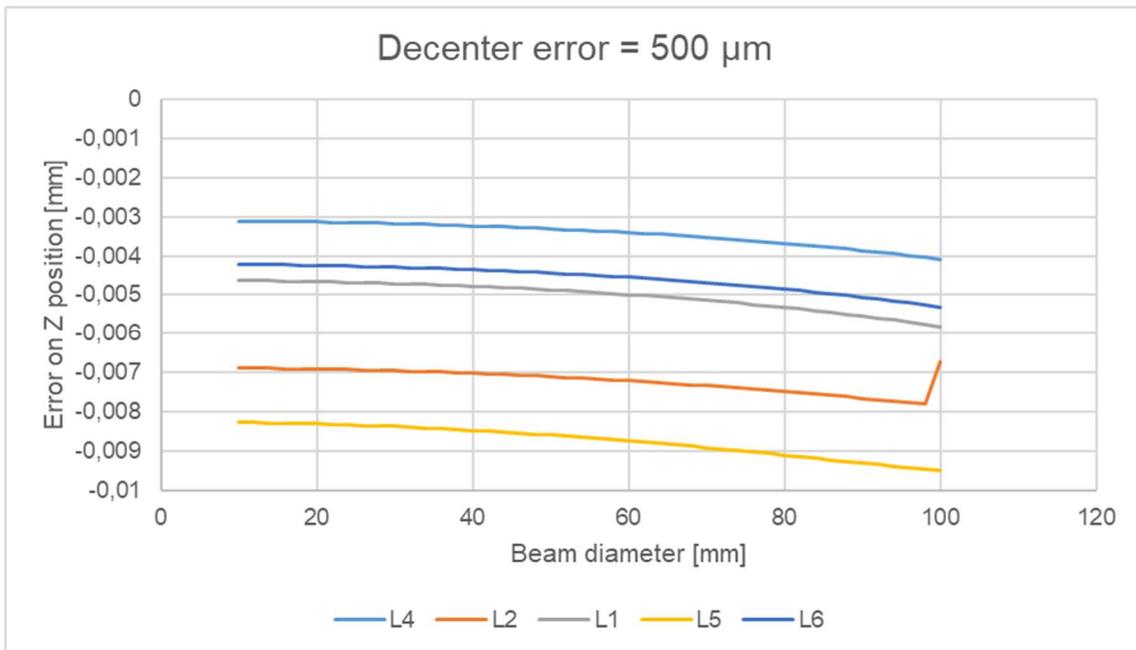

*Figure 59. The simulation figuring out the error in the z-axis position due to a decentering of the beam mask of 500 μm.*

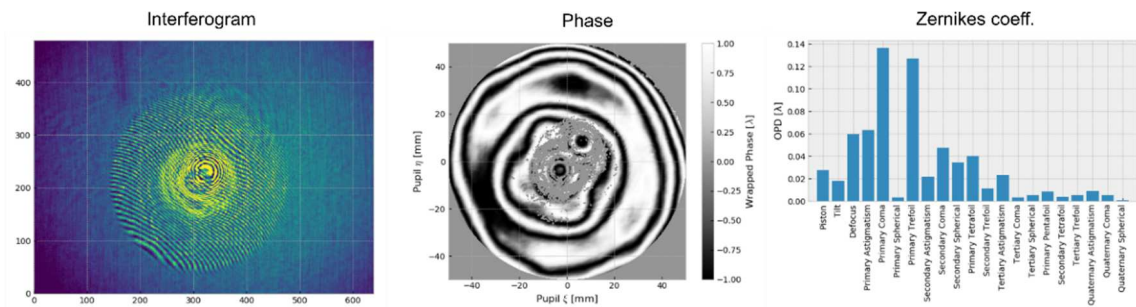

*Figure 60. One example of interferograms analysis from PRYSM for L3 alignment in direction 180 degrees.*

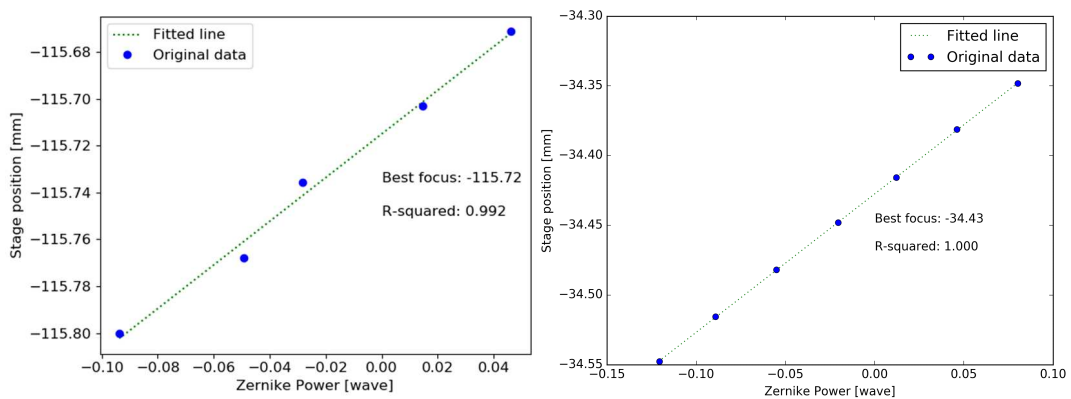

*Figure 61. The stage position for the best focus of L3 illuminated with a beam of 22mm and performing an interferometric test by using a spherical mirror below the L3 lens. We acquired a sweep in focus measuring the Zernike power term and found defocus zero-position by the linear least-squares fitting. Left L3 at orientation 0 deg, right L3 orientation 180 deg.*

## 2.5 Opto-mechanical bench

In this section, we describe the setup used for the alignment. In an optical bench, we did install an interferometer, a laser beam system, the various GSEs needed, two motorized translation stages, and a Coordinate Measuring Machine (CMM). We recall all the main steps of the procedure, describing the optomechanical setup which has been used, the lenses assembly, the alignment and the main test which have been performed. Figure 63 shows a concept of the optical bench setup, while Figure 64, shows the overall optical bench. All the optical mounts of the bench are characterized by 2 connection points, as shown in Figure 62. This choice has been performed after some preliminary tests on the bench setup stability.

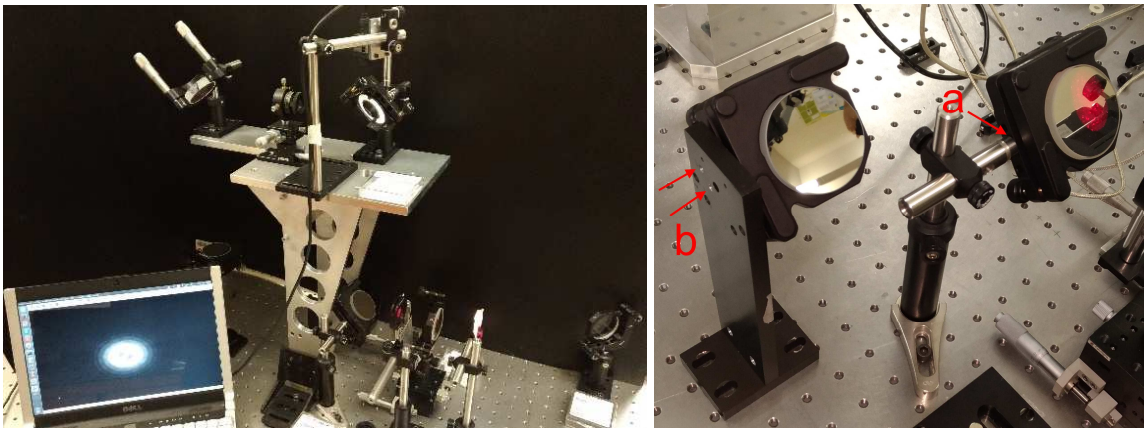

*Figure 62. On the left: the preliminary mechanical mount of the BB rebuild to check the final bench configuration. On the right: the old mount with one fixed optics mounting point [a] and the final choice with a minimum of two fixed mounting points [b].*

The laser system is composed of the laser itself, a beam expander, and an iris, creating a reference beam of the size of about 1 mm in diameter. FM1 reflects the beam coming from the laser system toward the upper part of the optomechanical setup, where a beamsplitter divides the beam into two parts: the reflected one is going toward FM3, while CCD-3 collects the transmitted beam to monitor the movements of the reference beam during the alignment.

FM3 is folding the light toward the lenses to be aligned, and CCD4 is used to check the transmitted beam, while CCD2 is used to check the light back-reflected from the lenses.

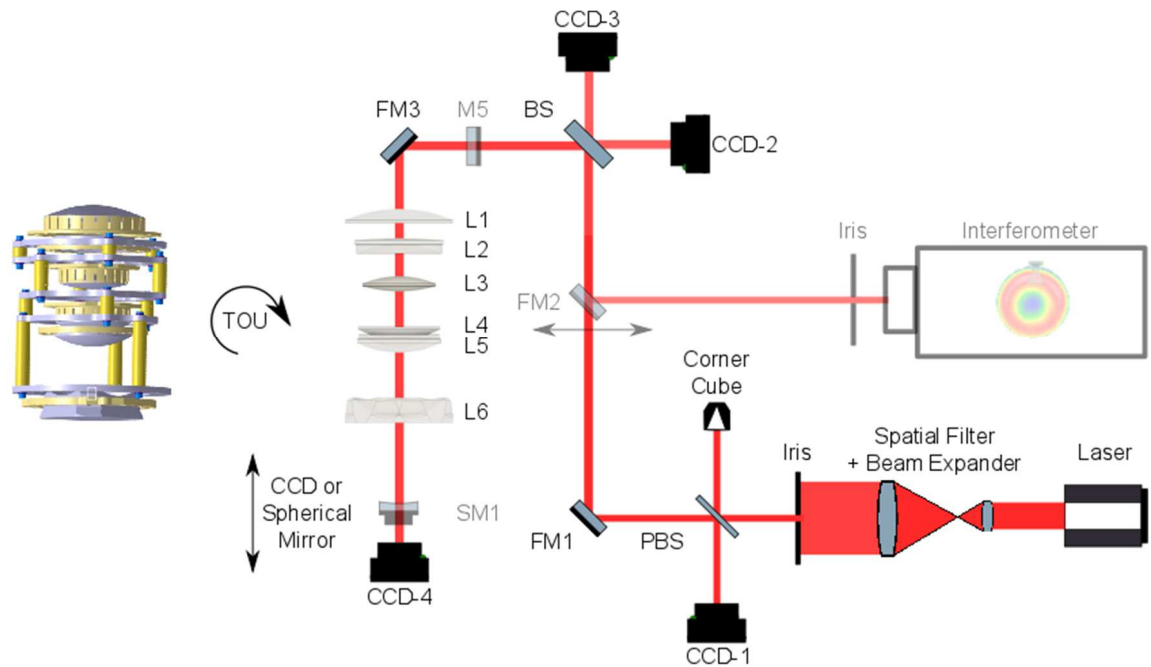

*Figure 63. The optical bench to perform the alignment. The folding mirror FM2 is mounted in a precision motorized translation stage and was inserted to select light coming from the laser or the interferometer. The M5 mirror was used to create a reference spot in the CCD-2. The motorized translation stage below the TOU moves in z-axis the CCD-4 or the spherical reference mirror for interferometric tests.*

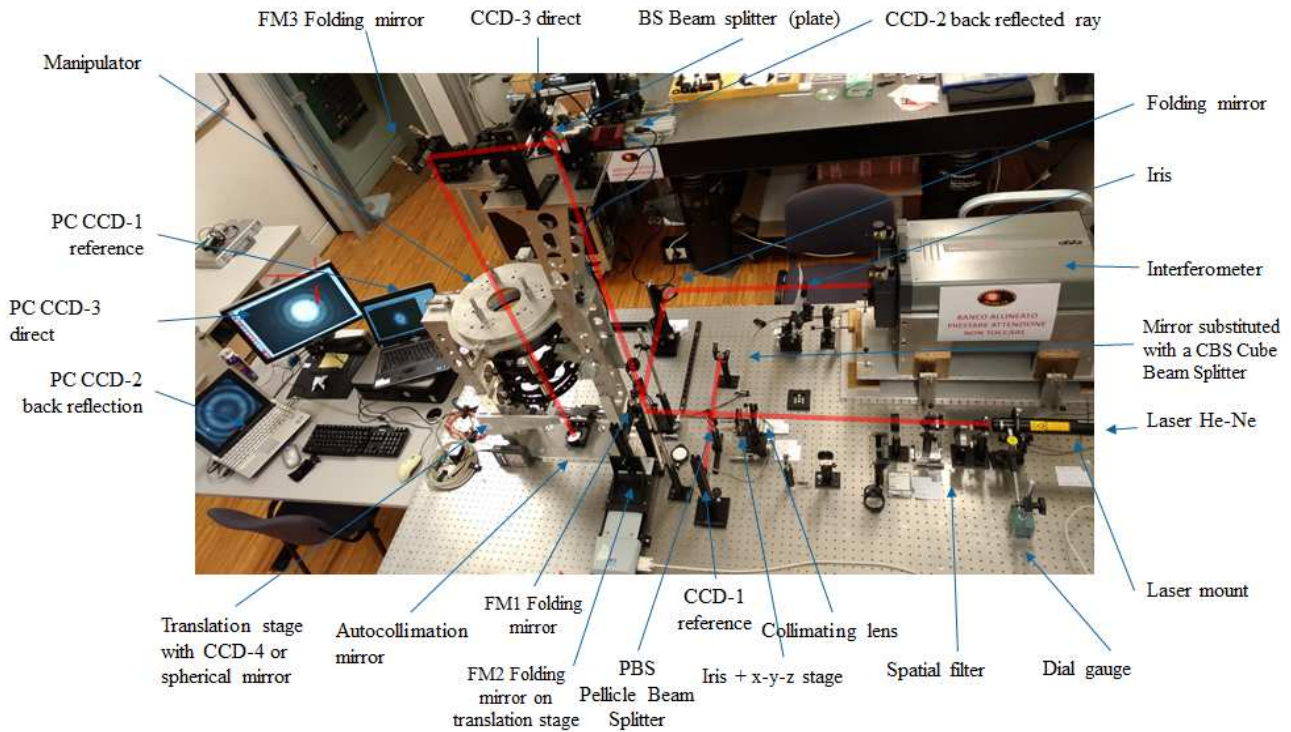

*Figure 64. The image of the optical bench and the GSE in the center. The laser ray coming from the right, an interferometer is co-aligned with the laser ray. By moving the FM2 folding mirror, mounted in a precise translation stage, the GSE catches the light from the laser or the interferometer, see the text description.*

## 2.5.1   Thermal control of the bench

Laboratory temperature changes may affect the operations of the alignment. The coefficient of linear thermal expansion (CTE, CLTE, α, or α1) is a material property that is indicative of the extent to which a material expands upon heating. Different substances expand by different amounts. Over small temperature ranges, the thermal expansion of uniform linear objects is proportional to temperature change. The CTE coefficient is defined as the change in length or volume of a material for a unit change in temperature, and it follows this formula $\alpha \equiv \frac{1}{L}\frac{dL}{dT}$ where $\alpha$ is the thermal expansion coefficient [1/K], L the length [m] and T the temperature [K]. Figure 3 shows the variation of the CTE with respect to the temperature for the AlBeMet alloy, used for the mount of PLATO, in comparison with aluminum and stainless steel. A change of one degree will cause (over 1 m length of an optical table made in stainless steel) a variation of its length of about 16 μm, and an equivalent variation of the length of the GSE (made in aluminum) of 22 μm. Due to the tolerances of optical alignment, it is very important to check accurately that the temperature of the materials and the air in the laboratory remains stable during the AIV operations.

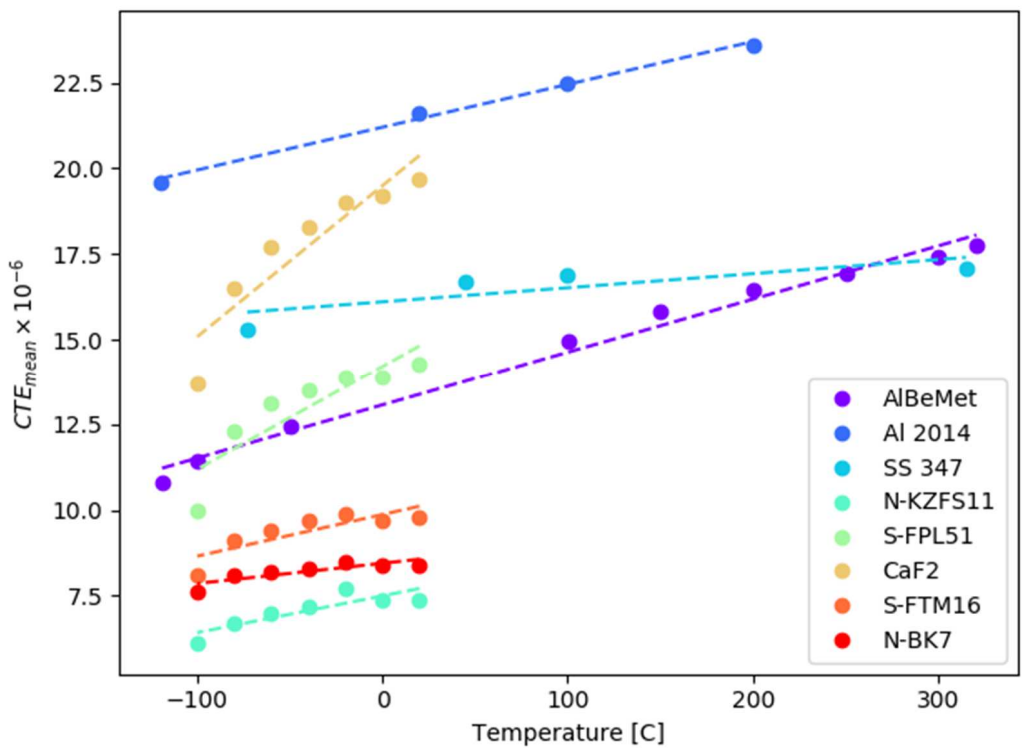

*Figure 65. The CTE characterizing the various materials of PLATO. Courtesy PLATO Consortium.*

We have placed several temperature sensors with an accuracy of ±0.15 °C, for monitoring the stability of the bench, of the laser, and of the GSE. First of all, we calibrated the signal of the four sensors (PT-100) by using a mixture of water and ice, a stable reference for zero degrees, by measuring the temperature during the time for the complete melting of the ice. So we apply a correction to the values read by the sensors to obtain the same value for all. After calibration, the value measured in the mixture of water and ice shows a PtV of 0.05 °C, six times lower the accuracy of the sensors.

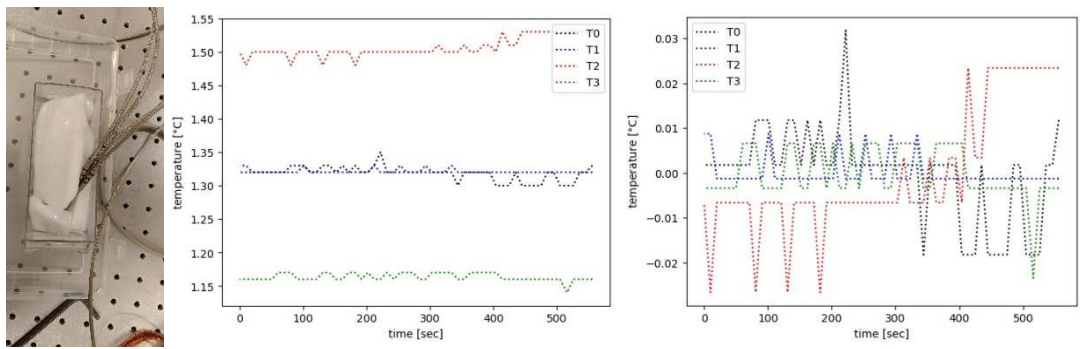

*Figure 66.The probes in the mixture of water and ice (on the left), the plot of the different temperature before the calibration (on the center), the residual after applying the correction (on the right)*

The probes are then positioned in a different part of the GSE, to help the temperature control of the laboratory. In parallel, we use a thermo-hygrometer PCE Instrument PCE-HT 110 with datalogger to control the air temperature of the laboratory with a resolution of 0.1 %RH  for the humidity reading, 0.1 degree resolution for the temperature reading.

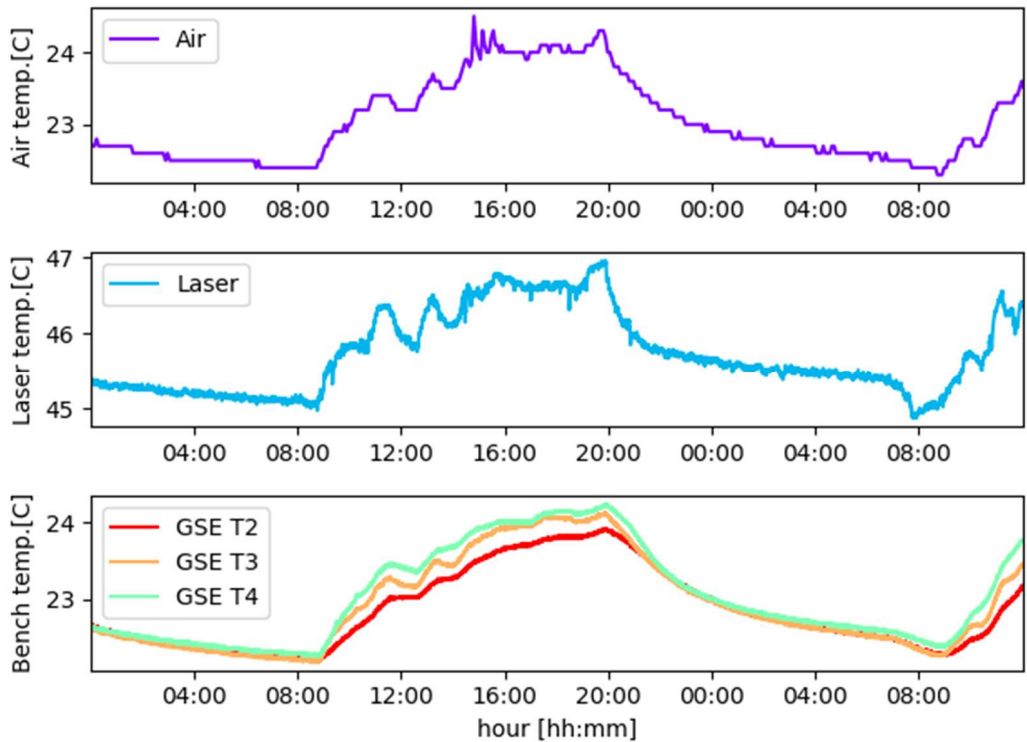

*Figure 67. The typical variation of the temperatures of the laboratory air, laser ad GSE over two days.*

Figure 67 shows the typical variation of the temperatures of the laboratory air, laser ad GSE over two days, showing the cycle day/night of the laboratory due to the conditioning system. The range of temperatures is always below two degrees. Figure 68 shows the typical value of the temperatures during an afternoon session and Table 11 resume that the PtV of all temperature is below 1 degree, variation small enough to ensure that no large variation due to the thermal expansion of the optomechanical components.

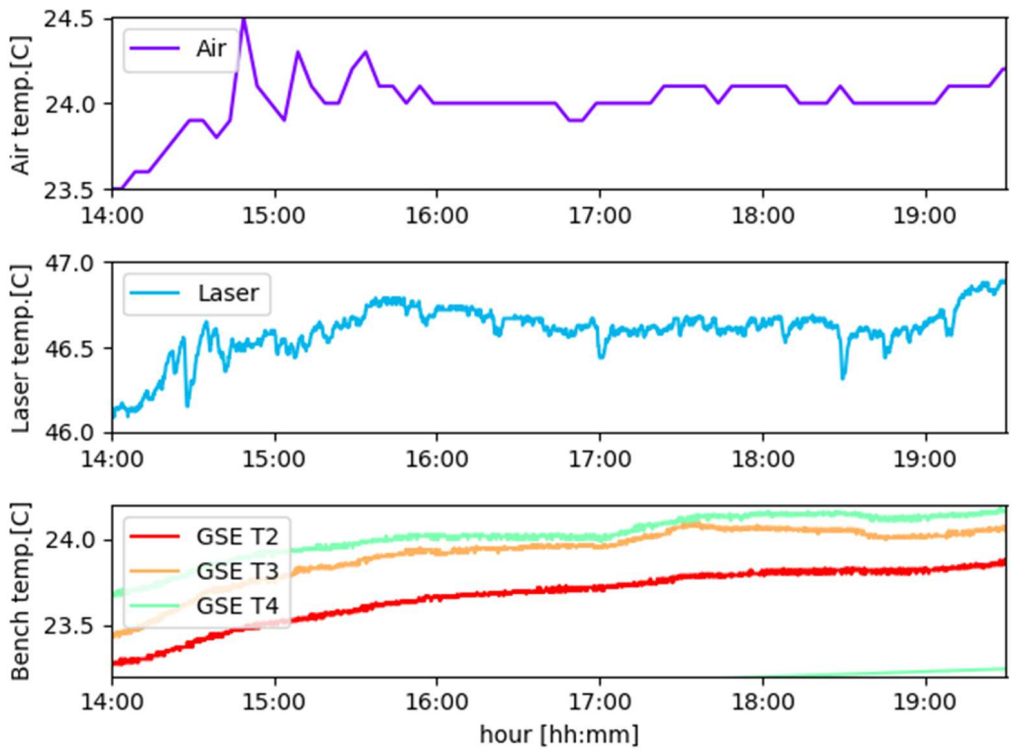

*Figure 68. The temperature during a normal work session in the optical alignment.*

*Table 11.The mean value and PtV of laboratory temperature during a working session.*

| Celsius | Air | Laser | GSE-T2 | GSE-T3 | GSE-T4 |
|---------|-----|-------|--------|--------|--------|
| **Mean** | 24,0 | 46,6 | 23,7 | 24,0 | 23,9 |
| **PtV** | 1,0 | 0,8 | 0,6 | 0,5 | 0,6 |
| **Dev.St.** | 0,2 | 0,2 | 0,2 | 0,1 | 0,2 |

To control the real thermal dependence of temperature in the lens positioning, we decide to measure the variation of the defocus with the temperature. We used a fizeau interferometer, a Zygo model, used in configuration with a lens system focused on infinity.

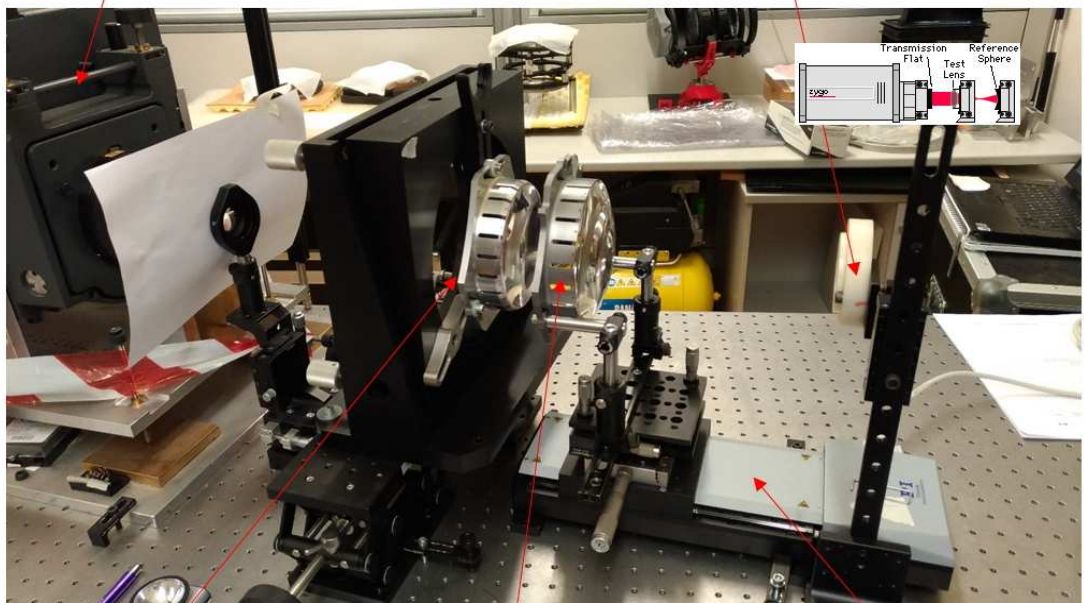

Interferometer with flat element    spherical mirror on a x-y-z mount, d = 22mm, R = 150mm

PLATO 1.0
Prototype L3 lens
Tip-tilt align error
≈ 160 arcsec

PLATO 1.0
Prototype L4 lens
Tip-tilt align error
≈ 130 arcsec

Motorizer linear stage
Tip-tilt align error
≈ 430 arcsec

*Figure 69. The interferometric test to evaluate the dependency in temperature of the defocus in a lens system.*

The interferometer software analyzes and plots all the Zernikes aberration coefficients, even if the power coefficient is the one we care about. For the test described in Figure 69, we used the PLATO 1.0 prototype L3 and L4 lenses, positioning them in front of the interferometer. The fizeau interferometer creates a flat reference beam that runs through the two lenses, passes the focus of the combined lens system and is back-reflected by a spherical reference mirror with a surface error of $\lambda/20$. The L3 lens is mounted in a tip-tilt mount to properly align it. By minimizing the defocus term, the focal position of L3 is characterized with an accuracy below 5 µm. L4 lens is mounted in a tip-tilt-x-z mount to properly align it with respect to L3. We did align the spherical mirror do minimize the defocus term gets from the interferometric analysis, and we performed some series of measurements over three days.

During the first test, we take a sweep through the focus position of the lens measuring the relative defocus term. We evaluated the slope of defocus power to L4 position in three different moments of the day with five different iterations, corresponding to three different temperature of the laboratory, see Figure 70. The resulting slope is -0.00082 [wave]/μm.

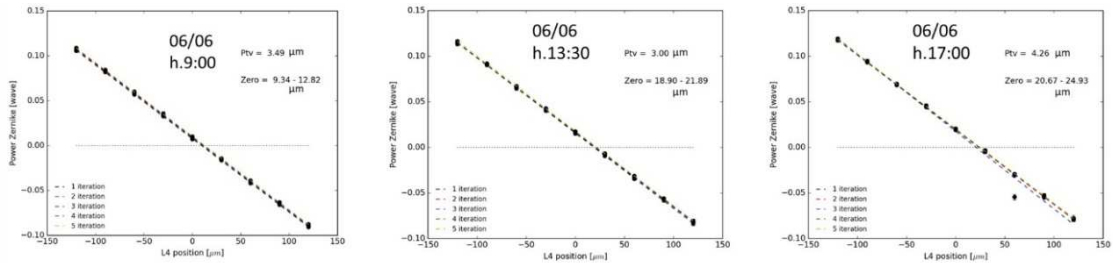

*Figure 70. The three different determination of the slope of focus power to L4 lens position.*

The second test measures continuously the changing of the defocus power and the air temperature of the laboratory for a set of 3 steps of three days. During this test, we don't move the optical bench parts. In Figure 71, we report only a set of measurements, on the left plot it is shown the temperature variation during days, while in the middle, the defocus term variation. The two signals are perfectly correlated, and the averaged slope is 0.0173 [wave]/°C.

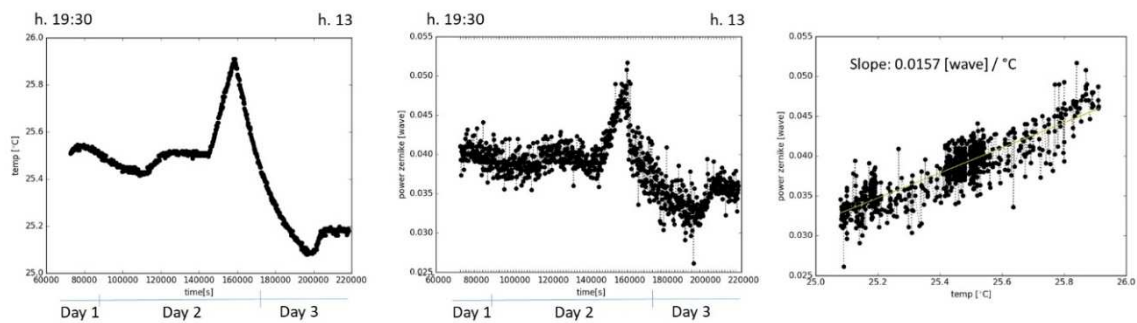

*Figure 71: On the left: the time variation of the temperature. On the middle: time variation of the defocus term. On the right: correlation between the previous data.*

Finally, we estimated the slope and direction of the thermal expansion on Z-axis to be 21 μm/°C, over two weeks of measurements. From this value, we have a constraint of one-degree C in the thermal control of the optical laboratory to improve the stability of the TOU setup. We assume that this expansion value don't change for all lens mount in the TOU due to low CTE of the AlBeMet alloy.

## 2.5.2   Laser stability

The He-Ne laser is a JDS Uniphase at 633nm with stability below 0.02 mrad, beam diameter of 0.68mm, beam divergence of 1.2 mrad, maximum power of 20mW, and tube length of 490mm. The laser tube is mounted in a stable mount with two fixed points, each composed by a three-screw adjustable ring mount and was monitored by two dial gauges, see Figure 72.

Two dial gauges (2μm of sensitivity) control tip-tilt movements of the laser, and they are placed toward the case ends. Monitoring the position of the gauges during a normal session of work, from morning to evening, reveals tip-tilt of the laser tube for thermal change with a typical pick to valley (PtV) of about 4 arcsec, considering also warm-up time and turn off transient of the laser. In the left part of Figure 73, the first two points corresponding to the warm-up time of the laser and thus they could be skipped. The plot shows that the laser position is stable, below the resolution of the gauge, in 70% of the working time of ~10 hours. We also perform a time series of images of the laser spot on a CCD with 5.2 μm pixel size for 10 hours and analyze the displacement located 1 meter away from the laser. The resulting displacement, computed with a Gaussian filter on the image before the centroid analysis, has a PtV of 7.3 μm. This corresponds to a tilt of the laser of 1.5 arcsec, which is in good agreement with the measurements performed with the dial gauges.

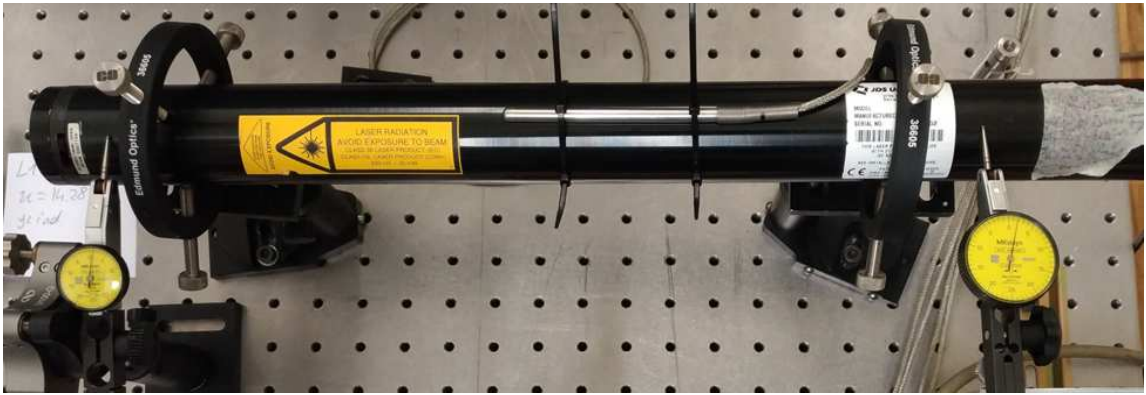

*Figure 72. The dial gauges control the position of the He-Ne laser, in the middle, the temperature sensor of the laser.*

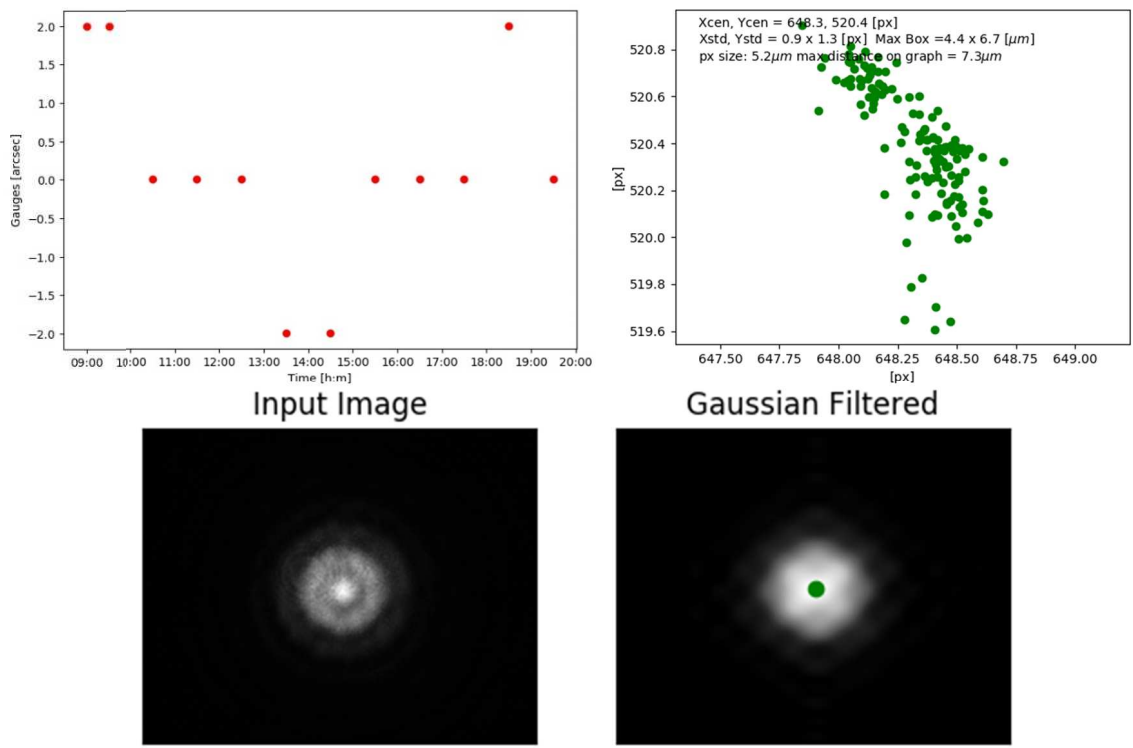

*Figure 73. On the left-top: a typical distribution of the values of gauges in arcsec. On the right-top: the stability of the laser beam. On the bottom: the image of a spot laser before the analysis (on the left) and after applying a Gaussian filter before finding the centroid of the spot (right).*

### 2.5.3   Spatial filter and beam expander

The laser beam is sent to a spatial filter and a beam expander, in a way to homogenize in the intensity profile and reduce the divergence of the beam. The beam expander is composed of a microscope objective lens with a focal length of 4.5mm and an NA of 0.65, which produces a spot of 7 µm in diameter. The spatial filter uses a pin-hole of 5 µm to ensure the correct spatial frequencies selection.

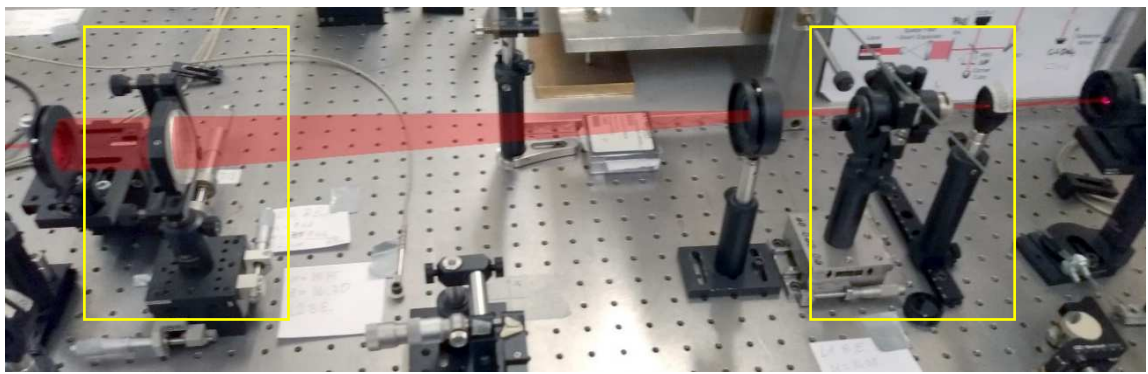

*Figure 74. On the left: the collimating lens in x-z-tip-tilt mount. On the right: the spatial filter in precise x-y mount.*

The second lens is an achromatic doublet with 500mm of focal length, and this ensures that the spot is diffraction limited seeing by the collimating lens and that the beam expansion is near 70x, and a consequent decreasing in beam divergence of the same value. The collimation is tuned by a collimation tester, the Newport 20QS20, and the minimum radius of curvature of the beam is ensured to be below 1000 m. An iris selects a beam of collimated light of ~1 mm in diameter.

### 2.5.4 Laser beam reference

A Pellicle Beam Splitter (PBS) reflects the ray toward a Corner Cube (CC) mirror and images it on the CCD-1 with 1.67 µm pixel size, see Figure 75. The spot realizes a reference for the laser position before the illumination of the GSE. We test the stability of the laser reflected by the PBS and after the PBS plus the CC, locating the CCD-1 1 meter away to the optics elements. For a time of 12 hours, the results are according to the stability of the laser head, evaluated the tilt in about 1.7 arcsec for the PBS and around 3.4 arcsec for the PBS+CC. In this second path we recognize a differential movement of the spot, due to double reflection, and a fast-displacement of the spot in the first ten images, corresponding to the starting image's sequence and depending on thermal stabilization[1] of CCD-1. This effect is also present with the CCD-2 of 4.65 µm pixel size as shown in Figure 76.

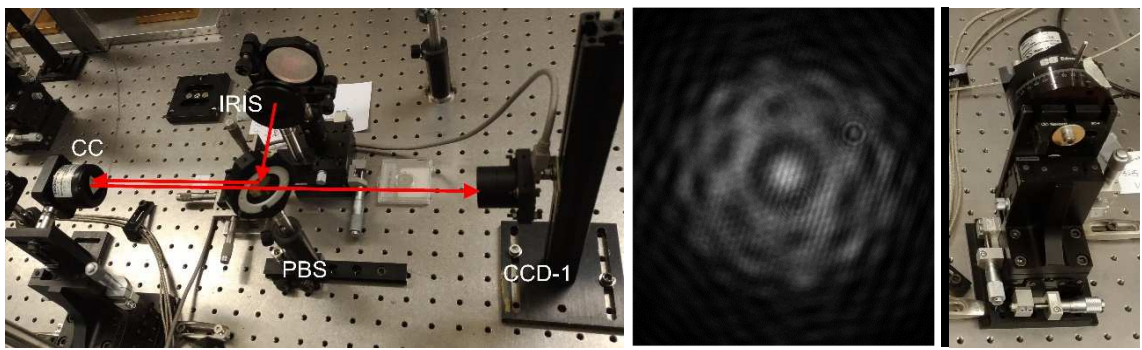

*Figure 75: On the left: the laser beam divided by the pellicle beam splitter reflected in the CC and back reflected on CCD-1. In the middle: the resulting spot shape on the CCD-1 after CC reflection. On the right: the CC mount in precise x-y-z and rotation mount.*

---

[1] The CCD-1 is a naked chip without fan or cooling, model IDS UI-1492LE-M pixel size of 1.67µm, resolution 3840 x 2748 px. During normal session the image is continuous reading, so the electronic board is heating. The time series of images is composed by a shot each five minutes, the electronic board between two images is not working and not generate heat. The passage from continuous reading and time series reading necessities of around 45 minutes to reach thermal stabilization and this explain the drift in the centroids, a drift of the electronic during a cooling phase.

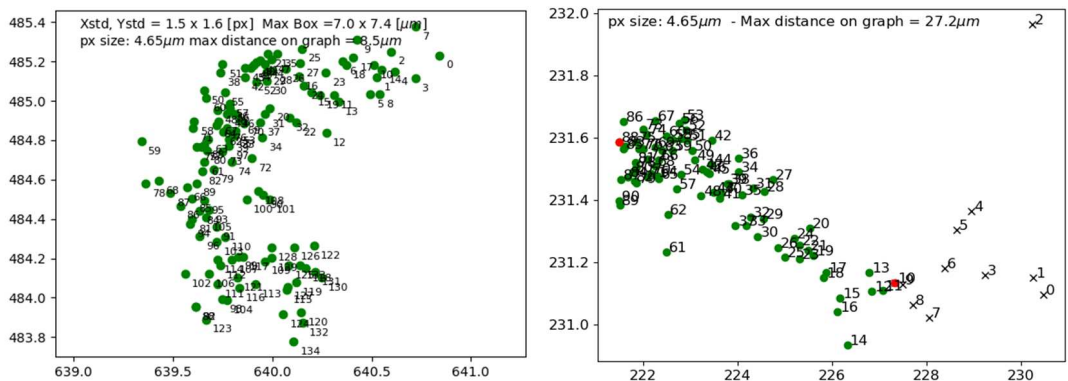

*Figure 76: On the left: stability over 12h for the ray reflected by the PBS only. On the right: stability over 7h of the path PBS + CC, we see the trend of the point in the lower part due to thermal stabilization of the CCD-2, the black dots are skipped, the green dots are used in the analysis, see note 1 page 80.*

## 2.5.5 Optical components adjustments

In the optical bench, we decided to move the key part with a micrometer or precise motorized stages. Any components with movable parts potentially introduce an issue for stability. For this reason, we used only micrometers with locking screw and stage with servo motors with brakes. We resume in Figure 77 the changes of the beam direction in the different parts of the bench. All CCDs are fixed, and the arrows represent the direction of the coordinate in the image, only CCD4 is movable in the z-axis. The movable parts are:

- the iris that is located after the laser system;

- the CC to be aligned with respect to the reference axis to reach a symmetrical image on CCD1;

- the folding mirror FM3 can be tilted with servo motors and translated by hand along the z-axis;

- the adjustable mirror M5 can be tilted to auto-collimate in CCD1 the chief ray.

To tune the z position of the lenses in the optical path, we can insert the beam coming from the interferometer by using two folding mirrors (see Figure 78): FM4 and FM2 movable along the z-axis. FM2 was mounted in a precision motorized translation stage to be inserted and removed from the optical path with high repeatability. The iris in front at the interferometer is adjustable in x,y to allow reaching on CCD-3 (control for decentering) and CCD2 (control for tip-tilt) the same position of the reference beam coming from the laser.

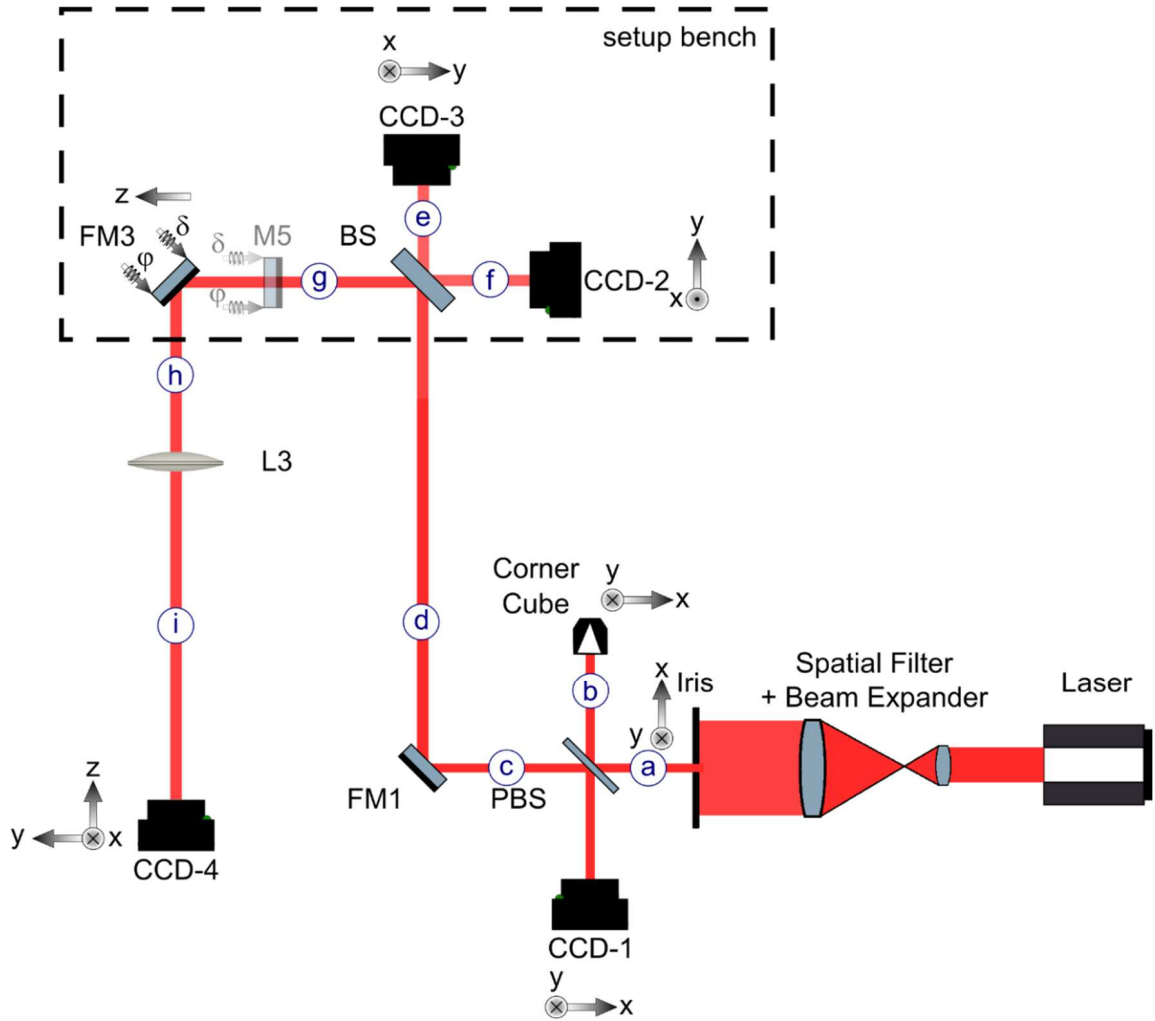

Figure 77. The alignment beam coming from the laser is shown in the diagram. The length of the arms are: a=20mm, b=480mm, c=200mm, d=1030mm, e=80mm, f=120mm, g=310mm, h=640mm (@ 0 deg) 570mm (@180 deg), i=300mm (@ 0 deg) 370 (@180 deg). The highlighted part is the small bench in the upper part of the GSE, also shown in Figure 61.

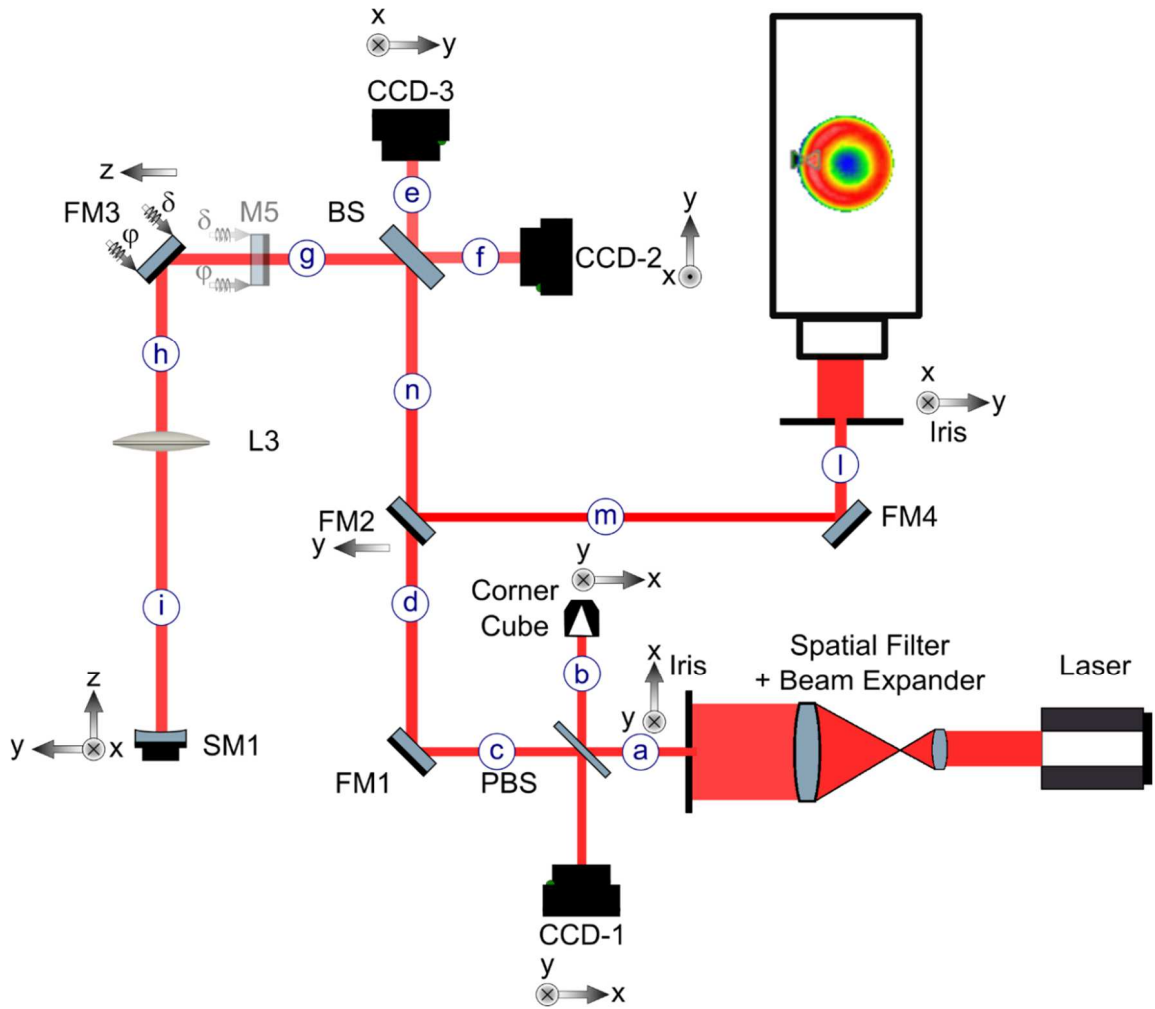

*Figure 78. The alignment beam coming from the interferometer. The length of the arms: a=20mm, b=480mm, c=200mm, d=110mm, e=80mm, f=120mm, g=310mm, h=640mm (@ 0 deg) 570mm (@180 deg), i=375mm (@ 0 deg) 445 (@180 deg), l=470mm, m=410mm, n=920mm. The selection between the beam coming from the laser or the one coming from the interferometer is done by inserting or removing FM2.*

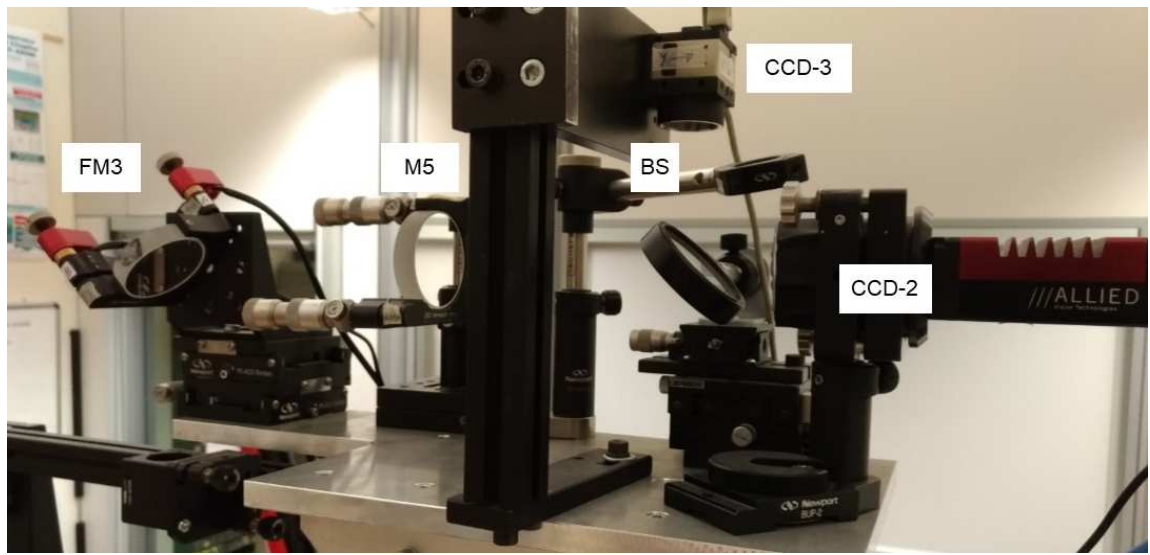

*Figure 79. The setup bench in the top of the GSE; the mirror M5 was mounted in the magnetic base for fast repositioning.*

We recall that, in the baseline procedure, the lenses position along the z-axis was foreseen to be performed with the interferometer. To do this, it is necessary to place a spherical reference mirror below the lens, in the correct position to perform a double pass interferometric measurements. The spherical reference mirror has been mounted on an Aerotech® PRO165SL translation stage, with a travel range of 150 mm, standard accuracy of ±6 µm, bidirectional repeatability of ±0.5 µm, and low pitch-roll-yaw of 6 arc seconds.

The Aerotech® translation stage had, of course, to be aligned with the optical beam, the laser beam, in the two possible Prototype orientations. Since the alignment has been performed by shimming, we decide to shim the stage in a way to move on an axis in between the two axis corresponding to the two prototype orientations, being the tilt difference very small (of the order of 30"), with essentially no impact on the position of the spherical mirror and, consequently, no impact on the interferometric measurements to tune the lenses focus.

## 2.6  Error budget

The optomechanical setup used for the Prototype AIV has been explained in the previous sections. It is mostly deployed on a single optical bench holding both the AIV setup and the prototype GSE, and it requires a certain number of GSE and optomechanical elements. Before starting the Prototype AIV, a full characterization of the AIV setup has been performed, has partly already explained, with the purpose to identify the accuracy of the alignment operations. To achieve this, we analyze the following items:

- characterize the setup stability over time;

- characterize the setup stability temperature dependence;

- characterize the mechanical repeatability and the effect on the alignment beam of a few crucial GSE (such as the rotation one - which is needed to be used during the AIV process - and the centering GSE);

- characterize the accuracy of the centroiding operations on CCD4, which is collecting the transmitted light used during the alignment process for the lenses centering adjustment;

- the tip-tilt alignment of the lenses requires that the Newton rings observed on CCD2 are uniformly illuminated, while their position on the detector is related to the lens centering (information the latter which is redundant with the transmitted spot position). For this reason, we characterized the accuracy of these operations on CCD2, which is collecting the back-reflected light used during the alignment process for the lenses tip-tilt adjustment.

In Figure 64, page 71, there is a photo of the alignment setup, in which the two channels that will be used for the alignment are visible:

- the laser one, that will be essentially used for the lenses centering and tip-tilt alignment;

- the interferometric one, that was supposed to be used for the lenses alignment in focus while, as described in section 3.2, it has not been used.

## 2.6.1  Laser's spot analysis

The laser generates a coherent beam with a divergence of 1mrad, reduced by the beam expander optics. When imaging a laser spot on a detector, the image obtained is characterized by an interference pattern. Another source of disturb is the fringing inside the CCD, which consists of a pattern of fringes generated inside the optical window of the CCD (see Figure 80) and it is superimposed to all the images.

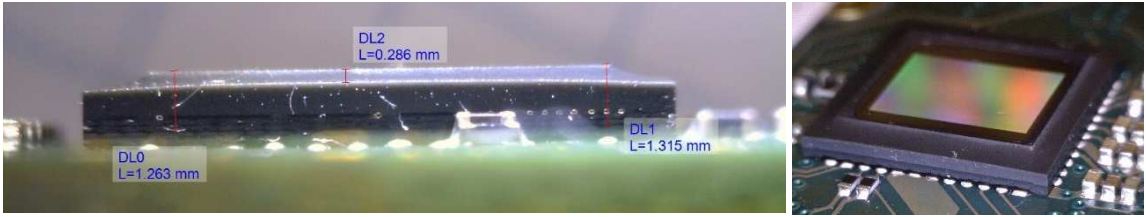

*Figure 80. The optical window of the CCD, measured in thickness in the laboratory*

In fact the temperature variation of the window, induced by the readout current, increases its thickness and consequently changes the position and the number of the fringes.

In the following subsections, we describe how we attached the problem for the 4 CCDs used during the alignment.

### 2.6.1.1 Reference spot on CCD1

Figure 81 (on the left part) shows an image of the CC reflection inside the CCD-1 that shows a typical fringe pattern. In the space of Fast Fourier Transform (FFT), the fringes are located in high frequencies range, and their position does not change during the series. Applying a mask in these zones in the Fourier space, see Figure 81 central image, in the image coming from the inverse FFT the fringes disappear, and the images are cleaned, see Figure 81 right part.

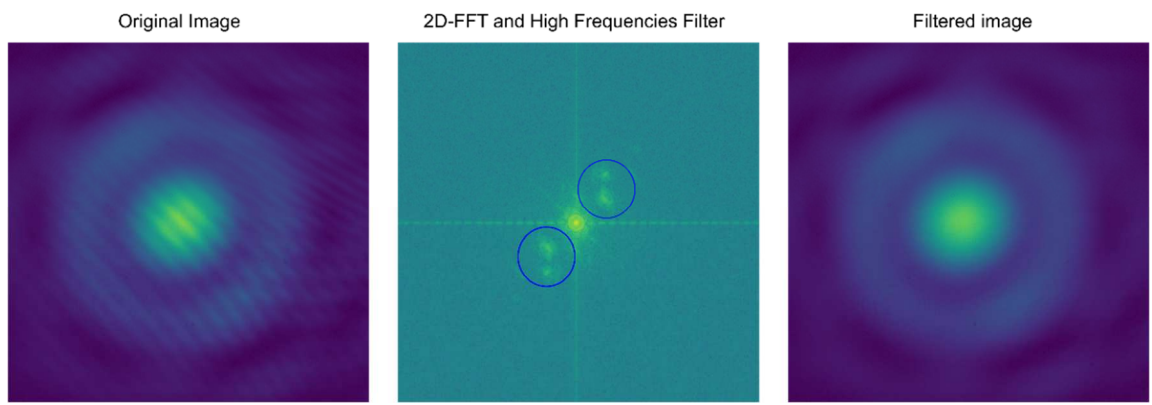

*Figure 81. On the left: fringes pattern over CCD images. Middle: FFT and the two high frequencies areas (in blue) removed. On the right: inverse FFT and images without fringes. The image is the reflection of the CC on the CCD-1.*

The final purpose is to compute a spot centroid very precisely. To achieve this, we suppose that the central part of all spots is modeled by a Gaussian function, which is the typical profile of laser beams, which is untouched when propagating through optics. To compute the centroid, we have evaluated different methods:

- finding the position of the maximum value over the pixel → poor value caused by pixel to pixel variation (fringes);

- compute the center of a circle fitting the isophotes values → if the image is not symmetrical the center is wrong;

- fitting two 1D-Gaussian over the extracted mean value along the x-axis and along the y-axis → fast computing but this could be affected by a gradient in image or fringes;

- removing the fringes by FFT and fitting with a 2D-Gaussian function → best value but with long computational time.

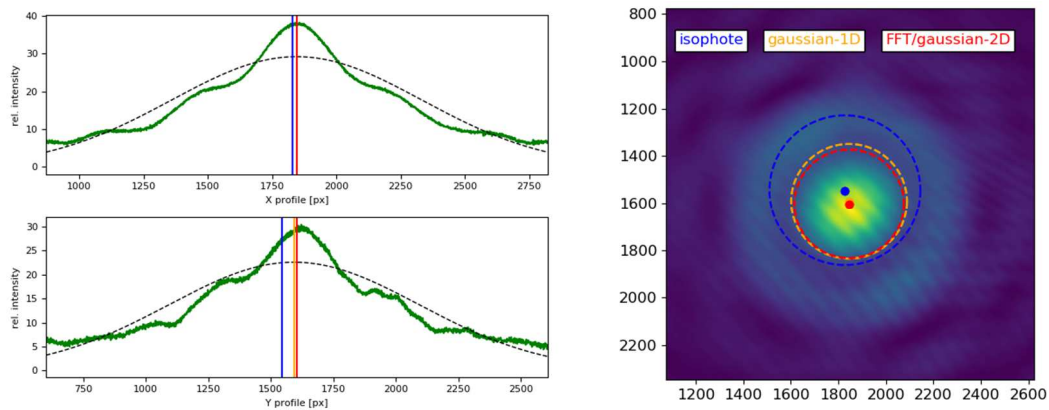

*Figure 82. On the left: mean profile along the x-axis and y-axis (green), Gaussian-1D fit (black), the position from the isophotes (blue line), from the Gaussian-1D fit (yellow line), from FFT filter and Gaussian-2D fit (red line). On the right: the same observable of the left plot but over a spot image. The image is the reflection of the CC on the CCD-1.*

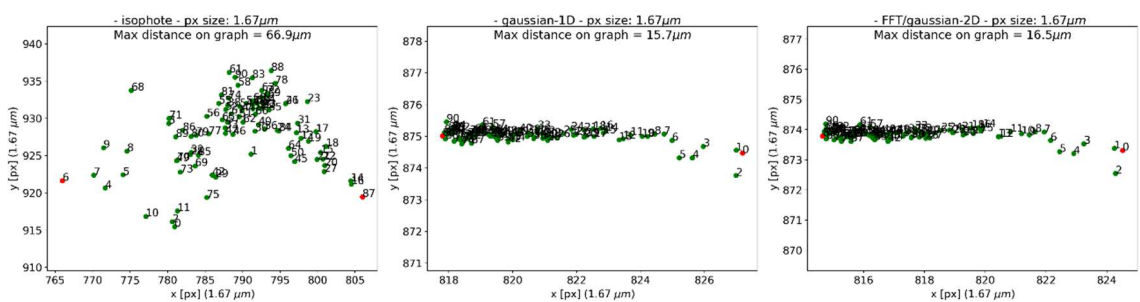

*Figure 83. Stability positions of the laser spot computed only with isophotes fitting (left), without FFT filtering and with Gaussian-1D fit (middle), and with FFT filtering and Gaussian-2D fit (right). The red dots represent the maximum pairwise distance between centroids.*

In Figure 83, we show the stability of a series of images taken in 7 hours computing the centroids by using: only isophotes fitting (left), without FFT filtering and with Gaussian-1D fit (middle), and FFT filtering and Gaussian-2D fit (right). The red dots represent the maximum pairwise distance between centroids. The plots show how the Gaussian fitting is more precise than the isophotal one, and the results of Gaussian fitting 1D and FFT/ Gaussian-2D are pretty similar. We suggest to use the FFT + Gaussian-2D method for precise measurements, i.e. for the position of the lens, and only Gaussian-1D fit for stability over time of the laser beam to reduce the computational time of a large number of files.

### 2.6.1.2   Reference spot on CCD2

CCD-2 was used to measure the tilt of the lenses by the BBR, in Figure 85 and Figure 86 is displayed all lenses BRR images and the related focus spot. We figure out the centroid position by the auto-collimation of the M5 mirror, and this defines the "Target position". After we extract the x-axis and y-axis profile from the center of a BRR image (in Figure 83 there is an example for L3). From the profile we identify the maximum and minimum position by a fitting: this allows us to define the center between the three couples of maxima. We also calculate the center derived from the isophotal fit of a circle. All values are resumed in a table below the images profile, and this procedure is working in real-time during the alignment process. Figure 84shows the panel of the python code for the L3 at the end of its alignment process.

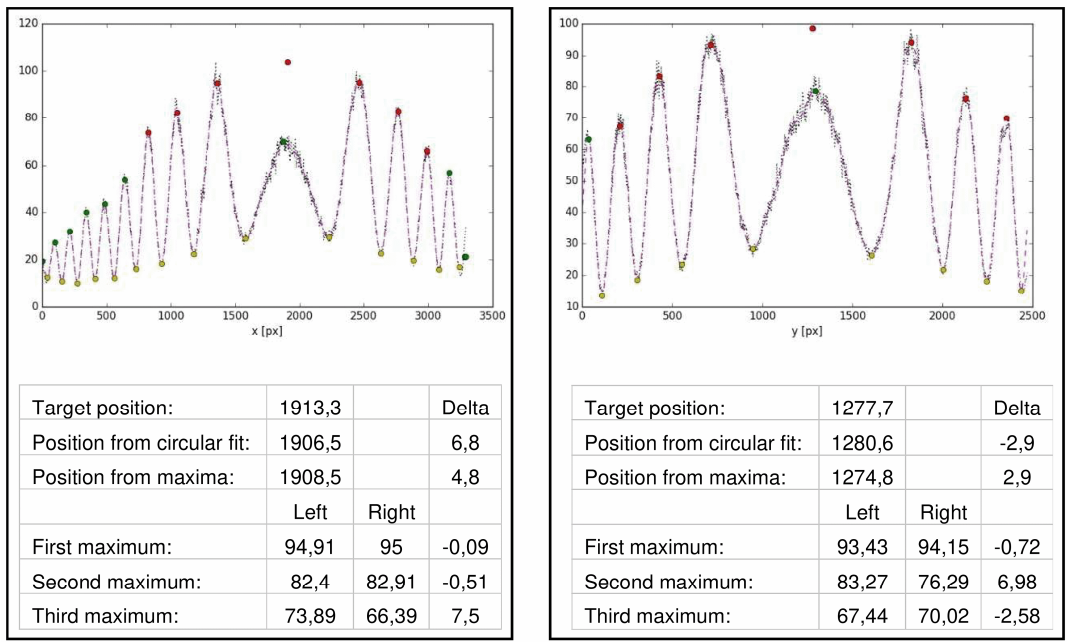

| Target position: | 1913,3 | | Delta |
|---|---|---|---|
| Position from circular fit: | 1906,5 | | 6,8 |
| Position from maxima: | 1908,5 | | 4,8 |
| | Left | Right | |
| First maximum: | 94,91 | 95 | -0,09 |
| Second maximum: | 82,4 | 82,91 | -0,51 |
| Third maximum: | 73,89 | 66,39 | 7,5 |

| Target position: | 1277,7 | | Delta |
|---|---|---|---|
| Position from circular fit: | 1280,6 | | -2,9 |
| Position from maxima: | 1274,8 | | 2,9 |
| | Left | Right | |
| First maximum: | 93,43 | 94,15 | -0,72 |
| Second maximum: | 83,27 | 76,29 | 6,98 |
| Third maximum: | 67,44 | 70,02 | -2,58 |

*Figure 84. The lenses observables defines the reference for tip-tilt and decenter by BRR.*

In Figure 85 are summarized the back reflected reference images acquired on CCD2 and the images of the transmitted beam are in merged in Figure 86.

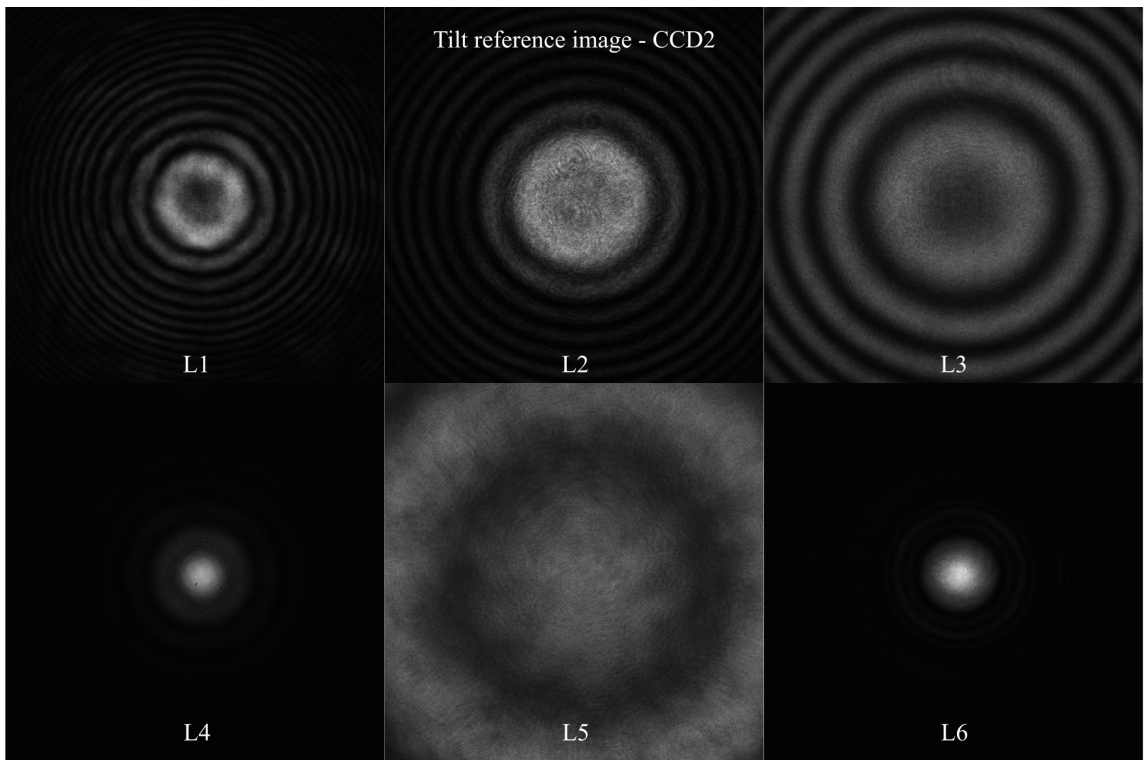

*Figure 85. A mosaic of images taken on CCD-2 from back-reflected light for each lens, is the tip-tilt reference image.*

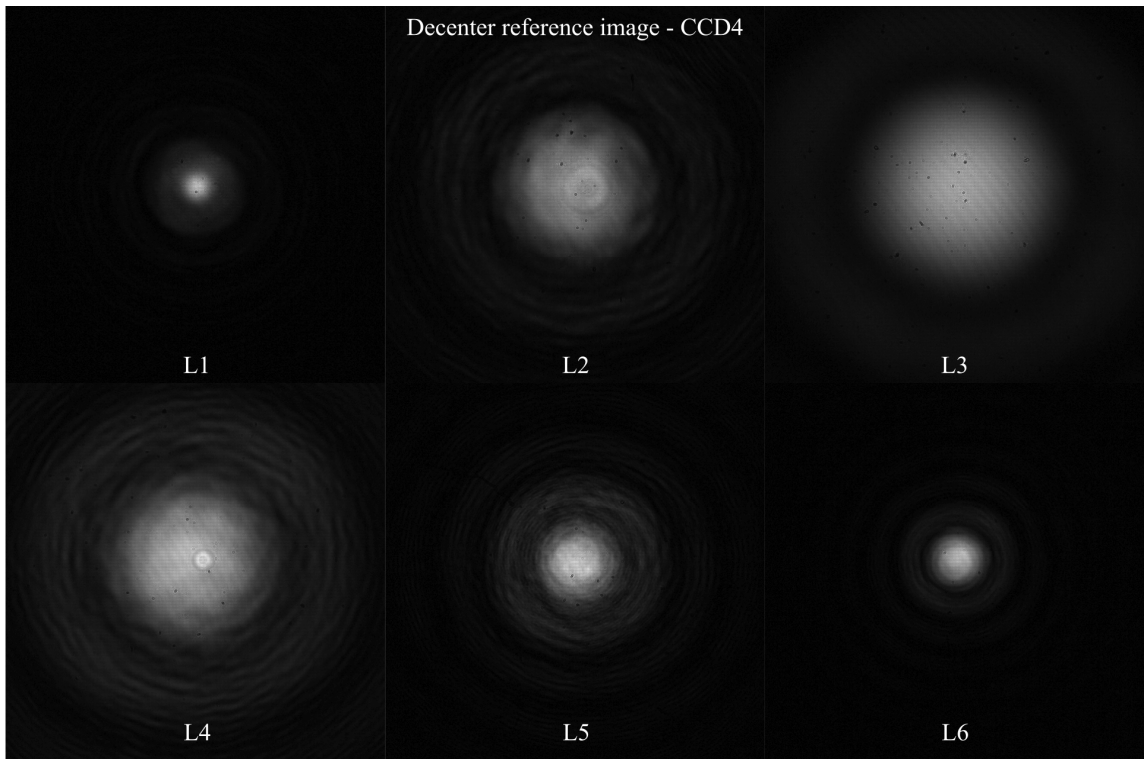

*Figure 86. A mosaic of images taken on CCD-4 from the transmitted beam for each lens, is the decenter reference image.*

### 2.6.1.3 Reference spot on CCD-3 and CCD4

As already mentioned, also on CCD-3 and CCD-4 the laser spot is affected by fringing, due to the change of thickness of the optical windows. In this case, the centroid's computation follows the FFT filtering and Gaussian-2D as previously described. In Figure 87 there shown the spot analysis for the CCD-4, where the white dots in the frequency space domain correspond to the high frequencies of the fringes.

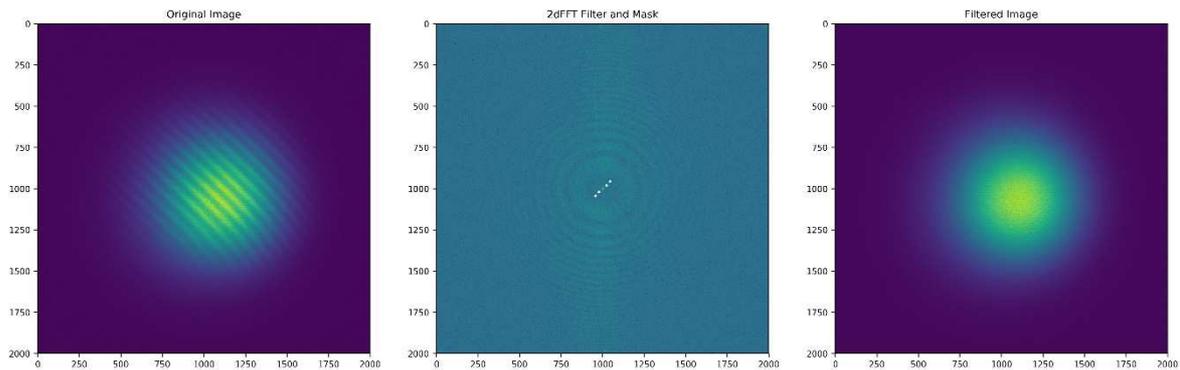

*Figure 87. The spot analysis with FFT filtering and Gaussian-2D fitting for CCD-4.*

For CCD-3 or CCD-4, the fringes dimensions are comparable with the diameter of the spot. In this case, applying a Gaussian filter before the Gaussian-2D fitting, the contribution of the fringes from the images was removed entirely, see Figure 88.

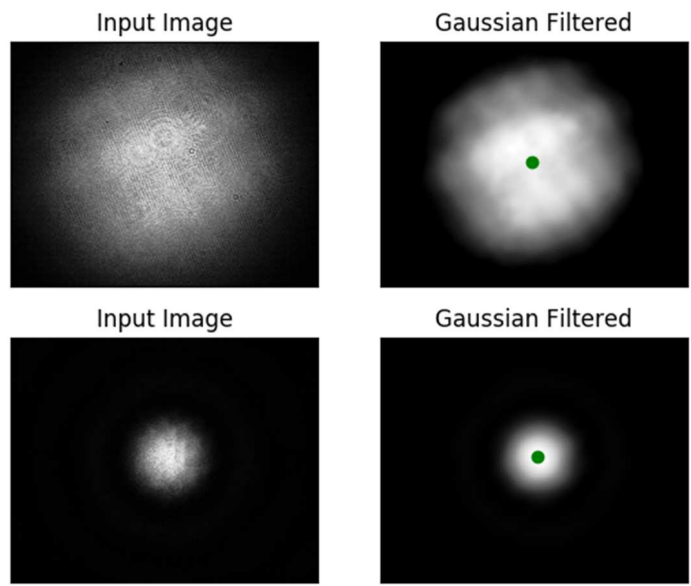

*Figure 88. The Gaussian filter efficiently reduces the fringing if their dimension is comparable with the spot size.*

## 2.6.2  Setup stability over time

The stability of the setup has been checked over different periods of time, by checking on CCD-2, CCD-3 and CCD-4 the movement of the spots created by the back-reflected beam, alignment beam, and transmitted beam respectively. The back-reflected beam movement on CCD-2 has been checked in two different configurations:

- with the CC sending back the light to CCD-2, having the beam not passing through the folding mirror FM3, Figure 90 arrow "a";

- with the flat mirror positioned where the lenses will be placed, also sending back the light which is passing back and forth through the flat mirror FM3, Figure 90 arrow "b";

In Figure 89 we report an example of a stability test, performed just on CCD-2 (in the 2 different configurations just mentioned) and CCD3. From this test, the maximum movement of the spot on CCD-3 over 3 hours is of the order of 1 pixel PtV, with a pixel size of 5.2μm, while on CCD-2 the maximum movement is of the order of 4 pixels (~20 μm) PtV with the CC, increasing to about 11 pixels (~60μm) PtV without CC, corresponding to about 3 arcsec of tilt PtV. The effect of the FM3 and of the mirror below the TOU is quite significant due to the long leverage effect of the setup (Figure 89 right). The strange movement of the spot follows the displacement of the GSE in the 3D space due to thermal variation during the test caused by the CTE of the material, mostly aluminum and stainless stell.

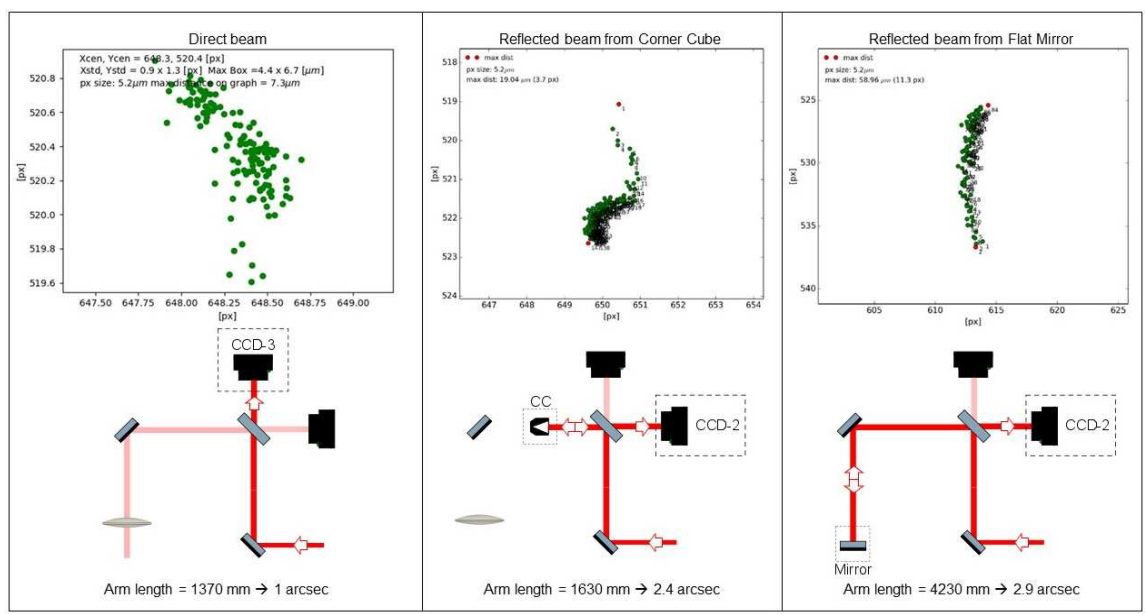

*Figure 89. An example of a stability test over 3 hours, performed on CCD-2 (in 2 different configurations, with and without CC) and CCD-3.*

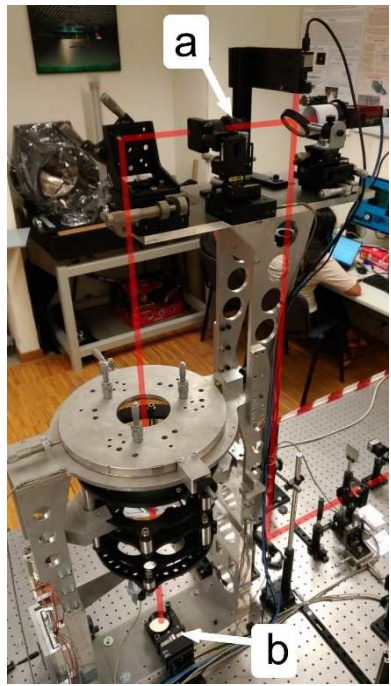

*Figure 90. The optical components for the stability test of the reference beam, a: the CC in the setup bench, b: the flat mirror below the TOU, the red ray is drawn over the image for better understanding of the optical path.*

Further test on a 10 hours time-scale have shown movement of the spot on CCD-2 in the setup without retroreflector of the order of 240µm PtV, correspond to about 12arcsec of tilt PtV, and movement of the spot on CCD-4 of the order of 24µm PtV, being the last about half of the overall alignment budget in centering of ±22µm for all the lenses but L6, which is looser.

For this reason and based on these test results, we tried to improve the setup stability by adding some metallic ropes connecting the GSE holding the prototype to the optical bench, see Figure 91. With such a modified setup, we did make several tests on different time scales and the result obtained has been better.

We emphasize again that we are mostly interested in time scales which are comparable to the time needed to align a single lens since normally the alignment beam is re-aligned at the beginning of each lens alignment. The alignment time for all the lenses ranges from 3-4 hours to a maximum of 7-8 hours, and thus this is the time scale we are mostly interested to explore.

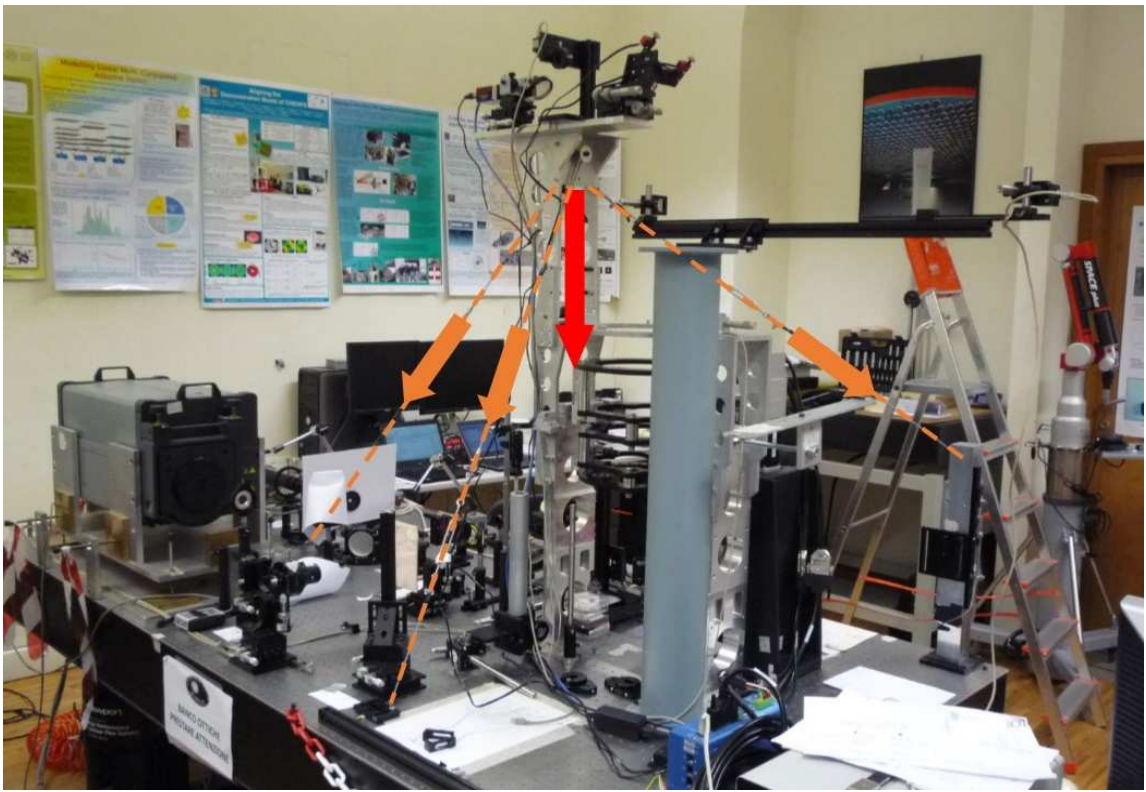

*Figure 91. The three metallic ropes added to the GSE holding the prototype, with the purpose to improve the setup stability; their preload can be adjusted individually and the resulting force is directed downwards, decreasing the oscillations of the GSE tower.*

The stability over, a bit conservative, 12 hours time frame with the improved setup supported by tests over several runs, is of the order of ~120 μm PtV on CCD-2, correspond to about 6 arcsec of PtV overall tilt, and ~12μm PtV of decentre on CCD-4, showing that the new setup is increasing the setup stability of about a factor 2.

All these values are accounting a few degrees of temperature variation, below 4, which were typical during the 10-12 hours test. We emphasize that the temperature was monitored in several points of the optical setup and prototype structure and that during the lens alignment we were able to maintain the temperature stable within 1 degree, by applying the right ventilation to the laboratory.

### 2.6.3   Interferometer accuracy

The interferometer Zygo FlashPhase GPI, equipped with a transmission sphere with an aperture of F/1.5, was used to perform a double pass test addressed at the end of the alignment. During the analysis, we chose the position and size of the aperture mask in the Metro Pro software, see Figure 94, but the mask position and size can affect the computation of the Zernike coefficients and, more generally, the analysis performed by the interferometer. For this reason, we carried out a sensitivity analysis concerning the translation and diameter variation of the mask, to obtain an estimate of the error on the final coefficients.

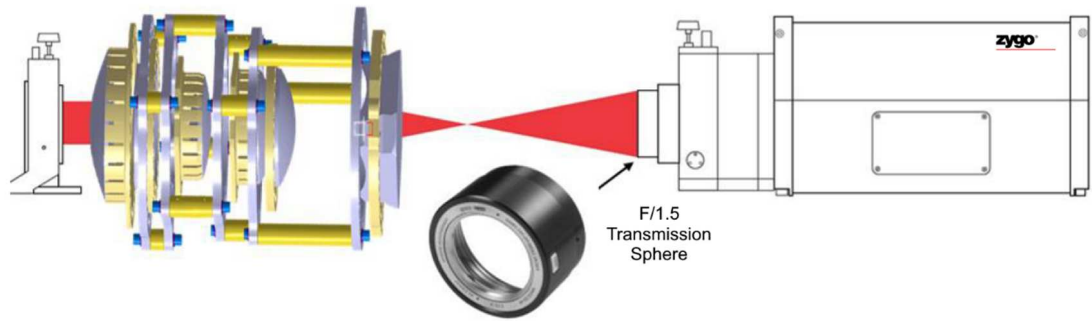

*Figure 92. The scheme of the double-pass interferometic test.*

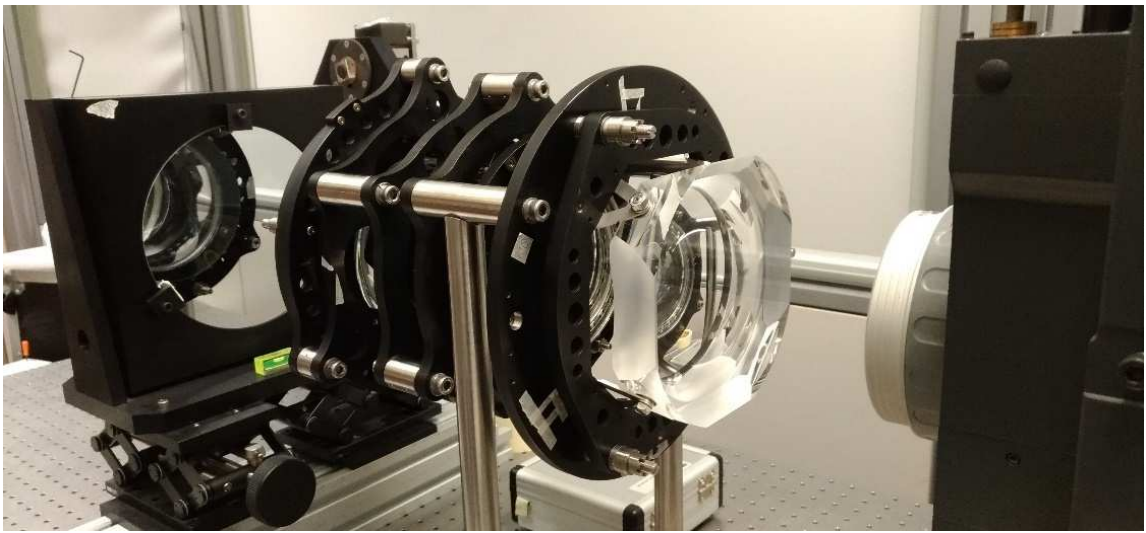

*Figure 93. The double pass test to measure the accuracy of the interferometer with respect to the position and dimension of the aperture mask. Plato is illuminated in direction L6 to L1.*

The aperture mask was selected to fill the 97% of the 120mm aperture of the lens, and its position was translated around a grid of 125 points of a matrix 12x12 to simulate the max error in positioning, see Figure 94. For each position, we measure the Zernikes coefficients and create an error map, Figure 95 resume the analysis. The results of such analysis have been compared with a simulation in ray tracing, considering the same aperture used in the test, which are summarized in Table 12. The values obtained (for spherical, coma and astigmatism) are within the simulated one, considering the best and worse case scenarios.

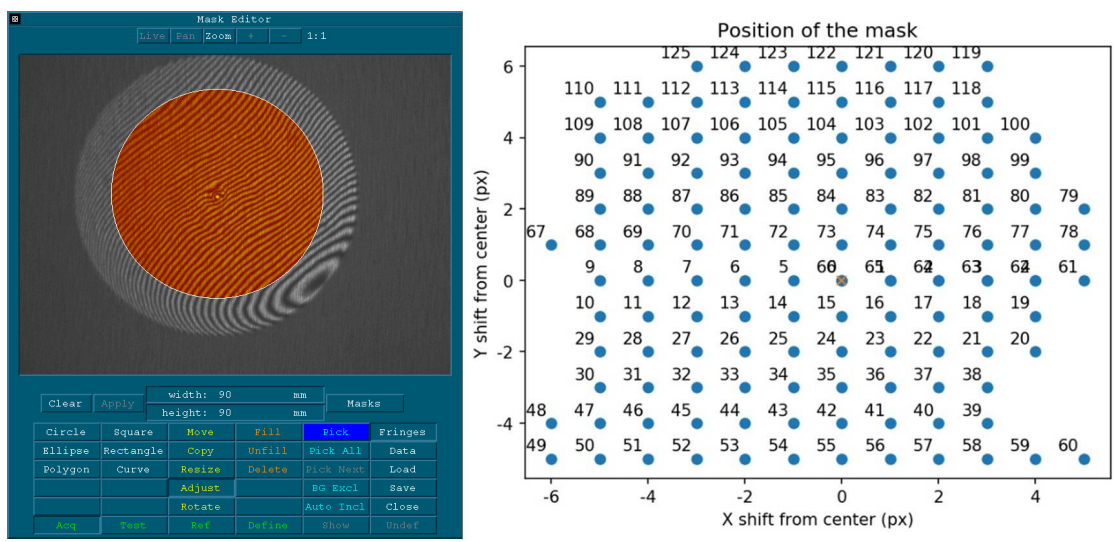

*Figure 94. On the left: the aperture mask selection in the Metro Pro software. The gray zone is the interferogram coming from the lenses of PLATO, the orange zone is an example of the aperture mask selection for the analysis. On the right: the matrix of decentering applied to the mask*

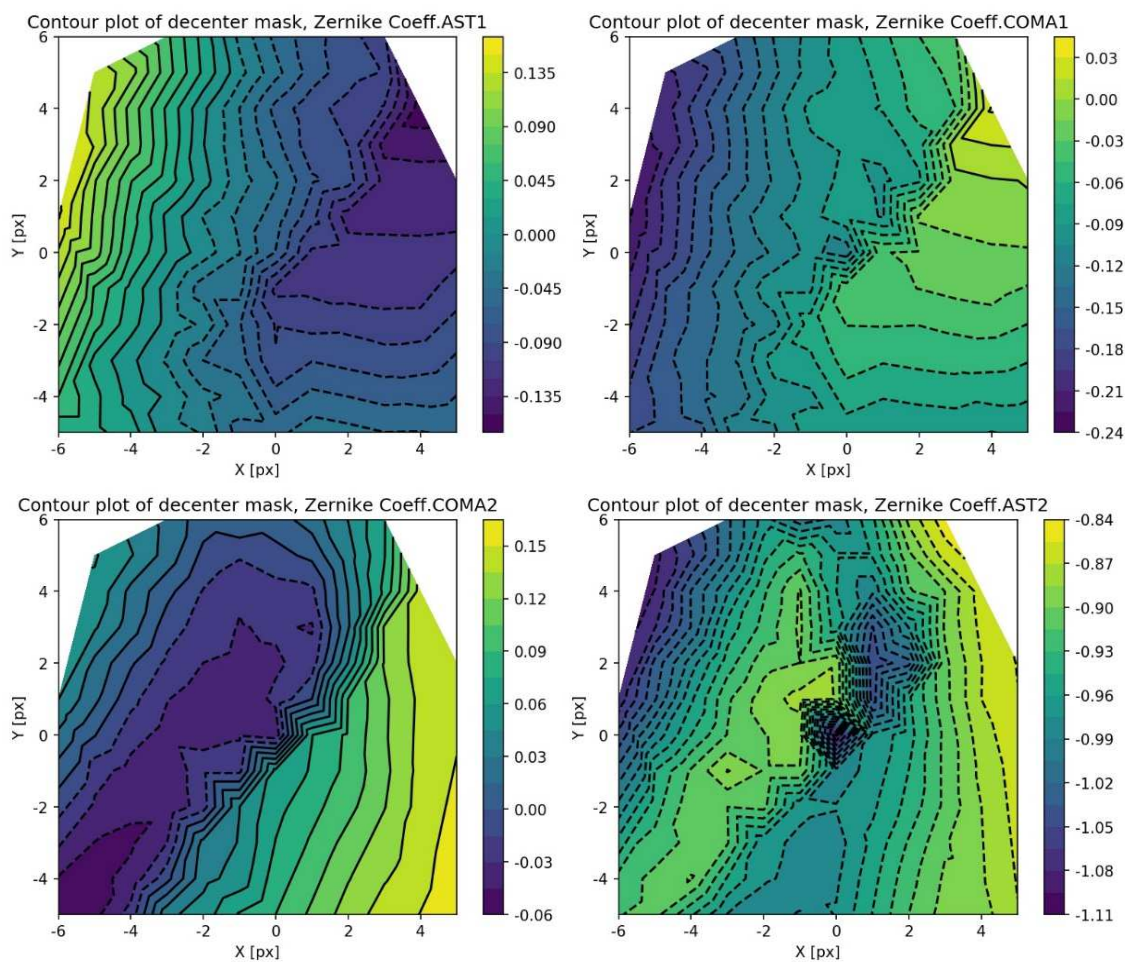

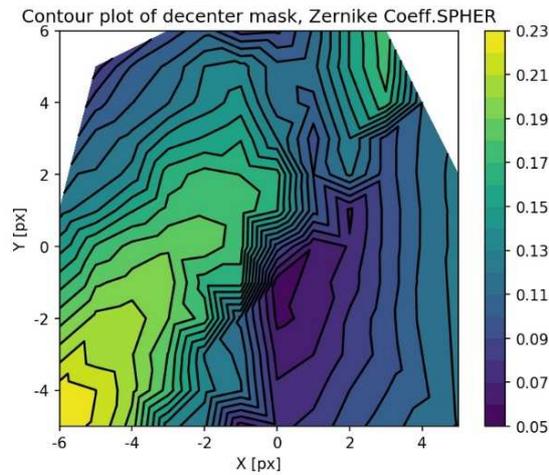

*Figure 95. The color map of the primary aberrations values moving the aperture mask on a matrix 3x3 points.*

*Table 12. The overall error for a decenter in the aperture analysis mask for the three primary aberrations. Left column: for a total decenter of ± 6 points 97% of the aperture lens. Middle column: same displacement but with a variation of the mask diameter of ± 3% with respect to the previous. Right column: the comparison with a simulation in Zemax respect the best case (mask aligned with the lens axis) to worst-case (mask with maximum displacement)*

| Zygo Coeff. | Decenter [wave] ± 6 pt (PtV) | Zoom mask [wave] ± 3% (PtV) | Simulation Zemax [wave] (best-worst case) PtV |
|---|---|---|---|
| SPH | 0.172 | 0.152 | 0.432 − 1.032 |
| COMA | 0.299 | 0.324 | 0.014 − 1.082 |
| AST | 0.400 | 0.270 | 0.010 − 0.787 |

To evaluate the impact of the mask diameter variation, we made the following test. The diameter has been changed from 91 mm to 98 mm and we obtained stable aberration coefficients in the range 91/94mm, see Figure 96.

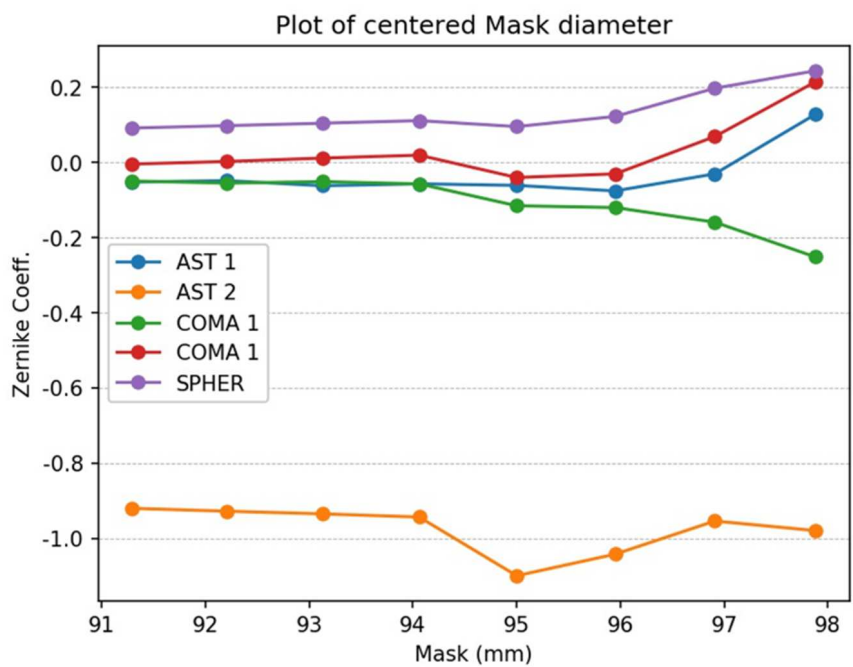

*Figure 96. The primary aberration values measured with a variation in diameter of the aperture mask, centered with respect to the optical axis of the TOU.*

We perform a second test concerning the mask decentering reducing the size of the grid to 3x3 points. The aberration values are of course less, and we performed this for 97% and 95% of the lens aperture, see Figure 97 and Figure 98.

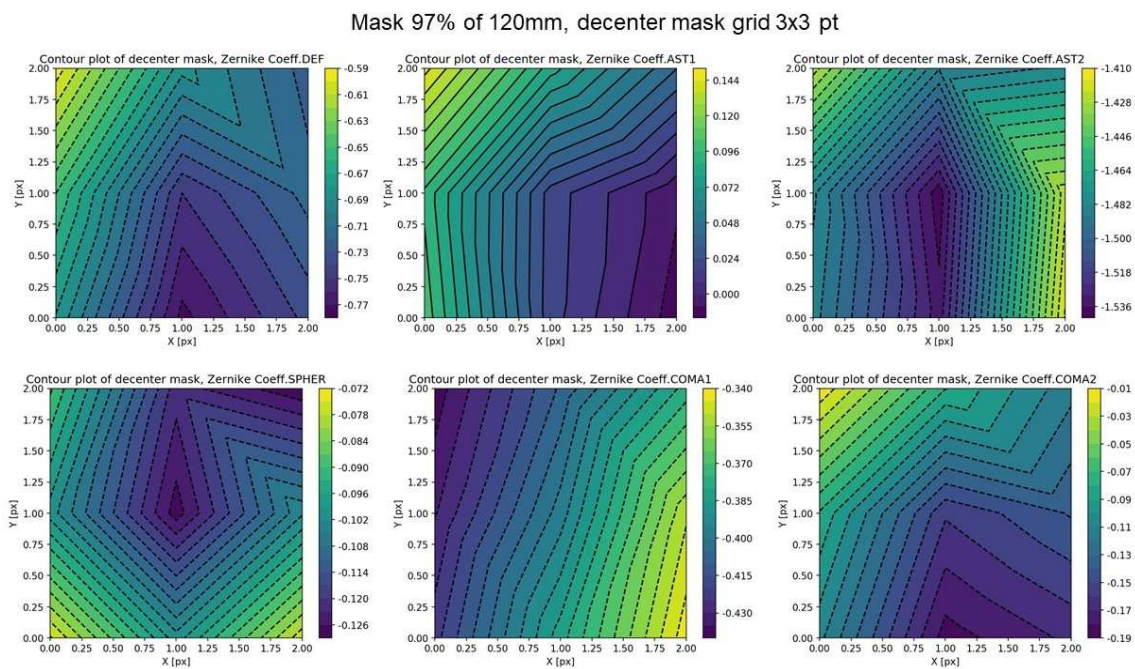

*Figure 97. The variability of aberration coefficients for a grid decenter of 3x3 points, and mask at 97% of 120mm.*

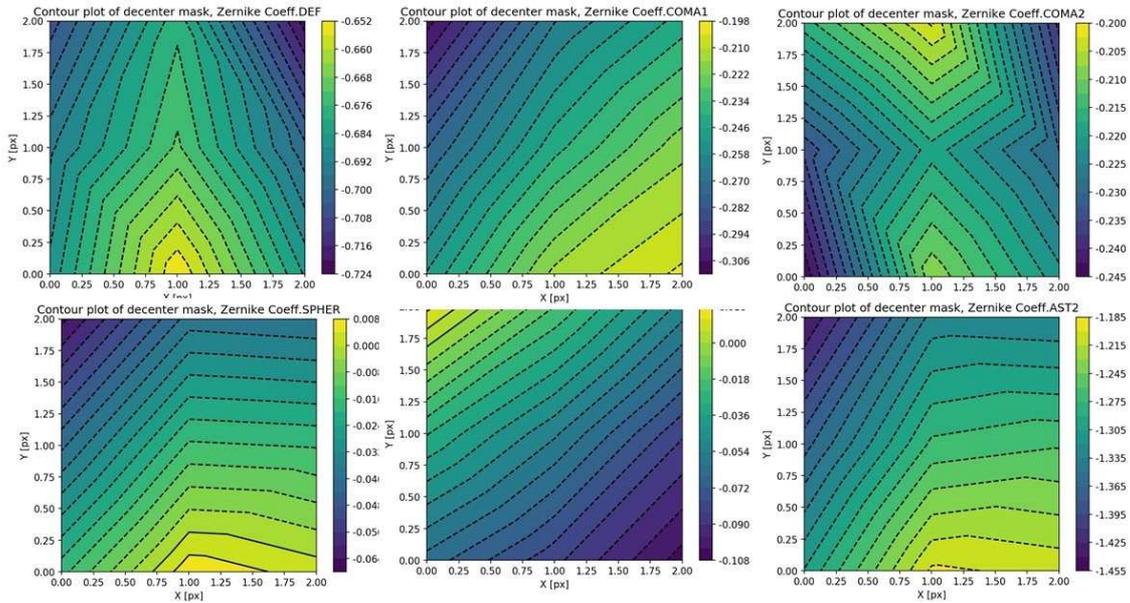

*Figure 98. The variability of aberration coefficients for a grid decenter of 3x3 points and mask at 95% of 120mm.*

*Table 13. Resume of aberration error considering a grid of displacement in an aperture of 3x3 points.*

| Mask 95% 120mm [wave] | | Mask 97% di 120mm [wave] | |
|---|---|---|---|
| PtV DEF: | 0.070 | PtV DEF: | 0.179 |
| PtV AST1: | 0.118 | PtV AST1: | 0.159 |
| PtV AST2: | 0.244 | PtV AST2: | 0.124 |
| PtV COMA1: | 0.106 | PtV COMA1: | 0.096 |
| PtV COMA2: | 0.042 | PtV COMA2: | 0.173 |
| PtV SPHER: | 0.072 | PtV SPHER: | 0.054 |

We finally estimated the error in applying the same mask to the warm and cold interferograms, due to the centering and to the diameter of the mask, and they have been realistically estimated in 1 pixel for the centering and ± 2% on the mask diameters, which leads to an uncertainty on the Zernike coefficients shown in Table 14.

*Table 14. The uncertainty on the Zernike coefficients (spherical, coma and astigmatism) due to errors in positioning (1 pixel in centering) and dimensioning (± 2% of the diameter) of the mask.*

| Error ± 1 px in mask centering and Error ± 2% in mask diameter [wave PtV] | | | | | |
|---|---|---|---|---|---|
| AST1: | 0.144 | COMA1: | 0.108 | SPHER: | 0.072 |
| AST2: | 0.270 | COMA2: | 0.180 | | |

### 2.6.4   Rotation GSE accuracy

When we aligned the beam to L3, to avoid degeneracy between centering and tilt, typical of all the spherical lenses, we decided to iterate between aligning the tip-tilt of the beam on L3 and the centring, the latter done by superimposing the transmitted spot (on CCD-4, which is collecting the transmitted beam) with and without L3. The last operation was performed by rotating the prototype of about 90° since the GSE structure allows the light to go through. When the prototype structure is moved back and forth to 90°, the spot detected on CCD-2 and CCD-4 was slightly moving. We performed about 100 rotations of the prototype structure recording the spots centroids on CCD-2 and CCD-4, and we computed the PtV values of the two clouds of points obtained, which correspond to an error to be added to the overall budget when aligning the beam to L3.

With L3 lens, the maximum diameter of the cloud on CCD-4 turned out to be ~26µm of decentring PtV, while on CCD2 the maximum diameter was of the order of 300µm, correspond to ~15arcsec of overall tilt PtV.

### 2.6.5   Lenses insertion and GSEs installation

Inserting each lens into the setup and installing each alignment GSE "manipulator" has a certain effect on the beam stability, both on CCD-2 and on CCD-4. We repeated the just mentioned operations, and the movement on CCD-4 turned out to be ~6 µm PtV, while on CCD-2 such a number is depending on the considered lens curvature and its distance from CCD-2. In the worst case, this value corresponds to ~10 arcsec PtV, and we decided to conservatively adopt the same number for all the lenses.

### 2.6.6   Centroid's accuracy

On CCD-4 the transmitted beam has been collected. To align each lens, the spot in the current configuration was compared with the reference spot of the previously aligned lens. The centroid has, of course, a certain accuracy, which has been computed and corresponds to ~6µm PtV. The last number account also for short term movements of the spot on CCD-4.

### 2.6.7   Newton rings uniformity accuracy

The accuracy of the operation of making the Newton rings uniform on CCD-2, which is collecting the back-reflected light used during the alignment process for the lenses tip-tilt adjustment, has been computed. Of course, such a number is depending on the considered lens curvature and distance from CCD-2. In the worst case, this value corresponded to ~11 arcsec PtV, and we decided to conservatively adopt the same number for all the lenses.

## 2.6.8   Alignment Axis re-definition at new lens insertion

When a new lens is inserted in a particular orientation, the alignment beam has to be double-checked by looking at the position of the transmitted spot on CCD-4 and the position of the back-reflected spots on CCD-2. If the alignment beam has to be tuned, this operation was performed within a precise accuracy, which we tested to be ~13 arcsec PtV of overall tilt on CCD-2 and ~6 µm PtV in centering.

## 2.6.9   Alignment Axis re-definition after prototype rotation

When the prototype structure has to be rotated to continue the lenses insertion, the alignment axis has to be re-aligned to L3 in the new orientation. This operation was performed within a certain accuracy, which we tested to be ~5 arcsec PtV of overall tilt on CCD-2 and ~26 µm PtV in centering.

## 2.7   Alignment process and results

We recall here the procedure foreseen for the Assembly and Integration of the Prototype, which was essentially based on the following main steps:

- each lens was aligned in tip-tilt by observing the BRR and in centering by observing reflected and transmitted spots positions (observed in two dedicated CCDs, named CCD-1 and CCD-3 respectively in Figure 63);

- each lens was aligned in focus by positioning the CCD collecting the transmitted light (CCD-3 respectively in Figure 63) in the theoretical focus position (computed by Zemax) and moving the lens to achieve proper focus on CCD-3.

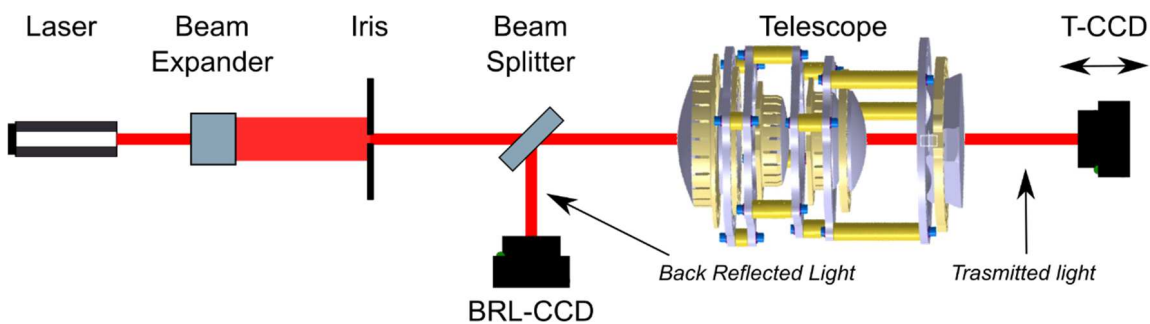

*Figure 99. The prototype alignment concept (in this figure, the TOU structure instead of the prototype one is shown).*

These are the general guidelines of the procedure, which is considering that a collimated beam is sent to the Prototype and used as a reference beam for the alignment, as it is shown in Figure 99.

We now specialize in the prototype the general procedure just described, considering all the constraints essentially due to the following issues:

- what shown in Figure 99 is a concept, the real procedure is considering to keep the prototype in a vertical position (as it is shown in Figure 100), in a way to simplify the insertion of the various lens assemblies, that shall always happen from the top of the structure;

- again for accessibility reasons to the inner part of the structure, the first lens to be inserted has to be one of the middle ones, and L3 has been selected to be the first one. This fact requires that the prototype structure has to be rotated at least one time to allow the insertion of all the lenses;

- given that the first lens to be inserted is L3, we decided to place L3 centering it mechanically to the prototype structure, and then align the collimated beam to L3 instead of doing the opposite. So, L3 is the reference to which all the other lenses are aligned;

- we identified the optimal sequence for the lenses insertion, given obvious constraint (after L3, we must insert either L2 or L4, and so forth), and following these guidelines:

    o the optimization of the visibility of the BRR (simulated with FRED Photon Engineering), since they are the observables used for the alignment;

    o the minimization of the travel required for the CCD #4 (T-CCD, the CCD collecting the Transmitted light) to make it possible to reach the focus position in the various configuration of inserted lenses;

    o the minimization of the number of rotations required to the prototype for the lenses insertion, since the rotation of the structure requires each time a re-alignment of the beam to the prototype;

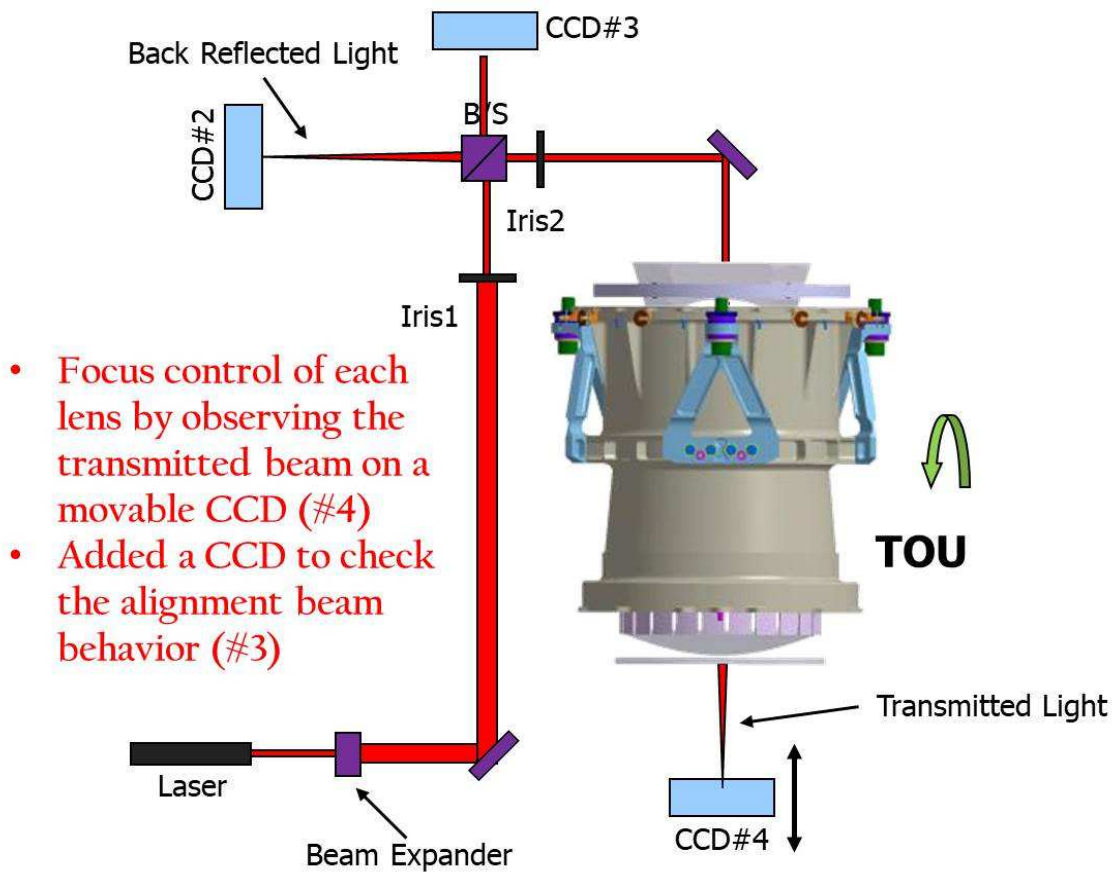

*Figure 100. The Assembly and Integration procedure foreseen for the PLATO prototype*

The identified sequence is shown in Figure 101.

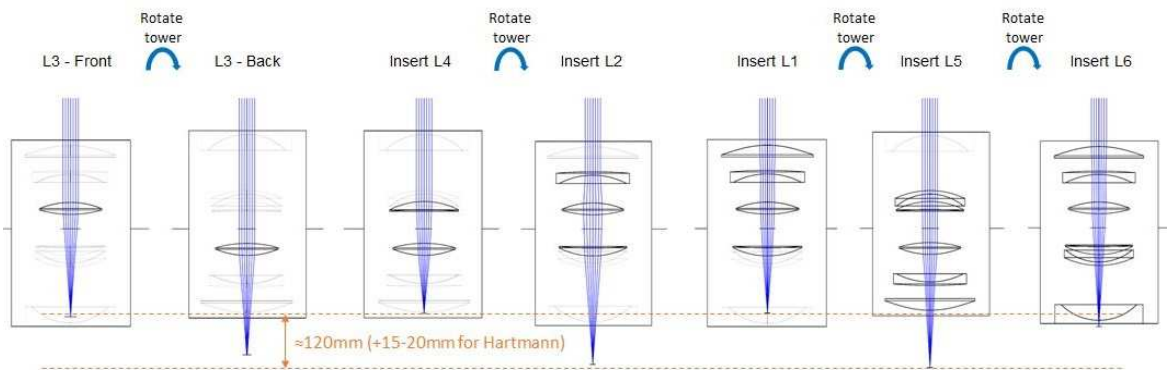

*Figure 101. The lenses insertion order, to allow their alignment along the optical axis*

## 2.7.1   L3 alignment, identification of the alignment axis

By the means of an optical setup realized on the same optical bench which is holding the prototype, we materialize a collimated beam (which will be later used for the lenses alignment) which has to be pre-aligned to the prototype structure and to L3.

The scheme of the just mentioned setup is shown in Figure 77, and the procedure followed for the alignment of the beam to L3 is the following:

1. the L3 lens-mount assembly is inserted into the prototype structure, mechanically aligned with the caliper in a way to be centered with respect to the fixing holes. The screws are then tightened to the foreseen 10Nm force per screw, bypassing through intermediate steps of 3Nm and 7Nm;

2. by rotating the prototype structure with the dedicated GSE, the beam can be materialized on CCD-4 with and without (prototype structure at 90° position) L3 in the optical path. We named the L6-L1 direction the "0° orientation", and vice versa the L1-L6 direction the "180° orientation". The Iris is moved till having the two positions on CCD4, with and without L3, coincident. In this way, the beam is centered to L3;

3. the CC is adjusted in a way to have it centered on the alignment beam, and the corresponding position on CCD-1 is recorded;

4. M5 was mounted on a magnetic base, which can be repositioned very precisely, is aligned in tip-tilt till having the spot on CCD-1 superimposed to the one coming from the corner cub. In this way, the ongoing beam is superimposed to the backward beam, and the position of the latter on CCD-2 is recorded. M5 is then removed;

5. the position of Newton's rings back-reflected by L3 surfaces is compared with the reference position of the beam on CCD-2, defined in the previous step in auto-collimation of the alignment beam with a flat mirror. Acting on the picomotors of FM3, the back-reflected spot from L3 is superimposed with the reference on CCD-2;

6. since the last operation will also affect the beam centering on L3, steps from 1 to 5 have to be iterated till achieving the required accuracy. Since the beam aligned to L3 is the reference for the alignment of the other lenses, the accuracy is reduced, since it will only determine how decentred and tilted it will be the final focal plane. We decided to consider accuracies <20μm in centering and <30 arcsec in tip-tilt.

In Figure 102 are shown a collage of these previous phases.

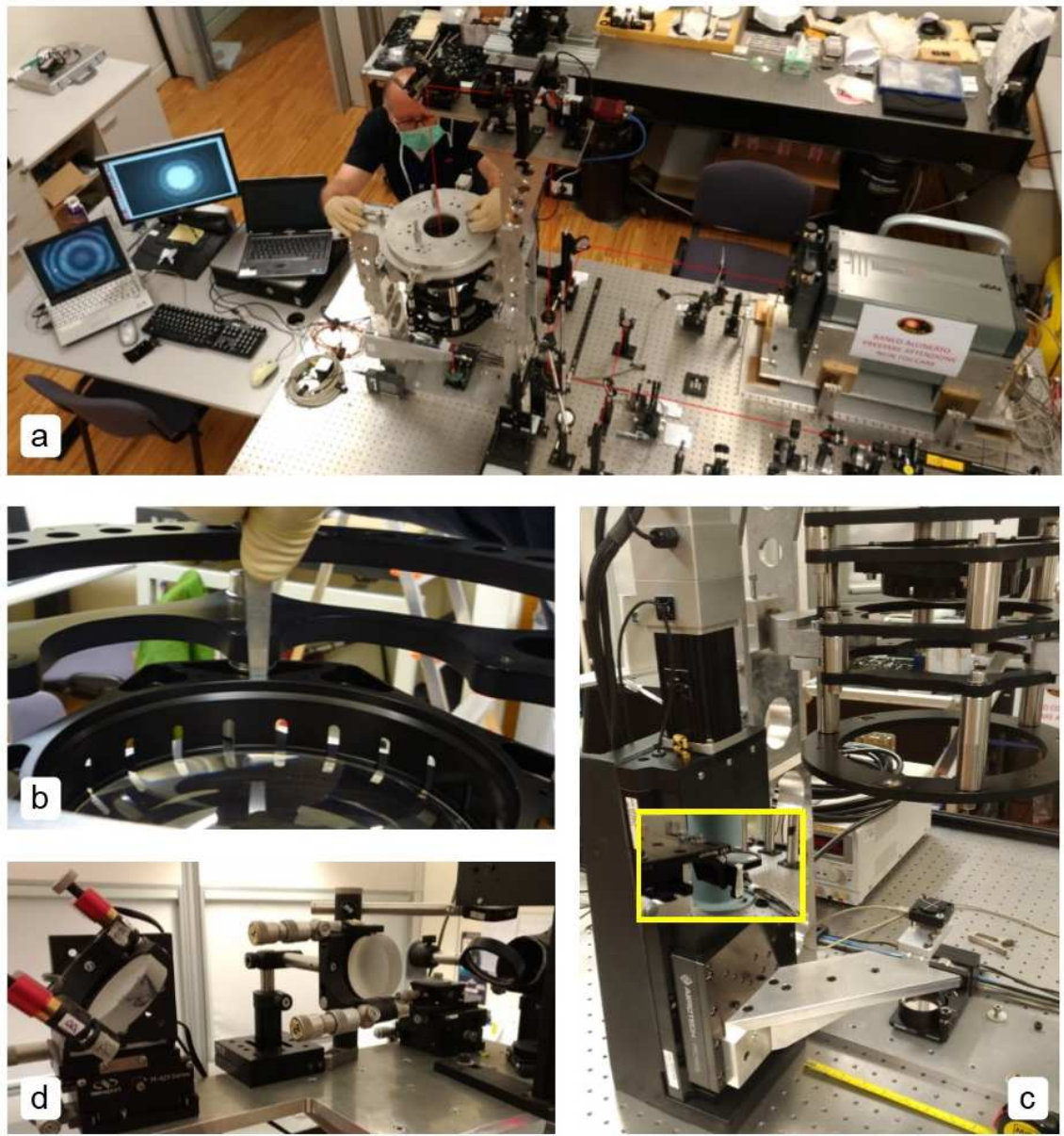

*Figure 102. The L3 alignment. Panel a: the overall bench with the interferometer and laser coaligned, panel b: the L3 mechanically aligned with the caliper, panel c: the CCD-4 and the spherical mirror in the yellow windows, below the CCD-4 there is a flat mirror to perform stability test, d: the M5 mirror in the magnetic base.*

## 2.7.2   L3 focus reference definition

As already mentioned, the focus position of L3 will be used as a reference to position the other lenses in their correct focus. In other words, given a certain orientation of the prototype, the L3 focus has to be correctly imaged on CCD-4, and the corresponding position of the focusing stage has to be recorded. When the next lens will be placed (L4 for example), by raytracing software Zemax (with the optical design updated with the measured lenses parameters) we provides the relative distance between the L3 focus and the L3+L4 focus, CCD-4 is positioned where predicted by Zemax starting from the L3 alone focus position previously recorded, and L4 has to be moved along the optical axis till having L3+L4 in their best focus on CCD-4. What just described has to be computed for both orientations of the prototype, to allow the proper focus of the lenses inserted both-ways into the prototype.

Due to the tight tolerances of the various lenses in focus (see Table 2), its measurement has to be done with very high accuracy. We initially thought to make a sweep in focus with the motorized stage moving CCD-4, but our characterization of such a method gave an accuracy of the order of ±15μm, which is comparable with the overall focus tolerance required for L1, and thus not acceptable.

We thus use also the interferometer, as it is shown in Figure 78, page 83. By using a deployable mirror FM2, the collimated beam coming from the interferometer can be sent to the prototype. The iris of the interferometry allows selecting a portion of the collimated beam, while a spherical mirror (equipped with centering adjustment through micrometers) can be placed on the moving part of the CCD-4 linear stage, just above CCD-4 itself. The spherical mirror is mounted on a kinematic magnetic base plate, which allows its removal and repositioning with high accuracy.

With this setup, the measurement of the best focus for example of L3 is done interferometrically looking at the beam sent back by the spherical mirror in auto-collimation. The position of the spherical mirror (which has to be placed at a distance from the nominal lens focus equivalent to its curvature radius) can be tuned very precisely thanks to the very high positioning accuracy of the CCD-4 linear stage (0.1 μm of bidirectional repeatability), and the achieved accuracy in focus by looking at the fringes on the interferometer is of the order of ±3μm, a factor 3 better than the tolerance of the tighter lens.

### 2.7.3   Changing in the focus procedure

When we did insert L4, the first lens in the sequence after L3, and performed the focus alignment as just described, we did a sanity check of the focus position of L4 with a CMM machine, and it turned out to be away from the foreseen position of about 150μm.

We started investigating what the reason could be, and it turned out that we underestimated the effect of the accuracy of the measurement of the primary lenses parameters, such as the radius of curvature, performed by the manufacturer. The radius of curvature measurement accuracy that was specified in 0.02% was causing an error on L4 positioning along Z of the order 40μm for the worst case, due to the uncertainties on the radius of curvature of both L3 and L4. Furthermore, this error propagates to the following lenses, due to procedure based on positioning the detector on the nominal focus, causing possible maximum displacements of the order of 2mm for the worst-case on L6, see Table 15.

*Table 15. The propagation error in the z-position of the lens due to 0.02% in error for the radius of curvature.*

|              | L1    | L2    | L3 | L4   | L5    | L6    | Focal plane |
|--------------|-------|-------|----|------|-------|-------|-------------|
| **Z error [mm]** | 0,192 | 0,074 | 0  | 0,04 | 0,313 | 1.719 | 0,284       |

Of course, such errors are far outside the tolerances specified in Table 2, and we have thus to modify the focus alignment procedure, by using a CMM machine to place the lenses mechanically in their nominal position. The CMM we have in Padova laboratories see Figure 103, has an accuracy in the determination of the distance between two points in the 3D space of about ±15μm. This error is comparable with the focus tolerance of the tighter lens, L1 lens, and therefore not optimal, but still much better than what we would achieve with the baseline procedure. Of course, this procedure requires particular care, since the operation has to be performed by touching each lens around its vertex, which has to be protected by using a double sheet of optical paper to prevent scratches. Furthermore, to take into account possible relative movements of the prototype structure and the CMM machine supporting structure, some intermediate tests have to be performed to double-check the position of a selected area in common between the two mechanical structures on the optical bench. The position was referred with respect to the least inserted mount, which has to be touched on three reference points 120° apart.

We also characterized what is the error that we perform in touching the lenses vertex, which is of the order 1mm in radius after an acquisition based on the average of about 50 measurements. We converted this radius in focus error for the lens with the largest radius of curvature that we have on the prototype. Thus a new focus error of ±2μm to be considered into the overall focus budget.

Of course, the change in the focus measurement procedure does not require any more to follow the sequence shown in Figure 5, and thus the more obvious insertion sequence becomes L3, L4, L5, L6, prototype structure rotated and then L2 and L1.

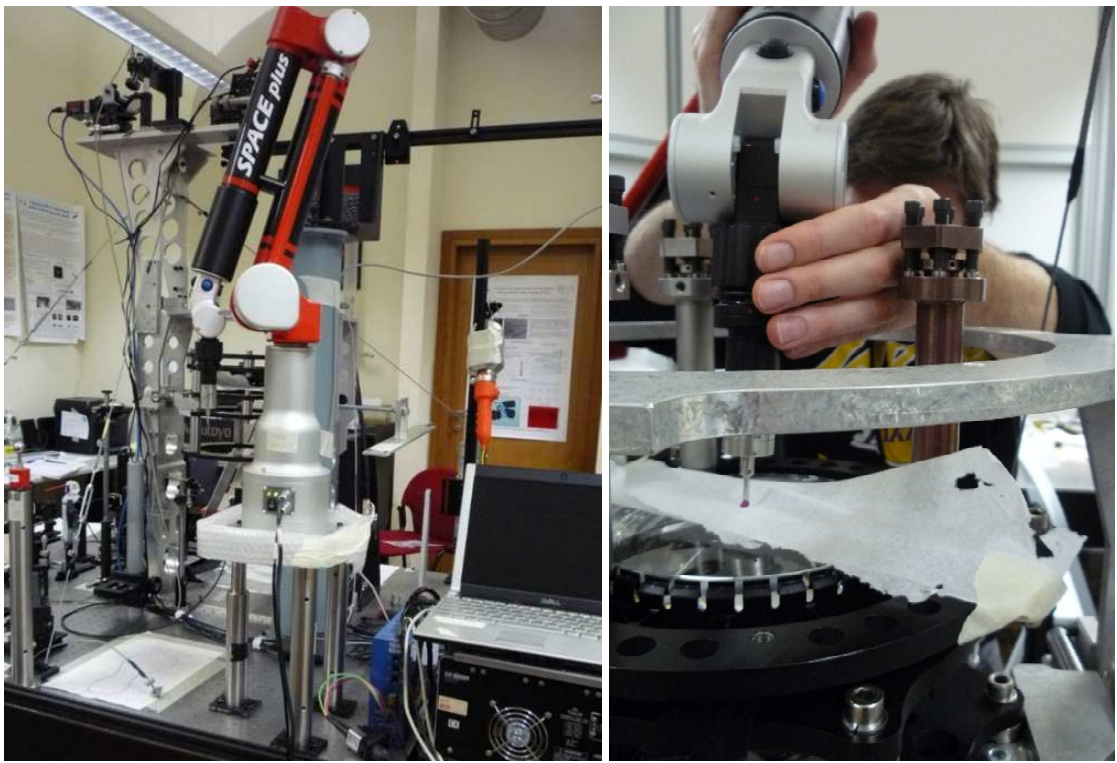

*Figure 103. The CMM used in Padova for the focus alignment of the TOU Prototype lenses and the vertex measurement of L3 lens.*

## 2.7.4   Other lens alignments

All the other lenses are integrated in a very similar manner, utilizing the beam aligned to L3 and with respect to the focus reference of L3, and following the new insertion order.

The procedure adopted is described in the following:

1. each lens is inserted into the prototype structure and positioned, with the nominal shims set, in the nominal focus position;

2. the tip-tilt of the lens is aligned, with a dedicated manipulator in the GSE for each lens, by superimposing the BRR on the reference on CCD-2;

3. centering is aligned, with dedicated GSE for each lens, by superimposing the transmitted spot on the reference on CCD-4;

4. since tip-tilt alignment is also affecting centering and vice-versa, steps 2 and 3 shall be iterated till reaching the required accuracy. At this point, shims thickness to realize the achieved tip-tilt are computed, pivoting around the lens center in order not to introduce any focus term with the shimming;

5. the lens assembly is removed, and shims for tip-tilt correction are inserted;

6. the lens assembly is re-positioned inside the prototype onto the shims and centering is again adjusted;

7. the fixing screws are now inserted and gently tightened;

8. the focus position of the lens is checked with the CMM machine, by comparing the measurement of the vertex of the previously inserted lens with the vertex measurement of the current lens;

9. If tip-tilt, centering and defocus are aligned with the required accuracy, the screws are tightened with the foreseen force, if not, shimming is recomputed and steps from 5 to 9 have to be repeated.

Before aligning each lens, we double-check that the alignment axis did not move with respect to the previously aligned lens. To perform this operation, we double-check the position of the transmitted spot on CCD-4 and the position of the back-reflected spots on CCD-2. If they are the same with respect to the previously aligned lens situation, we proceed with the alignment of the new lens. If not, we insert a mirror in the alignment setup before the last folding mirror (M5 of Figure 77) to disentangle if the beam movement is coming from FM3 or not, and we act either on FM3 tip-tilt through tip-tilt or on the beam centering to re-align the beam. Once we achieved this situation, we start with the alignment of the new lens.

After having aligned L6, the prototype structure has to be rotated and the alignment axis has to be re-aligned to L3 in the other direction, in a way that will be described in detail in section 2.8.

## 2.7.5   L3 results

The alignment of L3 was been performed in two directions of the TOU with respect the incoming beam from the laser from the top of the GSE, in direction L1 to L6, defined in the following text L1-L6 direction 180° orientation, and L6 to L1, defined as L6-L1 direction 0° orientation.

### 2.7.5.1  L3 alignment in L1-L6 direction (180° orientation)

We align the laser beam to L3 in the L1-L6 direction. This configuration is then recorded to be used for the later phase in which L2 and L1 will be assembled and aligned.

The observables are:

- Decenter: the relative position of two spots on CCD-4, taken with and without the lens in the optical path (to remove the lens from the optical path we rotate the TOU structure to 90deg position). During this operation, the CCD-4 is positioned at the bottom-end of its travel, at -150mm of the z-stage, about 160mm from the lens focus of 213mm. A Δ between the positions of the two spots is caused by a shift of the laser beam from the vertex of the lens, when tilt is aligned too, of ~0.6Δ. The goal: Δ<10um on CCD-4.

- Tilt: the position of Newton's rings back-reflected by L3 surfaces is compared with the reference position of the beam on CCD-2, defined in auto-collimation of the laser beam with a flat mirror. A Δ between the positions of the two spots is caused by a tilt of the laser beam with respect to the normal to the vertex of the lens when the lens is centered too, which is, of course, depending from a distance between the lens vertex and CCD-2, which is 1070mm. For L3 in this configuration, a Δ of about 100um, correspond to 18 pixels on CCD-2, gives ~0.01/1070/2 [rad] = 10 arcsec, that we assume to be a reasonable error considering that the overall tolerance for the L3 tilt is +/- 44 arcsec. The goal is thus to have Δ<100um on CCD-2 (18 pixels).

- Focus: we identified with the CMM machine the vertex of L3, to be recorded and used as a reference for the L2 positioning in focus. As already mentioned, this operation is performed several times, normally 50, to average all the results. The error due to the fact that we mistake the vertex position, which is also depending on the lens curvature, if of the order of 5μm for the most curved surface. We also emphasize that we also touch three positions on the mount of L3, to have access points when L2 is into double-checking possible movements between the CMM and the prototype.

Results for L3 are:

- decenter for x=8 μm and y=13 μm corresponding to 15μm of overall decenter;

- tip-tilt for δ=-6 arcsec and φ=8 arcsec corresponding to 10 arcsec of overall tilt.

Additional error sources to be considered for the decentring are:

- the effect of the setup stability on CCD-4 over the alignment time, see section 2.6.2;

- the spot movement on CCD-4 due to the TOU, see section 2.6.4;

- the error in the centroid computation and short time stability of the, see section 2.6.6.

Their linear sum is 44μm PtV, while their quadrature sum is 30μm PtV.

Additional error sources to be considered for the tilt are:

- the effect of the setup stability on CCD-4 over the alignment time, see section 2.6.2;

- The effect on CCD-2 due to the TOU rotation, see section 2.6.4;

- The error in evaluating the intensity profile of the Newton rings and short time stability of the measurements, see section 2.6.7.

Their linear sum is 32 arcsec, while their quadrature sum is 20 arcsec.

Concerning the focus, we measured the vertex position of L3 in the way we just explained and we recorded the position which is the reference for the alignment in focus, in this orientation, of L2 and L1. The error to be associated with this measure is the CMM accuracy, which is ±15μm.

### 2.7.5.2 L3 alignment in L6-L1 direction (0° orientation)

We rotate the TOU and align the laser beam to L3 in the L6-L1 direction, and we perform the alignment as the previous section, the observables are:

- Decentre: the relative position of two spots on CCD-4, taken with and without the lens in the optical path. To remove the lens from the optical path, we rotate the TOU structure to the 90deg position. During this operation, the CCD-4 was positioned at the bottom-end of its travel, at -150mm of the z-stage, about 160mm from the lens focus of 213mm. A Δ between the positions of the two spots is caused by a shift of the laser beam from the vertex of the lens of ~0.7Δ. Goal: Δ<10 μm.

- Tilt: the position of Newton's rings back-reflected by L3 surfaces is compared with the reference position of the beam on CCD-2, defined in auto-collimation of the laser beam with a flat mirror. A Δ between the positions of the two spots is caused by a tilt of the laser beam with respect to the normal to the vertex of the lens which is, of course, depending from the distance between the lens vertex and CCD-2 is 1170mm. We thus assumed the same goal of having Δ<100 μm, 18 pixels on CCD-2.

- Focus: we identified with the CMM machine the vertex of L3, to be recorded and used as a reference for the L4 positioning in focus.

The results are:

- decenter for x=10 and for y=-11 μm corresponding to 15 μm of overall decenter;

- tip-tilt for δ=8 arcsec and φ=6 arcsec corresponding to 10 arcsec of overall tilt.

Additional error sources to be considered for the decentring are:

- the effect of the setup stability on CCD4 over the alignment time, see section 2.6.2;

- the spot movement on CCD4 due to the TOU rotation, see section 2.6.4;

- the error in the centroid computation and short time stability of the measurements, see section 2.6.6.

Their linear sum is 44 μm PtV, while their quadrature sum is 30 μm PtV.

Additional error sources to be considered for the tilt are:

- the effect of the setup stability on CCD-4 over the alignment time, see section 2.6.2;

- the effect on CCD2 due to the TOU rotation, see section 2.6.4;

- the error in evaluating the intensity profile of the Newton rings and short time stability of the measurements, see section 2.6.7.

Their linear sum is 32 arcsec PtV, while their quadrature sum is 20 arcsec PtV.

Additional error source concerning the defocus is the accuracy of the CMM, which is ±15μm.

## 2.7.6   L4 results

We first double-check the alignment beam and tune it if necessary. After that, with the prototype structure in the 0° orientation, we insert L4 on the foreseen shims, setting the nominal L4 position in focus. We then align L4 to the laser beam, previously co-aligned with L3 in the 0° orientation.

The allowed adjustments are ΔX, ΔY on the GSE manipulator micrometers for centering adjustments, θ1, θ2 and θ3 on the GSE tilt micrometers, see Figure 104, whose conversion is 10 μm to ~16 arcsec of L4.

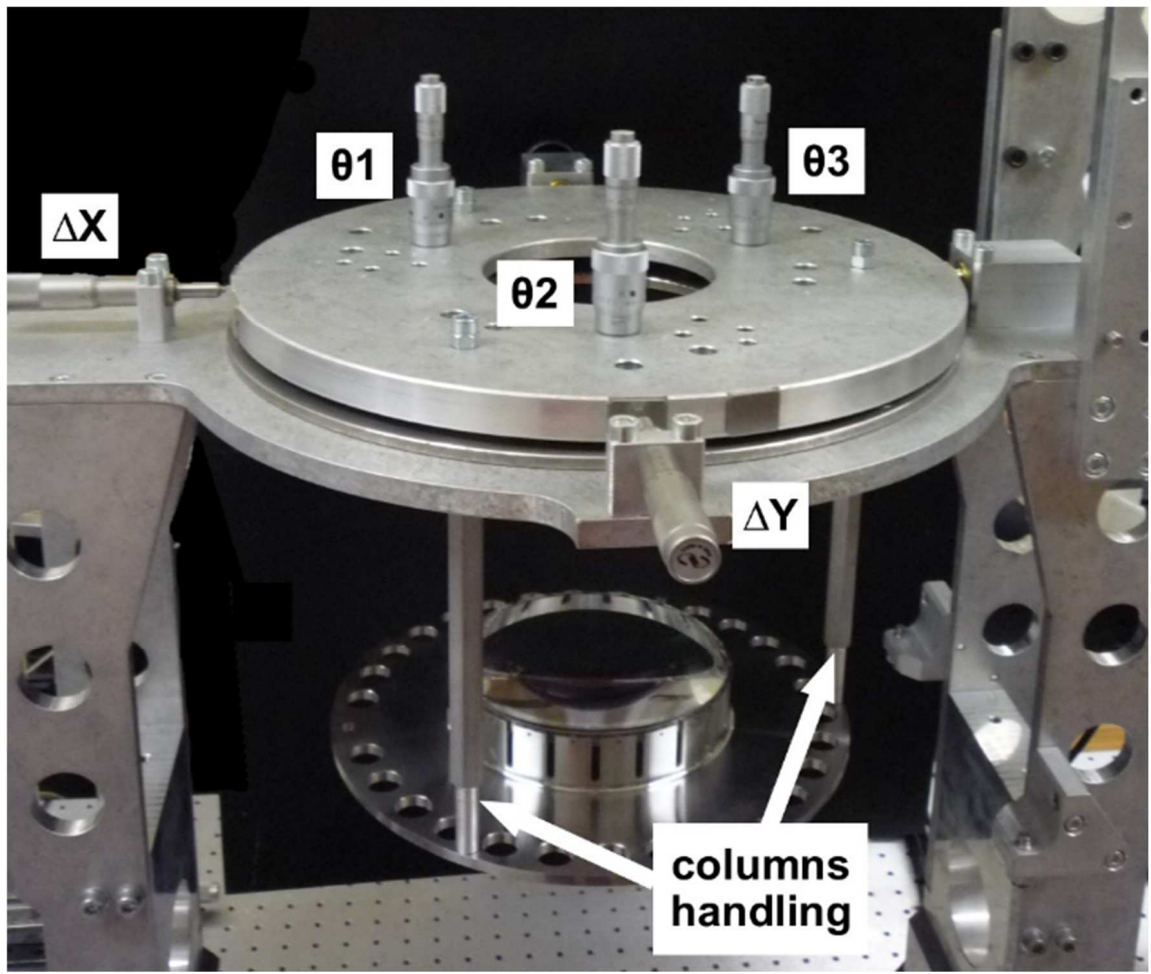

*Figure 104. The centering and tip-tilt alignment GSE 'manipulator'.*

The observables are:

• decenter: the recorded position of the spot with L3 only on CCD-4, which is the reference, and the position of the spot L3+L4, which must superimpose with the reference;

• tilt: the position of Newton's rings back-reflected by L4 surfaces is compared with the reference position of the beam on CCD-2, defined in auto-collimation of the laser beam with the flat mirror M5. The Newton rings have to be made uniformly illuminated as much as possible;

• focus: the focus position of the lens is checked with the CMM machine, by comparing the measurement of the vertex of the previously inserted lens with the vertex measurement of the current lens.

At this point, L4 is removed from the prototype structure, the right shimming to accomplish the measured tip-tilt and focus are inserted in the right position in the interface plate, L4 is placed back into the structure onto the shims, its centering is again performed by acting on $\Delta X$ and $\Delta Y$ micrometers, superimposing the transmitted spot on the CCD-4 reference. Finally, the screws are tightened with the bolt elongation tools.

Alignment results:

- decenter for x= 5.1 μm and for y=-0.3 μm corresponding to a total decentre of about 5 μm;

- tip-tilt for $\delta$=25 arcsec and $\varphi$=16 arcsec corresponding to about 30 arcsec of overall tilt;

- focus: ~5 μm.

Additional error sources to be considered for the decentring are:

- the error in the initial repositioning of the transmitted spot to the last lens aligned position, the operation performed in the preliminary phase before inserting the current lens, see section 2.6.8;

- the stability of the setup over the alignment time, see section 2.6.2;

- the movement of the spot due to the lens insertion and GSE installation, see section 2.6.5;

- the error in the centroid computation and repeatability of the measurements, see section 2.6.6.

Their linear sum is 30 μm PtV, while their quadrature sum is 16 μm PtV.

Additional error sources to be considered for the tilt are:

- the error in the initial repositioning of the back-reflected spot to the previous reference, the operation performed in the preliminary phase before inserting the current lens, see section 2.6.8;

- the stability of the setup over the alignment time, see section 2.6.2;

- the movement of the spot due to the lens insertion and GSE installation, see section 2.6.5;

- the error in evaluating the intensity profile of the Newton rings and short time stability of the measurements, see section 2.6.7.

Their linear sum is 40 arcsec PtV, while their quadrature sum is 20 arcsec PtV.

An additional error source concerning the defocus is the accuracy of the CMM, which is ±15μm.

## 2.7.7   L5 results

We first double-check the alignment beam and tune it if necessary. After that, with the prototype structure in the 0° orientation, we insert L5 on the foreseen shims, setting the nominal L5 position in focus. We then align L5 to the laser beam, previously co-aligned with L3 and L4 in the 0° orientation.

The allowed adjustments are ΔX, ΔY on the GSE manipulator micrometers for centering adjustments, θ1, θ2 and θ3 on the GSE tilt micrometers, whose conversion is 10 μm to ~12 arcsec of L5.

The observables are:

- decenter: the recorded position of the spot with L3 and L4 on CCD-4, which is the reference, and the position of the spot L3+L4+L5, which must superimpose with the reference;

- tilt: the position of Newton's rings back-reflected by L5 surfaces was compared with the reference position of the beam on CCD-2, defined in auto-collimation of the laser beam with a flat mirror. The Newton rings have to be made uniformly illuminated as much as possible;

- focus: the focus position of the lens was checked with the CMM machine, by comparing the measurement of the vertex of the previously inserted lens with the vertex measurement of the current lens.

At this point, L5 is removed from the prototype structure, the right shimming to accomplish the measured tip-tilt and focus are inserted in the right position in the interface plate, L5 is placed back into the structure onto the shims, its centering is again performed by acting on ΔX and ΔY micrometers, superimposing the transmitted spot on the CCD-4 reference. Finally, the screws are tightened with the bolt elongation tools.

Alignment results:

- decenter for x= 4.1 μm and for y=-0.1 μm corresponding to a total decentre of about 2 μm;

- tip-tilt for δ=5.4 arcsec and φ=7.9 arcsec corresponding to about 10 arcsec of overall tilt;

- focus: ~10 μm.

Additional error sources to be considered for the decentring are:

- the error in the initial repositioning of the transmitted spot to the last lens aligned position, the operation performed in the preliminary phase before inserting the current lens, see section 2.6.8;

- the stability of the setup over the alignment time, see section 2.6.2;

- the movement of the spot due to the lens insertion and GSE installation, see section 2.6.5;

- the error in the centroid computation and repeatability of the measurements, see section 2.6.6.

Their linear sum is 30 μm PtV, while their quadrature sum is 16 μm PtV.

Additional error sources to be considered for the tilt are:

- the error in the initial repositioning of the back-reflected spot to the previous reference, the operation performed in the preliminary phase before inserting the current lens, see section 2.6.8;

- the stability of the setup over the alignment time, see section 2.6.2;

- the movement of the spot due to the lens insertion and GSE installation, see section 2.6.5;

- the error in evaluating the intensity profile of the Newton rings and short time stability of the measurements, see section 2.6.7.

Their linear sum is 40 arcsec PtV, while their quadrature sum is 20 arcsec PtV.

An additional error source concerning the defocus is the accuracy of the CMM, which is ±15μm.

## 2.7.8   L6 results

We first double-check the alignment beam and tune it if necessary. After that, with the prototype structure in the 0° orientation, we insert L6 on the foreseen shims, setting the nominal L6 position in focus. We then align L5 to the laser beam, previously co-aligned with L3, L4, and L5 in the 0° orientation.

The allowed adjustments are $\Delta X$, $\Delta Y$ on the GSE manipulator micrometers for centering adjustments, θ1, θ2 and θ3 on the GSE tilt micrometers, whose conversion is 10 μm to ~10 arcsec of L6.

The observables are:

- decenter: the recorded position of the spot with L3, L4 and L5 on CCD-4, which is the reference, and the position of the spot L3+L4+L5+L6, which must superimpose with the reference;

- tilt: the position of Newton's rings back-reflected by L6 surfaces is compared with the reference position of the beam on CCD-2, defined in auto-collimation of the laser beam with a flat mirror. The Newton rings have to be made uniformly illuminated as much as possible;

- focus: the focus position of the lens is checked with the CMM machine, by comparing the measurement of the vertex of the previously inserted lens with the vertex measurement of the current lens.

At this point, L6 is removed from the prototype structure, the right shimming to accomplish the measured tip-tilt and focus are inserted in the right position in the interface plate, L6 is placed back into the structure onto the shims, its centering is again performed by acting on $\Delta X$ and $\Delta Y$ micrometers, superimposing the transmitted spot on the CCD-4 reference. Finally, the screws are tightened with the bolt elongation tools.

Alignment results:

- decenter for x= 18.1 μm and for y=-14.1 μm corresponding to a total decentre of about 23 μm;

- tip-tilt for $\delta$=16.4 arcsec and $\varphi$=13.9 arcsec corresponding to about 21 arcsec of overall tilt;

- focus: ~13 μm.

Additional error sources to be considered for the decentring are:

- the error in the initial repositioning of the transmitted spot to the last lens aligned position, the operation performed in the preliminary phase before inserting the current lens, see section 2.6.8;

- the stability of the setup over the alignment time, see section 2.6.2;

- the movement of the spot due to the lens insertion and GSE installation, see section 2.6.5;

- the error in the centroid computation and repeatability of the measurements, see section 2.6.6.

Their linear sum is 30 μm PtV, while their quadrature sum is 16 μm PtV.

Additional error sources to be considered for the tilt are:

- the error in the initial repositioning of the back-reflected spot to the previous reference, the operation performed in the preliminary phase before inserting the current lens, see section 2.6.8;

- the stability of the setup over the alignment time, see section 2.6.2;

- the movement of the spot due to the lens insertion and GSE installation, see section 2.6.5;

- the error in evaluating the intensity profile of the Newton rings and short time stability of the measurements, see section 2.6.7.

Their linear sum is 40 arcsec PtV, while their quadrature sum is 20 arcsec PtV.

Additional error source concerning the defocus is the accuracy of the CMM, which is ±15μm.

## 2.7.9   L2 results

The prototype structure has to be rotated in a way to allow the L2 insertion. The alignment axis has to be moved to be aligned with L3 in the new orientation. This operation is performed by setting the recorded position for the micrometers of the iris (decenter) and of the FM2 (tilt) for the 180° orientation, to go about in the correct position. We then tune the alignment axis, by adding a new error described in 2.6.9.

The reference axis did not move with respect to the previously aligned lens. To perform this operation, we double-check the position of the transmitted spot on CCD-4 and the position of the back-reflected spots on CCD-2.

After that, with the prototype structure in the 180° orientation, we insert L2 on the foreseen shims, setting the nominal L2 position in focus. We then align L2 to the laser beam, previously co-aligned with L3, L4, L5 and L6 in the 180° orientation.

The allowed adjustments are $\Delta X$, $\Delta Y$ on the GSE manipulator micrometers for centering adjustments, $\theta 1$, $\theta 2$ and $\theta 3$ on the GSE tilt micrometers, whose conversion is 10 μm to ~12 arcsec of L2.

The observables are:

- decenter: the recorded position of the spot with L3, L4, L5 and L6 on CCD-4, which is the reference, and the position of the spot L2+L3+L4+L5+L6, which must superimpose with the reference;

- tilt: the position of Newton's rings back-reflected by L2 surfaces is compared with the reference position of the beam on CCD-2, defined in auto-collimation of the laser beam with a flat mirror. The Newton rings have to be made uniformly illuminated as much as possible;

- focus: the focus position of the lens is checked with the CMM machine, by comparing the measurement of the vertex of the previously inserted lens with the vertex measurement of the current lens.

At this point, L2 is removed from the prototype structure, the right shimming to accomplish the measured tip-tilt and focus are inserted in the right position in the interface plate, L2 is placed back into the structure onto the shims, its centering is again performed by acting on $\Delta X$ and $\Delta Y$ micrometers, superimposing the transmitted spot on the CCD-4 reference. Finally, the screws are tightened with the bolt elongation tools.

Alignment results:

- decenter for x=-10.1 µm and for y= 14.1 µm corresponding to a total decentre of about 17 µm;

- tip-tilt for $\delta$=4.8 arcsec and $\varphi$= 5.5 arcsec corresponding to about 7 arcsec of overall tilt;

- focus: ~27 µm.

Additional error sources to be considered for the decentring are:

- the error in the initial repositioning of the beam after the prototype structure rotation, the operation performed in the preliminary phase before inserting the current lens, see section 2.6.9;

- the stability of the setup over the alignment time, see section 2.6.2;

- the movement of the spot due to the lens insertion and GSE installation, see section 2.6.5;

- the error in the centroid computation and repeatability of the measurements, see section 2.6.6.

Their linear sum is 50 µm PtV, while their quadrature sum is 30 µm PtV.

Additional error sources to be considered for the tilt are:

- the error in the initial repositioning of the beam after the prototype structure rotation, the operation performed in the preliminary phase before inserting the current lens, see section 2.6.9;

- the stability of the setup over the alignment time, see section 2.6.2;

- the movement of the spot due to the lens insertion and GSE installation, see section 2.6.5;

- the error in evaluating the intensity profile of the Newton rings and short time stability of the measurements, see section 2.6.7.

Their linear sum is 40 arcsec PtV, while their quadrature sum is 20 arcsec PtV.

An additional error source concerning the defocus is the accuracy of the CMM, which is ±15µm.

## 2.7.10 L1 results

We first double-check the alignment beam and tune it if necessary. After that, with the prototype structure in the 180° orientation, we insert L1 on the foreseen shims, setting the nominal L1 position in focus. We then align L1 to the laser beam, previously co-aligned with L3, L4, L5, L6 and L2 in the 180° orientation. The allowed adjustments are $\Delta X$, $\Delta Y$ on the GSE manipulator micrometers for centering adjustments, $\theta 1$, $\theta 2$ and $\theta 3$ on the GSE tilt micrometers, whose conversion is 10 µm to ~10 arcsec of L1.

The observables are:

- decenter: the recorded position of the spot with L3, L4, L5, L6 and L2 on CCD-4, which is the reference, and the position of the spot L3+L4+L5+L6+L2+L1, which must superimpose with the reference;

- tilt: the position of Newton's rings back-reflected by L1 surfaces is compared with the reference position of the beam on CCD-2, defined in auto-collimation of the laser beam with a flat mirror. The Newton rings have to be made uniformly illuminated as much as possible;

- focus: the focus position of the lens is checked with the CMM machine, by comparing the measurement of the vertex of the previously inserted lens with the vertex measurement of the current lens.

At this point, L1 is removed from the prototype structure, the right shimming to accomplish the measured tip-tilt and focus are inserted in the right position in the interface plate, L1 is placed back into the structure onto the shims, its centering is again performed by acting on $\Delta X$ and $\Delta Y$ micrometers, superimposing the transmitted spot on the CCD-4 reference. Finally, the screws are tightened with the bolt elongation tools.

Alignment results:

- decenter for x=0.1 µm and for y=-4.1 µm corresponding to a total decentre of about 4µm;

- tip-Tilt for δ=15.8 arcsec and φ=-1.4 arcsec corresponding to about 16 arcsec of overall tilt;

- focus: ~17 µm.

Additional error sources to be considered for the decentring are:

- the error in the initial repositioning of the transmitted spot to the last lens aligned position, the operation performed in the preliminary phase before inserting the current lens, see section 2.6.8;

- the stability of the setup over the alignment time, see section 2.6.2;

- the movement of the spot due to the lens insertion and GSE installation, see section 2.6.5;

- the error in the centroid computation and repeatability of the measurements, see section 2.6.6.

Their linear sum is 30 µm PtV, while their quadrature sum is 16 µm PtV.

Additional error sources to be considered for the tilt are:

- the error in the initial repositioning of the back-reflected spot to the previous reference, the operation performed in the preliminary phase before inserting the current lens, see section 2.6.8;

- the stability of the setup over the alignment time, see section 2.6.2;

- the movement of the spot due to the lens insertion and GSE installation, see section 2.6.5;

- the error in evaluating the intensity profile of the Newton rings and short time stability of the measurements, see section 2.6.7.

Their linear sum is 40 arcsec PtV, while their quadrature sum is 20 arcsec PtV.

An additional error source concerning the defocus is the accuracy of the CMM, which is ±15µm.

## 2.7.11 Alignment accuracy

We report in Table 16 the comparison between the prototype alignment tolerances and the achieved accuracies, which are given for each lens as the sum of the lens measured position, and as accuracy the quadrature sum of all the error sources described in the previous sections.

*Table 16. The prototype alignment tolerances (green background) and the achieved accuracies (yellow background), summing the errors of each lens in quadrature.*

| | Alignment tolerances | | | Alignment accuracy | | |
|---|---|---|---|---|---|---|
| Lens | Decenter | Tilt | Focus | Decenter | Tilt | Focus |
| L1 | ± 22 µm | ± 44 arcsec | ± 15 µm | -4 ± 8 µm | -16 ± 10 arcsec | -17 ± 15 µm |
| L2 | ± 22 µm | ± 44 arcsec | ± 30 µm | -17 ± 15 µm | -7 ± 10 arcsec | -27 ± 15 µm |
| L3 | ± 22 µm | ± 44 arcsec | ± 40 µm | -15 ± 15 µm | -10 ± 10 arcsec | 0 ± 15 µm |
| L4 | ± 22 µm | ± 44 arcsec | ± 20 µm | -5 ± 8 µm | -30 ± 10 arcsec | -5 ± 15 µm |
| L5 | ± 22 µm | ± 44 arcsec | ± 20 µm | -2 ± 8 µm | -10 ± 10 arcsec | -10 ± 15 µm |
| L6 | ± 42 µm | ± 44 arcsec | ± 20 µm | -23 ± 8 µm | -21 ± 10 arcsec | -13 ± 15 µm |

While for the tilt there is margin for all the lenses, concerning centering and focus some lenses are at the edge of the tolerances. L2 and L3 decenter may be out of specification depending on the way the errors sum up, and the same applies to the focus of L1, L2, L5, and L6.

The situation is becoming slightly worse if, to the measured lens position, the error sources described in the previous sections are summed up linearly, as shown in Table 17, even if it has to be considered the linear sum of the errors an improbable event to happen.

*Table 17. The prototype alignment tolerances (green background) and the achieved accuracies (yellow background), summing the errors of each lens linearly.*

| | Alignment tolerances | | | Alignment accuracy | | |
|---|---|---|---|---|---|---|
| Lens | Decenter | Tilt | Focus | Decenter | Tilt | Focus |
| L1 | ± 22 µm | ± 44 arcsec | ± 15 µm | -4 ± 15 µm | -16 ± 20 arcsec | -17 ± 15 µm |
| L2 | ± 22 µm | ± 44 arcsec | ± 30 µm | -17 ± 25 µm | -7 ± 20 arcsec | -27 ± 15 µm |
| L3 | ± 22 µm | ± 44 arcsec | ± 40 µm | -15 ± 22 µm | -10 ± 16 arcsec | 0 ± 15 µm |
| L4 | ± 22 µm | ± 44 arcsec | ± 20 µm | -5 ± 15 µm | -30 ± 20 arcsec | -5 ± 15 µm |
| L5 | ± 22 µm | ± 44 arcsec | ± 20 µm | -2 ± 15 µm | -10 ± 20 arcsec | -10 ± 15 µm |
| L6 | ± 42 µm | ± 44 arcsec | ± 20 µm | -23 ± 15 µm | -21 ± 20 arcsec | -13 ± 15 µm |

We report a few considerations on the performed alignment:

- the individual errors which were considered for the various lenses alignment are all PtV errors, which is already a conservative approach. Adding them up linearly seems to be an approach too conservative;

- looking at the results reported in Table 16, achieving better alignment numbers (for example, on the centering of L2 and L3) would significantly improve the probability to be on specification. We want to emphasize that the more significant errors associated to L3 and L2 are due to our specific setup since they are mostly coming from the need to define a reference for L3 and from the need to rotate the prototype structure to insert L2, and thus may be reduced a lot by using different GSEs;

- to achieve better alignment numbers, we strongly suggest that the procedure to tighten the screw will have to be improved; in fact, most of the uncertainty was coming tightening the screws, even if we have been using the dedicated tool. We believe that having a GSE which is holding the mount in place while tightening may minimize the movement of the mount itself during the process;

- even when looking at Table 17, considering different GSEs with the purpose both to minimize the errors associated to the centering of L2 and L3 and to improve the final alignment achievable accuracies, may leave margins to finalize the alignment being entirely sure to be within tolerances;

- the change in the focus measurement procedure obliged us to use a CMM not fully characterized and not optimal because of the need to physically touch the lens vertex. Such a procedure is not acceptable for the real TOUs, and thus a different approach has to be studied, touchless CMM for example;

• the time needed for each lens alignment is ranging from 4 to 8 hours, meaning that the overall alignment time is of the order of 4 to 6 days;

• the procedure has a margin for improvements and simplification and seems to be doable, even if it requires care and skill-full people continuously controlling several degrees of freedoms and it is probably of difficult implementation for a more industrialized approach.

## 2.7.12 Alignment lesson learned

There are many lessons learned that we briefly report here.

All the alignment GSEs that we have been using was acting on a plate, positioned on the top of the prototype structure, to tune both tip-tilt and centering, by using dedicated micrometers and holding the lenses through dedicated columns (see Figure 104). It turned out that this operation is challenging to be performed both because of the correlation between centering and tip-tilt movements and because friction is making the centering movement not smooth, also because we are acting with the micrometers on a plate very far the mount. We strongly suggest to tune the mounts centering directly on the side of the mount, by having dedicated holes on the TOU structure, and also to have a counteracting system, springs, for example, to allow mount movements in all directions.

The tightening process is, in reality, moving the mounts, decreasing in this way, the alignment accuracies that can be achieved. A system that constraint the mount position while tightening the screws may minimize such movements, allowing to improve the final alignment accuracy.

The method used for the prototype lens focus positioning, the CMM touching the lenses vertex, paying precautions to avoid damaging the lenses, did not show any sign of consequences on the lenses themselves. A touchless measurement of the lenses focus positioning will further reduce the positioning error.

An effort has to be also placed in ensuring a more repeatable tightening process since the achieved positioning accuracy of the lenses is sharply reduced.

The errors associated with the procedure are of the order of ±15 μm for all the lenses, but L2 and L3 may be easily improvable.

## 2.8  Quality of the alignment

The prototype mechanical structure is equivalent to the TOU final structure, in terms of thermal behavior, to mimic the relative movements of the lenses when going from "warm" to "cold" conditions. The purpose of such a prototype is to validate an alignment procedure and test, both in warm and in cold conditions, the on-axis and off-axis optical performance of the system.

In this section, we describe the result of the optical tests performed to assess the optical quality of the system both using interferometry and focal plane images analysis. The warm tests were conducted at the Padova laboratory, while the cold tests in the cryo-vacuum chamber Cryotec at the Leonardo facilities in Campi Bisenzio, Florence tests, where the warm test was re-iterated.

We performed three different tests on the TOU:

- a standard interferometric test, performed with an F/1.5 transmission sphere to achieve the proper TOU input F-number and a reference flat mirror with tip-tilt capabilities;

- a PSF test with monochromatic light; since the PSF is degrading quickly in warm conditions going off-axis, and since we have in Padova only available a 100mm diameter beam, being the pupil size of the TOU 120mm instead, only a few positions have been tested under certain conditions, see section 2.8.2;

- a Hartmann test, to measure the optical quality of the TOU. Again, this test is limited to a 100mm diameter beam at the Padova laboratory.

Due to the limitations in terms of the collimated beam diameter in Padova laboratory, we did repeat at Leonardo both the warm interferometric test and the warm Hartmann test, with precisely the same setup used for the cold test and with the right collimated beam diameter.

### 2.8.1  Warm Interferometric Test in Padova

The goal is to test the performance of the TOU with the Zygo FlashPhase GPI in the Padova laboratory and to practice with the TOU/Interferometer setup in differents angles and rotational positions. To do this we use two different interferometric tests, named setup #1 and setup #2.

Figure 105 shows a first simplified schematic of the experimental setup #1 that has been used. The interferometer is equipped with a transmission spherical element F/1.5, illuminating the TOU from its focal plane. A high quality, λ/20 PtV, reference flat mirror located on the other side of the TOU reflects the light back into the interferometer for wavefront analysis. Since the light is passing twice through the optical element under examination, the TOU in this case, this procedure is called a double-pass interferometric test. We point out that in test setup #1 the Zygo Spherical element (F/1.5) has a beam aperture of 36.9°, while the TOU (theoretically F/2) has a beam aperture of 28.1°. This means that the entrance pupil of 120mm of the TOU, near L3, is fully illuminated by the interferometer beam. Due to the fact that the focal plane is 6.345 mm away from the last surface of L6, only a small portion of L6 is illuminated (about 4.25 mm).

Since the interferometer is quite heavy and difficult to handle, in order to verify the system optical quality also off-axis, we rotated the TOU in 6 different positions, using a combination of tilting the TOU with respect to the interferometer beam by -16 and + 16 degrees and a rotation of the TOU around its optical axis by 120 and 240 degrees, see Figure 106 and Figure 107.

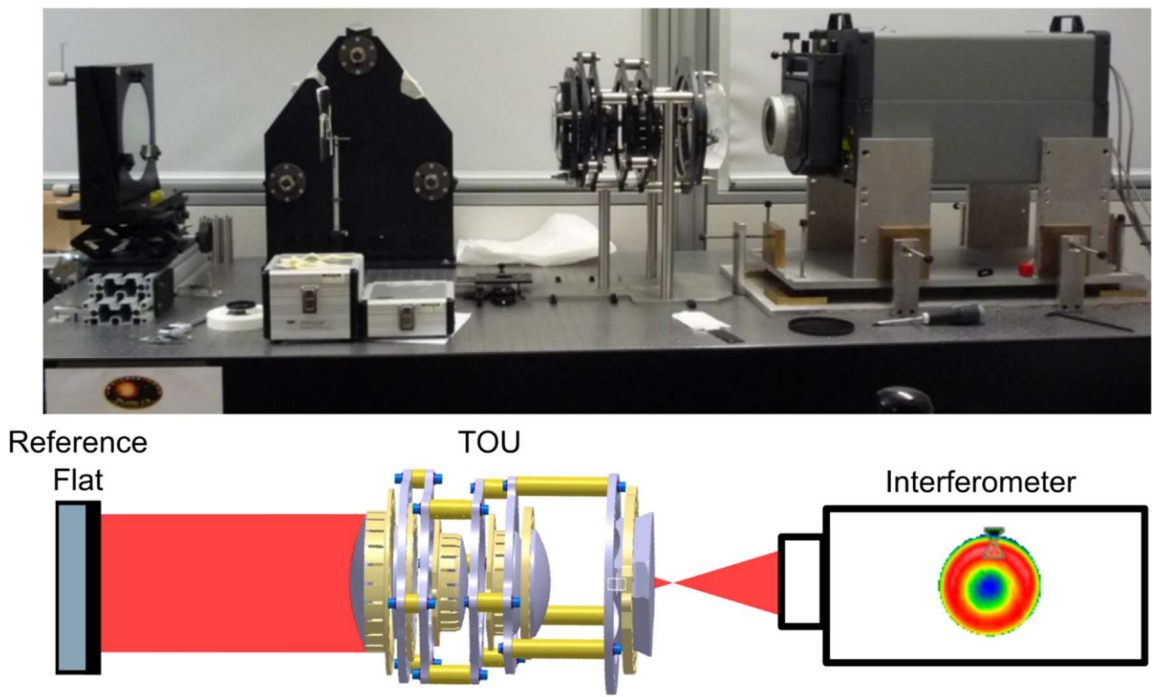

*Figure 105. The warm interferometric test of the TOU with the transmission sphere, setup #1.*

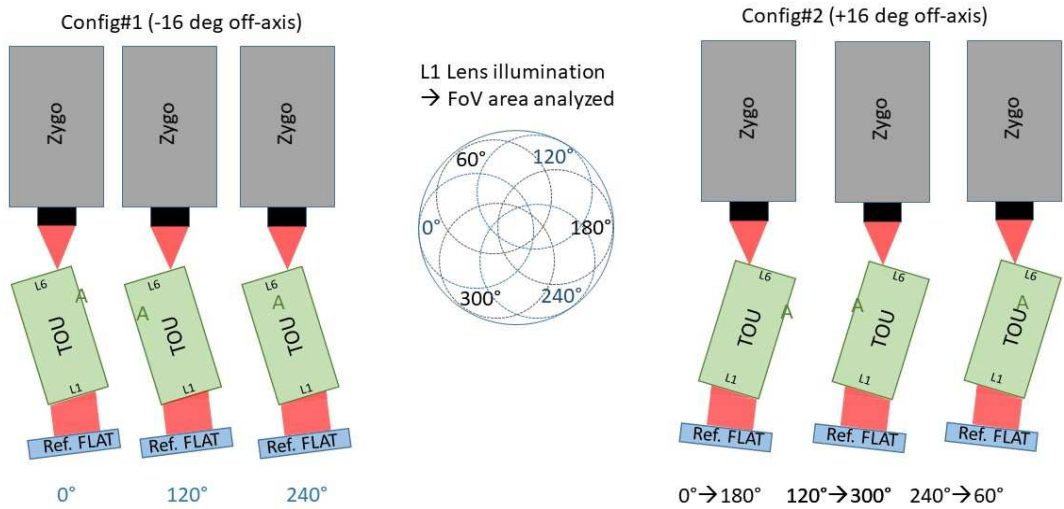

*Figure 106. The scheme of the TOU off-axis position and axial rotation, setup #1.*

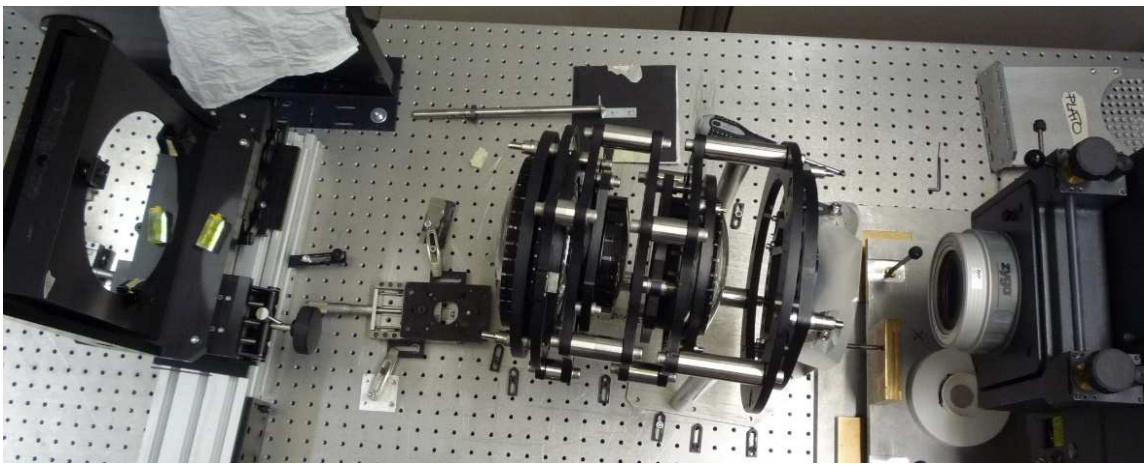

*Figure 107. The interferometric test in the off-axis position, setup #1. The manual translation stage performs a fine focus position of the TOU.*

Figure 108 shows the simplified schematic of the experimental setup #2. The interferometer is now equipped with a flat element that generates a collimated beam of 100mm in diameter. The double-pass interferometric test in this configuration has been possible using a reference spherical mirror. Since the center of curvature of this mirror should lie exactly where the TOU focuses the beam in order to send the light back into the interferometer without introducing high aberrations itself, the mirror was mounted on precise translation stages.

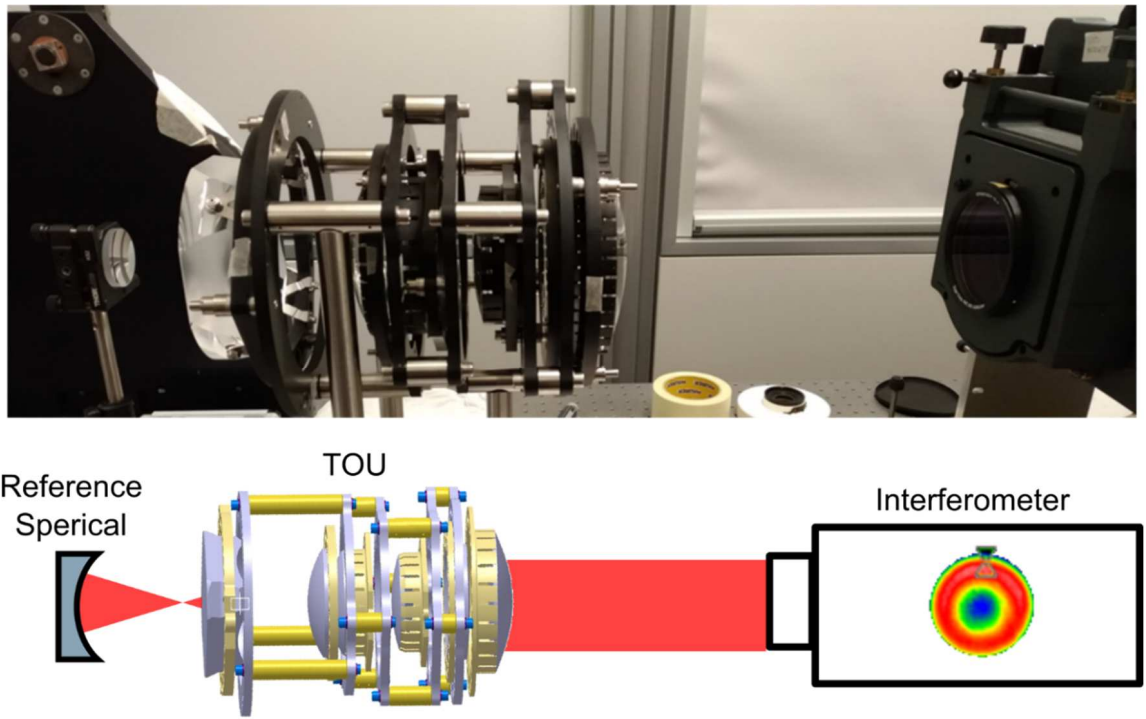

*Figure 108. The warm interferometric test of the TOU with the flat element and aperture of 100mm, setup #2.*

In setup #1 the off-axis angle was estimated from the angle between the axis orthogonal to the TOU optical axis and the reference flat mirror angle. Also, it can be estimated by the decentering of the outcoming axis with respect to the L6 center. With setup #1 we have conducted preliminary investigations for the complete tests in Leonardo:

- Test A: The TOU mainly decentered by 12mm;

- Test 0: On-axis;

- Test B: The TOU tilted by about ±2°.(B-2_0°; B+2_0°);

- Test C: The TOU was tilted and decentered (off-axis 12°+/-1°) for three configurations rotating the TOU around its optical axis, counterclockwise, looking from L1. (C-12_0°, C-12_120°, C-12_240°). C-12_360° is a double-check for C-12_0°.

With this same notation test, B+2_0° should be equivalent to B-2_180°.

A mask equivalent to 97% of the overall aperture was used, as shown in Figure 109, superimposed to the interferogram (with tilt inserted for Zygo measurements).

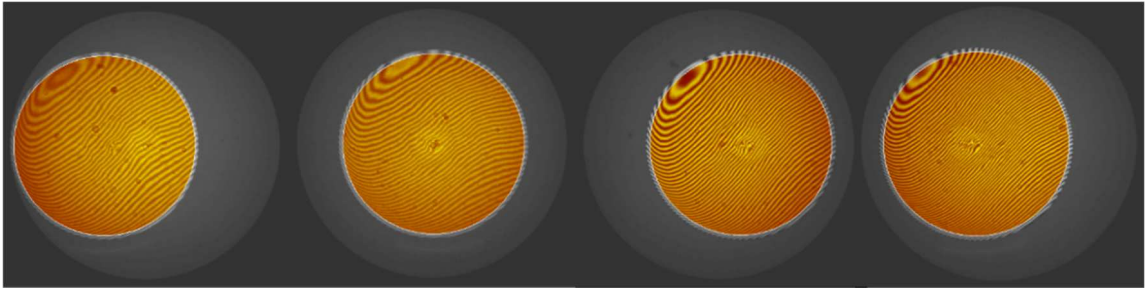

*Figure 109. The mask covering 97% of full aperture, from left to right, for configuration A, 0, B+2, B-2, with setup #1.*

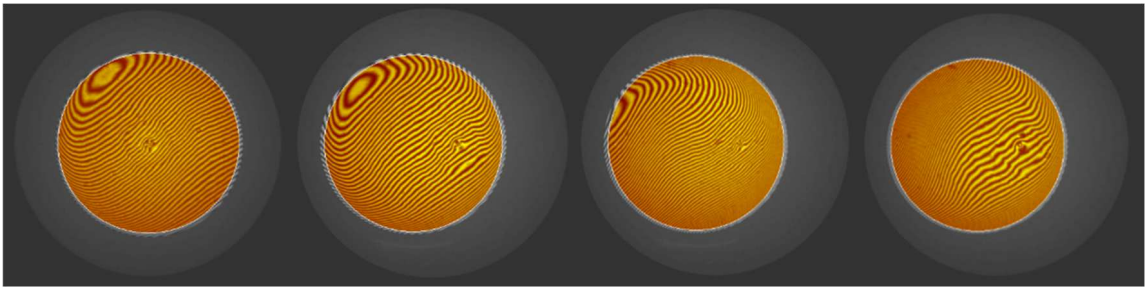

*Figure 110. The mask covering 97% of full aperture, from left to right, for the configuration 0°; C-12_120°, C-12_240°, with setup #1.*

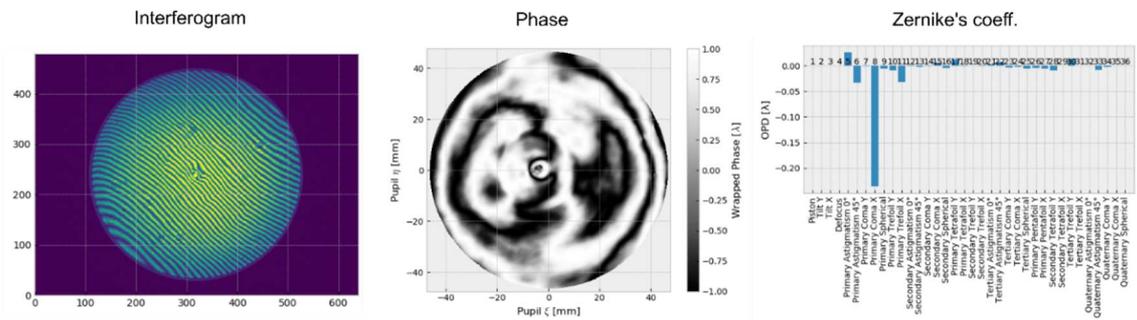

*Figure 111. The on-axis interferogram with analysis of Zernike's coefficients for 95mm mask centered on the TOU aperture.*

The RMS and PtV on the reconstructed wavefront given in Table 18 and Table 19 are subtracted by piston and tilt terms, which are just a result of residual misalignment between the reference mirror and the interferometer beam and are not related to the TOU intrinsic optical quality. The number of points used to sample the wavefront is also shown since Zernike values can be compared only if the amounts of points are comparable. The principal Zernike aberration coefficients are resumed in Table 18.

*Table 18. The Tilt, Astigmatism, and Coma are shown in the quadrature sum of the single Zernike terms for these aberrations, with test setup #1. In green low aberration values, in yellow medium values, and orange/red high aberration values.*

| TOU Z pos.-> | A | B+z | 0 | B-z | 0bis | C-12_0° | C-12_120° | C-12_240° | C-12_360° |
|---|---|---|---|---|---|---|---|---|---|
| **Tilt [wave]** | 0,249 | 0,178 | 0,375 | 0,348 | 0,525 | 0,155 | 0,193 | 1,339 | 0,116 |
| **Def [wave]** | 0,078 | -0,065 | -0,221 | -0,078 | 0,287 | 0,195 | 0,558 | 0,217 | -0,033 |
| **Ast [wave]** | 0,175 | 0,443 | 0,586 | 0,396 | 0,471 | 0,635 | 2,04 | 1,299 | 0,634 |
| **Coma [wave]** | 0,712 | 1,044 | 1,083 | 1,033 | 1,134 | 0,247 | 1 | 2,13 | 0,19 |
| **Spher [wave]** | 0,11 | 0,103 | -0,077 | -0,072 | 0,076 | 0,411 | 0,335 | 0,257 | 0,143 |
| **PtV [wave]** | 1,419 | 2,112 | 1,837 | 1,735 | 1,738 | 1,227 | 3,894 | 1,202 | 1,204 |
| **Rms [wave]** | 0,266 | 0,343 | 0,312 | 0,311 | 0,314 | 0,24 | 0,749 | 0,172 | 0,24 |
| ***Points*** | *57789* | *58279* | *58019* | *59446* | *56974* | *47016* | *45251* | *23291* | *46383* |

The beam diameter has been limited to 95mm in the analysis mask, to compare the results in the two different interferometric setup #1 and #2. For setup #1 we recalculated the aberration coefficients by applying a software mask (79%) to test 0 degrees, in order to have the same illuminated area on L1 (95mm), the result is listed in Table 19.

*Table 19. Results of setup #1 measurements, flat element + L1L6 TOU + spherical reference mirror, axial rotation of 240 deg.*

| PtV [wave] | Rms [wave] | Ast1 [wave] | Ast2 [wave] | Ast Angle [wave] | Coma1 [wave] | Coma2 [wave] | Coma angle [wave] | Spherical [wave] |
|---|---|---|---|---|---|---|---|---|
| 1.384 | 0.221 | 0.088 | -0.063 | -17.7 | -0.641 | -0.005 | 179.5 | -0.0077 |

We repeated the test with a different configuration (setup #2), using the Flat element (101.6 mm beam) on the interferometer, the TOU with L1 facing the interferometer and a Spherical mirror (2" diameter and f = 38.1 mm) as reference with its center of curvature placed in the TOU focus (Figure 108). We first aligned the TOU to the interferometer by looking at the light reflected by L6. Then we placed the spherical reference minimizing the fringes (with mirror tip-tilt) and Defocus Zernike term, by moving it with a stage.

We averaged 20 interferograms with a mask corresponding to 95% of the aperture (Figure 112). Since we did not change the rotation of the TOU along the optical axis, we compare the results to the previous setup #1, test 0 degrees, but applying a software mask in order have the same illuminated area on L1 (i.e. we consider a 95 mm beam illuminating L1 both in the flat element + L1L6 and in the spherical element + L6L1 configurations). The results are listed in Table 20.

*Table 20. Results of setup #2 measurements, flat element + L1L6 TOU + spherical reference mirror, axial rotation of 240 deg, 95mm beam illumination.*

| PtV [wave] | Rms [wave] | Ast1 [wave] | Ast2 [wave] | Ast Angle [wave] | Coma1 [wave] | Coma2 [wave] | Coma angle [wave] | Spherical [wave] |
|---|---|---|---|---|---|---|---|---|
| 1.192 | 0.231 | 0.011 | 0.093 | 41.6 | -0.679 | 0.006 | 179.5 | -0.0852 |

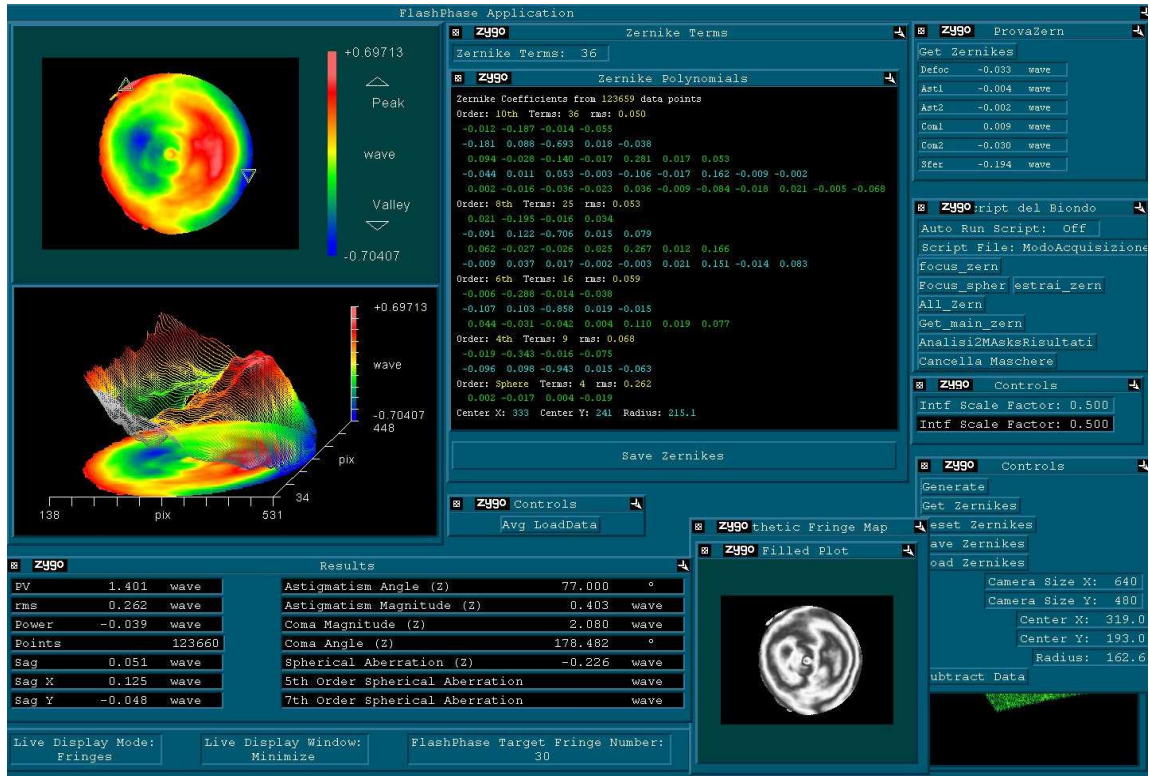

*Figure 112. Interferometric data for setup #2.*

*Table 21. The comparison between the simulation on axis, test setup #1 and #2. Coma aberration is slightly larger than expected from the simulation.*

| Interferometric Test | Simulated (Best/Worst case 1000 Montecarlo) | Measured Setup #1 (objective on interferometer) | Measured Setup #2 (flat element on interferometer) |
|---|---|---|---|
| Interferogram 95mm φ Overall PtV | 0,37/1,68 λ | 1,38 λ | 1,19 λ |
| Interferogram 95mm φ Overall RMS | 0,07/0,31 λ | 0,22 λ | 0,23 λ |
| Interferogram 95mm φ Spherical RMS | 0/0,13 λ | 0,08 λ | 0,09 λ |
| Interferogram 95mm φ Coma RMS | 0/0,45 λ | 0,65 λ | 0,68 λ |
| Interferogram 95mm φ Astigmatism RMS | 0/0,58 λ | 0,10 λ | 0,10 λ |

In Table 21, we report the comparison of the two interferometric tests, noticing that the coma aberration is slightly more significant than expected from the simulation. The simulation was done whit the following hypotheses:

- a collimated beam of 95mm diameter and warm condition (25°C);

- manufacturing tolerances: radius of curvature +/-0.02%, refraction index ±3×10$^{-5}$, Abbe number ±2×10$^{-5}$, thickness ±10µm, surface error from inspection report provided by manufacturer simulated with Zernike low order polynomials from 4 to 15;

- centering tolerances: decenter, tilt and defocus as in Table 2;

- results are computed in waves @ 633nm;

- Montecarlo simulations: 1000;

- Zernike polynomials: "Zernike Fringe Polynomials" i.e. "University of Arizona polynomials".

We conclude that the TOU appears having an asymmetry on optical quality in off-axis C-12 with best values on 0° interferogram, the on-axis analysis figures out coma aberration, see Figure 111. But this test has been then performed again at Leonardo premises using a collimated beam larger than the entrance pupil of the TOU.

## 2.8.2   Warm PSF Test in Padova

The PSF analysis in the warm condition in Padova anticipates the complete test at Leonardo laboratory and has been performed mainly to characterize the GSE and to make experience in advance both in terms of PSF analysis and possible problems to be faced while running the test at Leonardo. This test was not completed for all angles of investigation because the input beam was 101 mm in size and does not cover the entire aperture of PLATO.

However, this phase has been really useful to develop the coding in Python, necessary for the PSF investigation in cold test.

A problem to solve in the PSF analysis was the background subtraction of the images because the acquired images in the climate chamber were taken in monochromatic light, without the possibility of removing the ambient light. The laboratory of Padova is not a dark room, and the possibility of a gradient in the background was limited but not eliminated.

The short exposure time may introduce only read-out noise, but the CCD in warm conditions is different in size and efficiency with respect to cold conditions.

We similarly expect air turbulence due to the long beam path, and probably the PSF will be aligned with respect to a reference before averaged out to improve the signal to noise ration,

Due to the small pixel size of the CCD, 1.67μm with respect to the 18 μm of the PLATO CCD, we assume that the PSF will be oversampled.

To find the best focus position, we performed a sweep around the focal point moving the CCD in steps of 10 μm. For each position we saved several images (100), all images were coaligned with respect to the first image of the series and were combined in a datacube. We extracted from the datacube a median value along the z-axis, and this represented the reference image for the given CCD position. For each median image, we fit a gaussian-2D profile to the PSF, extracted the Gaussian RMS width (σ) and calculated the FHWM = 2.355 x σ. A fourth-order least-squares polynomial fit has been used to interpolate the FWHM values extracted for all the CCD positions. We identified the minimum of such a fit to locate the best focal position. We then moved the stage to this position to acquire an image of the PSF at the best focus. Figure 113 shows the focal point determination (on the left) and the FWHM extraction by the gaussian-2D fitting. Due to not symmetrical shape of the PSF, also confirmed from the simulation, we already know that the PSF FWHM gives only an indicative value of the real energy collected by the pixel size, while the EE computation typically achieved better results.

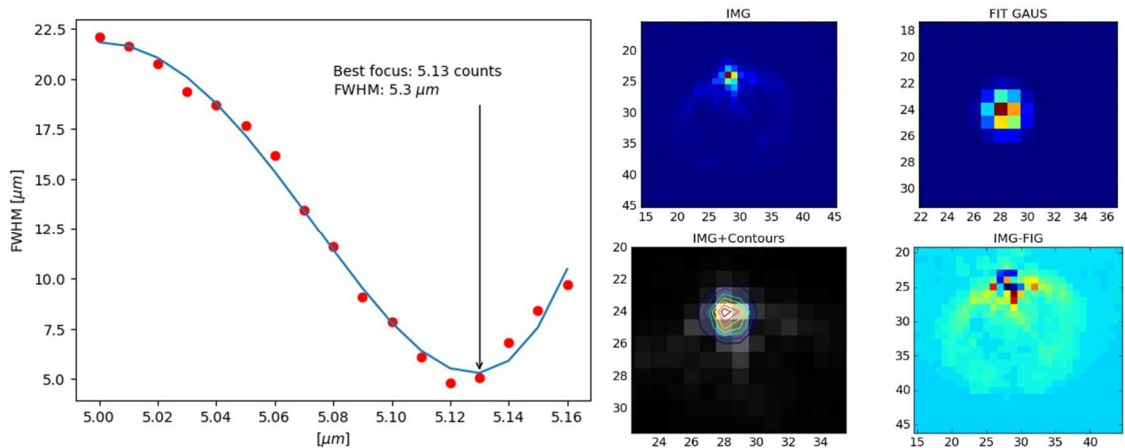

*Figure 113. The PSF on-axis TOU rotation 0°, beam size of 101mm. Left: determination of the PSF best focal position. Right: Gaussian-2D fit of the best focus image.*

For a short comparison between PSF images and interferometric test we collect the result data in a display, see Figure 114, to evaluate the consistency of the two tests. The PSF image is plotted with "spectrum" colormap to underline the low-level information from the outer part of the PSF, very few counts with respect to the center. The interferograms, due to the limited size of the beam to 100mm, are taken shifting the TOU around their optical axis. The interferograms represented with the border yellow on-axis, red for 3 degrees off-axis, and blue for 12 degrees off-axis. The PSF with gray border are taken at 16 degrees off-axis. We see a correspondence between the coma aberration of the PSF and the value measured in the interferogram.

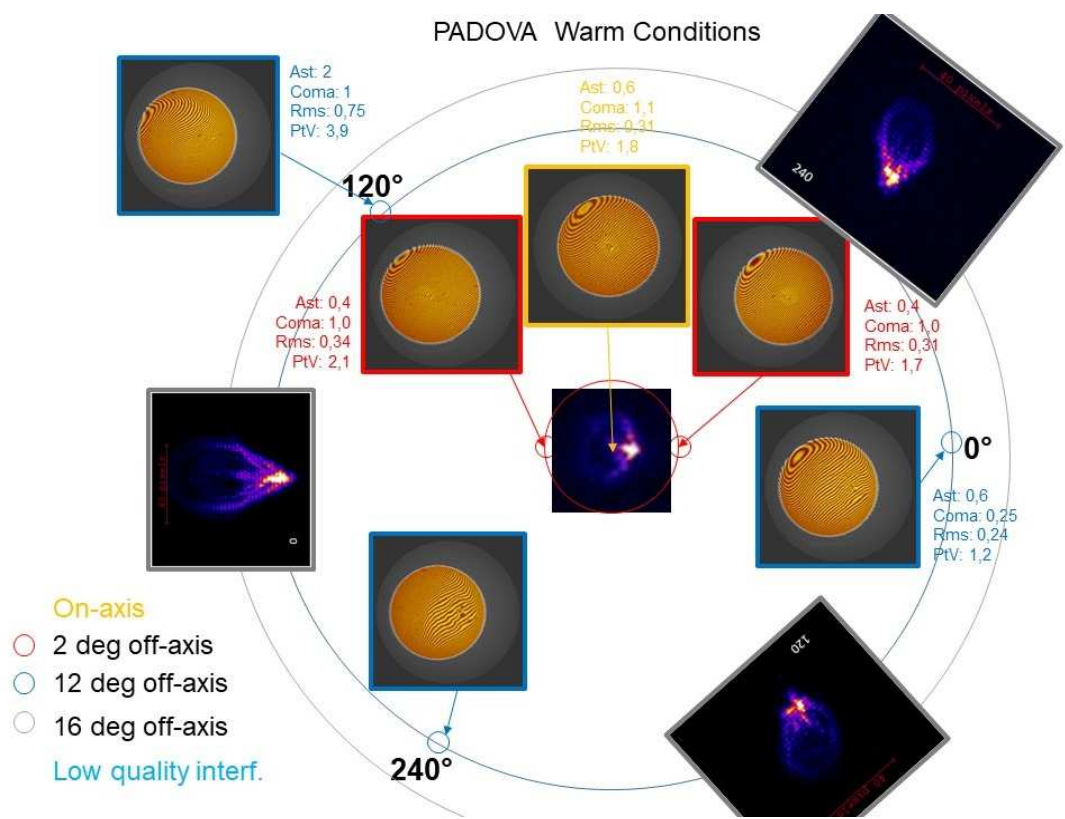

*Figure 114. The comparison in warm condition for the PSF images and interferograms analysis at 0°,120°,240°, and 2°,12°,16° off-axis, see legend and text.*

### 2.8.3   Warm Hartmann test in Padova

Hartmann test (Malacara-Hernández and Malacara-Doblado, 2015) is a screen test for quality examination of an optical system. This test needs a screen, Hartmann plate, which has several holes. The screen should be located just in front of the optical system under examination. The beam passing through a hole can be treated as a ray. For the ray tracing near the focus, we take a pair of images in front of and behind the focus and connect the pairs of points which are the images from the same hole. We calculate the coordinates of the rays on the focal plane by using linear interpolation. The mean distance between the rays and the optical axis on the focal plane is *Hartmann constant*.

In the case of an ideal optical system, *Hartmann constant* is zero. A real optical system, however, has somewhat aberrations; therefore Hartmann constant is greater than zero.

The coordinates of the rays on the focal plane are related to the aberrations of the geometrical optics. A relation between geometrical and wavefront aberration is the *Nijboer relation*, which is shown below.

$$\delta x = R \frac{\partial W}{\partial x} \, , \qquad \delta y = R \frac{\partial W}{\partial y}$$

Where $\delta x$ and $\delta y$ are the coordinates of a ray on the focal plane, $R$ is the distance between the exit pupil and the focal plane, $W$ is the wavefront equation, expressed with Zernike polynomial. By using the above relation, we get the coefficients of Zernike polynomial by the least square fitting method. Consequently, we get followings:

- the size of blur spot on the focal plane, Hartmann constant;

- the wavefront from Zernike polynomial.

We perform a Hartmann test to obtain the enclosed energy on the focal plane and the best focus position. The test was carried out on axis, with monochromatic light and in warm conditions (T ≈ 23.5 °C). The setup was composed by:

- Zygo FlashPhase GPI interferometer with 4" diameter transmission flat. λ = 632.8nm;

- 2748 X 3840 px CCD, px size: 1.67 µm;

- tip-tilt mount + Linear stage (10 µm resolution) to hold the CCD;

- Hartmann mask with 76 holes;

- TOU held in the horizontal configuration.

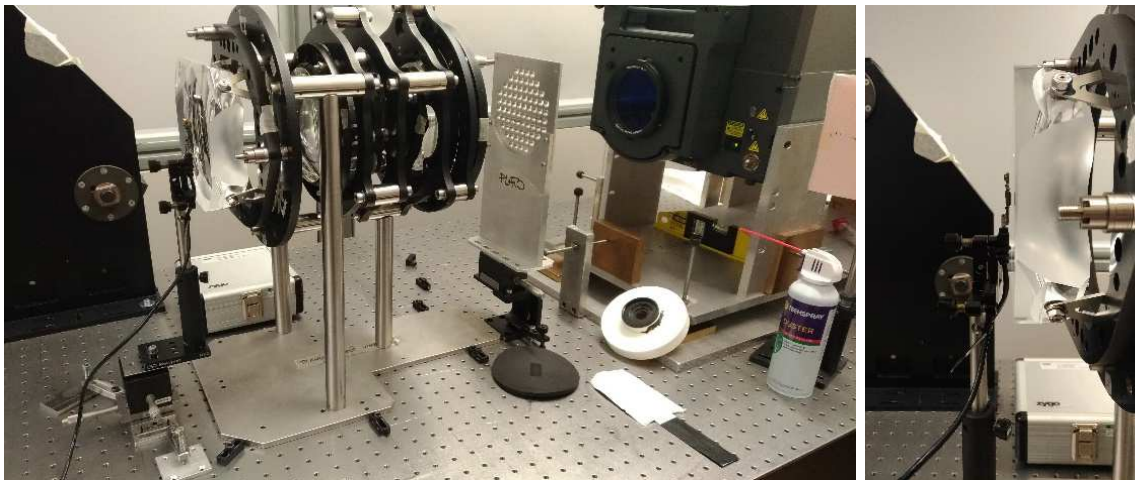

*Figure 115. On the left: the Hartmann test setup. On the right: the CCD in the focal plane, near L6 surface.*

The TOU was placed in front of the interferometer with L1 facing it, see Figure 115, and aligned to the interferometer. For the tip-tilt, we minimized the fringes of the interferogram by looking at the L6 back-reflected light. We aligned it in decenter by measuring the center of the transmitted beam with and without the TOU: the two centers should coincide.

For the Hartmann test, we used the interferometer to provide a collimated beam to materialize the TOU focus at 6.345 mm (nominal) behind L6. The CCD, on a tip-tilt mount, is placed behind the TOU on a linear stage to be able to move it in intra and extra focal positions. The Hartmann mask, Figure 116, is put between the interferometer and the TOU, roughly aligned in tip-tilt and aligned in decenter by looking at the quality of the spots on the CCD in extra-focal position.

We acquired 100 images for two intra and four extra-focal positions of the CCD and average them. In the zero position of the linear stage, the CCD is about 14 mm far from the rear surface of L6. The considered positions are: 0, 2, 3, 4 mm (extra-focal), 11, 12.92 mm (intra-focal).

With the 4 inches diameter beam (101,6mm), we are able to illuminate 52 mask holes, resulting in 52 spots on the CCD.

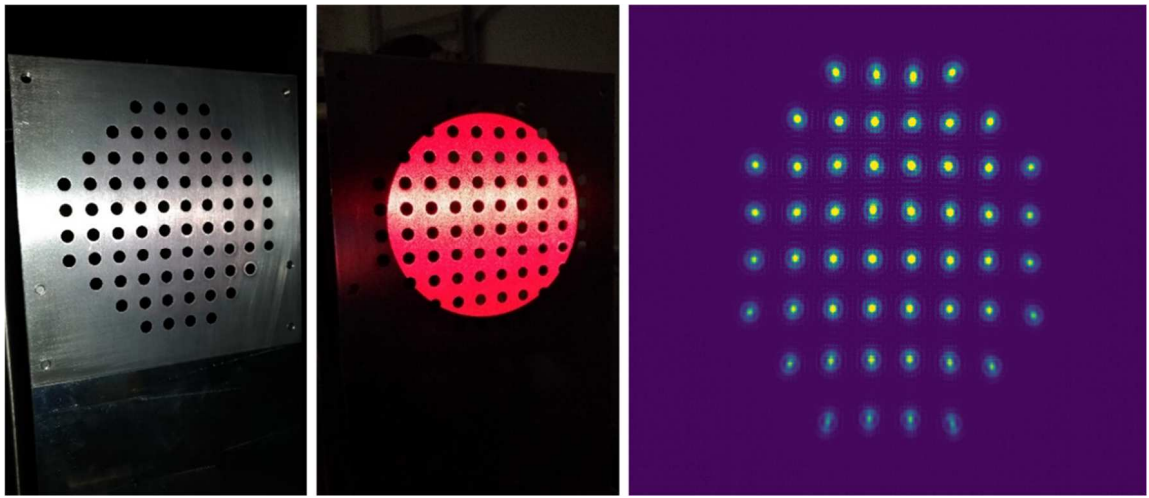

*Figure 116. On the left: the 76 holes Hartmann mask. In the middle: the 52 holes illuminated by 4 inches laser beam. On the right: the CCD image of the 52 spots.*

We apply the following steps to find the centroid positions:

- average the images creating a datacube and extracting the mean value along the z-axis;

- smooth the averaged image with gaussian 3 px filter;

- find 52 maxima on spots and compute the coordinates by a gaussian-2D fitting;

- store maxima coordinate in a reference file.

We calculate the best focal plane position recalling the position of the centroid, fitting the lines coefficient for each conjugated spot and making a dense matrix of lines passing through the focus. We plot the intersection of fitted lines with z-slicing planes, calculating the RMS of the intersections spatial distribution. In the z-axis we found the position where this RMS is minimum; this defines the best focal plane. In this position, we recalculate the intersection of the fitted lines with the best focal plane. Figure 117 shows an example of Hartmann code analysis, in this case with three intra focal positions and four extra focal positions. The intersection of all fitted lines is highlighted by a square.

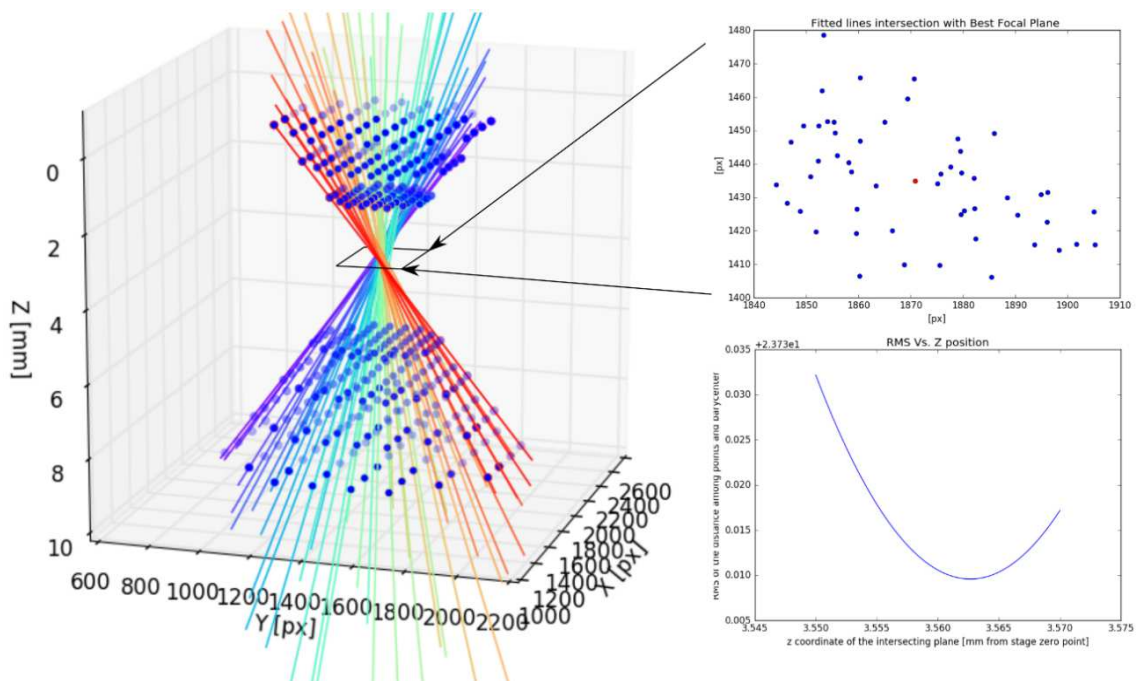

*Figure 117. On the left: line fitting on all 52 points in the on-axis Hartmann test. On the right-top: the intersection of the lines on the Best Focal Plane, red spot is the barycenter. On the right-bottom: The fitting of the minimum of the RMS fitting.*

To computing the Enclosed Energy, we calculate the dimension of the square that encloses 90% of the points in the retrieved best focus position. In this case, we used 47 points over 52. We build a sliding window of increasing size, moving on the image, and we plot the enclosed energy diagram, see Figure 118. We find that 90% of energy was collected in a square of about 21 μm side, i.e. 1.16 pixels. We compare the results with the analysis coming from considering squares and circles centered around the barycenter of the distribution, see Figure 119: the results are 1.30 px and 1.36 px respectively for 90% enclosed energy.

A Montecarlo simulation was performed to compare the test result and has been also useful in Leonardo laboratory. The simulation parameters are:

- wavelength: 632.8 nm;

- ambient conditions: T=23°C, P=1atm3;

- variable beam size, the fraction of ensquared EE and FoV, see table;

- Montecarlo numbers: 500;

- best focal position at Zernike defocus=0;

- tolerance manufacturing: the ray of curvature ±0.02 %, refraction index ±3e-5, Abbe ±2e-5, thickness ±10μm, surface error from inspection report simulated by Zernike's polynomial from 4 to 15;

- decenter, tilt and defocus tolerances follow the design requirements of Table 2;

- Output: Semi-width of the square enclosing n% of energy [μm].

In Table 22 are resumed the values coming from the Montecarlo simulation. Our results for the on-axis field and with the aperture limited to 52 spots (101.6mm) are located near the 80% percentile of the worst case.

*Table 22. The Montecarlo simulation in warm condition for square enclosing energy. The values from the real Hartmann test are marked in blue, between the 80% percentile and worst case for the on-axis field.*

| | **Field** | **on-axis** | | | | **off-axis 16.2°** | | | |
|---|---|---|---|---|---|---|---|---|---|
| **Warm** | **Beam diameter [mm]** | **101.6** | | **46.9** | | **101.6** | | **46.9** | |
| | **Fraction of ensquared EE** | **80%** | **90%** | **80%** | **90%** | **80%** | **90%** | **80%** | **90%** |
| **Semi-width of the square enclosing n% of energy [μm]** | **Nominal** | 4.87 | 8.94 | 2.90 | 6.25 | 27.91 | 39.72 | 10.29 | 14.34 |
| | **Best** | 4.2 | 5.93 | 2.89 | 6.22 | 19.89 | 29.34 | 8.14 | 12.26 |
| | **20% percentile** | 4.92 | 8.43 | 2.99 | 6.29 | 24.27 | 34.59 | 9.62 | 13.57 |
| | **50% percentile** | 5.31 | 9.45 | 3.13 | 6.34 | 26.06 | 37.39 | 10.34 | 14.41 |
| | **80% percentile** | 5.74 | 10.31 | 3.43 | 6.53 | 28.54 | 40.93 | 10.92 | 15.18 |
| | **worst** | 7.15 | 12.84 | 4.63 | 7.13 | 32.39 | 45.35 | 12.56 | 16.95 |

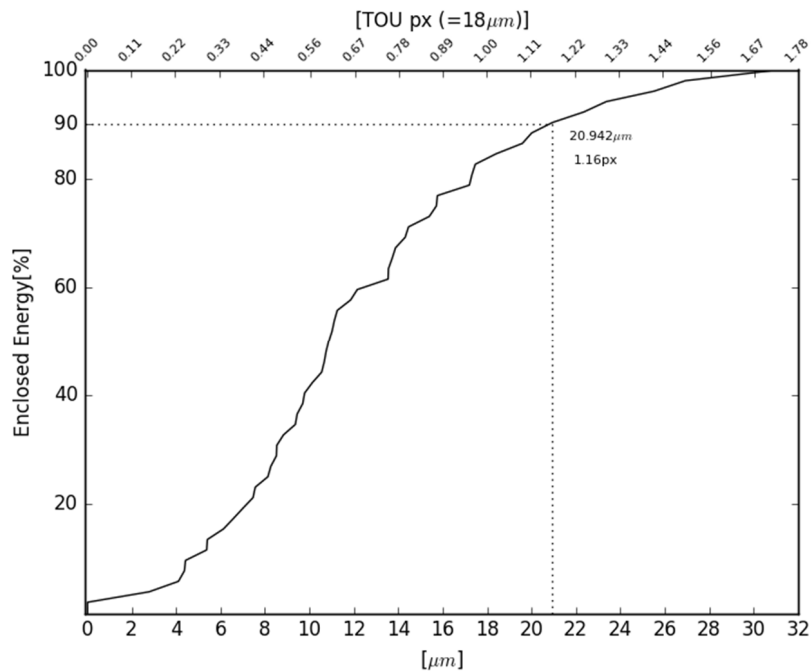

*Figure 118. The Enclosed energy diagram for the 52 points of the Hartmann mask.*

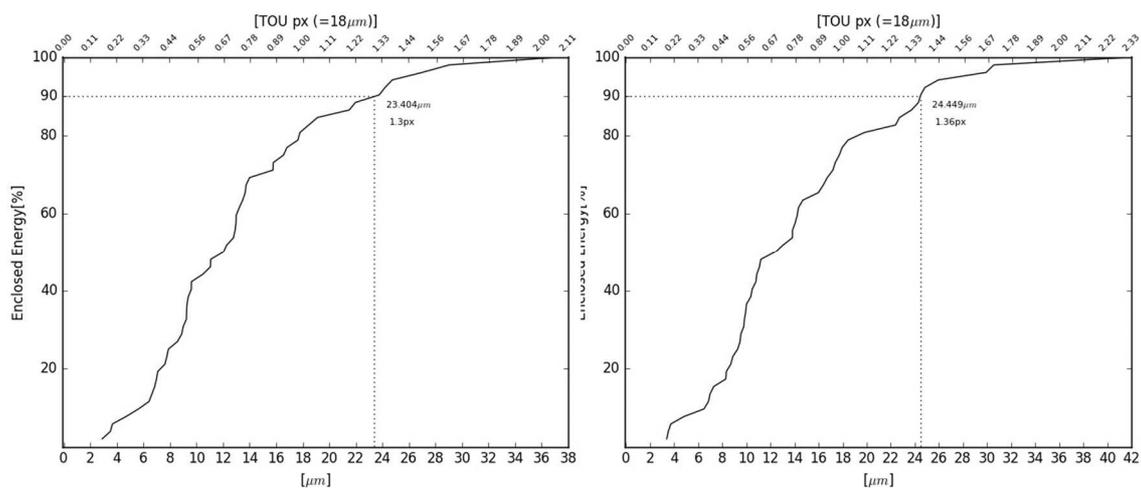

*Figure 119. The Enclosed energy diagram considering a growing square (on the left) and a growing circle (right) both centered on the barycenter of the distribution.*

## 2.8.4   Warm and cold test installation in Leonardo

At Leonardo the setup, described in the following, was the same for both the warm and cold test, being the only difference the Cryotec door open or close, with vacuum and proper temperature set. The Thermal Vacuum Chamber (TVC) "CRYOTEC_1" at Leonardo premises (see Figure 120) has the following characteristics:

· presence of a dedicated cold trap which shall be maintained at, at least, 50°C lower than the expected coldest temperature;

· presence of contamination sample;

· vacuum level up to $10^{-7}$ mbar [from 1000 mbar to $10^{-5}$ mbar in 2 hours]

· internal diameter: 1240 mm

· internal axial length: 1500 mm

· entrance door diameter:  1300 mm

· frontal optical window in quartz [diameter 320 mm]

· presence of lateral optical windows

· baseplate and TVC shroud with separate thermal control in the range:[ -200 °C; +100°C].

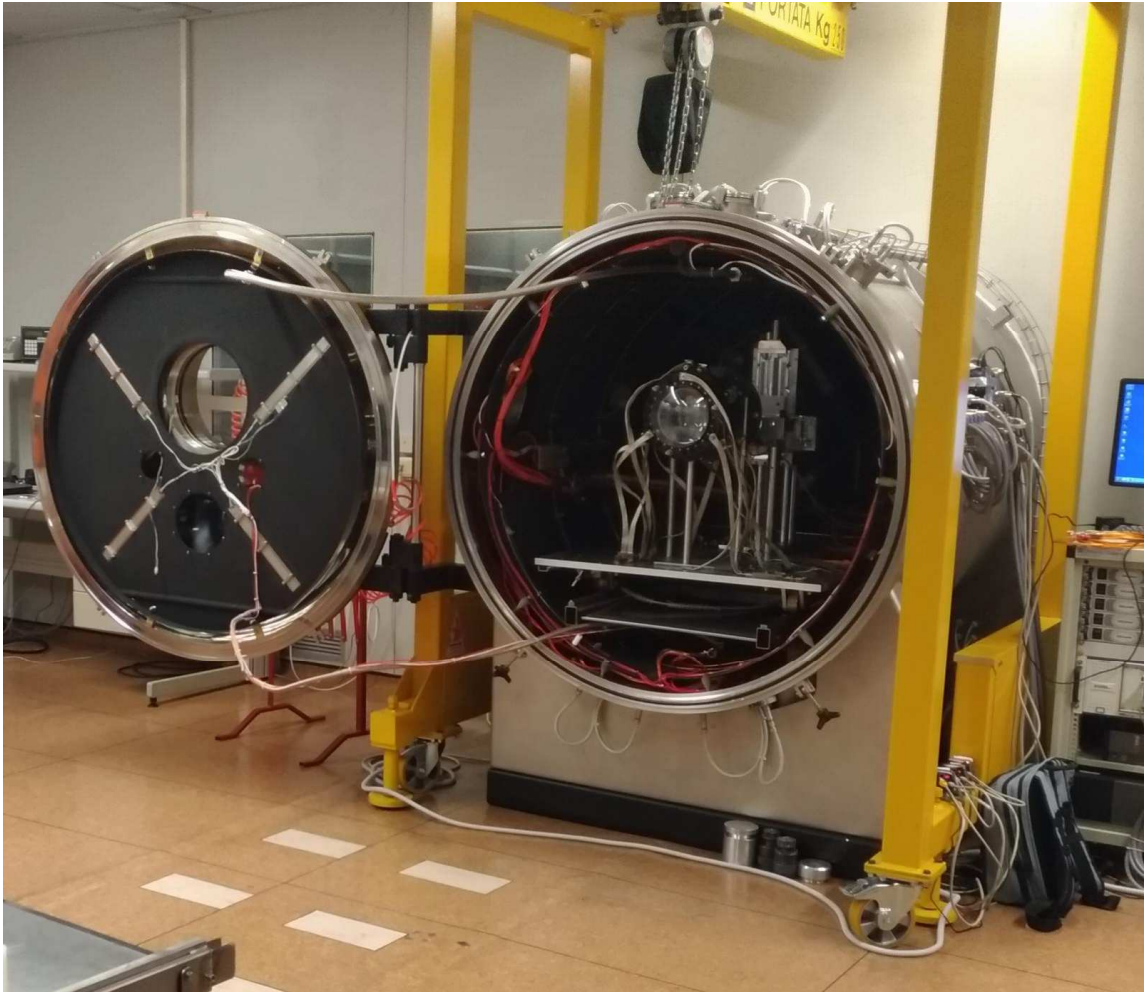

*Figure 120. TVC "Cryotec 1" in ISO-8 @Leonardo with the TOU installed inside (Firenze).*

The light source was a large collimated monochromatic beam of about 300mm in diameter and 632.8nm of wavelength, coming from an interferometer/beam expander optical setup. A folding mirror with 400mm in diameter has been used to fold the beam generated by the beam expander toward the Cryotec camera optical window.

The test foreseen are three separated tests, on and off-axis:

- PSF direct measurements see Figure 121;

- Hartmann measurements see Figure 122;

- Interferometric measurements with full illumination of the entrance lens.

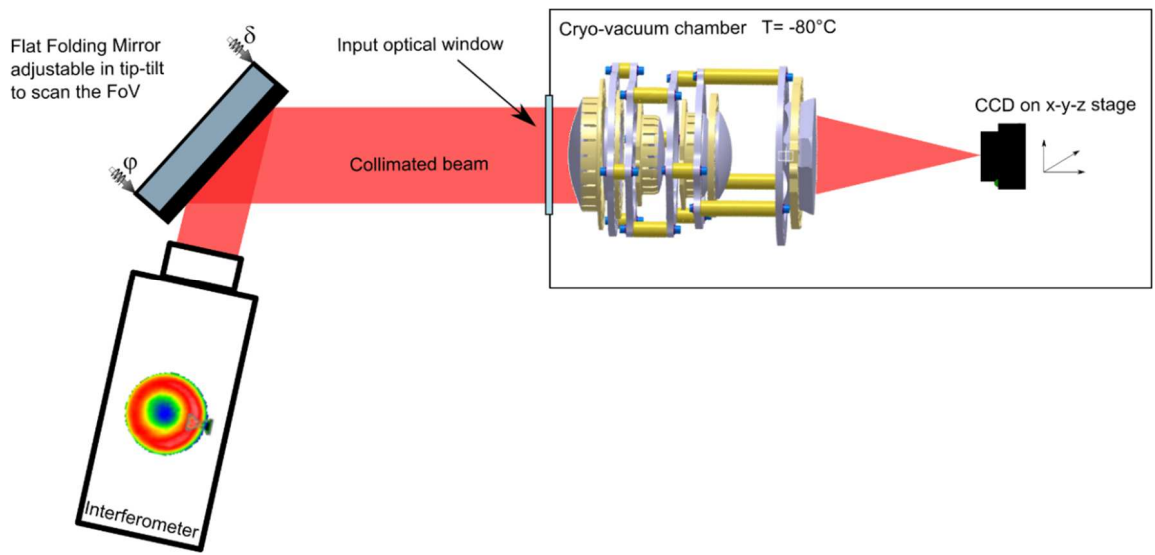

*Figure 121. The conceptual sketch of the cold PSF test*

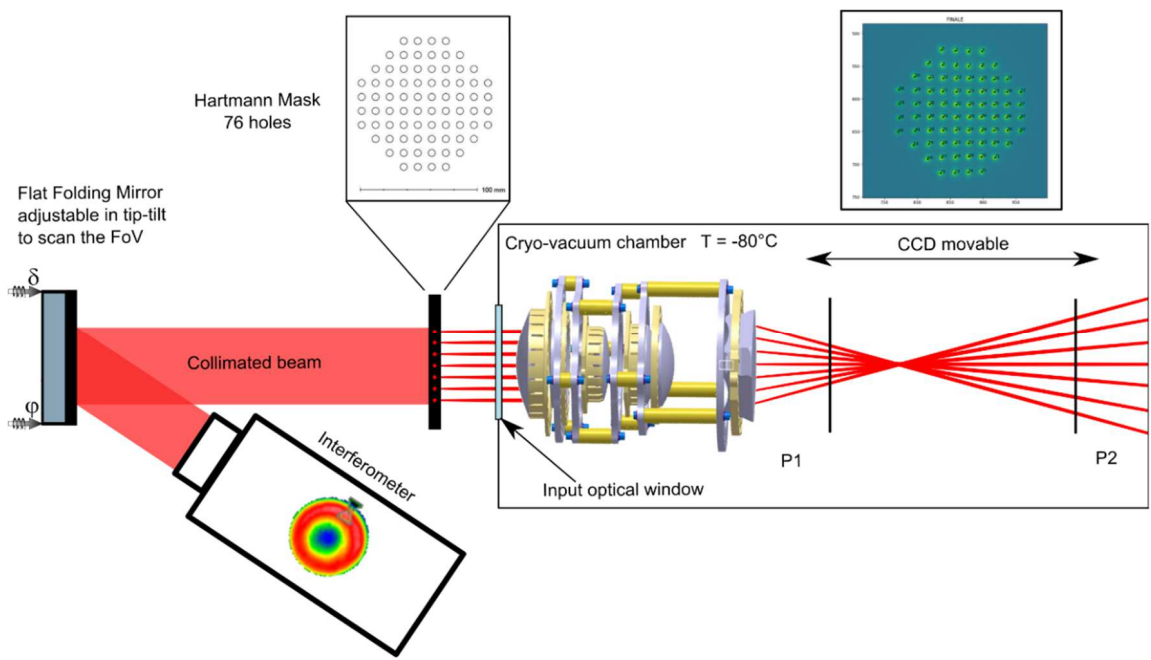

*Figure 122. The conceptual sketch of the cold Hartmann test, with the drawing of the Hartmann,
mask used for the test, each hole has 3mm diameter and the center to center distance between
the holes is 10mm. In total, there are 76 holes distributed over the 120mm pupil.*

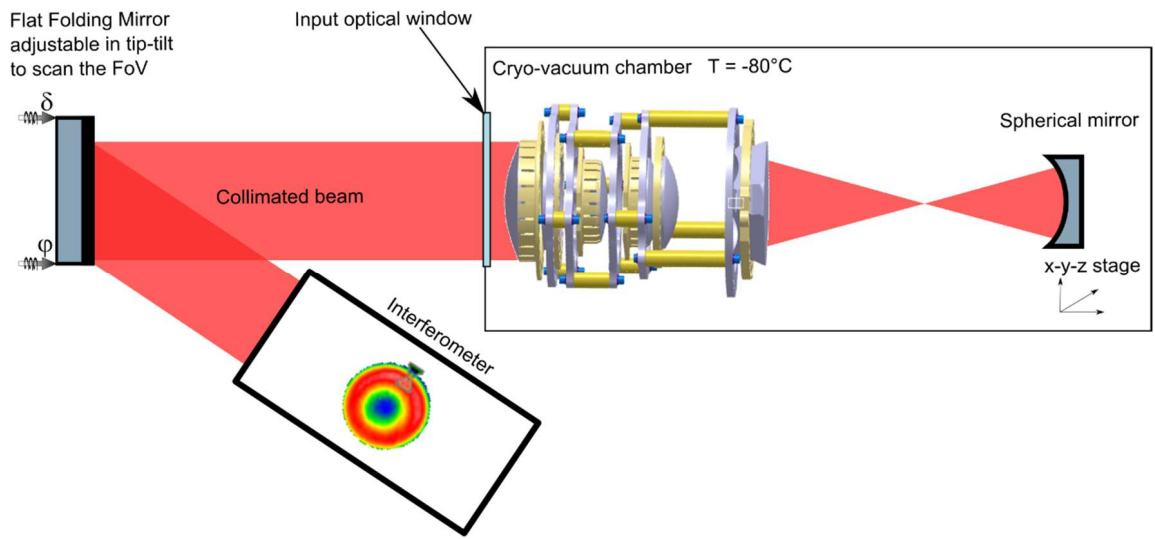

*Figure 123. The conceptual scheme of the cold interferometric test, using a spherical reference mirror in the x-y-z stage.*

The off-axis tests have been performed on six radii displaced by 60 deg and at three distances from the center, equivalent to off-axis angles of 5.6deg, 11.2deg, and 16.7deg, for a total of 19 positions over the FoV.

The off-axis performance has been tested by tilting the input beam on the TOU prototype by using a combination of tilt/shift of the folding mirror horizontally. Due to how the folding mirror support was realized, its shift in vertical position cannot be accomplished, and thus the foreseen off-axis test will require two 120 degree rotations (for a total of 3 test positions) along the optical axis of the TOU prototype, as we also did in Padova. After each rotation, a position in common with the previous TOU prototype orientation (the on-axis one) has been saved to measure any possible shift. Due to the rotation, along the TOU prototype optical axis, we use a hard stop to reposition the prototype after each rotation, in the same place along the optical axis, with a repeatability of about 10 μm, see Figure 124. Each prototype rotation will require the Cryotec to cycle between ambient and cold, which takes several hours. In the following, we intend the Z-axis to be parallel to the TOU prototype optical axis, while X (Horizontal) and Y (Vertical) both orthogonal to it.

As already mentioned, the prototype can be rotated on its support structure of 120 degrees, having thus three possible orientations. A reference screw was screwed on the L6 mount at the Padova laboratory to fix a reference in the TOU 3D space.

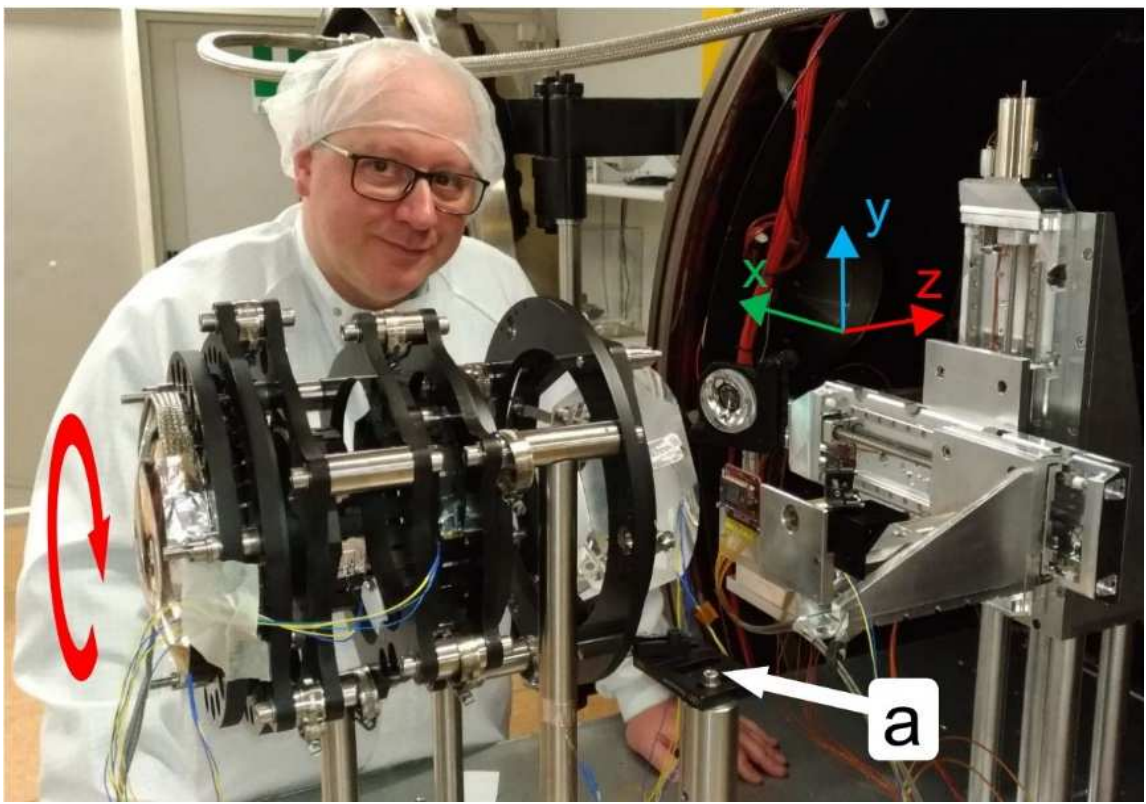

*Figure 124. The TOU in the optical bench of the climate chamber, "a" is the mechanical reference for the z-axis for each rotation of the TOU around its axis, in the three-position of 120° each. The multiple axis stage implements CCD and spherical reference mirrors.*

We define the prototype orientation as follows:

- Looking at the prototype from the L6 side, is the position with the reference screw at about 10 hours, defined position "A".

- Looking at the prototype from the L6 side is the position with the reference screw at about 2 hours, defined position "B".

- Looking at the prototype from the L6 side, is the position with the reference screw at about 6 hours, defined position "C".

The camera used for the cryo-vacuum tests is an Apogee Ascent 2000SL. Since this camera was not rated for cryo-vacuum operations, but the detector could work in such environment, the camera was modified to have the chip separated from the electronics.

The detector is a KAI2020, characterized by a pixel size of 7.4µm x7.4µm, by an active area of 1640x1214 pixels, equivalent to 11.84mm x 8.88mm (14.80mm on the diagonal), and it has a quantum efficiency of about 30% at 620nm.

Three linear stages have been used to span the full TOU FoV with the detector:

- Two Physik Instrument/Micos LS-110 for cryo-vacuum operations, with 190 mm travel range.

- One Micos PLS-85 for cryo-vacuum operations, with 32 mm travel range.

The performances of these stages in cryo-vacuum conditions are not available, but Physik Instrument estimates some of the main parameters to be, at -80 °C:

*Table 23: The main characteristics of the stages when used in cryo conditions. Values estimated by Physik Instrument.*

|  | Pitch [μrad] | Yaw [μrad] | Roll [μrad] | Flatness [μm] | Bidir. Rep. [μm] |
|---|---|---|---|---|---|
| **LS-110** | ± 240 | ± 40 | ± 1200 | ± 10 | ± 3 |
| **PLS-85** | ± 180 | ± 60 | ± 200 | ± 3 | ± 3 |

Prototype "Cold" test on vacuum has also been performed, with the purpose of double-checking the on and off-axis performance in the final working conditions at operating temperature of about -80°C.

The selection of the test to be performed in cold conditions has been made taking into account a few problems to be faced, mostly due to the low testing temperature and to the gradient between the inner and the outer parts of the Climate Chamber. In fact:

- the optical input window of the Climate Chamber is affected by aberrations, the so-called "lens" effect;

- a characterization of such aberrations should be included in the test error budget, and their effect could be minimized by using a parallel beam at the entrance window;

- if the test requires auxiliary optical components inside the chamber, they will introduce additional aberrations and particular care shall be given to their mounts to ensure the proper temperature compensation.

With these criteria, we have identified two tests to be performed on the prototype described in section 2.8.2.

The input optical window was tested with an interferometer to evaluate the lens effect during test operations. A flat reference mirror, with the surface quality of 16nm (RMS) was used in different conditions inside the TVC:

- with the door open at 1 atm, to evaluate the aberration of the mirror;

- with the door closed at 500mbar, $10^{-2}$ mbar, and $10^{-6}$ mbar, to evaluate the aberration of the window.

The heat of this window was lost only for radiation and the thermal gradient was small, as a result, the pressure gradient exceeds that the temperature gradient, see Figure 125.

Figure 126 resumes the 3rd-Order Seidel Aberrations, using 35 Zernike terms, during the depressurization of the TVC. A residual lens effect is present at $10^{-6}$ mbar, with the amount of 90nm in power, not negligible for absolute interferometric measurements. The optical quality tests performed with this TVC, evaluate the quality of the TOU by relative measurements, and the lens effect became negligible. Figure 127 shows the interferogram of the input optical window of the TVC at $10^{-6}$ mbar.

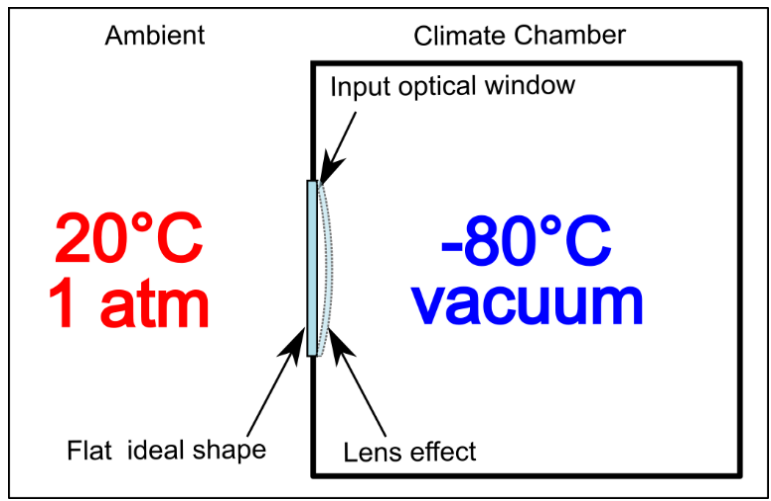

*Figure 125. The resulting lens effect at the entrance of the TVC "Cryotec 1".*

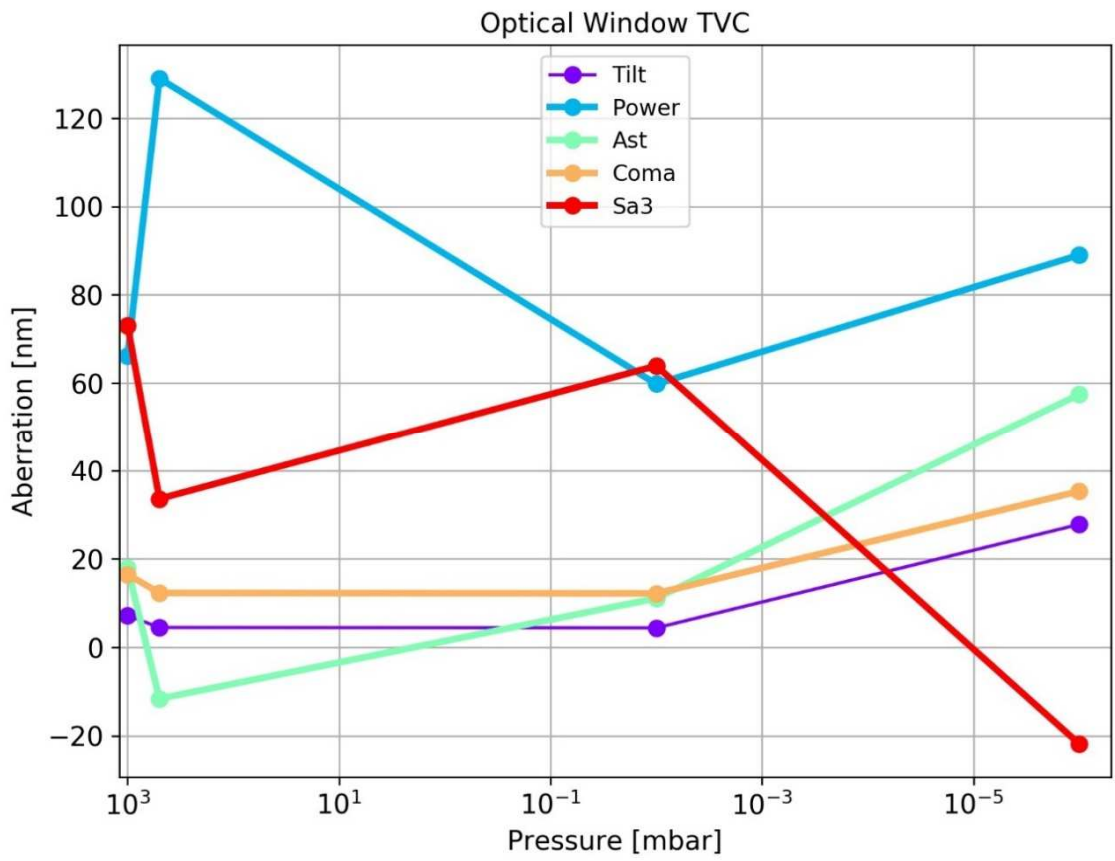

*Figure 126. The optical aberrations of the input window of "Cryotec 1" during the depressurization.*

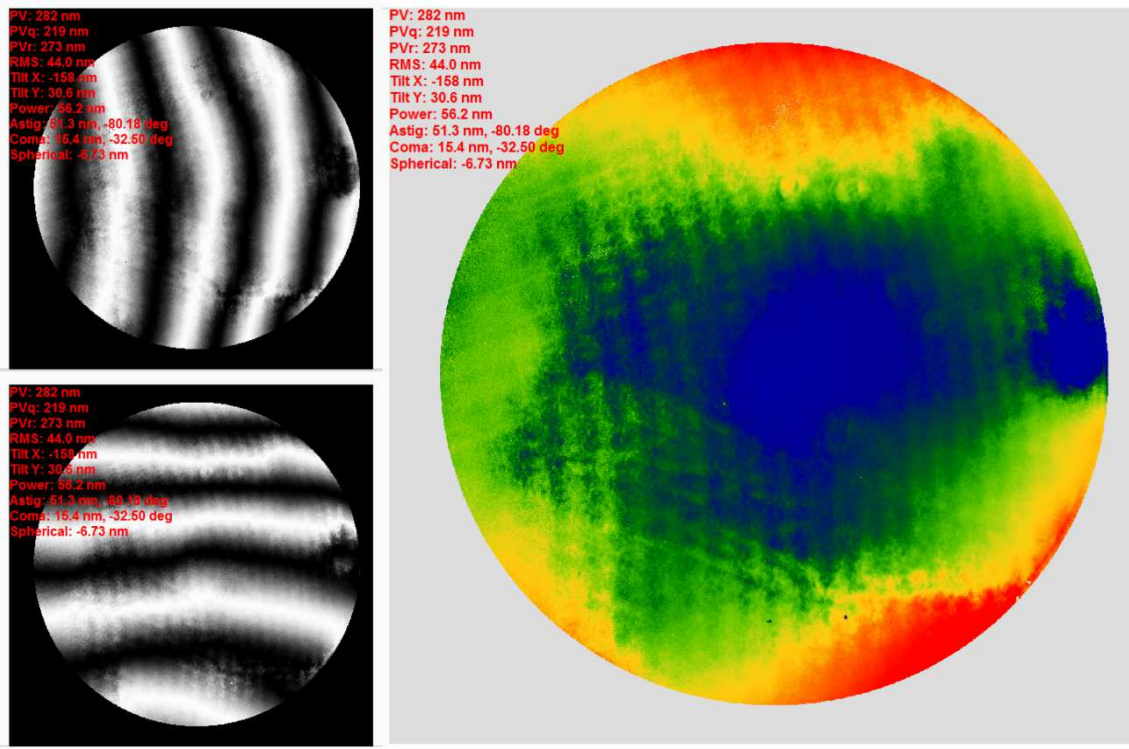

*Figure 127. Interferometric test of the input windows of the TVC at $10^{-6}$ mbar, courtesy Leonardo.*

### 2.8.4.1 Thermal stability of the Prototype

The cold inside the Cryotec camera was done by cooling down the optical bench and the shield which is surrounding the bench itself, called shroud. Several temperature sensors were installed in various positions, such on some lenses mounts, on the structure, on the motors. We needed a few cycles warm-cold to assess the right temperature to be imposed to the bench and the shroud, both to achieve the right temperature and to have a cooling down ramp not too fast for the lenses.

We finally were imposing the temperature of the bench to 163K and of the shroud to 148K, obtaining in this way a cooling down ramp for the lenses of the order of 0.25°/minute in average for the lenses. L6 was the fastest one with ~0.3°/minute, L1 was the slowest one with ~0.2°/minute, with the lens L3 achieving the right temperature of -80°C after about 17h.

We underline that we always experienced a significant temperature difference between the prototype lenses. Temperature sensors were attached to L1, L3, and L6, and they have shown an axial temperature gradient between L1 and L6. L1 is the closest lens to the optical input window and the warmest, L3 is in the middle and characterized by an intermediate temperature between L1 and L6, and L6 was the coldest. An example of this temperature gradient can be seen in Table 24, showing the values of the temperatures after the thermal stabilization. The difference between L3 and L1 is ranging from 22° to 25°, increasing when lowering the L3 temperature. The difference between L3 and L6 is of the order of 14°, quite stable in the nominal L3 working temperature range.

We have to consider that small differences, of the order of ±1°-2°, in the reported values may depend from the settling time for the temperature in the different moment of the test, but still the total gradient between L1 and L6 is of the order of 37°-38°, not decreasing with a more extended settling period.

*Table 24. The thermal gradient with L1 close to the optical input window; the lines in orange represent the situation with L3 at the TOU nominal working temperature (-75°/-85°).*

| L1 [°K] ([°C]) | L3 [°K] ([°C]) | L6 [°K] ([°C]) | Δ(L3-L1) [°K] | Δ(L3-L6) [°K] |
|---|---|---|---|---|
| 222 (-51) | 203 (-70) | 190 (-83) | 19 | 13 |
| 220 (-53) | 198 (-75) | 184 (-89) | 22 | 14 |
| 217 (-56) | 193 (-80) | 178 (-95) | 24 | 15 |
| 213 (-60) | 188 (-85) | 175 (-98) | 25 | 13 |
| 209 (-64) | 183 (-90) | 169 (-104) | 26 | 14 |

Nevertheless, this gradient between the lenses is explainable, considering that the prototype L1 was lens very close to the Cryotec input optical window, thus seeing the warm coming from the window. We did, at the end of the optical test, a couple of additional tests, with the purpose to check the origin of this temperature gradient. We first rotated of 180° the prototype, having in this way L6 instead of L1 close to the Cryotec window. The results are reported in Table 25, where it is visible that the situation is inverted, being L6 the warmest and L1 the coldest.

*Table 25. The thermal gradient with L6 close to the optical input window, in the situation with L3 in the TOU nominal working temperature range (-75°/-85°).*

| L1 [°K] ([°C]) | L3 [°K] ([°C]) | L6 [°K] ([°C]) | Δ(L3-L1) [°K] | Δ(L3-L6) [°K] |
|:---:|:---:|:---:|:---:|:---:|
| 187 (-86) | 198 (-75) | 232 (-41) | -11 | -34 |
| 183 (-90) | 193 (-80) | 231 (-42) | -10 | -38 |
| 179 (-94) | 188 (-85) | 229 (-41) | -9 | -44 |

We positioned the prototype away from the window, see Figure 128, and we shielded the optical input window, in Figure 129, with L1 looking toward the input window.

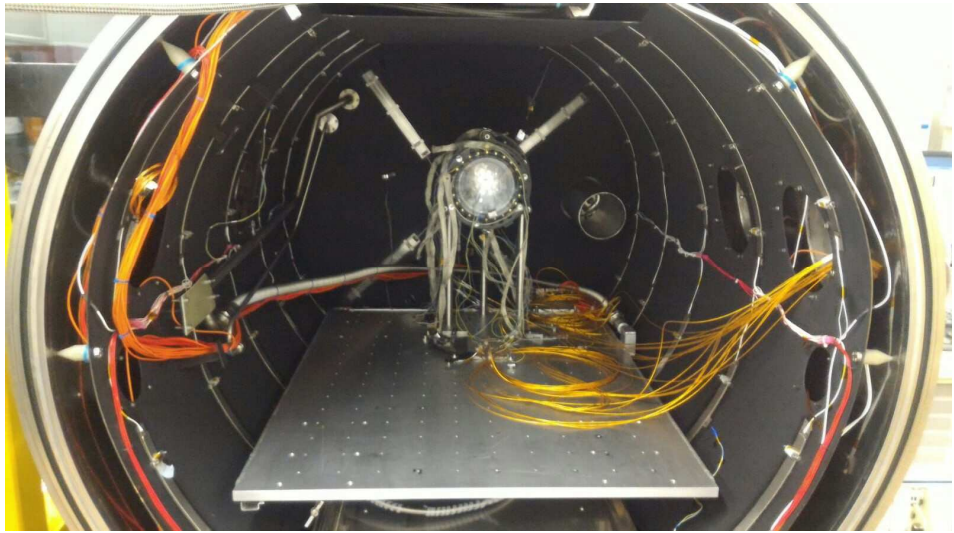

*Figure 128. The prototype positioned as far as possible from the optical input window.*

In these conditions, the difference of temperature between L1 and L6, and with L3 at the nominal temperature of -80°C, is about 10°, to be compared with the about 40° reported in Table 24. The maximum difference of temperature between L1 and L6 throughout the whole descending and ascending ramp has been 14°, thus again confirming that most of the problem is coming from the warm of the input optical windows and that the future test setup has to be appropriately designed taking into account this issue. One possible suggestion could be to use an input window on the side of the thermos-vacuum chamber, and to have a flat mirror, properly coated to minimize the thermal background, folding the light toward the TOU, if space is not an issue.

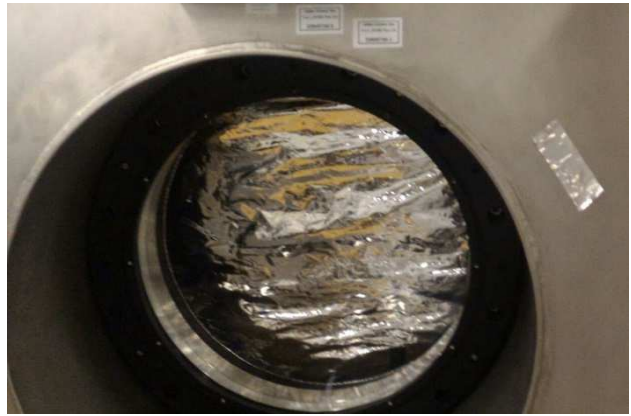

*Figure 129. The input window of the Cryotec camera shielded.*

We thus estimated the effect of this thermal gradient on the EE determination of this gradient in temperature. We simulated the experienced case with Zemax, using a ΔT between L1 and L6 of 36°C.

The simulation results are shown in Table 26, from which it looks pretty clear that the effect on the EE performance is small but not negligible, and moreover sometimes improving or decreasing the quality, depending on the considered position on the field.

*Table 26. The simulated effect of the thermal gradient on the EE performance, by using focus as the usual compensator.*

|  | Semi-width of the square enclosing 90% energy [μm] | | | |
|---|---|---|---|---|
|  | **On-axis** | **Field 5.6°** | **Field 11.2°** | **Field 16.7°** |
| **Warm** | 17.9 | 20.6 | 32.4 | 62.8 |
| **Cold** | 7.6 | 7.7 | 8.3 | 12.3 |
| **Axial gradient** 36° of ΔT L1-L6 | 8.1 | 8.4 | 7.1 | 9.1 |
| **Effect of axial gradient** | 7% worst | 9% worst | 17% better | 35% better |

## 2.8.5   Warm and cold test procedure in Leonardo

We emphasize that this procedure is valid both for the warm and cold test, being the only difference the Cryotec door open or closed (with vacuum and proper temperature set), and follows these items.

1.  The prototype has been installed on the Cryotec optical bench through its support structure. The position has to be as much as possible centered with the optical input window, the precision required 1-2mm), and the prototype optical axis has to be as much as possible parallel to the input window optical axis, around 1 degree.

2.  The XYZ motorized stages are installed after the prototype lens L6. Their installation requires that:

    •  With the focus stage at the travel end toward L6, the detector is positioned with a reasonable safety margin of minimum 1mm from L6 itself, and this condition has to be verified wherever in the FoV (moving X and Y stages of their complete travels).

    •  With the X and Y stages at mid-travel, the center of the detector is about corresponding to the FoV center (3-4mm of precision required). This can be accomplished by initial mechanical alignment, and double-checked when the folding mirror will be aligned to the TOU prototype in a way to define the FoV center.

3.  The detector and motors cables have been connected, and they have been functionally tested.

4.  The Cryotec Temperature sensors have been installed and they have been functionally tested.

5.  The reference to define mechanically the z position of the TOU prototype was installed on the bench; this ensures to minimize the prototype movement in Z when its rotation between A B and C orientations will be performed.

6.  The camera opening door is closed, and the folding mirror is positioned in front of the Cryotec input optical window.

7.  The folding mirror has been aligned with the prototype. To perform this operation, we look at the back reflections coming back into the interferometer. The two surfaces of L6 are the more visible ones, and they produce two sets of fringes. A decentering of the TOU wrt the incoming beam will shift rigidly the fringes coming from the two L6 surfaces, while a tilt of the TOU will increases/decrease the distance between the two sets of fringes coming from the two faces of L6. The procedure consisted of changing the tilt of the beam, by acting on the TT of the folding mirror, till nulling the distance between the two sets of fringes and having them centered to the beam diameter. Both accuracies are quite relaxed. In fact, due to the large beam oversize, the beam centering can be accomplished to 2-3mm of accuracy, while due to the large FoV of the TOU prototype, the tilt alignment can be accomplished with 0.5 degrees of accuracy, see Figure 130.

8. The CCD has been switched on and centered on the beam by moving the stages. The position of the spot on the CCD and the positions of the stage (in motor steps) have to be recorded, being the reference for the on-axis position.

9. We put references on the floor, pieces of tape indicating the positions of table legs, see Figure 130, in a way that the movable table carrying the folding mirror can be removed and successively placed back with a coarse accuracy. The fine-tuning of the folding mirror repositioning process can be done by centering the spot on the position recorded on the CCD.

The other foreseen off-axis positions have been achieved moving the detector (in the X direction) of the right amount corresponding to the desired angle, computed by Zemax considering also the field distortion, and successively tilting and moving the external flat folding mirror in order to position the spot on the desired reference pixel of the CCD. The positions in the FoV equivalent to the off-axis angles foreseen to be tested are in Table 27.

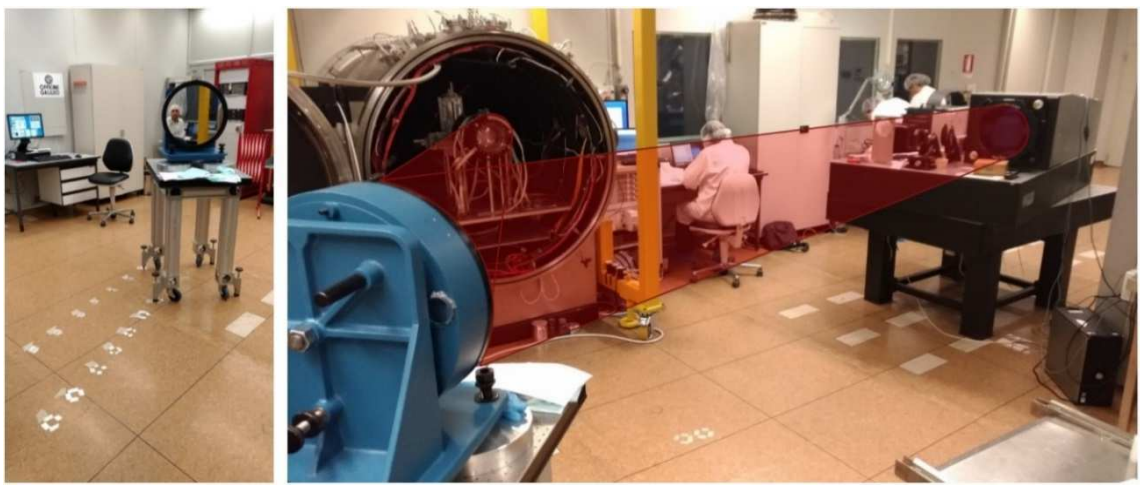

*Figure 130. The TOU installed in the climate chamber. Left: the folding mirror positions for simulating the off-axis beam. Right: the large beam expander folded by the mirror inside the climate chamber.*

*Table 27. The positions in the FoV corresponding to the off-axis angles in the entrance pupil and the focal plane with respect to the chief-ray.*

| Off-axis *input angle* | Off-axis pos. on L6 | Off-axis exit angle L6 |
|---|---|---|
| **5.6°** | 24.46 mm | 12.0° |
| **11.2°** | 49.87 mm | 24.3° |
| **16.7°** | 76.86 mm | 36.9° |

For interferometric tests, we did install a 0.6-inch F/1.6 spherical mirror to perform the on-axis interferometric test, mounting the mirror on the right side of the CCD. Further to the right, we install a second spherical reference mirror, 0.5 inch F/1.0 titled for C/+5.6. We also have installed a 2-inches diameter spherical mirror with a very fast F/# of 0.5, which made it possible to acquire also interferograms at the three off-axis positions C/-5.6, C/-11.2 and C/-16.7. Figure 132 shows the three spherical mirrors in the left and right of the CCD, mounted in the same x-y-z stage.

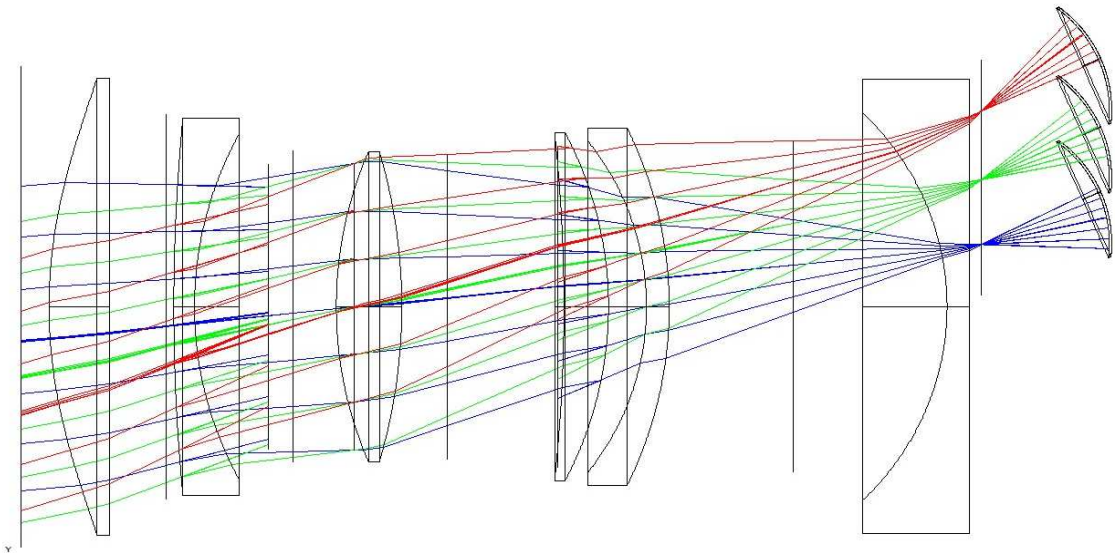

*Figure 131. Schematic view of the positions in the FoV consistent with the three off-axis input angles, and the position of the spherical mirror of 2 inches in diameter.*

In this way, we have the direct comparison, with the same setup, between cold and warm interferometric measurements, in the just mentioned prototype orientation and off-axis positions.

A Zygo Flash-Phase interferometer producing a 100 mm diameter collimated beam, coupled to a beam expander 3x, provided a 300mm diameter collimated beam.

Such a beam was sent to a 400mm flat folding mirror, positioned at about 45° in front of the Cryotec camera opening door; tilting and translating the flat-folding mirror, the full FoV of about 38° could be explored in one direction.

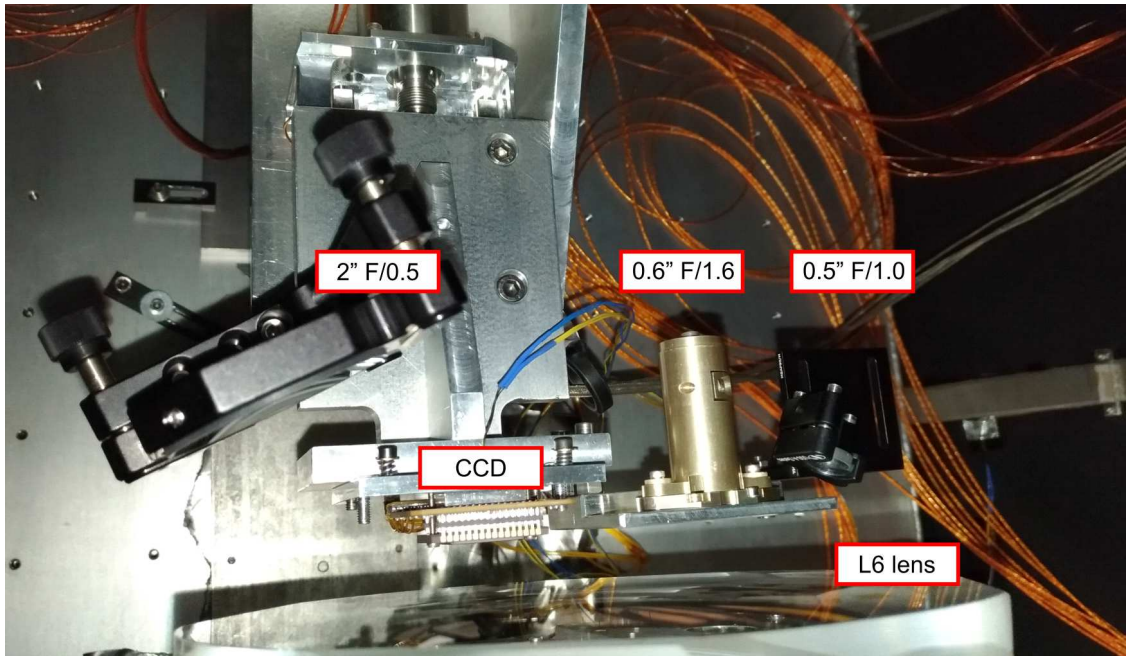

*Figure 132. The three spherical mirrors in the same frame of the CCD to perform interferometric tests.*

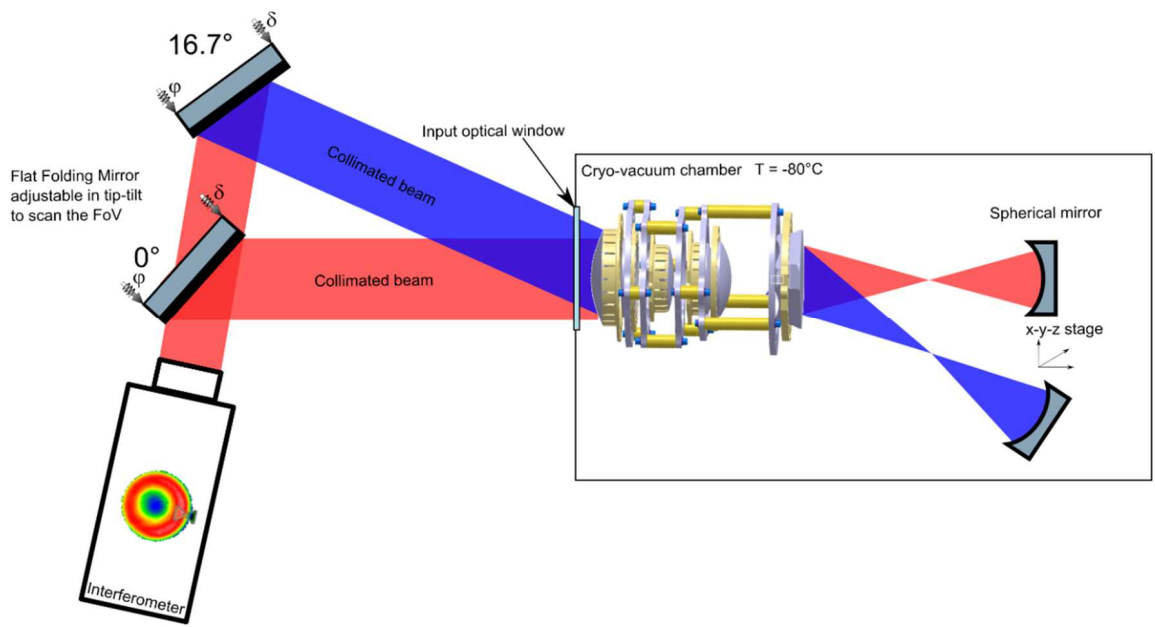

*Figure 133. The setup used for the cold interferometric test, the two extremal positions of the folding mirror are conjugated with the off-axis position described in the text.*

## 2.8.5.1 **Warm Interferometric Test On-Axis**

To acquire on-axis interferograms, we used the spherical mirror F/1.6, on the side of the detector. We first carefully aligned the mirror to the beam in this way:

- by looking at the references on the interferometer in alignment mode[2], see Figure 159, superimposing the back-reflected spots coming from the mirror to the interferometer reference;

- we then switched to the fringes mode and performed the focus alignment of the mirror by minimizing the number of visible fringes;

- a last iteration of the focus alignment has been done by acquiring interferograms and minimizing the power/defocus term.

Once the mirror was properly aligned, the XYZ position of the stages was recorded, and we acquired 50 interferograms. We discover that the dynamic range of the interferometer don't cover the analysis of the full aperture, then the mask used for the interferogram computation, see Figure 134, has been dimensioned and positioned to maximize the common sensed[3] area with the correspondent cold interferogram. In this case, an 82mm diameter mask has been used.

In Figure 135, there is the interferogram resulting from the average of 50 interferograms, taken with the prototype oriented at 240° (configuration C). In Table 28 we report the primary aberrations measured compared with the simulated ones over the same pupil diameter of 82 mm, the color-coding help to find if the determination was located near the best or worst case.

---

[2] In "alignment mode" of the interferometer a target (crosshair) appears on the fringe monitor, with in addition two bright spots. The brighter of the spots represents the reflection off the uncoated surface of the transmission flat, the second spot coming from the reference flat. By moving the tip and tilt knobs of the interferometer, we complete the alignment of the transmission flat. The reference concave mirror sreflect spot is centered with x-y movements of the stage until the complete superimpositions.

[3] The common sensed area, is the region of the interferogram were the distortions of the fringes due to aberration is well measured by the interferometer.

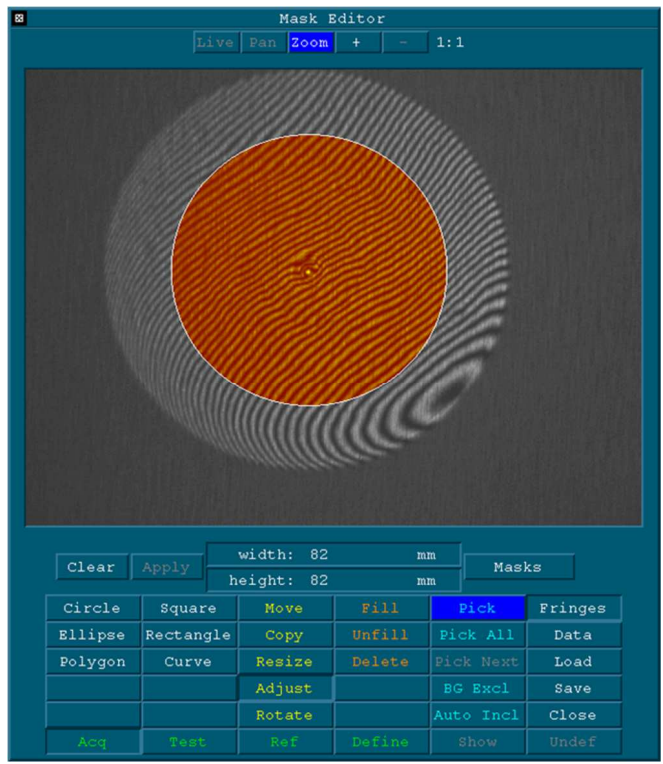

*Figure 134. The mask used to compute the warm interferograms on-axis with the prototype oriented at 240° (configuration C).*

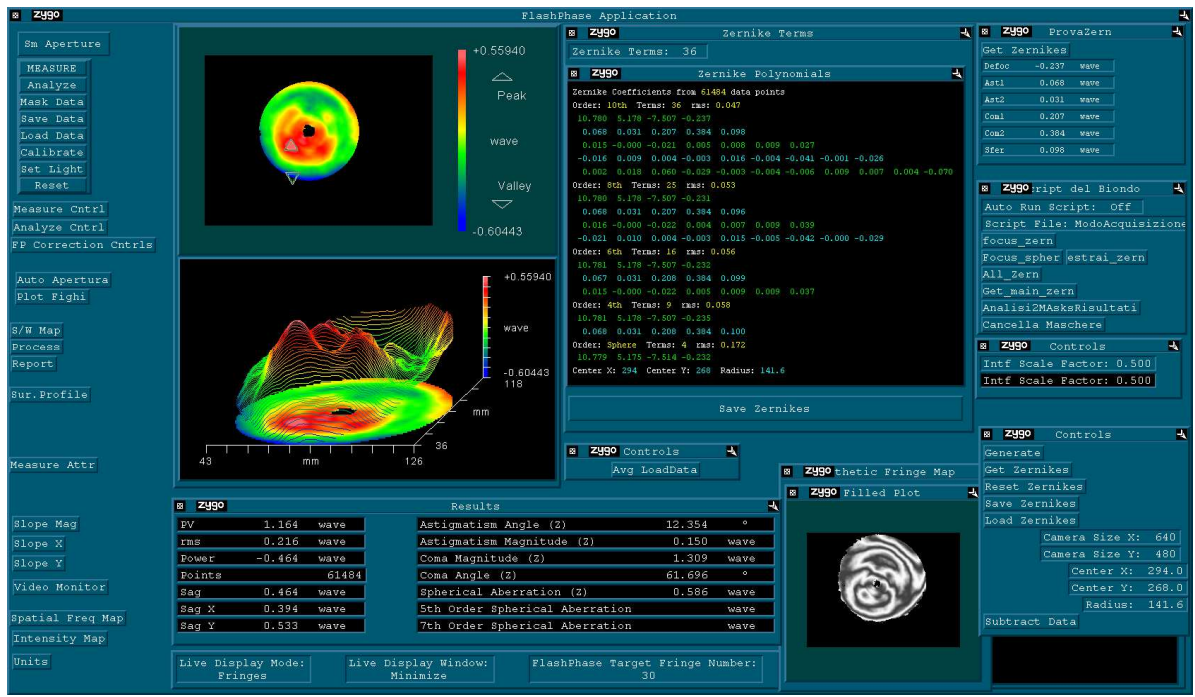

*Figure 135. The warm interferograms on-axis with the prototype oriented at 240° (config. C).*

*Table 28. The main aberrations measured (boldface) with warm interferograms on-axis with the prototype oriented at 240° (configuration C), compared to the simulated ones (Zemax) over the same pupil diameter. The defocus Zernike term measured for this data-set was -0.237 λ.*

| Rot. 240° on-axis 82mm diam. | Spherical (λ) | Coma (λ) | Astigmatism (λ) | Overall PtV (λ) | Overall RMS (λ) |
|---|---|---|---|---|---|
| **Measured** | 0,1 | 0,44 | 0,07 | 1,16 | 0,21 |
| **Sim. worst case  0°** | 0,18 | 0,09 | 0,49 | 1,27 | 0,48 |
| **Sim. best case 0°** | 0,07 | 0 | 0,01 | 0,27 | 0,12 |

## 2.8.5.2 Warm Interferometric Test at position C/-11.2°

We set the detector position to the one related to -11.2° and aligned the folding mirror to the detector reference pixel. At that point, we move the stages to have the spherical mirror sending back the light till having it appearing on the interferometer screen in aligning mode. At this point, we iterate the alignment by using the spherical mirror TT fine adjustment and the XZ movement of the stages to have both the spot reflected from the mirror aligned with the interferometer reference and the mirror centered as much as possible with the beam. In this way, at the two adjacent off-axis positions (-5.6° and -16.7°), the alignment of the beam can be performed only by remotely fine adjusting the XYZ stages, without having the mirror vignetting the beam.

From this point on, the same procedure of the previous section has been used. We underline that the focus adjustment has, in this case, to take into account that the chief ray corresponding to the 11.2° was tilted of about -24.3° with respect to the surface of L6, and thus any Z movement requires to move also the X stage, of an amount given by Z*tan(24.3°) in the right direction, to maintain the centering alignment.

Once the mirror was aligned correctly, the XYZ position of the stages was recorded, and we acquired 50 interferograms.

The mask used for the interferogram computation has been dimensioned and positioned to maximize the common sensed area with the correspondent cold interferogram. In this case, an 82 mm diameter mask has been used. In Table 29 we report an example of the primary aberrations measured compared with the simulated ones over the same pupil diameter.

*Table 29. The main aberrations measured (boldface) with warm interferograms at -11.2°off-axis position with the prototype oriented at 240° (configuration C), compared to the simulated ones (Zemax) over the same pupil diameter. The defocus Zernike term measured for this data-set was 0.098 λ.*

| Rot. 240° -11.2° off-axis 82mm diam. | Spherical (λ) | Coma (λ) | Astigmatism (λ) | Overall PTV (λ) | Overall RMS (λ) |
|---|---|---|---|---|---|
| **Measured** | 0,14 | 0,41 | 0,64 | 2,01 | 0,31 |
| **Sim. worst case** | 0,09 | 0,35 | 1,25 | 3,28 | 0,65 |
| **Sim. best case** | 0 | 0,14 | 0,17 | 1,22 | 0,29 |

### 2.8.5.3 Warm Interferometric Test at position C/-5.6°

From the previous position, we moved the X stage of the right amount relating to the off-axis of -5.6°, 49.87-24.46=25.41mm of X relative movement from the previous -11.2° position, and we tilt and decenter the flat mirror till having the spot appearing on the interferometer screen in aligning mode.

Then, we carefully aligned the mirror by following the same procedure described in the previous section, compensating any Z movement with the right X shift, accordingly to the chief-ray proper tilt due to this off-axis configuration. Once the mirror was aligned correctly, the XYZ position of the stages is recorded and we acquired 50 interferograms.

The mask used for the interferogram computation has been dimensioned and positioned to maximize the common sensed area with the correspondent cold interferogram. In this case, an 82mm diameter mask has been used. In Table 30 we report the primary aberrations measured compared with the simulated ones over the same pupil diameter.

*Table 30. The main aberrations measured (boldface) with warm interferograms at -5.6°off-axis position with the prototype oriented at 240° (configuration C), compared to the simulated ones (Zemax) over the same pupil diameter. The defocus Zernike term measured for this data-set was -0.128 λ.*

| Rot. 240° -5.6° off-axis 82mm diam. | Spherical (λ) | Coma (λ) | Astigmatism (λ) | Overall PTV (λ) | Overall RMS (λ) |
|---|---|---|---|---|---|
| **Measured** | 0,12 | 0,46 | 0,28 | 1,35 | 0,23 |
| **Sim. worst case** | 0,19 | 0,16 | 0,68 | 1,72 | 0,37 |
| **Sim. best case** | 0,07 | 0,01 | 0,01 | 0,42 | 0,07 |

### 2.8.5.4 Warm Interferometric Test at position C/-16.7°

From the previous position, we moved the X stage of the right amount related to the off-axis of -16.7°, 76.86-24.46=52.40mm of X relative movement from the previous -5.6° position, and we tilt and decenter the flat mirror till having the spot appearing on the interferometer screen in aligning mode.

Then, we carefully aligned the mirror by following the same procedure described in the previous section, compensating any Z movement with the right X shift, accordingly to the chief-ray proper tilt due to this off-axis configuration. Once the mirror was aligned correctly, the XYZ position of the stages is recorded, and we acquired 100 interferograms.

The mask used for the interferogram computation has been dimensioned and positioned to maximize the common sensed area with the correspondent cold interferogram. In this case, an 80x83mm elliptical mask has been used. In Table 31, we report the primary aberrations measured compared with the simulated ones over the same pupil diameter.

*Table 31. The main aberrations measured (boldface) with warm interferograms at -16.7°off-axis position with the prototype oriented at 240° (configuration C), compared to the simulated ones (Zemax) over the same pupil diameter. The defocus Zernike term measured for this data-set was 0.178 λ.*

| Rot. 240° -16.7° off-axis 82mm diam. | Spherical (λ) | Coma (λ) | Astigmatism (λ) | Overall PTV (λ) | Overall RMS (λ) |
|---|---|---|---|---|---|
| **Measured** | -0,08 | 1,02 | 1,23 | 3,74 | 0,56 |
| **Sim. worst case** | 0,28 | 0,63 | 2,61 | 6,48 | 1,71 |
| **Sim. best case** | 0,18 | 0,38 | 0,93 | 3,93 | 0,8 |

### 2.8.5.5 Warm Hartmann test at Leonardo

Due to the limited beam diameter (101.6mm) we had in Padova during the warm test, we decided to repeat the full sweep of warm Hartmann test at Leonardo, with the same setup used for the cold Hartmann test, see section 2.8.6, performed for on-axis and off-axis positions, with the same holed mask.

In every field position, we saved between 6 and 10 locations intra and extra focal, and for each of those, we saved between 20 and 50 images, depending on the amplitude of the spot vibrations that we were experiencing in each case. To maintain the Cryotec cold, the LN2 is suddenly injected in the climate chamber generating vibrations that make often the spot movements becoming pretty large. In this case, a larger number of measurements to be averaged was necessary. We paid particular attention to atmospheric turbulence due to the large path of the collimated beam emerging from the interferometer, around 5 meters.

The analysis of the images was the same used at the Padova laboratory with minor differences. For this test we followed these steps:

- we average the 20-50 images for each position of the detector;

- we find the position of the 76 centroids, for each position of the detector;

- we find the 76 lines that fit the position of each centroid at different detector position;

- we intersect the 76 lines with planes orthogonal to the TOU optical axis;

- we identify the focal plane as the plane where the 76 intersection points are distributed minimizing their rms wrt their centroid;

- on the focal plane, we find the square, which encloses 90 % of the 76 points. This square is centered on an arbitrary point to minimize the size of the square itself.

The results are summarized in Figure 136 and Figure 137, where we report the semi-dimension of the square enclosing the 90% of the 76 points (i.e. 90% energy), comparing the results with simulations. As can be seen in Figure 136, we also estimate the associated error bar by a Monte Carlo analysis. First, we compute the error associated with the centroid determination.

We found the centroid coordinates in two ways: automatically by a DAOPHOT-based python routine, or by a "click input" and then a 2D Gaussian fit within a typical size box containing the single spot. We consider a set of realizations (n=50) of this computation (clicks) and compare the results among them and with the automatic routine, finding a maximum difference of 10 px. Given this number, we consider an on-axis Hartmann test and its centroid positions for each plane. We simulate 1000 sets starting from the considered one, applying random errors, within the 10 px, in x and y to each centroid. For each set, we re-calculate the best focal plane, the enclosed energy, resulting in a Gaussian distribution with a standard deviation of about 3µm, which is the dimension of the error bar in the plots.

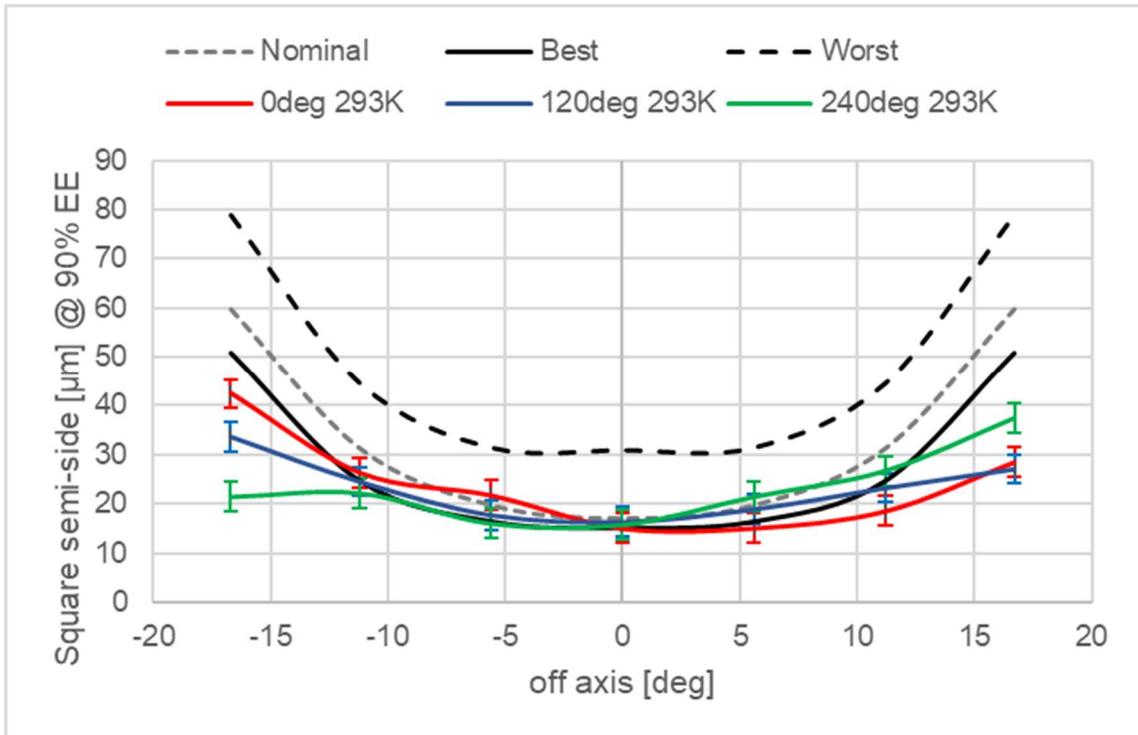

*Figure 136. The semi-width of the square enclosing the 90% of the energy at the focal plane, as a function of the off-axis angle, coming from the Hartmann test. Different colors (red, blue, green) refer to three orientations of the TOU (rotation along the optical axis). Black lines refer to worst, best (solid) and nominal (dotted) results (@296K) from ray-tracing simulations. Data acquired in warm conditions (see legend).*

We also consider the trend of the position of the focal plane at the different off-axis position and for the three orientation of the TOU, referring it to the position of the on-axis focal plane, and comparing it with simulation. The result of this analysis provides the tilt of the detector with respect to the actual focal plane. This error comes out when repositioning the prototype onto the structure to test it at different rotation angles, Figure 138, we re-arrange the data rotating them along their fitting straight line (Figure 139). The simulated positions of the focal plane on-axis and at the usual three off-axis positions are shown with the black dotted line.

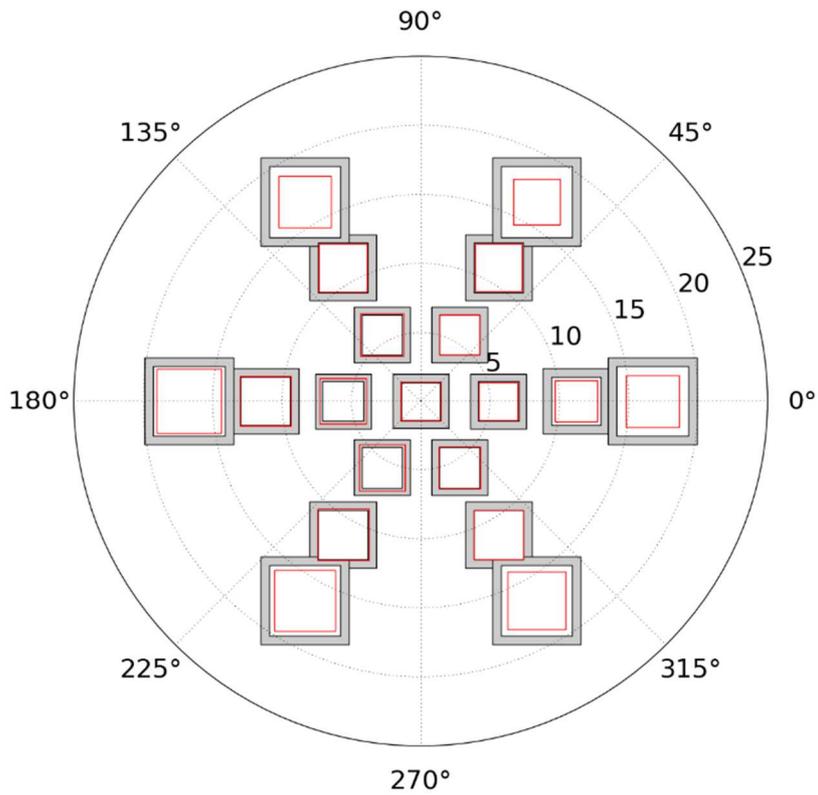

*Figure 137. A 2-D representation of result shown in Figure 136 with the off-axis angle in the radial coordinate and orientation of TOU in polar coordinate. Grey "square ring" is the area between the best and worst results coming from ray-tracing simulations (@ 296K). Red squares are the measured values. When red squares are enclosed in the grey area, data are in according with nominal results.*

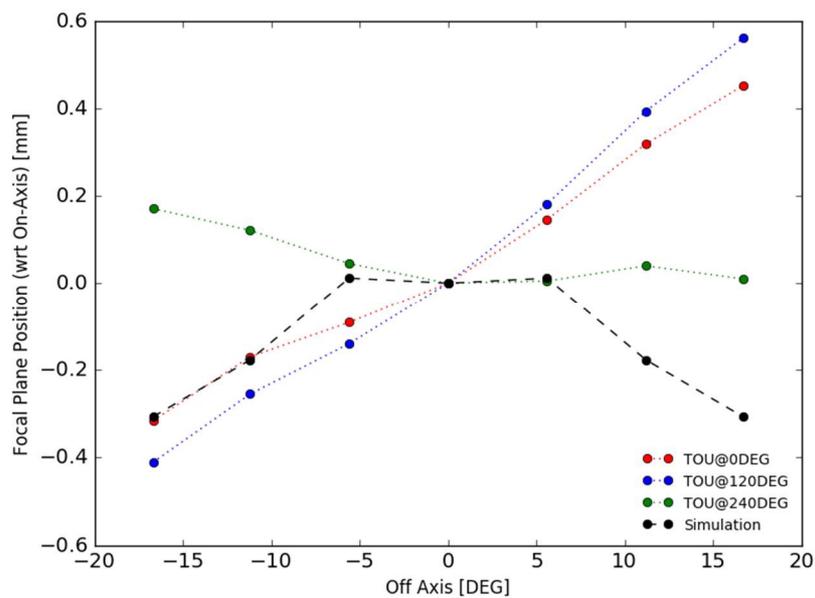

*Figure 138. The warm case. Assuming the same focal plane position on-axis for the three TOU configurations, here the trend of the focal planes for the off-axis positions and the three orientations of the TOU. The data carries the error in tip-tilt for the detector.*

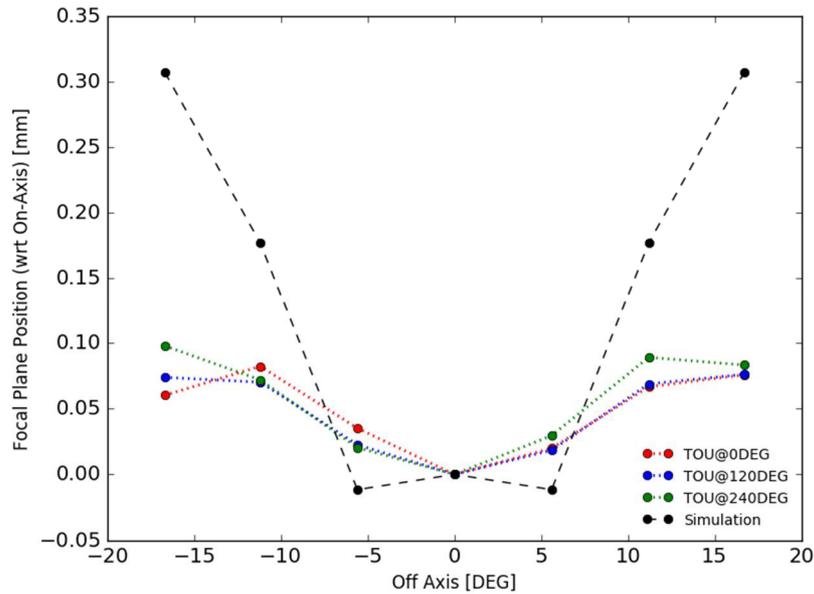

*Figure 139. As in Figure 138, but with data re-arranged by rotation. Warm case.*

## 2.8.6   Cold PSF test in Leonardo

For each foreseen position in the FoV, we were sweeping around the focus position with steps of 10 microns, doing ~10 steps intra and ~10 steps extra focal, and acquiring typically 20 images for each step. For this thes test the Hartmann mask was removed.

There was a fast and evident movement of the PSF, both due to the Cryotec vibrations and to the $N_2$ weak continuous flow we had to apply to the Cryotec input optical window, to avoid condensation and ice formation. As a result, averaging the 20 images taken for each step would have considerably increased the PSF size, and we thus applied a "shift & add" algorithm for each focus sweep position. The best focal position is obtained calculating the squared energy for every sweep position and by fitting the results with a parabola. The background was calculated from the median value over all pixels, the image size is 6Mpx, and the PSF image is of the order of 100px width. Figure 140 shows the values of the 90% of Enclosed Energy for position A of the TOU and on-axis illumination. Actual data are represented by the black line, fitting them with a parabolic function permits to determine the minimum, identifying the best focal position. Figure 141 shows the EE and the image in the best focal plane. Figure 142 and Figure 143 show the same result for the maximum FoV of 16.7°, where the PSF is pretty larger.

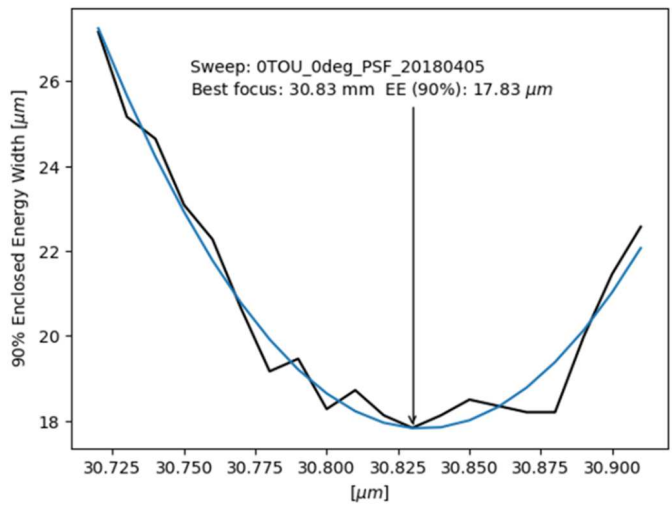

*Figure 140. The position A/on-axis. Best focal position computed by 90% Enclosed Energy in the sweep positions.*

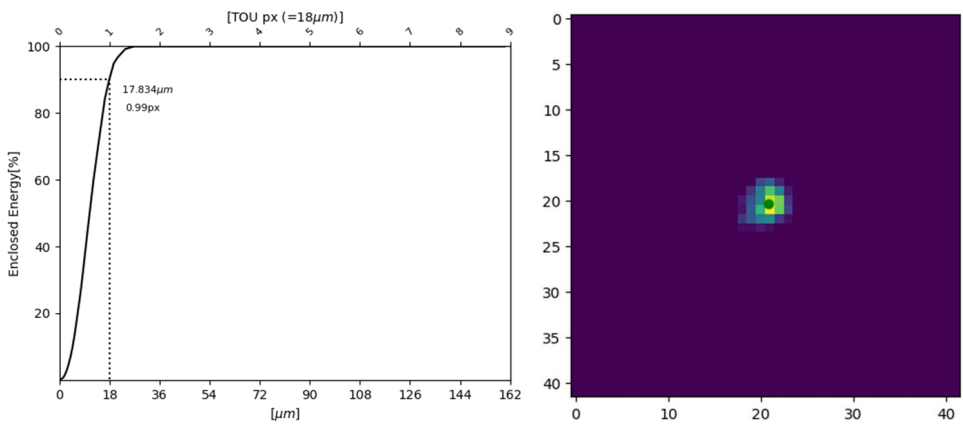

*Figure 141. The best focal position. Left: Enclosed Energy. Right: PSF image.*

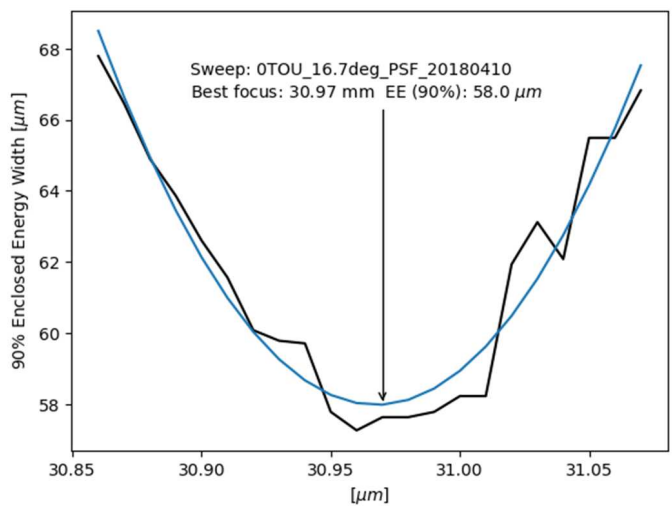

*Figure 142. The position A/+16.7°. Best focal position computed by 90% Enclosed Energy in the sweep positions.*

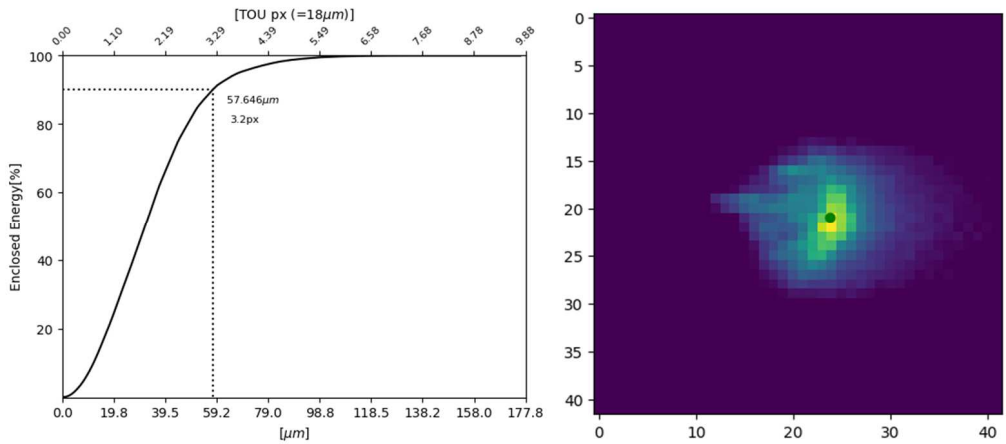

*Figure 143. The best focal position. Left: Enclosed Energy. Right: PSF image.*

The EE computed from the PSF measurements is pretty much dependent on the choice of the background to be removed. We analyze the background subtraction to reduce the error in the EE computation, which induces an error also in the best focus position. As an example, in Figure 144, we report the effect on the estimated EE related to different background subtractions, computed with σ=3 (on the left image) and σ=4 (on the right image). With respect to the median background of the original image, adding a value of σ=3, the background increases by 1.5%, while with σ=4 the background increases of 1.9%. The relative difference between σ=3 and σ=4 is 0.4%, equivalent to a change in the EE (with 90% of threshold) of about 17%, from ~43μm to ~35μm.

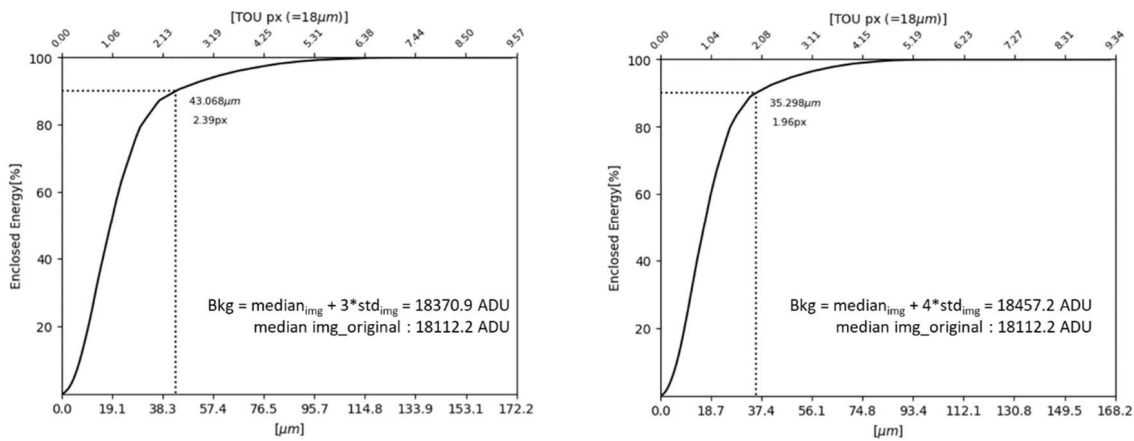

*Figure 144. The effect of the σ choice on the background estimation*

Figure 145 shown three different methods to compute the background in the PSF image:

a. $bkg = median + k \cdot std, k = 3\ (case\ "a"), 10\ (case\ "b"), 30\ (case\ "c")$

the background is computed as the sum of the median value in the box image and a number of times the standard deviation, all negatives values are put to zero;

b. $bkg = median$

the background is computed as the median value in the box image, no correction for negative values;

c. $bkg = average[^4_1 Box(5x5\ px)]$

the background is computed as the average value in four boxes of 5x5px in size located in the four corners of the image, no correction for negative values.

The best focal position determined with the three methods has a PtV of 30µm, the EE show similar values for the case "b" and "c" but twice the value compared to the case "a", closer to the values obtained with the Hartmann test, see in the next section Figure 154. For this reason, we investigate more accurately the case "a".

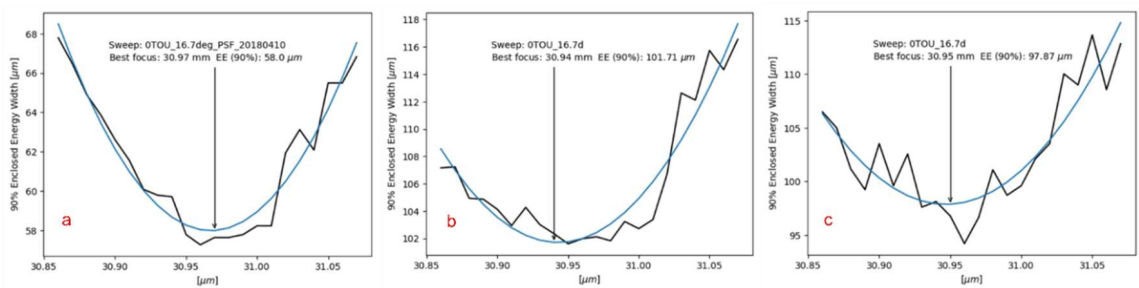

*Figure 145. The tree different analysis of the EE background subtraction, see text for description.*

For the case "a" we compute the variation of EE when the constant "k" of standard deviation in the formula $bkg = median + k \cdot std$ is increasing from 3 to 30, to find the best value of "k". We found that the EE reach asymptotic values over ten times the standard deviation, see Figure 146.

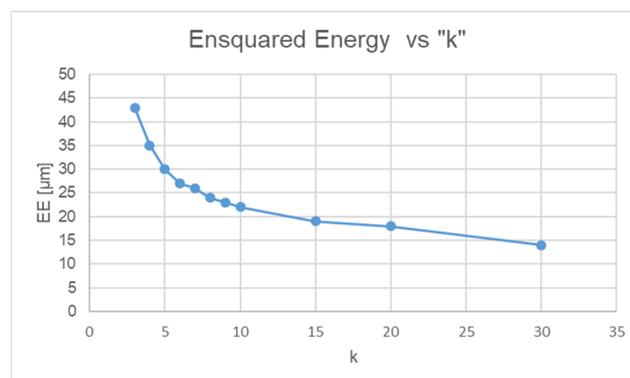

*Figure 146. The EE variation with the increasing of the "k" constant multiplied to the standard deviation to compute the background estimation, case "a".*

Figure 147 shows the resulting image of the PSF after the subtraction of the background in the case "a" for different value of "k" constant, again showing that the choice of the sky background to be subtracted is impacting a lot the results. We have computed the EE in configuration 0deg of the TOU for different values of constant "k", Figure 148 shows that the value of k=10 is comparable with the same determination of the Hartmann test in the same conditions.

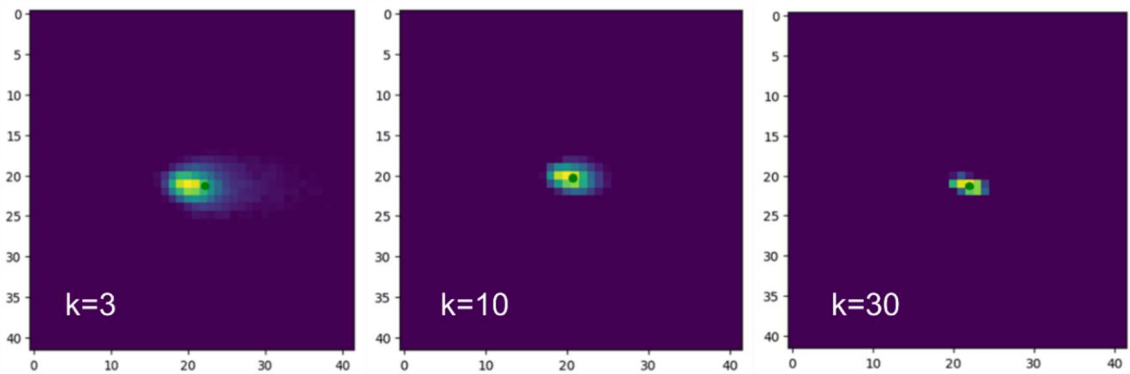

*Figure 147. The images represent the PSF for the case "a" after the subtraction for k=3,10,30.*

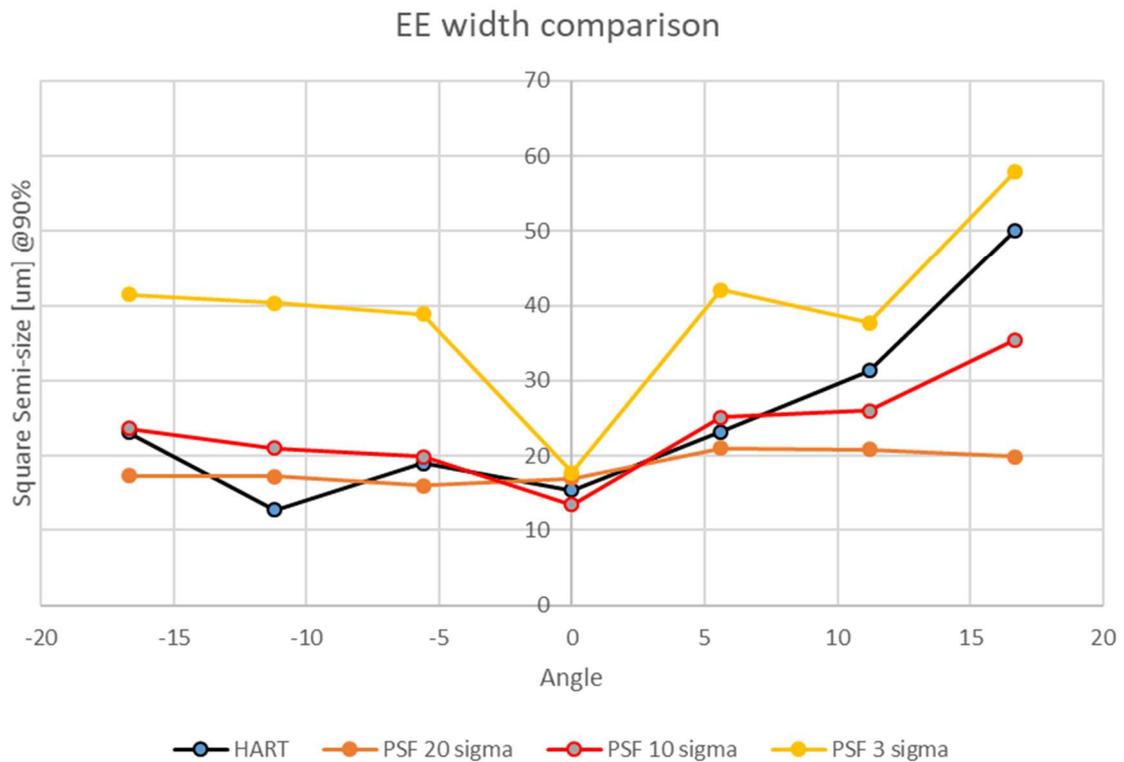

*Figure 148. The EE width comparison for different values of k in the case "a" for all FoV.*

After this empirical analysis of the error in the estimation of the EE due to background subtraction, we select the more consistent value of the background as $bkg = median + 10 \cdot std$. In Figure 149 we compute the EE values for all FoV and compared them simulation. We conclude that for all rotation of the TOU over the EE is below the worst case. These results are comparable with the EE computed from the Hartmann test, Figure 154. We can confirm that the background subtraction pays an essential role in the quality image analysis and this confirms that EE from PSF analysis is less robust than the Hartmann test, and thus not completely representative of the real situation.

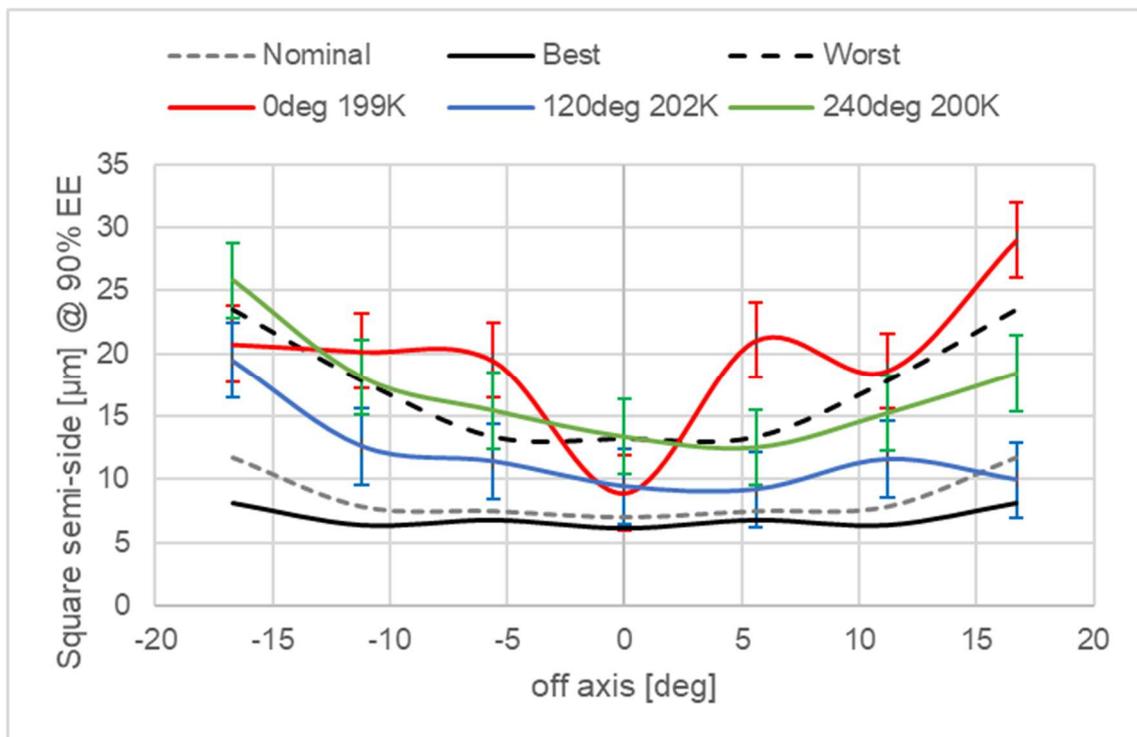

*Figure 149. The semi-width of the square enclosing the 90% of the energy at the focal plane, as a function of the off-axis angle, coming from the PSF test. Different colors (red, blue, green) refer to three orientations of the TOU (rotation along the optical axis). Data acquired in cold conditions. The semi-EE was computed subtracting the background estimation $bkg = median + 3 \cdot std$.*

We anyhow estimated what is the minimum value of σ that makes the background nearly zero over the field, in order to have a somehow objective criterium for the choice of σ. It turned out to be about σ=3. With such a choice, the estimation of the EE done through the PSF measurement is reported in Figure 150, from which it is clear that the estimation is in agreement only in half of the tested positions with the results from the Hartmann test, while in the worst case the difference can be as significant as 28 µm.

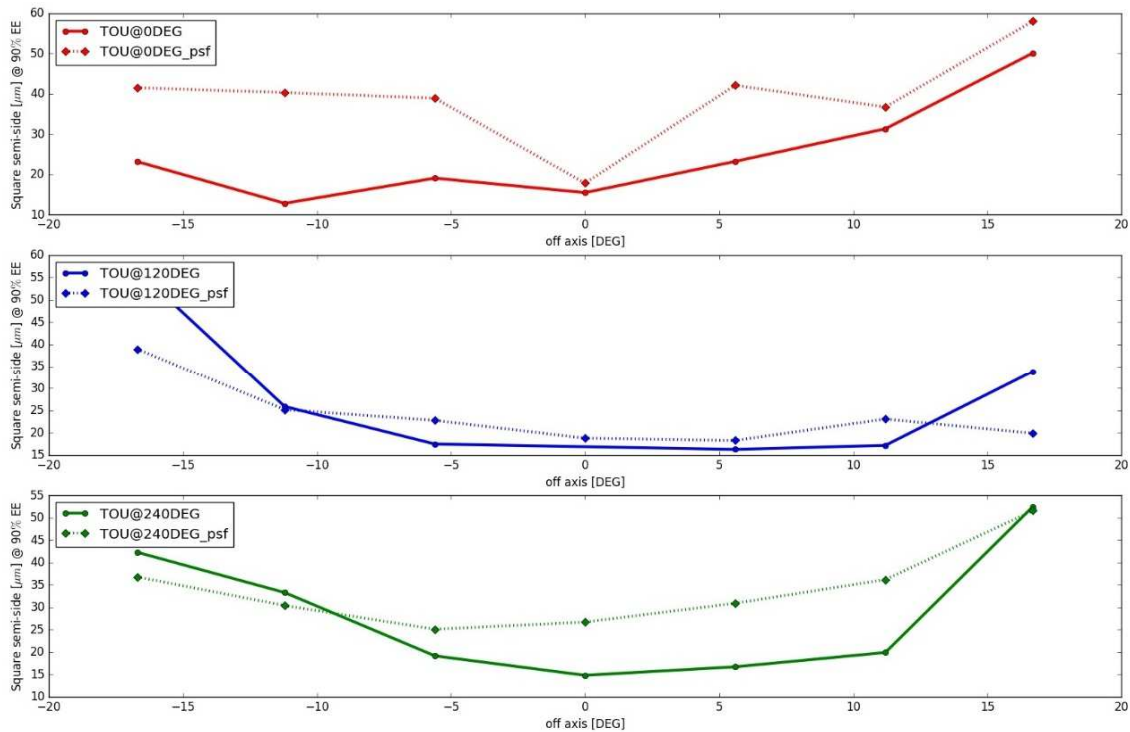

*Figure 150. The comparison between the enclosed energy results coming from the Hartmann test (solid line), and the PSF test (dotted line) with the background removed with σ=3.*

We simulate with Zemax the size of the PSF for the FoV of 0° (on-axis), 5.6°,11.2°,16.7°, for the nominal and worst case, see Figure 151. In Figure 152. The simulation Zemax of the PSF size in nominal and worst-case after a Monte Carlo simulation, based on ray tracing.we report the simulation with Zemax of the PSF size (based on ray tracing) in nominal and worst-case after a Monte Carlo simulation. The worst case is made by a Monte Carlo simulation having as degrees of freedom the alignment tolerances. Figure 153 shows the result of the PSF measurements; the images were de-rotated since the CCD axis is fixed with respect to the TOU mount, while the TOU has been rotated and tilted during the test. Visually the PSF appears between the nominal and worst case, but with a decreasing of quality in 240° with respect 120°, as confirmed by Hartmann analysis in Figure 150.

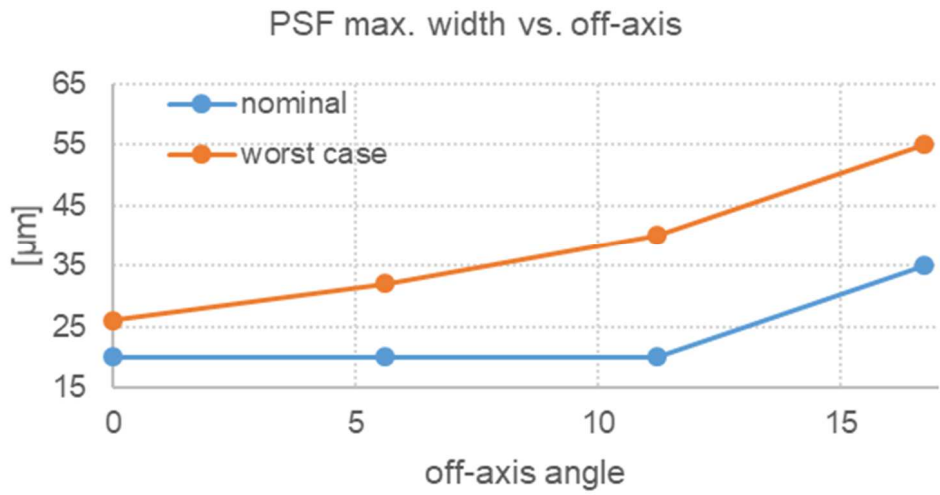

*Figure 151. The Zemax simulation of the PSF size in nominal and worst-case after a Monte Carlo simulation.*

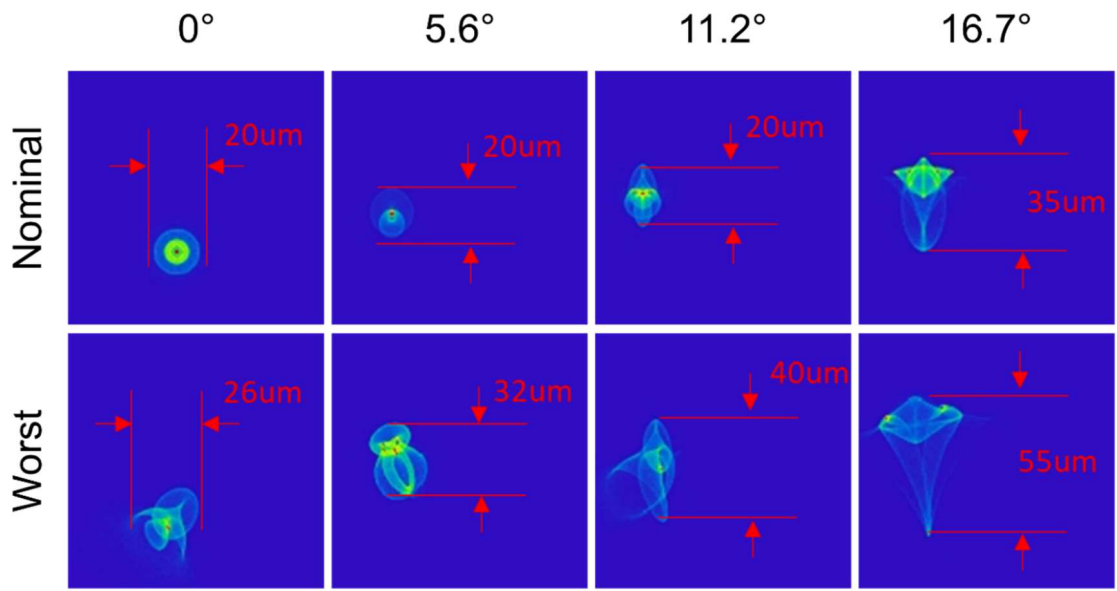

*Figure 152. The simulation Zemax of the PSF size in nominal and worst-case after a Monte Carlo simulation, based on ray tracing.*

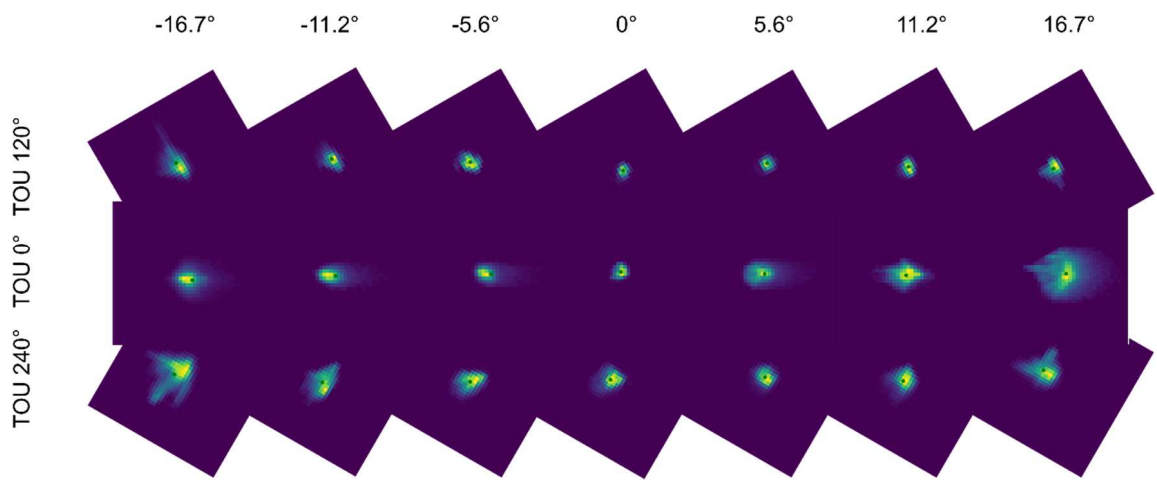

*Figure 153. A mosaic of all best focus PSF for all FoV angle and rotations of the TOU of 0°,120°, and 240°. The images are de-rotated since the CCD was fixed axis with respect to the mechanical mount of the TOU.*

### 2.8.6.1 Cold Hartmann test in Leonardo

One of the goals of the cold test was exploring the performance of the TOU over the full working temperature range ( -75°C/-85°C),

In Figure 154 we report, for one of the three prototype orientations (240°), the performance of the prototype in terms of EE at two different temperatures, ~200°K (-73°C) and ~190°K (-83°C), pretty much close to the boundaries of the working temperature range. The two curves are the green and yellow ones respectively. Considering the error bars, the two curves look pretty much similar, showing variations of the order of 10-15%, apart from the position at 16.7° off axis, where the difference is of the order of 30%. The results of the EE for all FoV are summarized in Figure 154 and Figure 155.

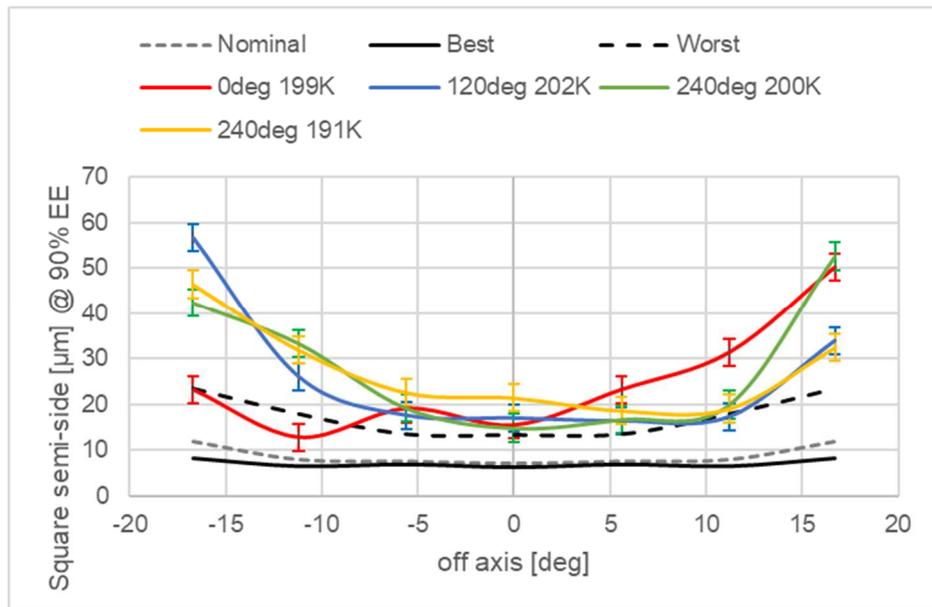

*Figure 154. The semi-width of the square enclosing the 90% of the energy at the focal plane, as a function of the off-axis angle, coming from the Hartmann test. Different colors (red, blue, green) refer to three orientations of the TOU (rotation around the optical axis). The yellow line represents a comparison at a different temperature for 240 degrees orientation. Black lines refer to worst (dashed), best (solid) and nominal (dotted) results @193K from ray-tracing simulations. Data acquired in cold condition, see the legend. The temperature shown is the one of L3.*

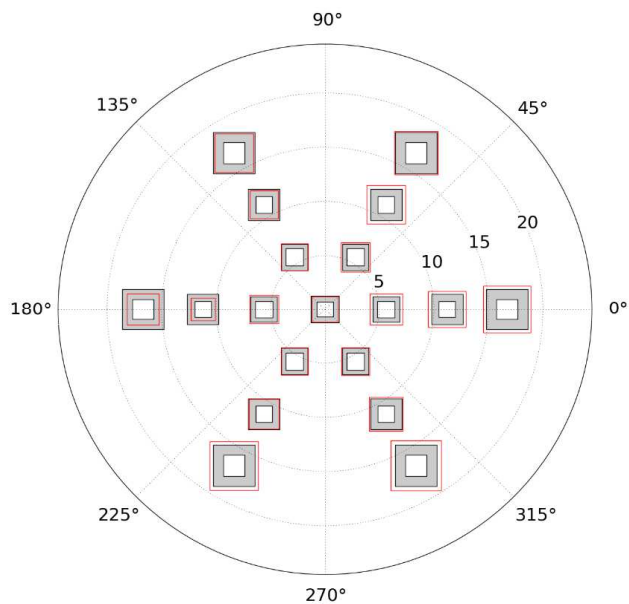

*Figure 155. A 2-D representation of result shown in Figure 154, with the off-axis angle in the radial coordinate and orientation of TOU in polar coordinate. Grey "square ring" is the area between the best and worst results coming from ray-tracing simulations @193K. Red squares are the measured values. When red squares are enclosed in the grey area, data are in according with nominal results.*

In Figure 156, we show the comparison of the results on the EE between the cold and warm case.

In Figure 157, we show the results on the trend of the focal plane position, like in the warm case. Also in this case, since the result of this analysis provides the tilt of the detector with respect to the actual focal plane, introduced when repositioning the prototype onto the structure to test it at different rotation angles, we re-arrange the data rotating them along their fitting straight line, Figure 158Figure 157. The simulated positions of the focal plane on-axis and at the usual three off-axis positions are shown with the black dotted line.

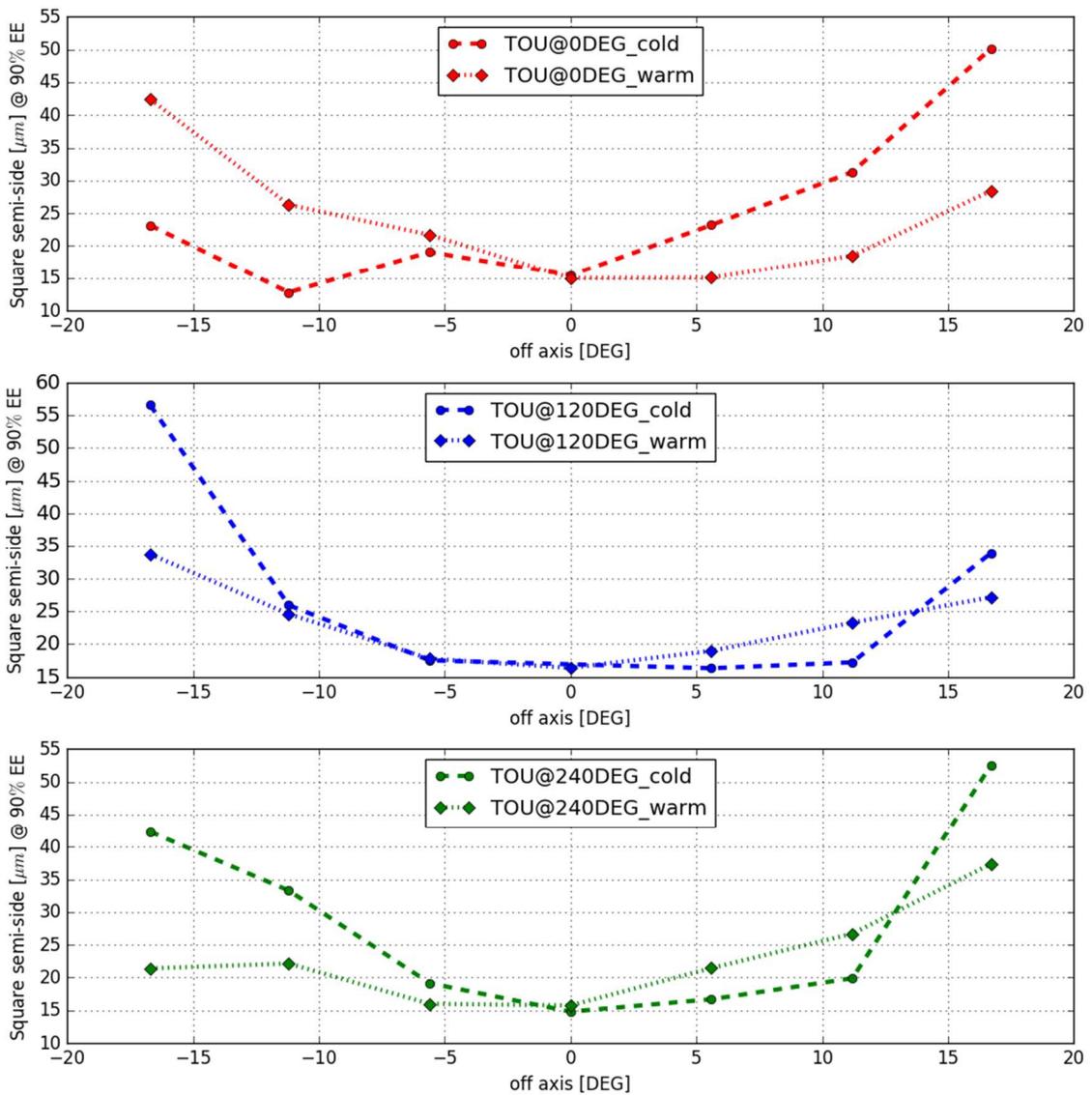

*Figure 156. The comparison between warm and cold results for the enclosed energy at different TOU rotation as a function of the off-axis position.*

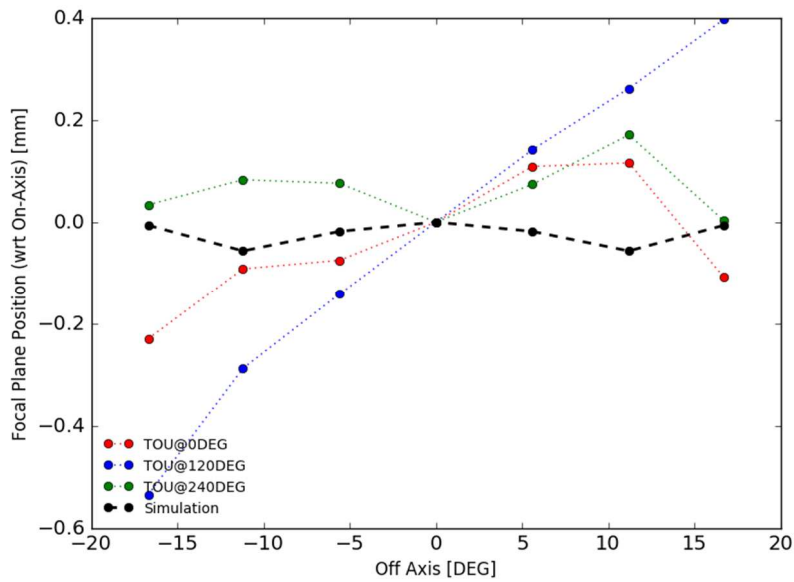

*Figure 157. Assuming the same focal plane position on-axis for the three TOU configurations, here the trend of the focal planes for the off-axis positions and the three orientations of the TOU. Cold case.*

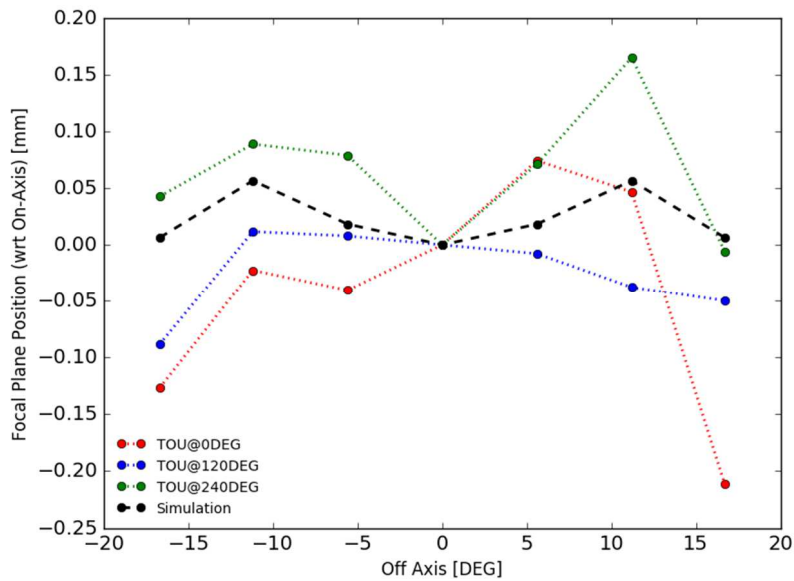

*Figure 158. As Figure 157, but with data re-arranged by rotation.*

## 2.8.6.2 Cold Interferometric Test in Leonardo

The cold interferometric test has the purpose of qualifying the TOU in full aperture and on a large FoV. To perform this test, we used three spherical reference mirror mounted on the CCD arm, as discussed in 2.8.2. We collected interferogram at the orientation A/+11.2, B/+5.6, C/-5.6, C/-11.2, and C/-16.7.

We have used here the same mirror positions used for the warm interferometric tests described in section 2.8.2. This was enough to make the spot appearing on the interferometer screen in aligning mode, even if of course, the ~100° of the temperature difference between the two tests did require an additional fine adjustment of mirror positions, above all in Z.

The fine alignment of the mirror to the beam was performed in this way:

- by looking at the references on the interferometer in alignment mode, superimposing the back-reflected spots coming from the mirror to the interferometer reference, Figure 159 a) and b);

- we then switched to the fringes mode and performed the focus alignment of the mirror by minimizing the number of visible fringes, Figure 159 c) show the back reflection in aligning mode;

- a last iteration of the focus alignment has been done by acquiring interferograms and minimizing the power/defocus term.

Once the mirror was adequately aligned, the XYZ position of the stages was recorded, and we acquired 35 interferograms.

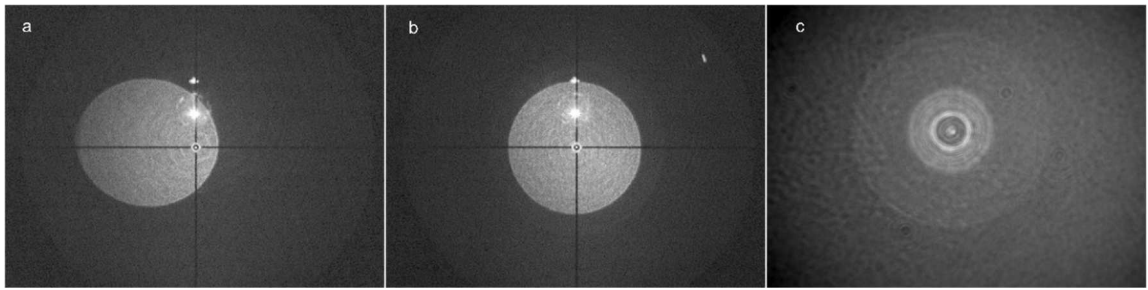

*Figure 159.The alignment of the spherical reference mirror inside the climate chamber with the interferometer in "align mode". Test on-axis a) the mirror not aligned, b) the mirror aligned, the bright spot is the back reflection from the flat reference in the beam expander, c) the mirror aligned and in correct z-position.*

The mask used for the interferogram computation (see Figure 160) has been dimensioned and positioned to maximize the common sensed area with the corresponding cold interferogram. In this case, an 82 mm diameter mask has been used.

As an example, we report in Figure 161 the interferogram resulting from the average of 35 interferograms, taken with the prototype oriented at 240° (configuration C) with the mask of 82 mm visible in Figure 160.

*Figure 160. The mask used to compute the cold interferograms on-axis with the prototype oriented at 240° (configuration C).*

*Figure 161. The cold interferograms on-axis with the prototype oriented at 240° (configuration C).*

### 2.8.6.3 Cold Interferometric Test at position C/0°, C/-5.6°, C/-10°, C/-16.7

We describe here the interferometric result for the position C (240° of rotation) since both in Hartmann and in PSF test this configuration appears more problematic. The other angles will be discussed in the conclusion of this chapter. The same procedure described in section 2.8.5.1 has been performed for all FoV angles. The mask used for the interferograms computation has been dimensioned and positioned to maximize the common sensed area with the corresponding cold interferograms. In all cases, an 82 mm diameter mask has been used.

In Table 32, we report the primary aberrations measured compared with the simulated ones over the same pupil diameter, for all angles included on-axis. The color code of the measured values is representative of the proximity to the simulated cases, blue is near the best case, and orange is near the worst case. We select the z-position of the spherical mirror to minimize the defocus term.

*Table 32. The primary aberrations measured (boldface) with cold interferograms at 0°,-5.6°,-10°,-16.7° position with the prototype oriented at 240° (configuration C), compared to the simulated ones (Zemax) over the same pupil diameter. The defocus Zernike term measured for this dataset was reported in the table.*

| 240° TOU rotation 82 mm pupil diameter | Spherical (λ) | Coma (λ) | Astigmatism (λ) | Overall PTV (λ) | Overall RMS (λ) |
|---|---|---|---|---|---|
| **On-axis  defocus -0.163 λ** | | | | | |
| **Measured** | 0,2 | 0,05 | 0,11 | 0,86 | 0,13 |
| **Sim. worst case  0°** | 0,42 | 0,27 | 0,46 | 2,22 | 0,43 |
| **Sim. best case 0°** | 0,31 | 0,01 | 0,01 | 0,96 | 0,16 |
| **Off-axis -5.6° defocus -0.409 λ** | | | | | |
| **Measured** | 0,38 | 0,57 | 0,24 | 1,71 | 0,38 |
| **Sim. worst case** | 0,41 | 0,45 | 0,52 | 1,78 | 0,4 |
| **Sim. best case** | 0,31 | 0,01 | 0,01 | 0,74 | 0,16 |
| **Off-axis -10° defocus 0.087 λ** | | | | | |
| **Measured** | -0,29 | 0,36 | 0,11 | 2,1 | 0,24 |
| **Sim. worst case** | 0,33 | 0,33 | 0,94 | 2,42 | 0,5 |
| **Sim. best case** | 0,24 | 0,01 | 0,03 | 0,65 | 0,14 |
| **Off-axis  -16.7° defocus 0.375 λ** | | | | | |
| **Measured** | -0,15 | 0,36 | 1,09 | 2,68 | 0,49 |
| **Sim. worst case** | 0,06 | 0,69 | 1,48 | 3,67 | 0,67 |
| **Sim. best case** | 0 | 0,07 | 0,14 | 0,99 | 0,22 |

## 2.8.7   Optical Quality conclusions

In this section, we try to summarize the test results, analyzing both the interferometric test, the Hartmann test, and the PSF measurements.

**Issue Solving:** the interferometric test was performed in Leonardo realizing on the fly a GSE to hold the spherical mirrors on the side of the detector. Because of space issues inside the cryo-vacuum chamber and of interface issues with the pre-existing GSE, we could select only a spherical mirror within a certain diameter and radius of curvature. We selected 3 different mirrors, whose geometrical characteristics allowed us to collect interferograms, both in warm and cold conditions, at four radial positions of the TOU FoV.

Additionally, to avoid frosting on the vacuum chamber input window, we have been obliged to blow a constant $N_2$ flow on it, causing much turbulence. The last fact, added to the vibrations of the camera for the normal flow of $LN_2$ to maintain the cold, was causing a very harsh environment to collect interferograms, while this is a minor problem for Hartmann and PSF test.

Moreover, one of the linear stages had an issue and stopped at temperatures below -70°C; this made very difficult the fine alignment of the mirror in cold conditions.

For all these reasons, the quality of the interferograms, in particular in cold conditions and the more off-axis positions was not optimal in the overall pupil diameter. Moreover, we have been obliged to use masks with a maximum diameter of about 82mm, that we used for all the possible configurations (warm/cold and all the positions in the FoV) to make a comparison as homogeneous as possible.

**PSF Ensquared Energy**: the EE computed from the direct PSF measurements is pretty much dependent on the background removal, giving to these measurements an intrinsic uncertainty that can reach values as high as the measurements themselves. To make an example, 0.5% of the variation in the choice of the background to be removed causes a 20% variation in the estimated EE.

**Hartmann Ensquared Energy**: the EE computed through the Hartmann test is independent both from the environment and from the background. Indeed every data set is, in reality, the average of a large number of measurements, and data have a good SNR since the results are depending on the spots barycentre.

For all these reasons, we believe that the Hartmann test gave the most reliable results.

We want to emphasize that:

- The various tests have been performed in slightly different temperature conditions, above all in the first phases of the campaign, with an orientation of the prototype at 0° and 120°. In fact, finding the optimal temperature to be applied to the Cryotec shroud and optical bench that allows keeping the TOU at the working temperature took longer than expected. The error associated with this fact has been evaluated in section 2.8.6.1, and it is of the order of 10%-30% of the estimated performance in terms of EE.

- We had to live with a relevant temperature gradient between the prototype lenses due to the optomechanical setup. The error associated with this fact has been evaluated with simulation in section 2.8.4.1, and it is of the order of 10%-30% of the estimated performance in terms of EE.

Considering the more significant error for both cases (30%), and summing them in quadrature, we conservatively envisage that the level of confidence of the estimated EE that will be shown in the following sections is of the order of 40%.

### 2.8.7.1 Interferometric Test comparison

Recalling the comparison in Table 32, we show a few comparisons between warm and cold interferometric test, performed with the same mask of diameter 82 mm centered on the interferogram, considering the error described in section 2.6.3 and recalled in Table 33, resulting from 1 pixel for the centering error and ±2% on the mask diameters error. The values highlighted in Table 34are worse than what simulated, also considering the error. Significant differences are observed in off-axis value and in cold conditions.

*Table 33. The Uncertainty on the Zernike coefficients due to errors in positioning and dimensioning the mask.*

|              | ASTIG.1 | ASTIG.2 | COMA1 | COMA2 | SPHER. |
|--------------|---------|---------|-------|-------|--------|
| Error (PtV)  | 0.144   | 0.270   | 0.108 | 0.180 | 0.072  |

We have shown in Figure 162 the difference between the measured performance with Zygo and the expected one with Zemax in Warm conditions. The differences should all tend to zero, while we experience a relevant difference in astigmatism at the most significant off-axis positions (11.2 and 16.7 degrees). Small differences are evident also in the coma at all off-axis positions.

*Table 34. The comparison between warm and cold interferograms over the same surface (mask diameter 82mm), unit in wave @633 nm.*

| APERTURE 82mm | | | | | | | | |
|---|---|---|---|---|---|---|---|---|
| | on-axis | | | | off-axis 5.6° | | | |
| | Warm | | Cold | | Warm | | Cold | |
| | Zemax | Measured | Zemax | Measured | Zemax | Measured | Zemax | Measured |
| Astig. 1 | 0,000 | **0,068** | 0,000 | **-0,094** | -0,217 | **0,251** | 0,014 | **0,244** |
| Astig. 2 | 0,000 | **0,031** | 0,000 | **-0,064** | 0,000 | **-0,121** | 0,000 | **-0,026** |
| Coma 1 | 0,000 | **0,384** | 0,000 | **-0,05** | 0,000 | **0,385** | 0,000 | **0,402** |
| Coma 2 | 0,000 | **0,207** | 0,000 | **-0,01** | -0,212 | **0,25** | 0,112 | **0,402** |
| Spher. | 0,125 | **0,098** | 0,358 | **0,2** | 0,129 | **0,122** | 0,363 | **0,384** |
| Defocus | 0,000 | **-0,237** | 0,000 | **-0,163** | 0,000 | **-0,128** | 0,000 | **-0,409** |
| | off-axis 11.2° | | | | off-axis 16.7° | | | |
| | Warm | | Cold | | Warm | | Cold | |
| | Zemax | Measured | Zemax | Measured | Zemax | Measured | Zemax | Measured |
| Astig. 1 | -0,720 | **0,588** | 0,287 | **0,034** | -1,826 | **1,215** | 0,763 | **1,028** |
| Astig. 2 | 0,000 | **-0,242** | 0,000 | **0,116** | 0,000 | **-0,222** | 0,000 | **0,349** |
| Coma 1 | 0,000 | **0,342** | 0,000 | **-0,258** | 0,000 | **0,239** | 0,000 | **-0,352** |
| Coma 2 | -0,719 | **-0,234** | -0,040 | **-0,238** | -1,441 | **-0,992** | -0,337 | **-0,096** |
| Spher. | 0,045 | **0,14** | 0,281 | **-0,303** | -0,225 | **-0,077** | 0,012 | **-0,151** |
| Defocus | 0,000 | **0,098** | 0,000 | **0,087** | 0,000 | **0,21** | 0,000 | **0,375** |

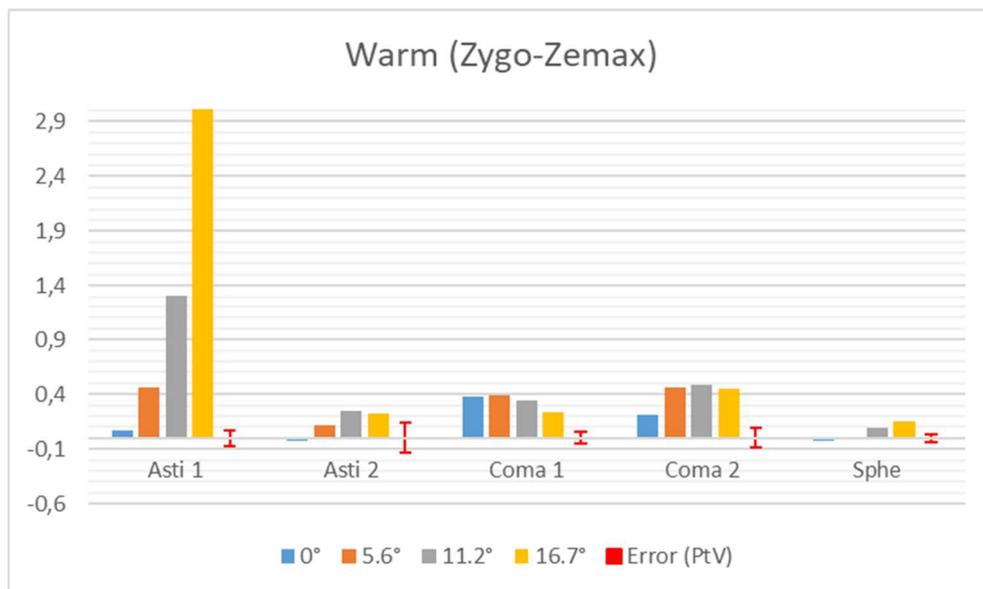

*Figure 162. The difference between measured performance with Zygo and the expected one with Zemax in Warm conditions. In red, is shown the relative error of the aberration.*

The differences between the measured performance with Zygo and the expected one with Zemax in Cold conditions are shown in Figure 163. Also, in this case, the differences should all tend to zero, and the discrepancies between the model and the measured data look to be smaller than in warm conditions.

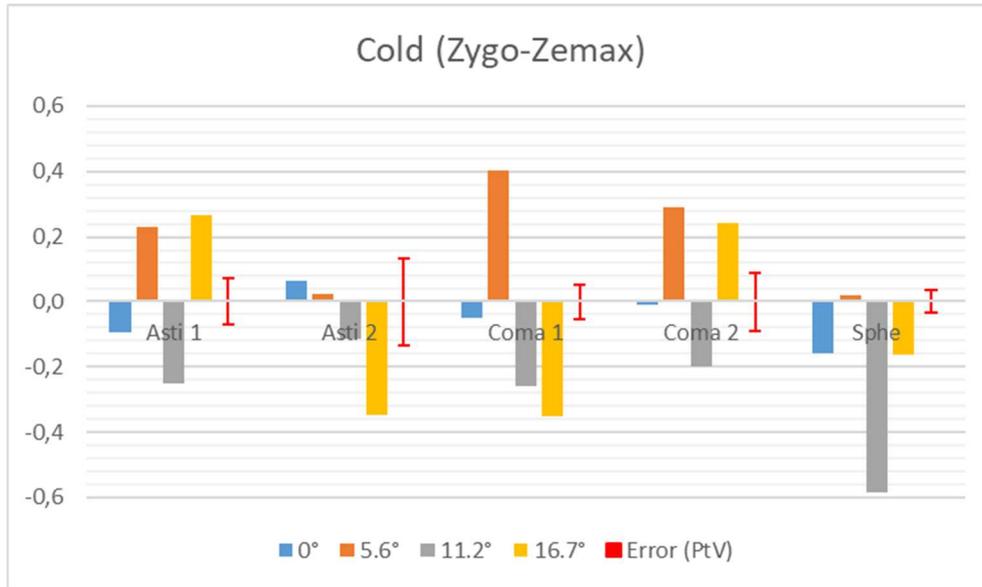

*Figure 163. The measured performance with Zygo and the expected one with Zemax in Cold conditions. In red, is shown the relative error of the aberration.*

In Figure 164, there is instead the trend of the Zernike coefficients in warm conditions computed with Zemax (on the left image) and measured with Zygo (on the right image). For the consistency between model and measured performance, the two images (left and right) should look the same, while the astigmatism trend seems to be inverted.

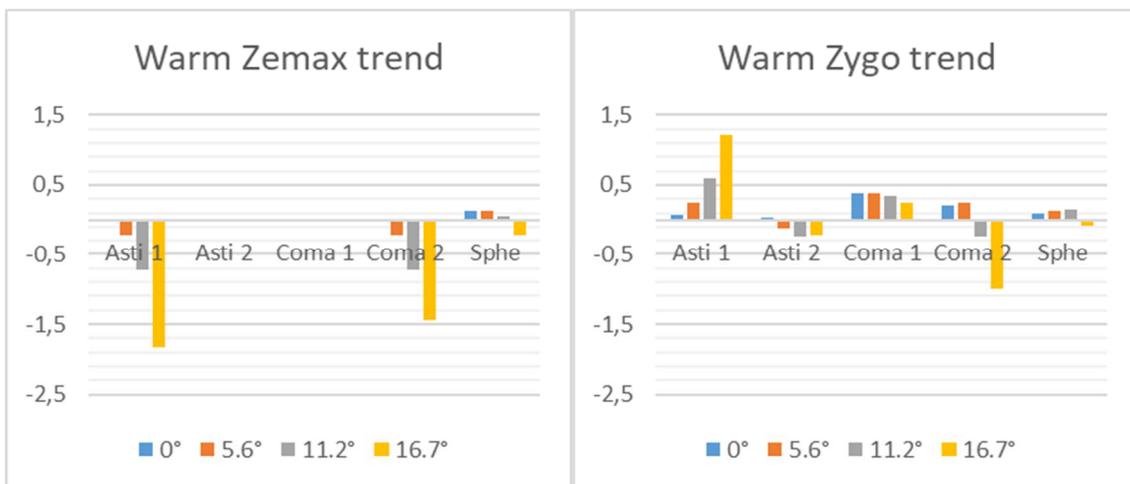

*Figure 164. The Zernike coefficients variations for different FoV positions as expected from Zemax (on the left image) and measured from Zygo (on the right image) in Warm conditions.*

The same comparison between the trends in cold conditions was shown in Figure 165. Also in this case, for the consistency between model and measured performance, the two images (left and right) should look the same, while the coma trend seems to be different, even if considering the uncertainties the variation is small, of the order of 200nm.

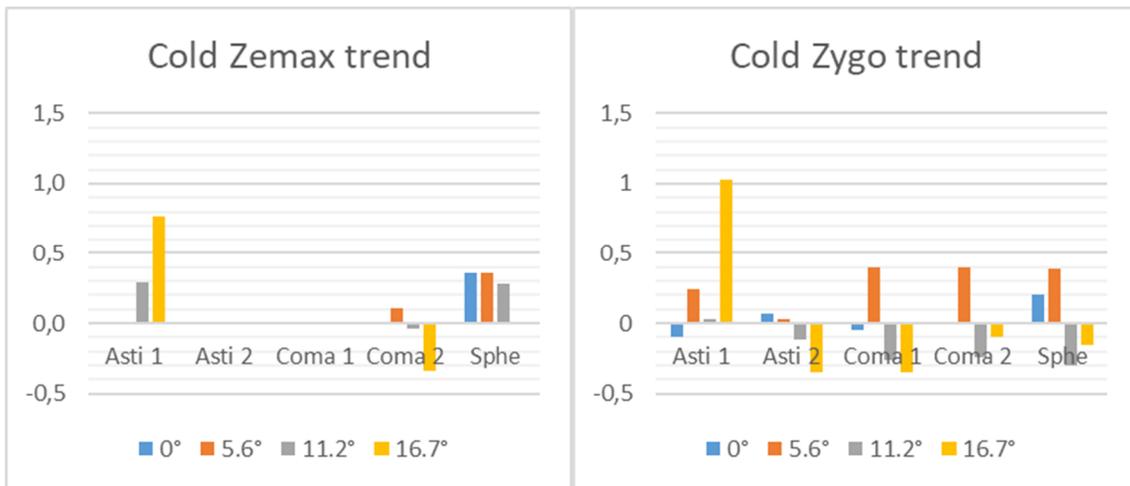

*Figure 165. The Zernike coefficients variations for different FoV positions as expected from Zemax (on the left image) and measured from Zygo (on the right image) in Cold conditions.*

In Figure 166, there is the difference between the cold and warm Zernike coefficients for different FoV positions as expected from Zemax, on the left image, and measured from Zygo, on the right image. For consistency between simulated and measured performance, the two figures should look the same, while there is a large discrepancy in the astigmatism coefficient at 11.2 and 16.7 degrees positions, and also the coma at 11.2 degrees is quite different, even considering the uncertainties.

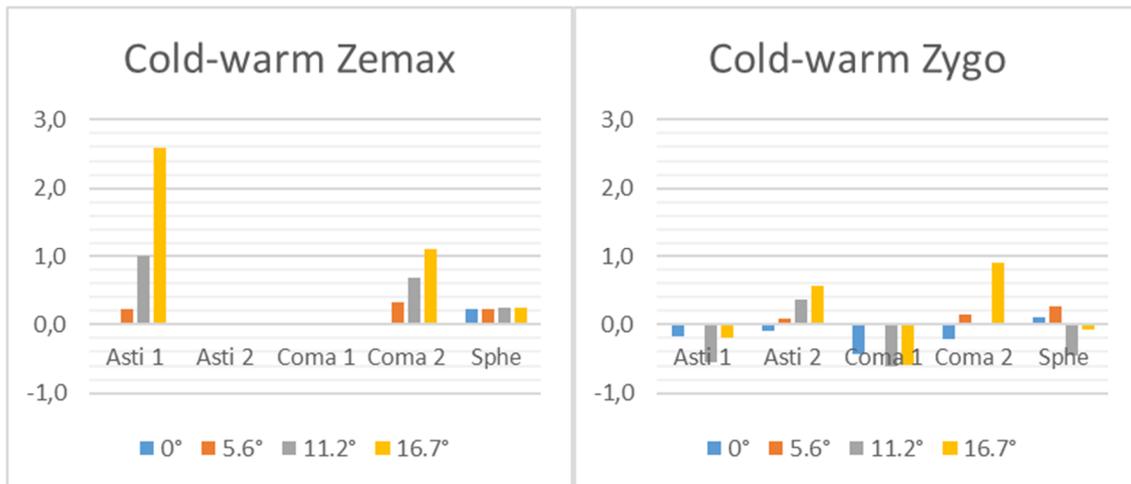

*Figure 166. The difference between the cold and warm Zernike coefficients for different FoV positions as expected from Zemax (on the left image) and measured from Zygo (on the right image).*

We have some doubts about the consistency of the signs of the coefficients computed by Zygo and simulated with Zemax, and we made a test with the interferometer which showed that the astigmatism coefficient comes with an inverted sign depending on the defocus coefficient sign. We thus recomputed all the plots in this case, based on the defocus coefficient sign shown Table 34, changing the signs of all the astigmatism coefficients with associated negative defocus. The results are shown in the plots of Figure 167, where we show the difference between the measured performance with Zygo and the expected one with Zemax, both in Warm (Figure 167 on the left image) and in Cold (Figure 167 on the right image) conditions. We then did the opposite exercise, changing the signs of all the astigmatism coefficients with positive defocus, and the results are shown in Figure 168.

These plots have to be compared with Figure 162 and Figure 163. The coma and spherical coefficients are the same while concerning astigmatism:

- in Figure 167, the situation is pretty much similar to the original plots, with small modifications, both in warm and in the cold, of the astigmatism coefficients at 0° and 5.6° off-axis positions;

- in Figure 168, the situation improves a lot in warm conditions, while it is getting worse in the cold, above all in the two most off-axis positions.

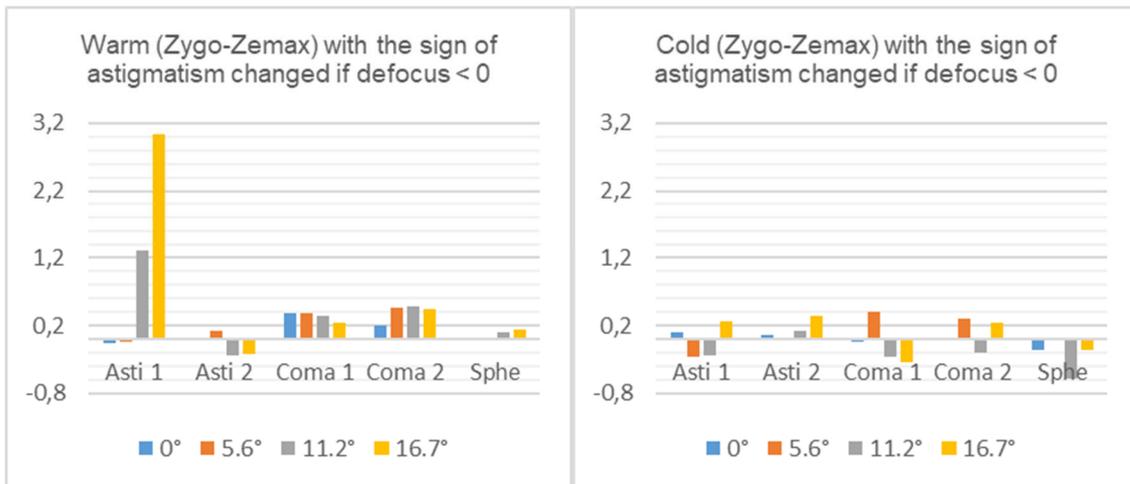

*Figure 167. The difference between measured performance with Zygo and the expected one with Zemax in Warm conditions (on the left image) and Cold conditions (on the right image) with astigmatism sign changed when the defocus term was negative.*

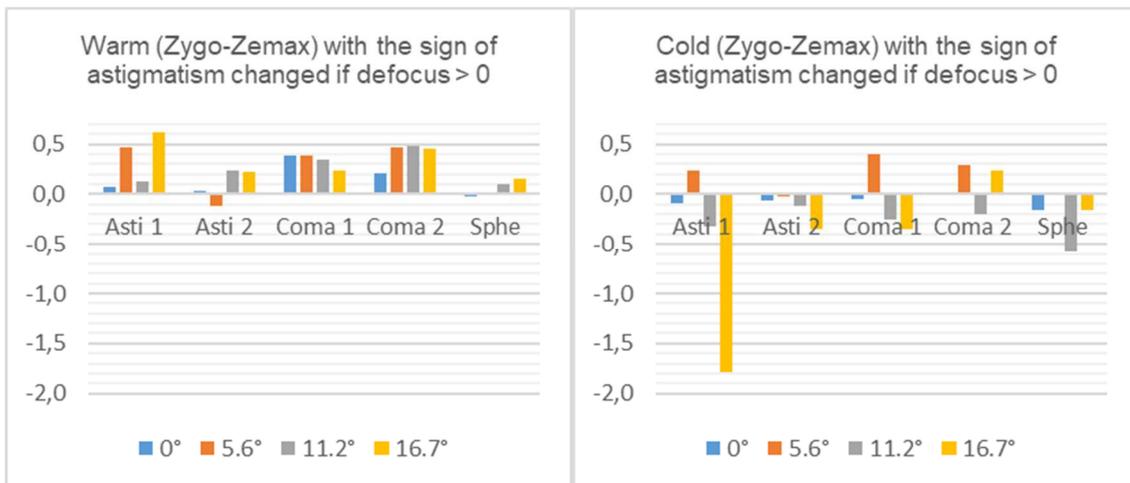

*Figure 168. The difference between measured performance with Zygo and the expected one with Zemax in Warm conditions (on the left image) and Cold conditions (on the right image) with astigmatism sign changed when the defocus term was positive*

## 2.8.7.2 Ensquared Energy performance comparison

We report here a summary of the performance obtained in warm (Figure 136, page 160) and cold (Figure 154, page 171) conditions, in terms of EE computed through the Hartmann test.

It has to be noted that:

- the warm performance is almost always better than both the best and the nominal simulated performance, and this behavior is even enforced when going off-axis;

- the cold performance is almost always worse than the worst simulated case and, also, in this case, this behavior is even enforced when going off-axis;

- while the expected/simulated nominal performance in cold conditions with respect to warm conditions should be better of a factor ranging from 2 on-axis to about six at the maximum off-axis tested (16.7 degrees), the measured performance in cold/warm condition is comparable, as it can be seen in Figure 169;

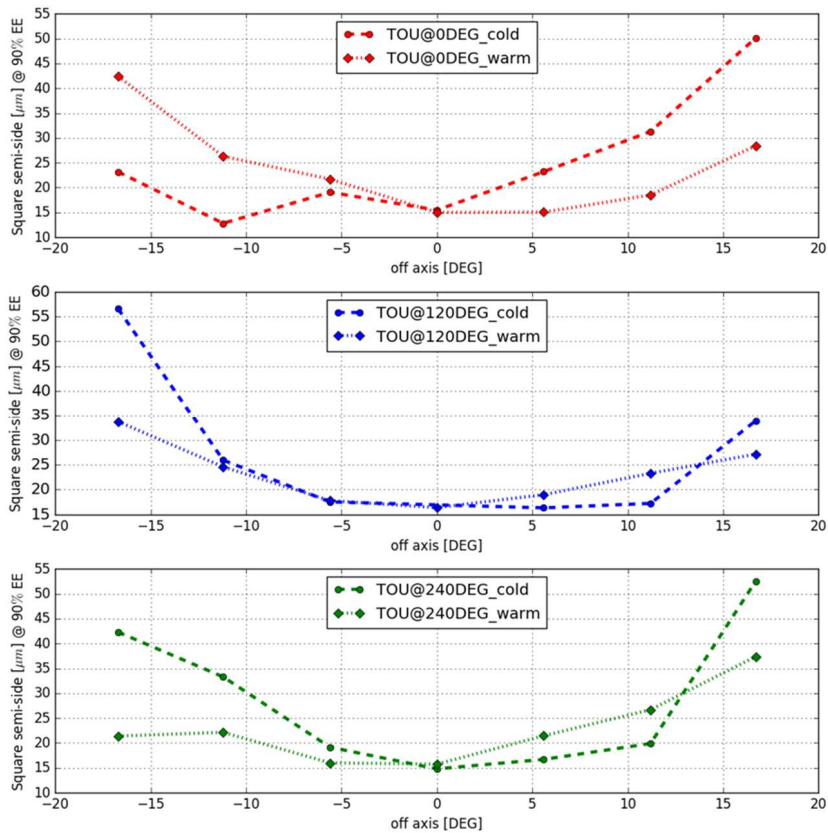

*Figure 169. The comparison between warm and cold results for the enclosed energy at different TOU rotation as a function of the off-axis position.*

- we also recall the massive discrepancy, in warm conditions, between the simulated best focal plane positions, and the measured ones (Figure 170), also pretty much field-dependent, similarly to the EE performance both in warm and in cold conditions.

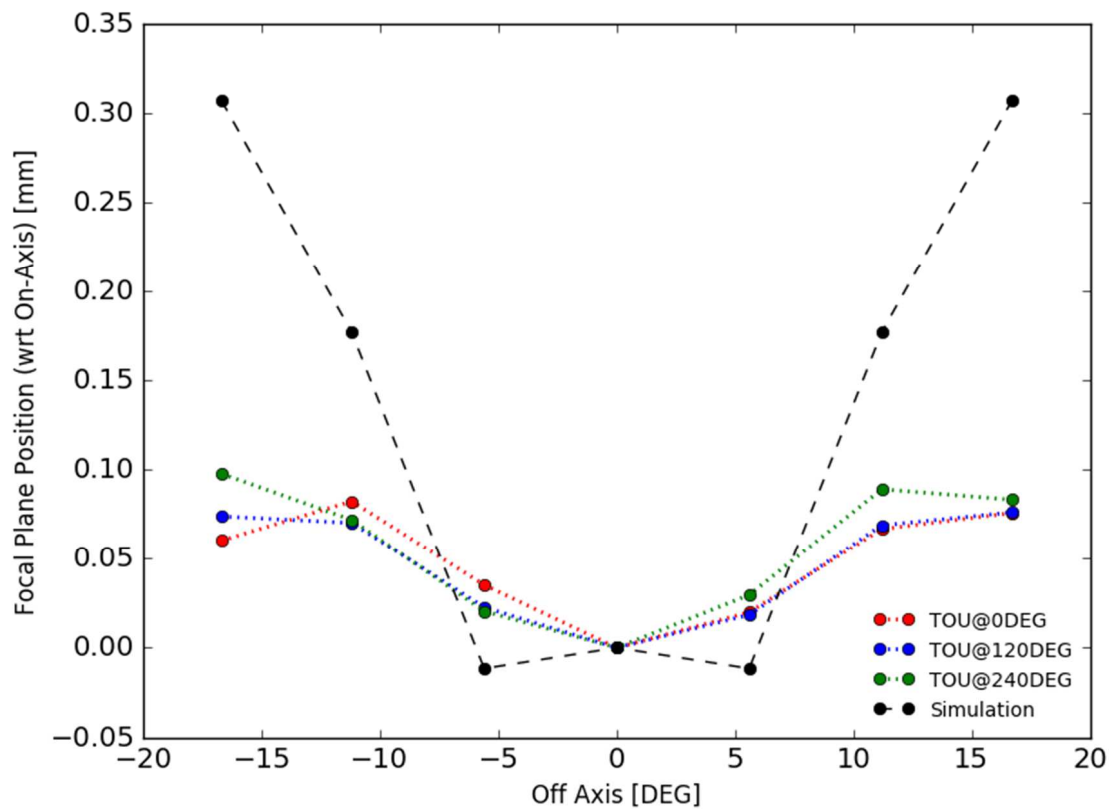

*Figure 170: Assuming the same focal plane position on-axis for the three TOU configurations, here the trend of the focal planes for the off-axis positions and the three orientations of the TOU. Warm case.*

Given these observables, our considerations are:

- there are a few issues in the warm and cold performance registered which are probably related to the alignment performed on the prototype, which is mostly related to a non-center-symmetric quality distribution, above all in the 0 degrees orientation in cold conditions. This fact can be seen in Figure 137, page 161 and Figure 155, page 171;

- nevertheless, there is a significant issue related to the overall quality obtained, which is better than expected in warm and worse than simulated in the cold, which is in our opinion challenging to be reconducted to the alignment performed;

- the fact that both the warm/cold performance and the "best focal plane position" in warm are field dependent suggest that it is very unlikely due to a problem in one of the spherical lenses, while it may be more probable that the problem may come from the aspherical lens, which by definition has a differential effect over the field.

### 2.8.7.3 Investigations of measured performance

We try in this section to find possible reasons explaining the difference between the measured and simulated (through Zemax) performance, evidenced in the previous section. The first thing we investigated is whether this effect may come from a problem deriving from prototype larger alignment errors. We thus simulated the case in which we did the alignment with tolerances exaggeratedly worse than the nominal ones, looking at the effect on the performance both in warm and in cold conditions. For this set of simulations, lenses manufacturing errors provided by the supplier have been considered. The results of the Montecarlo analysis are shown in Table 35 and Table 36, from which it can be clearly seen that the cold performance we measured may be compatible with the worst case of the 2x and 3x tolerances and the warm performances may be compatible with the best case of the 2x and 4x tolerances.

Given these results, we decided to reassess the errors associated with the alignment procedure in a more accurate way to get reasonable values to be used for further analyses.

*Table 35. The warm performance when simulating alignment tolerances 2x and 4x worse than the nominal ones.*

| Semi-width of the square enclosing n% of energy [um] | WARM WITH HARTMANN MASK | | | | | | | | | | | |
|---|---|---|---|---|---|---|---|---|---|---|---|---|
| | Field | on-axis | | | off-axis 5.6° | | | off-axis 11.2° | | | off-axis 16.7° | | |
| | Align tol. | nominal | 2x worse | 4x worse | nominal | 2x worse | 4x worse | nominal | 2x worse | 4x worse | nominal | 2x worse | 4x worse |
| | Nominal | 17.0 | 17.0 | 17.0 | 19.7 | 19.7 | 19.7 | 31.3 | 31.3 | 31.3 | 59.9 | 59.9 | 59.9 |
| | Best | 15.2 | 14.8 | 15.2 | 16.5 | 15.8 | 15.7 | 24.8 | 22.2 | 17.0 | 51.0 | 45.3 | 36.4 |
| | 20% percentile | 17.0 | 17.7 | 19.5 | 19.5 | 20.0 | 22.0 | 29.8 | 30.2 | 33.2 | 59.0 | 59.8 | 59.4 |
| | 50% percentile | 18.0 | 19.3 | 23.5 | 20.8 | 22.4 | 27.1 | 32.6 | 35.2 | 41.1 | 63.3 | 65.5 | 72.7 |
| | 80% percentile | 20.3 | 22.5 | 30.2 | 22.5 | 25.8 | 34.5 | 35.8 | 40.6 | 51.0 | 67.2 | 71.9 | 86.5 |
| | worst | 31.0 | 35.8 | 50.0 | 31.5 | 37.5 | 52.0 | 44.6 | 59.1 | 80.1 | 78.8 | 93.5 | 124.5 |

*Table 36. The cold performance when simulating alignment tolerances 2x and 3x worse than the nominal ones.*

| | Field | on-axis | | | off-axis 5.6° | | | off-axis 11.2° | | | off-axis 16.7° | | |
|---|---|---|---|---|---|---|---|---|---|---|---|---|---|
| **Semi-width of the square enclosing n% of energy [um]** | Align tol. | nominal | 2x worse | 3x worse | nominal | 2x worse | 3x worse | nominal | 2x worse | 3x worse | nominal | 2x worse | 3x worse |
| | **Nominal** | 7.8 | 7.8 | 7.8 | 7.7 | 7.7 | 7.7 | 7.6 | 7.6 | 7.6 | 12.2 | 12.2 | 12.2 |
| | **Best** | 5.9 | 6.3 | 5.6 | 6.7 | 7.1 | 7.4 | 6.2 | 6.6 | 7.7 | 9.8 | 11 | 10.6 |
| | **20% percentile** | 7.6 | 8.9 | 10.8 | 8.1 | 9.7 | 11.9 | 8.4 | 11.4 | 14.6 | 13.1 | 15.5 | 21.7 |
| | **50% percentile** | 8.6 | 11.3 | 15.1 | 9.1 | 12.2 | 15.9 | 10.4 | 15.2 | 20.2 | 15.8 | 21.3 | 34.9 |
| | **80% percentile** | 9.8 | 14.2 | 19.5 | 10.5 | 15.4 | 21.8 | 13.3 | 20.1 | 27.8 | 20.4 | 29.9 | 53.6 |
| | **worst** | 17.1 | 24.2 | 32.6 | 15.6 | 24.7 | 35.2 | 21.8 | 34.8 | 49.3 | 38.1 | 56 | 95 |

*Table 37. The alignment tolerances as from requirements (light blue), first estimation of the achieved accuracy during the alignment (green) and a revised and more conservative estimation (grey).*

| Lens | Alignment tolerances | | | Estimated alignment accuracy | | | Revised alignment accuracy | | |
|---|---|---|---|---|---|---|---|---|---|
| | Decenter | Tilt | Focus | Decenter | Tilt | Focus | Decenter | Tilt | Focus |
| L1 | ± 22 μm | ± 44'' | ±15 μm | ± 15 μm | ± 10'' | ±15μm | ± 20 μm | ±20'' | ±44 μm |
| L2 | ± 22 μm | ± 44'' | ±30 μm | ± 15 μm | ± 10'' | ±15μm | ± 20 μm | ± 20'' | ±44 μm |
| L3 | ± 22 μm | ± 44'' | ±40 μm | ± 15 μm | ± 10'' | 0 μm | ± 12 μm | ± 20'' | 0 μm |
| L4 | ± 22 μm | ± 44'' | ±20 μm | ± 8 μm | ± 10'' | ±15μm | ± 18 μm | ±20'' | ±44 μm |
| L5 | ± 22 μm | ± 44'' | ±20 μm | ± 8 μm | ± 10'' | ±15μm | ± 18 μm | ± 20'' | ±44 μm |
| L6 | ± 42 μm | ± 44'' | ±20 μm | ± 8 μm | ± 10'' | ±15μm | ± 18 μm | ± 20'' | ±44 μm |

Re-evaluating each error source both for tilt, decentre and focus and applying a certain number of conservative assumptions we devised the numbers shown in Table 37 which are:

• concerning tilt and decentre, slightly worse than the first estimation (shown in orange in Table 37), but still within alignment tolerances (shown in light blue in Table 37);

• concerning focus accuracy, due to several reasons explained in section 2.8.7.5, worst of a factor about 3 with respect to the initial estimation.

Comparing what we obtained after the revision of the accuracies with the Montecarlo simulations presented above, we believe that the achieved alignment tolerances are unlikely to be so loose (especially for what concerns tilts and decentre) and, for this reason, we evaluated other possible sources of error which are presented in the paragraphs below.

## 2.8.7.4 The problem in the L1 profile

Considering that the performance discrepancy is strongly field-dependent, we did evaluate the possibility that the only aspherical lens, L1, may have a problem. We decided to make a test redesigning the profile of L1 based on the warm measured performance, to check what the impact may be on the cold simulated performance. This attempt was furthermore justified from the fact that one of the worst offenders (in terms of EE deterioration and position of the best focus against FoV) is indeed the error in the radius of curvature of L1. Thus, we changed the first surface of L1 (radius, conic and aspheric coefficients) to match the observations in terms of encircled energy (Figure 136), best focal plane position (Figure 170) and the Zernike spherical aberration coefficient (which has been measured about 0.1 waves for 82mm aperture, see We have shown in Figure 162 the difference between the measured performance with Zygo and the expected one with Zemax in Warm conditions. The differences should all tend to zero, while we experience a relevant difference in astigmatism at the most significant off-axis positions (11.2 and 16.7 degrees). Small differences are evident also in the coma at all off-axis positions.

Table 34), all of them in warm conditions. The parameters of the aspherical surface that we obtained in this way are reported in Table 38, and the L1 sag that we obtained in this way is shown in Figure 171.

*Table 38. The parameters of the aspherical lens L1 derived from the warm performance.*

| L1 PARAMETERS (warm) | | |
|---|---|---|
| | nominal | modified |
| R | 185.817 | 184.292 |
| k | -3.857 | -3.612 |
| a4 | 2.917E-08 | 2.600E-08 |
| a6 | -3.947E-12 | -3.760E-12 |

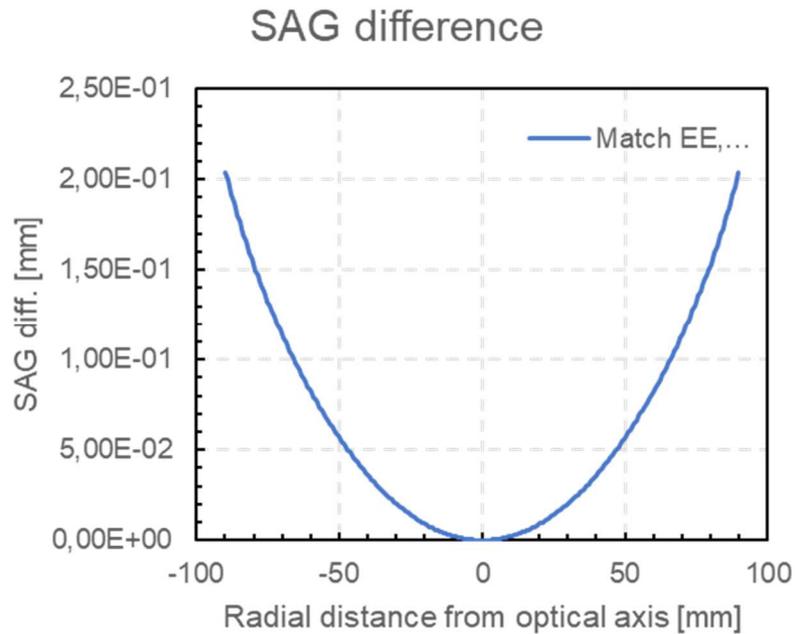

*Figure 171. The SAG difference between the theoretical L1 profile and the one modified based on the warm performance*

In Table 39 we report the simulated warm/cold performance by using the modified L1 profile, while in Table 40 we report the simulated best focal position in warm/cold again applying the modified L1 profile. In Figure 172 we show the nominal performance in warm conditions of the prototype model with the modified L1, which should be similar to the warm measured performance by definition, being L1 modeled on the warm performance.

*Table 39. The warm and cold performance correspondent to the modified L1*

| HARTMANN 90%EE SQUARE SEMI-SIDE | | |
|---|---|---|
| Field [°] | warm [um] | cold [um] |
| 0 | 14.1 | 7.8 |
| 5.6 | 18.3 | 17.6 |
| 11.2 | 18.2 | 33.4 |
| 16.7 | 34.7 | 51.7 |

*Table 40. The warm and cold best focal plane position correspondent to the modified L1*

| Best focal plane position [mm] | | |
|---|---|---|
| Field [°] | Warm | Cold |
| 0 | 0 | 0 |
| 5.6 | -0.029 | 0 |
| 11.2 | -0.075 | 0.049 |
| 16.7 | -0.082 | 0.232 |

In Figure 173 we show instead the cold estimated (with Zemax) performance of the prototype model with the L1 modified, derived from the thermal model (black solid line) compared to the cold measured performance.

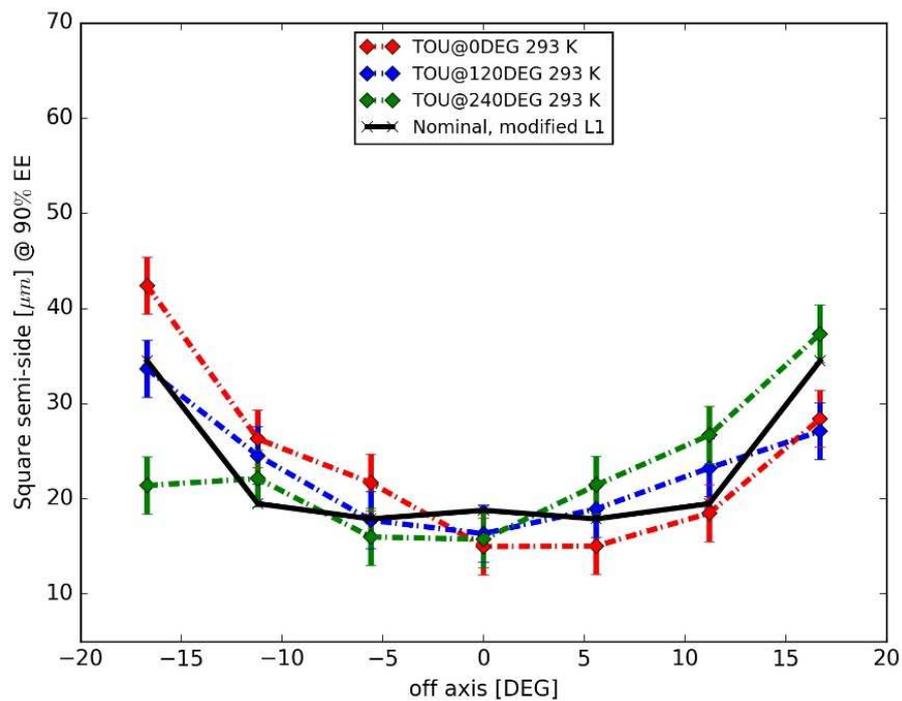

*Figure 172. The nominal warm performance in black solid line, compared to the measured warm performance, based on which L1 has been modelled; of course, the measured and the simulated EE performance are similar by construction in the warm case.*

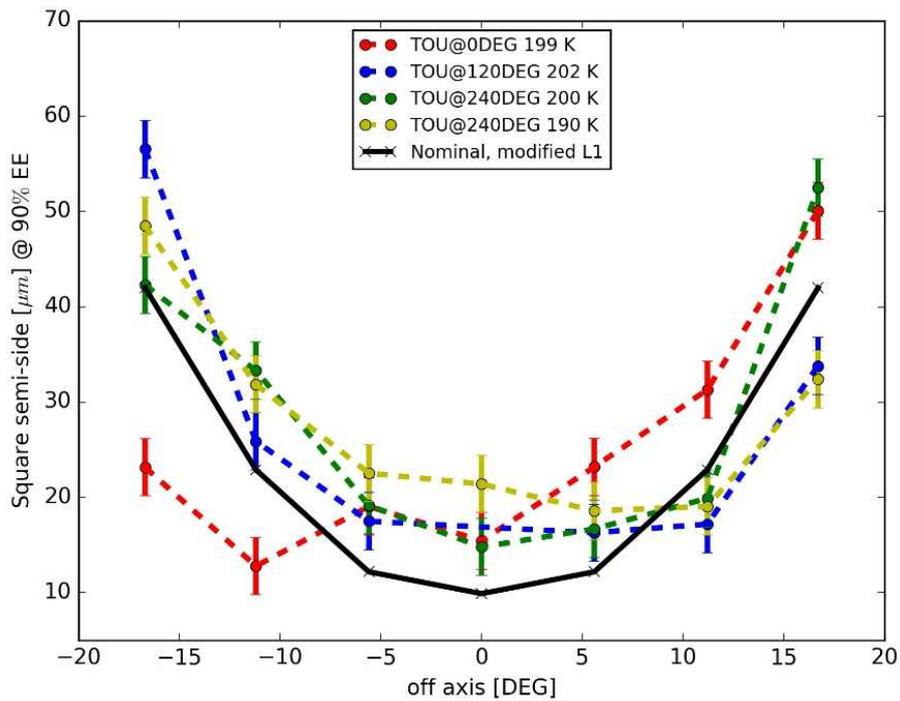

*Figure 173. The nominal cold performance (black line) by using the model with the modified L1.*

It has to be noted that the estimated nominal cold performance derived from the prototype model with the modified L1 is matching in a good way the prototype performance measured at LND, both on and off-axis.

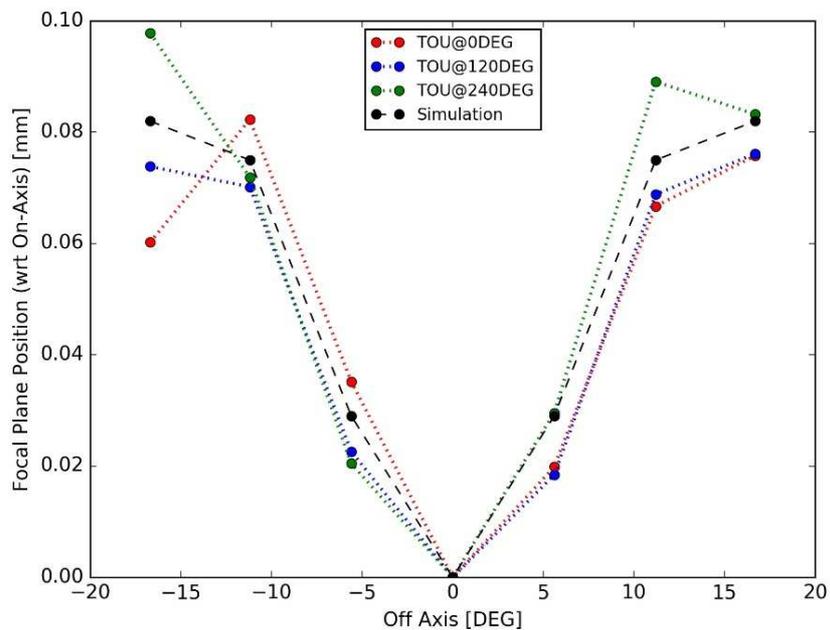

*Figure 174. The warm simulated vs measured best focal plane position, based on which L1 has been modeled; of course, the measured and the simulated positions are similar by construction in the warm case.*

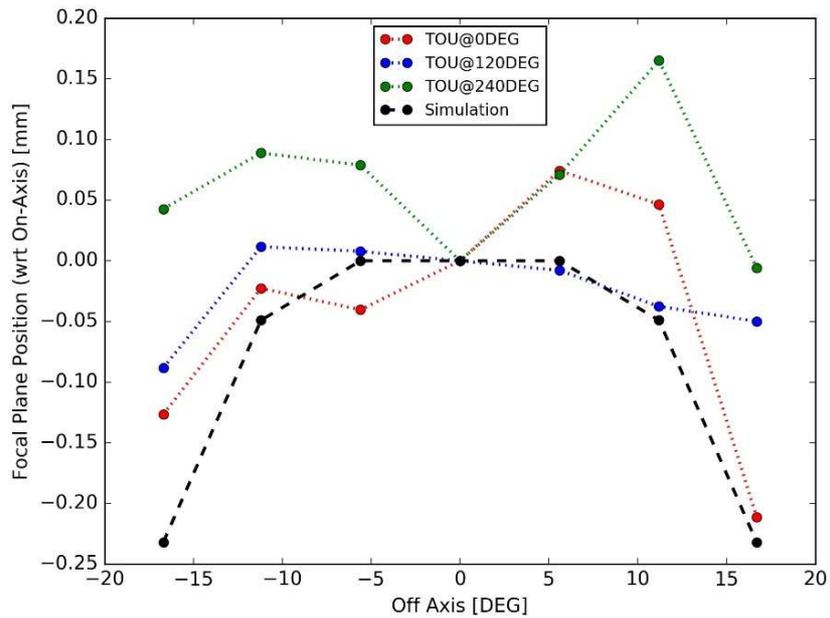

*Figure 175. The cold simulated vs. measured best focal plane position, by using the model with the modified L1.*

Also, the cold best focal plane position (see Figure 175) looks to be reasonably similar to the measured best focal plane positions, apart from the more off-axis position where the difference appears to be more significant. Notice that the plane position is in agreement with the expected field curvature describe in Figure 13.

Indeed, it has to be noted that the difference in the profile (shown in Figure 171) between the nominal L1 and the modified one is reaching about 200µm at the edge of the lens, which is quite a large discrepancy somehow challenging to foresee. The L1 profile has been measured by Leonardo-Medialario, giving a deviation of the L1 profile much smaller than the one expected from Figure 171, of the order of maximum deviation of 10 microns PtV as shown in Figure 176.

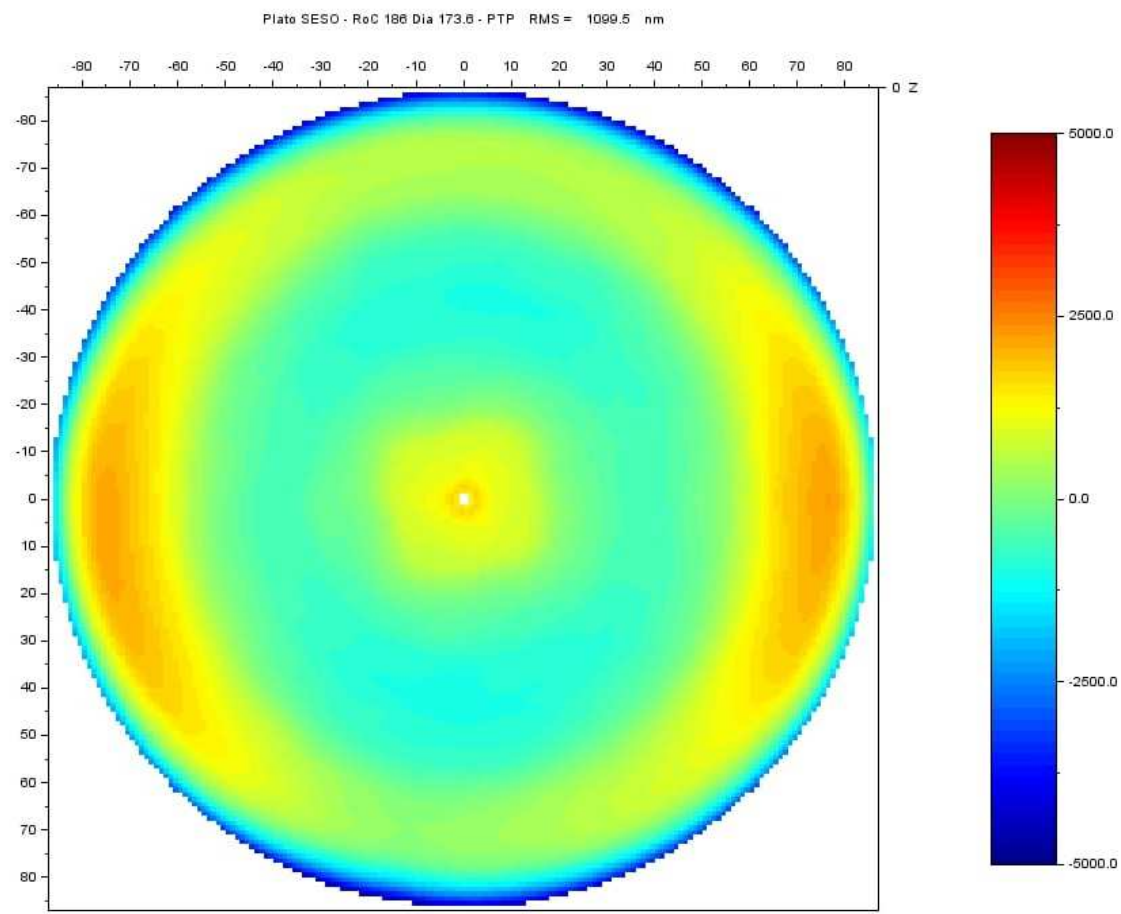

*Figure 176. The L1 surface figure deviation from the nominal profile measured by LND-Medialario.*

Thus, the slightly different profile of L1 absolutely cannot explain the discrepancy between simulated and measured performance.

### 2.8.7.5 Focus error during the alignment.

Another parameter that has a powerful impact, in terms of EE deterioration and position of the best focus against FoV, is indeed the inter-distance between the lenses.

We thus tried to simulate a condition in which the error in aligning the focus position of the lenses is much larger both than that aimed from the tolerance analysis and estimated during the AIV procedure.

The required tolerances in the focus of the lenses are recalled in Table 41. In order to decide new values for the focus alignment tolerances to be used in the simulation, we first revised the estimated precision that we could achieve during the alignment process. Our initial estimate was the one presented in the 3rd column of Table 41, which is essentially ±15 µm for all the lenses but L3, which is, in fact, the reference for the z alignment of all the other lenses.

*Table 41. The alignment tolerances of the 6 lenses (we recall that L3 is the reference for the focus alignment of the other lenses).*

| Lens | Focus tolerances | Focus achieved with AIV | Focus achieved with AIV revised |
|------|------------------|-------------------------|----------------------------------|
| L1   | ± 15 µm          | ± 15 µm                 | ±44 µm                           |
| L2   | ± 30 µm          | ± 15 µm                 | ±44 µm                           |
| L3   | ± 40 µm          | 0 µm                    | 0 µm                             |
| L4   | ± 20 µm          | ± 15 µm                 | ±44 µm                           |
| L5   | ± 20 µm          | ± 15 µm                 | ±44 µm                           |
| L6   | ± 20 µm          | ± 15 µm                 | ±44 µm                           |

This estimate was based on the CMM machine measurement accuracy, claimed to be ±10µm by the producer, which we always applied to every measured position during the AIV process. For every lens position, we had to touch the vertex both of the lens previously inserted and the vertex of the newly inserted, which leads to the quadrature sum of 2 times ±10µm, giving as error estimate about ±15µm, which has been reported in the 3rd column of Table 41 as estimate of the precision achieved during the lenses alignment.

We decided to precisely characterize the CMM machine measurement accuracy, and we ended up with a slightly larger number (about ±15µm, conservatively consider ±20µm) than the one reported in the CMM machine specifications.

Additionally, the focus alignment of each prototype lens has been done slightly differently from what initially foreseen, since we decided to characterize also the position of each lens with respect to one point of the CMM supporting structure, to trace also every possible relative movement between the prototype and the CMM machine itself. Thus, every lens estimated position was in reality performed by:

1. touching a mechanical reference on the optical bench where both the CMM machine and the prototype were placed;

2. touching a mechanical reference on the prototype to see any possible relative movements between the CMM and the prototype structure;

3. touching the vertex of the previously inserted and aligned lens;

4. touching the vertex of the newly inserted lens;

5. touching the same mechanical reference on the prototype just mentioned in item 2 to see if any movement of the structure itself happened during the insertion of the new lens.

All these considerations caused a revised estimate in error to be associated to the focus alignment process of each lens, which is coming from the quadrature sum of 5 times ±20µm, giving as revised accuracy ±44µm, reported in the 4[th] column of Table 41.

Given these considerations, we decided to evaluate the impact on the performance considering two possible different focus alignment accuracies: ±50µm and a largely exaggerated ±100µm. The adopted procedure is the same used for the modified L1 profile: the warm model has been optimized to match the observations in terms of EE and best focal plane position by using the inter-distance between lenses as compensators (the inter-distance was constrained to vary no more than 50um and 100um from the nominal position). Then, the model was transformed into cold and compared to the measurements.

In Table 42 we report the result of the Zemax analysis, in term of 90% EE semi-side square, in both cases (±50µm and ±100µm) and warm and cold conditions, while in Table 43 we report the same but in term of best focal plane position vs. FoV.

*Table 42. The simulated performance (90%EE of the semi-side square) considering more significant focusing alignment tolerance in two different cases, ±50µm and ±100µm of error for each lens.*

| | Hartmann 90%EE Square semi-side [µm] | | | |
|---|---|---|---|---|
| | Error Z ±50µm | | Error Z ±100µm | |
| Field [°] | Warm | Cold | Warm | Cold |
| 0 | 20.5 | 11.7 | 19.3 | 9.4 |
| 5.6 | 18.7 | 12.4 | 18.1 | 12.7 |
| 11.2 | 19.1 | 19.9 | 20.7 | 24.7 |
| 16.7 | 39.2 | 35.7 | 35.3 | 45.3 |

*Table 43. The simulated performance (in term of best focal plane position vs FoV) considering more significant focussing alignment tolerance in two different cases, ±50µm and ±100µm of error for each lens.*

| | Best focal plane position [mm] | | | |
|---|---|---|---|---|
| | Error Z ±50µm | | Error Z ±100µm | |
| Field [°] | Warm | Cold | Warm | Cold |
| 0 | 0.000 | 0.000 | 0.000 | 0.000 |
| 5.6 | -0.038 | -0.013 | -0.032 | -0.006 |
| 11.2 | -0.109 | -0.011 | -0.085 | 0.013 |
| 16.7 | -0.166 | 0.098 | -0.116 | 0.146 |

The results shown in Table 42 and Table 43 are graphically plotted in Figure 177, Figure 178, Figure 179 and Figure 180, for the case of ±50µm focusing accuracy.

They look to be in good agreement, with the exception of Figure 179, in which there is still a particular disagreement between the warm best focal plane position simulated and measured, above all going off-axis in the FoV, where the difference is of the order of factor 2 at 16.7° off-axis. This is due to the fact that ±50 µm of axial displacement is not enough to bring the warm system to the measured values.

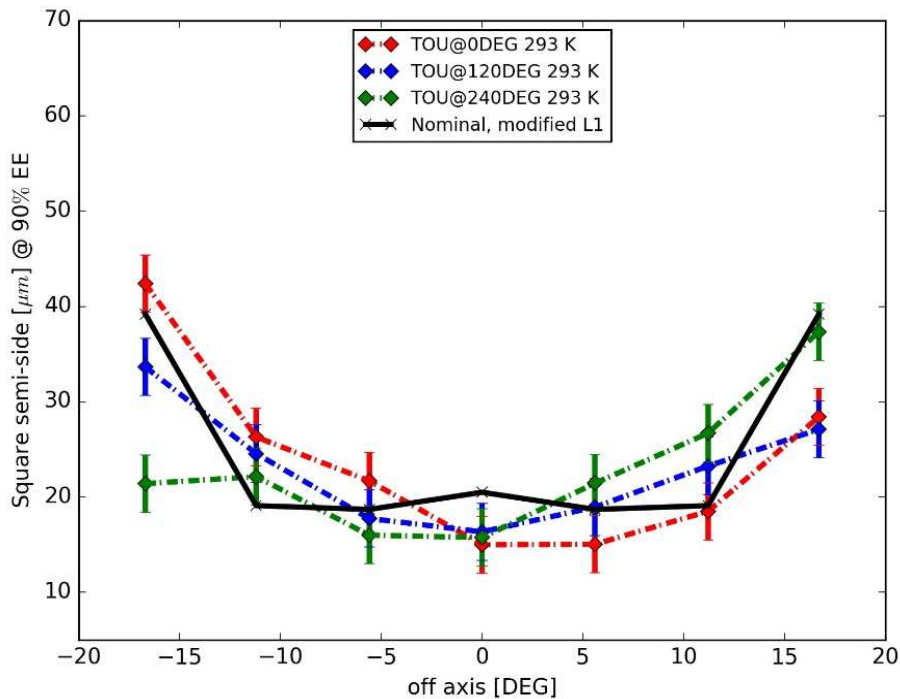

*Figure 177. The **warm** simulated performance (continuous black line) compared with the warm measured performance (colored lines in the usual prototype three different orientations and off-axis positions), in terms of **90% EE** of square semi-side, in the case of the **±50µm** accuracy assumption.*

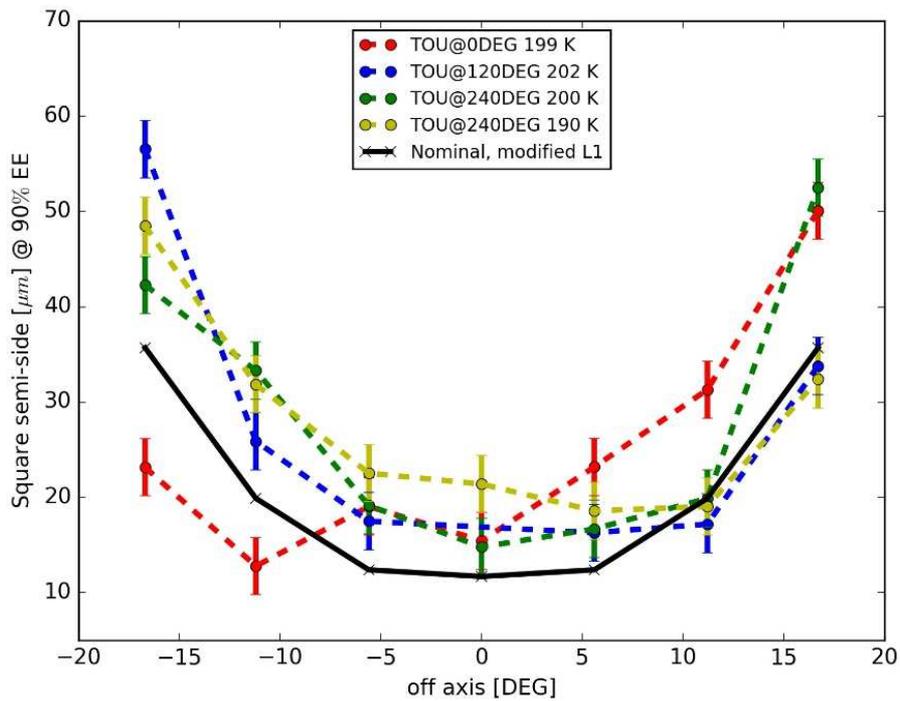

*Figure 178. The **cold** simulated performance (continuous black line) compared with the cold measured performance (colored lines in the usual prototype three different orientations and off-axis positions), in term of **90% EE** of square semi-side, in the case of the **±50µm** accuracy assumption.*

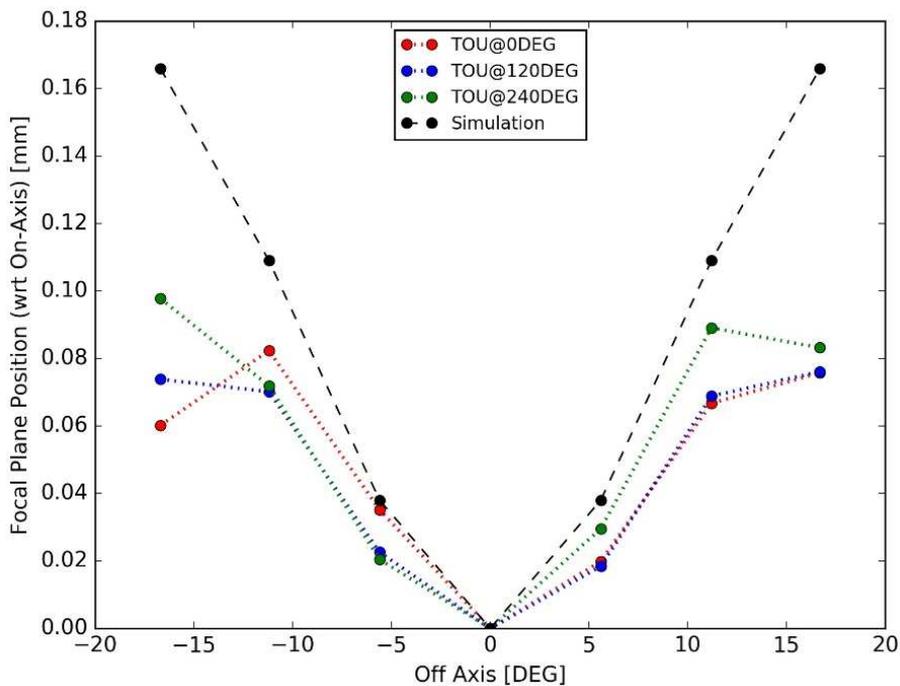

*Figure 179. The **warm** simulated performance (continuous black line) compared with the warm measured performance (colored lines in the usual prototype three different orientations and off-axis positions), in terms of **best FP position vs FoV**, in the case of the **±50µm** accuracy assumption.*

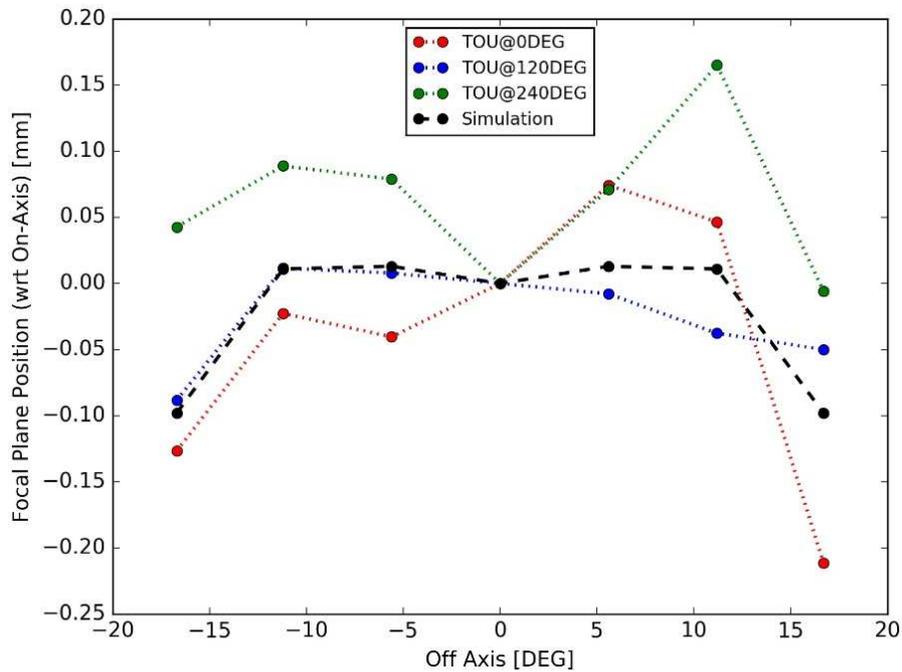

*Figure 180. The **cold** simulated performance (continuous black line) compared with the cold measured performance (colored lines in the usual prototype three different orientations and off-axis positions), in terms of **best FP position vs FoV**, in the case of the **±50µm** accuracy assumption.*

In Figure 181, Figure 182, Figure 183 and Figure 184 we plot instead of the results shown in Table 42 and Table 43, for the case of ±100µm focussing accuracy.

It looks like there is still good agreements concerning the performance in terms of warm and cold EE, while the warm discrepancy in term of best FP position is definitely getting smaller, and the cold best FP positions are still in good agreement with measured values (see Figure 184).

Let us emphasize again that ±100µm looks to be too loose accuracy, but we did try it anyhow to see if the trend was going in the right direction. It looks like ±50µm is already going in the right direction of explaining both warm and cold behavior, even if there must be a combination with something else, which originated the next attempt to investigate the causes of the discrepancies between measured and simulated performances, presented in the next section.

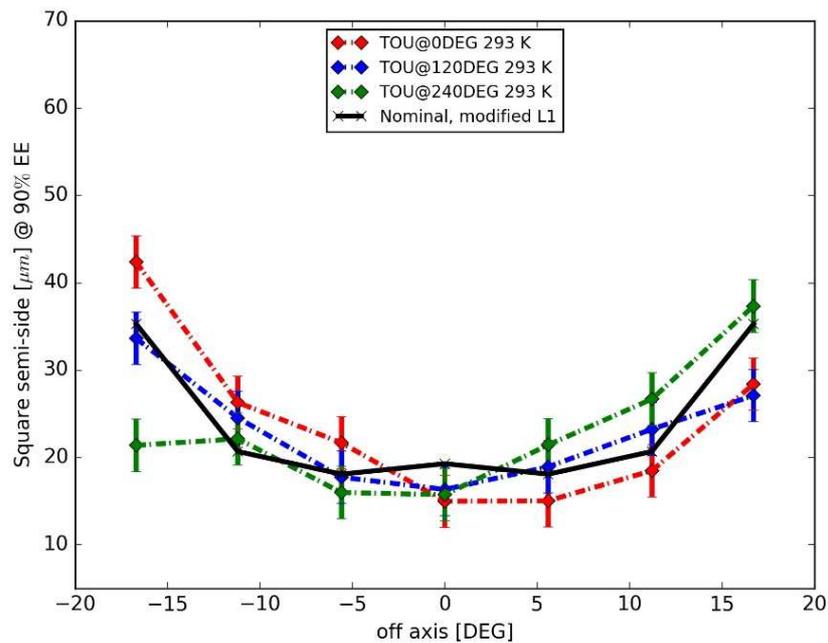

*Figure 181. The **warm** simulated performance (continuous black line) compared with the warm, measured performance (colored lines in the usual prototype three different orientations and off-axis positions), in terms of **90% EE** of square semi-side, in the case of the **±100μm** accuracy assumption.*

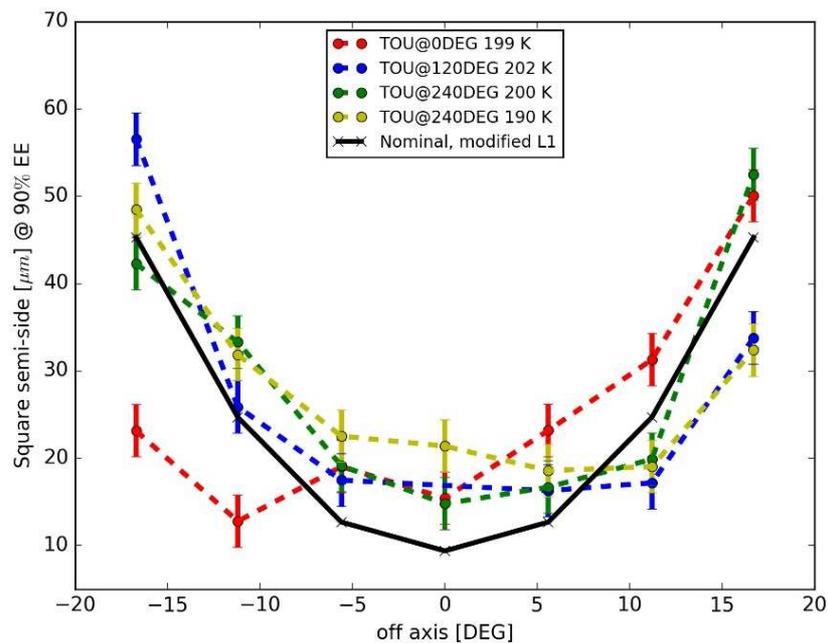

*Figure 182. The **cold** simulated performance (continuous black line) compared with the cold measured performance (colored lines in the usual prototype three different orientations and off-axis positions), in terms of **90% EE** of square semi-side, in the case of the **±100μm** accuracy assumption.*

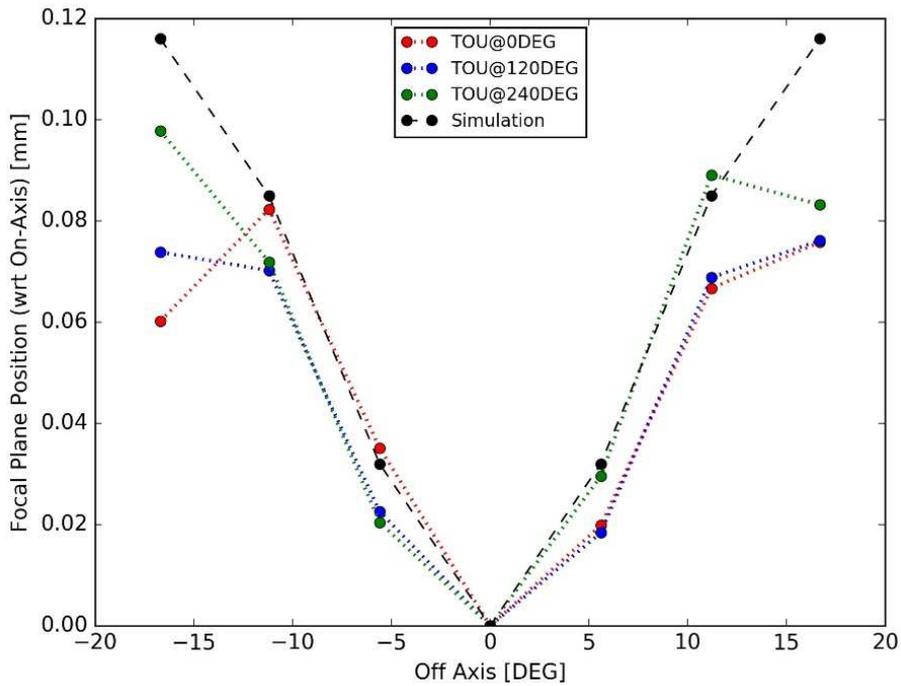

*Figure 183. The **warm** simulated performance (continuous black line) compared with the warm, measured performance (colored lines in the usual prototype three different orientations and off-axis positions), in terms of **best FP position vs. FoV**, in the case of the **±100μm** accuracy assumption.*

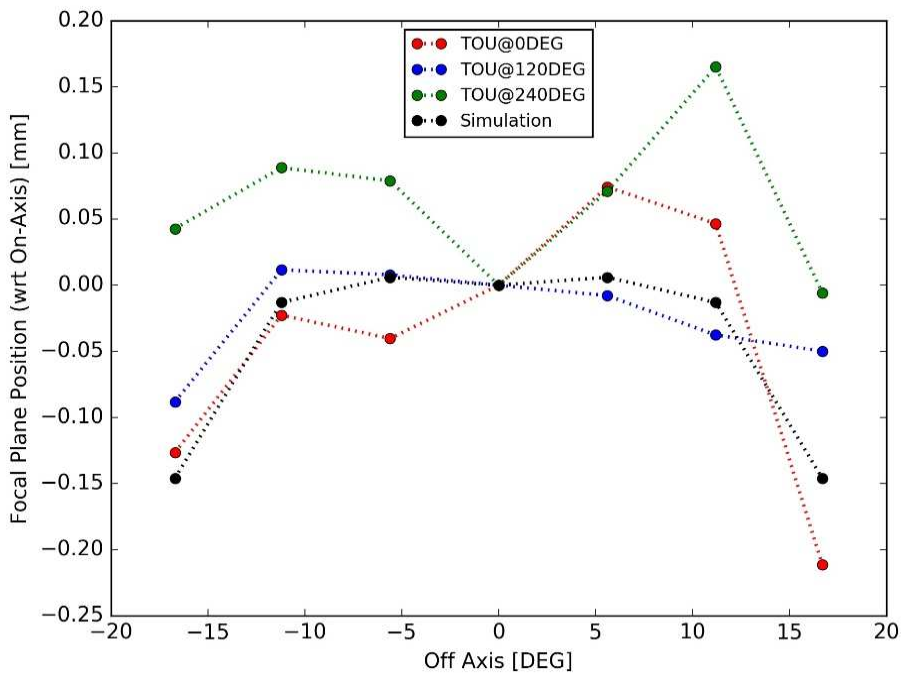

*Figure 184. The **cold** simulated performance (continuous black line) compared with the cold measured performance (colored lines in the usual prototype three different orientations and off-axis positions), in terms of **best FP position vs FoV**, in the case of the **±100μm** accuracy assumption.*

### 2.8.7.6 Focus error from lens thickness and radius of curvature

From the previous sections, it looks clear that the discrepancy between the measured and the simulated performance is probably due to a combination of effects, one of which being for sure a more significant error in the lens focus alignment performed during the AIV of the prototype. Setting the last value to ±50μm, which looks more realistic accordingly to the considerations done in the previous section, 2 other parameters belonging to the worst offender list are the lens thickness and the error in the Radius of Curvature (RoC) measurement.

We decided then to re-evaluate the prototype estimated performance by considering an error on the RoC measurement which is doubled with respect to what specified by SESO (+/-0.02%) and an error on the thickness measurement accuracy of ±30μm, which are of course adding to the ±50μm focusing accuracy, and the results are shown in Table 44. Again the system is optimized to match the measurements in warm condition and then transformed into the cold to check that also the cold measurements are in agreement with the model.

*Table 42. The warm and cold simulated performance in terms both of 90% EE square semi-side and best focal plane position, assuming a focusing accuracy of ±50μm, a thickness error of ±30μm and an RoC accuracy measurement of +/-0.04%.*

| Field [°] | HARTMANN 90%EE SQUARE SEMI-SIDE (μm) | | Best focal plane position [mm] | |
|---|---|---|---|---|
| | Warm | Cold | Warm | Cold |
| 0 | 18.8 | 9.9 | 0.000 | 0.000 |
| 5.6 | 17.9 | 12.2 | -0.033 | -0.007 |
| 11.2 | 19.5 | 22.9 | -0.087 | 0.011 |
| 16.7 | 34.5 | 42.0 | -0.119 | 0.144 |

The results are shown in Table 44 are graphically plot in Figure 185, Figure 186, Figure 187, and Figure 188.

The results plot, in this case, looks in reasonable agreement when comparing the simulated and measured results (both in warm and cold conditions and both in term of EE and best focal plane position), and pretty similar to the case analyzed in Sec. 2.8.7.5, with the wicked, largely overestimated focus accuracy of ±100μm.

These results tend to suggest that what observed during the warm/cold test prototype campaign has a reasonable explanation, coming from the combination of errors that may have happened both during the lens characterization and during the alignment phase.

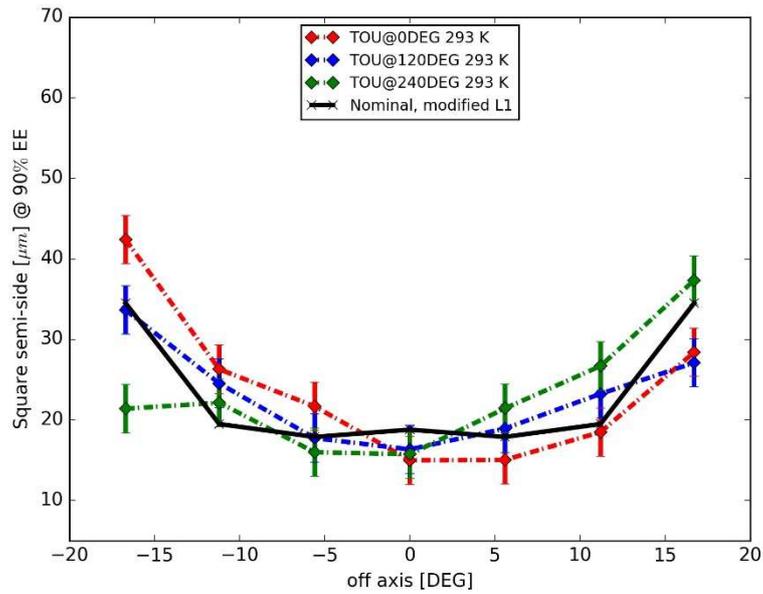

*Figure 185. The **warm** simulated performance (continuous black line) compared with the warm measured performance (colored lines in the usual prototype 3 different orientations and off-axis positions), in terms of **90% EE** of square semi-side, assuming a **focusing accuracy of ±50μm, a thickness error of ±30μm and a RoC accuracy measurement of +/-0.04%**.*

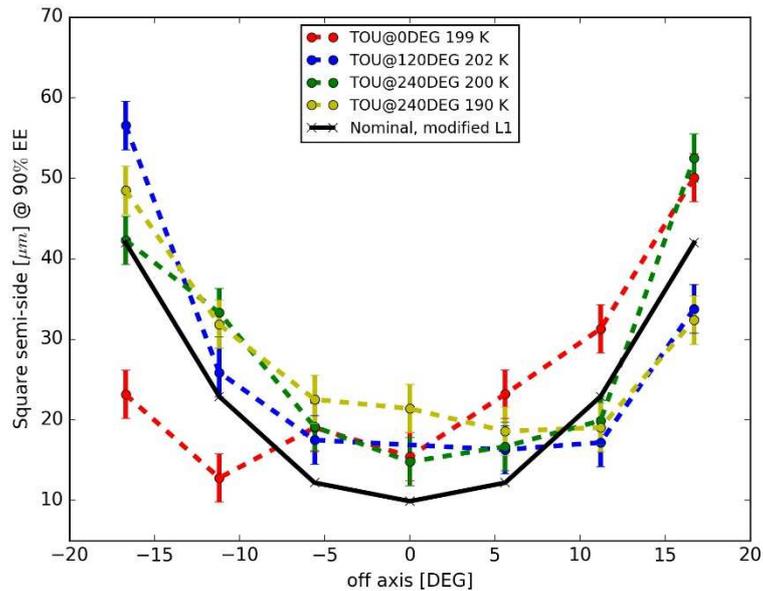

*Figure 186. The **cold** simulated performance (continuous black line) compared with the cold measured performance (colored lines in the usual prototype 3 different orientations and off-axis positions), in term of **90% EE** of square semi-side, assuming a **focusing accuracy of ±50μm, a thickness error of ±30μm and a RoC accuracy measurement of +/-0.04%**.*

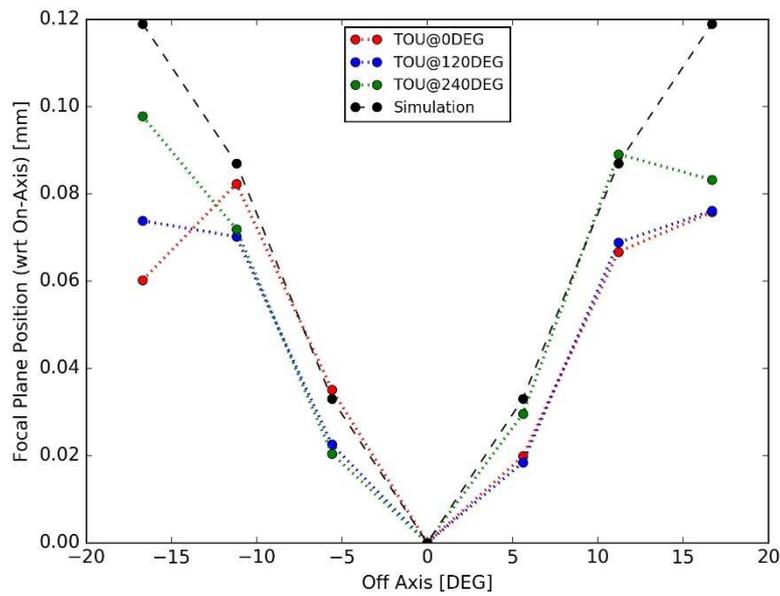

*Figure 187. The **warm** simulated performance (continuous black line) compared with the warm measured performance (colored lines in the usual prototype 3 different orientations and off-axis positions), in terms of **best FP position vs FoV**, assuming a **focusing accuracy of ±50μm, a thickness error of ±30μm** and an RoC accuracy measurement of +/-0.04%.*

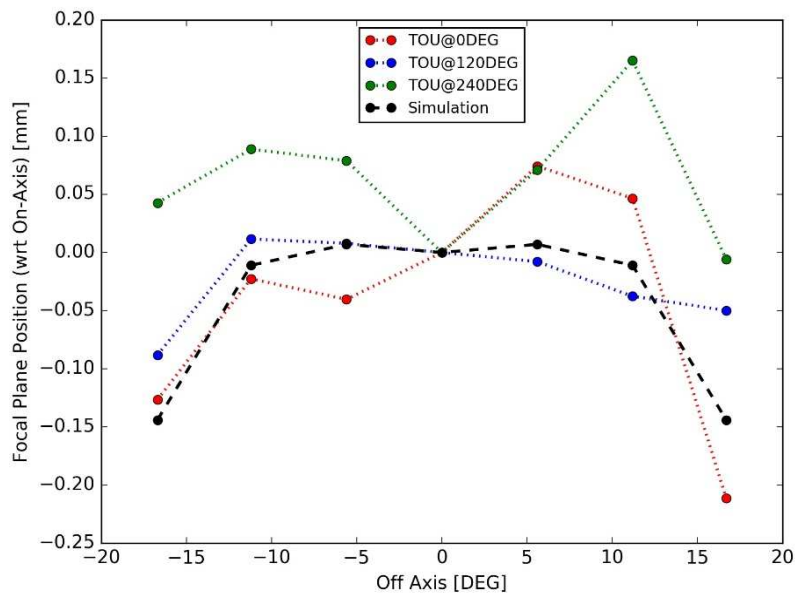

*Figure 188. The **cold** simulated performance (continuous black line) compared with the cold measured performance (colored lines in the usual prototype 3 different orientations and off-axis positions), in terms of **best FP position vs FoV**, assuming a **focusing accuracy of ±50μm, a thickness error of ±30μm** and an RoC accuracy measurement of +/-0.04%.*

We emphasize that the primary error source is the low accuracy in aligning the lenses in defocus, an operation that we have been obliged to perform with our CMM machine, which has performances far from what can be obtained with similar commercial devices, such as the Zeiss CMM machine, available at Leonardo.

## 2.9  Conclusions

The ESA mission PLATO (Planetary Transits and Oscillations of stars), which will detect many exoplanets down to Earth-size in the habitable zone, is composed of 26 Telescope Optical Units. In my Ph.D. I have been involved in the AIV phase of the Prototype of the TOU, whose primary goal was to demonstrate and validate a procedure of alignment to achieve optimal performance within the project tolerances. I performed the optical alignment by defining the strategy of alignment, the observables, the methods of measurement, and their accuracy.

I developed a Python library focused on image analysis to evaluate, with high accuracy, centroid position, spot intensity profile and the reference position of the alignment beam, with the possibility of real-time analysis with the acquisition. In this manner, many procedures are also now automatized and optimized to reduce errors in the computation of parameters fundamental for the TOU alignment and performance evaluation. I dedicated particular attention to the thermal stability of the AIV laboratory, controlling the temperature down to 1 °C, due to the tight tolerances required for the alignment, the presence of several micrometers and materials with different CTEs used in the setup. I interacted with the optical group to correlate instrumental measurements with simulation results, redefining techniques and accuracies.

After the alignment, I tested in "flight condition" (-80C and vacuum) the optical quality of the TOU, by using Hartmann, Ensquared Energy, and interferometric techniques. The cryogenic temperature tests were conducted in a cleanroom at Leonardo premises, in an international environment with a high technological level.

After the integration and verification, long and detailed analysis has been run to evaluate the error budget and solve results discrepancies, which showed that measured and simulated performances of the TOU can be matched if considering relaxed tolerances on some of the worst offender parameters. The Prototype thus proved to be very useful in giving indications on important parameters to be kept under control in the Manufacturing Assembly Integration and Verification (MAIV) phase of PLATO. The industry will define a manufacturing serial process for the 26 TOU composing the optical part of PLATO, taking a considerable advantage from the many lessons learned during the AIV Prototyping phase.

The mechanical structure of the GSE was investigated to solve stability and repeatability issues, with a gain in terms of overall accuracy, total AIV process time and final design solution of the GSE adopted for industrialization. The choice of non-contact techniques for the lenses focus alignment, such as low coherence interference, are the results of the investigation performed in this Ph.D. work. Also, the interferometric test, the most accurate today available for optical quality determination, was unsuitable due to the manufactured uncertainties in the radius of curvature and refraction index of the lenses.

In terms of qualifying the optical performances of all TOUs, the Hartmann test has proved to provide a fast and precise assessment of EE in the whole FoV of the TOU, more efficient than the analysis of the PSF, useful instead for the focal plane determination. The use of a precise x-y-z translation stage to span the entire TOU FoV with a detector is desirable also for easy robotization, but it is tricky in cold conditions for the emerged issues during the tests.

The experience acquired during tests in the cryo vacuum chamber raised a warning on the thermal stabilization of the TOU. Inside the cryo vacuum chamber, the external optical window radiated heat toward the TOU, which showed a thermal gradient along the optical axis, a problem that we were not able to eliminate during the tests, but we envisaged possible solutions to overcome it.

# 3  SHARK-NIR

The discovery of an exoplanet is going together with a detailed study with a ground-based telescope. Large mirror and eXtreme Adaptive Optic (XAO) are necessary to collect many photons and reach a high Strehl ratio, permitting direct imaging of evanescence luminosity of exoplanets through sophisticated optical techniques as the coronagraphy. SHARK (System for coronagraphy with High order Adaptive optics from R to K band) is an instrument proposed for the Large Binocular Telescope (LBT) in the framework of the "2014 Call for Proposals for Instrument Upgrades and New Instruments", now in the integration phase and will be installed in 2020 at the telescope, a binocular telescope with 8.4m of diameter each. It is composed by two channels, a visible and a near-infrared arm, to be installed one for each LBT telescope, and it will exploit unique, challenging science from exoplanet to extragalactic topics with simultaneous spectral coverage from B to H band, taking advantage of the outstanding performances of the binocular XAO LBT capability. The instrument is composed of two channels, a visible one (SHARK-VIS) operating from 0.5 μm to 1.0 μm, and a near-infrared one (SHARK-NIR) operating between 0.96 μm to 1.7 μm. This document refers only to the SHARK-NIR channel.

In this thesis I am involved in the AIV phase of the infrared channel, studying the procedure for the best alignment of the optical bench, testing components accuracy and making software coding to help the integration phase in the cleanroom.

## 3.1  The coronagraphy

In exoplanet research, direct imaging of an exoplanet is a crucial point to understand many astrophysical processes of these systems and represents a considerable technical challenge because of the  very small angular separation and high star-to-planet brightness ratio. In this observing situation, the contrast from a star and the nearby planet is a parameter that easily overcomes the dynamic range of the best image sensor. The coronagraphy is necessary to achieve dynamic range, a technique born with Lyot for observation of the solar corona in 1930, employing mask in the focal plane of the telescope to dither the luminosity of the star. The draw in Figure 189 shows the scheme of the first Lyot coronagraph, used to reach the $10^6$ at 2 arcmins from the surface of the Sun. Impurity-free glass, highly polished lenses, and a high altitude site to minimize atmospheric pollution were all essential (Lyot, 1939).

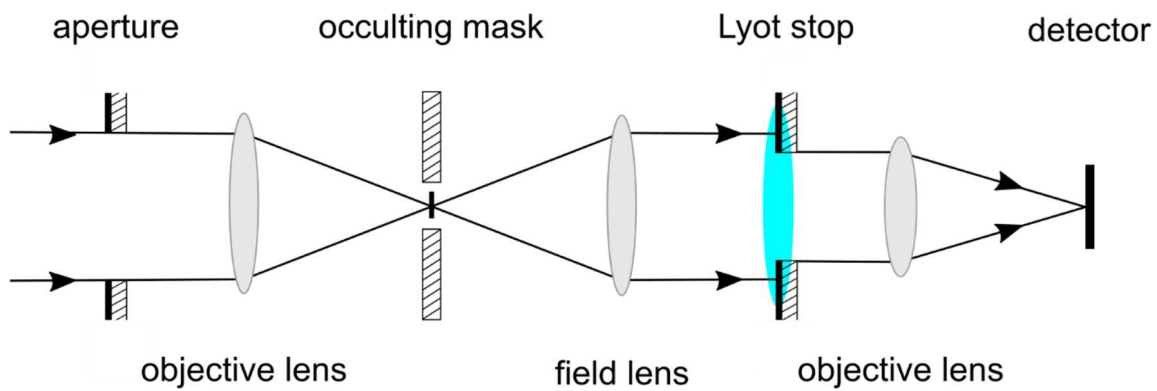

*Figure 189. The scheme of the Lyot coronagraph, the lens in cyan color is the image of the first objective lens by field lens in the Lyot stop.*

In the scheme of the Lyot coronagraph, Figure 189, the first objective lens forms an image of the disk and corona in the focal plane. The occulting mask is not enough to eliminate the diffracted light, part of this light reaches the focal plane, for this reason, a Lyot stop (LS) is mandatory. By using the field lens, the image of the objective lens and its diffraction pattern are created in the LS stop position, cyan color in Figure 189. The LS intercepts the diffraction ring and allows most of the light from the surrounding structure to pass. The second objective lens relays the filtered image onto the detector. In a perfect Lyot coronagraph, about 50% of the light from a nearby source might be lost, compared to the suppression of about 99% of the stellar light.

Advanced stellar coronagraphy, see Figure 190, uses a similar concept of Lyot, but with some innovations. The star incoming the coronagraph is focused by an AO system, in some cases with high Strehl ratio, the PSF is near to the diffraction limited resolution. Several Airy rings are superimposed on a widely scattered light halo, containing several percents of the total flux, and we have access to a region within few $\lambda/D$. The apodizer mask, which is located in the entrance pupil plane, rejected the light of the star from the area of interest in the instrumental focal plane, improving the contrast. In the focal plane, there is the Fourier Transform of the pupil plane, the peculiar shape of the apodizer mask changes the profile of the PSF. These masks typically change the amplitude of the original wavefront, as in the original Lyot coronagraph, but some masks change the phase, as the Four-Quadrant Phase Mask (FQPM). In the intermediate focal plane, the contrast could be improved by applying an occulter mask reducing the residual light of the star.

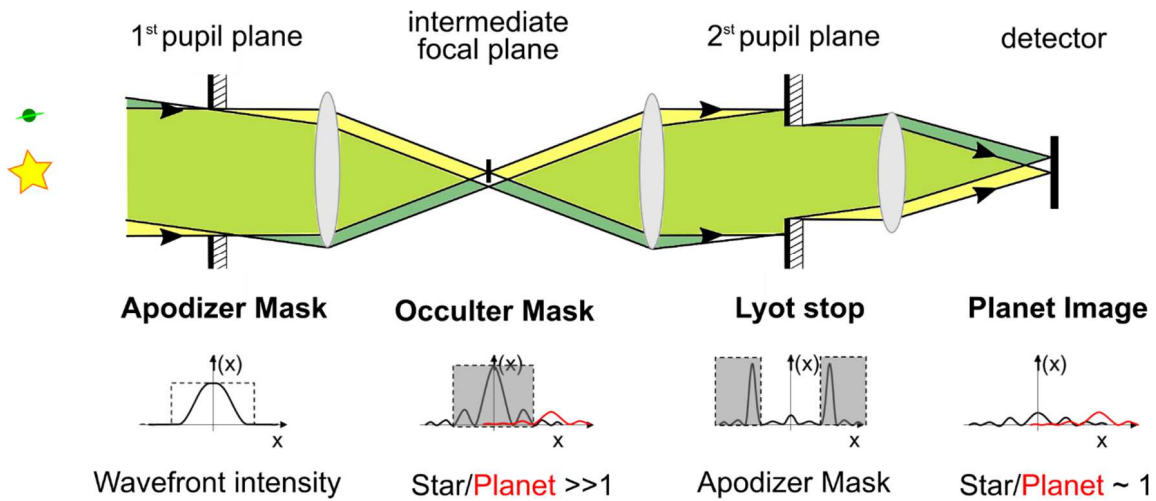

*Figure 190. New coronagraph uses an Apodizer mask to vary the intensity of the wavefront and an occulter mask combined with a LS to decrease the starlight below the exoplanet intensity level.*

The focal plane region of interest is characterized by a discovery space defined by a minimum and a maximum angle from the star, similar to a circular crown around the occulted PSF of the star. The smaller radius, the Inner Working Angle (IWA), is either defined as the minimum angular separation at which the total energy throughput reaches 50% of its maximum value or the raw contrast reaches a threshold value. The largest radius, the Outer Working Angle (OWA), is either defined as the maximum angular separation where throughput is greater than 50% of its maximum or the raw contrast increases above a threshold.

A second imaging system creates the image of the pupil in a secondary plane, where is located the LS. The typical diameter of the LS is 80% of the first pupil, the residual light is removed and the resulting image on the scientific camera shows the region nearby the star at high contrast. From an optical design perspective, modern coronagraphs are able to reach up to $10^{10}$ in contrast within 5 $\lambda$/D. They are classified into four main groups:

- **interferometric coronagraphs**: create multiple beams from a single telescope beam, producing a destructive interference on-axis and constructive interference off-axis;

- **pupil apodization coronagraphs**: their designs are characterized by a modification of the pupil complex amplitude, yielding a PSF suitable for high-contrast imaging. The apodization can be performed by a pupil plane amplitude mask which can be continuous or binary, or by a phase mask;

- **improved Lyot amplitude masks**. The mask adapts its amplitude by using band-limited transmission profiles in the focal plane;

- **improved Lyot phase masks**: a phase mask can be used to introduce phase shifts in the focal plane to create self-destructive interference, the FQPM as an example, the related achromatic phase knife coronagraph, and the optical vortex coronagraph, comprising both scalar and vector variants.

For a complete description, see Guyon et al., 2006.

SHARK-NIR will provide for 3 pupil apodizer masks, a FQPM, and a Gaussian-Lyot coronagraph.

## 3.2  Science with SHARK

An essential property of the LBT AO system is the capability of realizing a high Strehl Ratio (SR) at moderately faint magnitude (R>10). The AO upgrade SOUL (Pinna et al., 2016) will provide an additional 1-2 magnitude gain to obtain the same level of correction as given by FLAO (see Figure 191), opening the field of high-contrast AO imaging to targets much fainter than feasible using SPHERE and GPI. This feature will allow deep search for planets around targets like, e.g., M dwarfs in nearby young associations and solar-type stars in nearby star-forming regions (e.g., Taurus-Auriga, located at d=140 pc).

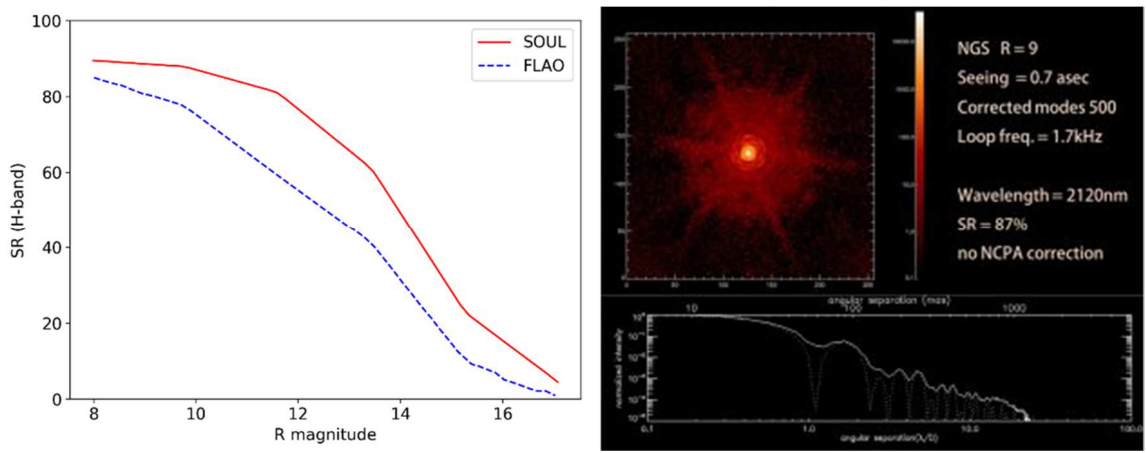

*Figure 191. On the left: Strehl ratio values obtained at 1650nm via E2E simulation for the case of SOUL and FLAO. On the right: First NIR PSF obtained on LUCI with the SOUL AO system (Pinna et al., 2016).*

The SHARK@LBT binocular configuration gives the possibility to face direct imaging and study these main scientific topics.

**Exoplanets**: modern astrophysics pushes new technologies towards the direct detection of exoplanets, providing access to information on giant gaseous planets outside the Solar System, similar to ours Jupiter, Saturn, Uranus, and Neptune. Due to the very demanding resolution and contrast required on the science images, direct detection of extrasolar planets is a very challenging task to be achieved. Since massive exoplanets cool and fade with time, direct-imaging surveys have been most successful around young stars. In order to build-up the spectral energy distribution of the sub-stellar object, the coronagraphy helps to perform spectroscopy and photometry in differents band, or precise astrometry over several years, to better characterize the physical properties, for example, exoplanets masses, density, surface temperature, and atmospheres. Spectroscopic mode with a modest spectral resolution is currently foreseen in SHARK-NIR through a long slit positioned into the intermediate focal plane.

**Proto-planetary disks:** to understand how matter aggregates to form the building blocks of planetary bodies, there is the need to investigate not only the evolution of the disk itself but also the role of jets in shaping its structure. The study of proto-planetary disks is fundamental to comprehend the formation of our solar system as well as of extra-solar planetary systems. This requires observing the system at a high angular resolution as close as possible to the parent star, occulting its light to enhance the area where the interplay between the accretion and ejection of matter dominates the dynamics. For example, applying the power of SHARK-NIR to well-studied Taurus-Auriga star-forming region, we can study the interaction between planets and disks, explaining the formation mechanism of exoplanet systems, now poorly accessible with Spectro-Polarimetric High-contrast Exoplanet REsearch instrument (SPHERE) and Gemini Planet Imager (GPI). Considering the distance of the Taurus-Auriga region (Martinez and Kraus, 2019), 140 pc, this science case motivates the instrument design with an IWA as small as 100 mas. The characteristics of the LBT AO system allow reaching 1 or 2 magnitudes fainter targets with respect to SPHERE and GPI, permitting to observe many more nearby small mass stars and solar-type stars in star-forming regions at close distance. The requirements for the IWA are a minimum of $4 \lambda/D$ with the goal for $2\text{-}3 \lambda/D$, for the contrast they are $10^{-5}$ -$10^{-6}$ in the range 300-500 mas and $10^{-5}$ -$5\text{x}10^{-6}$ for IWA below 300 mas.

**Brown dwarfs in young open clusters:** brown dwarfs (Béjar and Martín, 2018) are objects with intermediate-mass between stars and planets, adding a new spectral type object L and T, the L type with very low effective temperature (500K) characterized by spectrum with atomic and molecular absorptions, the T type with spectral shapes explained by dust grains suspended in the photosphere. These coldest brown dwarfs contain a historical record of the star formation process at very low masses and at epochs, many billions of years before the stellar birthplaces we see today. Using SHARK to this object will help to define the actual hypotheses on the formation of gas-giant planets, the core accretion, and disk instability. These formation models, therefore, predict two populations of giant planets segregated by orbital distance, with the closer planets formed by core accretion (3 to 10 AU) and the outer ones by disk instability (10 to 100 AU). Pleiades is the richest of these clusters with its about 800 known members, with direct imaging we can detect self-luminous companions placed on wide orbits (> 1 AU), to characterize their atmosphere and constrain their orbital parameters. The coronagraphic requirements for IWA are a maximum of 4 $\lambda$/D with the goal for 2-3 $\lambda$/D, for the contrast they are of $10^{-5}$ for angular separation below 300 mas.

**Giant planets around low-mass stars**: the member populations of moving groups remain largely incomplete and searches to find the missing members are ongoing. The low-mass populations of moving groups are crucial for studies of recent local star formation, disks and exoplanets, and the astrophysics of young and low-mass stars. Direct imaging survey for giant planets around young moving groups a relatively new research field doable with a large telescope equipped with AO (Biller et al., 2013). These observations can infer key constraints on multiple systems and planet formation in the low-mass regime. For SHARK there are many potential targets for a search for planets in wide orbits. The coronagraphic requirements for IWA are a maximum of 4 $\lambda$/D with the goal for 2-3 $\lambda$/D, for the contrast they are of $10^{-5}$ -$10^{-6}$ in the range 300-500 mas and $10^{-5}$ - $5\times10^{-6}$ for angular separation below 300 mas.

**Stellar jets and disks**: Coronagraphic imaging of stellar jets play a crucial role in the comprehension of the environment of young stars, that are often associated with accretion disks and stellar jets. The stellar jet physic was a research field only for space telescopes, but with XAO and large telescopes, we can achieve more significant results (Cox et al., 2012). AO observations of the base of jets disclose a clear picture of the jet very close to the launching point and of the instabilities and shocks in the flow. A coronagraph with an IWA of 150 - 200 mas and modest contrast in J and H bands ($10^{-2}$ - $10^{-3}$) would constitute a big leap in the observational modes available today: it would allow sampling the jet as close as 10 AU from the source in nearby systems like Lupus and Taurus. Images of the stellar disk are now standard in the millimeter range, supported by large array of radio telescopes. An optical and near-infrared instrument, equipped with a coronagraph, could achieve an angular resolution of 20-40 mas with contrast of $10^{-4}$ - $10^{-5}$, probing the small dust grains in the disk through scattered light imaging down to a few AU from the star, figure out in the visible range the structure and dynamics of planet-forming regions (Wagner et al., 2015). The coronagraphic requirements are 2-3 $\lambda$/D (0.15"-0.20") with FoV below 1"x1" and for the contrast are between $10^{-4}$ and $10^{-5}$ for disks, $10^{-3}$ for jets at separations of 200-300 mas.

**Active Galactic Nuclei (AGN)**: black hole-powered 'central engines' producing bright emission at various wavelengths, and jets extending far beyond the galaxy are common to many galaxies but show different properties when observed. Those differences led to a variety of names, such as quasars, blazars, or Seyfert galaxies. Astrophysicists construct a unified model to explain the differences, including a central black hole, a rotating disk of infalling material surrounding the black hole, and the jets speeding outward from the poles of the disk, calling this standard set of features an active galactic nucleus (AGN) (Antonucci, 1993). Direct imaging with optical telescopes opens the doors to new science, until now only doable with radio astronomy. High-resolution images are required in order to be able to disentangle the faint extended light of the host galaxy from the halo of the bright nucleus. The capabilities of SHARK-NIR in terms of spatial resolution and contrast enhancement may be applied to study the AGN-host relations as well as Dumped Ly-α systems (DLAs), to constrain the Black Hole feeding mechanism and to trace, in bright quasars, molecular outflows powerful enough to clean the inner kilo-parsec and quench the star formation. SHARK will offer the unique opportunity for a breakthrough in these fields, thanks to:

i) improved SR in the NIR at faint magnitudes with respect to present NIR instrumentation,

ii) AO correction in the optical band which will allow achieving resolutions a factor three better than HST,

iii) integral field spectroscopy in the optical band,

iv) the fundamental synergy with the coronagraphic modality of LMIRCAM in K-band.

To reach the scientific goals, both SHARK visible and near-infrared channels are required. The coronagraphic requirements are 2 to 8 $\lambda/D$ with FoV of 5"x5" and 15"x15", ideal respectively for DLAs and AGN inner morphology. For contrast, the requirements are $10^{-5}$ - $5x10^{-6}$ at 2000 mas and z=0, $10^{-4}$ at 700 mas and z=0.1 and z=1.1.

## 3.3 Scientific Requirements

We reported in Table 44 a summary of the top-level scientific requirements, which have been used as drivers for the system design. The design specification for IWA and contrast, in the last column of the table, are resulting from simulations (E. Carolo et al., 2018).

*Table 44. Summary of the top-level scientific requirements, compared with the design specifications.*

| Parameter | Requirement | Design Specification |
|---|---|---|
| IWA | 4 $\lambda$/D (goal: 2 $\lambda$/D) | as low as 2.6 $\lambda$/D |
| Contrast | $10^{-5}$ - $10^{-6}$ (goal: $10^{-7}$) | 2 x $10^{-6}$ with seeing 0.4" at 500 mas radial distance |
| Spectral resolution | Low: R ~ 100 Medium: R ~ 700 | Low: R = 100 @ $\lambda$ = 1.33 $\mu$m Medium: R = 100 @ $\lambda$ = 1.33 $\mu$m |
| LSS minimum SNR per pixel | 5 | >10 |
| FoV | 15" (goal: 18") | 18"x18" |
| PSF position stability on the SCICAM | 0.25$\lambda$/D in field stabilized mode (goal) 3 mas in pupil stabilized mode | |

| Target | Mode | Requirements |
|---|---|---|
| Exoplanets | CI, DBI | IWA: maximum 300 mas, goal: 100 – 200 mas Contrast: $10^{-5}$– $5\times10^{-6}$ for separations < 300 mas; $10^{-5}$–$10^{-6}$ in the range 300 − 500 mas; $R_{mag}$: 5 – 10 FoV: < 1 × 1 arcsec |
| | LSS | Resolving power: R ~ 100 and R ~ 700 SNR: > 5 |
| Disk | CI | IWA: 150 – 200 mas Contrast: $10^{-4}$– $10^{-5}$ at separations of 200 – 300 mas $R_{mag}$: 6 – 12 FoV: < 1 × 1 arcsec |
| Stellar jets | LSS | IWA: 150 – 200 mas Contrast: $10^{-3}$ at separations of 200 – 300 mas $R_{mag}$: 6 – 12 FoV: < 1 × 1 arcsec Resolving Power: R ~ 100 |
| AGN | DI, CI | IWA: 200 – 600 mas Contrast: $10^{-5}$– $5\times10^{-6}$ at 2000 mas for z = 0; $10^{-4}$ at 700 mas for z = 0.1 ÷ 1.1; $R_{mag}$: 9 – 14 FoV: 5 × 5 arcsec and 15 × 15 arcsec for, respectively, DLAs and AGN |

## 3.4  SHARK-NIR description

The SHARK-NIR basic idea is a camera for direct imaging, coronagraphy, and spectroscopy, using the corrected wavefront provided by the LBT Adaptive Secondary Mirror (ASM), operated through one the existing AO WaveFront Sensing (WFS). SHARK-NIR will be installed at the entrance of LBTI (see Fig. 3), very close to the WFS that is dedicated to LBTI itself, which will be used to sense and drive the ASM, providing the corrected wavefront to SHARK-NIR. A dichroic, deployable in front of the entrance window of LBTI, shall pick-up and re-direct the wavelength range between 0.96 and 1.7 microns toward SHARK-NIR, letting the VIS light to go to the WFS. The graphics in Figure 192 show the path of the light incoming from the telescope to SHARK position:

1. the light from the astronomical source is collected by the two mirrors, 8.40m of diameter each, one for blue and the other for the red part of the electromagnetic spectrum;

2. the AO in prime focus correct at Earth's atmosphere distortions of the light, moving with 672 magnets on the back of the secondary mirrors of 910mm that change their shapes 1500 times per second;

3. the tertiary mirror deflects the corrected light towards the center of the telescope where various instruments, which combine the beams, are installed;

4. the dynamic balancing system compensates for movements of the telescope and helps it remain fixed on one spot in space;

5. the light enters in the LBT Interferometer (LBTI) canceling the light from a bright star and allowing astronomers to look for faint, orbiting planets. SHARK-NIR will installed before LBTI;

6. the position of the interferometric camera is used to combine the light in phase, allowing high-resolution images. The telescope has the equivalent sharpness of a 22.8m instrument.

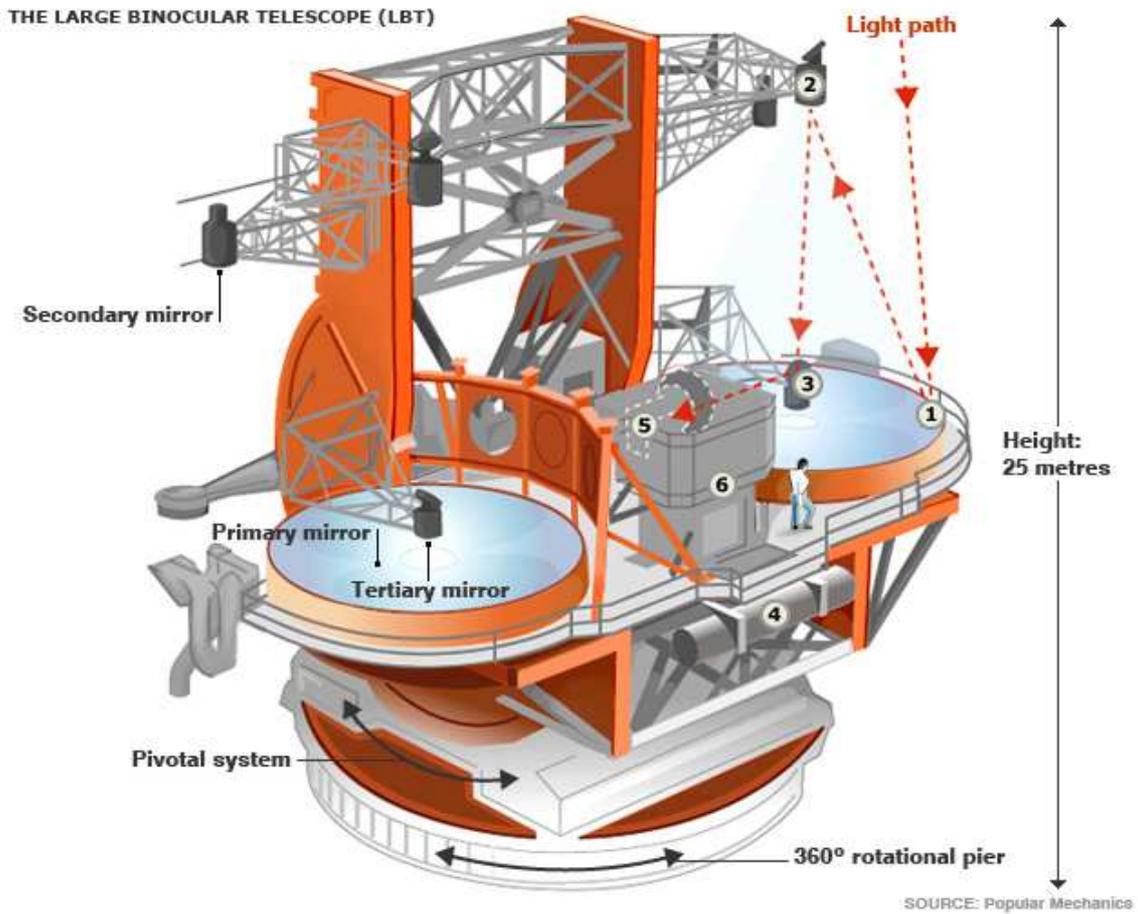

*Figure 192. A graphics of LBT with the path of the incoming light from an astronomical source to the scientific instrument (Source POPULAR MECHANIC).*

The main characteristics of SHARK-NIR are resumed in Table 45.

*Table 45. The main characteristics of SHARK-NIR*

| PARAMETER | SPECIFICATION |
|---|---|
| Observing modes | Imaging/Coronagraphy/Spectroscopy/DBI |
| Waveband | 0.96 – 1.7 µm |
| Field of view | 18 x 18 arcsec |
| Diagonal Field of View | 25.5 arcsec |
| Airy Radius @ λ = 0.95 µm | 29 mas / 2 px |
| Plate scale | 14.5 mas/px |
| Working F/# | 31.4 |
| Detector format | 2048 x 2048 px |
| Pixel pitch | 18 µm (square) |
| # of mirrors in the camera | 8 ( 3 flat, 1 DM, 4 OA parabolas) |
| Nominal Strehl within 18 arcsec FoV diameter in all bands | > 98% |

### 3.4.1 Instrument modes

Being SHARK-NIR a coronagraphic instrument, the camera has to be designed to accomplish an extreme performance, ideally not to decrease the correction provided by the AO system. All the coronagraphic techniques that may be implemented need a Strehl Ratio as high as possible to provide excellent contrast. High contrast needs optical elements of high accuracy manufacturing, moreover these components must be installed and aligned with high precision. The whole instrument mechanics has to be very stiff and designed to minimize the effect of the flexures. To maintain the performance at every observing altitude, it is necessary to implement an Atmospheric Dispersion Corrector (ADC) to compensate for the atmospheric dispersion.

Some of the foreseen science cases need to perform the field derotation, to realize which the whole instrument has to be mounted on a mechanical bearing.

A NIR camera, based on a Teledyne H2RG detector, cooled at about 80°K to minimize the thermal background, will provide an FoV of the order of 18"x18", with a plate scale foreseeing two pixels on the diffraction limit PSF at 1μm.

A few subsystems have been introduced in the instrument design with the purpose of optimizing the instrument performance. A Deformable Mirrors (DM) used as a fast tip-tilt (TT) mirror, will be used to correct both the NCPA and the undesired PSF movements during a scientific exposure. The last correction requires a dedicated fast TT sensor, which has been placed after the first pupil plane, into the collimated beam by a beam splitter will pick-up 5% of the light to be sent to the sensor.

Between the TT mirror and the beam splitter feeding the TT sensor, a filter wheel positioned at 50mm from the pupil plane carries the apodizing masks. These kinds of masks are typically placed precisely into the pupil plane, which is occupied by the DM in our design. The impact of the slight displacement of the masks with respect to the pupil plane was taken into account during the design phase of the masks. SHARK-NIR has mainly three observing modes, described in the following sections.

#### 3.4.1.1 Direct Imaging mode

In observing mode, SHARK will provide an unobstructed FoV of 18"x18", with a correction which is nominally nearly over the full 18" diameter with SR>93% at 1μm, and perfect on-axis achieving R=99.7%, see Figure 193. Considering very relaxed tolerances for the alignment of the optical elements such ±200μm for the off-axis parabola decenter, ±3' for their tilt, the final performance achieves SR>96% at 1μm over almost the full 18" diameter.

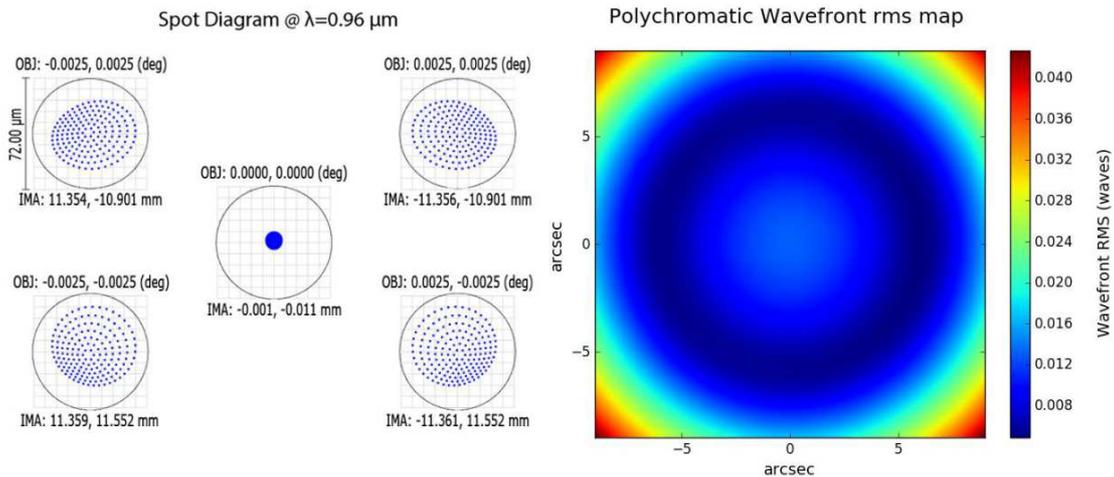

*Figure 193. On the left: spot diagram at the scientific focal plane (FP-SCI) for five field positions at λ=0.96 μm. The circle represents the Airy disc at the same wavelength, the box represents a 4x4 pixel area. On the right: polychromatic wavefront rms map of the full field for the nominal design.*

By using the ADC, the optical performance remains excellent, ensuring, for example, an on-axis SR >98% for every zenith angle. The total throughput of the instrument is about 50%.

The field rotation can be compensated through the mechanical bearing, which will accomplish a tracking precision of about 7", the requirement is 20", and perform a maximum rotation of 185°.

The scientific filters available for the direct imaging mode are 10, distributed in the two filter wheels positioned in the collimated beam after the 2nd pupil plane.

This mode assumes no coronagraphs nor slits are inserted along the optical path. Ten scientific filters, distributed in the two filter wheels positioned in the collimated beam after the 2nd pupil plane, are available to the observer in Classical Imaging (CI) configuration, while a limited number of filters couples can be used in combination with a Wollaston prism, to achieve Dual Band Imaging (DBI) with an image separation of 14". The CI and BDI FoV are 18x18 arcsec (square) and 1" (radial FoV), respectively.

### 3.4.1.2  Coronagraphic mode

The current design of SHARK-NIR foresees two intermediate pupil and focal planes, to implement a large variety of coronagraphic techniques. Three techniques will be implemented:

6. **Gaussian Lyot**, which requires a gaussian stop into the 1st focal plane and a pupil stop on the 2nd pupil plane;

7. **Shaped Pupil (SP),** which requires an apodizing mask into the 1st pupil plane and an occulting mask into the 1st focal plane;

8. **FQPM**, which requires a four-quadrant mask into the 1st focal plane and a pupil stop into the 2nd pupil plane.

All these techniques dim the light of the central star, in order to improve the contrast in the vicinity of the star itself, allowing to detect much fainter companions (exo-planets case for example) or to explore the morphology of the investigated object (Jets/Disks case and AGN/QSO case). These faint structures are characterized by different operating distances from the central star (IWA), and by different contrast that can be reached at a certain distance, by different throughput, and by different FoVs.

The design also includes a simple dual-band imaging channel, which can be selected by inserting a Wollaston prism into the second pupil plane. The prism divides the input beam into two output beams inversely polarized. Different combinations of filters can be applied to the two beams, which are focused and recorded on the H2RG detector displaced by about 14". Each of the two-beam is characterized by an FoV of about 2".

The coronagraphic mode will provide the possibility to observe both in the field and pupil stabilized mode. Since SHARK is located at the Gregorian focus of LBT, there is no optical rotation of the pupil during the observation. The field stabilized observations thus need a mechanical derotation, so SHARK-NIR is supplied with a derotator, adding considerable complexity to the instrument. In this mode, only the Gaussian Lyot coronagraph and the FQPM can be used because SP apodizers are designed taking into account the spider shadows of the secondary mirror. On the other side, in pupil stabilized mode, the whole set of coronagraphs could be exploited.

The coronagraphic FoV will depend on the coronagraphic techniques selected and its range will be from about $8\lambda/D$ for the SP, to the whole FoV (18"x18") in the Gaussian Lyot case. The IWA is also depending on the technique and from 4 to 5 $\lambda/D$.

The total throughput of the instrument depends on the coronagraphic technique, and it is about 25% with the Gaussian Lyot technique, and ~7% with the SP coronagraph.

### 3.4.1.3 Spectroscopic mode

Long-slit spectroscopic (LSS) coronagraphic mode will be implemented in SHARK, with two resolutions:

- a low-resolution mode (R~ 100, through a prism), in order to target faint targets;

- a medium-resolution mode (R~700, through a VPH grating) to get spectral information for the brightest objects.

In the focal plane wheel, two positions will be dedicated to two coronagraphic long slits, which are characterized by different IWA. The total throughput with the prism is of the order of 25%, while is about 15% with the grism.

## 3.4.2 Optomechanical description

### 3.4.2.1 Direct imaging and coronagraphic mode

The optical design of SHARK-NIR is characterized by two intermediate pupil planes and an intermediate focal plane and will allow the implementation of advanced coronagraphic techniques, such as the SP coronagraph. The nominal optical quality is characterized by a very low wavefront error, and an error budget has been allocated to keep NCPA of the science channel within few tens of nanometers RMS, allowing to reach contrast better than $10^{-5}$ at 500 mas separation, when observing bright target in good seeing conditions.

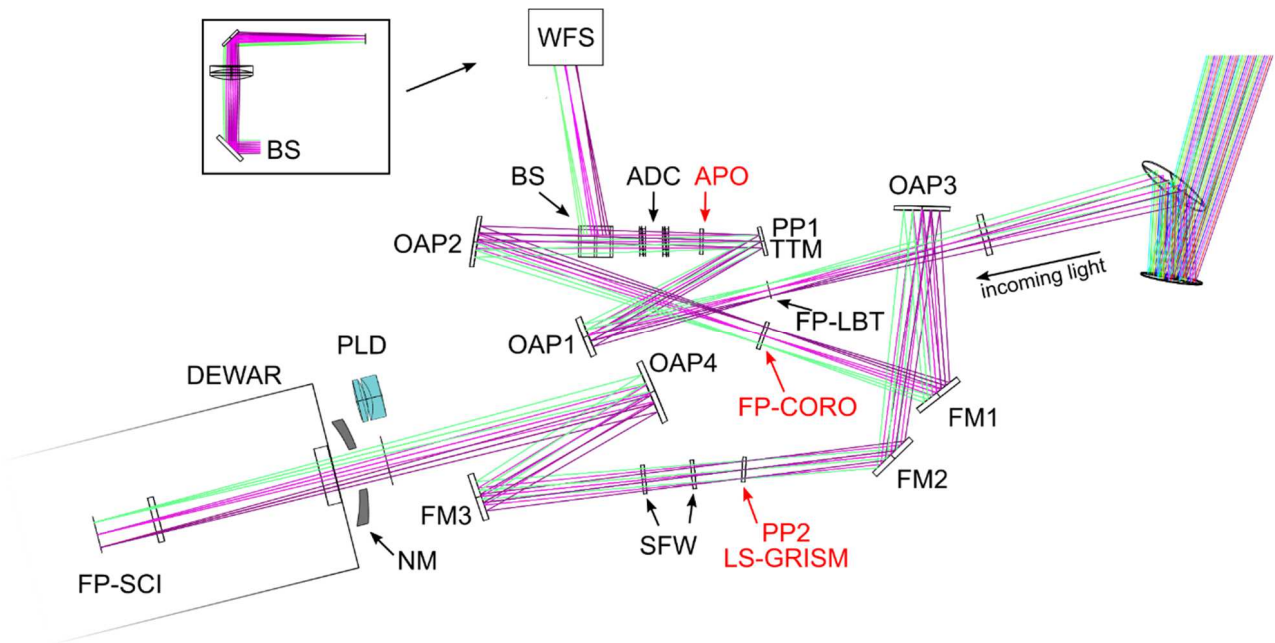

*Figure 194. The optical layout of SHARK-NIR. In red the coronagraphic planes, for component description see Table 46.*

*Table 46. The name definitions of the fundamental planes in SHARK-NIR.*

| Principal Optical Components | Acronym | Pupil diameter | Plate scale | Pixel scale |
|---|---|---|---|---|
| Incoming F/15 Focal Plane | FP-LBT | - | 1.67 mas/μm | - |
| First Pupil Plane (TT mirror) | PP1 - TTM | 11.15 mm | - | - |
| Folded Focal Plane (local WFS) | WFS | - | 1.4 mas/μm | 21 mas/px |
| Intermediate Focal Plane (occulters, slits) | FP-CORO | - | 1.14 mas/μm | - |
| Second Pupil Plane (Lyot masks, LSS dispersing elements, Wollaston prism) | PP2 | 14.92 mm | - | - |
| Final Focal Plane (scientific) | FP-SCI | - | 0.8 mas/μm | 14.4 mas/px |
| Lyot Stop and Grism | LS-GRISM | | | |
| Science Filter Wheels | SFW | | | |
| Apodizer Wheel | APO | | | |
| Atmospheric Dispersion Corrector | ADC | | | |
| Off-axis Parabola | OAP1...4 | | | |
| Pupil Lens Deployable | PLD | | | |
| Narcissus Mirror | NM | | | |

The optical design is shown in Figure 194, and the main components are the following:

- an off-axis parabola (OAP1) is creating a pupil plane, called PP1, of about 11.15mm of diameter onto the deformable mirror TTM. The DM to be used has the ALPAO 97-15, characterized by a maximum pupil size of 13.5mm in diameter, by 97 actuators (usually the NCPA correction is limited to the first 15-20 modes), the PtV stroke of the actuators is 40-60um and a bandwidth of about 750Hz;

- a filter wheel (APO) will select between different apodizing masks, positioned 50mm after the pupil plane.

- in the collimated beam is placed the Atmospheric Dispersion Corrector (ADC). The ADC is deployable, in a way to optimize the system performance at observing altitudes that didn't require the correction, typically for zenithal distances smaller than 25°-30°;

- between the ADC and the second off-axis parabola (OAP2), a beam splitter is placed to send a small portion of the light (~5%) to a very simple tip-tilt sensor, mark as WFS, which is placed in vertical position with respect to the plane of the drawing, composed of a lens and a commercial detector sensitive to J band. The TT sensor gives the advantage to monitor, at low frequency once every minute, for example, possible drifts of the spot during a single exposure, to be then compensated with the local DM, ensuring in this way to maintain the proper mask alignment;

- OAP2 is refocusing the beam on an intermediate focal plane (FP-CORO) with an F/# = 22, where a filter wheel can select between different occulting masks. The same wheel accommodates a couple of low spectral resolution grism (R~100 and R~700) to perform spectral characterization of the science targets;

- after a folding mirror FM1, a third off-axis parabola (OAP3) is creating the 2nd re-imaged pupil plane (PP2), where a filter wheel can select between different pupil stops used to properly mask the spiders and the secondary mirror, to minimize diffraction effects. On the same collimated beam, two additional filter wheels (SFW) will allow the insertion of seven scientific filters each;

- after a folding mirror (FM3), the fourth off-axis parabola (OAP4) is creating the final focal plane onto the detector, where the diffraction limit PSF is Nyquist sampled at 1µm;

- a deployable pupil lens (PLD) can be inserted between OAP4 and the cryostat window, with the purpose to create an image of the pupil onto the detector, which can be used before each scientific exposure to properly calibrate and compensate pupil shifts;

- the scientific camera (FP-SCI) is Cryogenic and will be used a Teledyne HAWAII-2RG detector. The liquid nitrogen tank shall ensure a hold time of about 28 hours.

The whole bench is installed on a mechanical bearing, see Figure 195, allowing the field rotation whenever required from the science cases.

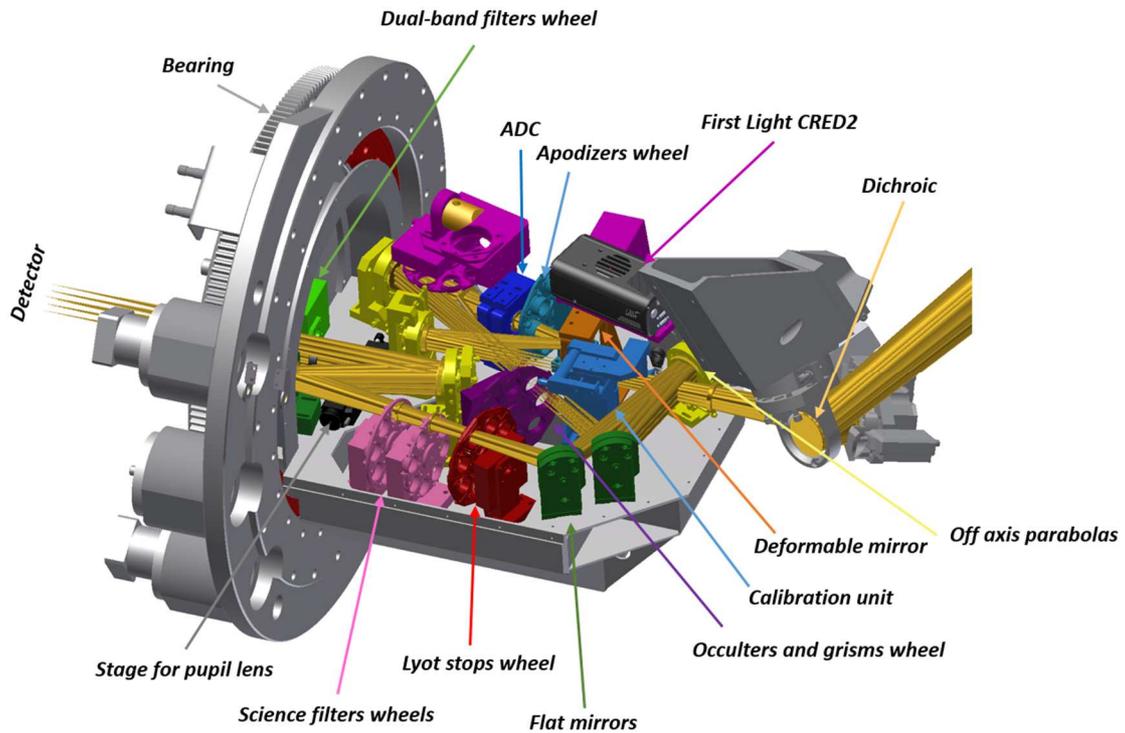

*Figure 195. A detailed lateral view of the optical bench without carters.*

The entrance window of the camera dewar is kept at 200mm of distance from the detector, in order to limit the solid angle through which the detector sees the external environment to limit thermal background. Moreover, a gold-coated annular Narcissus mirror (NM) is placed 40mm before the dewar window to limit thermal background, see Figure 196. The mirror has a central hole (Ø=43.4mm) to transmit the scientific FoV and the surface looking at the detector has a spherical shape with the center of curvature placed on the detector (R=250mm). The solid angle subtended by the hole on the narcissus mirror, as seen from the detector center, is Ω=0.023 sr.

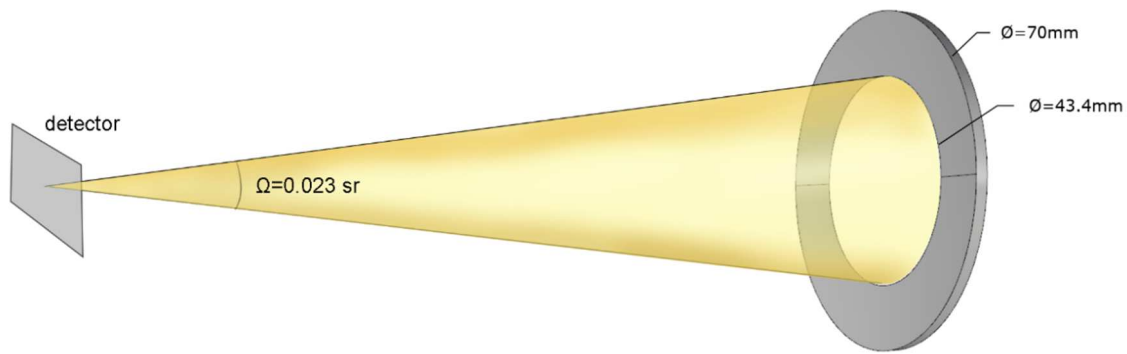

*Figure 196. The scheme of the Narcissus mirror.*

The properties of the coronagraphic planes are resumed in Table 47.

*Table 47. The properties of the coronagraphic planes.*

| CORONAGRAPHIC PLANES SPECS | |
|---|---|
| **Intermediate FP** | |
| F/# | 22.01 |
| Plate scale | 1.14 arcsec/mm |
| **Scientific FP** | |
| F/# | 31.44 |
| Plate scale | 0.80 arcsec/mm |
| Pixel size | 18 μm |
| Plate scale | 0.014 arcsec/px |
| **First Pupil plane** | |
| Pupil diameter | 11.15 mm |
| Mirror tilt | 15 deg |
| Projected pupil size | 11.15 x 11.55 mm |
| **Second pupil plane** | |
| Pupil diameter | 14.92 mm |
| Mask tilt | 10 deg |
| Projected pupil size | 14.92 x 15.24 mm |

## 3.4.2.2 Spectroscopic mode

The slits can be inserted in the intermediate focal plane by the FP-CORO wheel. Two slits are foreseen:

- one with 100 mas width and 100 mas central occulter

- one with 100 mas width and 200 mas central occulter.

The dispersion is produced on the second pupil plane. An Amici prism is used for the LR mode and a Grism or a double VPHG is used for the MR mode.

### 3.4.2.3 Dual-band imaging mode (DBI)

Dual-band imaging mode is characterized by the same instrument configuration of the direct imaging mode with the two following differences:

1. in the second pupil plane (PP2) a Wollaston prism angularly separates the "s" and "p" polarized light in two beams. The corresponding image separation on the scientific FP-SCI is 14". Each image is characterized by an FoV diameter of 2";

2. a couple of filters (dual-band filter), placed in front of the dewar window, selects the waveband for each image (see Figure 5).

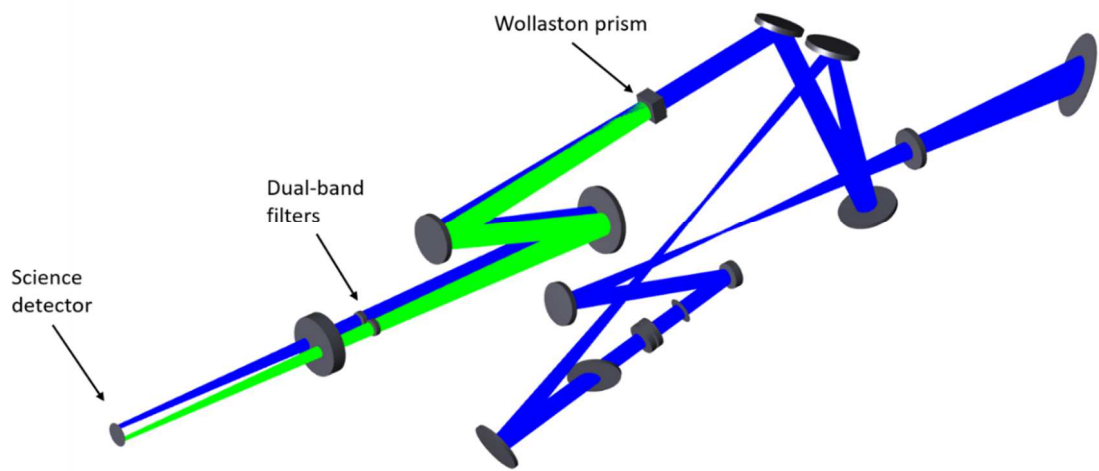

*Figure 197. The layout of the DBI observing mode.*

### 3.4.2.4 NCPA measurement in DBI mode

SHARK-NIR is positioned very close to the WFS area and is a very compact instrument characterized by very relaxed alignment tolerances and by very small internal flexures. All these facts should help to minimize the NCPA between the instrument and the WFS. Nevertheless, the characterization of NCPA and their removal may need to be performed after every new telescope pointing, to maximize the optical quality achieved by the NIR camera. The baseline approach for the measurement of NCPA aberrations is to use "phase diversity" (Robert et al., 2008). Phase diversity requires 2 images to reconstruct the phase aberrations in the pupil plane: one in focus and the other one out-of-focus. Both images are generated simultaneously on the science camera by using the dual-band imaging mode of SHARK. One of the dual-band filters will be a flat filter (as the other narrowband filters used for the DBI mode, but with broadband transmission), while the other one will be a plano-convex lens with very low optical power (f=6480 mm, $R_{curv}$=2900 mm, 15mm diameter) to produce a defocus of 370nm rms (1.9 radians of phase for the NarrowBand_J filter and 1.4 radians for the NarrowBand_H filter). The surface Sagitta (SAG) of the lens is shown in Figure 198 and the maximum deviation from the planar surface is 8-9 µm. Defining R the radius of curvature and D the diameter of the lens, the formula for the SAG profile is:

$$SAG = R - \sqrt{R^2 - \left(\frac{D}{2}\right)^2}$$

In order to perform a good reconstruction of the wavefront with phase diversity, it is required to observe with narrowband filters. The waveband selection will be performed with the scientific filters.

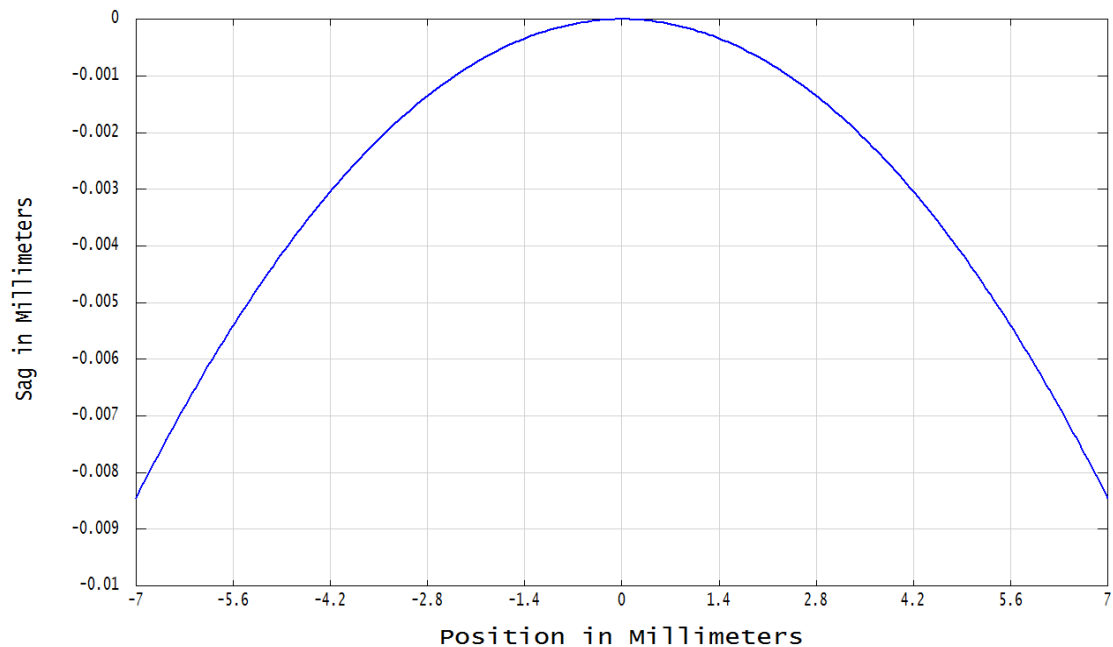

*Figure 198. The surface SAG of the lens used to introduce de-focus for phase diversity.*

### 3.4.2.5 Pupil re-imager

The pupil re-imager, PLD in Figure 194, is an achromatic doublet that can be inserted between OAP4 and the dewar window to make an image of the pupil on the scientific detector to be used for alignment of the pupil masks, Apodizer and LS, before observation. The lens is a 1-inch commercial achromatic doublet by Thorlabs. The focal length is f=300 mm, and it is optimized over the wavelength range λ=1050 − 1700 nm. The diameter of the pupil image is ≈11.8 mm and is sampled by the scientific camera with approximately 650 pixels.

The pupil re-imager could also be used, together with a π/2 phase mask on the intermediate FP (FP-CORO), for low-order wavefront sensing accordingly to the Zernike phase mask concept described in N'Diaye et al., 2013. The working principle consists in using a π/2 phase mask in the intermediate FP to transform the phase aberrations of the wavefront into intensity variations in the next pupil plane that can be measured by the scientific detector. The drawback of the phase mask concept would be the use of a position in the intermediate focal plane filter wheel, which is already entirely used for the occulting masks and the slits for the spectroscopic mode.

### 3.4.2.6 Tip-tilt wavefront sensor

The tip-tilt WFS module, WFS in Figure 194, is made up of three components:

9.  a beam splitter to send a small portion of the light (≈5%) to the WFS. The beam splitter is always inserted in the science channel;

10. an achromatic doublet to image the star on the detector with the proper sampling;

11. a detector used to record the spot.

The imaging lens is a commercially available achromatic doublet produced by Thorlabs. It is a Ø2" cemented doublet optimized for the wavelength range 1050 – 1620 nm. The focal length is f=200 mm and images a field of view of 13.6x10.8 arcsec on the detector with an F/17.7 focal ratio. The resulting plate scale is 21 mas/px and the spot is sampled with ≈3 pixels across the diameter.

The detector selected for the design is the C-RED2 InGaAs camera produced by FirstLight. The specifications of the product are summarized in Table 48.

*Table 48. The specifications of the C-RED2 camera.*

| PARAMETER | VALUE |
|---|---|
| Imaging Device | InGaAs sensor |
| Pixels (H x V) | 640x512 |
| Pixel Size, H x V (µm) | 15x15 |
| Image acquisition options | Windowing, region of interest |
| Frame Rate | 400 Hz (Full frame) – 14 KHz (32x32 window) |
| Waveband (µm) | 0.8 – 1.7 |
| Average Q.E. | >70% |
| RON | 25 electrons (lower with non-destructive reading) |
| Cooling | Air or liquid |
| Bit depth | 14 bits |
| Mount interface | C-Mount |

### 3.4.2.7 Wollaston prism and Dual-band Filters

The Wollaston prism is placed in the LS wheel (PP2). It is designed to achieve a beam deviation of δ=2.15° which corresponds to 14" on-sky. The material of the birefringent crystal is YVO4 selected for its low chromatic elongation in the wavelength range of SHARK-NIR. At λ=1.55μm and considering a Δλ=26nm, the lateral chromatic elongation is 4 mas on-sky. The parameters of the prism are summarized in Figure 199, on the left.

Considering an FoV of 2" for each dual-band image, the two beams generated by the Wollaston prism are completely spatially separated before the Dewar window. Therefore, it is possible to position the dual-band filters outside the cryostat. At the position of the dual-band filters, the two beams are separated by ≈ 6mm allowing some margin for the mounting of the filters, see Figure 199, on the right.

An error of ±0.5° on the Wollaston prism angle corresponds to an error on the beam deviation of ±0.2°. The same error translates to a movement of 780μm on the dual-band filter. Concerning the alignment of the Wollaston prism, a tilt error of ±1° produces negligible effects on both the beam deviation and beam position on the dual-band filters.

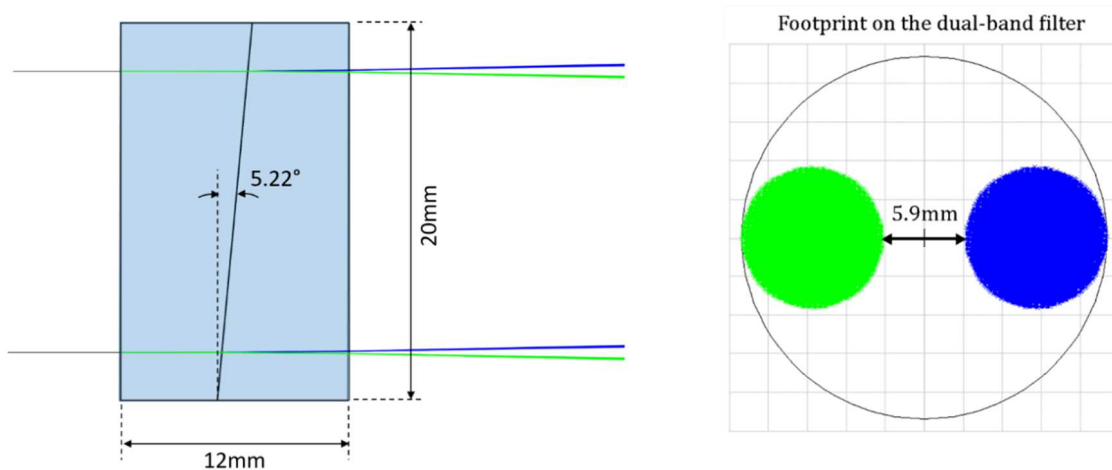

*Figure 199. On the left: parameters of the Wollaston prism. On the right: the footprint of the beams at the position of the dual-band filter.*

### 3.4.3   Apodization and Coronagraphic masks

The coronagraphic SP masks of SHARK, both apodizers and occulters are designed at Institut de Planétologie et d'Astrophysique de Grenoble (IPAG). They are designed considering the diffraction of the LBT spiders. Figure 200 shows an image of LBT arm and a similar entrance pupil shape with the spiders.

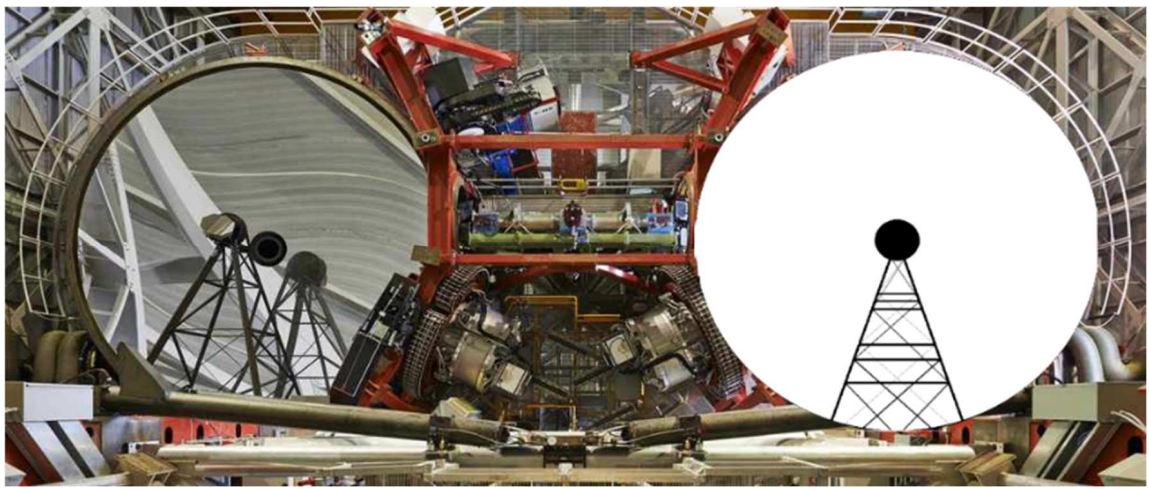

*Figure 200. The arms supporting the secondary and the tertiary mirrors of LBT generate a shadow in the primary mirror that design the entrance pupil of the telescope. Left: An image of LBT arms, right: the superimposed equivalent pupil.*

In Table 49 are shown the three SP masks of SHARK-NIR. The first is symmetrical and the resulting PSF has a working angle from 100 to 320 mas achieving a contrast of $10^{-5}$. The other two have more contrast $10^{-7}$ but are characterized by a discovery space of 220°. The light diffracted outside the discovery space is masked by the occulters.

The FQPM suppresses on-axis starlight by using a phase mask in the focal plane. The mask divides the focal plane into four quadrants and induces a π phase shift on two of them on one diagonal. Provided that the image of the star is perfectly centered on the common vertex of the quadrants, then the four outcoming beams combine destructively at infinity, and the stellar light in the downstream pupil plane is totally rejected outside of the pupil area (Rouan et al., 2000). This light is then easily blocked by means of LS.

The components realized for SHARK-NIR have been designed by Observatoire de Paris, Laboratoire d'études spatiales et d'instrumentation en astrophysique (LESIA) at the wavelength of 1.6μm and have been proven to provide rejection of more than a factor 500 in monochromatic light, see Figure 201. The PSF peak intensity, with and without the coronagraph, is a function of wavelength for this phase mask optimized to work at 1.6μm. In our case, we have a local minimum also at a wavelength around 550nm, and we can test this mask in the visible spectrum, for several reasons much easier.

*Table 49. The design of the SP masks. In the first column the characteristics of the three apodizers are presented. In the second column the coupled occulters are shown. In the last column there is the resulting PSFs. From the top the symmetrical mask (SP1_H), and the two asymmetrical ones (SP2a_H, SP2b_H).*

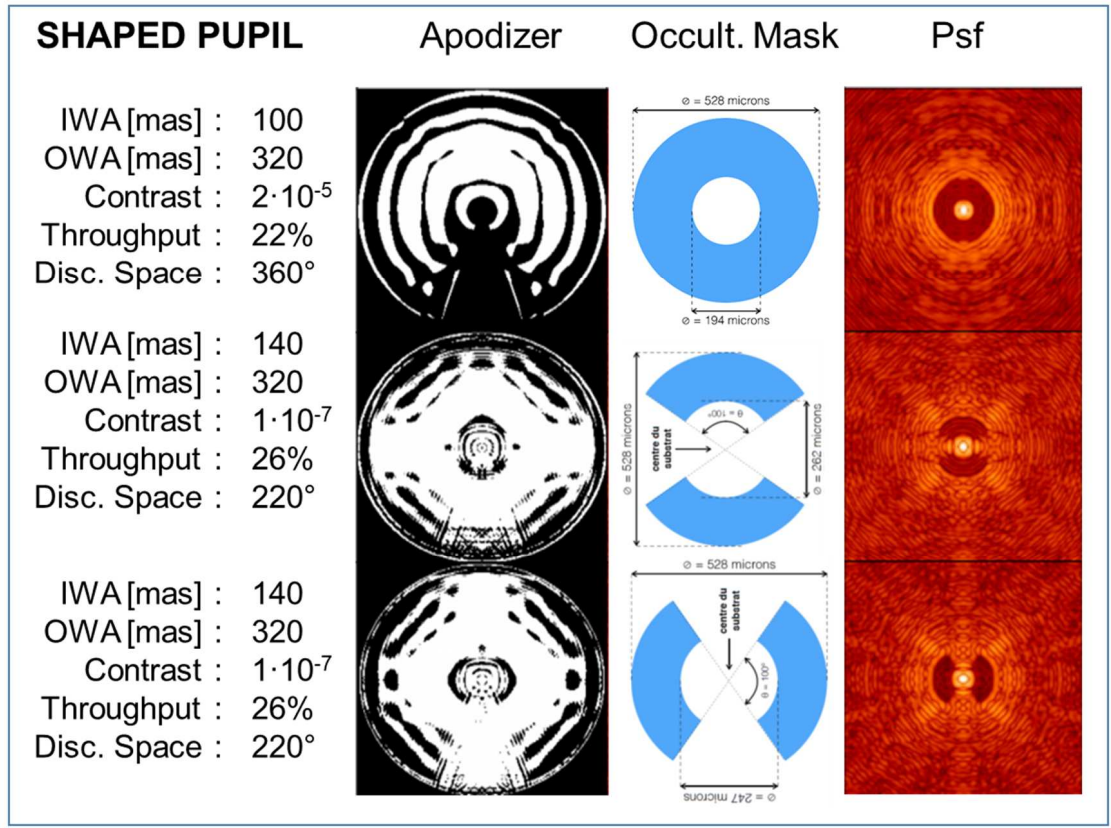

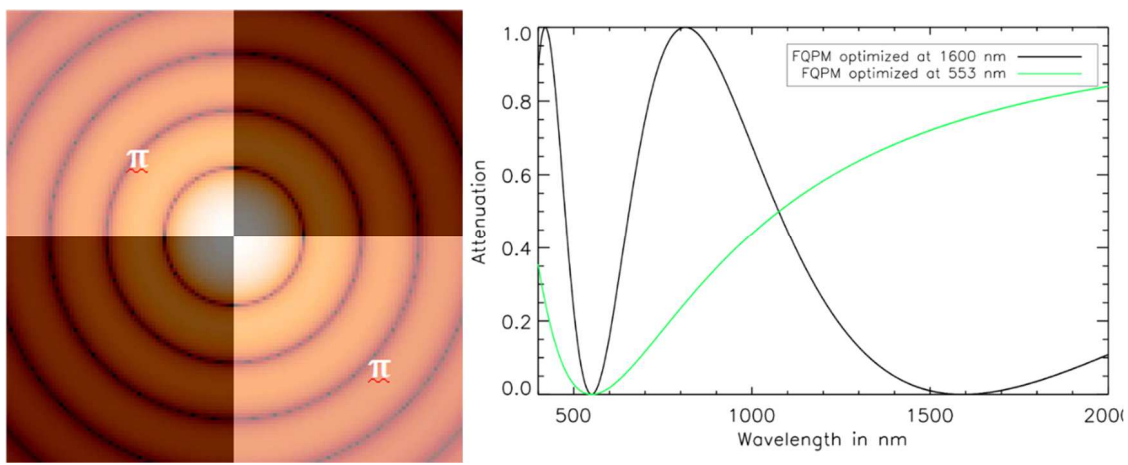

*Figure 201. On the left: Four Quadrant Phase Mask. On the right: Comparison between theoretical PSF attenuation as a function of wavelength for a FQPM optimized at 1.6μm and 553nm. The former can cancel a star at 550nm (sharp), but its higher sensitivity to the bandwidth makes it worse in performance with respect to the latter at the end.*

## 3.5 Alignment of SHARK-NIR

For each component of SHARK-NIR, the project management defined a specific work package to perform a quality test of the optomechanical components, final alignment procedure, and relative error budget to implement all necessary tools to characterize the final alignment of SHARK-NIR in the cleanroom. I am responsible for the characterization and alignment procedure alignment for the SP and the FQPM.

### 3.5.1 Shaped Pupil alignment

The alignment procedure was performed in an optical bench that simulates the final one of SHARK-NIR but without the off-axis parabolas. The design constraints of this bench are:

- an entrance pupil of diameter 11.15mm and located 50mm before the apodizer wheel;

- the same focal plane scale of SHARK-NIR, 35 μm at 1 λ/D;

- the working wavelength of 633nm, 550nm, and 1 μm;

- the LS diameter of 80% of pupil diameter;

- a system to reference for the optical axis.

Figure 202 illustrates the optical bench parts: starting from the laser, the bench develops in a linear fashion by using different folding mirrors. Working in different wavelengths we need a setup that can change the focal plane scale changing minimal parts position. We decided to localize the main PSF image in the surface of the mirror, like the cat's eye[4] reflection, to have a robust reference position of the focal plane. From this, we re-imaged the PSF with a second lens in the coronagraphic plane. A collimator (L4) and focusing lens (L5) create in a fixed camera (CCD-1) both the pupil and the image PSF image, only inserting or removing the focusing lens (L5).

---

[4] Cat's eye reflection occurs when a collimated beam is focused by a lens in the surface of a mirror, and the back reflected light beam assumes the same aperture of the incoming rays.

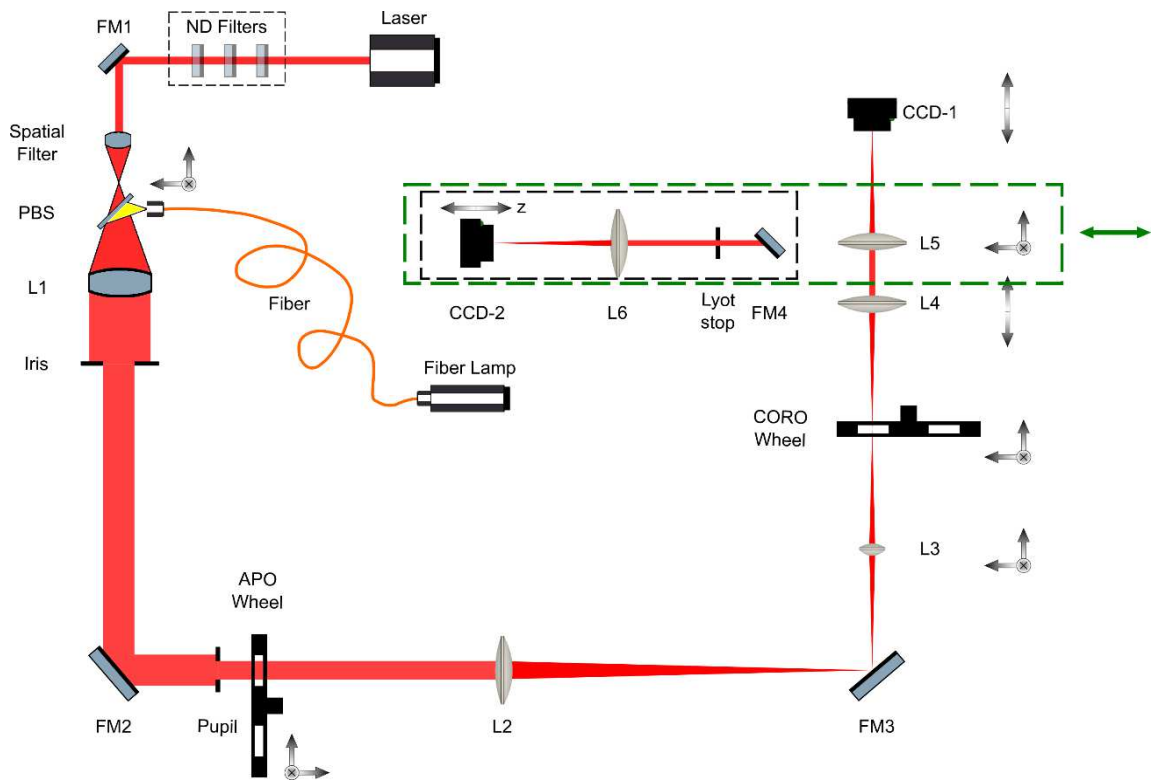

*Figure 202. The optical drawing of the Test Bench. In green is highlighted the deployable path of the LS for the Four Quadrant mask, with the folding mirror FM4 inserted to reflect the light in the LS arm. The CCD-2 is movable to register an image of the pupil or the PSF.*

A collimated beam from the laser is stabilized by a spatial filter and beam expander. In the inner part of the beam expander, a pellicle beam splitter can be inserted to collect light from a fiber optic coaligned with the laser. Folding mirror FM3 is used to compensate the wedge of the apodizer mask. An iris located 50mm before the apodizer plane is illuminated with a collimated beam and simulates the entrance pupil of SHARK-NIR with 11.15mm of diameter. The lens L2 focuses the beam cat's eye on the mirror FM3, this ensures that the PSF is located in a specific position. We used a deployable Foucault slit in a translation stage to move it on the surface of FM3 until half of the PSF is covered. This permits to have a sharp feature in the coronagraphic focal plane to thoroughly define its position in z. Lens L3 is slightly movable in the z-axis to change the plate scale with different working wavelengths and matching 35 μm at 1 λ/D. The coronagraphic FP is finally reimaged on a camera using two lenses, L4 as collimator, and L5 as a focusing lens. L5 is deployable and when removed in the CCD is imaged the entrance pupil or the apodizer mask.

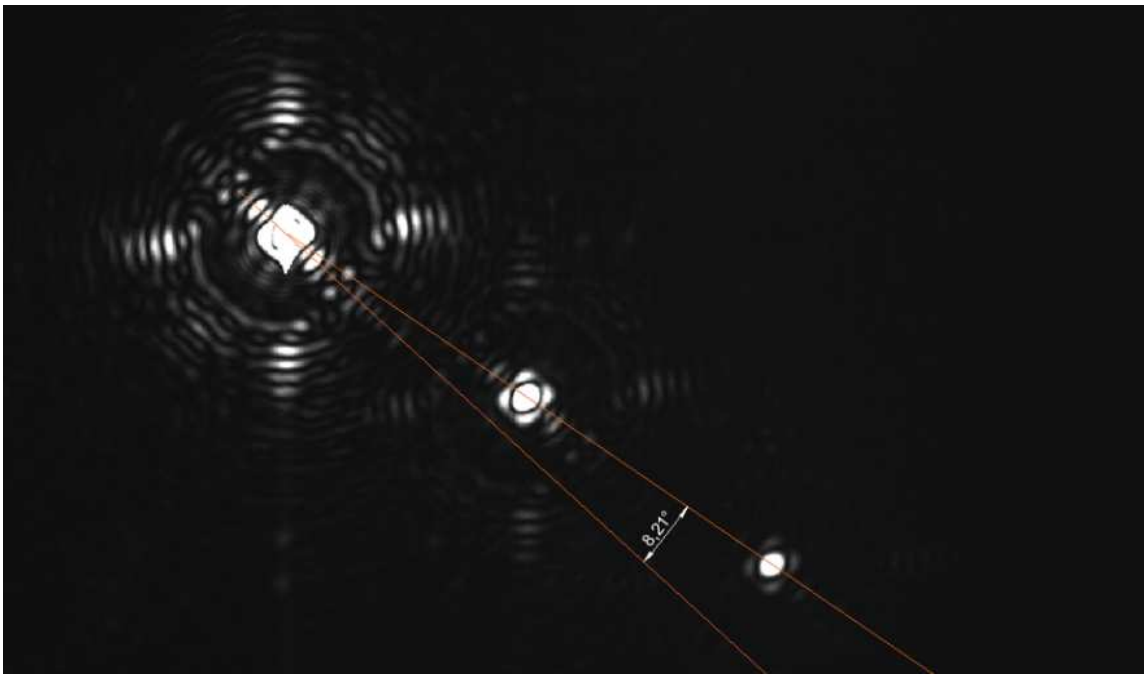

*Figure 203. The multiple reflections of the SP2a_H Apodizer mask, falling outside the discovery space due to the presence of a wedge in the glass substrate of 2 arcmin. 8.21 degrees is the angle between the wedge direction (where the ghosts located) and the direction of the luminous wings of the PSF.*

The optical bench for the coronagraphic masks tests and alignment included this main parts: 20mW He-Ne laser (633nm), ND filters, Plane mirror [FM1], spatial filter, Pellicle Beam Splitter [PBS], Lens (BB2-E02 D50.8 FL=300mm) [L1], Plane mirror [FM2], Iris [Pupil], Apodizer wheel [APO Wheel] on PI stage + Mercury controller, Lens (026-1730 D50.80 FL=700mm) [L2], Plane mirror [FM3], Lens (SINGLET D25.4 FL=70mm) [L3], Coronagraphic FPM wheel [CORO wheel] on PI stage + Mercury controller, Lens (D50.80mm F100mm P5630) [L4], Lens (D50.80mm FL=150mm) [L5] on linear stage PI M511.DG, pos. rep. 0.1µm, Camera (AVT Pike F-145B pixel size 6.45µm resolution 16bit) [CCD].

Figure 204 describes the optical layout with distances of the single parts, and in Figure 205 a photo of the bench is presented. Section a) is the illuminating system composed by the laser and the coaligned fiber-optic by a pellicle beam splitter. Section b) is the following path where the entrance pupil is a hole of 11.5 mm in diameter. The FM3 is cat's eye on the focus of L2. The lenses L2, L3, L4, and L5 are made with fused silica and feature a standard polish of PtV at 632.8 nm. L5 lens is deployable. The CCD camera with 6.45 µm pixel size is in a fixed position on the bench. All distances are in millimeters. The optical rays design the path from the pupil to the focal plane in the CCD, where the PSF is imaged. Section c) is the same as the previous, but the rays design the path from the pupil to the pupil image in the CCD, with L5 removed. Between the L2 lens and the FM3 a fast rotating diffuser (12000 RPM) can be inserted to remove the interference in the pupil and focal plane when necessary.

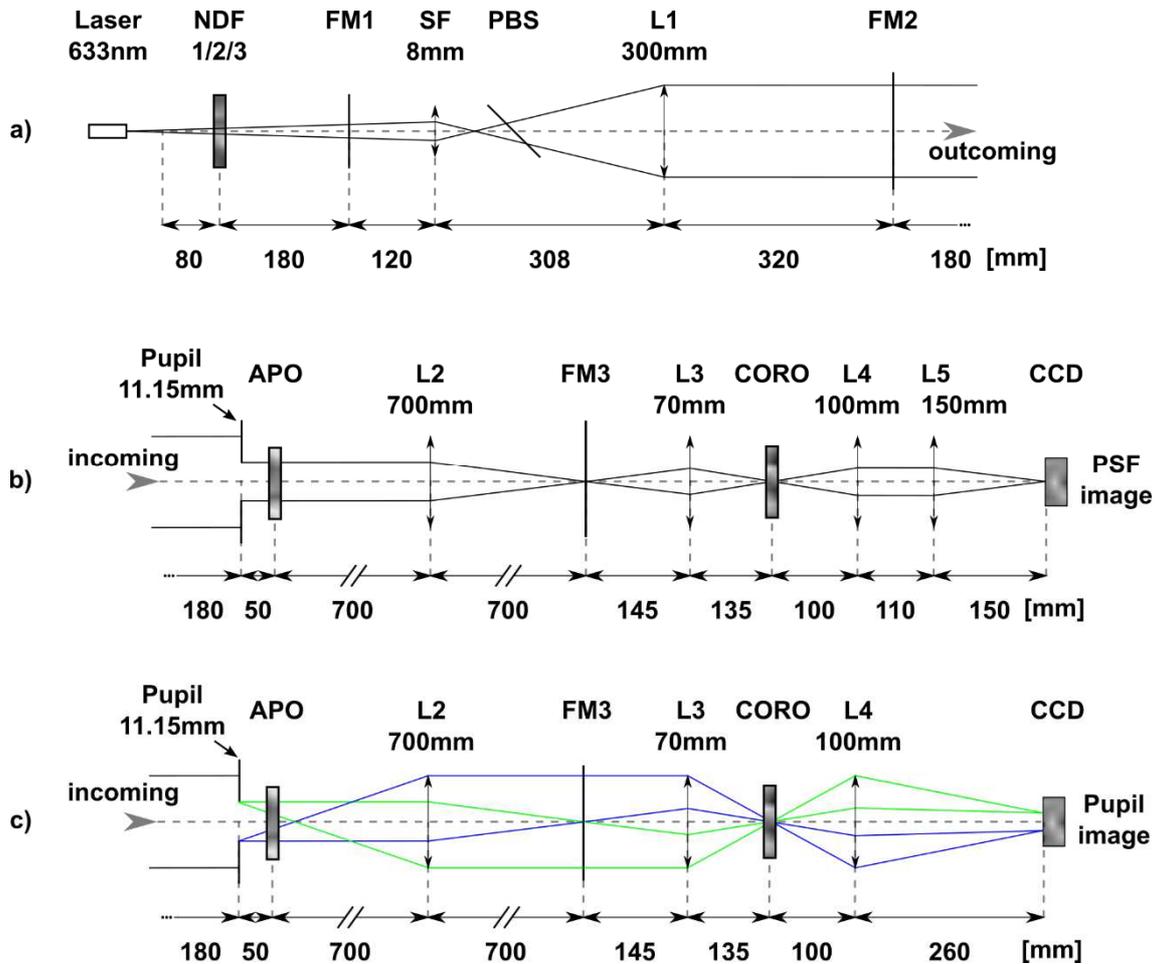

Figure 204. The optical layout of the SP coronagraph experiment. Section a) is the illuminating part composed by a laser and beam expander and a coaligned fiber-optic by a pellicle beam splitter. Part b) is the following path where the entrance pupil is a hole of 11.5 mm in diameter. The FM3 is cat's eye on the focus of L2. The lenses L2, L3, L4, and L5 are made with fused silica and feature a standard polish of PtV at 632.8 nm. L5 lens is deployable. The CCD camera with 6.45 μm pixel size is in a fixed position on the bench. All distances are in millimeters.

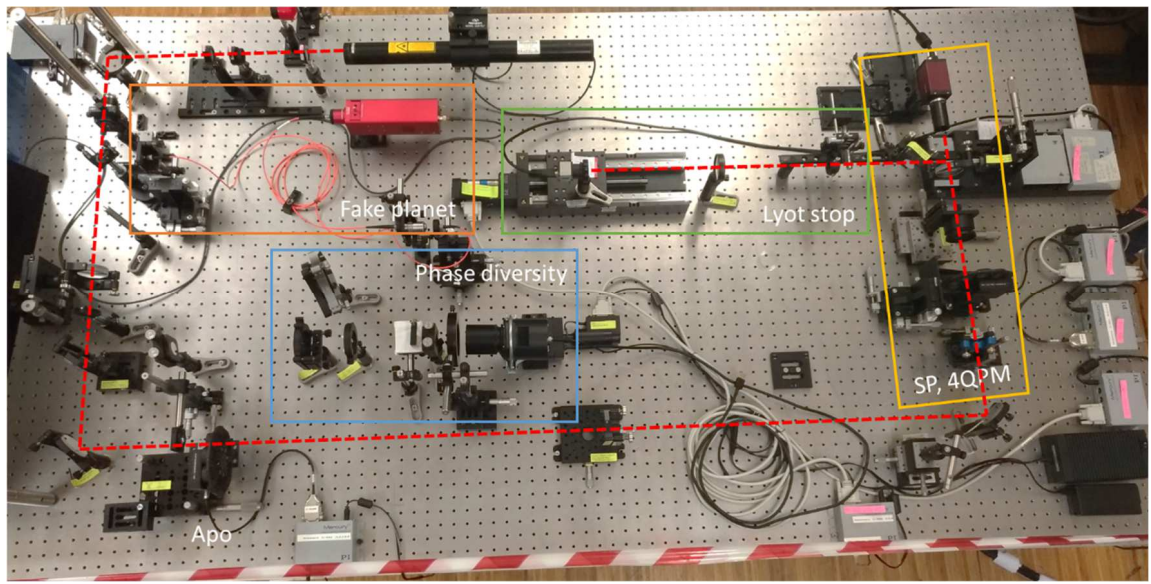

*Figure 205. The image of the optical bench with highlighted the part of the fiber optic illuminator, the arm of the coronagraphic masks, and the arm of the LS.*

The decenter and tip-tilt of the optical components were performed as in the PLATO case, with transmitted and back-reflection spots.

To test the alignment along the z-axis, parallel to the light beam, we started without masks and illuminating L2 with large size beam of diameter 30mm, in order to ensure less depth of focus with respect to the pupil diameter. FM3 was mounted in autocollimation, and we checked the back-reflected beam with a collimation tester, Newport 20QS20, installed before L2. Moving L2 we achieved the best-collimated beam. We pivoted FM3 without translating the position of the focus of 45 degrees in the direction of the lens L3. A Foucault slit is inserted in the surface of the FM3 until half PSF is occulted. In the conjugated plane of L3, we installed a CCD to image this feature. When the edge of the slit appears sharp, without a diffraction pattern, we are confident that the focal plane is projected by L3 in the CCD. Removing the slit and mounting the fixed iris of 11.15 mm in diameter, we measured the $\lambda/D$ of the PSF, performing steps in z-position of the CCD and L3 until we achieved the desired value. We need a PSF of 35μm on the focal plane because this is the physical dimension of the spot on SHARK-NIR coronagraphic FP.

Defined the position of the focal plane, L4 was aligned in the z-axis with the help of the collimation tester. By this, the CCD-1 reached the final position when the image of the pupil was brought into focus by L2/L3/L4 system. We took images moving the CCD at the step of 10μm and to find the best focus position with a Laplacian of Gaussian[5] (LoG) filter which amplifies the edge contrast in the image, following these steps:

- removing high frequencies with FFT filtering;

- compute the Laplacian of Gaussian of the image (LoG) using `scipy.ndimage.gaussian_laplace` Python module;

- do the FFT of the LoG, High Frequencies in Fourier Space increase (HFFS), and sum the counts in HFFS, which was defined by a distance, r, from the center;

- fitting the resulting data with a polynomial of 5[th] order and find the maximum;

- compute the position error with a variation of the inner radius of HFFS, which defines the accuracy of this method.

As an example, we reported in Figure 206 and Figure 207 the efficiency of the LoG that increases the intensity in the region of the HFFS. In Figure 208 the sweep around the focus is performed for a range of 350 μm and the position is determined with a precision of 26 μm PtV, varying the radius of the HFFS from 16% to 33%, 0% is only the central point of the Fourier space, 100% is all frequencies of the images.

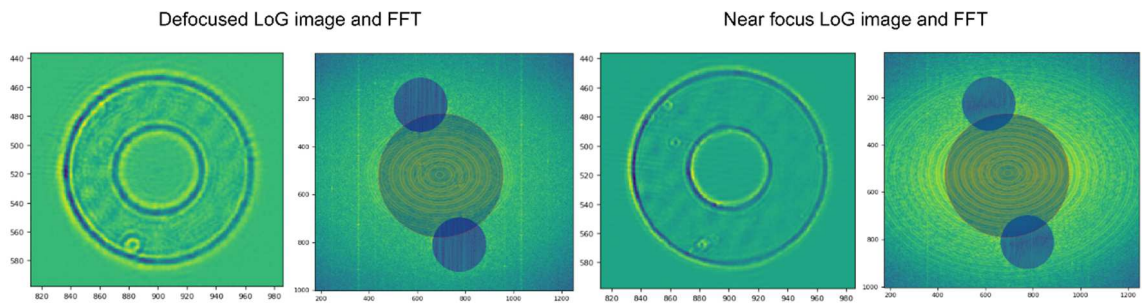

*Figure 206. The Laplacian of Gaussian of the image and their Fast Fourier Transform. The sum of counts in the High Frequencies in Fourier Space (outside blue regions), increases and is used to fit the best focal image.*

---

[5] Multidimensional Laplace filter using gaussian second derivatives (A. Huertas and G. Medioni, 1986)

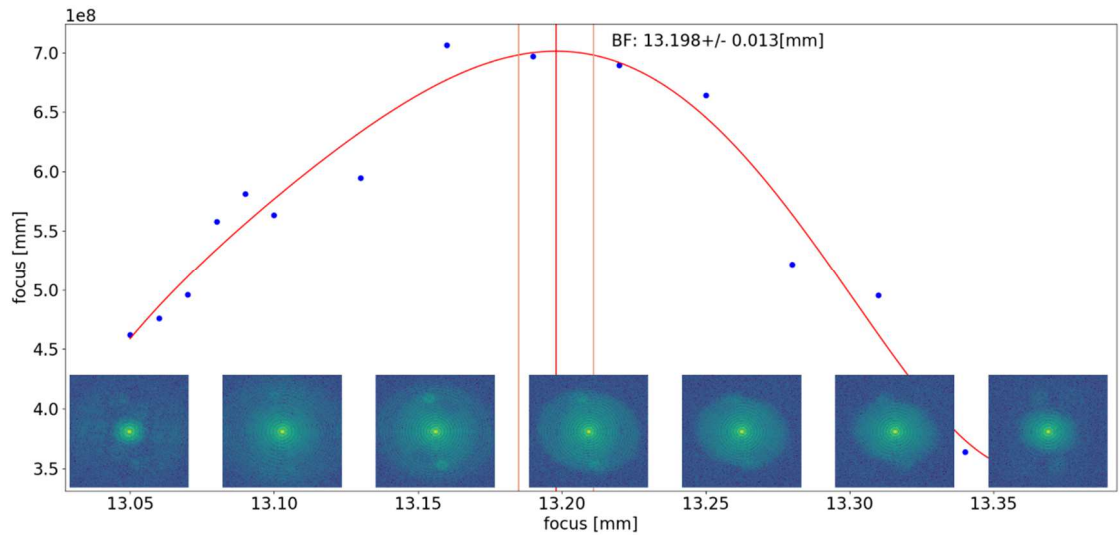

*Figure 207.The variation of the HFFS passing through the focus applying the LoG method compared with the FFT of the LoG image.*

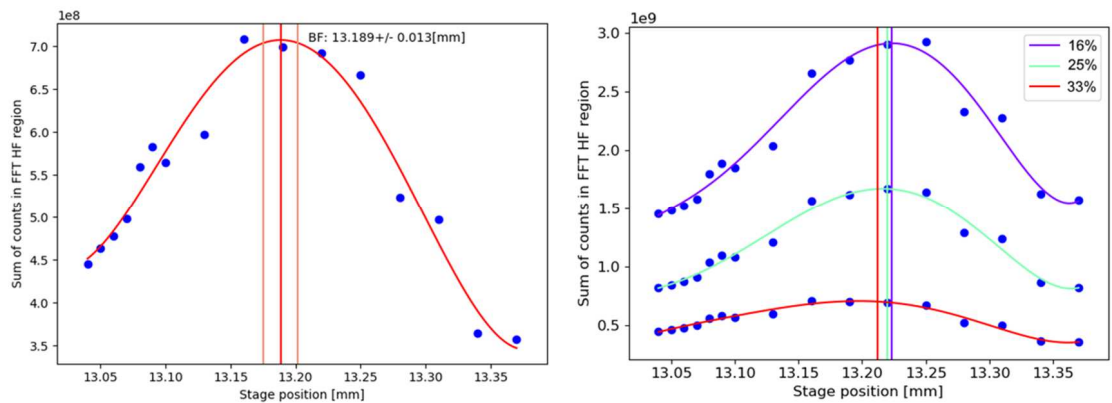

*Figure 208. On the left: to define the best focal position of the coronagraphic mask, we fitted the maximum of the sum of counts in the High Frequencies in Fourier Space of Laplacian of Gaussian of the images. On the right: accuracy determination varying the radius of the HFFS from 16% to 33%.*

The last action was inserting the L5 lens, and the z-position was achieved with a fit of the minimum value in the standard deviation of the Gaussian model of the image the PSF versus the L5 position.

Figure 209 shows the radial profile of the PSF in the CCD-1 at their final position, after the L4 and L5 lens. CCD-1 looks at the focal plane with a 1.5x magnification, magnification. Combined with the connection between FWHM and $\lambda/D$ of an Airy profile, the resulting FWHM of 54.2μm on CCD-1 leads to 1 $\lambda/D$ = 36μm on the FP.

To measure the spot diameter on the FP we followed these steps:

- acquiring 30 images and co-aligned with a cross-correlation to reduce jitter of air turbulence of the laboratory;

- fitting the PSF with a 2D gaussian and extracting the coordinate of the center $(x_0, y_0)$;

- performing a radial profile centered in $(x_0, y_0)$;

- find the local maxima and minimum of the profile;

- we used the formula to translate the information in px to micron:

$$size \ on \ FP \ [\mu m] = \ size \ on \ CCD[px] \times \frac{6.45[\mu m]}{1.5 \times 1.028} \qquad ,$$

where 6.45µm is the pixel size of the CCD-1, 1.5 is the optics magnification of L4-L5, and FWHM =1.028 λ/D.

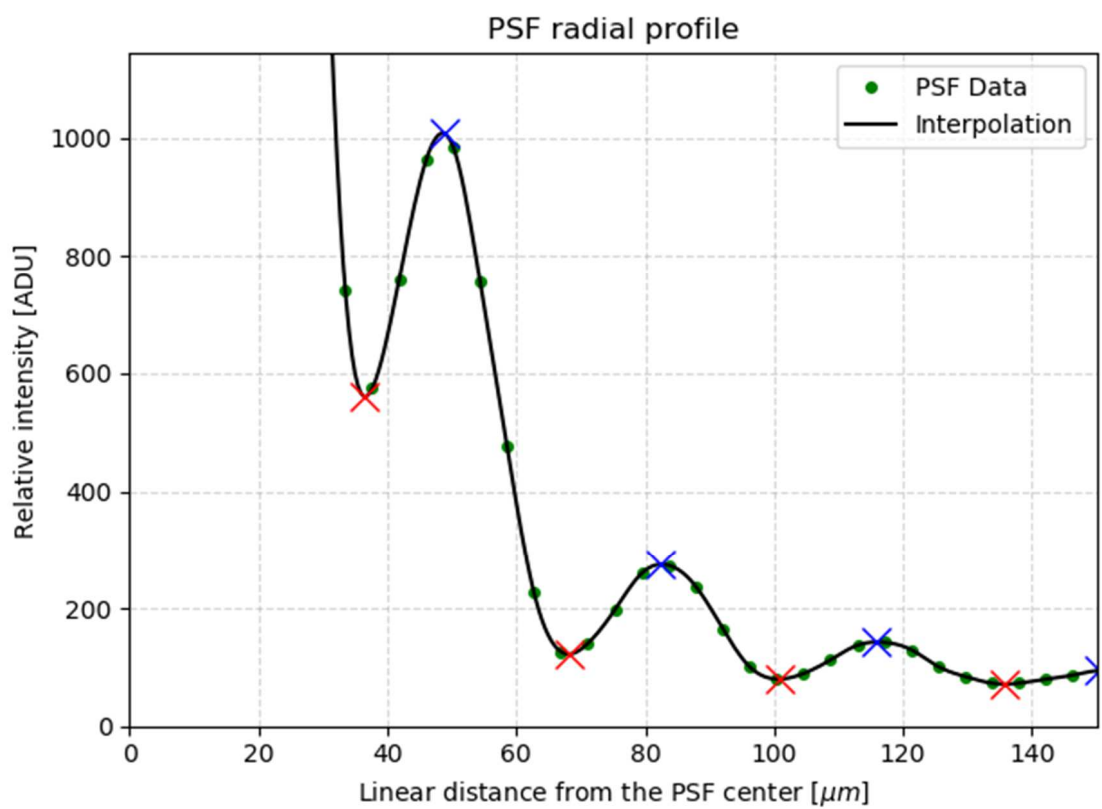

*Figure 209. The spot dimension on CCD in the best focus configuration. On the FP the spot dimension is 36µm corresponding at the first minimum.*

The SP masks realized by the company Optimask followed an optical inspection that confirms their nominal dimensions. There are three different apodizer designs, each of them coupled to a specific occulter, in Table 50 I show the resulting images taken in the optical testing bench at 550nm and 633nm wavelength and using the diffuser for the focal plane images.

*Table 50. The names and images of the SP of SHARK-NIR.*

| Name | SP Image @633nm | SP Image @550nm | CORO Image @633nm |
|------|-----------------|-----------------|-------------------|
| SP1_H and SP1_FPM_H | 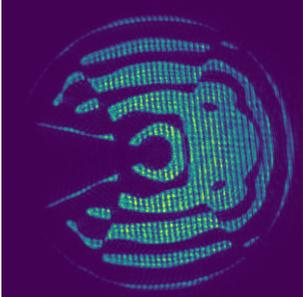 | 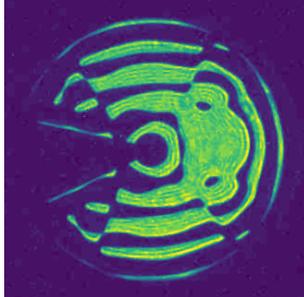 | 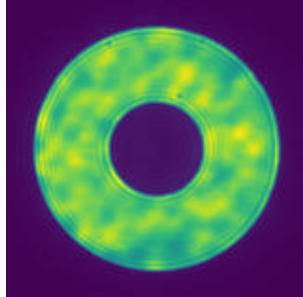 |
| SP2a_H and SP2a_FPM_H | 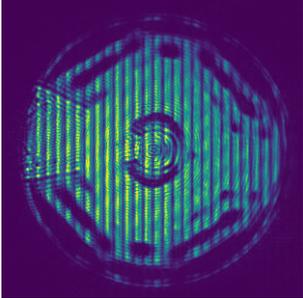 | 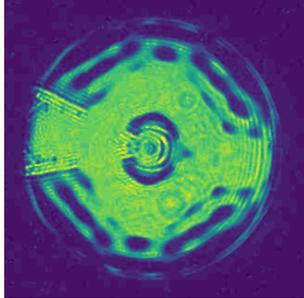 | 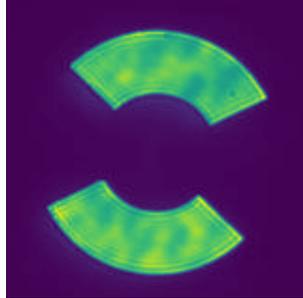 |
| SP2b_H and SP2b_FPM_H | 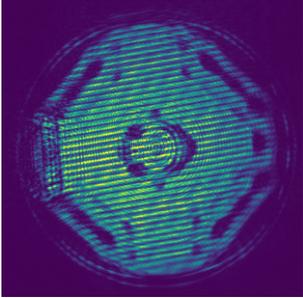 | 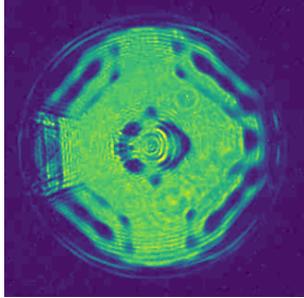 | 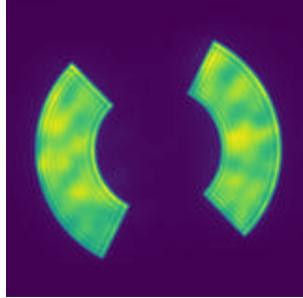 |

By using a laser with monochromatic light, we have experience of interference fringes which can make the precise measurements of the features in the images difficult. For example, we discussed the case of measuring the center of the symmetrical focal plane mask, see Figure 210. Panel a) shows the mask with the PSF generated by the lens system in the center. In the transmissive area, we can see Airy's rings but is not possible to measure the inner and outer radius of the occulter. The light of the PSF is scattered in multiple directions introducing a diffuser before the FM3, the IWA and OWA appear, but the speckles are dominant, and the measurement is still tricky, panel b). A fast rotating diffuser mitigates the speckles, and only the interference fringes due to the thickness of the mask remain. We easily remove them with a Fourier filtering, panel c).

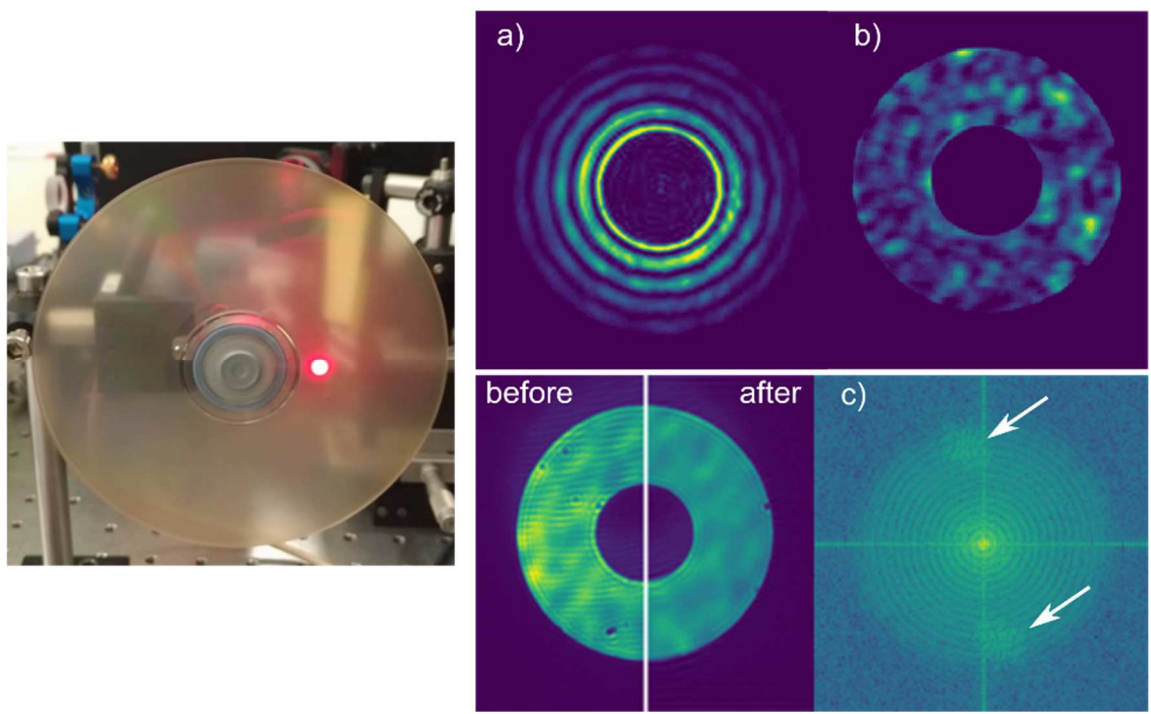

*Figure 210. The fast-rotating diffuser and the improvement in the quality of images.*

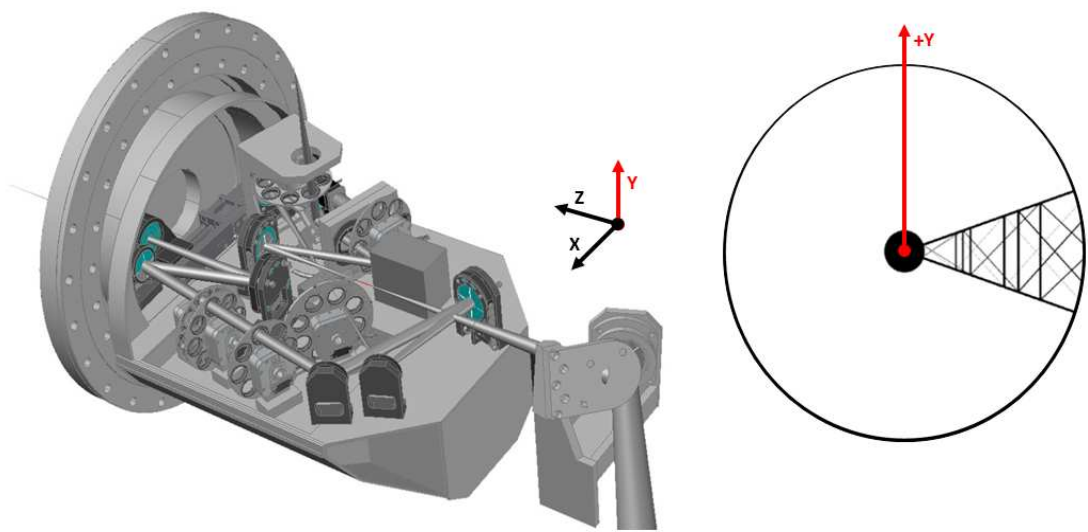

*Figure 211. The global coordinates of SHARK-NIR mechanics. The image on the right shows the position of the arm of LBT. The apodizers are mounted on the wheel such as the arm drawn on the mask is oriented as in the figure when the mask is illuminated by the collimated beam.*

The rotation angle of the asymmetric mask follows the global coordinates of SHARK-NIR and is correlated with the position of the supporting arm of the secondary mirror of LBT, see Figure 211.

### 3.5.1.1 Accuracy of optical axis reference

We measured the error in the positioning of the optical components in such order:

- The entrance pupil became the reference for the following parts;

- PSF on CCD after L2/L3/L4/L5;

- Shaped Pupil Apodizer mask, 50mm after the entrance pupil;

- Coronagraphic mask in the simulated focal plane of SHARK-NIR;

**Entrance Pupil**: without L5 we have in the CCD the image of the pupil, a python code measures the center of the aperture to realize the first reference of the optical axis. We have some noise to be reduced as the interference pattern generated by the monochromatic laser beam and the disuniformity in the illumination of the path. Figure 212 shows an extreme situation and the iteration steps:

a) is the original image with the interference pattern;

b) is the filtering in the spatial frequency domain corresponding to the high frequencies of the interference;

c) is the Fourier filtered image;

d) is the background map from the ASTROPY (Price-Whelan et al., 2018) routine *astropy.bkg_estimator*, used to create a normalized flat reference;

e) is the final image corrected for the background gradient, and the resulting intensity of the pupil border is more uniform than the original image.

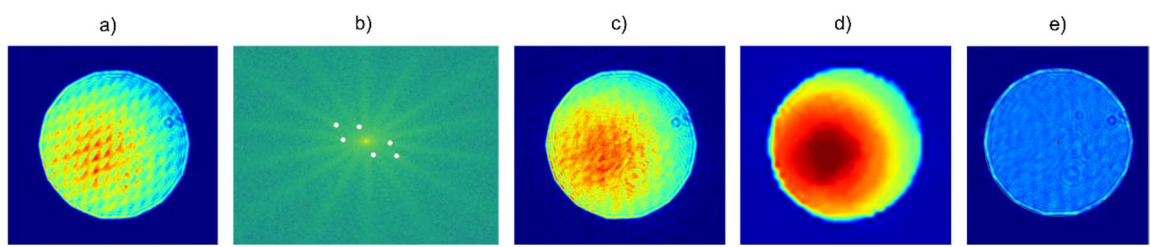

*Figure 212. The software procedure to analyses the pupil image.*

The resulting image of the pupil shows the diffraction pattern of the border, distributed over 10 pixels, around 60μm, with 30x10⁴ levels of intensity. We defined as the center of the pupil, the center of a circle that fits an isophote to a precise value of the border. Moreover, the inner part of the profile shows a typical Fresnel diffraction amplitude profile. In optics, a solution of the Fresnel diffraction for an aperture is calculated in terms of integrals that cannot be obtained by analytical methods. These integrals contain arguments in terms of the Cartesian coordinate system and are known as Fresnel integrals (Sandoval-Hernandez et al., 2018). The Fresnel integral $\hat{F}$ are generally composed by real cosine and sine Fresnel integrals:

$$C(P) \;=\; \int_0^P \cos\left(\frac{\pi x^2}{2}\right) dx\,, \qquad S(P) = \int_0^P \sin\left(\frac{\pi x^2}{2}\right) dx$$

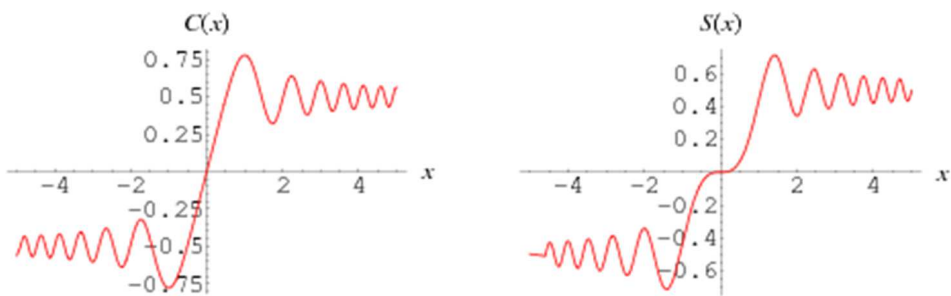

$$\hat{F}(P) \;=\; C(P) + i\,S(P) \;\equiv\; \int_0^P \exp\!\left(i\,\frac{\pi}{2}\,x^2\right) dx$$

Where $P \geq 0$ is a real number and $x$ is a real variable. When $P$ tends to infinite, $C = S = 1/2$. The Fresnel integral $\hat{F}$ is writable in terms of Gaussian Error Function (Korn and Korn, 2000).

$$\text{Erf}(x) = \frac{2}{\sqrt{\pi}} \int_0^x \exp\left(-\tau^2\right) d\tau$$

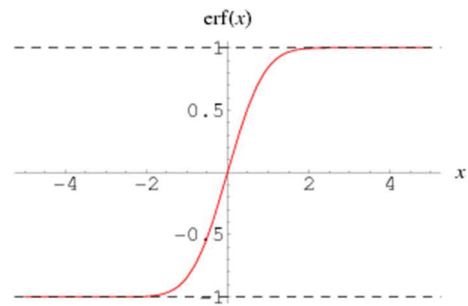

We fitted the Fresnel profile of the pupil edge with a Gaussian Error Function to extract the mean value of the rising part and perform isophotal interpolation for $\pm 1000$ intensity levels with steps of 10. We obtained 200 estimations of the center from which we extracted the mean value and the associated accuracy.

In Figure 213 we reported the center of the pupil $x_0 = 726.51$ +/- 0.09 [px] and $y_0 = 513.68$ +/- 0.04 [px]. The fitted diameter of 287.69 +/- 0.31 [px] was used to determine the scale factor and the root sum squared (RSS) associated to the center, equal to **4.4µm**.

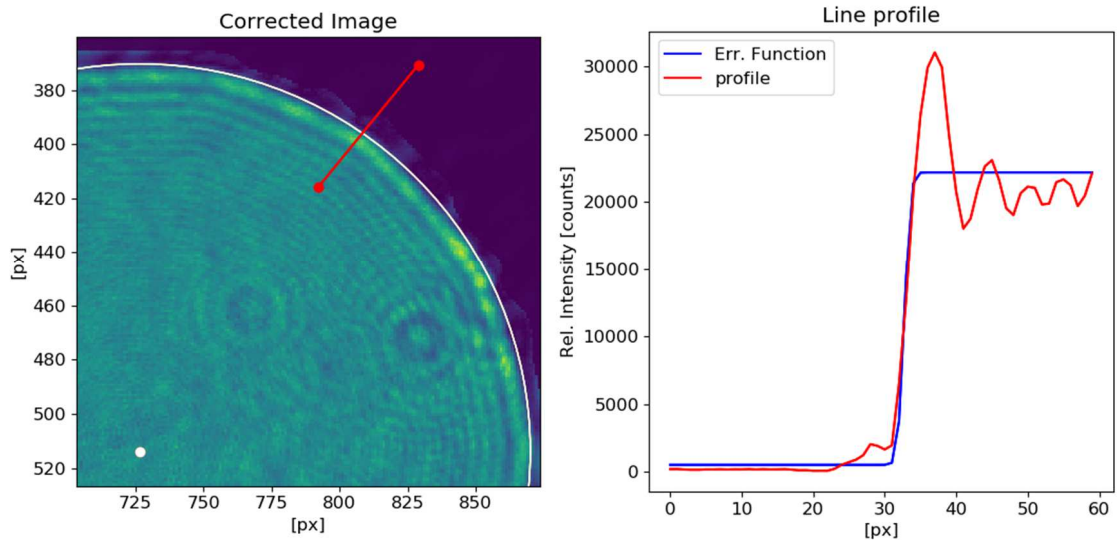

*Figure 213. On the left: the center of the pupil is determined by an interpolation of the isophotal level of the border with a circle. On the right: the line Fresnel profile of the border, from outer to the inner part, fitted by the Error Function.*

**PSF on CCD**: L5 lens was decentered in x and y to minimize the distance from the center of the PSF and the reference point of the pupil. We extracted the center of the PSF with the gaussian 2D fit. The resulting parameters of the Gaussian model were passed to a non-linear least-squares fit of a parametrized Gaussian 2D model using LMFIT-py code (Newville et al., n.d.), which associates at the center also the accuracy. The computed accuracy is very high, of 0.02%, that means in the CCD plane 0.02 µm. The error is then related to the accuracy of the stage supporting L5 lens: x-axis with linear stage PI M511.DG position repeatability of 0.1µm, y-axis manual translation stage with repeatability of 1µm. Remembering the magnification factor 1.5x of the L4/L5 system, a conservative value is **3.0µm** in the coronagraphic plane, half of the pixel size.

**Shaped Pupil Apodizer**: the requirement is that the apodizer mask is aligned better than the 0.5% of its diameter. The SP image is not perfectly sharp because it is positioned 50mm after the pupil, but we can use symmetrical features to define the center of the mask. The outer part of the three masks is perfectly circular by design. We selected many points by hand and perform a circle interpolation. Iterating this method ten times we derived the conservative accuracy of the measurements. From this, we also measured the diameter of the mask. The result is 0.9px in standard deviation, which with the 285 px in diameter means 0.3%, inside the requirement.

**Coronagraphic FP mask**: we used the same procedure applied to the pupil, deriving the precise plate factor (μm/px) in the coronagraphic plane and the accuracy of the center. We used the rotating diffuser to have a more uniform illumination of the masks. After ten determinations the estimated uncertainty is 0.018 px corresponding to 0.071 μm in the FP. A conservative value adopted is **0.1 μm**. For the asymmetric mask, we also need to measure the rotation angle. We selected by hand many points at the edge of the mask, to define two straight lines by the linear least-squares fitting. The crossing point of the lines defines the center, and we efficiently computed the rotation angle from the slope of the bisector. The accuracy in position is 0.16px, corresponding in the FP to 0.67 μm, with a conservative value of 1μm. The angle accuracy is 0.13 deg (7.8 arcmins).

In Figure 214, I resumed the fitting point for the PSF, Apodizers and FP masks.

a)        b)        c)        d)

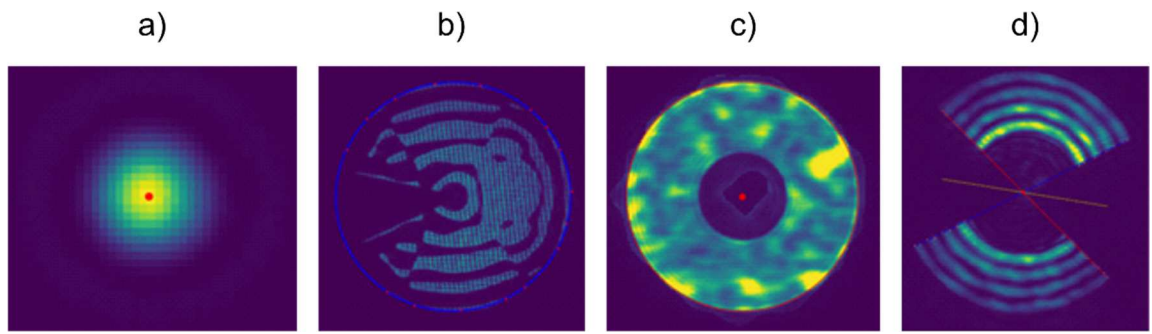

*Figure 214. The image analysis: a) PSF center, b) SP center, coronagraphic mask, c) symmetrical and d) asymmetrical.*

### 3.5.1.2 Coronagraphic wheel

The coronagraphic wheel is a crucial tool to move the masks in the correct positions, by a rotational movement of the wheel and a radial movement of the mask mount in the wheel, see Figure 215. The rotational movement is entrusted to a motor with high precision, the M-116.DGH of PI-Instrumente. The budget allocated for the occulting mask wheel positioning repeatability is 8 arcsec (40 μrad). Since during the instrument life all the motors will be initialized several times, we tested in our laboratory the repeatability of all M116 rotary stages in reaching a well-determined position after homing.

For alignment purposes, it is required to rotate the focal plane coronagraphic masks in steps not wider than 2 μm. Since the distance of the center of these masks from the pivot point of the wheel is 43 mm, this translates into a minimum incremental step for the rotary stage of 9.5 arcsec (47.5 µrad). From the Physik Instrumente datasheet, we have unidirectional repeatability of 10 µrad and minimum incremental step of 25 µrad.

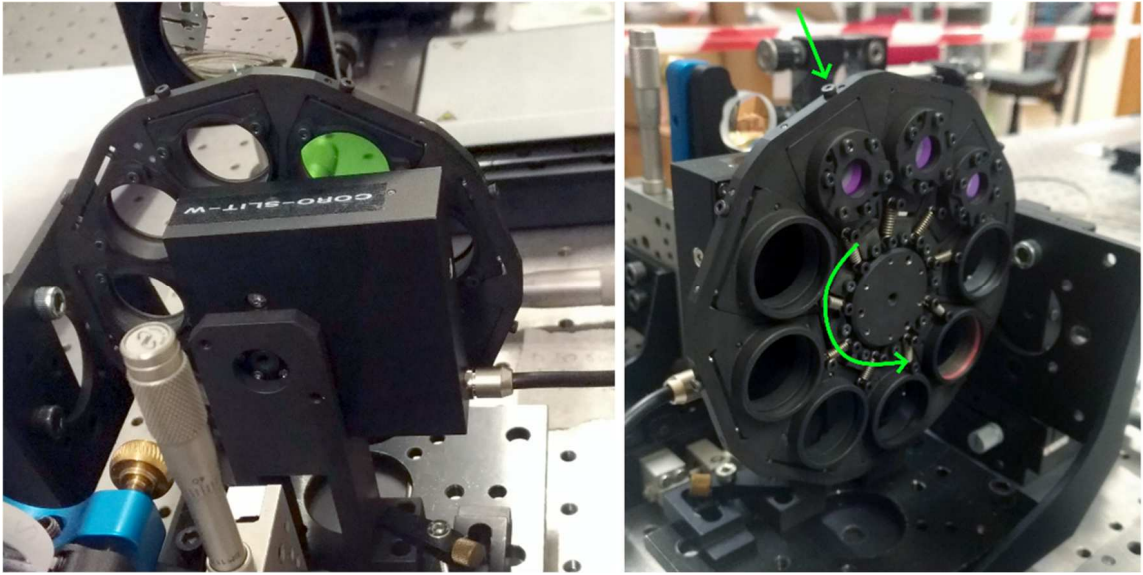

*Figure 215. The masks wheel, the arrow indicates the rotational and radial movements.*

Both these parameters of the motor have been tested with both native controllers (Mercury) and with the SHARK-NIR electronic and software, resulting in a minimum incremental step value of **3.6 arcsec**. Considering the PtV and the worst scenario, the repeatability in positioning the stage in a given position is **6.4 arcsec.**

We optically tested the position repeatability of the coronagraphic wheel by acquiring a sequence of images of SP1 occulter in diffused light. Before each acquisition, we moved the wheel at home position and back on the occulter. We repeated the procedure 9 times. To obtain the position repeatability we measured the shift of the centroids. The occulter diameter on the wheel is 528µm, on the CCD this value is 832µm. The shift of the center of the coronagraphic mask for each iteration is the ratio of these two values times the pixel value of the CCD (pix size = 6.45µm), the real magnification factor during the test. The radius of the coronagraphic wheel (from the center of the mask to the rotation center) is 43mm. The centroid shift values are the shift of the coronagraphic mask divided for the radius of the focal plane wheel, resumed with this formula:

$$shift_{x,y}[arcsec] = shift_{x,y}[px] \times 6.45[\mu m] \times \frac{528\,[\mu m]}{832\,[\mu m]} \times \frac{206265}{43 \cdot 10^3[\mu m]}.$$

The resulting repeatability from the PtV values of the shift movement is **3 arcsec**.

The screw pitch is 400μm/360° as declared by the company. When we turn the screw, the coronagraphic mask moves in a radial direction with respect to the center of its wheel. For this reason, to quantify the amount of shift when the screw is turned, we fixed a goniometer on the wheel to screw of a known amount, as in Figure 216. We can consider a sensitivity in radial centering of the FPM of about **10μm.**

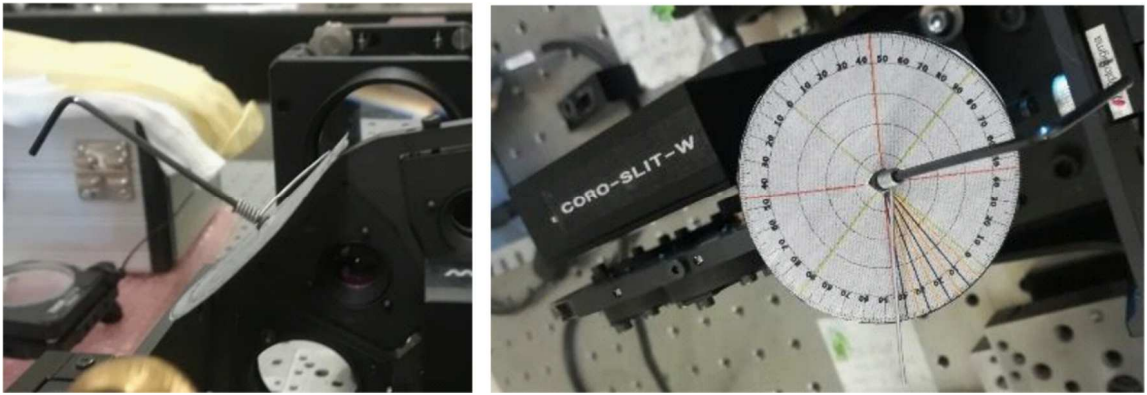

*Figure 216. The goniometer fixed on the wheel to screw of a known amount the mask mount.*

With this accuracy, we wrote a Python code that, given the coordinates of FPM and PSF centers on CCD-1, calculates the number of screws and the rotation angle of the wheel necessary to align the mask in the radial and azimuthal direction, respectively. The Python procedure converts the distance between the FPM and the PSF from Cartesian to polar coordinates (ρ, θ) relative to the center of the wheel.

The natural choice for the origin of the polar reference frame is the position of the FPM wheel center on the CCD. With this choice, the differences Δρ and Δθ between FPM and PSF centers directly translate, respectively, into a number of screws and a number of degrees of rotation of the wheel.

This operation thus requires a preliminary calibration step, in which the coordinates of the FPM wheel center are determined. To do that, it is necessary to acquire four images of the FPM (with the diffuser) in four different positions. An example of the four images (assembled together via software) is shown in Figure 217. Position 2 is obtained from position 1 by turning the radial screw five times (the same holds for images 3 and 4). The line passing through the centers of 1 and 2 intercepts the one passing through the centers of 3 an 4 in a point which is the position of the FPM wheel center. It is preferable to do this procedure with a SP occulter since it is much easier to determine its center compared to the FQPM.

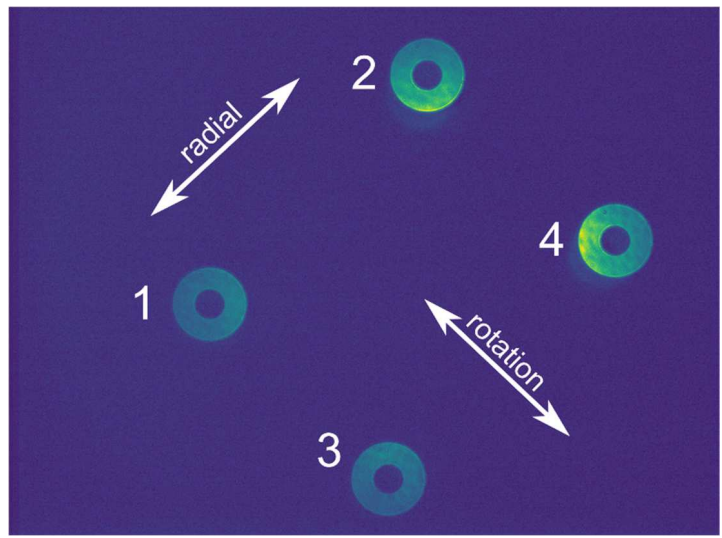

*Figure 217. To calibrate the radial and rotation movement of the mask wheel, a series of four images are taken with known separation.*

The accuracy is the sum squared root of the centroids and wheel accuracy, 10μm radial, 1μm rotation, and 1μm centroid achieving 10 μm. This significant error is due principally to the worst accuracy of the radial movement performed with an Allen wrench. A new tool that fit better into the bolt and equipped with a precise goniometer could reach the desired accuracy of 3μm, we proposed one like in Figure 218.

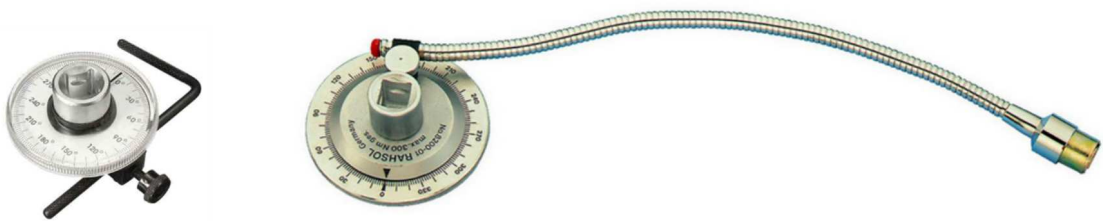

*Figure 218. The proposed torque goniometer to achieve better accuracy in radial movement.*

This accuracy reaches a confidence level only if compared with the sensitivity in defocus and decenter of the mask position in the computation of the final raw contrast of the coronagraph, defining the confidence interval of the contrast.

The z-position of the wheel changes the FP mask position with respect to the PSF, affecting the final contrast of the coronagraph. We applied the same procedure used for the CCD-1 best focal position, and we estimated the related tolerance of a defocus in the mask. The tolerance to focus alignment is calculated by moving the FPM in z (with the micrometer) around the best focus at steps of 10-30μm and acquiring, for each position, an image of the FPM in diffusion and a coronagraphic image. The center of the FPM is used as the pole for computation of the azimuthal average of residual light in the corresponding coronagraphic image. We repeated the test also for a larger amount of defocus, by moving the FPM in z (with the micrometer) around the best focus at steps of 500um, both in intra and extra-focal direction. Figure 219 shows the result of the contrast computation, showing small sensitivity with respect to the defocus position.

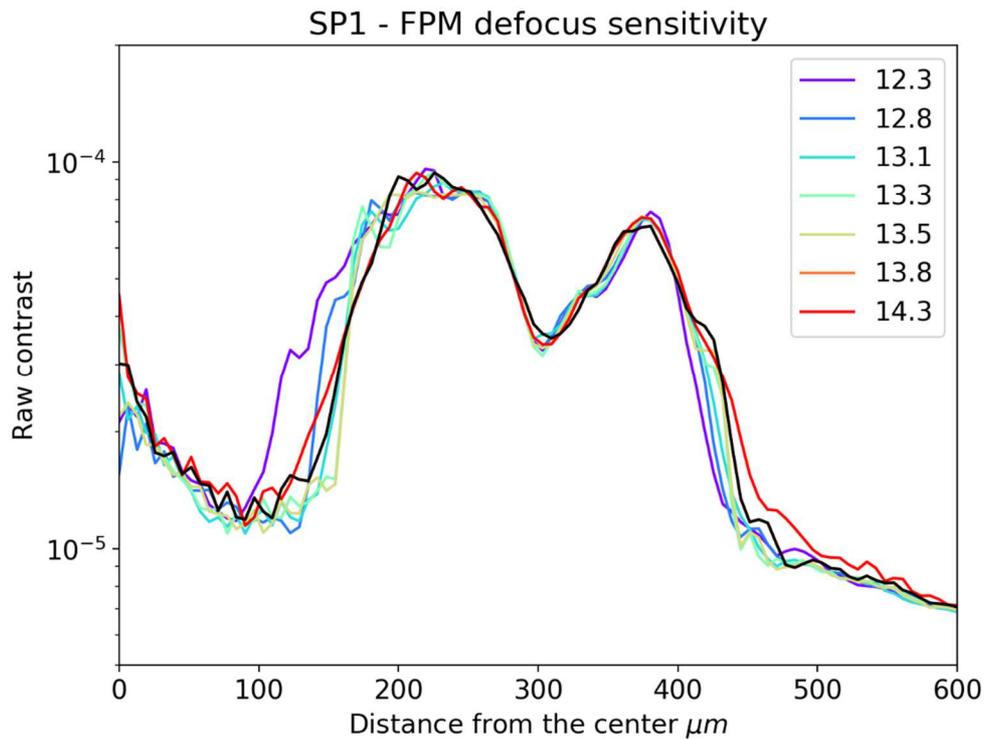

*Figure 219. The contrasts achieved. The best focus is shown in black, while in rainbows color the curves obtained at all the other positions in z, from 12.3 to 14.3mm position with the best focus in the middle.*

We also investigated the sensitivity to movements of the FPM in decentering (x,y). In this test, we moved the FPM from its aligned position (24.13mm) with the micrometer at steps of 10μm in both x and y direction and acquired, for each position, an image of the FPM in diffusion and a coronagraphic image. The formers, as for the defocus test, are needed for contrast calculation. Figure 220 shows the contrasts achieved. We are sensitive for 10 μm decenter. We can also quantify the amount of light passing for different decenter value. This amount slightly similar to the accuracy in centering of the mask, releasing a bit the constraint in the screw rotation, however the new Allen key with goniometer is necessary.

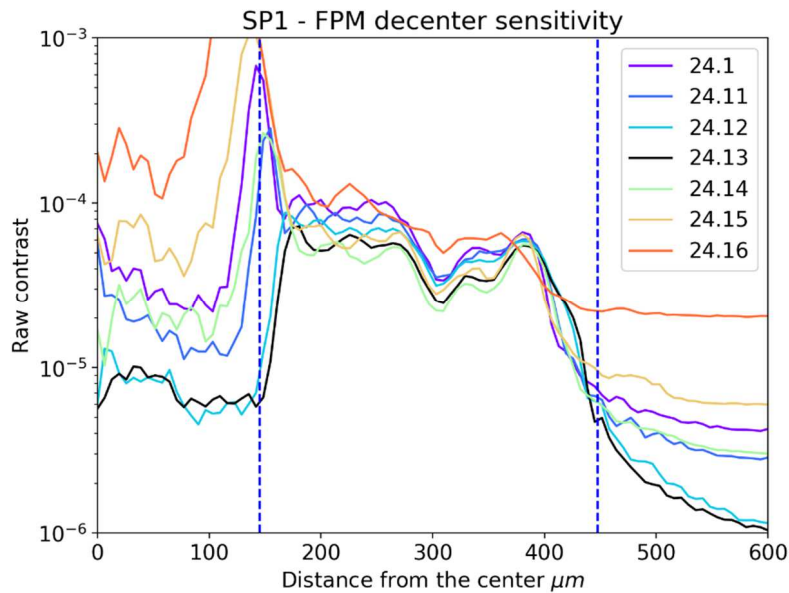

*Figure 220. The contrast curves achieved at different decenter positions from the on-axis position, in black, decenter position in mm. Blue dotted lines are the position of IWA and OWA.*

### 3.5.1.3 Raw Contrast

The scientific yield of a coronagraph depends on performance metrics that are a fundamental part of the design of an instrument. The most widely used metric is the **raw contrast**, defined as the fraction of collected planet and starlight detected in the discovery space. The **total rejection** measures the ratio of the total intensity of the direct image to that of the coronagraphic image. The **peak attenuation** is the ratio of the maximum of the direct image on-axis to that of the coronagraphic image. The **minimum detectable planet-to-star flux ratio** is an observational metric that limits the detection in an observation, related to stellar photon noise and signal to noise ratio for the detection. Coronagraph could be designed to minimize the **limiting flux ratio**, considering the photon noise, the speckle noise due to wavefront stability, and the interaction between the wavefront error and the diffracted starlight, in the un-aberrated case.

With the mechanical and software tools previously described, we defined a procedure to obtain a well-aligned coronagraph and compute the final raw contrast. Following the optical layout of Figure 202, the steps are:

1. First, we defined the position [$x_0$, $y_0$] of the optical axis on the camera by taking an image of the pupil, without L5 and masks. We inserted OD2 and/or OD3 for the correct exposure time on the CCD. The ND filters were inserted orthogonally with respect to the laser axis, by autocollimation of the back-reflected ray from the ND.

2. Insert L5 and find the best position of PSF image in the FPM plane by moving L3 lens, with rotating diffuser and Foucault slit. Capture 100 images to measure the sharpness of the Foucault slit.

3. Remove Foucault slit, insert OD3, and check the PSF diameter (need 32$\mu$m), if it is not ok we repeated the item 2 by slightly moving L3.

4. Moving L5 on a motorized and manual translation stage for centering the PSF in [$x_0$, $y_0$], note the accuracy.

5. Insert the FPM and perform the best focus position of the mask moving the occulters wheel in z-position.

6. Light up with the rotating diffuser the FPM and capture 100 images to check the center position and calculate the magnification factor L4-L5: if necessary centering the FPM in [$x_0$, $y_0$] using the manual stage, note the accuracy.

7. Remove the FPM, remove L5 and OD3

8. Capture 1 image of the pupil, check the center of the pupil [$x_0$, $y_0$].

9. Insert L5, insert OD3

10. Check the PSF of the iris, the center [$x_0$, $y_0$] and the FWHM.

11. Insert the apodizer, remove L5, remove OD3, insert OD2.

12. If the apodizer has defocused, we found the best focus by moving the camera in the z-direction.

13. Remove the apodizer and check the position of the center of the pupil [$x_0$, $y_0$].

14. Insert L5, insert FPM, remove OD2.

15. If we perform a movement of the CCD-1, we moved L5 by the same amount of the z-direction of the camera and re-calculated the magnification factor L4-L5, if the magnification factor is different by the previous one, we repeat the item, 2,3,4.

16. Remove L5, remove FPM.

17. Check the pupil image with the apodizer inserted: centering of the apodizer by rotating the wheel, by radial movement of the apodizer and by moving in the x-direction the stage. Measure the center and note.

18. Insert L5, remove apodizer.

19. Check PSF image to check the center [$x_0$, $y_0$].

20. Insert apodizer.

21. Move FM2 to shift the PSF center to the center of the optical axis [$x_0$, $y_0$].

22. Insert FPM.

23. Check the position of the FPM with respect to the center of the PSF.

24. Now with the optical bench aligned, we acquire the coronagraphic images and compute the raw contrast.

The efficiency of a coronagraph resides in its capability to suppress the incoming light selectively with respect to the position of the source in the field: it has to block on-axis light while transmitting off-axis radiation as much as possible. Raw contrast is the ratio between these two components. Hereafter, we will use the terms "off-axis PSF" and "coronagraphic PSF" to refer to:

- **Off-axis PSF:** the PSF of an off-axis source obtained with the apodizer in the optical path, but no FPM. Here "off-axis" has to be intended inside the discovery region of the coronagraph.

- **Coronagraphic PSF:** the PSF of a source on-axis obtained using the combination apodizer+FPM.

Raw contrast is defined as:

$$C(r) = \frac{I(r)}{\hat{I}_0 \cdot M(r) \cdot t_r}$$

Where:

- $\hat{I}_0$ is the peak intensity of the off-axis PSF.

- $M(r)$ is the FPM intensity transmission curve. In the simplest case of a hard-edge amplitude mask, the case of SP, M is 1 in the transmissive region of the mask and zero elsewhere.

- $t_r$ is the ratio of exposure times (coronagraphic to off-axis)

- $I(r)$ is the intensity in the coronagraphic PSF, computed by taking the azimuthal average of the counts at steps of 6.45μm, the pixel size of the camera. The center used for the calculation is one of the FPM.

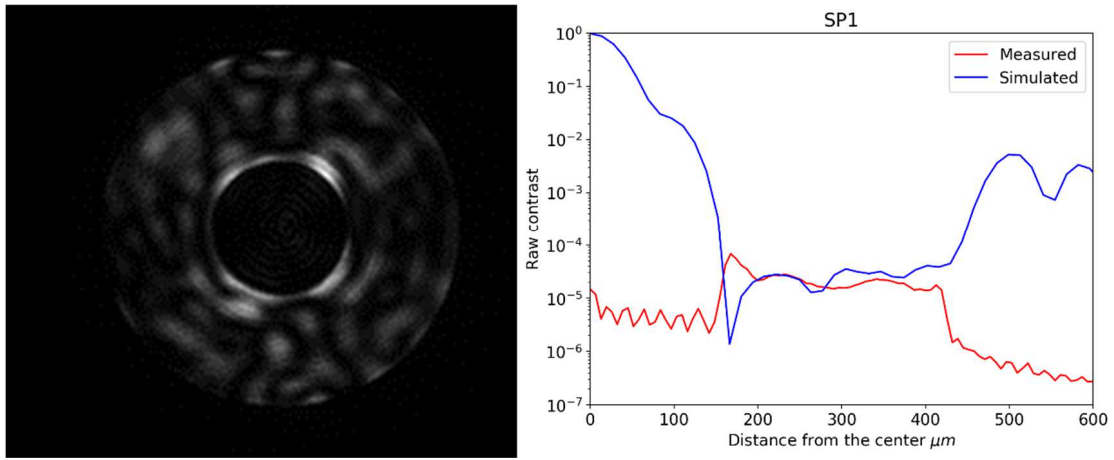

*Figure 221. The SP1_H PSF (on the left) and raw contrast (on the right).*

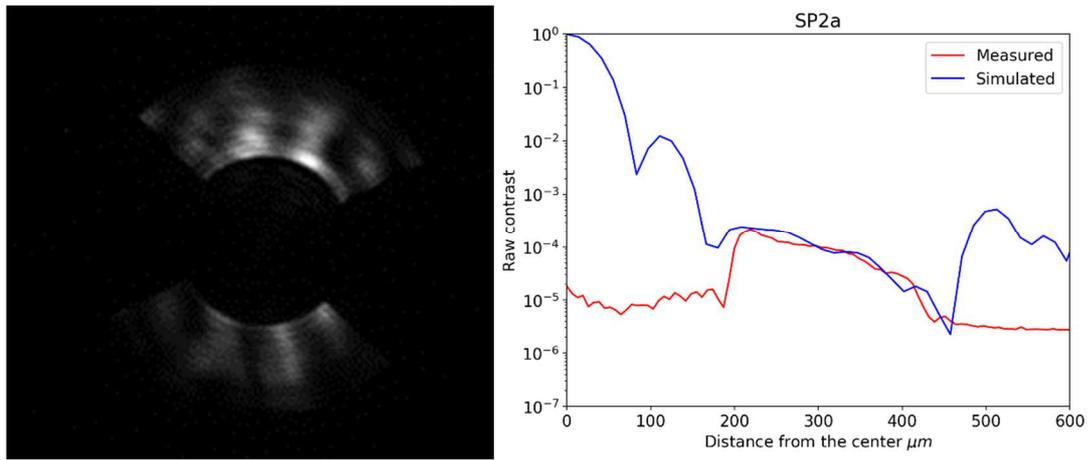

*Figure 222. The SP2a_H PSF (on the left) and raw contrast (on the right).*

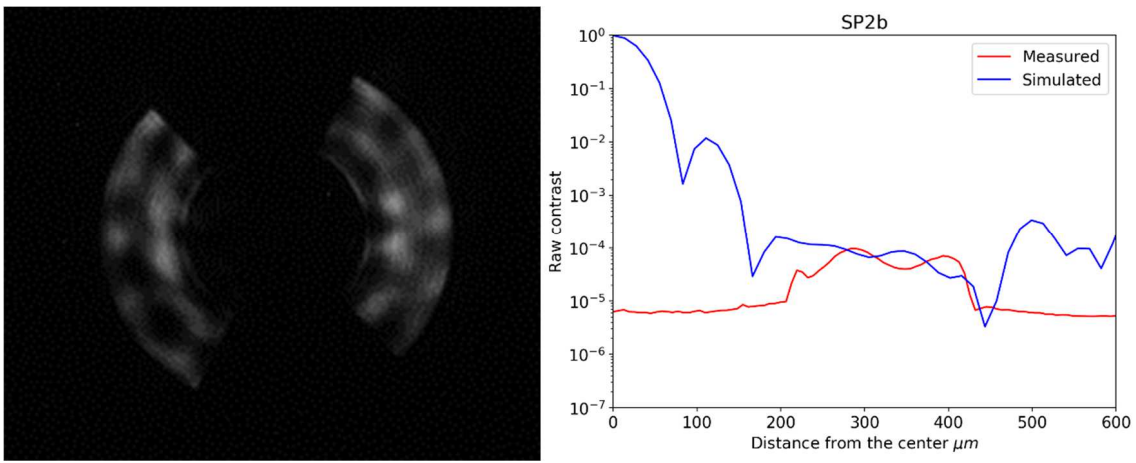

*Figure 223. The SP2b_H PSF (on the left) and raw contrast (on the right).*

Measured raw contrasts for the three SP designs are reported in Figure 221, Figure 222 and Figure 223, where they are compared with simulated ones. For displaying purposes, simulated contrasts are not zero outside the transmissive region of the FPM. Inside the transmissive regions (whose boundaries are reported in Table 4), we found a perfect match between measurements and simulations for all three designs.

For the simulated curve for the SP1 in Figure 221 the level contrast less than $10^{-5}$ near the IWA is a fake effect, the nominal raw contrast for this coronagraph is $2x10^{-5}$. For both the asymmetric SP, the measured raw contrasts are consistent with the simulations obtained in the same conditions of the laboratory test:

- the wavelength of 633nm monochromatic (laser) different from the working SHARK-NIR one (1600nm uncoherent);

- the entrance pupil of the bench setup is realized by an iris, while in the real case of SHARK-NIR the entrance pupil is a reimaging from the LBT telescope.

Because of these two effects the incoming electric field generates is different with respect to the one used to design the apodizer patterns.

For the SP2b_FPM_H, it seems that the IWA is larger than the nominal one by a not negligible amount, after some investigation, we reported in Table 51 the IWA and OWA values for the whole SP masks.

*Table 51. The IWA and OWA values for the SP coronagraphic masks.*

| Coronagraphic mask technique | IWA [μm] | | OWA [μm] | |
|---|---|---|---|---|
| | theoretical | measured | theoretical | measured |
| SP1_FPM_H | 196 | 198 | 528 | 521 |
| SP2a_FPM_H | 262 | 266 | 528 | 521 |
| SP2b_FPM_H | 247 | 250 | 528 | 521 |

### 3.5.1.4 Fake planet

The simulation of a fake planet in the discovery space is an alternative test to qualify the raw contrast of the coronagraph. Figure 224 and Figure 225 show the planet simulator: is composed of a Thorlabs lamp (SSL201L/M) with a narrow band He-Ne filter positioned inside. The light is injected into the optical path by means of a Pellicle Beam Splitter (PBS) and two optical fibers in cascade:

- M28L01 Multi-Mode, Ø400 μm, 400 to 2200 nm, 0.39 NA

- P1-630A-FC-2 Single Mode, 633 - 780 nm, 3.6 - 5.3 μm @633 nm

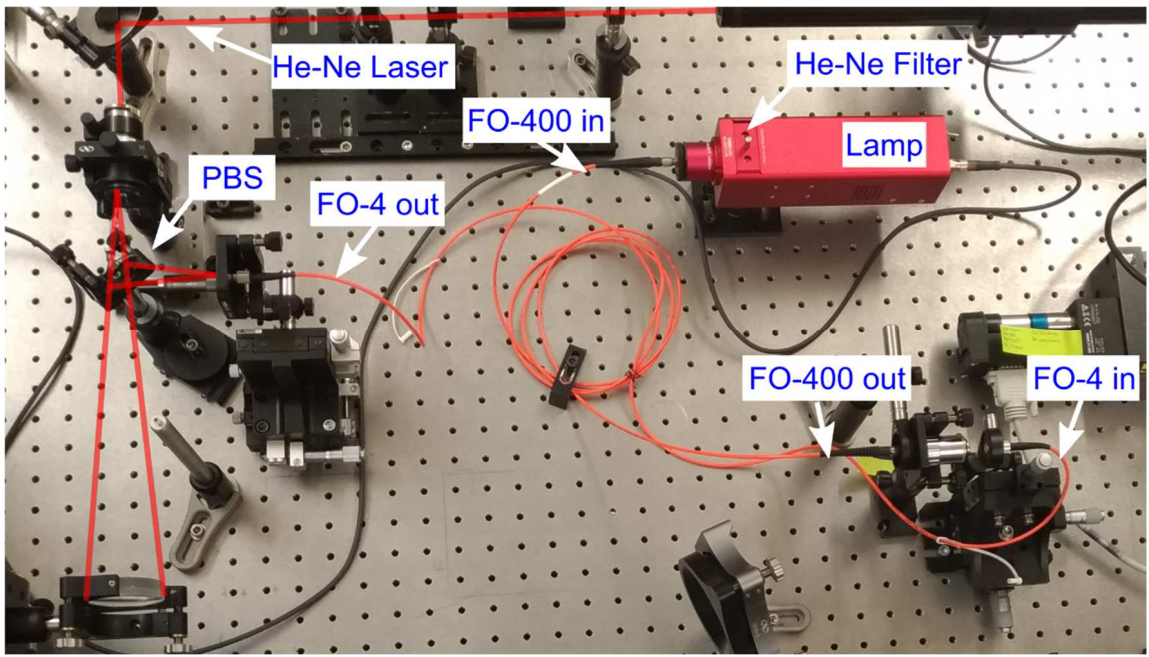

*Figure 224. The fake planet simulator, generated by a fiber optics.*

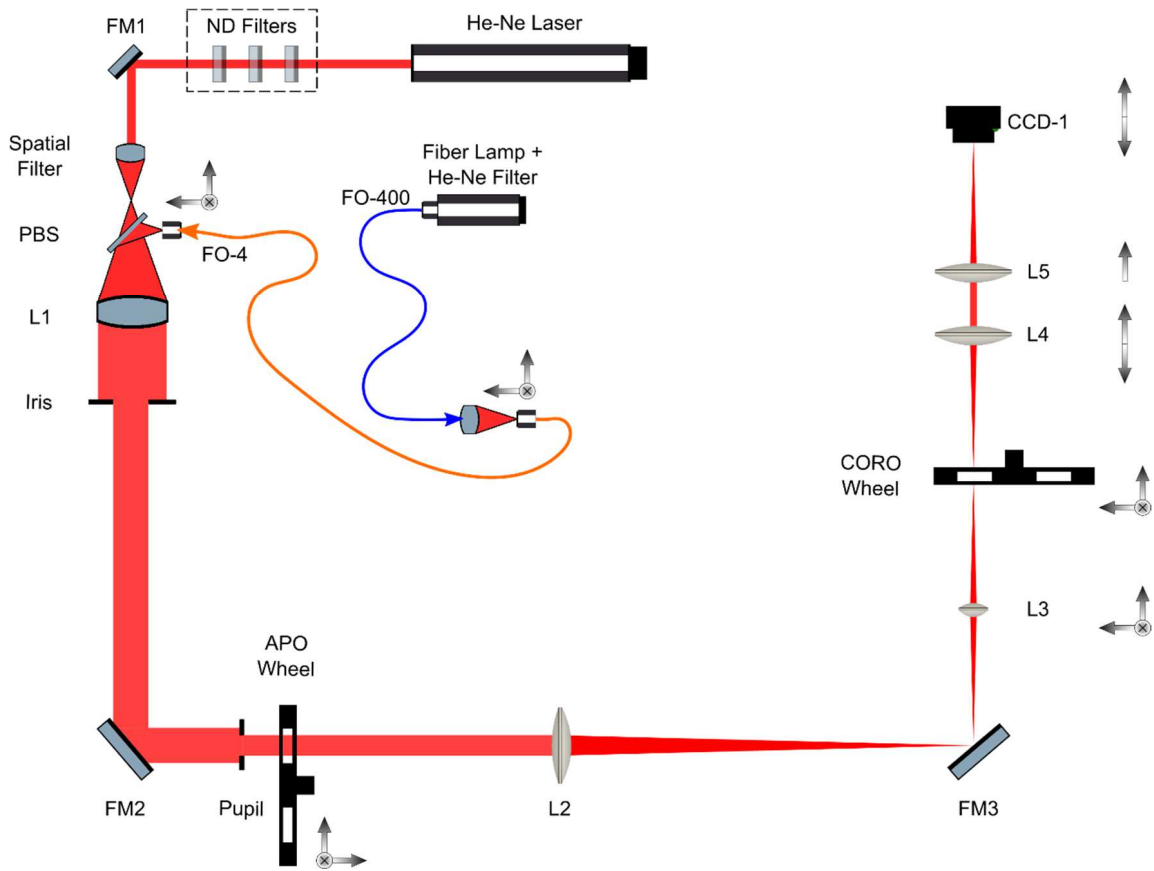

*Figure 225. The optical bench to simulate a planet near the IWA.*

The first fiber with a large core was used to collect as much light as possible, and the output was coupled with an objective lens with 40X power. The input of the second fiber was mounted in an x-y-z translation stage to reach with high accuracy the focal position of the objective lens. The alignment was performed measuring with a power meter + photodiode the maximum intensity at the fiber output.

The He-Ne laser beam passes through a spatial filter and is collimated by a second lens. Positioning a mirror after the beam expander in autocollimation, the back-reflected light is deviated by the PBS in the direction of the fiber optic. Here too the fiber is aligned and focused with an x-y-z translation stage, and the focus is reached observing the maximum flux in a power meter. We observed, in CCD-1, the two PSF in the focal plane of the coronagraph, one from the laser and one from the lamp. The same wavelength guarantees the absence of chromatism and focus dragging. The core of the FO-4 has the same dimension of the pin-hole of the spatial filter in the beam expander, generating a second unresolved source.

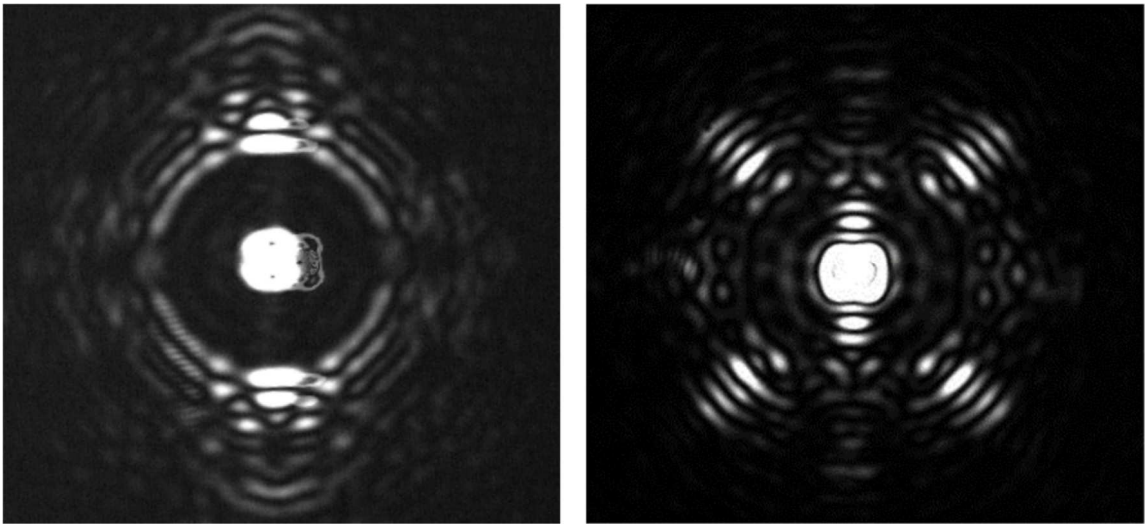

*Figure 226. On the left: PSF of the apodizer SP1. On the right: PSF of the apodizer SP2. The discovery space is the dark inner region, and the excess of the PSF light is diffused outside the OWA.*

Inserting the apodizer SP1 the discovery space appears around the principal PSF, see Figure 226, and we dislocated the PSF of the fake planet close to the IWA, acting on the translation stage of the fiber optic output FO4-out. Then we removed the SP1, and we measured the EE of the two PSF until they reach the same value, inserting OD2 and OD3 in the laser path and slightly decentering the FO-4 input. At this point, we boosted the exposure time of the CCD to see as best as possible the Airy's rings. The intensity of each ring is well known, and we decentered the FO-4 input with respect the objective lens axis to realize the same intensity of the 8th ring, corresponding to a contrast of $10^{-4}$, see Figure 227 left. We inserted the apodizer SP1 and the corresponding focal plane mask obtaining the high contrast image of the fake planet near the IWA, with a Signal to Noise Ratio of 36 with a single image, see Figure 227 right, confirming the previous results of Figure 221.

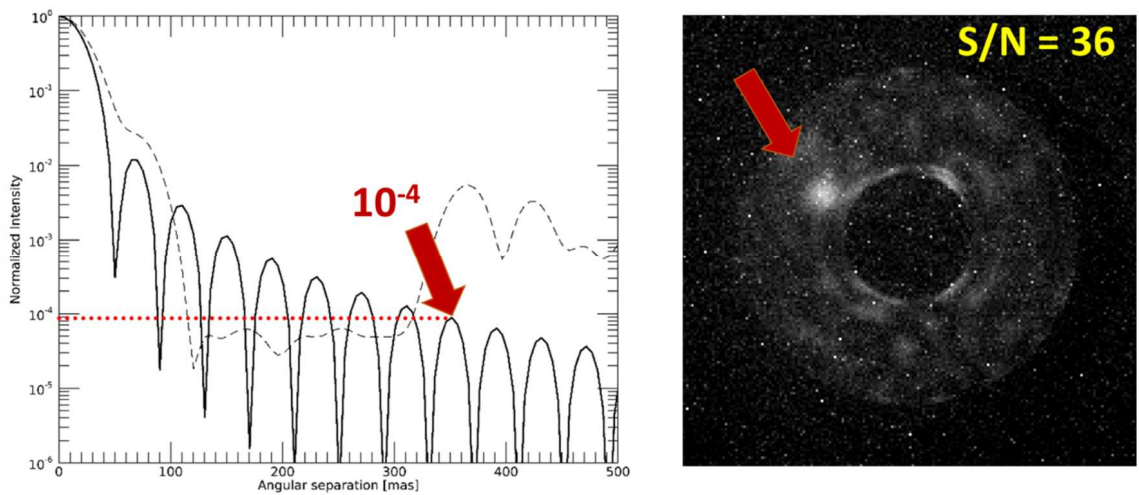

*Figure 227. On the left PSF intensity to the 8th Airy annulus, with superimposed the contrast measured for SP1. On the right: the fake planet a $10^{-4}$ in contrast near the IWA.*

## 3.5.2   Four Quadrant alignment

We defined the alignment procedure of the coronagraph both for the phase mask in the FP and the LS in the downstream pupil plane and measured its performance in terms of raw contrast and rejection. SHARK-NIR FQPM is monochromatic and its working wavelength is 1.6um. Looking at the ratio of the PSF peak intensity with and without the coronagraph (i.e. the attenuation) as a function of wavelength for such a phase mask, we see a local minimum at a wavelength around 550nm, see Figure 201 on the right. For this reason, we performed the procedure in visible light, using the lamp coupled with a filter and optical fiber. Figure 228 shows the optical bench for the tests; it is arranged with the same components used for the SP tests, but we use a stabilized tungsten-halogen Light Source coupled with the Thorlabs FB550-40 filter with central lambda at 550nm and FWHM equal to 40nm, see Figure 230. The FB550-40 filter has a spectral resolution R=$\lambda/\Delta\lambda$ ~14, different from the tests realized at LESIA (R~100). Moreover, the detrimental effect of the spectral bandwidth is magnified at 550nm with respect to 1600nm because of the stronger variations in the glass refractive index at optical wavelengths. The typical FQPM  rejection is proportional to $R^2$ (Riaud et al., 2001). For these reasons, just because of the operating wavelength difference, we expected to measure rejections a factor 50-100 smaller than the ones measured at LESIA laboratories.

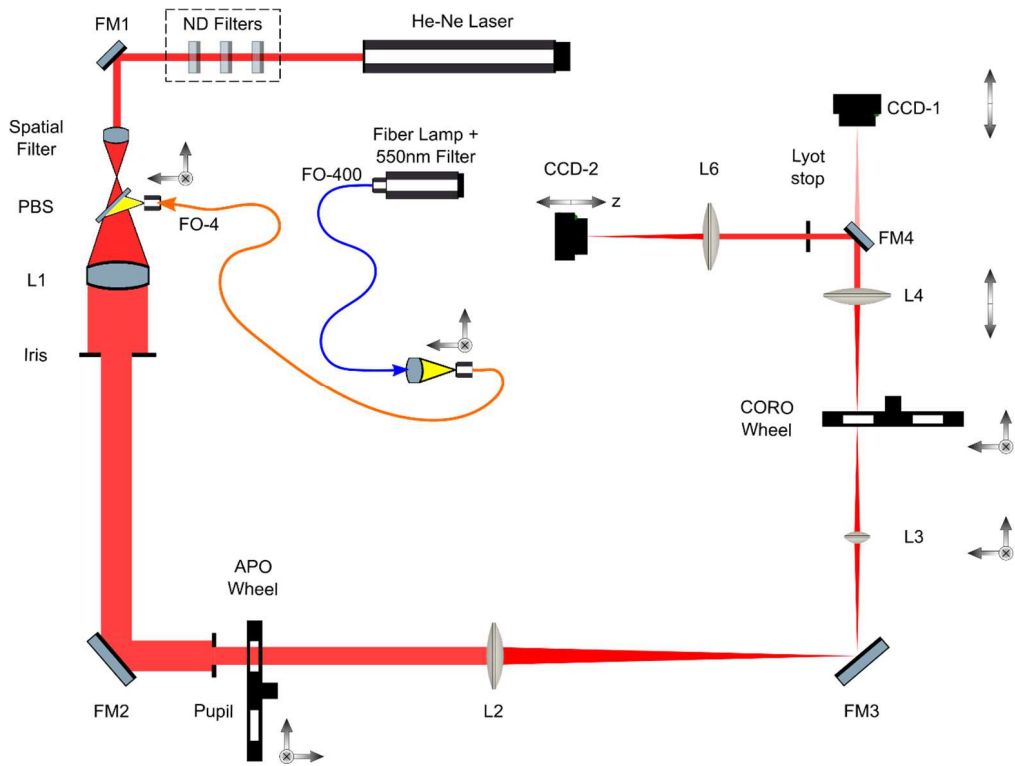

*Figure 228. The optical design of the bench for testing the FQPM. After L4 the folding mirror FM4 is inserted to reflect the light in the LS arm. The CCD-2 is on al linear stage to register an image of the pupil or the PSF.*

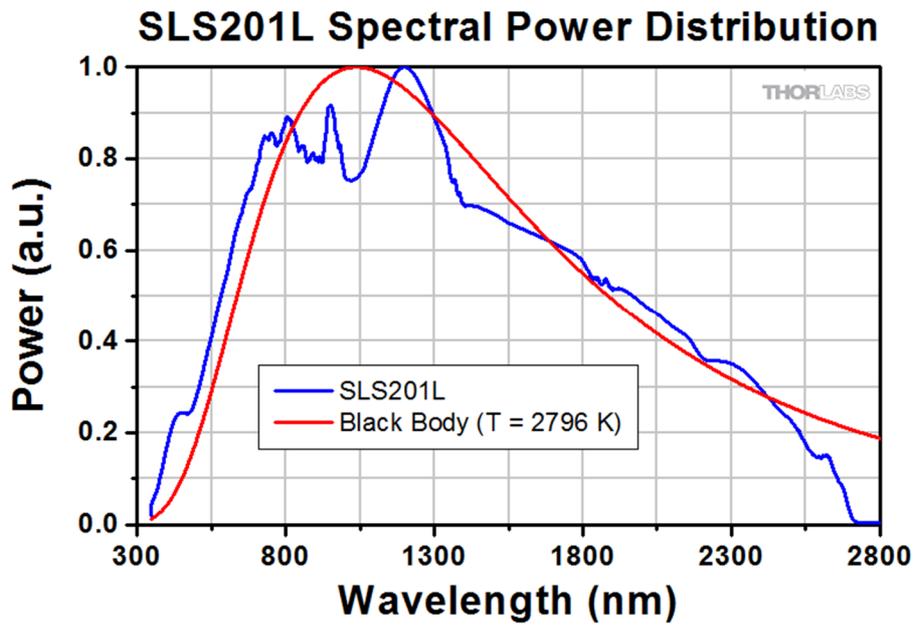

*Figure 229. The spectral power distribution of the selected light sources for SHARK-NIR, as from Thorlabs website.*

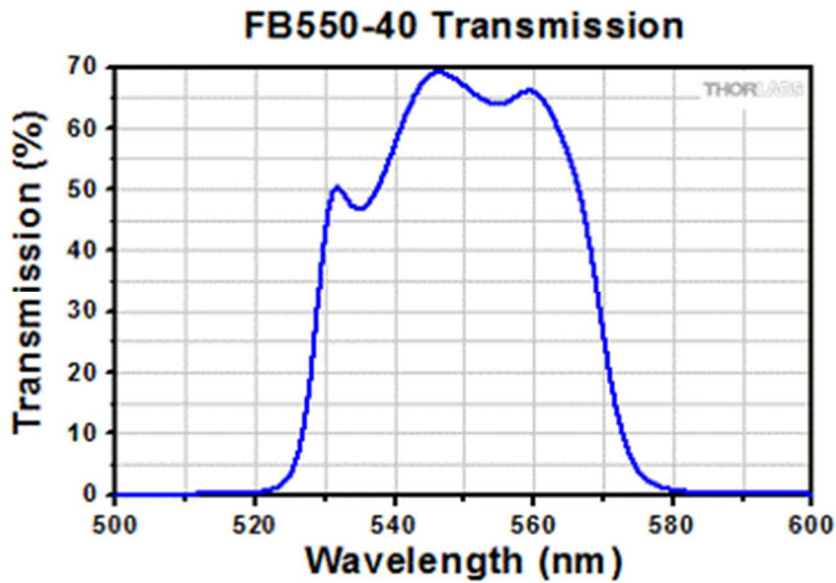

*Figure 230. The transmission curve of the filter used for the tests, catalog Thorlabs.*

We started from the optical set-up used for SP tests with the planet simulation, but in the lamp (SSL201L/M) we inserted the FB550-40 filter, and the optical fiber was re-aligned to ensure maximum transmission of light. The optical layout is shown in Figure 231.

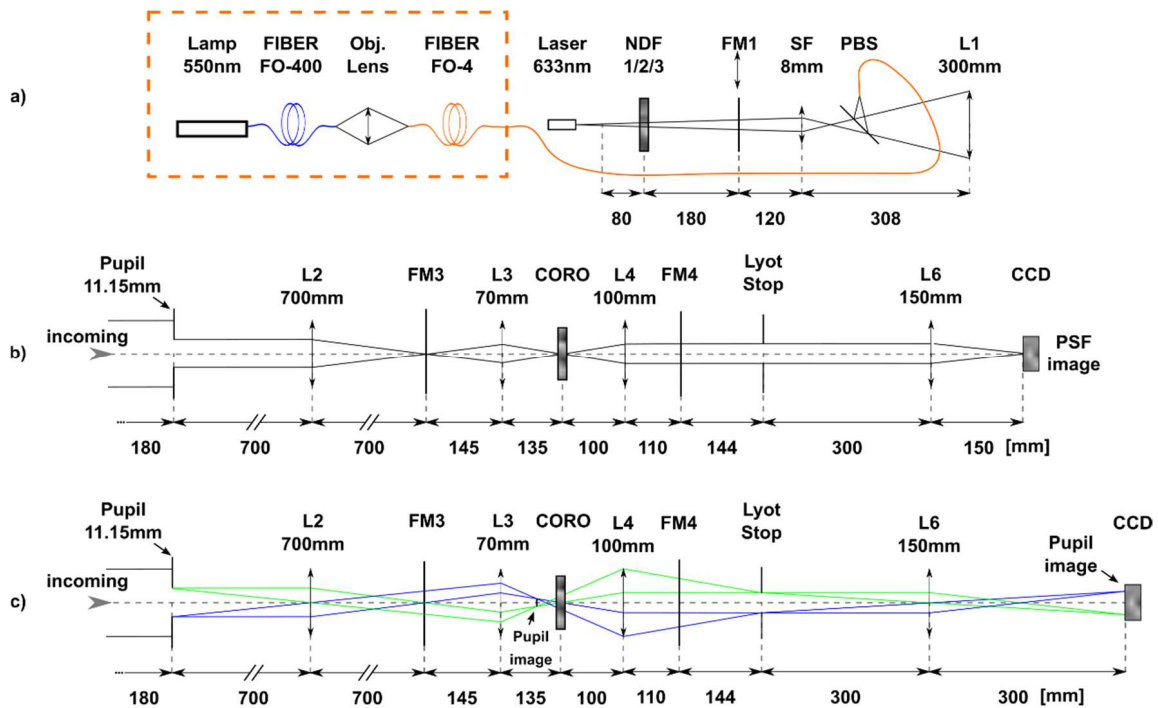

*Figure 231. The optical layout of the FQPM test bench. a) is the illuminating system with two optical fibers coaligned with the laser, b) is the path to generating the PSF image on the CCD-2, c) the path for the acquisition of the pupil image.*

The change in wavelength imposed to reset the z-position of L3, CORO, and L4. This process has been made following the same procedure described in section 3.5.2.

The LS is mandatory for this coronagraph to work, being the component that physically blocks the light. In fact, the upstream phase mask just diffracts it outside the pupil. In order to materialize a pupil plane after the phase mask, we introduced a deployable mirror (FM4) which folds the light towards the center of the bench, Figure 231 c). The beam folds after being collimated by L4 so that no additional lens is needed to form the pupil. After the pupil, a lens (L6) has been placed in front of a camera Prosilica GT-3300, hereafter CCD-2. Depending on the position of CCD2, L6 creates the image of the source or the image of the pupil. By removing FM4, it is possible to take images of the phase mask at high resolution using CCD-1, with the same setup of the SP, see section 3.5.2.

Normally, the LS is slightly undersized with respect to the pupil in order to increase the coronagraph rejection. For our experiment, we used an iris of diameter 20% smaller than the pupil one. In order to derive the correct stop size, we preliminarily positioned CCD-2 in the LS plane and acquired an image of the pupil, see Figure 232. We then used the well defined Python code to measure its diameter. Afterward, with CCD-2 in pupil imaging position (almost 1:1 magnification with respect to LS plane), we optically measured the dimension of the stop and verified that its diameter was 80% of the pupil one.

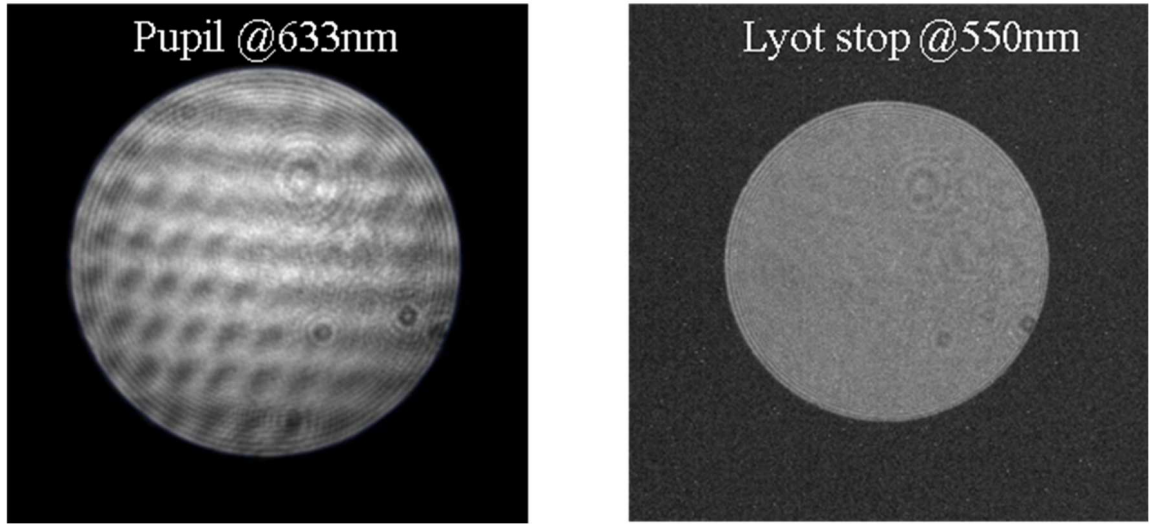

*Figure 232. Representation of the pupils without (on the left) and with (on the right) the LS.*

In the following, I reported the coronagraph alignment procedure:

- First of all, we had to re-align the arm (L3+L5+CCD1), since we worked at a different wavelength with respect to the tests with the SP masks. We did it in three steps:

    o First, we moved CCD1 in z to re-conjugate it to the pupil plane. We looked at the image of the pupil (without L5) cut with a piece of paper to guide the eye. We performed fine-tuning with the LoG method.

    o Then we inserted L5 and the FQPM. We moved L5 in z-direction until we obtained a sharp image of the transitions between the mask quadrants on CCD-1. A Python code computed the width of the discontinuity and performed the best fit to the minimum size versus the sweep position. In this way, we defined the coronagraphic FP.

    o Then we moved L3 in the z-direction to find the best focus of the PSF on the focal plane by re-imaged it on CCD-1 with L4+L5. We used the same code to fit the best focal position of the PSF.

- We co-aligned the laser light and the fiber spot passing through the BE by observing the PSF on the camera. Of course, the spot dimensions weren't the same, because of the different wavelengths, but we were just interested in the decentering of the two sources.

- We inserted the FQPM and looked at the PSF on CCD-1. We moved the FP wheel, by acting on the micrometers on its mount to center the mask on the PSF. We considered the mask well centered at the time we saw the shadows of the quadrants cutting the PSF in four parts almost symmetrical. A refined procedure, based on the measurement of the center of the FQPM cross, has been developed afterward, and it is described in section 3.5.3.1.

- We removed the FQPM and inserted FM4 to fold the light towards the CCD-2 arm.

- We moved CCD-2 on its linear stage to conjugate it to the pupil plane by utilizing a single lens (L6) to relay 1:1 magnification. We calculated the coordinates of the pupil center on CCD-2 and its diameter with an ad-hoc Python code. The measured diameter was 1.916mm @633nm; accordingly, we made a conic iris of 1.533mm diameter;

- We inserted the LS and adjusted its position in z to focus it on the camera. We acquired images of the LS and calculated the coordinates of its center to properly align it in (x, y) to the reference position set by the pupil. We measured for the LS a diameter of 1.498mm @633nm and 1.493mm @550nm, corresponding to around 78% of the iris diameter, near 80% as desired, see Figure 232.

- Without moving CCD-2 from its position, we inserted the FQPM. The characteristic diffraction pattern of the FQPM appeared on the camera, it was composed of 4 bright corners and an almost dark pupil (the coronagraphic pupil, hereafter, Figure 235).

- We switched CCD-2 to the PSF position. This position in z-direction has been found by fitting with a parabola.

- We took images of the coronagraphic (on-axis) PSF, and then we off-centered the spot by rotating the FQPM wheel and acquired some off-axis PSF, necessary to compute raw contrast.

I report in Figure 233 some images from the Python pipeline showing pupil and PSF center determination.

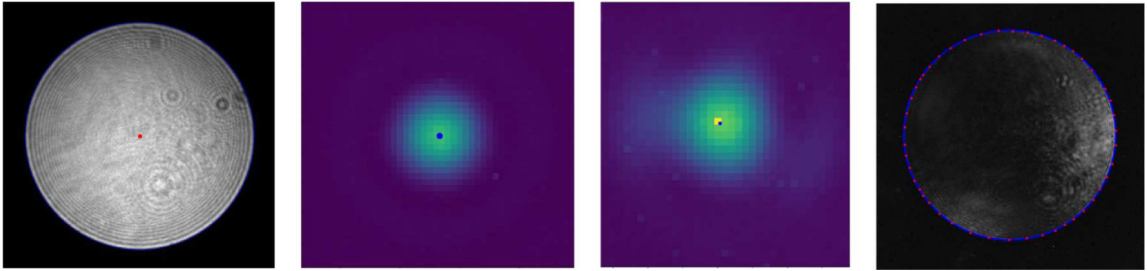

*Figure 233. From left to right: calculation of the pupil center with isophotal fitting; calculation of non-coronagraphic PSF center with Gaussian fitting; calculation of coronagraphic PSF center with Gaussian fitting; fitting of the pupil with interpolation of several points manually selected on the image.*

### 3.5.2.1 Results

As for the SP, the raw contrast is defined as:

$$C(r) = \frac{I(r)}{\hat{I}_0 \cdot M(r) \cdot t_r}$$

Where:

- $\hat{I}_0$ is the peak intensity of the off-axis PSF.

- $M(r)$ is the FPM intensity transmission curve. In the following, is it assumed that the FQPM transmits 100% of the light of any off-axis source (M=1 ∀r). This actually holds true only if the off-axis source is not too close to the center (say further than ~1.5 λ/D) and, of course, if the source does not fall on one of the transitions.

- $t_r$ is the ratio of exposure times (coronagraphic to off-axis)

- $I(r)$ is the intensity in the coronagraphic PSF, computed by taking the azimuthal average of the counts at steps of 5.5μm, the pixel size of the camera GT3300. The center used for the calculation is one of the FPM.

To measure the λ/D scale in the PSF off-axis image on CCD-2, we implemented the fitting model `AiryDisk2D` in Astropy (a Python library), which is able to extract the radius of the first zero of the PSF, and the λ/D value. We compared this result with the interpolated radial profile of the PSF to obtain the positions of the local minima. We applied a linear regression of these measurements in pixel versus their theoretical values in λ/D, and we obtained 1 λ/D = (9.76 +/- 0.04) pix. Figure 234 shows the residual image of the Airy 2D fitting with a 3% amount of the initial level range (PtV) and the interpolation of the radial profile of the original (green) and modeled (red) data. The differences between the models are due to high order aberrations in the system, but the first minimum is well fitted.

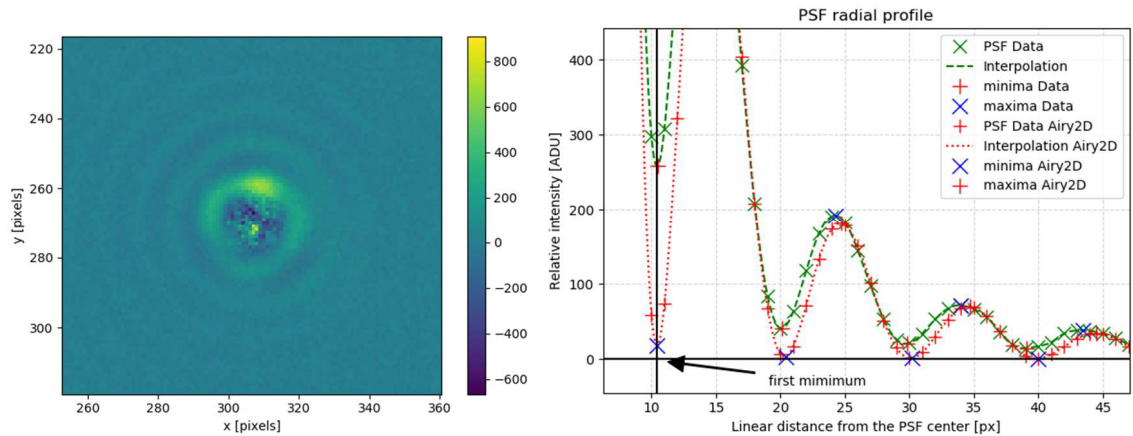

*Figure 234. On the left: the residual image of the PSF Airy fitting. On the right: the interpolated radial profile of original and modeled data.*

One more metric commonly used with FQPM is the total rejection. It is calculated by considering the integrated flux in the image plane with no coronagraph, neither phase mask or Lyot stop, and dividing it for the same quantity measured on the coronagraphic image. Usually, a square area of ~10-20 $\lambda$/D centered on the star is used in order not to include too much noise. It has been shown that the rejection $\tau$ is proportional to the square of the spectral resolution R=$\lambda$/$\Delta\lambda$: $\tau \propto R^2$.

Figure 235 shows the coronagraph pupil and PSF with two different sources, in the top with a resolved fiber optic, and bottom with unresolved one, simulating worst and best case of performance in AO. The improvement in light rejection from the pupil is visible.

Pupil with FQ                      Preliminary PSF

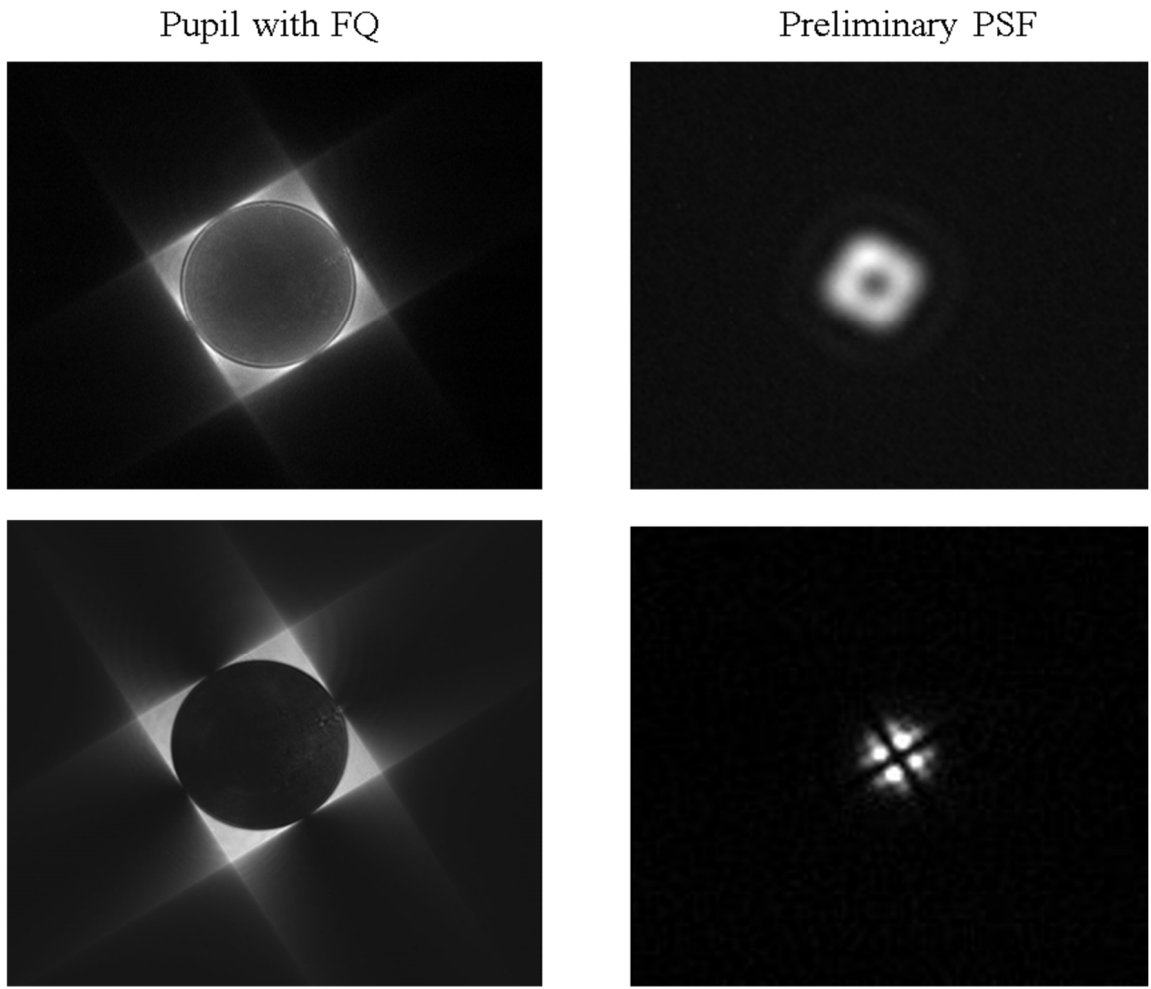

*Figure 235. On the top: the image of the coronagraphic pupil (left) and PSF (right) before fine alignment. For that preliminary test, we used a fiber optics of 25μm core (resolved source). On the bottom: the best pupil and PSF after alignment. The coronagraph was illuminated with an unresolved source (a fiber of 5μm core). The PSF shown here is the one acquired with the Pike camera (CCD-1) and used for the alignment of the phase mask. The images are taken without the LS.*

In Figure 236 and Figure 237 is shown, respectively, the contrast curves calculated at LESIA and the one obtained during this test. There is almost one order of magnitude difference between the two, which is totally expected due to the non-optimal working wavelength.

In the LESIA test, the parameters were R~100 at $\lambda$=1591nm and resulted $\tau$ ~ 400. We expected for the rejection of the coronagraph in ratio with LESIA, with R~ 14 and $\lambda$=550nm, $\tau$ ~ 8, see section 3.5.2. The test value has been calculated on a square window of size 25 $\lambda$/D and results to be $\tau \approx$ 40, 5 times more the expectation and only a factor ~10 smaller than what measured at LESIA. This number is in agreement with the contrast that we expect to have. The effect of the wavelength has thus been partially compensated by some other effects, such a possible explanation is that we used a much smaller pupil with respect to LESIA tests (11mm vs 30mm). That might have helped in mitigating the impact of system aberrations and bench turbulence.

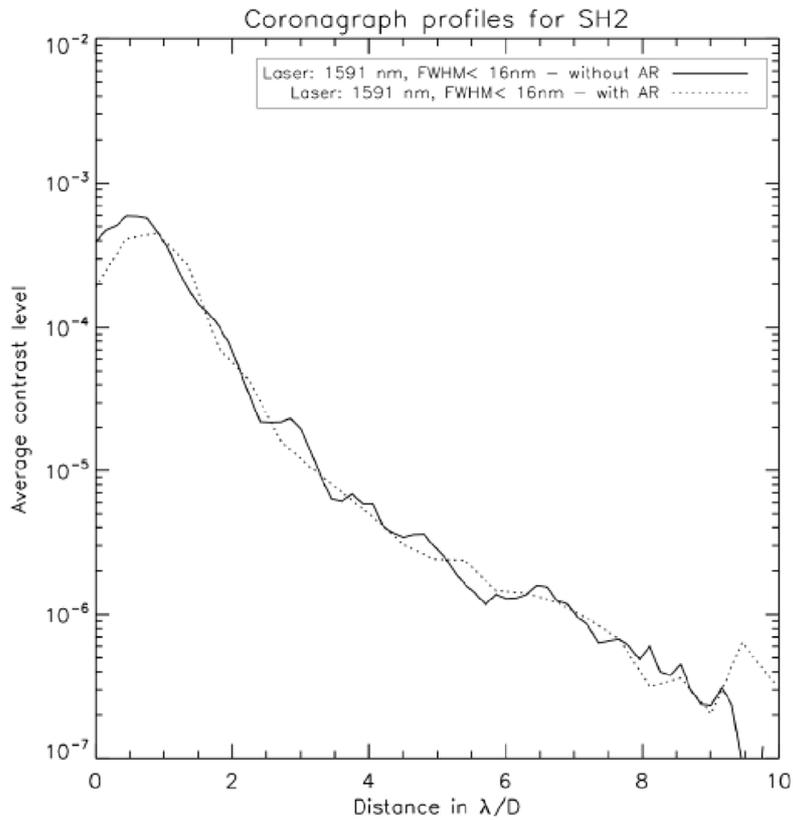

*Figure 236. The contrast level for the FQPM obtained at LESIA with a laser source @1591nm.*

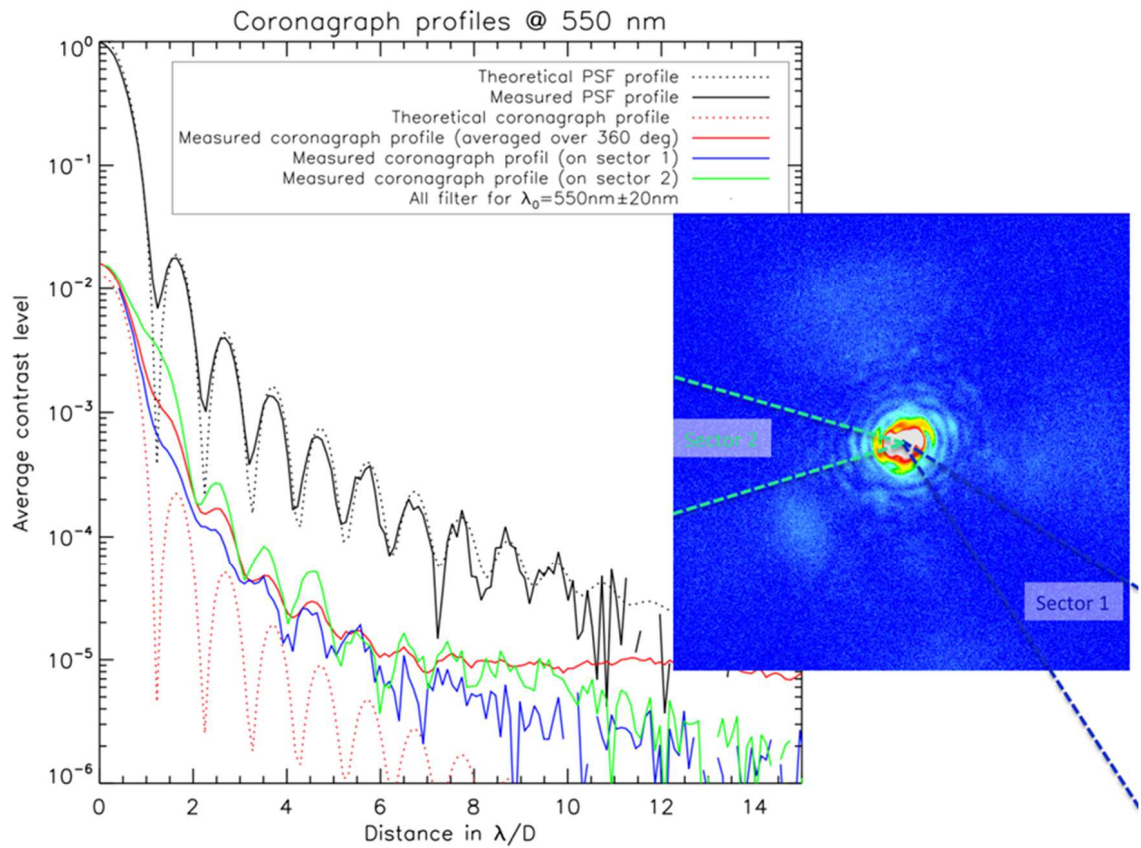

*Figure 237. The FQPM achieved contrast compared to the theoretical one. After the alignment tests, the contrast curves were computed into two dark sections 1 and 2, and by averaging over 360 degrees. Computation courtesy LESIA.*

## 3.5.3   Error budget of alignment

The error budget was computed considering the contributes of the center determination of FQPM and PSF, and the minimum radial and azimuthal movement of the wheel.

### 3.5.3.1 FQPM center determination

The center of the FQPM is determined with a semi-automatic procedure which is able to identify the transition between the quadrants, in order to have an image of the mask with diffused illumination of the FP. In Figure 238 shows the efficiency of the fast rotating diffuser in illuminating the whole surface of the FQPM a revealing the transitions. The transitions are unresolved, and their profiles appear sharp on-axis and larger off-axis because of field aberrations. However, in both cases, the extracted profile across the transition shows a clear drop in transmission, see Figure 238 on the right.

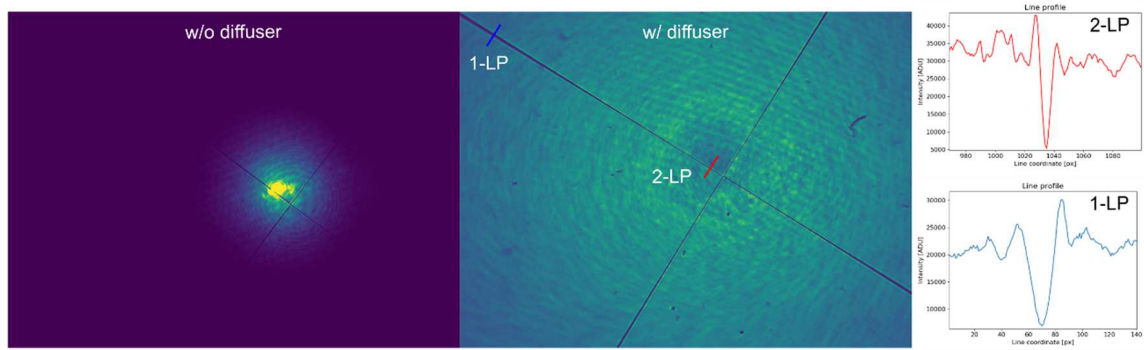

*Figure 238. We stressed the large utility of the fast rotating diffuser to measure with high precision the center of the FQPM. 1-LP and 2-LP represent the linear profile along the segments.*

For every line of the image, has been selected the minimum values of the horizontal line profile which define a series of points. All these points have been fitted with linear regression, in order to trace one line for each quadrant transition. The intersection of the four lines defines the center of the FQPM. The fitted lines didn't intersect in a single point, but defined a square, see Figure 239. This is due to the intrinsic misalignment between the quadrants: the pixel size of CCD-1 was small enough, to appreciate it. We defined the maks center as the geometrical center of the square.

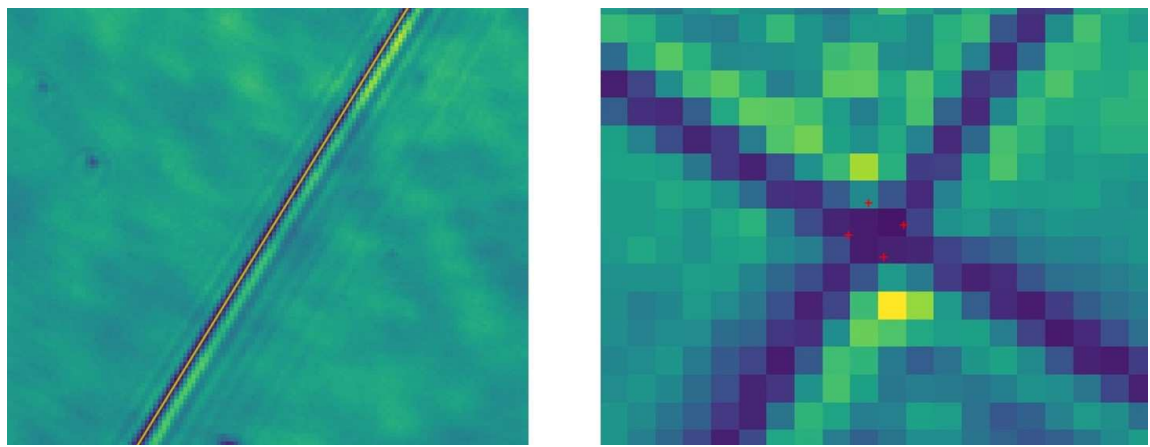

*Figure 239. On the left: a detail of the transitions fitting. On the right: zoom of the center of the mask. Red crosses indicate the points of intersection of the four lines. The misalignment of the quadrants is visible.*

Since the CCD-1 see the FP by the L4 and L5 re-imaging system, we measured the precise magnification factor inserting a SP with well know physical dimension. The measured magnification of L4+L5 was 1.64x, little more than the theoric 1.5x. The error in the FPM center determination stems from the goodness of fit of the transitions. If the misalignment of the quadrants is appreciable, as in our case, then the center of the FPM can be determined with accuracy as small as +/- 0.03 pixels. This translates into 0.1µm in the FQPM plane, considering that the measured magnification of L4+L5 is 1.64.

### 3.5.3.2 PSF center determination

To calculate the standard error on the PSF center we started with the preliminary determination of the center with a Gaussian2D fit. Then we passed this coordinates as a parameter of a non-linear least-square fit (LMTIF library of Python) that computed the center of the PSF and its error, corresponding to 0.02 μm in the FP, is negligible.

### 3.5.3.3 Minimum radial movement

The radial movement of the mask is controlled by using the micrometric wheel screw. Dedicated tests led to an estimated sensitivity of 10μm in radial centering. This is quite a conservative value. Measures of the Δρ between FPM and PSF is a way to understand which is the proper rotation value of the screw. The tests that have been made can demonstrate that we are able to match the required rotation with a precision of few degrees. In our consideration, 5° is a reasonable estimate of the error on the rotation of the screw. It corresponded to 5μm in the FQPM plane.

### 3.5.3.4 Minimum azimuthal movement

The movement is motorized, therefore it is more precise with respect to the radial one. The minimum rotation step of the motor translates into an azimuthal step of 24μrad, which is 1μm because the radius of the CORO wheel is 43mm.

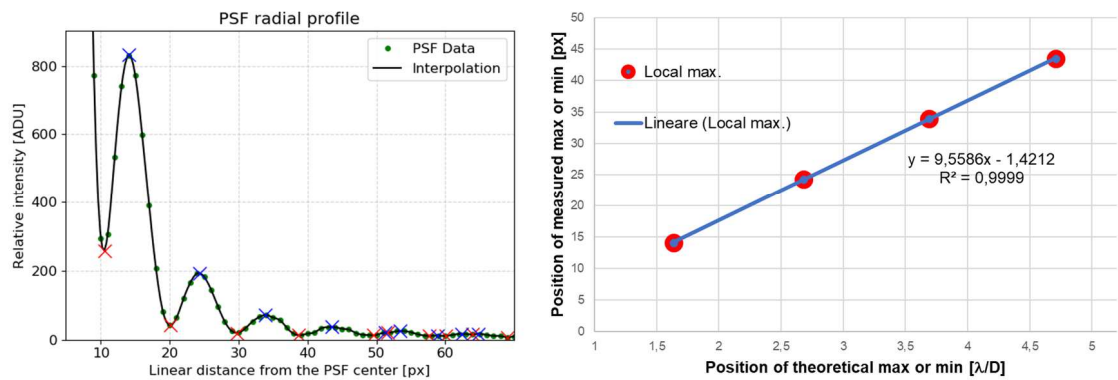

*Figure 240. On the left: the radial profile of the off-axis PSF. The blue crosses indicate the positions of the maxima. On the right: the linear fit of the measured maxima compared with the theoretical ones.*

### 3.5.3.5 Error budget summary

To have an estimate of the quality in the alignment, we simulated the misalignments of the focal plane masks. We analyzed a guide star mag$_R$=8 and seeing is 0.4" (high-Strehl regime), the result is plotted in Figure 241. The misalignment is a lateral shift of the mask with respect to the star. This shift is assumed to remain constant throughout the ADI observation. With a 4 μm misalignment, in the worst case, the detection limit decreases by 0.3 magnitudes.

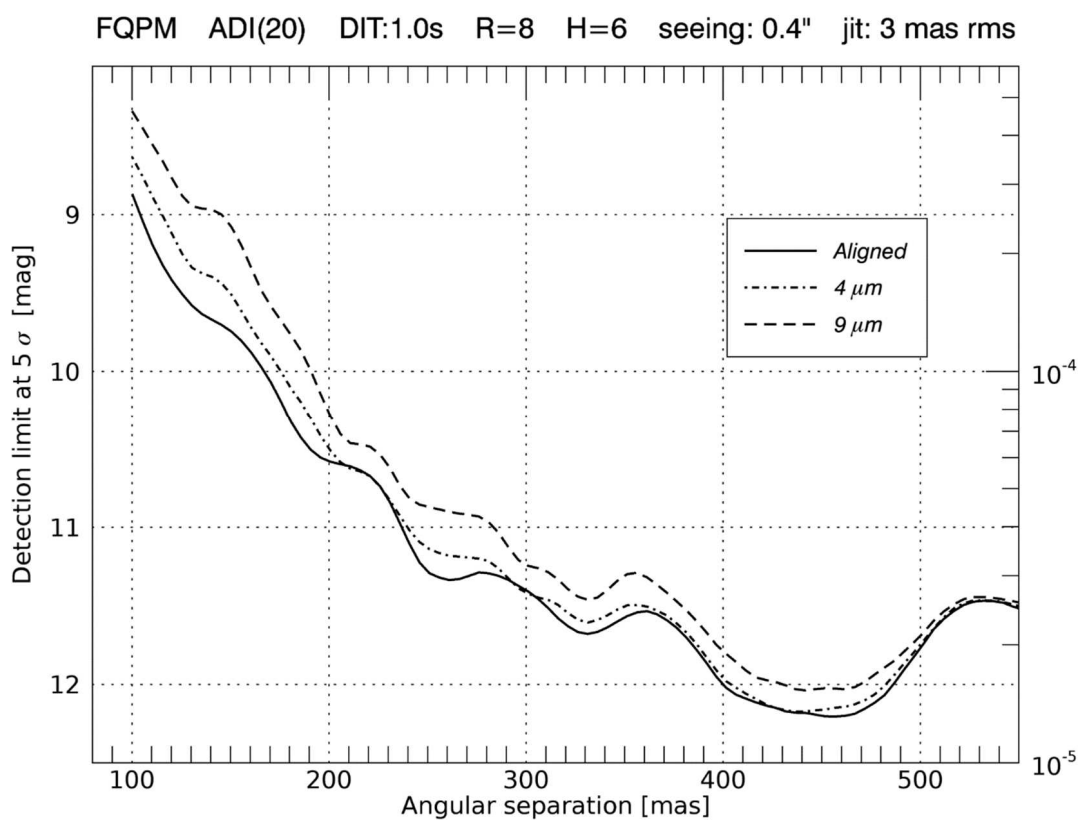

*Figure 241. The impact of misalignments of the FQPM, for a guide star of mag$_R$=8 and seeing of 0.4" with FLAO. Jitter is 3 mas rms. We considered 20 simulated images and a 40° of FoV rotation.*

The total error budget is 5 μm, as summarized in Table 52, which is compatible with a decrease in the detection limit of 0.3 magnitudes in the worst case.

*Table 52. The amount of the error budget.*

| Determination | Uncertainties [±μm] | Method |
|---|---|---|
| FQPM center | 0.1 | With fit of the mask transitions |
| PSF center | 0.02 | Gaussian LMFIT fit |
| Accuracy in radial centering | 5 | By hand (conservative value) |
| Accuracy in azimuthal centering | 1 | By piezo |
| CORO wheel repeatability | 0.6 | By piezo |
| **Total** | **5.1** | |

## 3.6 Non-Common Path Aberrations

In the design of AO systems, a part of the optical path is not corrected by the WFS. This path includes the optics that are after the BS separating the AO channel from the instrument channel. In the discovery and characterization of exoplanets with coronagraphy, the planet signal could be mistaken with the speckles background, the residual signal generated by NCPA. The final strategy adopted in SHARK-NIR will depend on the real optical stability, that will be characterized during the on-sky validation phase.

SHARK-NIR will use the already corrected wavefront coming from the LBT Adaptive Secondary Mirror, but there is the necessity to correct the residual jitter of the telescope and the Non-Common Path Aberrations, i.e. all the aberrations that are generated inside the instrument, due to temperature and gravity variation or misalignments of the optics.

The SHARK-NIR AO channel is composed of a DM and an IR camera as a wavefront sensor. From an operational point of view, the first thing to do is estimate the NCPA by a phase diversity technique, on the scientific detector. Phase diversity is a focal plane wavefront sensing technique that is able to retrieve the phase aberration introduced by a camera starting from two images of whatever object, one of which (the diverse image) is intentionally corrupted by a know aberration (D. Vassallo et al., 2018).

Then, the DM is deformed in order to correct the NCPA, up to 20-30 modes: these two operations are performed before the observation. Finally, during the observation, the AO loop works in order to correct the tip-tilt, by evaluating the centroid of the star on the detector through a real-time computer which manages also the DM correction. To test this system, we built a Test Bench composed of 3 channels:

12. CRED-2 channel, for the acquisition of the fast tip-tilt images.

13. Basler channel, for slower tip-tilt control purposes.

14. Shack-Hartmann channel, to check if all the aberrations but tip-tilt, are preserved during the fast tip-tilt loop.

SHARK-NIR is equipped with an ALPAO 97-15 DM for a static correction of the NCPA and a dynamic correction of the tip-tilt. A technical camera, the CRED-2 from First Light, acquires the images for the tip-tilt loop. These images are analyzed by an RTC from Microgate, whose output is the commands for tip-tilt compensation to the deformable mirror. In this field, I am involved in the construction of an optical test bench in order to test RTC before the integration of the DM and the CRED-2 into SHARK-NIR in a closed-loop layout.

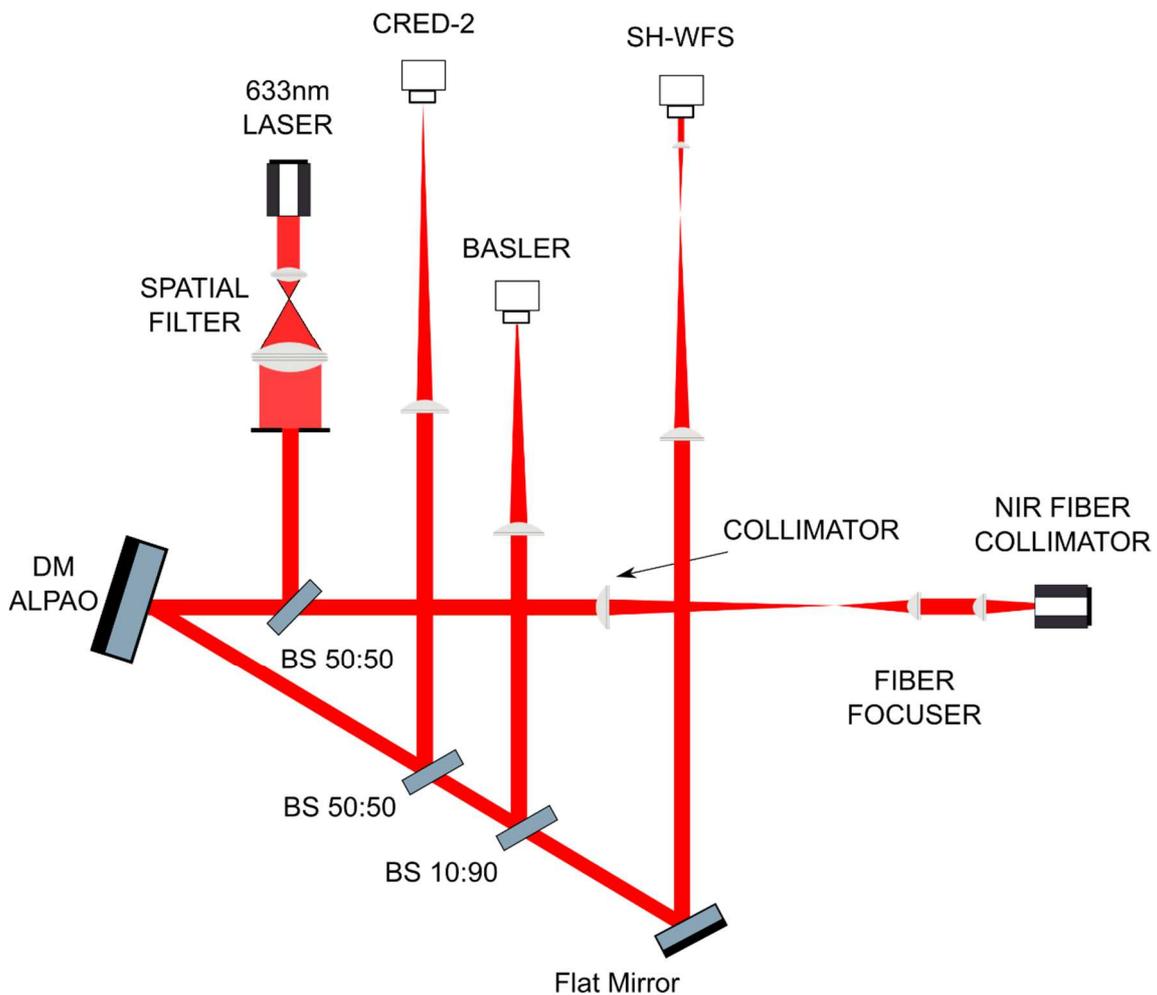

*Figure 242. The optical layout of the RTC test bench.*

The optical layout has been studied to closely match the SHARK-NIR pupil dimension onto the DM (11.15 mm), to closely match the incidence angle of the beam on the DM surface (15°) and to have the same F# on the technical camera.

In addition to this, the other two channels have been introduced in the optical layout of the bench:

15. The Basler channel, whose scope is just a slower control of the tip-tilt loop (maximum framerate 60 fps).

16. The Shack-Hartmann (SH) channel, whose scope is to check if all the aberrations, but tip-tilt, are preserved during the fast tip-tilt loop.

The optical layout is showed in Figure 242.

There is a telescope simulator, whose light is collimated to produce a beam with a diameter close to 11.15 mm on the DM. A 50:50 beam splitter folds half of the light toward the CRED-2 camera. This beam is focused by the same lens that will be used in SHARK-NIR, in order to have the same F# onto the camera.

The rest of the light continues toward a second beam splitter (10R:90T). 10% of the light is reflected toward the Basler camera, while 90% of the light continues toward the Shack-Hartmann WFS. Due to the small dimensions of the lenslet array, the beam is compressed by two lenses to fit the array.

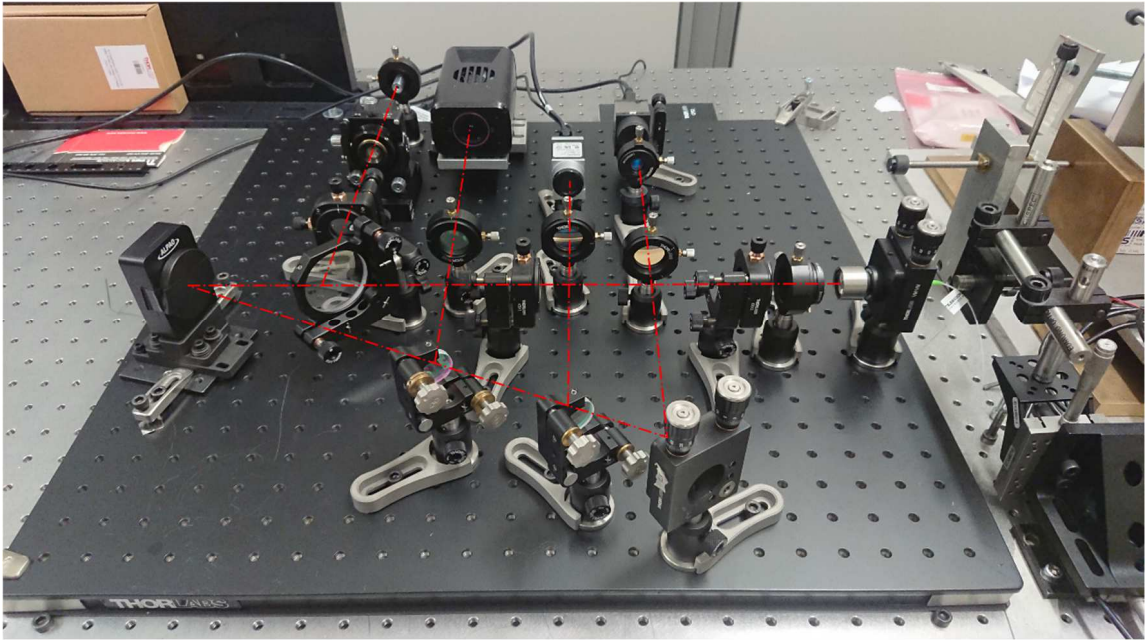

*Figure 243. A picture of the RTC Test Bench, fully integrated and aligned.*

The SH is not sensitive to the NIR light, and the lenses of the telescope simulator are coated with anti-reflection coating in the 1.05 – 1.70 μm wavelength. In order to have a decent amount of light on the SH camera, an auxiliary channel fed by a He-Ne laser is introduced in the path by a beam splitter just before the DM. This beam splitter is coated to be 50R:50T in the visible light, while the infrared light should be almost completely transmitted. In this way, the laser light will illuminate the SH, while the CRED-2 should be almost blind to the laser wavelength and thus should not be affected by this light. The laser light is spatially filtered with a 5 μm pinhole to reduce the aberrations.

### 3.6.1 Test Bench parts

The mechanical layout of the Test Bench has been conceived to use as much as possible commercial elements and to provide all the degrees of freedom needed for the alignment of the optics. The optical bench is an aluminum breadboard 600 x 600 x 25 mm by Thorlabs, with M6 taps, and the components used are listed in the following.

### 3.6.1.1 Fiber collimator

The fiber collimator F810APC-1310 is interfaced to a Tip-Tilt mount VM1/M from Thorlabs through an AD15NT adapter from Thorlabs. The VM1/M mount allows tip-tilt adjustments up to 3° with a resolution of 18 arcsec. The height of the collimator is tuned manually, as the VM1/M mount is interfaced with a 30 mm post from Thorlabs, inserted into a 34.7 mm post holder from Thorlabs. The post holder is secured to the optical bench with a CF125 clamping fork from Thorlabs.

### 3.6.1.2 Iris

In order to select the diameter of the beam to feed the telescope simulator and thus to illuminate the DM with an 11.15 mm beam, a variable iris is inserted just after the collimator. The iris is an SM1D12-SM1, screwed to an XY mount for centering purposes. The mount is an LM1XY/M from Thorlabs, with a travel range of ± 1 mm and a resolution of 250 μm per revolution of the centering screws. The mount is screwed to a 30 mm post from Thorlabs, inserted into a 34.7 mm post holder from Thorlabs.

### 3.6.1.3 Fiber Focuser

The Fiber Focuser lens has 5 degrees of freedom to be aligned: X, Y, Z, Tip and Tilt. For this reason, the lens is mounted, through an S1TM09 ring adapter, on a K5X1 5 axis mount from Thorlabs. This mount allows a travel range along X and Y axis of ± 1 mm with a resolution of 254 μm/revolution, a travel range of ± 3.2 mm along Z-axis with a resolution of 318 μm/revolution and a ± 4° Tip-Tilt with a resolution of 8 mrad/revolution.

The mount is screwed to a 30 mm post from Thorlabs, inserted into a 34.7 mm post holder from Thorlabs.

### 3.6.1.4 Collimator

The Collimator lens has 5 degrees of freedom to be aligned: X, Y, Z, Tip and Tilt. For this reason, the lens is mounted on a K5X1 5 axis mount from Thorlabs. This mount allows a travel range along X and Y axis of ± 1 mm with a resolution of 254 μm/revolution, a travel range of ± 3.2 mm along Z-axis with a resolution of 318 μm/revolution and a ± 4° Tip-Tilt with a resolution of 8 mrad/revolution.

The mount is screwed to a 30 mm post from Thorlabs, inserted into a 34.7 mm post holder from Thorlabs.

### 3.6.1.5 Beam Splitter #1

The role of this beam splitter is to insert into the main path the light from the auxiliary laser and to co-align in Tip-Tilt the laser beam to the fiber beam. As said, the light from the laser will feed the Shack-Hartmann WFS and the Basler camera.

The beam splitter is mounted into a 2 inches Tip-Tilt mount KM200 from Thorlabs. It provides an angular range of ± 3° with a resolution of 5 mrad/revolution.

The mount is supported by a 50 mm post by Thorlabs, inserted into a 34.7 mm post holder by Thorlabs.

### 3.6.1.6 Laser

The auxiliary laser source, a CPS635R laser diode from Thorlabs, is mounted onto an LM1XY/M centering mount from Thorlabs, through an AD11F ring adapter.

The mount is screwed to a 30 mm post from Thorlabs, inserted into a 34.7 mm post holder from Thorlabs.

### 3.6.1.7 Spatial Filter

In order to reduce the aberrations in the laser beam and stabilize their position, the laser light is spatially filtered. The spatial filter is constituted of a 910A mount from Newport, with 5 degrees of freedom. An M20-X microscope objective is mounted in the spatial filter, and the light is spatially filtered by a 910PH-5 five μm pin-hole. The pin-hole can be translated in XY with a resolution of 254 μm/revolution, while its position along the Z-axis (optical axis) can be adjusted with a resolution of 318 μm/revolution. The whole mount has gimbal tip-tilt capabilities.

A 20 mm shim below the 910A mount allows having the optical elements of the spatial filter at the correct height. The 20 mm shim is constituted of a pile of 2 BA2T2/M standard bases from Thorlabs.

### 3.6.1.8 Spatial Filter Collimator

The output beam from the spatial filter is collimated by an AC254-125-B lens from Thorlabs. This lens has 5 degrees of freedom to be aligned: X, Y, Z, Tip and Tilt. For this reason, the lens is mounted on a K5X1 5 axis mount from Thorlabs. This mount allows a travel range along with X and Y axis of ± 1 mm with a resolution of 254 μm/revolution, a travel range of ± 3.2 mm along Z-axis with a resolution of 318 μm/revolution and a ± 4° Tip-Tilt with a resolution of 8 mrad/revolution.

The mount is screwed to a 30 mm post from Thorlabs, inserted into a 34.7 mm post holder from Thorlabs. The output beam is a collimated beam with a diameter of ~ 23.5mm, enough to completely fill the DM optical surface, even considering the 15° incidence angle on the DM surface.

### 3.6.1.9 Deformable Mirror

This is one of the key elements of the bench. In order to maintain the system as similar as possible to the SHARK-NIR design, the mount of the ALPAO 97-15 is exactly the one that will be used in the instrument. The mount is composed of a metal frame, directly connected to the DM with two M3 screws. This metal frame has 4 slotted holes to tune centering of the mirror along one axis.

Under the metal plate, a base plate with 4 threaded holes allows securing the metal frame to it. The baseplate has also 4 slotted holes, rotated by 90° to allow tuning along the orthogonal axis. Since the clearance holes to fix the DM to the metal frame are completely hidden by the baseplate, if it is required to separate the DM from the mount, the metal frame should be removed from the baseplate first. Once the DM + metal frame are in your hand, the DM can be taken off by simply removing the 2 M3 screws. To install a new mirror, simply reverse the process. The DM mount is positioned above a 5 mm thick shim, in order to have the center of the DM at the correct height (about 70 mm).

Firstly we tested the optical axis with the dummy flat mirror mounted in place of the DM, then the dummy was replaced with the DM.

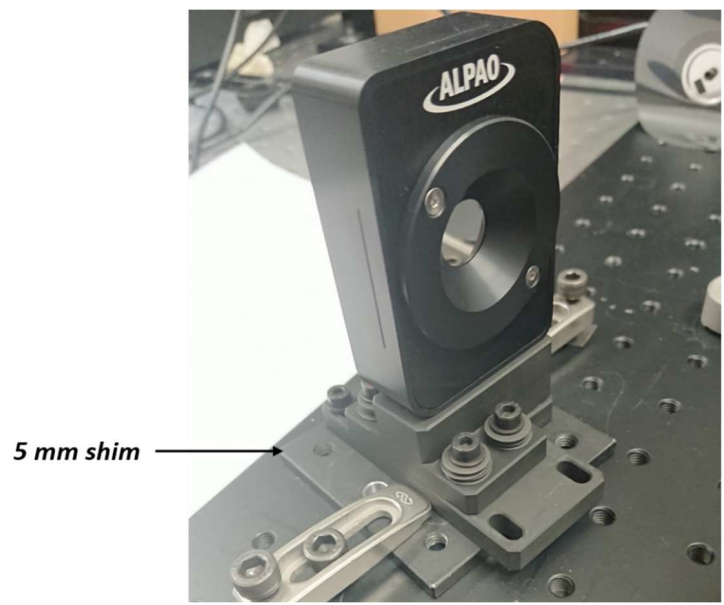

*Figure 244. The dummy flat mirror mounted on the test bench. A 5 mm shim below the mount allows having the mirror at the correct height.*

### 3.6.1.10 Beam Splitter #2

The scope of this beam splitter is to fold part of the light towards the CRED-2 camera. This is a beam splitter BSW23 from Thorlabs, 50:50 in the 0.8 – 2.6 µm wavelength range. It is seated into a U100-A tip-tilt mount from Newport. The mount is screwed to a 30 mm post from Thorlabs, inserted into a 34.7 mm post holder from Thorlabs.

### 3.6.1.11 CRED-2 focuser

A commercial AC254-300-C from Thorlabs focuses the beam onto the CRED-2 camera. The lens holder is an LM1XY/M from Thorlabs, with a travel range of ± 1 mm and a resolution of 250 µm per revolution of the centering screws. The mount is screwed to a 30 mm post from Thorlabs, inserted into a 34.7 mm post holder from Thorlabs.

### 3.6.1.12 CRED-2

This is the second key item of the test bench. In this case, we could not use the final mount of the camera, as it will be much higher than the rest of the optical elements on the bench. We thus designed and produced an aluminum plate to interface the CRED-2 camera to a manual linear stage M-UMR5.25 from Newport, whose scope is to translate the camera along the y-axis. The linear stage is actuated by a BM11.25 micrometer head from Newport, with a sensitivity of 2 µm. The linear stage is then interfaced with a baseplate with a slotted hole. This baseplate lies on 10 mm shims to have the focused spot close to the center of the detector. The shims are two standard bases from Newport. This mount is attached to the optical bench with a PI M-126.DG1 Precision Translation Stage, with a bidirectional repeatability 1 µm, performs z-axis movement to focus the camera.

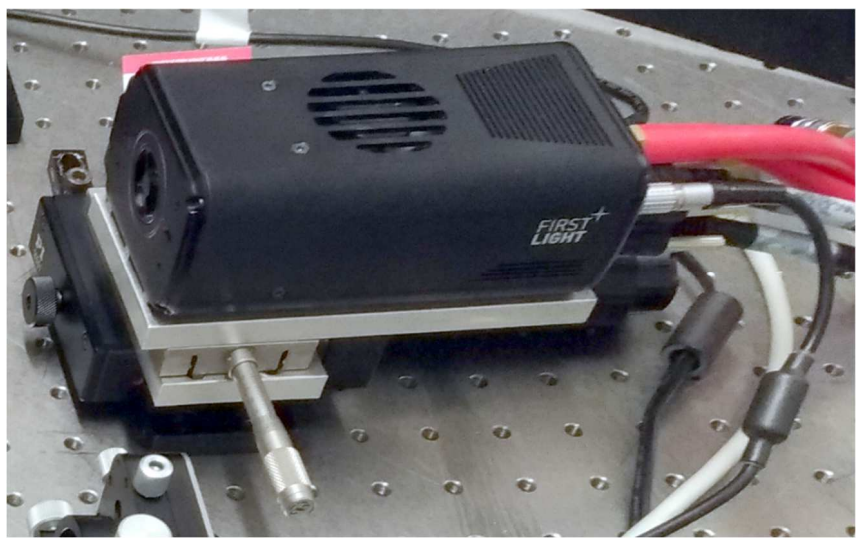

*Figure 245. The CRED-2 mounted in the optical bench with two translation stages.*

### 3.6.1.13 Beam Splitter #3

Scope of this Beam Splitter is to fold a small percentage of the light (around 10%) to the Basler camera, for a check of the spot centroid position at a framerate much slower than with the CRED-2 camera (maximum 60 Hz), but with a higher SNR, taking advantage of the laser light. This is a BSN11 beam splitter from Thorlabs, with a 10R:90T behavior in the 0.65 – 1.2 µm wavelength range. It is seated into a U100-A tip-tilt mount from Newport. The mount is screwed to a 30 mm post from Thorlabs, inserted into a 34.7 mm post holder from Thorlabs.

### 3.6.1.14 Basler Focuser

This is a commercial lens to focus the beam onto the Basler camera. The lens is a LA1509-B from Thorlabs, seated into an LM1XY/M centering mount from Thorlabs.

The mount is screwed to a 30 mm post from Thorlabs, inserted into a 34.7 mm post holder from Thorlabs.

### 3.6.1.15 Basler Camera

This camera is a CMOS sensor NIR enhanced, so in principle, it could observe light up to 1.1 µm. However, the NIR light will be greatly exceeded by the laser light, which will provide much higher SNR on the images with shorter exposures. The exact model is the Basler acA1300-60gmNIR. It is powered by its own power supply and it has a GigE interface for acquisition, the main characteristics are listed in Table 53.

*Table 53. Main informations on Basler acA1300-60gmNIR.*

| | |
|---:|:---|
| **Sensor Type** | CMOS |
| **Sensor Size** | 6.8 mm x 5.4 mm |
| **Resolution** | 1280 x 1024 px |
| **Pixel Size** | 5.3 µm x 5.3 µm |
| **Frame Rate** | 60 fps |
| **Dark Noise (typical)** | 23.2 e⁻ |
| **Saturation (typical)** | 7.4 ke⁻ |
| **Pixel bit depth** | 12 bits |

A 30 mm post from Thorlabs into a 34.7 mm post holder from Thorlabs holds the camera, through a homemade aluminum interface flange. The camera is fixed to the interface flange with 3 M3 screws, and the interface flange is secured to the post with 1 M4 screw.

### 3.6.1.16 Flat Mirror

This is a 1-inch lab mirror folding the beam toward the Shack-Hartmann. It is interfaced with a Tip-Tilt mount VM1/M from Thorlabs. The VM1/M mount allows tip-tilt adjustments up to 3° with a resolution of 18 arcsec. The height of the collimator is tuned manually, as the VM1/M mount is interfaced with a 30 mm post from Thorlabs, inserted into a 34.7 mm post holder from Thorlabs.

### 3.6.1.17 Shack-Hartmann Focuser

Since the beam is now too large to be sampled by our SH WFS, it has to be compressed. This is done by a couple of lenses. The first lens is a LA1509-B from Thorlabs, which focuses the beam. Its holder is an LM1XY/M from Thorlabs, screwed to a 30 mm post from Thorlabs inserted into a 34.7 mm post holder from Thorlabs.

### 3.6.1.18 Shack-Hartmann Collimator

The diverging beam is finally collimated by a LA1304-B lens from Thorlabs, which produces a beam with a diameter of ~ 5 mm, fitting the lenslet array of the SH WFS.

Its holder is an LM05XY/M from Thorlabs, screwed to a 30 mm post from Thorlabs inserted into a 34.7 mm post holder from Thorlabs.

### 3.6.1.19 Shack-Hartmann WFS

The collimated beam is collected by a commercial Shack-Hartmann WFS, WFS300-14AR, from Thorlabs. This is characterized by a lenslet array with a maximum number of active lenslets of 19x15, with a lenslet pitch of 300 μm. It is powered via USB, the main characteristics are listed in Table 54.

*Table 54. Main characteristics of the Shack-Hartmann WFS300-14AR from Thorlabs.*

| | |
|---:|:---|
| **Sensor Type** | CCD |
| **Sensor Size** | 5.95 mm x 4.76 mm |
| **Resolution** | 1280 x 1024 px |
| **Pixel Size** | 4.65 μm x 4.65 μm |
| **Frame Rate** | 15 fps |
| **Exposure range** | 0.079 ms – 65 ms |
| **Wavelength Range** | 0.4 – 0.9 μm |
| **Pixel bit depth** | 8 bits |
| **Number of Active Lenslets MAX** | 19x15 |
| **Lenslet Pitch** | 300 μm |

The WFS seats on a KM100WFS mount, with Tip-Tilt capabilities, to align the lenslet array to the beam. The mount is screwed to a 20 mm post from Thorlabs, inserted into a 25 mm post holder from Thorlabs.

### 3.6.1.20 NIR light source

Externally to the optical bench, we have the light source. This is a high power Tungsten-Halogen lamp, 75 W, near black body source with color temperature 3000 K. It has a driving voltage knob to tune the luminosity of the output. The lamp is an ASBN-W075F with FC/PC connector from Spectral Products. A custom fiber is connected to this lamp at one end, and to the F810APC-1810 Fiber Collimator to the other end. Since the Fiber collimator has an FC/APC connector, the fiber is connectorized FC/PC at one end and FC/APC at the other end. The fiber model is precisely the same that will be used in SHARK-NIR to simulate an on-axis point source, i.e. the Nufern 980-HP, a single-mode fiber. The main characteristics are listed in Table 55 and Table 56.

*Table 55. Main characteristics of the ASBN-W075 light source.*

| Item | ASBN-W075 |
|---|---|
| **Color Temperature** | 3000 K |
| **Power** | 75 W |
| **Light Output** | 1400 lumens (nominal) |
| **Voltage** | 12.0 V (nominal) |
| **Average Life** | 2000 hours |
| **Filament Size** | 1.6 mm x 5 mm |
| **Fiber Coupling** | FC/PC |
| **Tunable luminosity** | Yes |

*Table 56. Main characteristics of the Nufern 980-HP.*

| Item | Nufern 980-HP |
|---|---|
| **Core** | 3.6 µm |
| **Cladding** | 125 µm |
| **Core NA** | 0.2 |
| **Core Attenuation** | 3.6 dB/km @ 980 nm |
| **Operating Wavelength** | 980 – 1600 nm |
| **Cutoff** | 920 nm |
| **Short Term Bend Radius** | ≥ 4 mm |
| **Long Term Bend Radius** | ≥ 9 mm |

## 3.6.2   Test Bench Alignment

Before starting the alignment and even buying the mechanical parts of the bench, an alignment tolerance analysis for the main optical elements has been performed in Zemax, using Montecarlo simulations. The results of the analysis are summarized in Table 57.

*Table 57. Alignment tolerances derived from Zemax Montecarlo simulations.*

| | Decenter (x & y) | Defocus | Tilt (x & y) |
|---|---|---|---|
| **Fiber collimator** | ± 100 um | ± 500 um | ± 0.5° |
| **Fiber Focuser** | ± 50 um | ± 500 um | ± 0.5° |
| **Collimator** | ± 250 um | ± 500 um | ± 0.25° |
| **DM** | < 500 um | - | - |
| **Beam Splitter#1** | - | - | - |
| **Beam Spiltter#2** | - | - | - |
| **CRED-2 Focuser** | ± 250 um | - | ± 0.5° |
| **Basler Focuser** | ± 150 um | - | ± 0.5° |
| **Shack-Hartmann Focuser** | ± 400 um | - | ± 1° |
| **Shack-Hartmann Collimator** | ± 200 um | ±500um | ± 0.5° |

The tolerances are quite loose, especially regarding the tilt. For this reason, all the lenses have been aligned in Tip-Tilt looking at the back reflections through a hole in a piece of paper located before the lens.

The centering of the lenses has been done observing at the position of the beam before inserting the lens and after inserting the lens, using a CCD or a sheet of graph paper in case of loose decenter tolerance. Moving the lens in decentering, we made coincide the position of the beam with the position recorded before inserting the lens.

The focus of the lenses in 633nm was done checking the collimation of the output beam with a wedge plate Newport 20QS20 and minimizing the defocus term measured by the Shack-Hartmann WFS. In infrared light, the beam was collimated with a two holes mask, acquiring images moving the CRED-2 near and far the Test Bench, for a path of 680mm path, measuring the distance of the double spot, and slightly focusing the collimating lens. Few iterations and the beam is collimated, Figure 247 shows the last step, with a difference in centroid position of 0.7px, corresponding to 3 arcsec of beam divergence.

### 3.6.2.1 Alignment procedure

The test bench works with two different wavelengths: 633nm from the laser and 1000-1600nm, the waveband of SHARK-NIR, from the lamp ASBN-W075 plus Nufern 980-HP. A multi-wavelength filter is used in combination with this lamp. Its transmission curve is plotted in Figure 246. The WFS works in visible light, permitting a more accurate sampling of the wavefront with respect to the infrared channel.

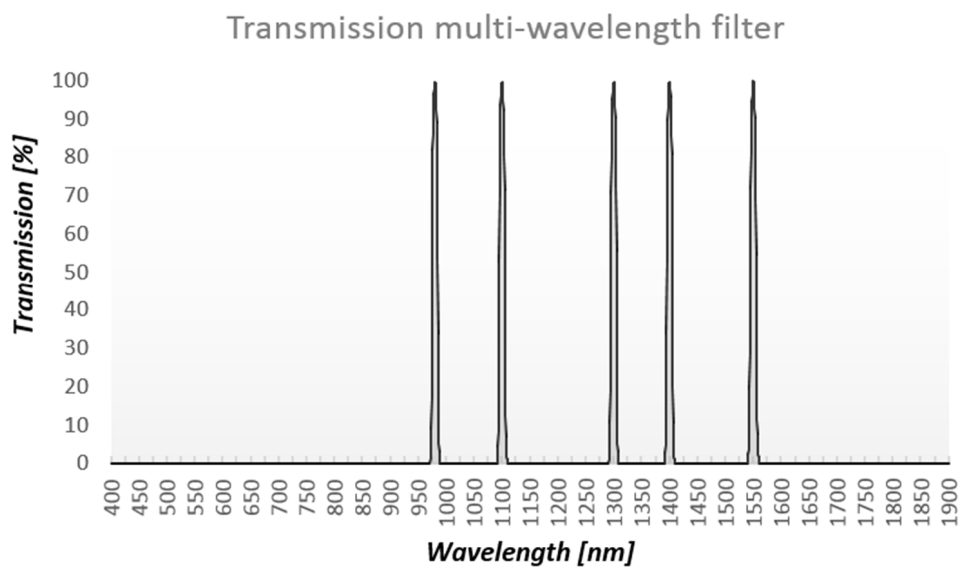

*Figure 246: transmission curve of the multi-wavelength filter from Chroma measured data.*

We started the alignment from the auxiliary laser, centering it to the M20-X microscope objective of the spatial filter using the mount of the laser, and adjusting the tip-tilt using the gimbal mount of the spatial filter. We then aligned the 5 μm pinhole to the converging beam using the 3 degrees of freedom provided by the pinhole mount, till we maximized the output flux from the spatial filter.

In order to align the Spatial Filter Collimator lens in decentering and tip-tilt, we had to remove the M20-X microscope objective. Once this operation was completed we re-introduced the M20-X into the path and aligned the collimator in focus using a wedge plate to check the collimation of the beam. Once this was done, we inserted the Beam Splitter #1 in the path. The tip-tilt of the BS#1 was adjusted in order to maintain the beam height and lateral shift within 1 mm on a 1-meter distance, moving a CCD back and forth on a rail on the optical bench, external to the RTC Test Bench.

We then started aligning the fiber channel. For alignment purposes, in this phase, we used a setup laser also for this channel. We placed the laser external to the RTC Test Bench, on an XY-Tip-Tilt mount, and we co-aligned the laser beam to the beam of the auxiliary laser, folded by the BS#1. This was done making the spots of the two-beam on the CCD on the rail coincident at two different positions of the CCD on the rail (about 1 meter away).

Using also a mirror placed on a tip-tilt mount on the position of the ALPAO, we autocollimated the first ray generate by an iris from the beam expander and we have checked the superimposition of the back-reflected laser spot from the second laser to the one. This ray became the temporal reference for the alignment of the lenses.

The 1SM1D12-SM1 iris was installed on the bench. The iris was as close as possible, and it was centered by eye in order to have the hole at the center of the footprint of the laser beam.

We then aligned the Fiber Focuser and Collimator lenses in tip-tilt and decenter as described before, maintaining the spot fixed on the CCD on the rail, and in focus using a wedge plate. At this point, we positioned a flat mirror in front of this CCD and aligned it in tip-tilt in order to have the beam going back into the iris.

We installed the Fiber Collimator, with the fiber plugged and aligned it in tip-tilt looking at the back-reflected light and in decenter maximizing the flux passing through the fiber, measuring it with a calibrated photodiode.

We positioned the dummy flat of the ALPAO 97-15, in order to have its surface with an inclination of ~ 15° with respect to the incoming beams. We then aligned the Beam Splitter #2, Beam Splitter #3 and the 1-inch flat mirror in order to have 3 beams mutually parallel and as parallel as possible to the lines of holes on the bench, by using a screen of graph paper moved back and forth along a rail.

All the remaining lenses have been aligned in tip-tilt and decenter using the light from the setup laser, as this beam was still very narrow. For the focus alignment, needed just for the Shack-Hartmann collimator, as for the other 2 channels, we only have 1 lens, we used the beam from the auxiliary laser, which is expanded to 15 mm by the Spatial Filter.

Finally, we positioned the detectors in the path and connected the fiber to the Fiber Collimator.

The CRED-2 has been positioned in order to have the focused spot more or less at the center of the detector, and it was focused by using the M-126.DG1 translation stage. Several intra and extra focus images are acquired to calculate the FWHM of the spot by a Gaussian2D fitting.

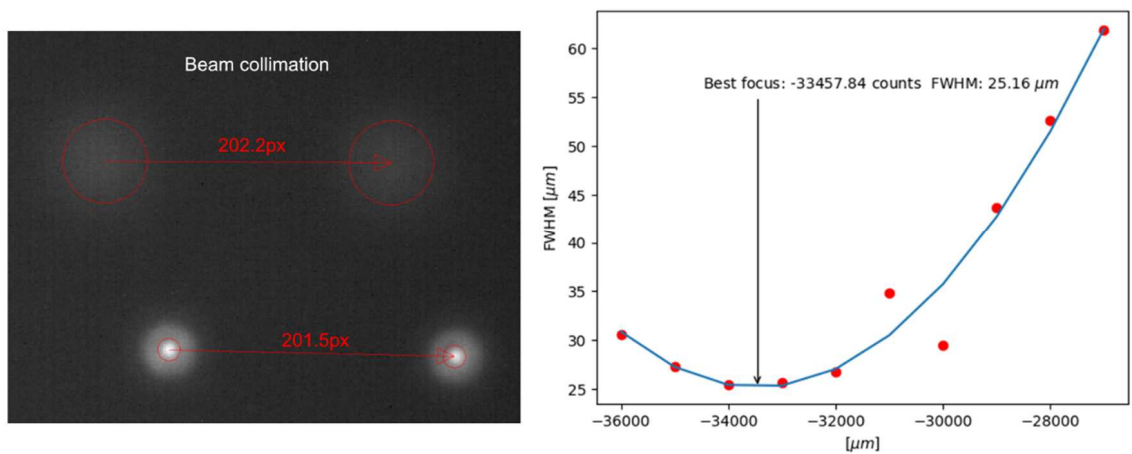

*Figure 247. On the left: the double spot image from the two holes mask acquired in the CRED2 far (top) and near (bottom) at the Test Bench. On the right: the sweep in focus to define the best focal position.*

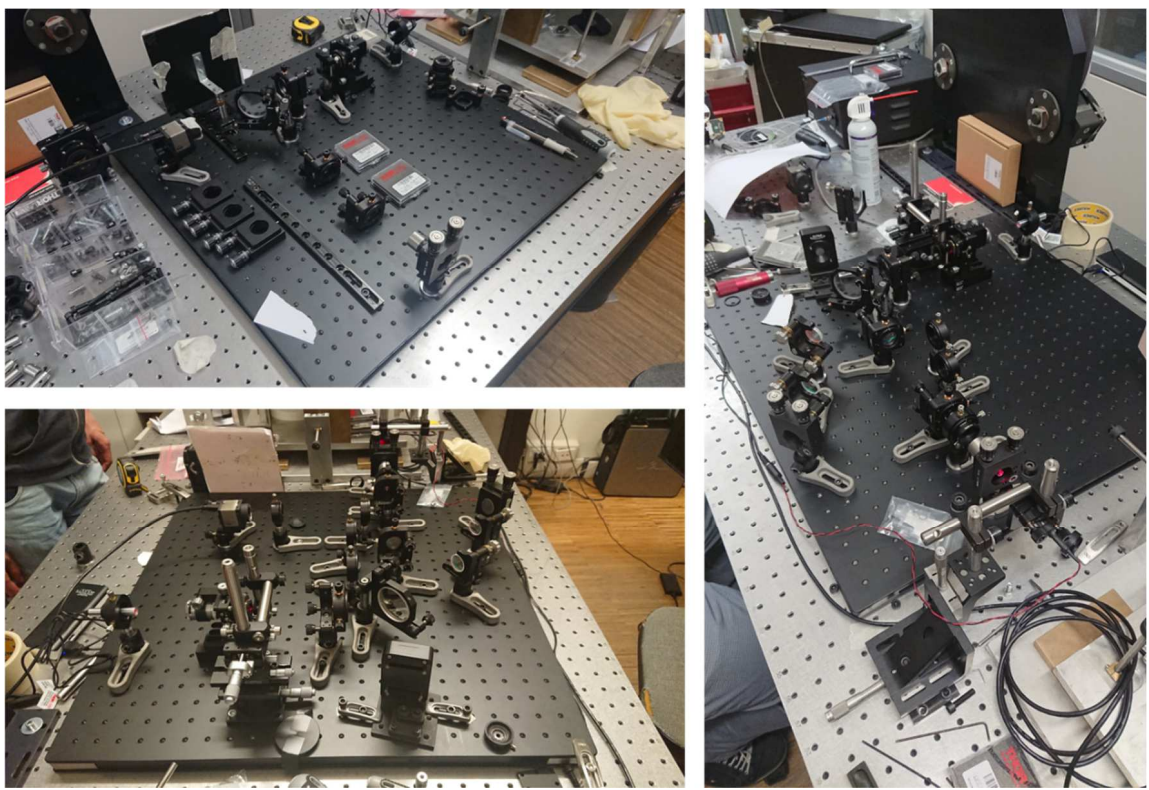

*Figure 248. Some alignment phases.*

The focus position was assumed in the minimum of a polynomial fitting of the points (FWHM, z-position), see Figure 247. The Basler camera has no linear stage for focusing, so it was positioned by hand as close as possible to the focus of its beam. The Shack-Hartmann WFS is positioned in a collimated beam, so we only had to center it to the beam.

We checked the images on the 3 detectors, verifying that we see the light in all of them and that for all the cameras we are not saturating, nor we have too low signal. For the CRED-2 camera, this was trivial as we could play with the voltage of the lamp to tune the amount of light. For the other two detectors, as the laser has a fixed power output and the two detectors have a different dynamic range, we used a Neutral Density filter.

As already said, one of the filters is immediately after the auxiliary laser, and this filter is enough for the SH WFS but not for the Basler, which is still saturating at its minimum exposure time (10 μs). A second filter was then positioned in front of the Basler Focuser lens. In this way, we are within the dynamic range of the two detectors.

### 3.6.2.2 Infrared beam diameter

The knowledge of the beam diameter in the arm of SHARK-NIR is crucial for a reasonable estimation of NCPA, as it is directly correlated with the number of modes used in the DM, and also for the phase diversity approach, that needs the f/number of the focused beam. We approached the measurement of the diameter from a different point of view with respect to the measure with the isophotal fitting. In particular, the Knife-Edge Diffraction (Kumar et al., 2007) of the pupil diameter depends on the wavelength and it is higher in the NIR than in visible light. The diameter of the beam in NIR needs a more accurate approach to be determined. We decided to model the defocused image of a lens by changing the diameter of the simulated beam, and by calculating the difference with the real image.

The test was performed in these steps:

- We illuminated the whole DM surface, which presents an optical full aperture of 14.00 mm, refers to Figure 249;

- We aligned the CRED focuser, described in 3.6.1.11, and found the best focus position of CRED camera, with PI M-126.DG1 translation stage to assure high accuracy. The best focus position is on count 144929, where each count is 1 μm;

- We moved the CRED in intra and extra focal positions, corresponding to 88000 (-483±1μm) and 200000 (468 ±1μm) counts, the images are shown in Figure 250;

- We performed a model of the focuser with PROPER, an optical propagation library of IDL, and simulated the intra and extra focal images. We parametrized the input diameter and simulated a series of defocused images in the interval from 13.0 to 15.0 mm;

- For each diameter determination, we calculated the sum of the residual image, computed as the difference between the normalized original image and the computed defocused image. In Figure 251 the images and the residual are shown;

- By a parabolic fitting of the diameter determination with respect to the sum of the residual counts, we found the best diameter of the input beam and the relative error.

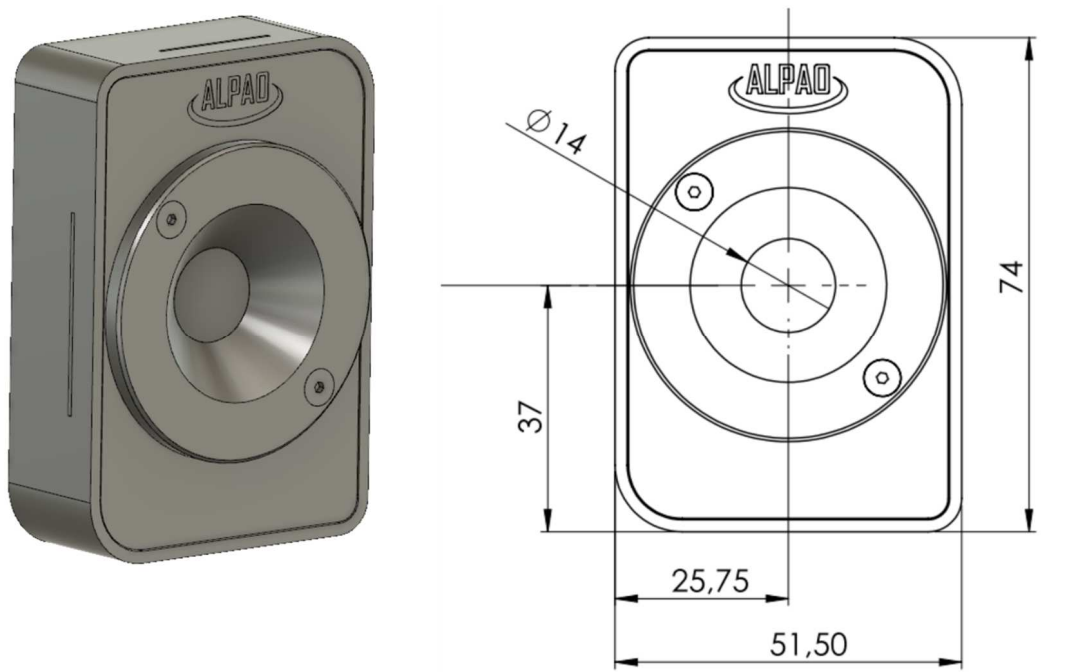

*Figure 249. The technical design of the ALPAO DM97-15, with full aperture of 14mm.*

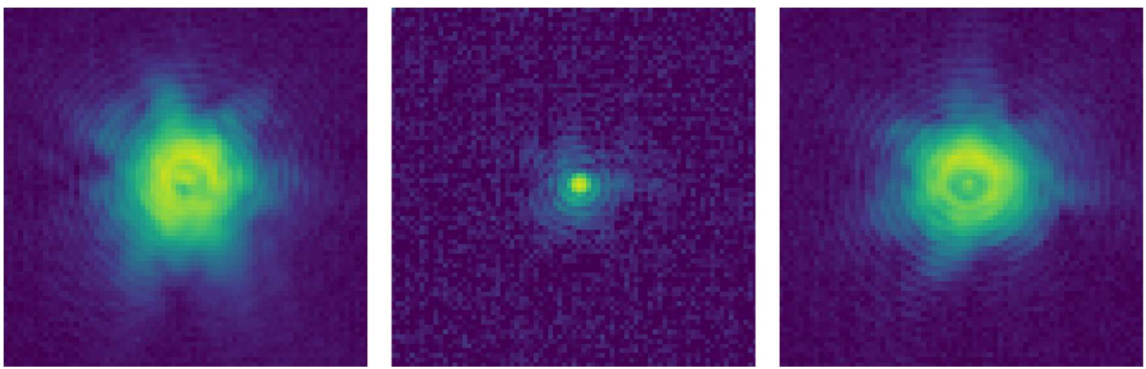

*Figure 250. The NIR images on CRED from intra focal position (on the left) to the extra focal position (on the right), and the best focal image (in the middle). The images are shown with log stretching to see the fine detail of the diffraction features.*

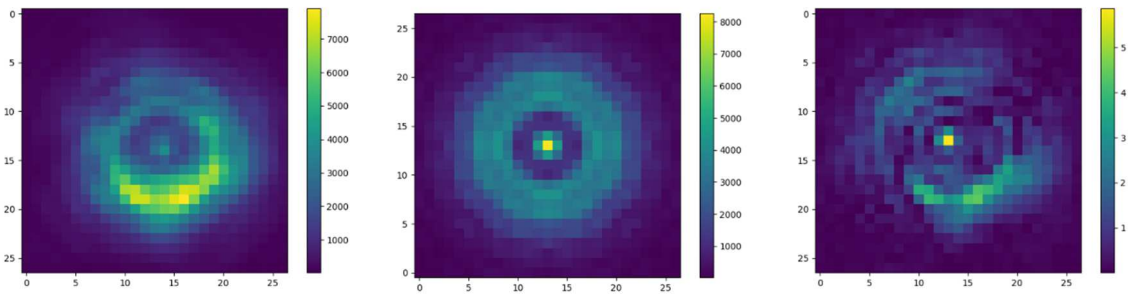

*Figure 251. From left to right: original image, reconstructed images for defocus position, residual.*

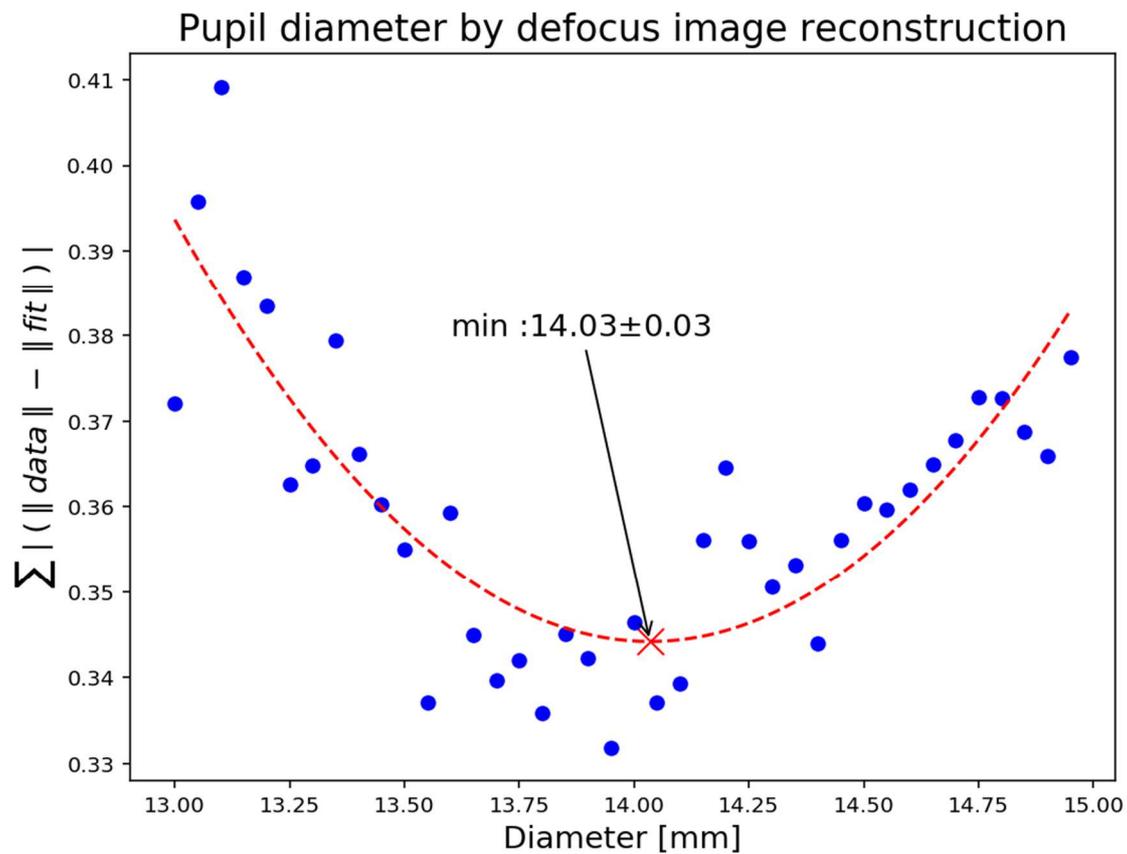

*Figure 252. The pupil diameter determination by defocused image reconstruction with respect to the intra focal image.*

The value of the diameter after the procedure is 14.03 ± 0.03 mm, very near the real value of the beam. By combining the full aperture of the DM (14.00 mm), the measured divergence in NIR of the collimated beam (3 arcsec) and the distance between DM and lens (350 mm), the beam diameter in the lens position is 14.01 ± 0.1 mm diameter. The values of diameter extracted from intra and extra focal images are identical, validating a robust method to characterize the beam also with an aperture below the full aperture of the DM.

### 3.6.2.3 Test bench results

The bench was then used to characterize the system simulating R magnitude from 8 to 10 and try to close the TT loop at 500 and 1000 Hz framerate. We injected into the system a fake TT history and try to correct it, with the same SW provided by Microgate along with the RTC. The typical amplitude of the TT is of the order of 5 to 15 mas: the resulting correction is presented in Table 58.

*Table 58. Results for TT correction expressed in mas.*

| R magnitude | Input time history / Frame rate [Hz] | 15 mas | 10 mas | 5 mas | No input Time History |
|---|---|---|---|---|---|
| Mag 8 | 500 | 5.7 | 4.6 | 2.1 | 0.9 |
| | 1000 | 3.2 | 1.5 | 1.2 | 0.4 |
| Mag 9 | 500 | 5.9 | 4.1 | 2.3 | 1.6 |
| | 1000 | 3.3 | 3.2 | 2.9 | 2.4 |
| Mag 10 | 500 | 6.8 | 4.3 | 2.8 | 1.4 |
| | 1000 | --- | --- | --- | --- |

Then we applied a typical NCPA correction to the DM and checked if the correction is still maintained during the TT loop, Table 59.

*Table 59. NCPA correction during the TT loop.*

| Maximum PtV values 500Hz | | Maximum PtV values 1000Hz | |
|---|---|---|---|
| **Flat rms** | **\|OL-CL\| [nm]** | **Flat rms** | **\|OL-CL\| [nm]** |
| 5 mas | 5.3 | 5 mas | 4.1 |
| 10 mas | 4.9 | 10 mas | 8.5 |
| 15 mas | 1.2 | 15 mas | 9.1 |
| **NCPA 100nm** | | **NCPA 100nm** | |
| 5 mas | 4.5 | 5 mas | 10.3 |
| 10 mas | 2.5 | 10 mas | 7.0 |
| 15 mas | 4.0 | 15 mas | 8.7 |

My personal contribution was in the phase of installation and maintenance of the optical alignment of the bench.

## 3.7 Conclusion

SHARK-NIR is a coronagraphic camera with spectroscopic capabilities for the LBT telescope currently in its AIV phase. The XAO of LBT pushes Strehl Ratio to the top-level regime and permits high contrast imaging of exoplanets, for new discoveries and characterizations. The coronagraphic components selected for SHARK-NIR, the SP masks, the Gaussian Lyot, and the FQPM, need high accuracy in optical alignment and high repeatability to achieve the goals required by the scientific cases.

In my Ph.D. I have been involved in the AIV phase of a sub-system of this instrument for LBT. In particular, I designed and implemented a test bench for analyzing and integrating the coronagraphic masks, by reproducing the focal plane scale, the entrance pupil and the coronagraphic techniques as similar as possible to the final assembly. I defined the alignment procedures of the coronagraphic masks and developed code in Python language to evaluate if these methods are accurate enough to reach the design tolerances. I employed robust mathematical algorithms and estimated the raw contrast achieved by the coronagraphs. The accuracy of the results contributed to the overall instrument error budget computation.

I also developed and aligned a second test bench used to outline the strategy of evaluation and correction of NCPA aberrations, to characterize the stability of the DM-TT correction and the efficiency of the WFS, which is part of SHARK-NIR for the correction of the residual jitter, to realize coronagraphic images with high stability and very low aberrations. I worked in a motivated research group, by collaborating not only in the alignment but also in the WF reconstruction by the phase diversity approach and on the coronagraphic field.

# 4 General conclusion

During my Ph.D. I have been involved in two international complementary projects on exoplanet research. I worked on the optical alignment of the telescope optical units prototype of PLATO Mission. This telescope is characterized by a very large field of view and a full refractive design. PLATO will detect transiting exoplanets in front of nearby stars with high accuracy (ppm) only if the alignment tolerances will be respected. The alignment was performed in warm conditions in order to achieve the best optical performance in "flight conditions" (temperature of -80°C and vacuum).

In PLATO I have been responsible for the bench setup and its qualification during the AIV phase of the TOU prototype, in terms of thermal stability, alignment observables selection and optical tests, such as the Hartmann test , the interferometric characterization and the PSF quality. I investigated the role of the CTE of different materials in the control of the "breath" of the optical bench, with the optical path length of the laser ranging from about 3 to 10 meters. I explored some optical alignment procedures to select the most reliable in terms of accuracy and alignment working time. The goal achieved in this part of my Ph.D. was to deliver to the industry a robust alignment procedure and demonstrate to ESA the alignment feasibility. PLATO is expected to fly in 2026.

I characterized the GSE of the PLATO-TOU used to manipulate the lenses, interacting with the optical design team to compare laboratory results with theoretical investigations and to achieve the expected accuracy of the optical design approved from ESA.

I developed a Python software library to analyze several shapes of the transmitted and back-reflected spot. An innovative script of the library was applied to analyze the corner cube reflection spot as an optical axis reference. The spots generated by the laser source are affected by an interference pattern, which has been removed by applying a spatial filter in the Fourier space, in order to improve the accuracy of the centroid analysis. This approach will be useful in any other future project that involves spot laser source as a reference for the alignment.

The evaluation of the alignment error budget of PLATO-TOU takes into account all the elements in the optical bench that contribute to the final error, combining them in quadrature. These results were shared with the other PLATO consortium sub-systems groups to compute the final error budget, which finally let us know the complete accuracy on photometry and astrometry of PLATO. This process is expected to be concluded in 2020.

The many lessons learned during the alignment procedure were a fundamental contribution to the industrialization of the TOU, that will of course include the alignment and qualification of the 6x26 lenses of the PLATO telescopes. The investigation of non-contact techniques (section 2.3.4) led Leonardo to select, for the production chain, a centering machine that uses low coherence interferometry techniques. This innovative approach could be used also in future projects to align all refractive systems for space or ground-based instruments.

The second project was SHARK-NIR, the high contrast imager for LBT, devoted to characterizing exoplanet with a ground-based telescope in near-infrared light, by using extreme adaptive optics in the regime of SR > 90%. I worked in the optical characterization and alignment of the coronagraphic masks, during its integration phase. The commissioning of the instrument is expected in 2020.

I was responsible for the definition of the alignment procedures of the many special optics composing SHARK-NIR coronagraphs and acting on either the phase or amplitude of the incident wavefront. I realized a test bench that simulates the real alignment case of the final instrument and a second bench used to outline the strategy for the correction of NCPA aberrations and for the removal of the residual jitter implementing a fast TT correction via a deformable mirror and a wavefront sensor. In particular, I developed software tools to find the center of the pupil apodizers, focal plane masks, and Lyot stop masks, with respect to the center of the entrance pupil, achieving very high accuracy. The FQPM was aligned in decentering and focus by using a fast rotating diffuser and a dedicated Python code. The error budget reaches the value of 5 μm, corresponding to 5.7 mas in the scale of the focal plane of SHARK-NIR. With this result, we have a decay of 0.3 magnitudes in the detection power of the FQPM, and negligible impact in SP because is sensitive over 10 μm. A useful innovation was the possibility to characterize the FQPM in the visible light (550nm) with respect to the nominal working wavelength of 1.6 μm, to have a simpler way to collect information about phase mask rejection. The optical bench was implemented with a double source simulator to characterize the efficiency of the coronagraph with a simulated fake planet, achieving a signal-to-noise ratio of about 36 near the IWA with a contrast between the planet and the source of about $10^{-4}$.

The use of the fast rotating diffuser was an innovative tool to properly illuminate the coronagraphic mask during the alignment process, becoming crucial in the final integration of SHARK-NIR, thus being able to avoid the need for an integration sphere to do so.

The NCPAs were characterized in a dedicated bench, working both in 633nm and in 1.6µm. The knowledge of the infrared beam diameter is crucial for the estimation of the NCPAs. I defined a software procedure, based on the propagation of light through an optical system using Fourier transform algorithms, to measure the beam diameter with high accuracy (0.2%). This innovative approach will be very useful in the integration phase of SHARK-NIR when we need to estimate the beam after the collimation through four off-axis parabolas. The TT correction in closed-loop at 1000Hz achieved the value of 3.3 mas residual jitter for a simulated star of mag 9 and starting from an open loop jitter of about 15 mas.

During my Ph.D. I have acquired skills on:

- Optical bench stabilization;

- Alignment of different optical instruments to a common path, such as interferometer co-aligned with the GSE in the case of PLATO or the bench to evaluate the NCPA for SHARK-NIR;

- Alignment in the visible and infrared domain;

- Use of interferometers and wavefront sensor to evaluate optical aberrations, and their analysis;

- Development of software tools to evaluate thermal stability and the accuracy of the optomechanical parts, the analysis of transmitted and back-reflected beams of a lens and the optical alignment of coronagraphic masks;

- Simulation of the light propagation through a lens system and comparison with laboratory results;

- Investigation on non-contact techniques to measure lenses position;

- Optical qualification in a climate vacuum chamber at working temperature of -80°C;

- Error budget analysis;

- Technical reports in the environment of an international project;

- Use and calibration of the pCCM touch machine.

The new instruments for ground-based telescopes that will take place in the next future, such as a 30/40-meter class telescope, need a huge specialization in optical alignment and will be developed in an optical benches of large size. The increasing of Extreme Adaptive Optic systems applied to this telescopes adds a new opportunity to use high contrast imaging in any telescope. Space mission composed by full refractive optics, i.e. Earth Observation, CubeSat and small astronomical missions, are usually aligned with laser spot based configuration and/or computer-assisted touching machines. The experience acquired in my Ph.D. can be applied to the integration and qualification of these instruments or telescopes, also provided by non-contact optical techniques.

# List of references

# Software Credits

- Astropy a core package for astronomy using the Python programming language. Astropy is licensed under a three-clause BSD license. https://www.astropy.org/

- Component Library by Alexander Franzen is licensed under a Creative Commons Attribution-NonCommercial 3.0 Unported License. http://www.gwoptics.org/ComponentLibrary/

- Inkscape, a professional vector graphics editor for Linux, Windows, and macOS. It's free and open source. https://inkscape.org/

- LMFIT Non-Linear Least-Squares Minimization and Curve-Fitting for Python is licensed under a three-clause BSD license. https://lmfit.github.io/lmfit-py/

- PRYSM: an open-source library for physical and first-order modeling of optical systems and analysis of related data. Under MIT License (MIT), copyright (c) 2017-2019 Brandon Dube https://prysm.readthedocs.io/en/stable/

- PROPER: an Optical Propagation Library, © 2006-2019. California Institute of Technology ("Caltech"), software designated as Technology and Software Publicly Available ("TSPA") http://proper-library.sourceforge.net/

- Uncertainties Python package, a free, cross-platform program that transparently handles calculations with numbers with uncertainties (like 3.14±0.01). is licensed under the Revised BSD License (© 2010–2016, Eric O. LEBIGOT) https://pythonhosted.org/uncertainties/

- VOSA, developed under the Spanish Virtual Observatory project supported by the Spanish MINECO through grant AyA2017-84089 http://svo2.cab.inta-csic.es/theory/vosa/

# Acknowledgments

I would like to express my enormous gratitude to my wife, that has supported me during these three years, with patience and personal support, to my family and my friends. You helped me understand what to do, and how to solve the problems that I encountered with strength and moral support.

Thanks to my supervisors Prof. Roberto Ragazzoni and Prof. Jacopo Farinato, first of all my friends, which led me in two complementary international projects, from space to ground-based environments.

I thank them for giving me their time and allowing me to develop my ideas and applying them to my projects, aware that the success of my work would also be partly a success for the whole group.

Very special thanks to everyone in the Adaptive Optic Group of INAF-Padova, Daniele Vassallo, Davide Greggio, Elena Carolo, Elisa Portaluri, Federico Biondi, Kalyan Kumar Radhakrishnan Santakumari, Luca Marafatto, Luigi Lessio, Marco Dima, Maria Bergomi, Valentina Viotto, Eleonora Alei, Carmelo Arcidiacono, Simonetta Chinellato, and Demetrio Magrin. Thanks for supporting me in learning sophisticated optical techniques, and for giving me access to a fantastic optic laboratory. Thanks for the unique opportunity I had to travel with you around the world and to live outstanding experiences. I could not imagine finding better people, both professionally than humanly. Each of you made me feel part of the group, I have met new special friends in my life.

I thank the Ph.D. school coordinator, Prof. Giampaolo Piotto, that completed my technological preparation with the activities of the school, such as the courses, and student activities, in particular, the possibility to teach in the courses of optics of the Master Degree Course.

Gabriele Umbriaco
*Università di Padova*
November 2019